\PassOptionsToPackage{dvipsnames}{xcolor}
\documentclass[a4paper,12pt]{book}
\usepackage{amsmath,amssymb,amsthm,mathrsfs,bbm}
\usepackage{latexsym,amscd,amsbsy,amsfonts,dsfont}
\usepackage{graphicx}
\usepackage{subfigure}  
\usepackage{float}
\usepackage{slashed}
\usepackage{cancel}
\usepackage{textcomp}
\usepackage{multirow}
\usepackage{soul}
\usepackage{physics}
\usepackage{comment}
\usepackage[font=small,labelfont=bf,margin=10pt]{caption}
\usepackage[margin=1in]{geometry}
\usepackage{bm}
\usepackage{pdfpages}
\usepackage[dvipsnames]{xcolor}
\usepackage[
      colorlinks=true,
      linktocpage=true,
      breaklinks=true,
      linkcolor=Blue,
      urlcolor=Purple,
      filecolor=black,
      citecolor=Purple,
      menucolor=black,
      pdfstartview=FitV,
      bookmarksopen=true
      ]{hyperref}
\usepackage{cleveref}     
\usepackage{fancyhdr}
\usepackage[backend=bibtex8,
            sorting=none,               
            citestyle=numeric-comp,     
            bibstyle=PhDThesisBibStyle, 
            giveninits=true,            
            isbn=false,                 
            maxbibnames=5               
           ]{biblatex}

\newcommand{\LCg}{\mathring{\Gamma}{}} 
\newcommand{\CDg}{\mathring{\nabla}{}}   
\newcommand{\Rg}{\mathring{R}{}}   
\newcommand\sT{{\scriptscriptstyle \mathrm{T}}}
\newcommand\eqFP{\mathcal{E}^{\scriptscriptstyle \text{FP}}}
\newcommand\eqWTDiff{\mathcal{E}^{\scriptscriptstyle \text{WTDiff}}}
\newcommand\ddh{(\partial^2\!\cdot\!h)}
\newcommand{\jordanf}{\text{(J)}}
\newcommand{\einsteinf}{\text{(E)}}

\definecolor{emerald}{rgb}{0.31, 0.78, 0.47}

\newcommand\Sext{\Sigma_{+}}
\newcommand\Sint{\Sigma_{-}}
\newcommand\Rext{L_{+}}
\newcommand\Rint{L_{-}}

\newcommand\Text{\tau}
\newcommand\Tint{\tau}

\newcommand\aext{\alpha}
\newcommand\aint{\alpha}

\newcommand\bext{\beta}
\newcommand\bint{\beta}

\newcommand\vpar{\vartheta}

\AtBeginDocument{\addtocontents{toc}{\protect\thispagestyle{empty}}}

\begin{document}

\thispagestyle{empty}

\includepdf[pages=1, pagecommand={}, fitpaper=true]{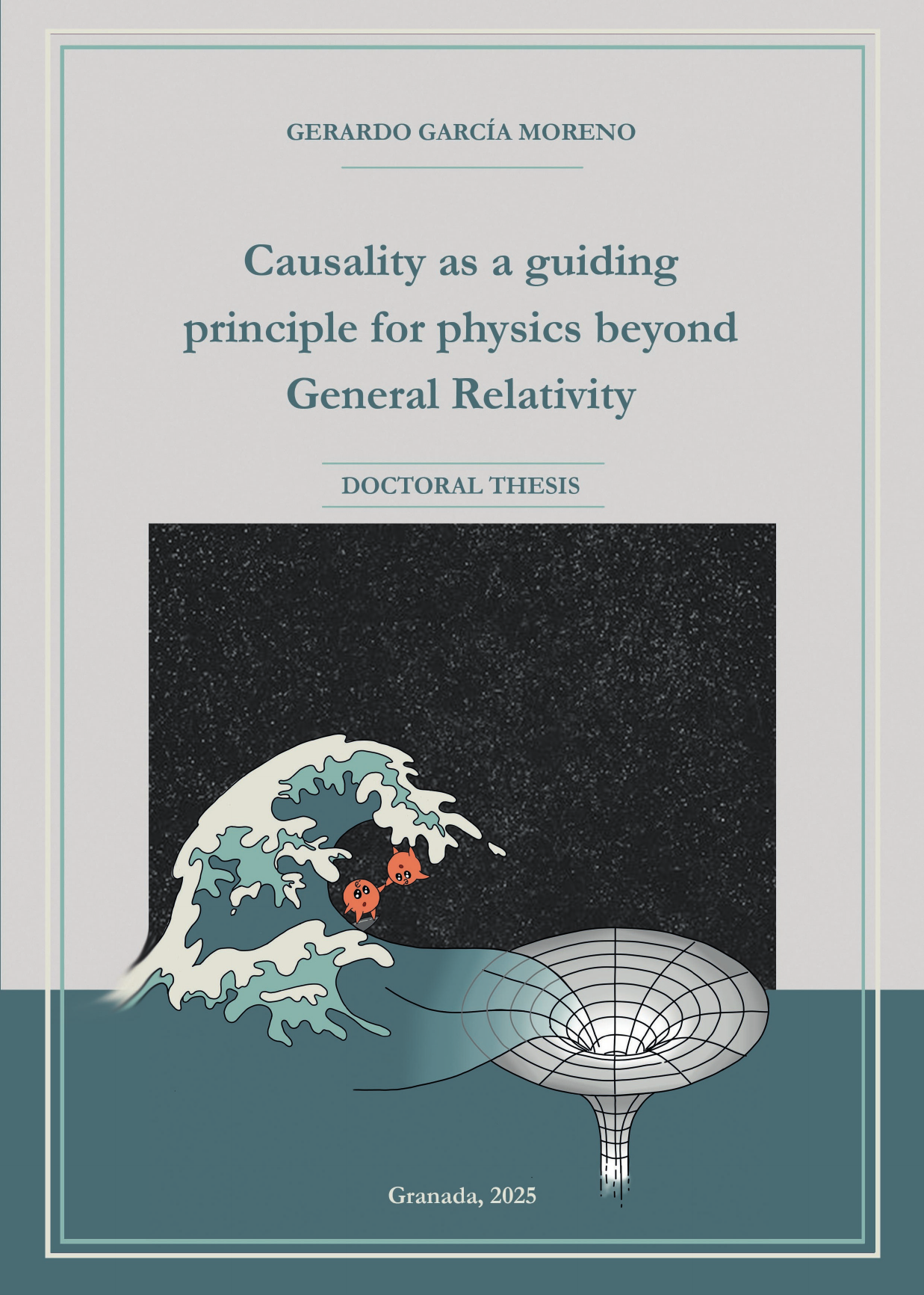}

\thispagestyle{empty}

\begin{titlepage}

\pagestyle{empty} 

\begin{center}

\vspace{10mm} \includegraphics[height=0.25\paperheight]{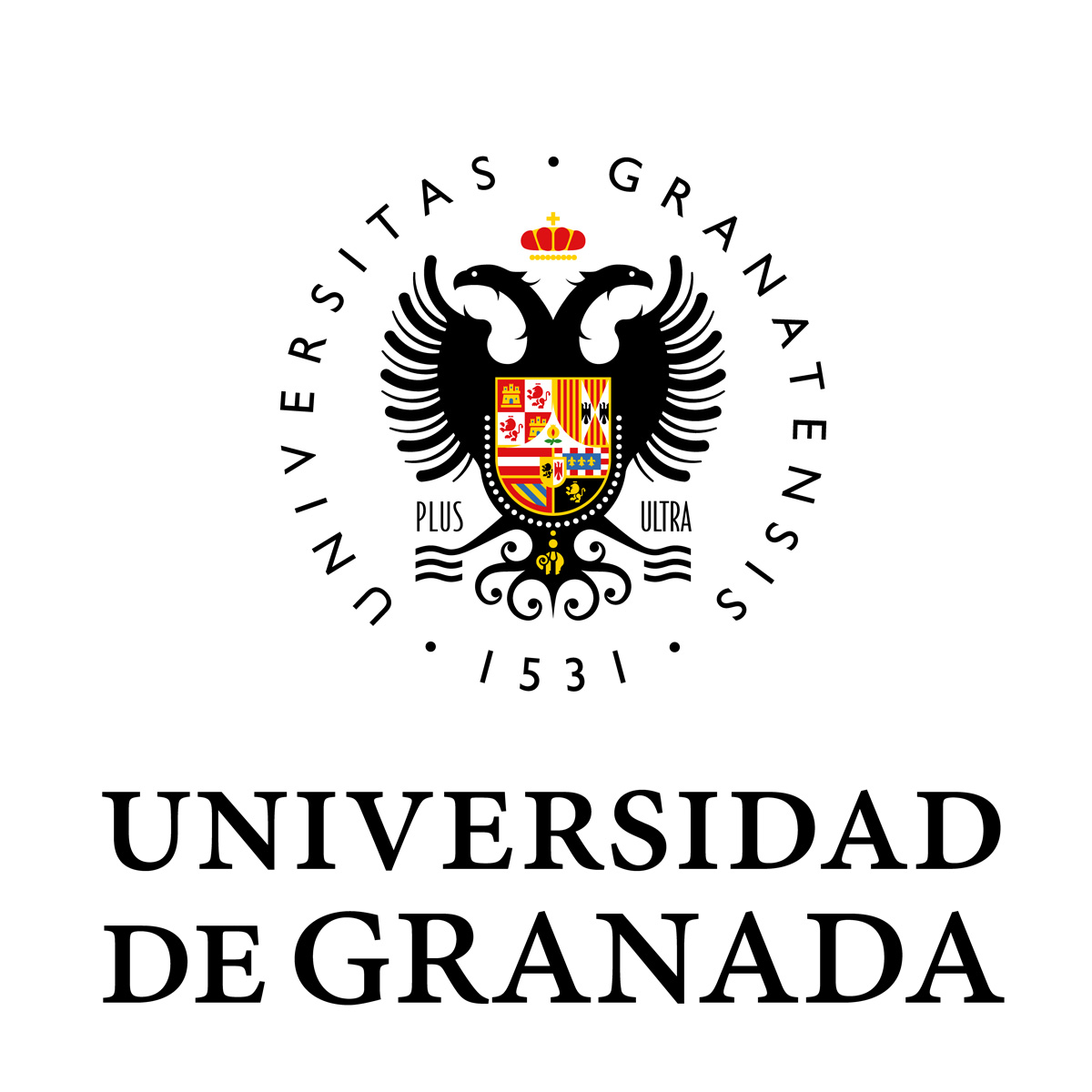}

\vspace{20mm} \textbf{\LARGE Causality as a guiding principle for physics}

\vspace{5mm}\textbf{\LARGE beyond General Relativity}

\vspace{15mm} {\Large G}{\large ERARDO} {\Large G}{\large ARC\'IA} {\Large M}{\large ORENO}

\vspace{15mm} {\large \textit{Thesis submitted for the degree of}}

\vspace{8mm}  {\large \textbf{DOCTOR OF PHILOSOPHY}}

\vspace{8mm}  {\Large  \textbf{Programa de Doctorado en F\'isica y Matem\'aticas}}

\vspace{5mm} {\large Supervisors:}

\vspace{5mm} {\large Carlos Barcel\'o Serón}

\vspace{5mm} {\large Luis Javier Garay Elizondo}

\vspace{5mm} {\large Ra\'ul Carballo Rubio}

\vspace{8mm}  {June 2025}

\vspace{8mm}  {Defended on 12 September 2025}

\end{center}

\cleardoublepage

\end{titlepage}

\pagenumbering{Roman}

\fancyhead[LE,RO]{\thepage}
\fancyhead[LO,RE]{Abstract}
\fancyfoot[C]{}

\thispagestyle{plain}

\noindent
    \begin{center}
        \textbf{ {\large Abstract}}
    \end{center}
    
This thesis is situated within the context of quantum gravity, understood broadly as any effort to explore the interplay between gravitation and the quantum realm, without necessarily entailing the quantization of the gravitational field itself. In particular, we focus on emergent theories, specifically, those in which the causal structure and geometric concepts underlying the gravitational field in General Relativity are not fundamental, but instead emerge from more basic underlying degrees of freedom. Moreover, our attention is directed toward emergent theories inspired by condensed matter physics contexts. 
    
Rather than constructing a full-fledged emergent theory and conducting a detailed analysis of its consequences, a task far too ambitious for a single thesis, this work offers a humble roadmap of analyses and reflections relevant to emergent frameworks, without committing to any specific theory. The aim is not to provide definitive answers, but rather to challenge certain established ideas and highlight the features that a potentially successful theory should possess. 

This thesis is divided into two distinct parts, reflecting the significantly different tools and analyses employed in each. The first part addresses fundamental and conceptual aspects of emergent theories, with a particular focus on the role of background structures, both in terms of their implications for the theory and their relevance from a constructive perspective. The second part begins with the assumption that singularities and horizons are absent, a feature often anticipated in emergent frameworks. However, the analyses in this part can also be interpreted as largely agnostic and independent of specific theoretical commitments.

The first part of the thesis comprises four chapters. The first two are situated within the framework of analogue gravity, which involves studying systems whose excitations, in certain regimes, can be described as relativistic fields propagating over an effectively curved geometry. In Chapter 1, we investigate whether closed timelike curves can be simulated within these analogue models. Since such systems are not constrained by the Einstein field equations, it is, in principle, possible to engineer configurations that would not naturally arise in general relativistic scenarios. However, we find that the system tends to avoid configurations associated with causally pathological emergent geometries: the analogue model breaks down before such configurations can be realized. We argue that this behavior is primarily due to the underlying more fundamental laboratory causality, although for observers restricted to probing only the emergent geometry, this reason may not be readily apparent.

In Chapter 2, we turn our attention to analogue systems with a quantum substratum and investigate the possibility of engineering superpositions of states, each associated with its own emergent geometry. We demonstrate that when the resulting causal structure becomes excessively blurred, the system destabilizes, rendering such superpositions unsustainable. We argue that although quantum gravity is expected to involve superpositions of spacetimes, configurations that significantly depart from a well-behaved causality are dynamically suppressed. This observation parallels arguments in the literature suggesting that, rather than quantizing gravity, it may be necessary to modify the quantum laws themselves to make them compatible with the general relativistic framework. We revisit several of these proposals, compare them with our findings, and draw lessons relevant to emergent gravity frameworks.

Chapter 3 addresses gravitational theories from a constructive perspective. It has long been argued that the only consistent non-linear extension of the Fierz-Pauli theory is General Relativity. We revisit this issue by examining the analogue problem for higher-derivative and metric-affine gravity theories. We demonstrate that any background-independent theory can be reconstructed from its linear counterpart. Furthermore, we show that GR is not the only consistent extension of Fierz-Pauli theory in dimensions greater than four, emphasizing that the constructive process involves making choices at each step, and does not uniquely lead to a single nonlinear theory. Finally, we also demonstrate that Unimodular Gravity can be reconstructed from its linearization, \mbox{WTDiff}, and, more generally, that any WTDiff-invariant background-dependent theory can be obtained from its linearization.

Chapter 4 offers a systematic comparison of General Relativity and Unimodular Gravity, as well as, more generally, Diff-invariant theories and their WTDiff-invariant counterparts. We examine these theories at the classical and semiclassical level, and some aspects of the quantum theory, concluding that the only distinction between them lies in the cosmological constant, which in General Relativity is a coupling constant, and in Unimodular Gravity, an integration constant. We critically review the implications of this distinction for model building and for string theory, particularly in relation to whether its low-energy sector can only be described by General Relativity.

The second part of the thesis focuses on horizonless and non-singular ultracompact objects and comprises three chapters. Chapter 5 revisits the no-hair theorems within the specific context of static and axisymmetric spacetimes. In particular, we relax the usual assumption of the presence of a horizon and instead impose the existence of a finite maximum redshift surface and bounded scalar curvatures. We find that, under these conditions, the multipole structure of the configuration must parametrically approach that of a black hole as the maximum redshift is taken to infinity. As a byproduct of this analysis, we apply the same tools to black holes in astrophysical settings and show that, for any given external gravitational environment, there exists a unique horizon shape compatible with it that does not exhibit singularities. In that sense, the environment is not able to grow hair to the black hole and the horizon deforms to adapt to the environment. 

Chapter 6 deviates slightly from the central themes of the thesis, as it presents the first example of asymptotically flat black holes with locally vacuum toroidal horizons in four spacetime dimensions featuring a non-singular exterior region. These are toy models constructed using thin shells and interpolating regions, each with a clear physical interpretation. As expected from classic results on black holes, their existence necessitates violations of energy conditions, which we explicitly characterize. Although this chapter is self-contained, it employs the same tools developed in the previous chapter and was directly inspired by the analyses conducted there. 

The final chapter of the thesis addresses compactness bounds, that is, the ratio of mass to radius, for spherically symmetric stellar-like objects in General Relativity. We focus on two classical results: the Buchdahl and Bondi bounds. In particular, we examine the assumptions underlying these results and argue that, when certain assumptions are relaxed, the bounds can be circumvented. We illustrate this with physically simple toy models that offer a clear physical interpretation. We emphasize that these assumptions are expected to be violated in physically realistic scenarios, and recall the similarities of our toy models with some models in the literature. Finally, we explore how modifying these assumptions, particularly through the imposition of various energy conditions, affects the bounds, and we review alternative, less stringent limits previously discussed in the literature.

The thesis concludes with a final chapter that summarizes the main results, outlines possible directions for future research, and situates the work within a broader context. In particular, we critically reflect on the findings, discuss their relevance to the emergent gravity framework, and explore several open questions, offering a more personal perspective on their significance. 

\cleardoublepage

\fancyhead[LE,RO]{\thepage}
\fancyhead[LO,RE]{Resumen}

\thispagestyle{plain}

\begin{center}
    \textbf{ {\large Resumen}}
\end{center}

Esta tesis se enmarca en el contexto de la gravedad cuántica, entendida en un sentido amplio como cualquier esfuerzo por explorar la relación entre la gravitación y el mundo cuántico, sin que esto implique necesariamente la cuantización del campo gravitacional. En particular, el análisis se centra en teorías emergentes, entendidas como aquellas en las que la estructura causal y los conceptos geométricos que aparecen en la descripción del campo gravitacional no son fundamentales, sino que emergen a partir de otros grados de libertad subyacentes. Además, nos centramos en teorías emergentes que están inspiradas en modelos de física de la materia condensada. 

En lugar de construir una teoría emergente y realizar un análisis detallado de sus consecuencias, una tarea demasiado ambiciosa para una sola tesis doctoral, este trabajo ofrece una hoja de ruta a teorías emergentes sin comprometerse con ninguna en particular, analizando algunos aspectos relevantes dentro de este contexto. El objetivo no es tanto proporcionar respuestas definitivas a algunas preguntas, sino cuestionar ciertas ideas establecidas y destacar algunas características que una teoría emergente debería poseer para ser exitosa. 

La tesis se divide en dos partes bien diferenciadas, que utilizan herramientas y enfoques diferentes. La primera parte aborda aspectos fundamentales y de índole más conceptual de las teorías emergentes, con un énfasis particular en el rol de las estructuras de fondo. Se analizan tanto las implicaciones para la teoría como su relevancia desde un punto de vista constructivo. La segunda parte de la tesis, parte de la suposición de que las geometrías que describen el campo gravitacional no tienen ni singularidades ni horizontes, una característica que se espera que tenga cualquier teoría emergente exitosa. No obstante, esta segunda parte también puede interpretarse como un análisis agnóstico partiendo de esta premisa sin comprometerse con ninguna teoría en particular. 

La primera parte de la tesis consta de cuatro capítulos. Los dos primeros se sitúan en el marco de la gravedad análoga, que consiste en estudiar sistemas cuyas excitaciones en ciertos regímenes pueden describirse como campos relativistas propagándose en una geometría lorentziana efectiva. En el Capítulo 1 investigamos si es posible simular curvas cerradas temporales en estas geometrías efectivas que aparecen en análogos gravitacionales. Dado que las ecuaciones de Einstein no restringen estas geometrías dinámicamente, es posible, en principio, diseñar configuraciones que no surgirían de forma natural en escenarios relativistas. Sin embargo, encontramos que el sistema tiende a evitar las configuraciones en las que la geometría emergente exhibe patologías causales. De hecho, la descripción análoga deja de ser válida antes de que dichas configuraciones se alcancen. Este comportamiento parece deberse principalmente a la presencia de una causalidad subyacente más fundamental: la causalidad del laboratorio. Sin embargo, esta razón no tiene porque ser evidente para observadores que solo pueden sondear la geometría emergente.  

En el Capítulo 2 se analizan sistemas análogos con un sustrato cuántico y se investiga la posibilidad de generar superposiciones de estados tales que, cada uno por separado, dan lugar a una geometría emergente concreta. Cuando la estructura causal se difumina demasiado, el sistema se desestabiliza, haciendo que dichas superposiciones sean extremadamente inestables. Aunque en gravedad cuántica se espera que las superposiciones de espaciotiempos tengan cabida dentro de la teoría, esto parece indicar que las configuraciones que se desvían demasiado de una causalidad bien comportada serían suprimidas dinámicamente. Esta observación guarda un paralelismo con argumentos similares en la literatura que sugieren que, más que cuantizar la gravedad, podría ser necesario modificar las leyes del mundo cuántico para hacerlas compatibles con el marco de la relatividad general. Se revisan varias de estas propuestas, se comparan con los resultados del modelo análogo y se extraen lecciones valiosas para el marco de la gravedad emergente. 

El Capítulo 3 aborda las teorías gravitacionales desde un punto de vista constructivo. Es habitual encontrar en la literatura la afirmación de que la única extensión no lineal consistente de la teoría de Fierz-Pauli es la relatividad general. En el capítulo se revisa esta cuestión examinando también el problema análogo para teorías alto derivativas y teorías métrico afines. Se demuestra que cualquier teoría que no tiene estructuras de fondo puede ser reconstruida a partir de su linealización. Además, se prueba que la relatividad general no es la única extensión consistente de la teoría de Fierz-Pauli en dimensiones espaciotemporales mayores que cuatro, enfatizando que el proceso constructivo requiere tomar decisiones en cada paso y distintas elecciones dan lugar a distintas teorías no lineales. Finalmente, se demuestra también que la gravedad unimodular puede ser reconstruida a partir de su linealización y, más en general, también cualquier teoría que exhibe estructuras de fondo pero es invariante bajo transformaciones WTDiff. 

El Capítulo 4 presenta una comparación sistemática entre la relatividad general y la gravedad unimodular, así como, de manera más general, entre teorías invariantes bajo difeomorfismos y sus contrapartes invariantes bajo difeomorfismos transversos y transformaciones de Weyl. Se comparan estas teorías a nivel clásico y semiclásico, y se analizan algunos aspectos de sus formulaciones cuánticas, concluyendo que la única distinción entre ellas radica en la naturaleza de la constante cosmológica. Mientras que en la relatividad general actúa como una constante de acoplo, en gravedad unimodular aparece como una constante de integración. Se revisan críticamente las implicaciones de esta diferencia tanto para la construcción de modelos como para la teoría de cuerdas. En particular, se analiza si el sector de bajas energías de teoría de cuerdas puede ser descrito únicamente por la relatividad general.   

La segunda parte de la tesis consta de tres capítulos y se centra en objetos compactos sin horizontes ni singularidades. El Capítulo 5 se centra en los teoremas de no pelo en el contexto específico de espaciotiempos estáticos y con simetría axial. En particular, no se asume la presencia de un horizonte, que es una suposición habitual, y se impone en su lugar la presencia de una superficie en la que la norma del vector de Killing temporal es mínima pero finita y los invariantes de curvatura escalares permanecen acotados. Se encuentra que, bajo estas condiciones, la estructura multipolar del objeto debe aproximarse paramétricamente a la de un agujero negro a medida que esa norma tiende a cero. Como subproducto de este análisis, aplicamos las mismas herramientas a los agujeros negros en entornos astrofísicos y mostramos que, para cualquier entorno gravitacional dado, existe una única forma posible para el horizonte que no da lugar a singularidades y es compatible con la relatividad general. En ese sentido, el entorno no puede crecerle ``pelo'' al agujero negro. 

El Capítulo 6 se desvía ligeramente de los temas principales de la tesis, y en él se presenta el primer ejemplo de agujeros negros asintóticamente planos, con horizontes toroidales en cuatro dimensiones espaciotemporales, que están localmente en vacío y con una región exterior no singular. Son modelos sencillos que presentan materia en algunas regiones, alguna de ella materia distribucional, es decir, capas finas. Como se espera en función de los teoremas clásicos sobre agujeros negros, la existencia de horizontes toroidales require violaciones de condiciones de energía en la región externa, que se caracterizan explícitamente en el capítulo. Aunque es autocontenido, emplea las herramientas desarrolladas en el capítulo anterior y fue directamente inspirado por esos análisis. 

El último capítulo de la tesis aborda los límites de compacidad, es decir, los límites para el cociente entre la masa y el radio de objetos esféricamente simétricos en relatividad general. Se analizan dos resultados clásicos: los límites de Buchdahl y Bondi. En particular, se examinan las hipótesis detrás de los resultados y se argumenta que, cuando se relajan algunas de esas suposiciones, los límites pueden sobrepasarse. Se ilustra esto con modelos simplificados y sencillos pero que ofrecen una interpretación física clara. En particular se enfatiza la similitud de estos modelos con algunos ya presentados en la literatura. Finalmente, se revisa cómo modificaciones de las hipótesis que conducen a estos límites, en particular a través de imponer condiciones de energía menos restrictivas, afectan al valor de las cotas que se derivan.

En las conclusiones se resumen los resultados principales, se discuten posibles direcciones de trabajo futuras y se sitúa el trabajo dentro de un contexto más amplio. En particular, se discuten de forma crítica los resultados de la tesis y su relevancia dentro del marco de la gravedad emergente. Finalmente, se exploran algunas preguntas abiertas surgidas a raíz de esta investigación y se ofrece una perspectiva más personal sobre su importancia. 

\cleardoublepage

\fancyhead[LE,RO]{\thepage}
\fancyhead[LO,RE]{Agradecimientos}

\thispagestyle{plain}

\noindent
\begin{center}
    {\LARGE \bf Agradecimientos}
\end{center}

\vspace{0.2cm}

Realizar esta tesis doctoral ha sido una aventura apasionante en la que ha habido momentos de todo tipo, desde la tristeza al júbilo, pasando por la desesperación y el agotamiento. Es el resultado de varios años de trabajo en los que he ido madurando y evolucionando, no solo a nivel científico, sino sobre todo a nivel personal. Me es imposible echar la vista atrás sin que me venga a la cabeza una lista interminable de anécdotas, lugares y momentos compartidos pero, sobre todo, de personas. Habéis dejado una huella imborrable en mí, y en mi forma de estar y ver el mundo durante estos años. Este proyecto no habría tenido sentido sin vosotros.

En primer lugar quiero agradecer a mis directores de tesis por su apoyo y confianza en mí durante estos años, a pesar de ser probablemente el estudiante de doctorado más disperso que han tenido. Agradezco a Carlos por las discusiones sobre física, filosofía, cine, música y una innumerable lista de temas: siempre consigues darle una vuelta de tuerca a mis reflexiones que me deja pensando. A Luis por abrirme esta ventana al mundo académico, por tu forma tan particular de entender y transmitir las cosas, de la que tanto he aprendido, y por los ratos por los bares de Malasaña. A Raúl por sus valiosos consejos sobre cómo desenvolverme en el entorno académico.

A Alfredo, porque el germen de esta tesis y mi interés por el mundo académico están sin duda enraizados en las infinitas conversaciones contigo. Gracias por tu amistad, por la pasión que desprendes por la óptica y la física, y por ofrecer tanta paz, no solo a mí, sino a todos los que nos sentamos a hablar contigo. A Laura por ser mi primera colaboradora y por ser una buena amiga durante tantos años, aunque como bien dijiste estemos condenados a no entendernos.

A toda la gente con la que he tenido la oportunidad de trabajar estos años: a Miguel, Marc y Raül por \textit{alabear} mi visión de la Física. A Bert por su entusiasmo, su pasión por la docencia y tantas buenas charlas sobre gravitación. A los otros dos generadores de nuestro SU(2), Alberto y Jose, y a Javi por ser nuestro subgrupo normal, gracias por las risas durante nuestros particulares juegos del calamar. A Gil por ese humor tan ácido y por ser un científico con el que se puede hablar de filosofía. Gracias a Ana por desprender ese entusiasmo y por hacerme de cicerone en Berlín, prometo ir saldando mis infinitas apuestas perdidas. A Edu, por las discusiones sobre física siempre tan estimulantes. A Jose Beltrán por sus preguntas siempre en el clavo, de las que han nacido tantas ideas. A Aitor y Di por su interés en trabajar conmigo y enseñarme tanto. 

To those who hosted me during my doctoral stays. Tomi, Manuel and Laür, thanks for giving me the opportunity to learn from you and and making my time in Estonia so interesting. Thanks Misha for letting me learn so much about gravity and quantum field theory during my stay in Switzerland. Gracias a Pepe y Valeria por las risas en el Barlova. Gracias Alejandro por acogerme en Suiza y hacerme sentir en casa, tenemos pendiente nuestra visita a Florencia. Eskerrik asko Ander, por las aventuras por los bares de Ginebra y Lausanne. Merci Alice pour les voyages à travers la Suisse et pour m'avoir fait découvrir les coins de Dijon. Gracias a José Luis Jaramillo por compartir raíces conmigo, por acogerme en Dijon y por ayudarme siempre en todo lo que has podido durante estos años, aún cuando tienes más tareas en un día de las que una persona normal podría hacer en un año. 

A Alejandro Jiménez: mi amigo, colaborador y apasionado de la tortilla. Me alegro infinitamente de que nuestros caminos se hayan cruzado. Te admiro como científico pero sobre todo como persona: gracias por tu amistad durante todos estos años y por la dedicación que pones en todo lo que haces. Estoy deseando leer lo próximo que escribas e ir a la caza del próximo pincho de tortilla, sea en Granada, Madrid, Salamanca, Bilbao o donde toque. 

A Alejandro Vivas, por tu amistad durante tantos años desde nuestro primer encuentro en la Complutense, desde el que nuestras vidas han discurrido en paralelo. Gracias por tu contagiosa e infinita inquietud intelectual.  

A todos mis compañeros de piso durante estos años: a Julio y Valentín por las infinitas risas y el caos compartido en aquella casa encantada. A Anthony por ofrecerme una visión particular sobre muchos temas y por estar tan cuajado como yo. A José (Water-Polo) por su amistad durante tantos años desde aquellas Olimpiadas de Física. A Pipo porque aunque estuvieras muy brevemente, aún me acuerdo de tu paso por casa y no puedo parar de reírme (y siguen apareciendo cosas tuyas por ahí, sinvergüenza). A Roberto e Ixaka por aguantarme en casa en la última parte de esta tesis. 

A mis compañeros del IAA que han hecho que el instituto sea mucho más acogedor. A Angela por los ``caffè correttos'' y por arreglar el mundo juntos con una barra de bar debajo del brazo. A Pedro por siempre querer ayudarme y llegar incluso a convencerme a veces de que, haga lo que haga, me va a salir bien. A Víctor por los tés y por ser el gurú de las simetrías. A Sebastiano por nuestras conversaciones furtivas de ciencia en los baños. A Gabri por tu apoyo, tu amistad y por contagiar ese amor por la Naturaleza (y a Momo también, claro). A Dani por compartir conmigo la pasión por la ciencia, la música y los bares, todo a la vez. A Bea por los buenos ratos, vino en mano, y los intercambios musicales. A Emilio por interesarse por mis delirios de físico teórico e invitarme a hacer divulgación en tantas ocasiones (por supuesto, de forma no remunerada). A Alicia por su amistad y por enseñarme a desenvolverme  en tantos contextos sociales (aunque quizá sin saberlo). A Teresa por las tardes de canciones pop y por ser la mejor (y peor) co-IP que he tenido hasta el momento. A Clara por ser el agente del caos que el IAA necesita para funcionar. A Julio por compartir su pasión por la guitarra y la música. A Celia por su amistad y apoyo y su infinita paciencia ayudándome a hacer animaciones. A mis compis de despacho durante estos años: Zuri, Adrián, Luis, Fran y David, por tantas risas y tantos momentos de desquicie colectivo. También gracias a mis matemáticos de la UGR: Fidel y Juansi por el apoyo mutuo, la amistad y tantas conversaciones interesantes durante estos años. A José Ignacio Illana por nuestros intercambios como aficionados a la Cosmología. 

A todos los amigos que os encontráis repartidos por el mundo, por vuestra amistad en la distancia. A mis amigos de la Complutense: a Víctor, a Isa, a Mario, a Adri y a Alex. A mis amigo de Córdoba: Molina, Lucía y Rocío, por aguantarme durante tantos años. A Jaime y Laura, por su amistad desde nuestras andanzas por tierras catalanas. A Marta por los vermús entre clases. A Héctor por su amistad desde las pachangas teóricas y sus clases de mecánica. A Merce por ser una gran profesora de grupos y por sus infinitos intentos por explicarme el formalismo de Loops. A Silvia y Antonio por su compañerismo y por su iniciativa de organizar nuestro magnífico QFTCS Workshop. A Adrià por convencerme continuamente de que lo raro es encajar en el molde. A los amigos del mundo académico que hacen de los EREP y los IberiCOS un encuentro familiar: Araceli, Sara, Eneko, Javi, Asier, Valle, Gabri y David.

A Migue por colgarme mi primera guitarra cuando no distinguía un Do de alguien cayéndose por las escaleras, gracias a ti ahora la música forma una parte enorme de mi vida. A mis Rayamanta: David, Pablo y Miguel, por su infinita paciencia conmigo, por las risas y por las canciones. A mis 11K: Ana y Quique por los buenos ratos en el local, aunque solo nos juntemos cuando pasa el cometa Haley. 

Quiero agradecer especialmente a los amigos que he hecho en Granada, por ser, en el fondo, los culpables de que esta ciudad me guste tanto. A Zoraida por su amistad durante estos años y por ser mi compañera de conciertos. A Jose, Rosa y la recién llegada Candela, aunque nos veamos menos de lo que me gustaría, me he sentido arropado por vuestro cariño y amistad desde el primer momento. A mis devotes del dedo corazón porque el meñique de oro está cerca, ¡creedme! En particular, a Iris por su amistad y por presentarme a tanta gente genial, a Iñaki por nuestros miércoles tontos en El Acorde y a Fagota y Alba por ser nuestra avanzadilla en Sevilla.  Quiero agradecer especialmente a mis habituales de los domingos: a Cristina, a Paula, a Santos (aka Patinete) y a Irene por su amistad, su apoyo, su visión del mundo y por ser un lugar al que siento que siempre podré volver. Gracias por aparecer. 

Quiero terminar agradeciendo a mi familia por su apoyo y preocupación durante todos estos años y, en especial, a mi madre por haberme animado a seguir una carrera académica. Te agradezco mucho el apoyo en las decisiones que he ido tomando durante estos años, el esforzarte por intentar entenderme aunque a veces sea imposible y por último, aunque no por ello menos importante, por ser la culpable de mi pasión por los Beatles y Antonio Vega. Gracias mamá. 

Esta tesis ha recibido financiación por parte del Ministerio de Universidades, a través de la ayuda FPU20/01684 y los proyectos nacionales PID2020-118159GB-C43 y PID2023-149018NB-C43.    

\vspace*{0.2cm}

\noindent

\cleardoublepage

\fancyhead[LE,RO]{\thepage}
\fancyhead[LO,RE]{List of Publications}

\thispagestyle{plain}

\noindent
\begin{center}
    {\LARGE \bf List of Publications}
\end{center}

\vspace*{0.2cm}

This is the list of articles published by the candidate with the original results on which this thesis document is based. 

\vspace*{0.2cm}

\begin{center}
    {\large {\bf Peer-reviewed journals}} 
\end{center}

\noindent \textbf{1.} \textit{Chronology protection implementation in analogue gravity}

\noindent Carlos Barcel\'o, Jokin Eguia S\'anchez, \underline{Gerardo Garc\'ia-Moreno}, Gil Jannes

\noindent \href{https://iopscience.iop.org/article/10.1088/1475-7516/2023/12/006}{{\bf Eur. Phys. J. C 82 (2022) 4, 299}}

\noindent \textbf{2.} \textit{Analogue gravity simulation of superpositions of spacetimes}

\noindent Carlos Barcel\'o, Luis J. Garay, \underline{Gerardo Garc\'ia-Moreno}

\noindent \href{https://link.springer.com/article/10.1140/epjc/s10052-022-10702-5}{{\bf Eur. Phys. J. C 82 (2022) 8, 727}}

\noindent \textbf{3.} \textit{Unimodular gravity vs general relativity: a status report}

\noindent Ra\'ul Carballo-Rubio, Luis J. Garay, \underline{Gerardo Garc\'ia-Moreno}

\noindent \href{https://iopscience.iop.org/article/10.1088/1361-6382/aca386}{{\bf Class. Quant. Grav. 39 (2022) 24, 243001}}

\noindent \textbf{4.} \textit{Bootstrapping gravity and its extension to metric-affine theories}

\noindent Adri\`a Delhom, \underline{Gerardo Garc\'ia-Moreno}, Manuel Hohmann, Alejandro Jim\'enez-Cano, Tomi S. Koivisto

\noindent \href{https://iopscience.iop.org/article/10.1088/1475-7516/2023/12/006}{{\bf JCAP 12 (2023) 006}}

\noindent \textbf{5.} \textit{Embedding Unimodular Gravity in string theory}

\noindent Luis J. Garay, \underline{Gerardo Garc\'ia-Moreno}

\noindent \href{https://link.springer.com/article/10.1007/JHEP03(2023)027}{{\bf JHEP 03 (2023) 027}}

\noindent \textbf{6.} \textit{Nonexistence of a parent theory for general relativity and unimodular gravity}

\noindent \underline{Gerardo Garc\'ia-Moreno}, Alejandro Jim\'enez-Cano

\noindent \href{https://journals.aps.org/prd/abstract/10.1103/PhysRevD.109.104004}{{\bf Phys. Rev. D 109 (2024) 10, 104004}}

\noindent \textbf{7.} \textit{No-hair and almost-no-hair results for static axisymmetric black holes and ultracompact objects in astrophysical environments}

\noindent Carlos Barcel\'o, Ra\'ul Carballo-Rubio, Luis J. Garay, \underline{Gerardo Garc\'ia-Moreno}

\noindent \href{https://iopscience.iop.org/article/10.1088/1361-6382/adc233}{{\bf Class. Quant. Grav. 42 (2025) 7, 075020}}

\noindent \textbf{8.} \textit{Beyond Buchdahl's limit: bilayered stars and thin-shell configurations}

\noindent Julio Arrechea,  Carlos Barcel\'o, \underline{Gerardo Garc\'ia-Moreno}, Jos\'e Polo-G\'omez

\noindent \href{https://doi.org/10.1103/dqbd-9z5f}{{\bf Phys. Rev. D 111 (2023), 124017}}

\noindent \textbf{9.} \textit{Toroidal black holes in four dimensions}

\noindent Carlos Barcel\'o, \underline{Gerardo Garc\'ia-Moreno}, Alejandro Jim\'enez Cano

\noindent \href{https://doi.org/10.1088/1361-6382/adf0e1}{{\bf Class. Quant. Grav. 42 (2025) 15, 155015}}

\newpage

Other articles that have been published during this thesis but are not directly related to the results presented in this document are:

\vspace*{0.2cm}

\noindent \textbf{1.} \textit{Hawking radiation from an analogue bouncing geometry}

\noindent Alberto Garc\'ia-Mart\'in Caro, \underline{Gerardo Garc\'ia-Moreno}, Javier Olmedo, Jose M. S\'anchez-Vel\'azquez

\noindent \href{https://journals.aps.org/prd/abstract/10.1103/PhysRevD.108.L061701}{{\bf Phys. Rev. D 108 (2023) 6, L061701}}

\noindent \textbf{2.} \textit{Classical and quantum field theory in a box with moving boundaries: A numerical study of the dynamical Casimir effect}

\noindent Alberto Garc\'ia-Mart\'in Caro, \underline{Gerardo Garc\'ia-Moreno}, Javier Olmedo, Jose M. S\'anchez-Vel\'azquez

\noindent \href{https://journals.aps.org/prd/abstract/10.1103/PhysRevD.110.025007}{{\bf Phys. Rev. D 110 (2024) 2, 025007}}

\vspace*{0.2cm}

\noindent

\cleardoublepage

\fancyhead[LE,RO]{\thepage}
\fancyhead[LO,RE]{Contents}

\thispagestyle{plain}

\tableofcontents

\cleardoublepage

\fancyhead[LE,RO]{\thepage}
\fancyhead[LO,RE]{List of Figures}

\thispagestyle{plain}

\noindent
\listoffigures

\vspace*{0.2cm}

\cleardoublepage

\fancyhead[LE,RO]{\thepage}
\fancyhead[LO,RE]{List of Acronyms}

\thispagestyle{plain}

\noindent
\begin{center}
    {\LARGE \bf List of Acronyms}
\end{center}

\vspace*{0.5cm}

\begin{center}
    \begin{tabular}{l l} %
      AdS & Anti-de Sitter \\
      BEC & Bose-Einstein Condensate \\
      BCFW & Britto-Cachazzo-Feng-Witten  \\
      BKL & Belinski–Khalatnikov–Lifshitz  \\
      BRST & Becchi-Rouet-Stora-Tyutin \\
      CTC & Closed Timelike Curve \\
      CFT & Conformal Field Theory \\
      DEC & Dominant Energy Condition \\
      EFT & Effective Field Theory \\
      FLRW & Friedmann-Lema\^{i}tre-Robertson-Walker \\
      GHP & Gibbons-Hawking-Perry \\
      GR & General Relativity \\
      NEC & Null Energy Condition \\
      RG & Renormalization Group \\
      SEC & Strong Energy Condition \\
      TDiff & Transverse Diffeomorphism \\
      TOV & Tolman-Oppenheimer-Volkoff \\
      UG & Unimodular Gravity \\
      WEC & Weak Energy Condition \\
      WTDiff & Weyl Transverse-Diffeomorphism \\
    \end{tabular}
\end{center}
    
\vspace*{0.2cm}

\cleardoublepage

\fancyhead[LE,RO]{\thepage}
\fancyhead[LO,RE]{Notation and Conventions}

\noindent
\begin{center}
    {\LARGE \bf Notation and Conventions}
\end{center}
\vspace*{0.2cm}

\noindent \textbf{\large{Units}}

\begin{itemize}
    \item Throughout this thesis, we set the speed of light in vacuum to unity, $c = 1$.
    \item In the first part of the thesis, we set Planck's constant to unity, $\hbar = 1$. The only remaining dimensionful fundamental constant is the Planck mass $M_P$ or, equivalently, $\kappa_{D}$ or Newton's constant, $G_N$. They are related through:
    \begin{align*}
          M_P = G_N^{- \frac{1}{D-1}}, \qquad \kappa_{D} = \sqrt{4 \pi G_N}.
    \end{align*}
    \item In the second part of the thesis, we set Newton's constant to unity, $G_N = 1$, so that the only remaining dimensionful fundamental constant is Planck’s constant, defined as:
    \begin{align*}
        \ell_P = \sqrt{\hbar}. 
    \end{align*}
\end{itemize}

\bigskip

\noindent \textbf{\large{Indices}}

\begin{itemize}

\item Type of indices appearing in the text:

\begin{itemize}
    \item[$\bullet$] Greek indices, $(\mu, \nu, \ldots )$ denote the spacetime components of tensors in a holonomic (coordinate) basis of the $D+1$-dimensional manifold. 
    \item[$\bullet$] Lower case latin indices from the middle of the alphabet, $(i,j, \ldots )$ denote the components of tensors on $D$-dimensional submanifolds defined in the text.
    \item[$\bullet$] Lower case latin indices from the beginning of the alphabet, $(a,b,\ldots)$ represent the components of tensors defined on $D-1$-dimensional submanifolds defined in the text. 
    \item[$\bullet$] Upper case latin indices from the beginning of the alphabet $(A,B,\ldots)$ represent the components of tensors in an anholonomic (non-coordinate) frame of the manifold.
    \item[$\bullet$] Upper case latin indices from the middle of the alphabet $(I,J,\ldots)$ represent a placeholder for all the fields living in the manifold, including both their spacetime and internal indices. 
\end{itemize}

\item Einstein summation convention is used throughout unless explicitly stated otherwise.

\item For tensors, upper indices correspond to the contravariant part (vector) and lower ones to the covariant part (covector).

\item Symmetrization and antisymmetrization of indices are defined as
\begin{align*}
    H_{( \mu_{1} \ldots \mu_{k} ) } := \frac{1}{k!} \sum_{\sigma \in S_{n}} H_{\sigma(\mu_1) \ldots \sigma (\mu_k)}, \qquad H_{[\mu_{1} \ldots \mu_{k} ]} := \frac{1}{k!} \sum_{\sigma \in S_{n}} \text{sgn} ( \sigma) H_{\sigma(\mu_1) \ldots \sigma (\mu_k)}, 
\end{align*}
where $S_n$ is the symmetric group on $n$-elements and $\text{sgn}$ the parity of the permutation. 

\end{itemize}

\bigskip

\noindent \textbf{\large{Metric and curvature conventions}} 

\begin{itemize}
    \item We work with the mostly plus convention signature for the metric $(-, +, \cdots, +)$. 
    \item For general considerations about structural aspects of theories, we work in $D+1$ spacetime dimensions, although for specific features that depend on the spacetime dimensions we focus on $D=3$. 
    \item Convention for the covariant derivatives:
        \begin{align*}
            \nabla_{\mu } B_{\nu}{}^{\rho} := \partial_{\mu} B_{\nu}{}^{\rho} + \Gamma_{\mu \nu}{}^{\sigma} B_{\sigma}{}^{\rho} - \Gamma_{\mu \sigma}{}^{\rho} B_{\nu}{}^{\sigma}.
        \end{align*}
    \item Conventions for the non-metricity $Q_{\mu \nu \rho}$ and the torsion $T_{\mu \nu}{}^{\lambda}$:
        \begin{align*}
            T_{\mu \nu}{}^{\lambda}:= \Gamma_{\mu \nu}{}^{\lambda} - \Gamma_{\nu \mu}{}^{\lambda}, \qquad Q_{\mu \nu \sigma} := - \nabla_{\mu} g_{\nu \sigma}.
        \end{align*}
    \item Conventions for the Riemann tensor, $R_{\mu \nu \rho}{}^{\sigma}$, the Ricci tensor $R_{\mu \nu}$, and the Ricci scalar, $R$:
        \begin{align*}
            & R_{\mu \nu \rho}{}^{\sigma} := \partial_{\nu} \Gamma_{\mu \rho}{}^{\sigma} - \partial_{\mu} \Gamma_{\nu \rho}{}^{\sigma} + \Gamma_{\mu \rho}{}^{\lambda} \Gamma_{\lambda \nu}{}^{\sigma} - \Gamma_{\nu \rho}{}^{\lambda} \Gamma_{\lambda \mu}{}^{\sigma}, \\
            & R_{\mu \nu} := R_{\mu \rho \nu}{}^{\rho}, \qquad R:=g^{\mu \nu} R_{\mu \nu}. 
        \end{align*}
\end{itemize}

\bigskip

\noindent \textbf{\large{Some notational comments}}

\begin{itemize}
    
    \item Boldface symbols, e.g., $\boldsymbol{v}$, $\boldsymbol{g}$), indicate abstract tensors or vectors, distinguishing them from their components, which are written with indices, e.g., $v^{\mu}$, $g_{\mu \nu}$.

    \item In Chapter~\ref{Ch3:SelfCoupling}, barred quantities, e.g., $\Bar{g}^{\mu \nu}$, denote background fields, around which perturbations are considered.

    \item $\Tilde{\boldsymbol{g}}$ denotes an auxiliary metric constructed from the metric tensor $\boldsymbol{g}$ and the background volume form $\boldsymbol{\omega}$, and its components are given by:
    \begin{align*}
        \tilde{g}_{\mu \nu} = \left( \frac{\omega^2}{\abs{g}} \right)^{\frac{1}{D+1}} g_{\mu \nu}
    \end{align*}
    The Levi-Civita connection compatible with $\Tilde{\boldsymbol{g}}$ is denoted as $\tilde{\nabla}$, to distinguish it from the one associated with $\boldsymbol{g}$, that we denote as $\nabla$. 
    \item Curvature tensors constructed from the Levi-Civita connection of a metric $\boldsymbol{g}$ are denoted as $R_{\mu \nu} \left( \boldsymbol{g} \right)$ when multiple metrics are involved, in order to distinguish between them. 

    \item The symbol $\Delta$ used at several places in the text does never represent the Laplacian, which is always denoted by $\nabla^2$. It is commonly used to represent a difference, unless explicitly stated otherwise. 

    \item The symbol $G_{\mu \nu}$ represents a metric defined in terms of the metric $g_{\mu \nu}$ and the Stueckelberg fields $Y^{\mu}$, and should not be mistaken for the Einstein tensor. 
\end{itemize}

\bigskip

\noindent \textbf{\large{Exceptions to the general notation}} 

\medskip

\noindent In a few places, a specific notation is introduced that deviates from the general conventions used throughout the thesis. This is done for clarity or to avoid an unnecessarily cumbersome notation. 
\begin{itemize}
    \item In Chapter~\ref{Ch2:AnalogueSuperp}, we use lower case latin indices from the middle of the alphabet, $(i,j, \ldots )$ as labels representing the different elements of a set, not the components of a tensor.
    \item In Subsection~\ref{Subsec:Constructibility}, we also use lower case latin indices from the middle of the alphabet, $(i,j, \ldots )$ as labels representing the different elements of a set, not the components of a tensor, and upper case latin indices from the middle of the alphabet $(I,J,\ldots)$ as representing subsets of elements.
    \item In Section~\ref{Sec:Strings}, latin indices from the beginning of the alphabet $(a,b, \ldots )$ are used to represent tensor indices in the worldsheet of strings. The target space metric is represented as $\mathcal{G}_{\mu \nu}$, with the Riemann and Ricci tensor and scalar represented as $\mathcal{R}_{\mu \nu \rho}{}^{\sigma}$, $\mathcal{R}_{\mu \nu}$ and $\mathcal{R}$, respectively. 
    \item In Appendix~\ref{App:Scattering}, we use latin indices from the middle of the alphabet, $(i,j, \ldots )$ as labels representing the different elements of a set, Greek indices from the beginning of the alphabet that are undotted $(\alpha, \beta, \ldots)$ and dotted $(\dot{\alpha}, \dot{\beta}, \ldots)$ to represent the $(1/2,0)$ and $(0,1/2)$ representations of $SL(2,\mathbb{C})$, respectively, and latin indices from the middle of the alphabet $(a,b,\ldots)$ to represent internal indices.  
\end{itemize}

\noindent

\makeatletter\@openrighttrue\makeatother   

\cleardoublepage

\setcounter{page}{0}

\pagenumbering{arabic}
\pagestyle{fancy}

\setcounter{secnumdepth}{-1}
\pagenumbering{arabic}

\chapter{Introduction}\label{Introduction}
\fancyhead[LE,RO]{\thepage}
\fancyhead[LO,RE]{Introduction}

General Relativity (GR) is the most successful theory that we have nowadays to describe gravitational phenomena. Its range of applicability is wide and spans from solar system phenomena, where it leads to small deviations from classic Newtonian gravity, to the description of the large-scale structure of the universe in cosmology. It also encompasses the description of compact objects with strong gravitational fields, such as astrophysical black holes and neutron stars, for which the nonlinearities of GR play a significant role. A natural question that arises then is why one would seek alternative descriptions of the gravitational phenomena beyond GR. 

One of the main reasons is the presence of singularities, which make it an undesirable theory. A singularity can be roughly and informally described as the breakdown of the spacetime structure itself. Given that the spacetime is the arena in which everything happens, it is not easy to ``locate'' singularities as events on spacetime. The most accepted operative definition of a singularity is that of geodesic incompleteness (although even such a broad definition leaves aside examples of spacetimes that one could consider as pathological~\cite{Beem1976}): a spacetime is said to be singular if it is geodesically incomplete and inextendable~\cite{Hawking1973,Wald1984}. This means that there are geodesics that cannot be further extended in their proper (affine) time (parameter) and the manifold is such that it cannot be smoothly extended either. 

A plethora of classic results which go under the name of singularity theorems, ensure that singularities form in GR under physically well-motivated and significantly general conditions~\cite{Hawking1970,Senovilla2014,Witten2019}. Different versions of the theorem make slightly different assumptions, but they all fall into two main categories: those that apply to gravitational collapse, demonstrating that once a trapped region forms, a singularity in the future of some observers becomes inevitable; and those that apply at the cosmological level, showing that the universe must have originated from an initial singularity. Although these theorems do not necessarily guarantee the formation of a curvature singularity, i.e., those for which the curvature invariants along the incomplete geodesics blow up, we will always be referring to such singularities along this thesis, which are commonly formed in gravitational collapse. Whereas there are claims that singularities may contain some physical content and are not necessarily pathologies of the theory~\cite{Curiel1999} once they are properly interpreted, the philosophical stand of this thesis is that, at least, the curvature singularities that typically form inside black holes in classical GR are signs of the breakdown of the theory. 

The nature of singularities in GR is not clear either. Singularities can be classified according to whether they are: spacelike, which means that they are in the future (or past) of timelike observers; timelike, which means that they are both in the future and past of timelike observers; and null which is a limiting case in which they are in the future (or past) of timelike observers but the singular surface can be regarded as a limit of null hypersurfaces. Timelike singularities are the most pathological in a sense, since they spoil the possibility of evolving initially regular data, which leads to a breakdown of predictability. This led Penrose to conjecture the so-called weak cosmic censorship~\cite{Penrose1999}, which roughly states that naked singularities (which are timelike singularities not covered by horizons) do not form in GR if physically realistic initial data are given. One of the main challenges in (dis)proving it lies in determining what constitutes physically realistic initial data, but even sharply formulating the conjecture remains a subtle mathematical problem with much ongoing work. 

When a black hole forms, its interior can be very complex and contain singularities covered by a horizon, in agreement with the weak cosmic censorship conjecture. In contrast, the exterior region of the final state of any black hole resulting from gravitational collapse is remarkably simple. In fact a folk interpretation of the no-hair theorems~\cite{Chrusciel2012} suggests that the configuration must settle to the Schwarzschild metric (if nonrotating) or to the Kerr metric (if rotating), as these represent the only two solutions for stationary vacuum black holes. The structure of the black hole's interior remains much less understood.

For nonrotating geometries, it is widely accepted that a spacelike singularity is formed in the interior, with the Oppenheimer-Snyder model~\cite{Oppenheimer1939} serving as the archetypal example. The situation becomes significantly more complex for rotating geometries. Although the singularity inside a Kerr black hole is timelike, classical analyses suggest that the singularity inside a dynamically formed rotating black hole is instead spacelike. Moreover, these analyses indicate that it is approached in a chaotic manner, as described by the Belinski–Khalatnikov–Lifshitz (BKL) conjecture~\cite{Belinskii1970}. However, this picture is complicated by the presence of a Cauchy horizon in the Kerr interior. Cauchy horizons are generically unstable due to the mass inflation mechanism, which is such that any generic perturbation triggers an instability, seemingly leading to the formation of a null singularity that replaces the horizon~\cite{Poisson1990,Ori1999,Marolf2011}. It remains uncertain whether this singularity is strong or weak, namely, whether geodesic trajectories can extend beyond it~\cite{Dafermos2017}.

Adding to this complexity, there are examples of gravitational collapse that produce naked singularities from initial data that are, in principle, well-behaved. This phenomenon, first identified in Choptuik's studies on critical collapse~\cite{Choptuik1992} and later rigorously demonstrated by Christodoulou~\cite{Christodoulou1994}, challenges or qualifies the weak cosmic censorship conjecture~\cite{Penrose1999} by permitting the formation of timelike naked singularities in certain fine-tuned situations. In fact, Christodoulou subsequently showed that these naked singularities arise within a subset of initial conditions that is non-generic~\cite{Christodoulou1997}. 

This discussion about singularities highlights two key aspects. First, the limitations in our understanding of GR, particularly regarding the formation of black hole singularities in generic scenarios without additional symmetries. Significant work remains to be done within classical GR to achieve a complete understanding of gravitational collapse in that framework. Second, although a complete understanding of singularities and their nature within GR remains out of reach, the substantial evidence for their formation within the theory strongly motivates the search for alternative frameworks where such issues might be resolved. Notably, some aspects examined in this thesis through the lens of beyond-GR frameworks also enhance our understanding of GR itself. Investigating theories beyond GR not only offers more general frameworks that address its foundational challenges but also provides valuable new insights into GR phenomena themselves. 

Another significant challenge GR faces is its difficulty in reconciling with the quantum realm. Even when considering the matter content from a quantum perspective while maintaining a classical spacetime structure for the gravitational sector, i.e., the framework of quantum field theory in curved spacetimes~\cite{Birrell1982}, the situation becomes highly complex. This approach reveals phenomena like Hawking radiation, which exemplify how nontrivial gravitational fields can give rise to excitations of the quantum fields and in fact to a rich new set of phenomena. These phenomena are often regarded as a foundational ``step zero'' in the quest for a theory of quantum gravity. The subsequent step, accounting for the backreaction of these excitations of the quantum matter on the classical gravitational field, which constitutes the framework of semiclassical gravity, is even more challenging. The resulting equations are so intricate that solving them is practically impossible. Although this framework remains poorly understood, heuristic analyses have uncovered problems like the information-loss paradox~\cite{Hawking1976}, which has sparked significant interest for decades. 

Furthermore, semiclassical gravity cannot be regarded as a final theory, as it exhibits several inconsistencies that highlight the need for a more fundamental framework. Various analyses indicate that coupling a classical system to a quantum one introduces theoretical challenges, as the dynamical evolution blurs the distinction between the two systems~\cite{Diosi1997,Barcelo2012,Garcia-Moreno2019,Brizuela2023}. Depending on the specifics of the coupling, different phenomena may arise, but a common outcome is the mutual exchange of properties: classical systems can acquire quantum traits, and quantum systems can exhibit characteristics of classical systems. To explore the nature of the gravitational field, several experiments have been proposed to determine whether gravitational interactions can generate entanglement between quantum systems~\cite{Bose2017,Marletto2017}. However, even if such experiments confirm the generation of entanglement, they do not necessarily imply the existence of quantum degrees of freedom inherent to the gravitational field~\cite{Martin-Martinez2022}.

Some proposals suggest that these and other inconsistencies cannot be resolved simply by quantizing the gravitational field. Instead, they argue for a fundamental revision of quantum mechanics to align it with the principles of GR, as proposed by Penrose with his idea of the \emph{gravitization} of quantum mechanics~\cite{Penrose1996,Penrose2014}. A key reason for the incompatibilities between GR and quantum mechanics lies in their differing approaches to background structures. GR eliminates background structures from the outset, treating the spacetime metric, the fundamental ingredient encoding the causal structure of the system\footnote{Actually only the causal structure is completely encoded in the light cones, the conformal factor is irrelevant from the causal point of view.}, as a dynamical degree of freedom. In contrast, quantum mechanics is only clearly formulated in the presence of background structures, particularly a fixed causal structure.

In summary, these issues call for a theory beyond GR, one that eliminates singularities and is, in some way, compatible with the quantum realm. Broadly, I will understand that any attempt to develop a theory that extends beyond the framework of GR to address such challenges falls under the umbrella of what is called quantum gravity. It is important to note that these considerations are primarily conceptual and not directly driven by empirical evidence. Nevertheless, the hope is that such a theory will yield observable implications in regimes of strong gravitational fields.

In particular, I believe that black hole physics, in particular gravitational wave observations due to the coalescence of compact objects to form a black hole, constitute the best candidates for observing beyond-GR phenomena. While the early universe is another strong-gravity regime where such phenomena might be necessary for a complete description, cosmological observations require extensive modeling of the universe's history. This makes it, in my opinion, a less favorable context for isolating and analyzing deviations from GR predictions.  

\section*{Conceptual framework of the thesis} 

Overall, this thesis is devoted to the problem of building new theoretical tools and models that help in the endeavor of having a consistent and predictive theory beyond GR. At this point, it is worth remarking that a theory of quantum gravity does not necessarily entail a framework in which the geometric degrees of freedom of the gravitational field are quantized. Approaches to quantum gravity are diverse, originating from a wide range of starting points and theoretical perspectives. For practical purposes, in this thesis we classify the quantum gravity approaches according to whether the causal structure and the geometric notions that describe the gravitational field are fundamental or not. We call emergent approaches to those for which at a fundamental level, geometrical notions such as spacetime itself are absent, and they arise as an effective description in some regimes.

Example of non-emergent theories are for instance canonical quantum gravity or its modern formulation: loop quantum gravity~\cite{Thiemann2007}, causal dynamical triangulations~\cite{Loll2019} and causal sets~\cite{Surya2019}. In all these approaches, the geometric spacetime structure itself is taken to be fundamental and different quantization procedures are applied to it. Examples of emergent approaches include for instance the most fashionable approach to quantum gravity: string theory~\cite{Polchinski1998a,Polchinski1998b}, where the fundamental degrees of freedom of the theory are taken to be relativistic strings, and all of the known interactions arise as the effective interactions among its excitations. 

In this thesis, we focus on a much less explored set of emergent theories: those inspired by condensed-matter-like systems~\cite{Volovik2008,Volovik2009,Carlip2012,Barcelo2014a}. The main observation that motivates the analysis in depth of this program is the observation that Lorentz invariance is ubiquitous in many physical systems that admit massless excitations. In fact, it can be even argued that Lorentz invariance is a straightforward consequence of having finite speed signals. Actually, it is also quite common to find systems (even nonrelativistic) for which the excitations of the system in some regimes can be described by a collection of weakly coupled relativistic fields propagating on top of an effectively curved metric, which mimics the behavior of a gravitational field. These systems are often called \emph{analogue gravity} systems~\cite{Barcelo2005}. 

For the systems known to date, the metric governing these equations does not satisfy a deformed version of Einstein’s equations or, more generally, any relativistic equations. Thus, this analogy just holds at a kinematical level, meaning that the effective geometry emerging in those systems does not display the dynamical features characteristic of GR. 

The central challenge for this program is to identify whether there exists a class of physical systems (a universality class, for instance) that, in certain regimes, can be described by a set of fields, including a Lorentzian metric, where both the fields and the metric obey the Einstein equations at leading order, potentially with additional corrections. In that sense, whereas the kinematical properties of a geometry are common in many systems, the gravitational dynamics seems hard to reproduce in analogues. The idea is that such equations, along with their corrections, will have phenomenological implications and offer a resolution to the singularity problems that plague GR. Additionally, given the inherently quantum nature of the substratum (the condensed-matter-like system) giving rise to the gravitational phenomena as an effective description, it would also dilute any problem in reconciling the gravitational and the quantum realms. 

The emergent gravity paradigm constitutes the framework in which this thesis has been developed. Rather than presenting a full-fledged emergent theory of gravity that is far beyond the scope of a single thesis, it constitutes a humble exploration of different generic aspects of the causal behavior that emergent theories beyond GR display: both foundational and phenomenological aspects, offering a road map to properties that a successful emergent theory should exhibit. As these aspects are well differentiated, the thesis is structured in two different parts. 

The first part of this thesis examines foundational aspects, particularly the role of background structures in emergent contexts and their influence on gravitational dynamics. For this purpose, different frameworks beyond GR are considered to extract lessons on how the causal structure of GR can be inserted or even diluted within them. 

The second part analyzes some phenomenological implications of such a theory. We begin with the assumption that an emergent theory of gravity prevents the formation of singularities. The removal of singularities suggests that event horizons might also be eliminated, leading to final states of gravitational collapse without any horizons. This aligns with the findings from the first part of the thesis, which indicate that extreme causality situations, such as chronological pathologies or undefined causalities due to superposing several of them, are not allowed in emergent contexts. Thus, it can also be expected that eternal (stationary) event horizons, which represent an extreme causality case, would also be prohibited in such a theory. 

\section*{Structure of the thesis}

The first part, which focuses on fundamental aspects of causality for emergent theories beyond GR, contains four chapters. Chapters~\ref{Ch1:ChronologyAnalogue}-\ref{Ch2:AnalogueSuperp} are framed within the context of analogue gravity. In Chapter~\ref{Ch1:ChronologyAnalogue}, the question that is addressed is whether it is possible to find analogue systems such that the emergent geometry contains causal or chronological pathologies such as Closed Timelike Curves (CTCs). Although such spacetimes are kinematically allowed in GR, they are generally considered unphysical: they require exotic matter, i.e., matter that violates energy conditions, to exist, and are typically unstable under perturbations, a feature closely tied to their exotic matter content. Furthermore, at the semiclassical level the situation is even worse, since the quantum vacuum cannot be ``turned off'' and, generically, semiclassical gravity predicts a large backreaction from the quantum fields on these spacetimes that destabilizes them. In analogue gravity one is not limited by the gravitational dynamics, since the metric does not obey Einstein equations, as discussed before. In fact, to some extent that depends on the freedom encoded in the specific analogue model under consideration, it is possible to engineer a large class of geometries that from a GR point of view would not be physical, either because they are unstable or because they would require exotic matter to support them. Thus, it is specially interesting to analyze whether these pathologies can happen (or can be engineered) in analogue systems. In that way, this chapter analyzes what are the restrictions that the background causality of the analogue frame imposes on the effective causality that emerges in the system. Then, we analyze the lessons that can be drawn for beyond-GR theories. 

Analogue systems can be classified according to whether its substratum has a classical or a quantum nature. Examples of a classical substratum would be a classical fluid, but one can also think of quantum fluids as analogue systems. In particular, Bose-Einstein Condensates (BECs) have been proven to be extremely useful for analogue gravity experiments. A theory of quantum gravity allows, at least in principle, superpositions of quantum states. For instance, we can think of two states in the theory that are the closest notion of a classical smooth spacetime, in the same way that coherent states are the closest notion to classical states of light in quantum optics. Hence, if we have a quantum analogue system, it is sensible to ask the same question: Is it possible to superpose two states that generate two different analogue metrics? The in-depth analysis of this question makes Chapter~\ref{Ch2:AnalogueSuperp}. Our findings in these two chapters unveil the consequences that background structures have on the emergent geometries. We find that pushing the emergent causality of analogue systems to limits such as developing chronological pathologies or attempting to superpose two such causalities, conflicts with the causality settled by the background structure of the analogue. The system's dynamics prevent it from acquiring an ill-defined causality, instead steering it toward a well-defined and consistent causal structure. In particular, the systems under consideration here display a Lorentzian/Galilean background structure which is the one that forbids the presence of chronological pathologies and blurred causalities due to superpositions.

In fact, it is remarkable that very little attention has been paid to background structures in the context of gravitational phenomena since the birth of GR. As discussed above, GR gets rid of all background structure in its construction and, probably because of its success in explaining experiments and observations, most of the analyses of beyond-GR theories have kept this background independence as a guiding principle. Emergent gravity contexts are antagonistic to this point of view, given that an emergent theory of gravity would naturally retain some of the background structures of the underlying system, at least in some regimes. Actually, beyond aesthetic or philosophical reasoning, there is no direct observation or experiment that points toward the fact that background structures must be forbidden in gravitational theories. The simplicity in which causal pathologies are avoided due to the presence of such structures and the lack of evidence for such pathologies could even be used as an argument in their favor. It is also worth remarking the similarity between these analysis and the semiclassical gravity analyses which illustrate how the quantum vacuum also tries to avoid the formation of extreme causal situations. Thus, the following chapters analyze the role of background structures in gravitational theories with an special emphasis on the analysis of the minimum background structure that can be added to GR without substantially modifying the dynamics: a spacetime volume form.

For that purpose, Chapter~\ref{Ch3:SelfCoupling} is dedicated to an analysis of the structure of gravitational theories from the point of view of a self-interacting theory of a massless spin-2 field, i.e., a graviton field, propagating on top of the Minkowski spacetime. Particular attention is given to the role of background structures. For a long time, it has been part of the collective thinking that GR is the unique consistent theory of self-interacting gravitons. This question has a long story with different attempts to prove it, and it was supposed to be settled by Deser in the seventies~\cite{Deser1970}. However, Deser's proof evades many subtleties of the construction, and it actually hides the potential nonuniqueness of the construction, leading to some criticism years later~\cite{Padmanabhan2008,Butcher2009}. Furthermore, the classic analyses leave aside from the beginning the possibility that the theory resulting from the self-coupling contains some background structure. In this chapter, we revisit the nonuniqueness of the construction as well as the role that background structures play in the construction. Furthermore, given that the generalization to higher derivative theories containing additional degrees of freedom is straightforward with the tools we develop, and the previous literature is not entirely clear in the subject~\cite{Ortin2017,Deser2017}, we also present it in detail. The analysis reported here suggests that if a system develops graviton excitations, it is not straightforward to conclude that its dynamics will be that of GR or that background structures need to be absent of the resulting theory. In fact, Unimodular Gravity (UG) is an example of how one can construct a self-consistent interacting theory of gravitons that still displays a background structure. 

Chapter~\ref{Ch4:UG} closes this first part precisely with a systematic study of unimodular gravity theories, i.e., theories with a background volume form, the minimum background structure that can be added to GR without significantly changing the dynamics. A systematic comparison of such theory with GR (as well as their higher derivative generalizations) is carried on, both at the classical, semiclassical, and perturbative quantum level. Our conclusion is that both theories are completely equivalent in their phenomenological consequences, presenting only a difference at a theoretical point: the nature of the cosmological constant. Whereas in GR the cosmological constant is simply a coupling constant, in UG the cosmological constant arises as an integration constant, i.e. a global degree of freedom. As a result, these theories are structurally different and are not simply different gauge fixings of a single theory. The implications of this different nature for the so-called cosmological constant problem are discussed in detail.   
 
The second part of the thesis is devoted to the analysis of compact objects beyond GR. It is inspired by the results from the first part, which suggest that emergent frameworks avoid any situation of extreme causality. Hence, we take the absence of horizons and singularities as a starting point in this second part for our investigation. The second assumption that we make is that GR is valid to describe the gravitational physics at large distances when the configuration has relaxed to a stationary state. It is not always true that beyond-GR phenomena need to be linked with high curvature. For instance, semiclassical effects can potentially become huge at horizons, where the curvatures do not necessarily acquire huge values. Nevertheless, our expectation is that once an object has formed, significant deviations from the GR dynamics occur only in the interior where high curvatures appear. Whereas our main motivation for making these assumptions and studying horizonless compact objects are the results from the first part of the thesis, these analyses can be regarded as agnostic explorations of alternative to GR black holes on their own, without invoking any specific quantum gravity framework. In fact, going beyond the framework of GR also turns out to be useful for understanding GR better.

We first revisit aspects of the so-called no-hair theorems. These theorems constitute a collection of classical results in GR that constrain the form of stationary, asymptotically flat vacuum solutions of Einstein's equations with an event horizon~\cite{Chrusciel2012}. It is important to note, however, that their proofs rely on several mathematical assumptions and it is not clear whether eventually some of them will turn out to be unnecessary.

When restricted to static configurations, the relevant result is Israel's theorem, which imposes additional constraints on the metric. Specifically, the theorem establishes that the metric must also be spherically symmetric and, in fact, must correspond to the Schwarzschild solution. This conclusion hinges on a combination of local conditions, such as the regularity of the horizon, and global ones, such as the assumption of a vacuum spacetime and asymptotic flatness. The interplay between these local and global assumptions complicates the task of disentangling their individual contributions to the theorem’s proof.

Furthermore, these constraints limit the theorem’s applicability to astrophysical scenarios, where black holes are typically immersed in external gravitational fields (there is often matter in the external region) or to horizonless objects. Chapter~\ref{Ch5:Nohair} addresses these limitations by disentangling aspects of the no-hair theorem in the static and axisymmetric case, a simplified yet rich scenario. We present a version of the no-hair theorem that accommodates the presence of external gravitational fields and horizonless objects. This is done replacing the presence of a horizon by a surface that is close to being one (a maximum redshift surface).

Our results demonstrate that, if an event horizon is present, its geometry is uniquely determined by the external matter content. Specifically, for a given external matter configuration, there exists only one nonsingular horizon geometry consistent with Einstein's equations. It reduces to the spherically symmetric Schwarzschild geometry in the absence of external matter, recovering the Israel's theorem scenario. For horizonless objects, we show that as the maximum redshift surface is approached, the multipole structure of the configuration converges to that of a Schwarzschild black hole.

Chapter~\ref{Ch6:ToroidalBHs} is an immediate follow-up of the previous chapter. Static and axisymmetric black holes need to be topologically either spheres or tori. Whereas the spherical configuration can exist in vacuum, with the intrinsic geometry deviating from the spherical symmetry according to the external matter content, the toroidal configuration necessarily requires some kind of external matter content that supports it. In fact, it is required that such external matter be exotic, i.e., that it violates energy conditions. The tools that we developed in the previous chapter allow us to present, to the best of our knowledge, the first families of locally vacuum toroidal black holes in four-dimensional spacetimes that are nonsingular in the external region. They are simple and they allow for an exhaustive analysis of energy conditions violations. 

Finally, starting from the premise that the final state of any gravitational phenomenon must be horizonless, a natural question arises: what are astrophysical black holes then, and what is their true geometry? This question is particularly relevant given the classic results in GR that constrain the compactness (mass-to-radius ratio) of spherically symmetric objects under certain assumptions about the behavior of matter. Broadly, these assumptions involve various forms of energy conditions. Chapter~\ref{Ch7:Buchdahl} addresses this question by revisiting two foundational results in the literature that establish upper bounds on the compactness of objects composed of nonexotic matter: Buchdahl’s~\cite{Buchdahl1959} and Bondi’s theorems~\cite{Bondi1964}.

First, we conduct an in-depth analysis of the assumptions underlying these theorems, demonstrating how relaxing each assumption independently allows one to circumvent the compactness bounds. To illustrate this, we introduce several toy models that violate these assumptions in isolation making them valuable examples for understanding how compact can an object become when the traditional assumptions are relaxed, exploring the resulting implications for compactness. These models are particularly insightful, as some of them replicate many of the key properties of semiclassical stars proposed in~\cite{Arrechea2021} as black hole alternatives. 

The thesis concludes with a section that summarizes the main results, outlines potential directions for future research, and places the work within a broader context. Additionally, it provides critical reflections on the relevance of the findings and offers a personal perspective on open questions.

\subsection*{How to read this thesis} 
Each chapter in this thesis has been written to be self-contained, and some of them are accompanied by an appendix where some technical details are provided. This structure allows each chapter to be read independently of the others. Moreover, every chapter begins with an introductory section that presents the problem being addressed and concludes with a summary of the main results, along with a discussion of how these results fit into the broader context of the thesis. We believe that reading the general introduction and conclusion of the thesis, together with the individual introductions and conclusions of each chapter, provides a comprehensive understanding of the key findings, their interplay with the general framework of the thesis and how they relate to the existing literature.

\addtocontents{toc}{\protect\thispagestyle{empty}}

\renewcommand{\theequation}{$\mathscr{I}$.\arabic{equation}}

\setcounter{secnumdepth}{2}

\part{Background structures and gravitational phenomena}\label{pt1}
\addtocontents{toc}{\protect\thispagestyle{empty}}

\renewcommand{\theequation}{\thechapter.\arabic{equation}}

\fancyhead[LE,RO]{\thepage}
\fancyhead[LO,RE]{Chronology protection mechanisms in analogue gravity}

\chapter{Chronology protection mechanisms in analogue gravity}
\label{Ch1:ChronologyAnalogue}


It is by now well known that many systems akin to condensed matter systems, in the sense of being composed by a large amount of elementary building blocks such as atoms or molecules, exhibit a behavior in certain regimes which can be characterized by the presence of some effective fields, classical or quantum, propagating on top of an effectively curved Lorentzian geometry. These behaviors are collectively called \emph{analogue gravity}\mbox{~\cite{Barcelo2005,Volovik2009}}. In its broadest description, the analogue gravity program intends to obtain new insights into gravitational behaviors by analyzing their equivalent counterparts within these analogue frameworks. The reverse direction also forms part of the analogue gravity realm: acquiring new ideas about laboratory systems by importing gravitational notions and techniques. 

The metric that emerges in the analogue system and controls the causal behavior of the fields does not obey Einstein equations. In fact, analogue systems often result from the splitting of a system, e.g., a fluid, into a background, such as the background density and fluid velocity, and some perturbations propagating on top of it, the sound waves. The components of the Lorentzian metric that the perturbations perceive are typically functions of these background specific properties and obey other kind of equations, e.g., Euler equations in the case of fluids. In that sense, given that the metric is not constrained through the gravitational dynamics, it is in principle possible to explore classes of geometries that would never happen in realistic situations since the gravitational dynamics forbids them. These geometries may not occur ``naturally'' either in the analogue systems, since the backgrounds variables do not tend to them without any external agent. However, they can be controlled experimentally, and as such, it is possible to think of simulating them in laboratory setups. In the case of fluids, for example, one can apply external forces to the system in order to engineer a desired metric. 

This chapter is devoted to the study of whether chronologically pathological spacetimes, for instance those containing CTCs, can be engineered in analogue systems. Although for some fluid analogue systems there exist results in the literature suggesting that chronological pathologies cannot arise~\cite{Visser1997b}, they do not dig into how the system forbids them from the dynamical point of view of the analogue system. The chapter is largely based on Ref.~\cite{Barcelo2022}, which is one of the articles published during this thesis.

\section{Analogue systems: the general picture}
\label{Sec:Analogues}

Analogue systems are systems whose equations of motion in some regimes correspond to relativistic fields propagating on top of an effectively curved metric. There are plenty of different analogue systems, from sound waves propagating on top of a fluid background, to light propagating inside an inhomogeneous dielectric medium characterized by some specific permittivity and permeability tensors. The most fundamental distinction that we can make among analogue systems is between classical and quantum analogues, though further refinements can be made. For many of the practical purposes, the classical or quantum origin is only relevant if one is examining phenomena close to the vacuum state of the emergent quantum fields. In particular, this is specially relevant to disentangle whether phenomena such as Hawking radiation or superradiance correspond to spontaneous or stimulated emissions. The classical or quantumness of the substratum for the analogue system will only play a relevant role in the next chapter. For the sake of concreteness, in this chapter and the following we are going to focus on sound waves propagating on top of fluids, although the discussion about chronology protection in analogue systems of Section~\ref{Sec:Chronology} will be quite generic. Furthermore, we will also exemplify the discussion with other analogue models. 

Let us consider an inviscid fluid, i.e., a fluid with a negligible viscosity, which is described by its velocity $\bm{v}$, its density $\rho$, and its pressure $p$. The dynamics of the system is governed by the continuity equation together with Euler equations:
\begin{align}
    & \partial_t \rho + \nabla \cdot \left( \rho \bm{v} \right) = 0, \\
    & \rho \left[ \partial_t \bm{v} + \left( \bm{v} \cdot \nabla \right) \bm{v} \right] = \bm{\text{\textflorin}}, 
\end{align}
where $\bm{\text{\textflorin}}$ represents the external force applied to the fluid. The system is closed once we consider a barotropic equation of state relating the density $\rho$ and the pressure $p$. Let us also assume that the fluid is irrotational and then the velocity field can be written as the gradient of a potential function $\phi$:
\begin{align}
    \bm{v} = - \nabla \phi.  
\end{align}
If we perturb the fluid with respect to a background value, i.e., we take $ \rho = \rho_0 + \delta \rho$, $p = p_0 + \delta p$, $\phi = \phi_0  + \delta \phi$, and linearize the equations, we find that $\delta p$ and $\delta \rho$ can be obtained algebraically from $\delta \phi$\footnote{A detailed derivation is presented in Sec.~2.3 of~\cite{Barcelo2005}.}:
\begin{align}
    & \delta p = \rho_0 \left[  \partial_t \delta \phi + \bm{v} \nabla \delta \phi \right] ,\\
    & \delta \rho = c^{-2} \delta p, 
\end{align}
where $c^{-2} = \frac{\partial \rho}{\partial p}$ represents the local speed of sound. The perturbation~$\delta \phi$ obeys a massless Klein-Gordon equation
\begin{align}
    \frac{1}{\sqrt{-g}} \partial_{\mu} \left( \sqrt{-g} g^{\mu \nu} \partial_{\nu} \delta \phi \right) = 0,
\end{align}
where the effective metric entering the equation can be expressed in terms of the background values $(\rho_0, p_0, \phi_0)$ as
\begin{equation}
    g^{\mu\nu}=\frac{1}{\rho_0 c} \left(
    \begin{array}{c|c}
    -1 & -v_0^i \\
    \hline \\
    -v_0^j & c^2 \delta^{ij}-v_0^i v_0^j
    \end{array}
    \right),
\label{Eq:Acoustic_Inverse_Metric}
\end{equation}
and $ v_0^i= - \partial_i \phi $. Inverting it, we get the acoustic metric $g_{\mu \nu}$ which provides the sound cones at each point of the spacetime:
\begin{equation}
    g_{\mu\nu}=\frac{\rho_0}{ c} \left(
    \begin{array}{c|c}
    - \left( c^2 - v_0^2 \right) & -v^i_0 \\
    \hline \\
    -v^j_0 & \delta_{ij} 
    \end{array}
    \right).
\label{Eq:AcousticMetric}
\end{equation}
Similar arguments can be applied to a BEC. In this case, we  find out that the perturbations propagating on top of the condensed wave function obey a Klein-Gordon equation. To put it explicitly, consider a many-body system of bosons described through a quantum field $\widehat{\Psi}$, whose dynamics is encoded in the Gross-Pitaevskii equation~\cite{Ketterle1999,Leggett2006}
\begin{align}
    i \frac{\partial}{\partial t} \widehat{\Psi} = \left(- \frac{1}{2 m} \nabla^2 + V (t,\Vec{x}) + g \widehat{\Psi}^{\dagger} \widehat{\Psi} \right) \widehat{\Psi},
\end{align}
with $g$ being the leading order four-body interaction among bosons and $V (t,\Vec{x})$ the external potential. At sufficiently low temperatures, these systems of weakly interacting bosons can develop condensation in the sense that a macroscopic number of particles populates the same state~\cite{Leggett2006}. To describe this condensation, it is customary to split the field into a macroscopic condensate and a fluctuation as $\widehat{\Psi} = \psi + \widehat{\delta \psi}$, where the condensation is expressed as $\langle \widehat{\Psi} \rangle = \psi$. Introducing the Madelung representation
\begin{align}
    \psi = \sqrt{n_c}  e^{- i \theta},
\end{align}
and considering a long wavelength approximation one can describe the perturbations in the phase $ \widehat{\delta \theta}$ as quantum acoustic perturbations obeying a Klein-Gordon equation in a curved background metric of the form~\eqref{Eq:Acoustic_Inverse_Metric}. The velocity of the background fluid is $\bm{v} = \nabla \theta  / m$ associated with the phase $\theta$ of the wave function $\psi$, and the background density is given by $\rho_0 = m n_c$, where $c$ represents the speed of the phonons in the medium 
\begin{align}
    c^2 = \frac{g n_c}{m}. 
\end{align}
As mentioned above, for this picture to remain consistent, we must require that the wavelengths of the sound excitations are sufficiently large. Specifically, they must be much larger than the so-called healing length of the condensate $\xi = 1/ (m c)$, i.e., \mbox{$\lambda \gg 2 \pi \xi$}. The healing length represents the characteristic scale at which the hydrodynamic approximation breaks down, revealing the microscopic granular structure of the system.
 
In more general terms, it is possible to consider BECs that display some kind of anisotropies, see~\cite{Barcelo2000} for a dedicated discussion. This is achieved by permitting the mass to become a 3-tensor: $m \to m_{ij}$. In fact, it is even possible to consider situations in which the tensor becomes space and time dependent, i.e., we have $m_{ij} (t, \vec{x})$. In this case, it is customary to introduce a scale $\mu$ carrying the mass dimensions and express the mass matrix as
\begin{align}
    m_{ij}=\mu h_{ij},
\end{align}
with $h_{ij}$ a dimensionless 3-metric encoding the anisotropic behavior. Although the appearance of an arbitrary dimensionful parameter $\mu$ might seem worrying, at the end of the day it is possible to see that it disappears from the computation of every observable~\cite{Barcelo2000}. Working out the hydrodynamic limit, we find that in the semiclassical limit, the phase obeys again a Klein-Gordon equation, but in this case we have the following line-element for the metric
\begin{equation}
    g^{\mu\nu}=\frac{g}{c^3} \left(
    \begin{array}{c|c}
    -1 & -v_0^i \\
    \hline \\
    -v_0^j & c^2 h^{ij}-v_0^i v_0^j
    \end{array}
    \right),
\label{Eq:Acoustic_Inverse_Metric_Anisotropic}
\end{equation}
where $c$ can formally be defined as the speed of sound of the fluid and it is given by
\begin{align}
    c^2 = \frac{g n_c }{\mu},
\end{align}
although the anisotropies make the actual speed of sound direction-dependent in this case. 

\section{A class of geometries amenable to simulation}
\label{Sec:Geometries}
In this section, we present some geometries containing CTCs. Most of them can be found somehow sparse throughout the literature. Here we revise them and present them in a unified way, so that they are amenable to be analyzed from the analogue gravity perspective. First, we describe the geometry that results from combining two warp drives to generate CTCs. Motivated by the properties of this geometry, we will introduce a family of spacetimes which are simpler but encapsulate their main geometric features. Finally, we will discuss probably the most paradigmatic example of spacetime containing CTCs: the so-called Misner spacetime. This spacetime is used as a proxy to more convoluted analyses, since the properties of its chronological horizon are considered to be generic. Although for most practical purposes this is true, we will put special emphasis here on the fact that Misner spacetime is topologically nontrivial as every spacetime containing CTCs in 1+1. In higher spacetime dimensions, it is possible to have spacetimes with CTCs exhibiting a trivial topology though, as we illustrate with specific examples. 

\subsection{CTCs engineered through warp-drive bubbles}
\label{Subsec:CTC-warp drives}
Warp drives were originally introduced by Alcubierre~\cite{Alcubierre1994}. They are based on perturbing a given spacetime within a compact region so that observers inside appear to move at superluminal speeds from the perspective of observers outside. They can be thought as ``tachyonic'' bubbles in spacetime that propagate faster than light. The simplest metrics representing warp drives can be written using N\'atario's line element~\cite{Natario2001}:
\begin{equation}
    \dd s^2 = - \dd t^2 + \delta_{ij} c^{-2} \left( \dd x^{i} - v^{i} \dd t\right) \left( \dd x^{j} - v^{j} \dd t\right).
    \label{Eq:Natario}
\end{equation}
It is possible to conceive warp drives with nonzero lapse function, see footnote 1 of~\cite{Alcubierre2017}, or with a non-Euclidean metric in the spatial sector, see~\cite{VanDenBroeck1999} for a conformally flat metric in the spatial sector, although we will not focus on them here for the sake of simplicity. In fact, the three most important examples of warp drives fall in this category: the original Alcubierre warp drive, which corresponds to an arbitrary element of N\'atario's line element~\cite{Alcubierre1994}, the zero-expansion warp drive of N\'atario which simply sets $\nabla \cdot \bm{v} = 0$, and the Lentz/Fell-Heisenberg~\cite{Lentz2020,Fell2021} zero-vorticity warp drive which uses a purely gradient flow, i.e., $\bm{v} = \nabla \phi$ so that $\nabla \times \bm{v} = 0 $. 

In this way, the metric of a warp drive is nicely adapted to be simulated with an acoustic metric, namely the one in Eq.~\eqref{Eq:AcousticMetric}. Essentially, we need to identify the velocity of the fluid with the shift functions entering the warp drive element. For concreteness, we can think of a warp-drive bubble moving in the $x$-direction whose profile acquires the following form
\begin{equation}
    v^{i} (t, \bm{x}) =  u (t) f \left( \sqrt{ \left(x-x(t)\right)^2 + y^2 + z^2 } \right) \delta^{i}_{\ x},
    \label{Eq:Velocity}
\end{equation}
where $f(x)$ is a function with compact support, describing the profile of the bubble which is peaked around the points of the trajectory 
\begin{equation}
    x(t) = x(0) + \int^t_0 \dd t' u(t'), 
    \qquad 
    y=z=0,
\end{equation}
and $u(t) = \dot{x} (t)$ represents the velocity of the bubble. 

\begin{figure}
\begin{center}
\includegraphics[width=0.45 \textwidth]{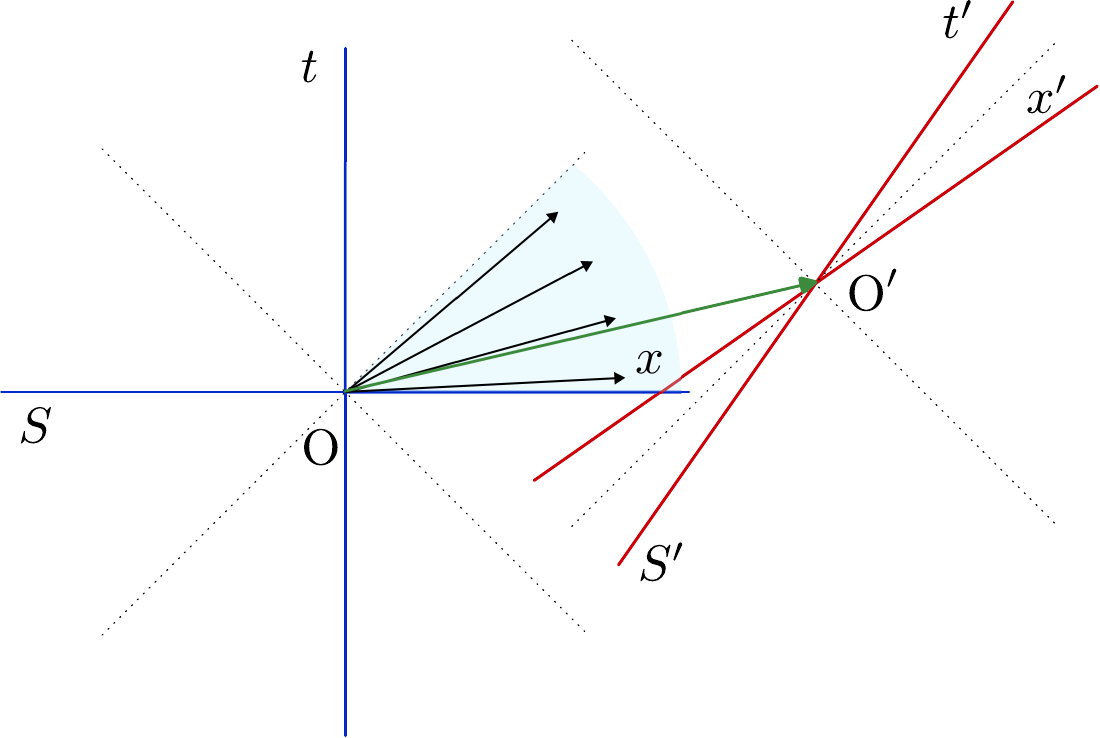}
\includegraphics[width=0.45 \textwidth]{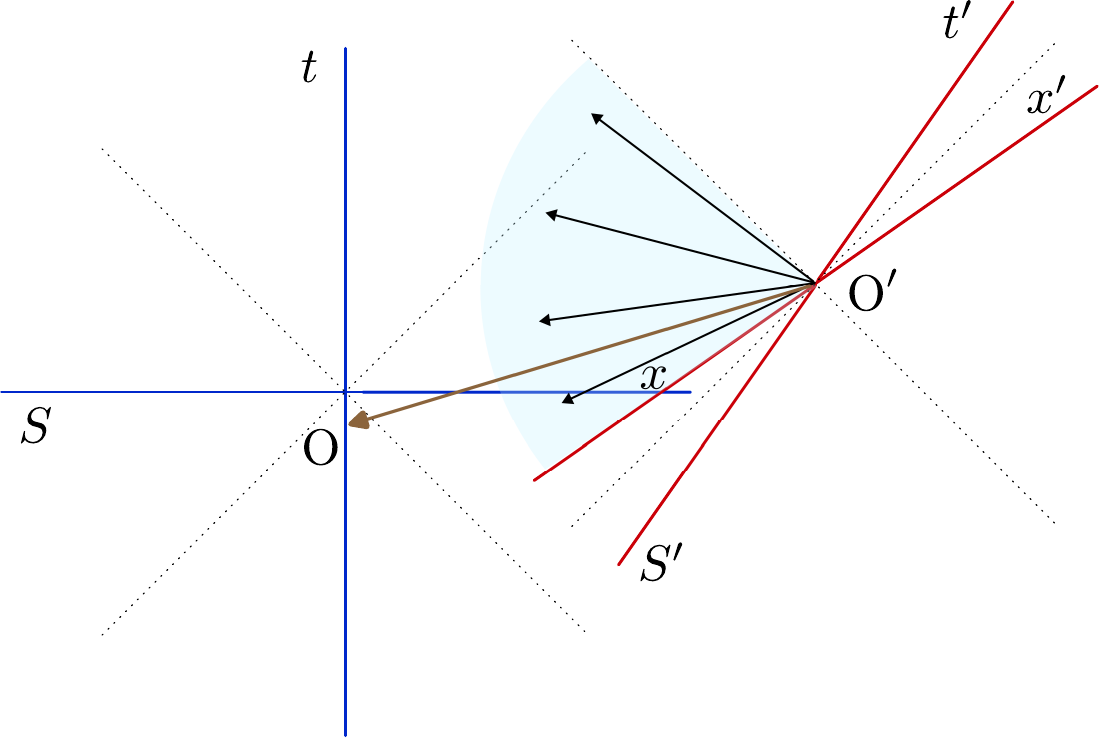} 
\includegraphics[width=0.45 \textwidth]{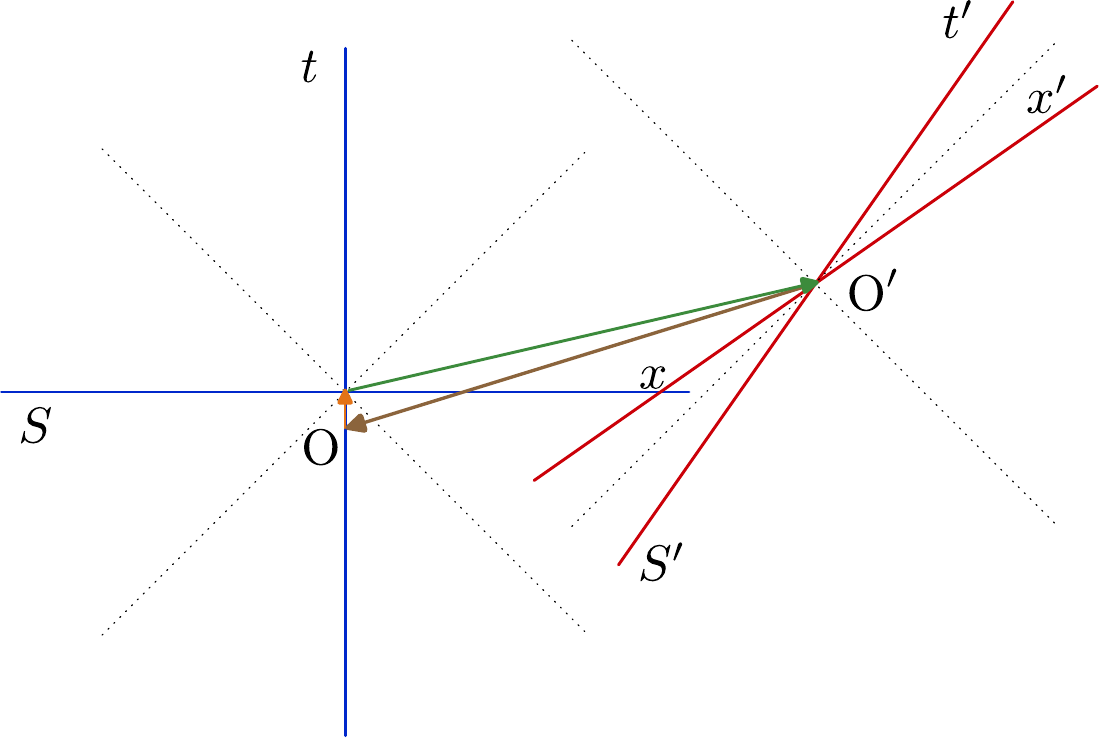}
\caption{The three figures above illustrate two inertial frames, $S$ (blue) and $S^{\prime}$ (red), with their respective origins at $\text{O}$ at $(t=0,x=0)$ and $\text{O}^{\prime}$ at $(t^{\prime}=0,x^{\prime}=0)$. The frame $S^{\prime}$ is translated and boosted with respect to $S$, as the tilt of their axes due to a hyperbolic transformation represents. The dotted lines at 45 degrees represent the light cones. If superluminal signals are allowed, the only restriction is that they must be future-oriented from the emitter’s perspective, even if they travel outside the light cones. This is shown in the upper left panel, where the shaded angular region represents the range of superluminal signals that can be sent from $\text{O}$ to the right. Notably, a signal can be sent to an observer passing through $\text{O}^{\prime}$, despite $\text{O}^{\prime}$ being spacelike separated from $\text{O}$. Similarly, in the upper right panel, the shaded region illustrates the possible superluminal signals that can be sent from $\text{O}^{\prime}$ to the left. The brown arrow represents a specific signal that ultimately reaches the past of $\text{O}$. The lower panel demonstrates how these two superluminal signals, combined with a third timelike signal depicted in orange, can create a causal paradox. In this setup, a signal originating from $\text{O}$ can be arranged to return to the past of $\text{O}$, i.e., it can reach $x = 0$  for $t<0$. In other words, an observer can receive a reply to a message before even sending the original message. This construction highlights how, in Lorentz-invariant theories, tachyons (superluminal particles) can be used to transmit signals into the past, leading to violations of causality. }
\label{Fig:Tachyons}
\end{center}
\end{figure}

The simulation of a single warp drive in a acoustic analogue system is straightforward~\cite{Barcelo2005}. One just needs to generate a region at which the velocity of the flow exceeds the speed of sound. From the internal perspective, this allows one to travel from one point of the fluid to another at velocities higher than the speed of sound. It is convenient to rewrite the warp drive metric in Eq.~\eqref{Eq:Natario} as a deformation of the flat spacetime metric $\eta_{\mu \nu}$ as
\begin{equation}
    g_{\mu \nu} = \eta_{\mu \nu} + b_{\mu \nu}, 
    \label{Eq:Natario2}
\end{equation}
with $b_{\mu \nu}$ having the following nonvanishing components:
\begin{align}
    & b_{00}  = u^2 (t) f^2\left( \sqrt{ \left(x-x(t)\right)^2 + y^2 + z^2 }\right), \nonumber \\
    & b_{01} = b_{10} = - u (t) f \left( \sqrt{ \left(x-x(t)\right)^2 + y^2 + z^2 }\right).
\end{align}
Now, although a single warp drive bubble by itself does not result in any chronology or causality violations, as already noticed in~\cite{Everett1995}, it is relatively easy to engineer a spacetime that contains CTCs by taking advantage of having two warp-drive bubbles. The idea is similar to the way in which one can send information to the past with a pair of tachyon particles in flat spacetime~\cite{Rolnick1969}, see Fig.~\ref{Fig:Tachyons} together with the description in the caption. What is then the clash, if any, between warp drive metrics and their analogue simulations? 
%

\begin{figure}
\begin{center}
\includegraphics[width=0.5 \textwidth]{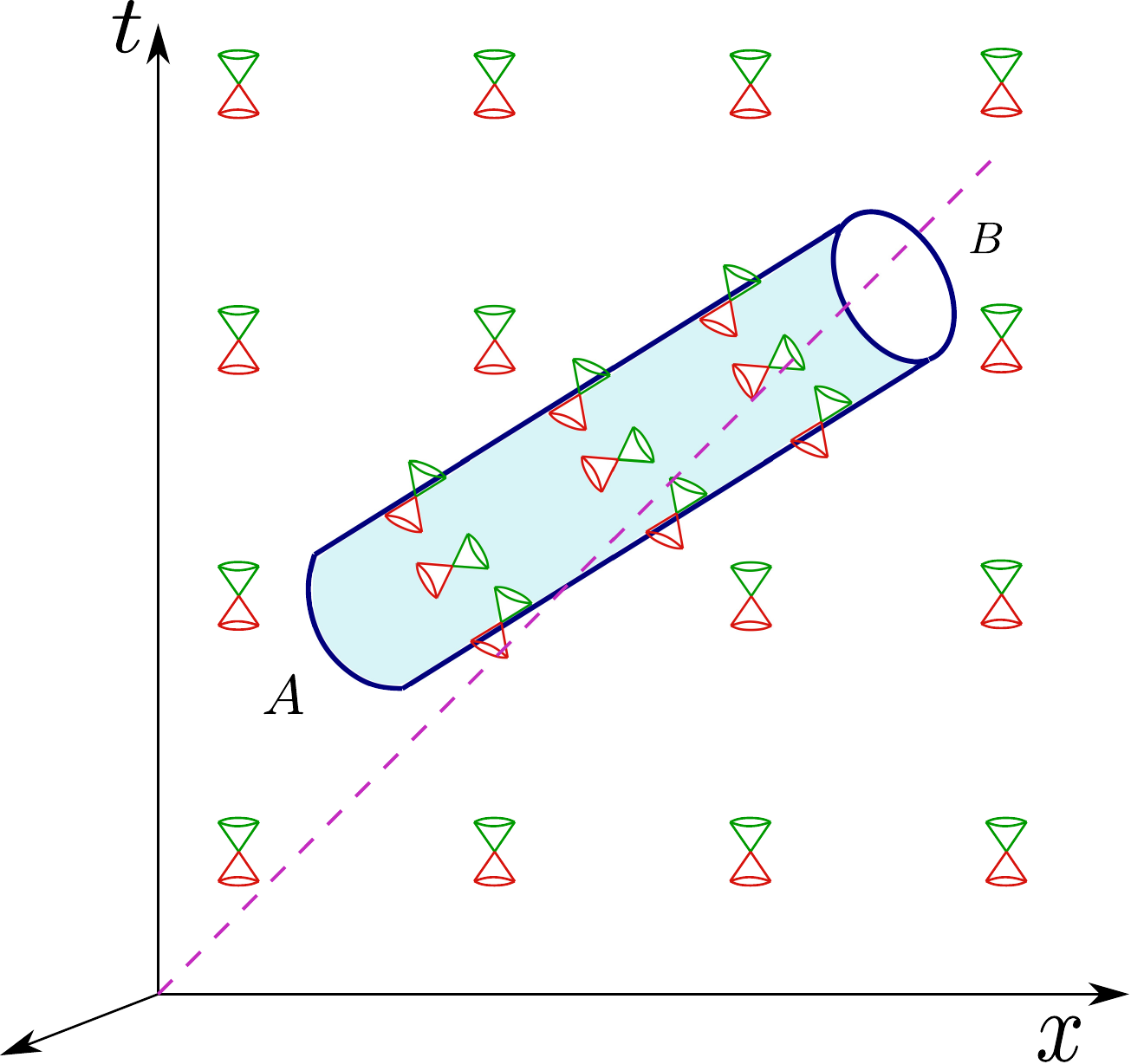}
\caption{The figure represents a warp drive tube that starts in a location A and ends in location B. The underlying spacetime can be considered $(D+1)$-dimensional with the warp-drive bubble moving in a straight line (along the $x$-coordinate, in the picture). This is the simplest construction of a warp drive and it can be simulated in an analogue gravity model without further problems.}
\label{Fig:warpdrive-forward}
\end{center}
\end{figure}

Let us begin with the most naive way in which one might try to make this configuration. Let us focus on the two-dimensional plane of a $D+1$ Minkowski spacetime, that we parametrize with coordinates $(t,x)$ of an inertial reference frame. Furthermore, let us choose two events $A$ and $B$ such that they are spacelike separated, with coordinates $(t_A,x_A)$ and $(t_B,x_B)$, respectively. Without loss of generality, let us assume that $t_B>t_A$. This setup is represented pictorially in Fig.~\ref{Fig:warpdrive-forward}. We can engineer a warp drive tube connecting the two events  (and their surrounding neighborhoods). Notice that the light cones inside the tube are modified with respect to the Minkowskian reference. The tube itself has thick walls where the light cones experience a tilting effect. It is precisely on those walls where the stress energy tensor supporting these configurations necessarily develops some energy conditions violations~\cite{Alcubierre1994}. From the perspective of an external observer, the trajectory connecting $A$ to $B$ along the warp drive appears spacelike. However, for observers within the bubble, the motion remains strictly timelike, as they continue to move forward in time. 

\begin{figure}
\begin{center}
\includegraphics[width=0.5 \textwidth]{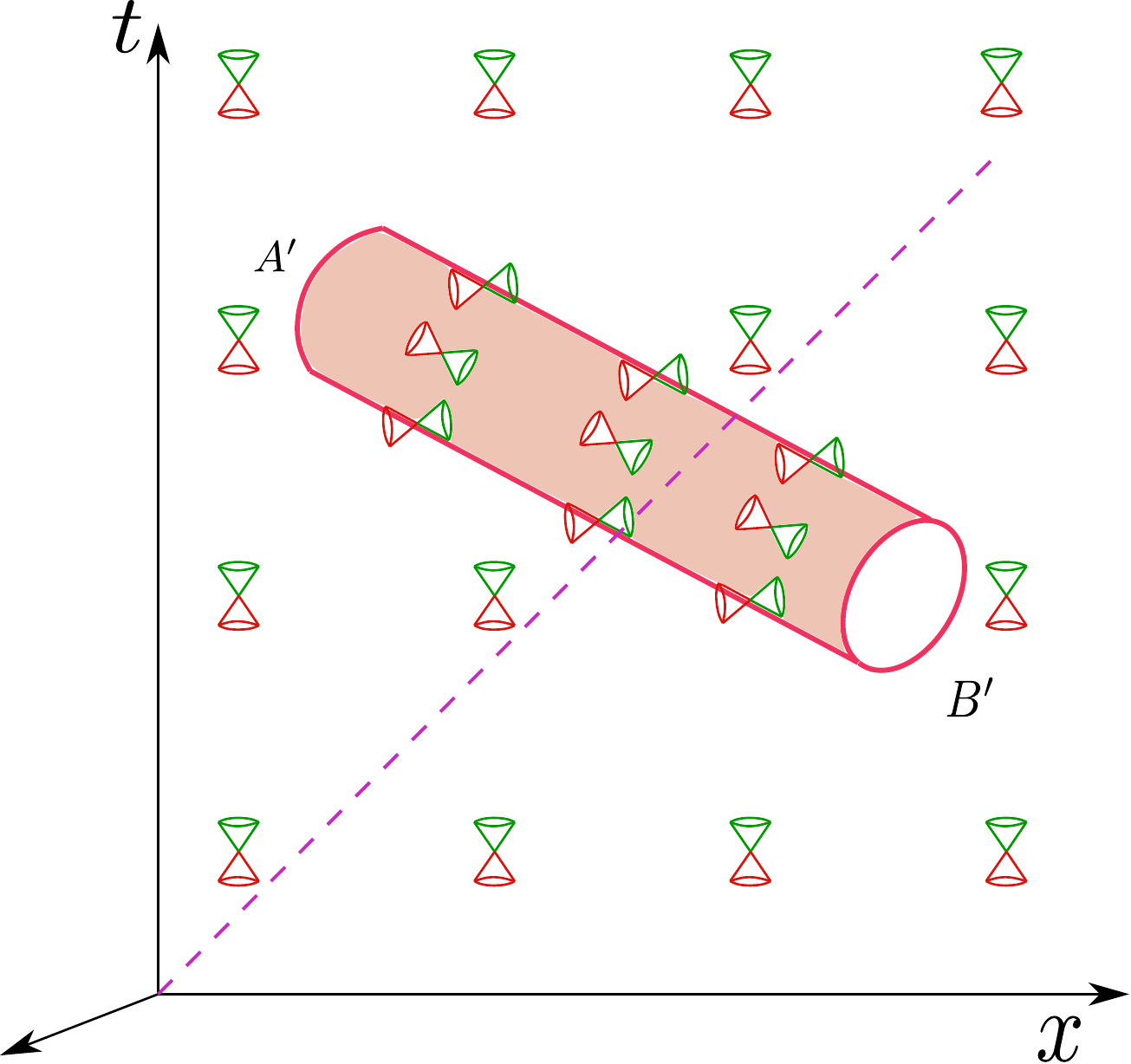}
\caption{The figure represents a warp drive tube that starts in a location $\text{A}^{\prime}$ and ends in location $B^{\prime}$. The underlying spacetime can be considered $(D+1)$-dimensional with the warp-drive bubble moving in a straight line (along the $x$-coordinate, in the picture). }
\label{Fig:warpdrive-backward}
\end{center}
\end{figure}

If such a warp drive can be constructed, then from a purely general relativistic perspective, it is also possible to create an equivalent configuration in which the coordinate time flows to the past instead of the future~\cite{Everett1995}. Let us write down such a metric explicitly. We start constructing a warp drive metric as the one just described but using another inertial reference frame $S'$ moving with velocity $v$ in the $x$ direction. Let us denote with a prime the cartesian coordinates of the reference frame $S'$, i.e., $(t',x')$. In these coordinates, the metric of the bubble takes the simple form $g'_{\mu \nu}$ of Eq.~\eqref{Eq:Natario}. To find the metric in the coordinates $(t,x)$ adapted to the inertial frame $S$, we simply need to perform a boost of velocity $-v$ in the $x$-axis, with $v$ the relative velocity between $S$ and $S'$. In this way, the resulting metric reads  
\begin{equation}
    g_{\mu \nu} = \eta_{\mu \nu} + c_{\mu \nu}, 
    \label{Eq:backwards_bubble}
\end{equation}
where $c_{\mu \nu}$ is transformed from the prime coordinates to the unprimed ones by an ordinary Lorentz transformation. We highlight that its functional form differs from the $b_{\mu \nu}$ tensor introduced in Eq.~\eqref{Eq:Natario2}. Actually, we can find its functional form explicitly, and it is described by the following linear transformation: 
\begin{equation}
  \left( \Lambda \right)^{\mu} _{\ \nu}  = \left(
  \begin{array}{cc} 
  \cosh \phi \ &  \sinh \phi  \\  
  \sinh \phi \ & \cosh \phi 
  \end{array}
  \right), \qquad \textrm{with} \qquad  \tanh \phi = v,
\end{equation}
Writing down the transformation we obtain the following $c_{\mu \nu}$ tensor
\begin{align}
    & c_{00} = u^2(t'(t,x),x'(t,x),y-y_0,z) \cosh^2 \phi - 2 u (t'(t,x),x'(t,x),y-y_0,z) \cosh \phi \sinh \phi , \\
    & c_{01} = c_{10} =  u^2(t'(t,x),x'(t,x),y-y_0,z) \cosh \phi \sinh \phi \nonumber \\
    & - u (t'(t,x),x'(t,x),y-y_0,z) \cosh^2 \phi -  u (t'(t,x),x'(t,x),y-y_0,z) \sinh^2 \phi, \\
    & c_{11} = u^2(t'(t,x),x'(t,x),y-y_0,z) \sinh^2 \phi - 2 u^2(t'(t,x),x'(t,x),y-y_0,z) \sinh \phi \cosh \phi.     
\end{align}
Notice that the functions $t'$ and $x'$ depend on the coordinates $(t,x)$ in a non trivial manner. In fact, the coordinates $ \{ x^{\mu} \}$ are related to the coordinates $\{ x^{\prime  \mu} \}$ through a Lorentz transformation from $S$ to $S'$ in which the Lorentz matrix is precisely $\Lambda^{\mu}_{\ \nu}$. Explicitly, the change of coordinates reads
\begin{align}
    & t' = t \cosh \phi  - x \sinh \phi, \\
    & x' = t \sinh \phi  + x \cosh \phi .
\end{align}
We just note that from these expressions one only needs to choose a profile for the bubble and the velocities to explicitly express the metric in global coordinates, since writing everything explicitly without specifying them would not be very illustrative. In general terms, we have constructed a warp drive traveling backward in coordinate time $t$, as depicted in Fig.~\ref{Fig:warpdrive-backward}. However, it is easy to verify that the Lorentz transformation we applied causes the new warp drive metric to differ from Natário's line element. For the arguments that follow, the precise form of the tubes will not be relevant.

\begin{figure}
\begin{center}
\includegraphics[width=0.5 \textwidth]{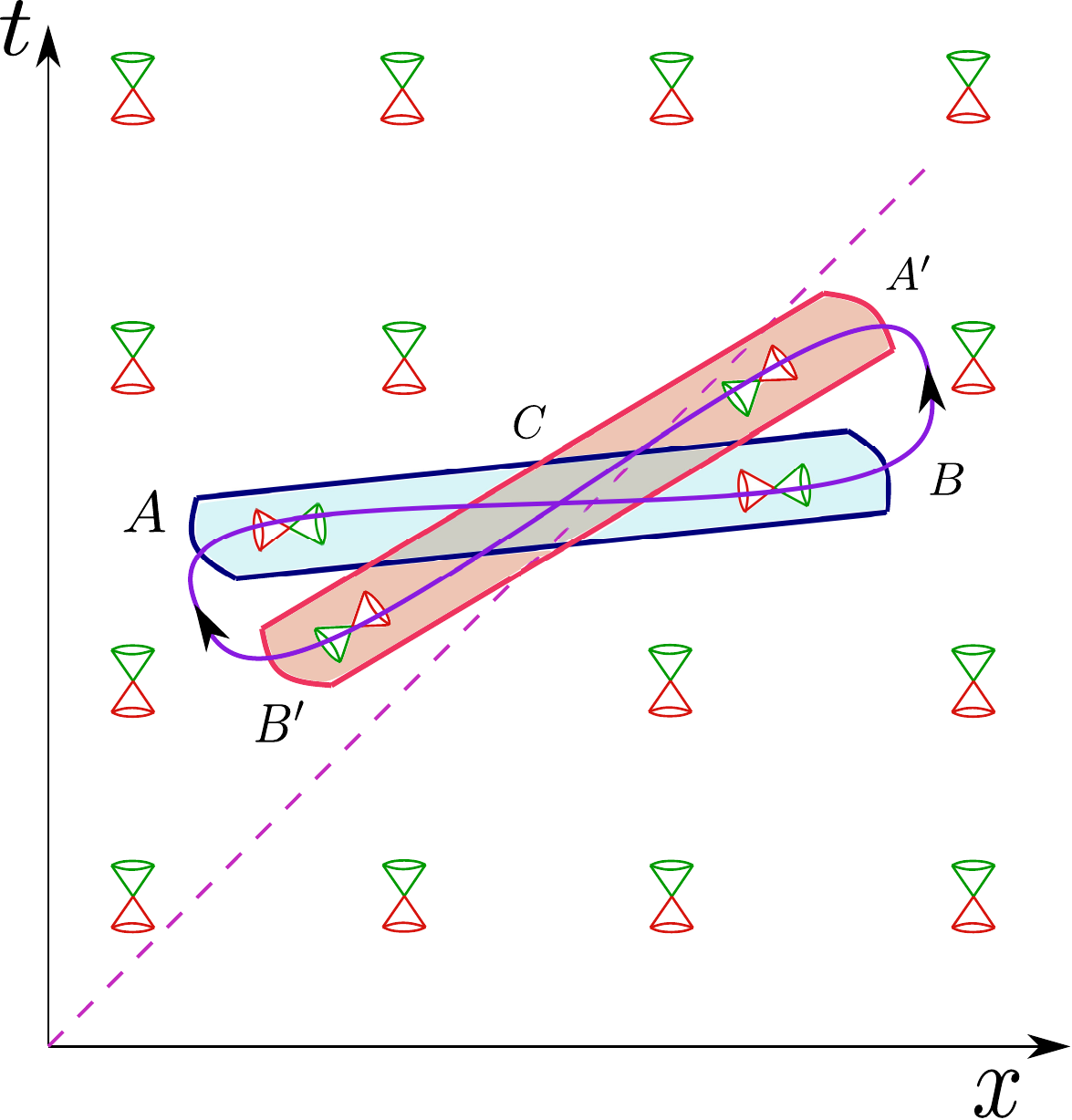}
\caption{We represent here the setup of two warp drives engineered to allow the presence of CTCs. The purple curve represents a generic CTC on this background. The problem with this configuration $1+1$ spacetime dimensions is that the metric in the region where the two warp-drive bubbles cross is ill-defined. In higher spacetime dimensions this problem is absent, we just need to engineer the two warp drives bubbles in different parallel planes to avoid the crossing.}
\label{Fig:warpdrive-crossing}
\end{center}
\end{figure}

By combining a ``forward" and a ``backward'' warp-drive bubble, one can attempt to construct a time machine, following the same logic described earlier for tachyons. First, we set up a ``forward" warp drive that allows faster-than-light travel from A to B. Upon exiting the bubble at B, the traveler could immediately enter a new warp drive, now of a ``backward'' type, and travel from $A'$ to $B'$, through another spacelike trajectory as seen from the external Minkowski spacetime. This second warp drive can take the time-traveler to an event $B'$ in the past of the initial event $A$. In this way a CTC is completed. This setup is pictorially represented in Fig.~\ref{Fig:warpdrive-crossing}. 

However, there is a fundamental issue with this construction in $1+1$ spacetime dimensions, as can be seen immediately from the figure. Specifically, at the crossing region $C$, the setup would require two different metrics simultaneously. This results in a region where the metric is not well-defined. Naturally, one might ask whether it is possible to resolve this issue by adjusting the starting and ending points of the bubbles, modifying their shapes, or fine-tuning their velocity profiles $v_1(t),v_2(t)$ to construct a completely regular Lorentzian metric that still allows for CTCs. The answer to this question is negative.

\begin{figure}
\begin{center}
\includegraphics[width=0.3\textwidth]{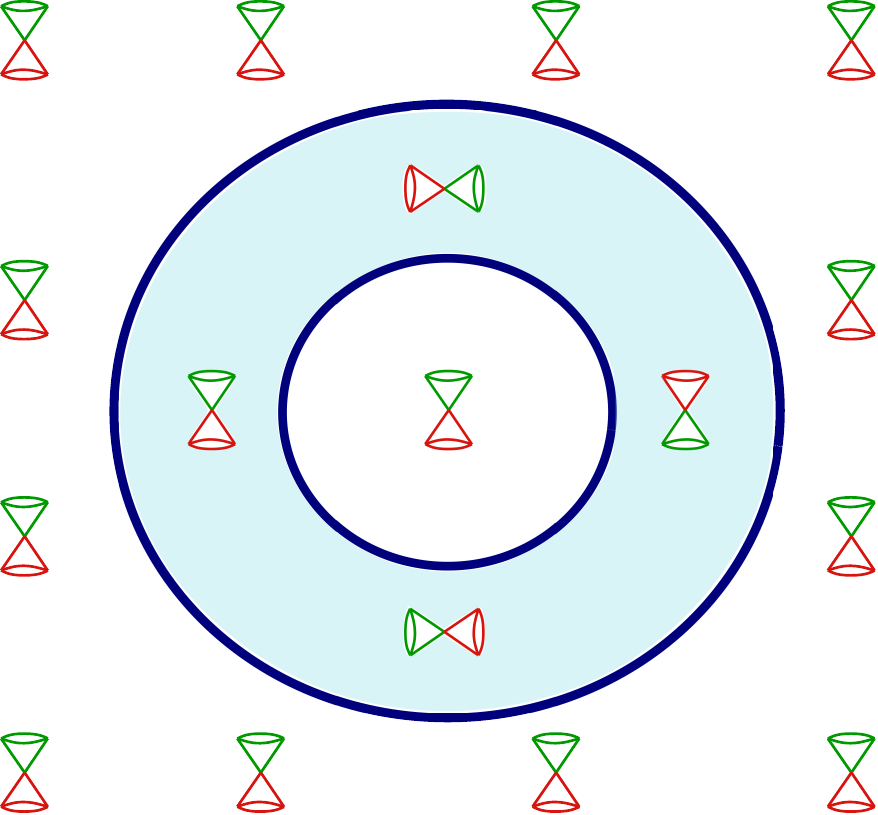}
\caption{We represent here the setup described in the text that already shows the difficulty present when trying to build CTCs in a spacetime with a trivial topology. The shaded region represents the region of abnormal behavior of the light cones. It is impossible to regularize the light cones without removing points from the spacetime, otherwise the metric would need to vanish at some point.}
\label{Fig:Toy_example}
\end{center}
\end{figure}

To understand why, it is useful to consider a toy geometry that effectively illustrates the obstruction. Specifically, imagine a spacetime that remains flat everywhere except along a circle of radius one around the origin in a given set of inertial coordinates $(t,x)$. At such circle, we make the light cones to make an angle of $45$ degrees with the tangent to the circle at every point. Clearly such geometry is singular since there is a jump in the metric. Even if we try to smooth such geometry by giving the circle a finite size and converting it into a disk, we can find a curve along which the light cones make a rotation of $360$ degrees and, hence, it is impossible to have a smooth metric in the region enclosed by such curve or regularize it in some way. This is depicted schematically in Fig.~\ref{Fig:Toy_example}.

This observation suggests that smooth two-dimensional metrics with the topology of $\mathbb{R}^2$ cannot exhibit CTCs. This aligns with existing literature, as all known smooth spacetimes that are pathological from a causal perspective in $1+1$ spacetime dimensions have a nontrivial topology. For instance, Misner's spacetime~\cite{Misner1967} (see Subsection~\ref{Subsec:Misner}), which serves as a model for spacetimes with chronological horizons, has the topology of a cylinder $\mathbb{R} \times \mathbb{S}^1$. This statement and actually even stronger constraints on the causality of two-dimensional spacetimes with trivial topology can be established.  

In Lorentzian geometry there exist a hierarchy of causality conditions such that each of them is stronger than the previous ones. The weakest of these conditions are the \emph{chronological condition} and the \emph{causality condition}. A spacetime is said to satisfy the \emph{chronological condition} if it does not contain any closed timelike curves, and it is said to obey the \emph{causality condition} if there are no closed nonspacelike curves.

Although the chronology condition is the weakest of such conditions, followed by the causality condition, and they are enough to rule out closed nonspacelike curves, one can still think of spacetimes that are ``arbitrarily'' close to containing closed causal curves. The \emph{strong causality condition} and the \emph{stable causality condition} attempt to formalize these notion of ``almost having closed curves''~\cite{Wald1984,Hawking1973}. The \emph{strong causality condition} is obeyed by a spacetime if for every point $p$ and every neighborhood $N$ of $p$, there exists a neighborhood $O$ contained in $N$ such that no causal curve intersects $O$ more than once. Finally, a spacetime is said to be \emph{stably causal} if there exists a timelike vector field $t^\mu$ such that the Lorentz metric defined as $\Tilde{g}_{\mu \nu} = g_{\mu \nu} - t_{\mu} t_{\nu}$ (which has larger light cones than $g_{\mu \nu}$ at every point) contains no closed causal curves. It can be proved that a spacetime is stably causal if and only if it admits a globally defined time-function, i.e., a function that is strictly increasing along each future directed causal curve~\cite{Hawking1973}. 

The following theorem which applies to two-dimensional spacetimes, formalizes the intuition described above that smooth two-dimensional metrics with the topology of $\mathbb{R}^2$ cannot display CTCs. 

\emph{Theorem:} Let $(\mathcal{M},\boldsymbol{g})$ be a two-dimensional simply connected spacetime (being $\mathcal{M}$ the smooth manifold and $\boldsymbol{g}$ its metric). Then, the chronology condition automatically holds. 

The idea of the proof is already contained in the observation that we have made above: having the structure of light cones enclosing a compact region, it is impossible to push them inwards or outwards that region without making them zero or singular. The formalization of this statement can be found in~\cite{Oneill1983}. Actually, it is possible to even prove a stronger result. Theorem 3.43 of~\cite{Beem1996} shows that every simply connected two-dimensional spacetime $(\mathcal{M},\boldsymbol{g})$ is indeed stably causal. In fact, two-dimensional spacetimes display further special properties from the causal point of view, as detailed in footnote 48 of~\cite{SanchezCaja2021}.

This concludes our discussion of two-dimensional spacetimes and warp drives. We have demonstrated that constructing CTCs while preserving the $\mathbb{R}^2$ topology is impossible. In $D+1$ dimensions the previous obstruction is no longer present. We can perfectly avoid the warp drive crossing by translating one of the tubes in an orthogonal direction: the forward and backward warp drives can be set up to live in different parallel planes. In fact, they do not need to be confined entirely to a plane, we just need to ensure that they do not intersect. 

\subsection{Generalized warp-drive regions}
\label{Subsec:Extended-Warpdrives}

As discussed above, a complete metric containing CTCs based on warp drives can be designed by just translating in a transverse direction one of the two warp drives. However, the explicit expression for such a metric is slightly convoluted and not so convenient for analogue gravity purposes. Therefore, we now construct a family of geometries that encapsulate the main properties of these spacetimes in a much simpler manner.  In addition to being useful for our discussion, these geometries are also suitable for further analysis of chronologically ill-behaved spacetimes, such as the investigation of semiclassical effects. 

The idea is to consider the manifold $\mathbb{R}^{3} \times \mathcal{N}$, with $\mathcal{N}$ a given manifold that we can think of as either compact or noncompact. Let us choose a set of cylindrical coordinates $(t,r,\phi)$ for $\mathbb{R}^3$ and another suitable set of coordinates $\{ x^{n} \}$ for $\mathcal{N}$. For the sake of simplicity, we will choose the metric to factorize into the Lorentzian geometric structure of $\mathbb{R}^{3} $ and a given Riemannian metric $g_{mn}$ on $\mathcal{N}$. Although realistic analogue gravity constructions will reproduce only 2+1 or 3+1 configurations, we will keep the discussion as general as possible for the moment. 

We introduce the following line element:
\begin{equation}
    \dd s^2 = - \dd t^2 + \dd r^2 - 2 r f(r) g(t) \dd t \dd \phi + r^2 \left( 1 - f^2(r) g^2(t) \right) \dd \phi^2 + g_{m n} dx^m dx^n,
\end{equation}
which is constructed in such a way that its determinant is $-r^2$ and hence the metric is Lorentzian independently of the values of $f(r)$ and $g(t)$. The orbits of $\partial_{\phi}$ are timelike for the values of $(t,r)$ for which $f(r) g(t) > 1$.

In addition to requiring $f(r) g(t)> 1$ somewhere, we must also impose that the function $f(r)$ satisfies $f(0) = f(\infty) = 0$ to have an asymptotically flat metric that is regular in the axis. We can choose it to be different from zero only in a compact region $ r \in (r_a,r_b)$. There are many different choices for $f(r)$ that do the job. For instance, a relatively simple choice is that of a $\mathcal{C}^{\infty}$ bump function
\begin{equation}
      f(r) = \left\{
\begin{array}{ll}
     0 & r \notin (r_a,r_b) \\
      N \exp \left[ \frac{-\sigma^2}{(r_b-r) (r-r_a)} \right] & r \in (r_a,r_b)\\
\end{array} 
\right. .
\end{equation}
For any $\sigma$ and $N>0$, the function is identically $0$ outside the interval $(r_a,r_b)$, and becomes positive inside of it. Actually, we can relax the compact support condition and choose a broader class of functions that might be useful for analytical analyses. For example, we can take a P\"osch-Teller like function
\begin{equation}
f(r) = \sigma^2 \sech^2 \left( r^* - r_0\right), \qquad  \textrm{with} \qquad r^*= r + r_0\ln(r/r_0),
\end{equation}
where $r_0$ represents the value of $r$ at which $f(r)$ reaches its maximum value which is equal to $\sigma^2$. Next, we must choose the $g(t)$ function according to the regions where we want the CTCs to exist: 
\begin{enumerate}
    \item \textbf{CTCs through every} $ \bf{t=\textrm{\textbf{constant}}}$ \textbf{surface:} In this case, we can simply choose to have $g(t) = 1$, with any of the $f(r)$ choices mentioned above, and we will have that $\phi$ is timelike for a whole cylinder of the spacetime $ \mathbb{R} \times I \times S^1$, with $I \subset \mathbb{R}$ an open interval. 
    
    \item \textbf{Gluing a flat region to a semi-infinite cylinder containing CTCs:} In this case, we are interested in choosing a function $g(t)$ that vanishes for $t< t_c$ and grows up to one for $t \rightarrow \infty $. A suitable choice for doing such an interpolation is an infinitely derivable bump function of the form
    \begin{equation}
          g(t) = \left\{
    \begin{array}{ll}
          0 & t < t_c \\
           \exp \left[ \frac{-\sigma^2}{t-t_c} \right] & t \in (t_c,\infty) \\
    \end{array} 
    \right. .
    \end{equation}
    Again, we can relax the condition of vanishing for $t<t_c$ by the condition of taking an almost zero value and replace the bump function with a sigmoid function, such as a hyperbolic tangent or an inverse tangent, i.e.,
    \begin{align}
        g(t) = \frac{1}{2} \left( 1 + \tanh \left( \frac{t}{\sigma} \right) \right), \qquad g(t) = \frac{1}{\pi} \left( \frac{\pi}{2} + \tan^{-1} \left( \frac{t}{\sigma} \right) \right),
    \end{align}
    where $\sigma$  modulates how quickly these functions approach 1, i.e., how fast the region where $\phi$ becomes timelike is reached as $t$ increases.
    
    \item \textbf{CTCs within a compact region:} In this case, we want to choose the function $g(t)$ such that it vanishes outside a compact interval $[t_a,t_b]$ and smoothly approaches one within a compact region. A convenient choice is:
    \begin{equation}
          g(t) = \left\{
    \begin{array}{ll}
          0 & t \notin (t_a,t_b) \\
          \exp \left[ -\frac{\sigma^2}{(t_b-t) (t-t_a)} \right] & t \in (t_a,t_b).\\
    \end{array} 
    \right. 
    \end{equation}
    Again, we might prefer to choose a function that is not exactly zero to ensure an analytic expression, such as a Gaussian or a P\"osch-Teller like function: 
    \begin{align}
    g(t) =  \exp \left[ - \frac{(t-t_0)^2}{2 \sigma^2}\right], \qquad g(t) = \sigma^2 \sech^2 \left( t - t_0\right),
    \end{align}
where we have introduced the center of the region at which $\phi$ becomes timelike, defined as $t_0 = (t_a+t_b)/2$. 
\end{enumerate}
\begin{figure}[H]
\begin{center}
\includegraphics[width=0.3 \textwidth]{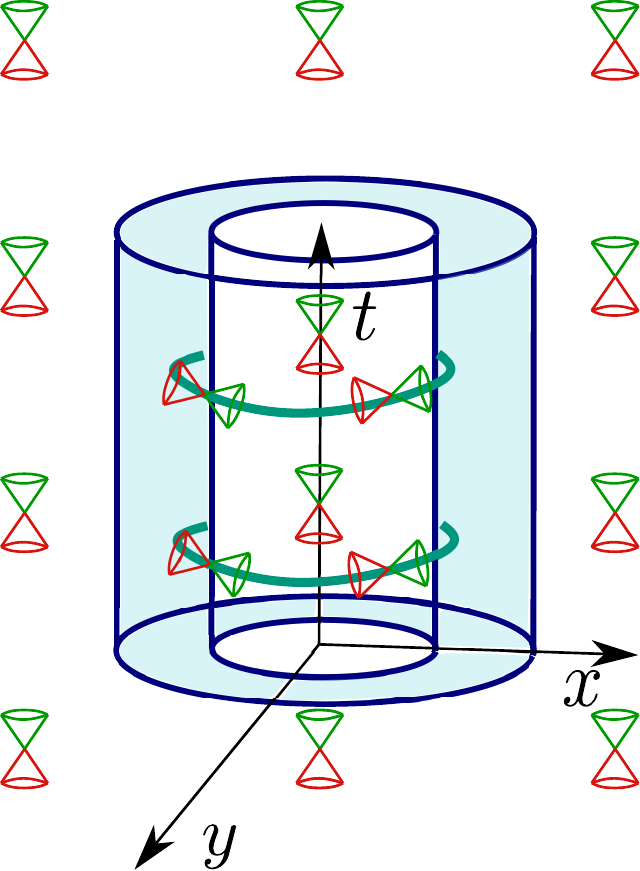}
\caption{Here, we depict the geometries described in the text. The different cases, corresponding to the finite, semi-infinite, and infinite cylinders, share the same essential structure, depending on whether the cylinder is confined to a compact interval $[t_a,t_b]$, a semi-infinite interval $[t_c, \infty)$ or the whole real line $(-\infty, \infty)$. The shaded region lying within the inner and outer surfaces of the cylinder, is the region where the periodic coordinate $\phi$ becomes timelike. Outside this region, the abnormal behavior of the light cones disappears. The presence of an additional spatial dimension is crucial for regularizing the light cones within the core of the cylinder, without requiring a change in topology, unlike the two-dimensional case.}
\label{Fig:Cylinder}
\end{center}
\end{figure} 
A pictorial representation of these geometries is provided in Fig.~\ref{Fig:Cylinder}. This completes our catalog of toy geometries containing CTCs which are inspired on those generated by the warp-drive tubes described in Subsection~\ref{Subsec:CTC-warp drives}. These geometries are significantly simpler, and thus easier to attempt to simulate within the framework of analogue gravity.

\subsection{Misner and Misner-like spacetimes}
\label{Subsec:Misner}

\begin{figure}
\begin{center}
\includegraphics[width=0.25\textwidth, height=0.35\textwidth]{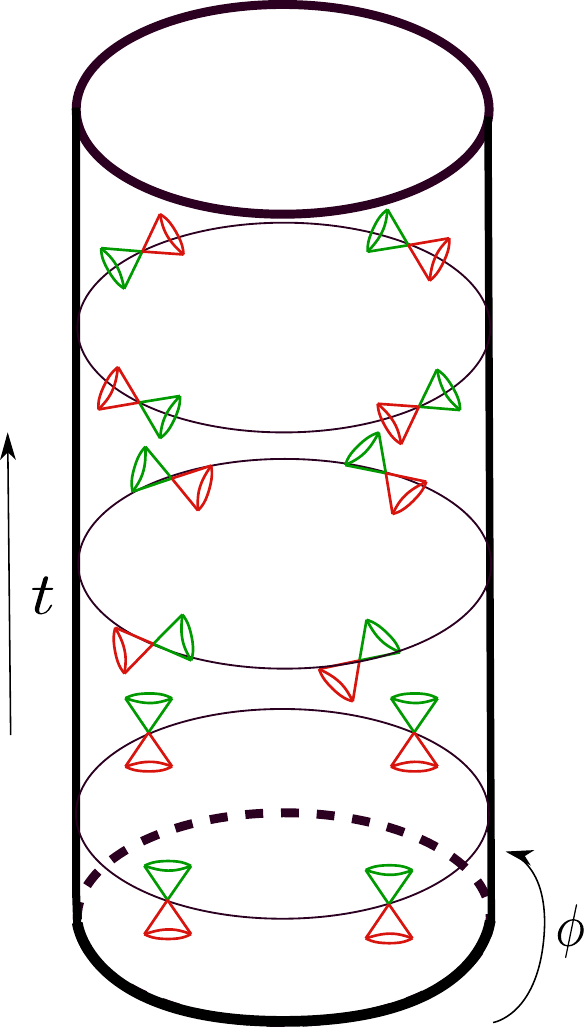}
\caption{ We represent here a cylinder $\mathbb{R} \times \mathbb{S}^1$ endowed with the light cone structure of Misner spacetime. At $t= - \infty$, the light cones seem to have a ``normal" causality in the sense of not allowing closed causal curves. They tilt as we move from $t=-\infty$ to $t=0$, where one of the generators of the light cones is tangent to the $\phi$ direction. Hence, at $t=0$, we got the first closed causal curve: a light ray that is confined to such circle. This $t=0$ slice corresponds to the circle in the middle of the three drawn. For $t \geq 0$, we enter into the regime of abnormal causal behavior containing closed causal curves. Notice that, although we are representing the light cones as cones for pictorial purposes, we are in 1+1 spacetime dimensions and they are not strictly ``cones''.} 
\label{Fig:misner_spacetime}
\end{center}
\end{figure}

The obstruction to build spacetimes with CTCs in a two-dimensional spacetime that we found in Subsection~\ref{Subsec:CTC-warp drives} was topological in nature. If we relax the condition of having a trivial $\mathbb{R}^2$-topology, we can build CTCs in $1+1$. Actually, the archetypal example of spacetime with a chronological horizon, i.e., a compactly generated Cauchy horizon~\cite{Hawking1991}\footnote{A compactly generated Cauchy horizon is a Cauchy horizon such that the past extension of its generators enters and remains within a compact subset of the spacetime, and they correspond to locally constructed time machines.} is precisely the so-called Misner spacetime~\cite{Misner1967,Hawking1973}. It has a $\mathbb{R} \times \mathbb{S}^1$ topology. This spacetime is commonly used as a proxy for studying chronological horizons. Given coordinates $(t,\phi)$ on the cylinder, its metric can be written as 
\begin{equation}
    \dd s^2 = -2 \dd t \dd \phi - t \dd \phi^2.
\end{equation}
It can be regarded as a simpler lower-dimensional version of the Taub-NUT spacetime~\cite{Hawking1973}. For $t<0$ it displays a normal causal behavior, but at $t=0$ we find a chronological horizon separating this chronologically ``safe" region from the chronologically ``sick'' region $t \geq 0$, which displays closed causal curves. A pictorial representation of this spacetime can be found in Fig.~\ref{Fig:misner_spacetime}. 

Imagine now that instead of endowing the cylinder with the Misner metric, we endow it with the flat metric. A well-posed problem in this spacetime could be given by simply considering data on a constant $t= t_0$ slice, that are $2 \pi$ periodic in $\phi$. After evolving forward and backwards such data, we would end up with a global solution. However, we should now recall that what we call ``time'' and what we call ``space'' in a two-dimensional spacetime is a choice. Pictorially, it corresponds to declaring which is the direction of time flow.

\begin{figure}
\begin{center}
\includegraphics[width=0.3 \textwidth]{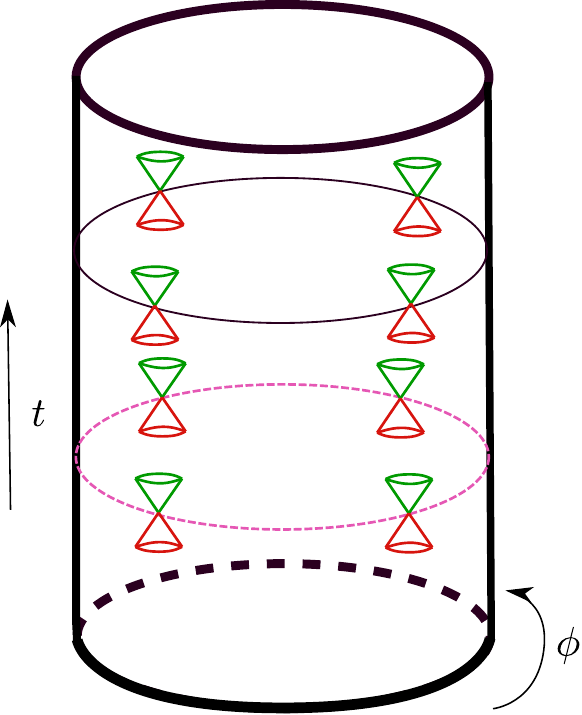}
\includegraphics[width=0.3 \textwidth]{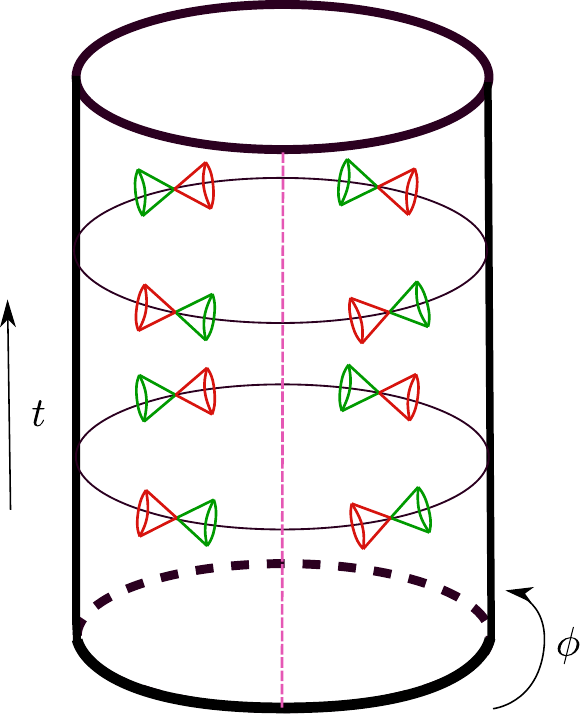}
\caption{The cylinder endowed with a flat metric allows two alternative perspectives regarding the causal description of the spacetime: a spacelike ring evolving in time or a spacelike line evolving periodically in time. The two alternatives are differentiated just by the choice of what we call time and what we call space. We have marked in dotted pink lines the putative surfaces on which we would place initial data, according to the choice.}
\label{Fig:flat_cylinder}
\end{center}
\end{figure}

We can also consider $\phi$ to be playing the role of time. Initial data given on a constant $\phi$-line would require to fulfill additional ``self-consistency'' constraints to have a well-defined evolution~\cite{Novikov1989b,Echeverria1991,Friedman1990b}. The existence of data fulfilling this criteria is guaranteed, since we know that global solutions exist. Thus, this illustrates that this configuration can be interpreted either as a spacelike ring evolving in time or as a line configuration evolving in a cyclic time. This is depicted in Fig.~\ref{Fig:flat_cylinder}

This example serves us to illustrate that, among the potential geometries displaying CTCs that one might consider, there are some of them which are mild in a sense. Whereas Misner spacetime is undoubtedly pathological since it develops CTCs from a chronologically well-behaved region, the second example of a cylinder with flat metric and CTCs falls in the second category of trivial CTCs, since it corresponds to a ``weird" choice of time direction. 

\section{Chronology protection in analogue systems}
\label{Sec:Chronology}

A property of the acoustic geometries described in Section~\ref{Sec:Analogues} is that they are, in principle, causally well-behaved. This was noticed in~\cite{Visser1997}, where it was shown that they automatically inherit the property of stable causality from the laboratory time. In fact, we just need to notice that the laboratory time $t$ behaves as global time function for the analogue system, since the the vector normal to the constant $t$ hypersurfaces, $\nabla_{\mu} t$, is always timelike
\begin{align}
    g^{\mu \nu} \nabla_{\mu} t \nabla_{\nu} t = - \frac{1}{\rho_0 c} < 0,
\end{align}
where the last inequality follows from the fact that the background density and the speed of sound need to be positive $\rho_0, c >0$. 

This appears to automatically rule out the possibility of simulating metrics with CTCs. Without further qualification this is not strictly true. It is interesting to realize that the flat cylinder with time $\phi$ periodically identified $\phi \sim \phi + 2 \pi$ can perfectly be simulated in an analogue system, for example, with a fluid inside an effectively 1-dimensional ring and declaring that the physical angular coordinate is a time coordinate. This simple configuration serves to illustrate that, contrary to the standard view, one can actually simulate spacetimes with CTCs in analogue systems. The same trick could be done in principle in $2+1$ and $3+1$ configurations although it would be slightly more complicated. The choice of time as an angular coordinate $\phi$ requires that the standard time and the additional spatial coordinates must acquire the same signature in the analogue model. In this way, the effective causality in the analogue system can have CTCs. 

However, as emphasized above, this kind of ``eternal" CTCs as seen from an external observer are somehow trivial. They correspond to declaring that an angular coordinate for the external observer is being used as a time coordinate for the internal observer. The price to pay is needing to impose extra ``self-consistency" conditions to obtain a well-posed problem. However, in the cases of eternal CTCs, those self-consistency constraints can be satisfied by some global solutions. This can be easily seen by noticing that, if we choose the nonperiodic time coordinate $t$, global solutions exist as the Cauchy development of some well-posed initial value problem on surfaces of constant $t$. Those global solutions automatically verify this self-consistency constraint after a trivial exchange of a space and the time coordinate. The analyses reported in~\cite{Sabin2016,Sabin2018} deal with this kind of CTCs. As such, their conclusions that these spacetimes can be simulated seem to be consistent with the results that we present here.

The most interesting case involves simulating geometries with a chronological horizon: spacetimes that feature a causally well-behaved region separated from a causally pathological region by a compactly generated Cauchy horizon. Here we point out that an acoustic realization of those configurations always involves the divergence of the velocity of the fluid. All the types of geometries described in Section~\ref{Sec:Geometries} involve tilting the sound cones until one is able to move infinitely fast with respect to the laboratory reference frame. But this is forbidden by the presence of the actual speed of light limit. Furthermore, to travel backwards in laboratory time would be forbidden even in a Galilean world (formally the $c \rightarrow \infty$ limit of a Lorentzian world). This means that, although an observer strictly following the acoustic metric would see the configurations in Fig.~\ref{Fig:warpdrive-forward}, a warp bubble moving to the future of some inertial time $t$, and Fig.~\ref{Fig:warpdrive-backward}, a warp bubble moving to the past of the inertial time, on equal footing, they are not truly equivalent due to the underlying background causality imposed by the laboratory frame. Hence, whereas the metric structure in Fig.~\ref{Fig:warpdrive-forward} can be simulated, the metric structure from Fig.~\ref{Fig:warpdrive-backward} cannot, if we understand the vertical time direction as the laboratory time. 

This situation seems to be generic in configurations in which a region with CTCs is separated from a region that does not have CTCs: the generation of CTCs is always accompanied by a divergence in the fluid velocity. The geometries explored in Subsection~\ref{Subsec:Extended-Warpdrives} illustrate this clearly as it can be seen by translating the line element into properties of the fluid background. The azimuthal velocity of the fluid always blows up at the chronological horizon. 

We now argue that the obstacle that we have identified in simulating spacetimes with chronological horizons applies to generic analogue systems, not only to the concrete acoustic example. The argument is based on the following property which holds, to the best of our knowledge, for all analogue systems explored in the literature so far. Independently of the specific form of the analogue physical metric\footnote{Here by ``physical'' we mean the analogue metric controlling the relevant fields, as opposed to an eikonal metric controlling just geometrical optics behavior.} of the particular analogue system, one can always take an eikonal approximation. The eikonal dispersion relation can always be written using a normalized inverse metric (one that obeys \mbox{$g_{\textrm{(E)}}^{tt}=-1$)}, resulting in
\begin{equation}
    g^{\mu \nu}_{\textrm{(E)}} p_{\mu} p_{\nu} = E^2 - 2 g^{ti}p_i + g^{i j} p_{i} p_{j},
\end{equation}
where $p_{\mu} = \left( - E, \boldsymbol{p} \right)$ and $\boldsymbol{p} = p^{i}$. Now, the central point is that the other components of the eikonal inverse metric are always identified with some properties of the substratum system~\cite{Barcelo2005}: in the fluid they can be identified with the velocities of sound and the fluid; in dielectric media they are identified with the permittivity and the permeability tensors $\epsilon_{ij}, \mu_{ij}$, etc. We expect something similar also to hold for Weyl-semimetals which seem to be good analogue systems~\cite{Volovik2016}. In such case, the identification of the divergent quantities with physical parameters is subtle, since one identifies the metric-components with some vectors characterizing the structure of the interacting Weyl-point~\cite{Volovik2017}, but the divergence still seems to represent a pathology of the system.

Consider now a spacetime as the ones we have discussed with a chronological horizon. One can always introduce a coordinate chart in which $\phi$, an angular coordinate in the causally well-behaved region, becomes a time coordinate in the causally pathological region. On the boundary, which constitutes the chronological horizon, it becomes null. In order to analyze what this implies for the eikonal metric components, let us write the metric and its inverse in matrix form. For the argument that follows it is sufficient to restrict ourselves to the $(t,\phi)$ coordinate sector (with $t$ the time coordinate of the well-behaved region). Then we have:
\begin{align}
    g = \left(
  \begin{array}{cc} 
  g_{tt} \ &  g_{t \phi}  \\  
  g_{t \phi} \ & g_{\phi \phi}
  \end{array}
  \right), \qquad \left( g\right)^{-1} = \frac{1}{g_{tt} g_{\phi \phi} - \left( g_{t \phi} \right)^2} 
  \left( \begin{array}{cc} 
  g_{\phi \phi} \ &  - g_{ t \phi}  \\  
  - g_{t \phi} \ & g_{t t}
  \end{array}
  \right).
\end{align}
The fact that $\phi$ becomes null at the chronological horizon means that $g_{\phi \phi}$ vanishes there. But this in turn implies that $g^{tt}$ becomes zero. To obtain the normalized eikonal metric with $g_{\textrm{(E)}}^{tt}=-1$ from the inverse physical metric, one needs to divide the inverse metric by a conformal factor $\Omega$ which must also tend to zero in order to keep the division normalized. But then, being divided by this same zero, the other components of the eikonal metric will blow up at the chronological horizon : $g_{\textrm{(E)}}^{ti}= g^{ti}/\Omega$, $g_{\textrm{(E)}}^{ii}=g^{ii}/\Omega \to \infty$. As these components are directly identified with physical properties of the system, the presence of this type of horizons would necessarily require a singular configuration of the analogue system. Even though the general relativistic metric components can be well defined, and so the hypothetical physical analogue metric components, divergences in the components of the inverse eikonal metric automatically translate in divergences on the physical properties of the analogue model and forbid their construction. The detailed physical properties that might diverge depend on the specific characteristics of the analogue system. In fluids, for instance, the velocity needs to blow up. This means that an infinite work from the external forces would need to be exerted, making thus not possible to simulate them.

\section{An example in detail: G\"odel spacetime}
\label{Sec:Godel}

As an explicit example, aside from those discussed in Section~\ref{Sec:Geometries}, we explore in detail the G\"odel geometry and some deformations of it in this section, since it serves as an archetypal example of a chronologically pathological spacetime containing CTCs~\cite{Godel1949,Hawking1973}. Furthermore, previous attempts to simulate it in an optic system have been presented in~\cite{Fiorini2021}. We find important to highlight that the primary limitation in analogue gravity lies in simulating chronological horizons, which are absent in the G\"odel geometry. Also, the identification of the metric components with the physical parameters of the analogue system does not seem to be done in the correct way, and hence the divergences that we observe in the horizons are absent in the analysis reported there.

G\"odel's spacetime is a solution of the Einstein equations with a negative cosmological constant $\Lambda$ and the energy-momentum tensor of a pressureless perfect fluid with a density: $  \rho = - \Lambda/(4 \pi)$. In suitable coordinates $(t,r,\phi,z)$, it can be written as follows~\cite{Kajari2004}: 
\begin{equation}
    \dd s^2 = -\dd t^2 + \frac{\dd r^2}{1 + \frac{r^2}{r_C^2}} + r^2 \left( 1 - \frac{r^2}{r_C^2}\right) \dd \phi^2 + \dd z^2 - \frac{2 \sqrt{2}}{r_C} r^2 \dd t \dd \phi,
    \label{Eq:Godel}
\end{equation}
where $r_C$ is a parameter characterizing the solution and related to the density and the cosmological constant through the equation $r_C^2 = - 4 \Lambda$. It can be seen that this geometry contains CTCs noting that for $r \geq r_C$, the (Killing) vector field $\partial_{\phi}$ becomes timelike. This vector needs to be periodically identified to avoid a conical singularity at $r=0$, i.e., $\phi \sim \phi + 2 \pi$. Therefore, it has closed orbits that are timelike for $r>r_C$. 

At first sight, it is not clear whether CTCs pass through the region $r < r_C$ since $\partial_{\phi}$ is spacelike there. However, we must take into account that this geometry is completely homogeneous, in fact it contains a group of five Killing vector fields acting transitively on the manifold~\cite{Hawking1973}. This means that every point of the manifold can be mapped to any other point in the manifold through a symmetry transformation. Hence, CTCs pass through every single point in this spacetime. However, there are no CTCs confined to the region $r<r_C$, in fact every CTC passing through the region $r<r_C$ crosses the cylinder $r = r_C$ an even number of times. 

From the point of view of the anisotropic acoustic metric for BECs introduced in Eq.~\eqref{Eq:Acoustic_Inverse_Metric_Anisotropic}, we realize that this precise system of coordinates allows for a direct realization of G\"odel's metric if we choose the fluid parameters conveniently. The nonvanishing components of the inverse G\"odel metric in these coordinates are
\begin{align}
& g^{tt}= -F(r), \nonumber \\
& g^{t\phi}= -\frac{\sqrt{2}}{r_C }    \frac{ 1 }{\left( 1-\frac{r^2 }{r_C^2} \right) } F(r), \nonumber \\
& g^{\phi\phi}=\frac{1}{r^2 }    \frac{ 1}{\left( 1- \frac{r^2}{ r_C^2} \right) } F(r), \nonumber \\
& g^{rr}= \frac{ \left(1+ \frac{r^2}{r_C^2} \right)^2 }{ \left(1- \frac{r^2}{r_C^2} \right)} F(r), \nonumber \\
& g^{zz}= \frac{ \left(1+ \frac{r^2}{r_C^2} \right) }{\left(1- \frac{r^2}{r_C^2} \right)}  F(r),
\label{Eq:InverseGodel}
\end{align}
where we have conveniently introduced the function $F(r)$ defined as
\begin{equation}
 F (r)= \frac{ \left(1- \frac{r^2 } {r_C^2} \right) }{ \left(1+ \frac{r^2 }{ r_C^2} \right)}.
\end{equation}
By comparing Eq.~\eqref{Eq:InverseGodel} with Eq.~\eqref{Eq:Acoustic_Inverse_Metric_Anisotropic}, we see that simulating the G\"odel metric requires implementing fluid motion only in the azimuthal $\phi$-direction; that is, we just need to take $v^i = v_{\phi} \delta^{i}_{\ \phi}$. First, we realize that from the $g^{tt}$ component of the G\"odel metric we have the following relation among the parameter $g$ encoding the quartic interactions of the BEC, the formal speed of sound $c$ of the BEC introduced above, and the function $F(r)$:
\begin{align}
    F(r) = \frac{g(r)}{c^3(r)}. 
\end{align}
We recall that $g(r)$ represents the four-body interaction coupling constant, while $c(r)$ formally corresponds to a speed of sound. However, due to the system’s anisotropic nature, this interpretation must be handled with care. The interaction can be made space-dependent, as it generally depends on external fields. By appropriately adjusting these external fields, the coupling constant can be shaped into any desired profile. Moreover, the formal speed of sound is directly linked to $g(r)$ and the background density $n_c(r)$, both of which can be readily controlled in experiments.

By inspecting the remaining components of the metric and comparing them with the line element in Eq.~\eqref{Eq:Acoustic_Inverse_Metric_Anisotropic}, we find that the azimuthal velocity of the fluid must be
\begin{equation}
    v_{\phi} = \frac{\sqrt{2}}{r_C} \frac{r}{1 - \frac{r^2}{r_C^2}},
\end{equation}
where we have taken into account the change to cylindrical coordinates. The matrix $h^{ij}$ is found to be
\begin{align}
    h^{rr}= g^{-2/3}   \frac{ \left( 1 + \frac{r^2}{r_C^2}\right)^{4/3}}{\left(1 -\frac{r^2}{r_C^2} \right)^{1/3}}, \nonumber \\
    h^{\phi\phi} = g^{-2/3}   \frac{\left(1 + \frac{r^2}{r_C^2}\right)^{1/3}}{\left(1 - \frac{r^2}{r_C^2}\right)^{4/3}}, \nonumber \\
    h^{zz}= g^{-2/3}  \frac{\left(1 + \frac{r^2}{r_C^2}\right)^{1/3}}{\left(1 - \frac{r^2}{r_C^2}\right)^{1/3}}.
    \label{Eq:Principal_directions2}
\end{align}
Furthermore, given that $h^{ij}$ is diagonal, the products of its components and the formal speed of sound $c^2$ can be physically interpreted as three anisotropic sound speeds $c_r^2, c_{\phi}^2$ and $c_z^{2}$. Explicitly they are
\begin{align}
    c_r^2 & = c^2 h^{rr}= \frac{ \left( 1 + \frac{r^2}{r_C^2}\right)^2}{\left(1 -\frac{r^2}{r_C^2} \right)}, \nonumber \\
    c_{\phi}^2 & = c^2 h^{\phi\phi} = \frac{\left(1 + \frac{r^2}{r_C^2}\right)}{\left(1 - \frac{r^2}{r_C^2}\right)^2}, \nonumber \\
    c_{z}^2 & = c^2 h^{zz}= \frac{\left(1 + \frac{r^2}{r_C^2}\right)}{\left(1 - \frac{r^2}{r_C^2}\right)}.
    \label{Eq:Principal_directions}
\end{align}
Let us analyze the properties of the fluid required to simulate the geometry. The first thing we notice is that the velocity of the fluid becomes infinite at $r = r_C$, where it also changes its sign. Hence the fluid system is singular, the $r<r_C$ and $r>r_C$ parts of the system are disconnected. However, there are CTCs living in the exterior region of the metric so it may still appear that the analogue system can locally simulate some trivial CTCs in the region $r>r_C$. However, looking at the speeds of sound we identify an additional issue. The speeds of sound also diverge at $r=r_C$, but moreover $c_r^2=c^2h^{rr}$ and $c_z^2=c^2 h^{zz}$ become negative for $r>r_C$, so $c_r$ and $c_z$ become purely imaginary. This means that we no longer have wave-like behaviors, or, in other words, causal signaling. The whole acoustic picture appears to break down. In that sense, the $r=r_C$ cylinder can be understood as a sort of ``domain wall":  it separates the interior region in which we have causal signaling from the region in which we have abnormal (exponentially amplified or attenuated) behavior of the putative sound-like excitations.

To design a realistic situation, imagine that we start with the fluid at rest and incite a rotation around the $z$-axis with the intention of evolving toward the G\"odel geometry. In order to do so, we need to obtain a configuration in which there is a separation between a clockwise and an anticlockwise rotating part of the fluid, separated by a surface at $r = r_C$ where the velocity needs to approach infinity and moreover the speed of sound also blows up. These requirements essentially imply that the hydrodynamic description of the BEC breaks down. Furthermore, because of the infinite fluid velocity at the $r=r_C$ surface, no signal can cross it. However, the physical parameters of the BEC are well defined for $r>r_C$, where one finds the CTCs. On the one hand, the sound speed becomes imaginary in that region, so it is not even sensible to talk about sound speeds nor any kind of propagating signals. In any case, the equations for the perturbations still hold, even if no causal interpretation is possible, and it turns out that these imaginary values can be attained in BECs with attractive interactions~\cite{Wieman2001,Duine2000}.

Another way to look at these imaginary speeds of sound is that two of the effective anisotropic masses of the BEC must be negative. The peculiarity of a particle with a negative mass is that it accelerates backwards when pushed forward\footnote{Notice that such negative masses are not fundamental, and thus need not result in tachyonic instabilities.}. However, it has been shown experimentally that it is indeed possible to achieve such strange behavior and create particles with negative effective masses~\cite{Khamehchi2017}.

Therefore, we have a surprising situation. The excitations of a quite strange BEC, with attractive interactions (which in principle would appear to forbid wave phenomena) combined with some negative anisotropic masses, end up behaving as if these excitations live in a perfectly Lorentzian world displaying CTCs. The situation would be equivalent in any other anisotropic fluid, not necessarily quantum\footnote{It is true, however, that engineering a classical fluid to display ``negative mass" excitations might be much more complicated, perhaps even impossible in practice.}. One would just need that $c_r^2$ and $c_z^2$ become negative in some region while $c_\phi^2$ stays positive. Roughly speaking, this ensures that the $r$ and $z$ coordinates acquire the same signature as the $t$ coordinate. Furthermore, it leaves the angular coordinate $\phi$ as the coordinate of different signature. From the perspective of the internal observers inside the fluid, ``time" would be what for an external (laboratory) observer is simply the angular coordinate. Thus, these CTCs would correspond to what we have called trivial CTCs, since they just correspond to a weird choice of time direction for the analogue system from the point of view of the laboratory system. 

Going back to G\"odel's metric, one could be tempted to modify the parameters of the analogue model and regularize the divergences. A simple example would be the following profiles where a suitable regulating parameter $\epsilon \ll 1$ is introduced (for simplicity we restrict our fluid to be effectively two-dimensional):
\begin{align}
    & v_{\phi} = \frac{\sqrt{2}r}{r_C}  \frac{1 - \frac{r^2}{r_C^2}}{ \left( 1 - \frac{r^2}{ r_C^2} \right)^2 + \epsilon^2}, \\
    & c_r^2 = \left( 1 + \frac{r^2}{r_C^2}\right)^2 \frac{1 -\frac{r^2}{r_C^2}}{\left( 1 -\frac{r^2}{r_C^2}\right)^2 + \epsilon^2}, \\
    & c_{\phi}^2 = \frac{1 + \frac{r^2}{r_C^2}}{\left(1 - \frac{r^2}{r_C^2}\right)^2 + \epsilon^2}.
\end{align}
However, it is straightforward to see that the associated acoustic metric is not a regular Lorentzian metric now, it degenerates at $r=r_C$. Still, strange as it may seem, this acoustic system does approach G\"odel metric for $r \gg r_C$. However, these CTCs are trivial again, and as such, do not conflict with our previous results.
   
To finish this section let us consider one final case: the possibility of simulating a geometry which approaches G\"odel's metric only in a finite range of the laboratory time $t$. To the best of our knowledge, this geometry does not have a general relativistic counterpart, in the sense that it is not a solution to the Einstein equations with the energy-momentum tensor of a known matter content. To construct such geometry, we introduce a modulating function $f(t)$, such that we can write a metric
\begin{equation}
    \dd s^2 = -\dd t^2 + \frac{\dd r^2}{1 + f(t) \frac{r^2}{r_C^2}} + r^2 \left( 1 - f(t) \frac{r^2}{r_C^2}\right) \dd \phi^2 + \dd z^2 - f(t) \frac{2 \sqrt{2}}{r_C} r^2 \dd t \dd \phi ,
    \label{Eq:Godel_finite_time}
\end{equation}
where $f(t)$ is chosen to have compact support. For instance we can take a bump function
\begin{equation}
    f(t) = \exp \left[ \frac{- \sigma^2}{(t_B-t) (t-t_A)}\right], 
    \label{Eq:Time_modulating}
\end{equation}
which is nonvanishing for $t \in (t_A,t_B)$. Hence, the metric represents a flat spacetime outside this interval, and develops CTCs within the region $(t_A,t_B)$. In order to also confine the CTCs to a compact region of space, one could force the metric component $g_{\phi \phi}$ to take negative values only within a finite interval of the $r$-component, for example through the following replacement 
\begin{equation}
    g_{\phi \phi} =  r^2 \left( 1 - f(t) \frac{r^2}{r_C^2}\right) \longrightarrow  r^2 \left( 1 - f(t) e^{-\frac{r^2}{\sigma^2}}\frac{r^2}{r_C^2}\right) ,
\end{equation}
where $\sigma$ must to be sufficiently large for the function $1 - e^{-\frac{r^2}{\sigma^2}} (r^2 /r_C^2)$ to exhibit two zeros. This new geometry exhibits CTCs  that are confined within a finite region of spacetime. However, for the same arguments explained above, it is not possible to simulate them as acoustic metrics, since the fluid would be required to develop a singular velocity. This again illustrates our more general point that it is impossible to generate metrics with nontrivial CTCs through an analogue metric.

\section{Conclusions}
\label{Sec:Conclusions}

In GR, all available evidence suggests that the construction of geometries with superluminal behavior\footnote{Defining superluminality in curved backgrounds is not straightforward. We simply mean signals that propagate faster than the speed of light in spacetimes with some flat regions.} requires violations of energy conditions~\cite{Penrose1993,Olum1998,Visser1998,Visser1999,Gao2000}. For instance, the construction of warp drives such as the ones described above requires exotic matter~\cite{Alcubierre1994}. Now, let us assume for a moment that exotic matter exists or could be created and manipulated at will for this purpose. The arguments that follow will be kinematic in character and not depend on dynamical considerations. Under the previous hypothesis we could in principle engineer a geometry containing a warp drive. The standard perspective provided by classical GR is that traveling faster than light entails the possibility of traveling to the past. For instance, the construction of a forward-in-coordinate-time warp drive is equivalent to the construction of a backward-in-coordinate-time warp drive due to the diffeomorphism invariance of GR. Under all these assumptions and tricks, it would be almost trivially possible to build a time machine.

Although this reasoning is of course logically correct, our analogue gravity analysis shows that even under the assumption of freely available and manipulable exotic matter, internal observers in an emergent gravity paradigm could face obstructions when trying to engineer CTCs in the presence of a fundamental background structure that is not observable to them in nonextremal situations. From their perspective, the inability to engineer such time machines could manifest itself differently depending on the specific ``high-energy" theory, for example as instabilities, singular behaviors or phase transitions. In all cases, this would constitute a dramatic breakdown of their internal description of physics in terms of low-energy effective fields, such as the metric description itself. For instance, the construction of the forward and backward warp drives would seem equivalent to these internal engineers\footnote{By internal engineers we mean observers only able to probe and manipulate the emergent causal structure, not the fundamental background causal structure.}. However, from the laboratory perspective, they are not equivalent at all, and in fact it is easy to see that the backward warp drive would be impossible to design. To sum up, traveling at speeds greater than the speed of sound is perfectly possible, but one cannot send signals backwards in time. 

Interestingly, one can also reverse the logic of the previous paragraph. When internal observers try to construct superluminal configurations, at the same time, they implicitly probe whether different inertial observers are indeed equivalent, or said in other words, whether there is a more fundamental causality underneath the GR description. It might happen that some inertial observers can construct a superluminal warp drive while others encounter a mysterious effect obstructing an equivalent construction. This nonequivalence would point at a breakdown of basic relativistic principles at this level of description. Our analysis also illustrates that beating the speed of light would not necessarily imply the possibility of traveling backwards in time and constructing a time machine. One might find insurmountable troubles in the way, caused by some fundamental substratum.

The starting point of this chapter was to investigate whether it is possible to simulate geometries with CTCs within the analogue gravity framework. Some claims in the literature suggest that reproducing spacetimes with CTCs is impossible, arguing that the background causality of the laboratory system must be inherited by the analogue model. We have shown that this argument requires further qualification. Specifically, we demonstrated that certain configurations with CTCs can indeed be faithfully reproduced in an analogue system; G\"odel's spacetime, for example, falls into this category. However, these configurations correspond to what can be considered trivial CTCs, as they extend throughout the entire spacetime rather than emerging from an initially well-behaved chronological region. In the analogue system, these CTCs can be interpreted as arising from an unusual identification of the time direction. The more intriguing cases involve spacetimes with CTCs that also feature chronological horizons, boundaries separating regions with CTCs from those without. It is this latter category of spacetimes that presents significant challenges and leads to pathologies when attempting to simulate them.

We have presented a simple catalog of analytical spacetime metrics containing CTCs that we have later attempted to simulate in analogue gravity setups. In all the examples analyzed, the presence of a chronological horizon leads to an insurmountable obstacle for its implementation in analogue gravity: one or more physical parameters of the analogue system must diverge at the chronological horizon. In some cases, these divergences can be smoothed out, but this destroys the regularity of the Lorentzian metric description which leads to the interpretation of the physical system as a gravitational analogue in the first place. We have focused on an acoustic system in which it is essentially the velocity of the fluid which must diverge in order to create the required tilting of the sound cones. We have shown that it is perfectly possible to simulate geometries allowing superluminal behaviors such as warp drives. However, this does not imply directly that one can build an analogue time machine. It is in fact the formation of a chronological horizon which is forbidden in the analogue implementation, since it is not possible to create a warp drive traveling backward in laboratory time.  

Furthermore, the obstructions found by exploring the analogue gravity implementation of CTCs resonate with Hawking's idea of a Chronology Protection mechanism in semiclassical GR~\cite{Hawking1991}. From the point of view presented here, such protection mechanisms arise naturally in frameworks in which the causal structure is emergent. In these emergent frameworks, there exists a background structure with a more fundamental underlying causality, which naturally prevents the type of manipulations required to create chronological pathologies.


\chapter{Superposing spacetimes in analogue gravity}
\label{Ch2:AnalogueSuperp}

\fancyhead[LE,RO]{\thepage}
\fancyhead[LO,RE]{Superposing spacetimes in analogue gravity}


Regardless of the specific framework chosen to develop a theory of quantum gravity, one fundamental requirement for any of them is the existence of states that represent the closest possible notion to a classical, smooth spacetime. These states would serve a role analogous to that of coherent states in quantum optics, which are the closest quantum counterparts to classical states of light. By extending the principles of quantum mechanics to gravity, it follows that one could, at least formally, generate superpositions of such states, i.e., superpositions of states describing (almost) classical spacetimes. Notably, the electromagnetic counterpart of these superpositions has already been demonstrated in laboratory experiments: Schrödinger-cat-like states of light have been successfully created~\cite{Brune1992}.

Building on the concepts introduced in the previous chapter, we now push the idea of emergent causality in analogue systems to its limits by exploring the superposition of two distinct causal structures. To this end, we consider an analogue system with a quantum substratum and investigate the superposition of two states, each of which gives rise to a different emergent geometry. While the analogy between a single state and a smooth classical spacetime is well established, here we undertake an exploratory analysis to determine whether it is possible to interpret the superposition of two such states from the perspective of analogue gravity, specifically, in terms of some sort of emergent blurred causality due to the superposition. This investigation forms the core focus of this chapter. To address this question, we concentrate on a specific quantum analogue system, the BEC model described in detail in Section~\ref{Sec:Analogues}. In particular, we examine a toy model consisting of a BEC in a double-well potential and analyze the states that could represent such a superposition of spacetimes.

The analysis we develop bears similarities to certain alternative approaches to quantum gravity, which propose that instead of blindly attempting to quantize the gravitational field, we should perhaps \textit{gravitize} quantum mechanics, to borrow Penrose's terminology~\cite{Penrose2014}. This perspective, though relatively niche but yet persistent, argues that some of the unique properties of GR, such as the fundamental causal structure embedded in the spacetime geometry, may not be amenable to be subject to the quantum rules. Blurring this structure, it is suggested, leads to pathological behaviors that the system's dynamics inherently avoid. In fact, one of the most striking challenges that fundamental physics faces, arises when we assume that the laws of quantum physics apply universally, across all scales and to all objects. However, we do not observe any phenomena associated with quantum superpositions of macroscopic states, despite such superpositions being conceptually permissible within quantum mechanics. The prevailing explanation for this absence involves a process known as environmental decoherence~\cite{Zeh1970,Zurek1981,Zurek1982}. However, Penrose has heuristically argued that gravity itself might underlie a more fundamental mechanism of decoherence~\cite{Penrose1989,Penrose1992,Penrose1996,Penrose2014}. The driving factor behind this mechanism is the lack of a well-defined causal structure: any deviation from a single, consistent causal framework effectively triggers a process of decoherence. As we shall see, our results within the framework of analogue gravity strongly resonate with Penrose's ideas, though notable differences also emerge.

The original motivation behind Penrose's proposal was to bridge the gap between the quantum and gravitational realms, with a particular emphasis on whether gravity could help resolve some of the conceptual challenges of quantum mechanics, such as the emergence of a classical macroscopic world. While Penrose's proposal is not the only one exploring this direction, his ideas are framed in a language closely aligned with the analyses carried out here in terms of analogue gravity. Therefore, in this chapter, we focus on comparing our results from analogue gravity with Penrose's proposals. However, for completeness, we provide an overview of the most prominent solutions to the foundational problems of quantum mechanics, along with other gravitationally inspired approaches similar in spirit to Penrose's, in Appendix~\ref{AppA}. This chapter together with such appendix are largely based on Ref.~\cite{Barcelo2021c}, which is one of the articles published during this thesis. 

\section{Analogue gravity: Superposing Spacetimes}
\label{Section:Superp_Spacetimes}

\subsection{Condensates in a double well}
\label{Subsection:Condensates_Double_Well}

Let us assume that we have scalar bosons in a symmetric double-well potential where we will label the wells by $i=1,2$. We will make the simplifying assumption for the moment that there is only one relevant state in each well, and particles within a well have a contact interaction controlled by the parameter $U$, which can be either repulsive ($U>0$) or attractive ($U<0$). Also, we will assume that we have a term describing tunneling between the wells controlled by the parameter $t$, which is the amplitude of probability for tunneling (not to be confused with time). The lower the potential barrier, the bigger the parameter $t$ becomes, making hopping more favorable. The Hamiltonian describing these particular interactions is the Bose-Hubbard Hamiltonian~\cite{Lewenstein2012} with just ``two sites'' in the lattice
\begin{equation}
    H = - t (a_1^{\dagger} a_2 + a_2^{\dagger} a_1 ) + \frac{U}{2}[n_1(n_1-1) + n_2 (n_2 -1) ],
    \label{Eq:BoseHubbard}
\end{equation}
where $a_i^{\dagger}, a_i$ represent the usual creation and annihilation operators that annihilate and create particles in the well $i$, with $n_i=a_i^{\dagger} a_i$ being the number of particles in the well $i$. Furthermore, we will work under the assumption of having a fixed number of particles \mbox{$N = n_1 + n_2$}. For the noninteracting Hamiltonian, i.e., Eq.~\eqref{Eq:BoseHubbard} with only the tunneling term, the solutions are explicit and can be found via an ordinary Bogolyubov transformation. In the regime in which the tunneling is highly suppressed, just the interacting term survives. It is diagonal in the number basis and the solutions are also explicit. Notice that we are omitting the hats to denote the quantum operators, to avoid an unnecessarily dense notation. 

We now describe the main properties of the states which will be interesting for our purposes: the ground state of the free Hamiltonian and the ground states for the Hamiltonian with $t=0$, i.e., the ones with no hopping between the two wells, either with attractive $U<0$ or repulsive interactions $U>0$. They will serve as approximate ground states for the regimes of hopping domination $t\gg \abs{U}N$ and interaction domination \mbox{$t \ll \abs{U}N $}, respectively. The intermediate ground states between both regimes are not so useful for our purposes and can be found numerically. However, understanding the evolution of the system from one regime to the other as we adiabatically vary the parameters sheds some light onto the behavior of the model and it can be well captured by the ansatz provided in~\cite{Mueller2006}.

\begin{figure}
    \begin{center}
        \includegraphics[width = 0.3 \textwidth]{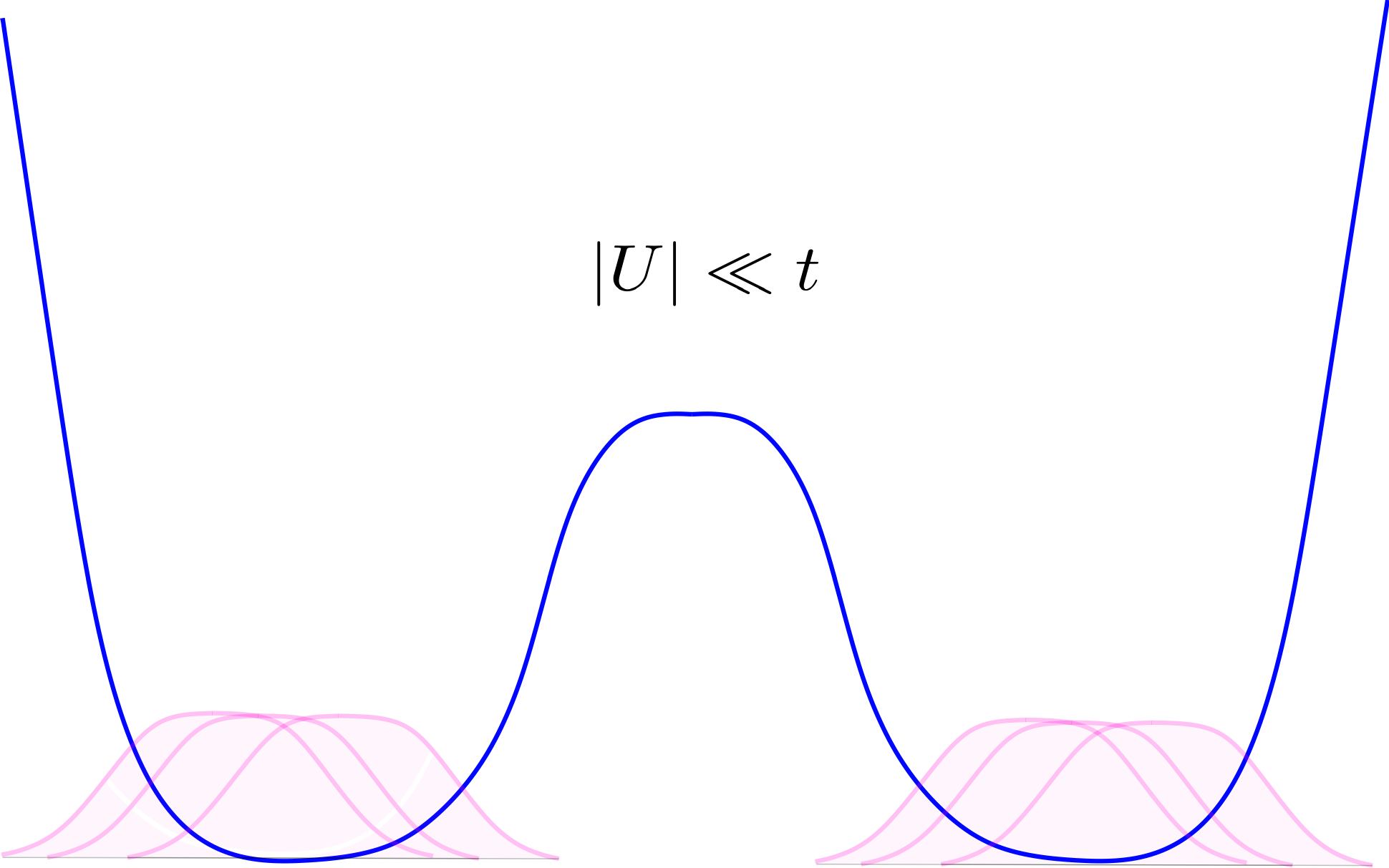}
        \includegraphics[width = 0.3 \textwidth]{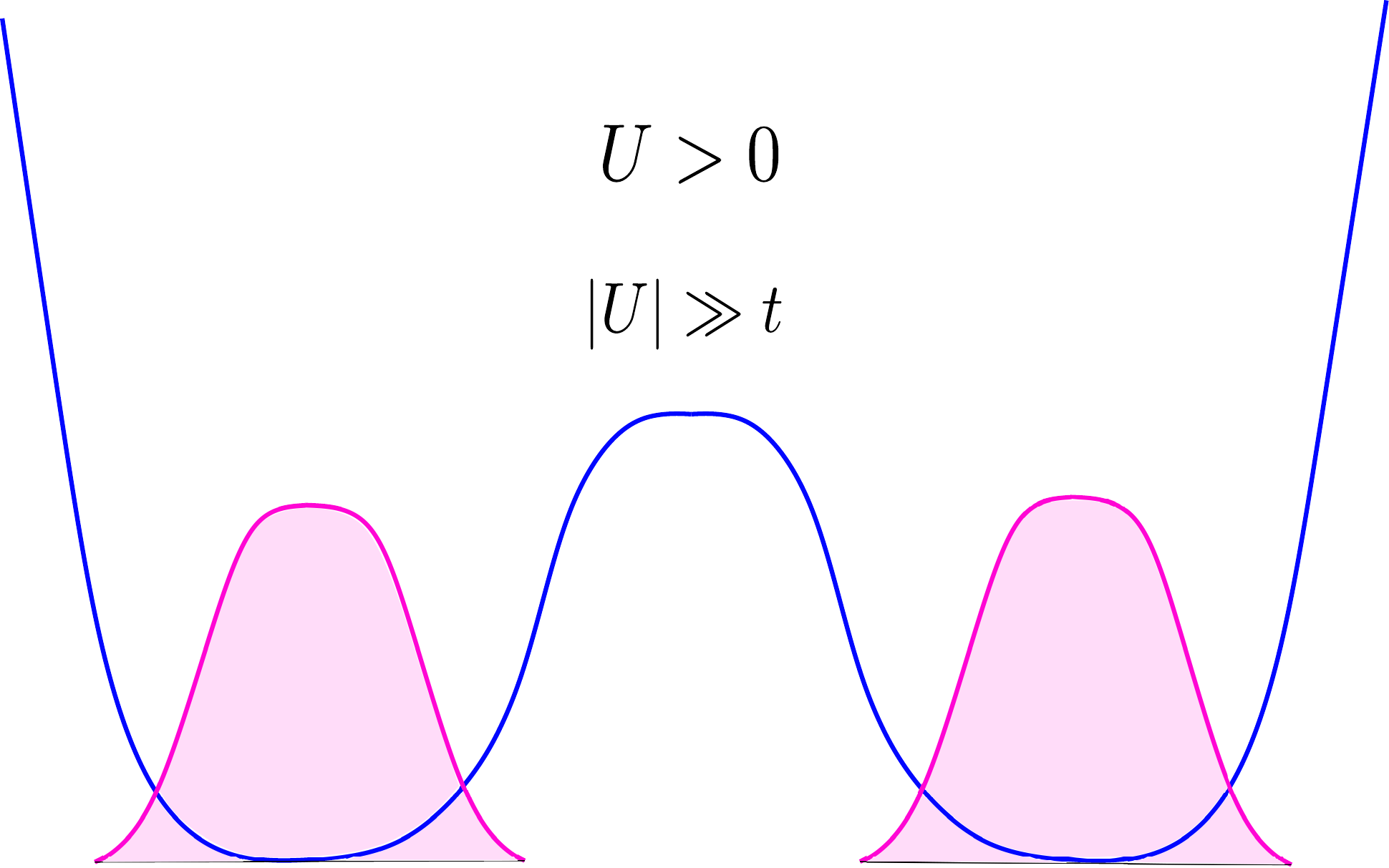}
        \includegraphics[width = 0.3 \textwidth]{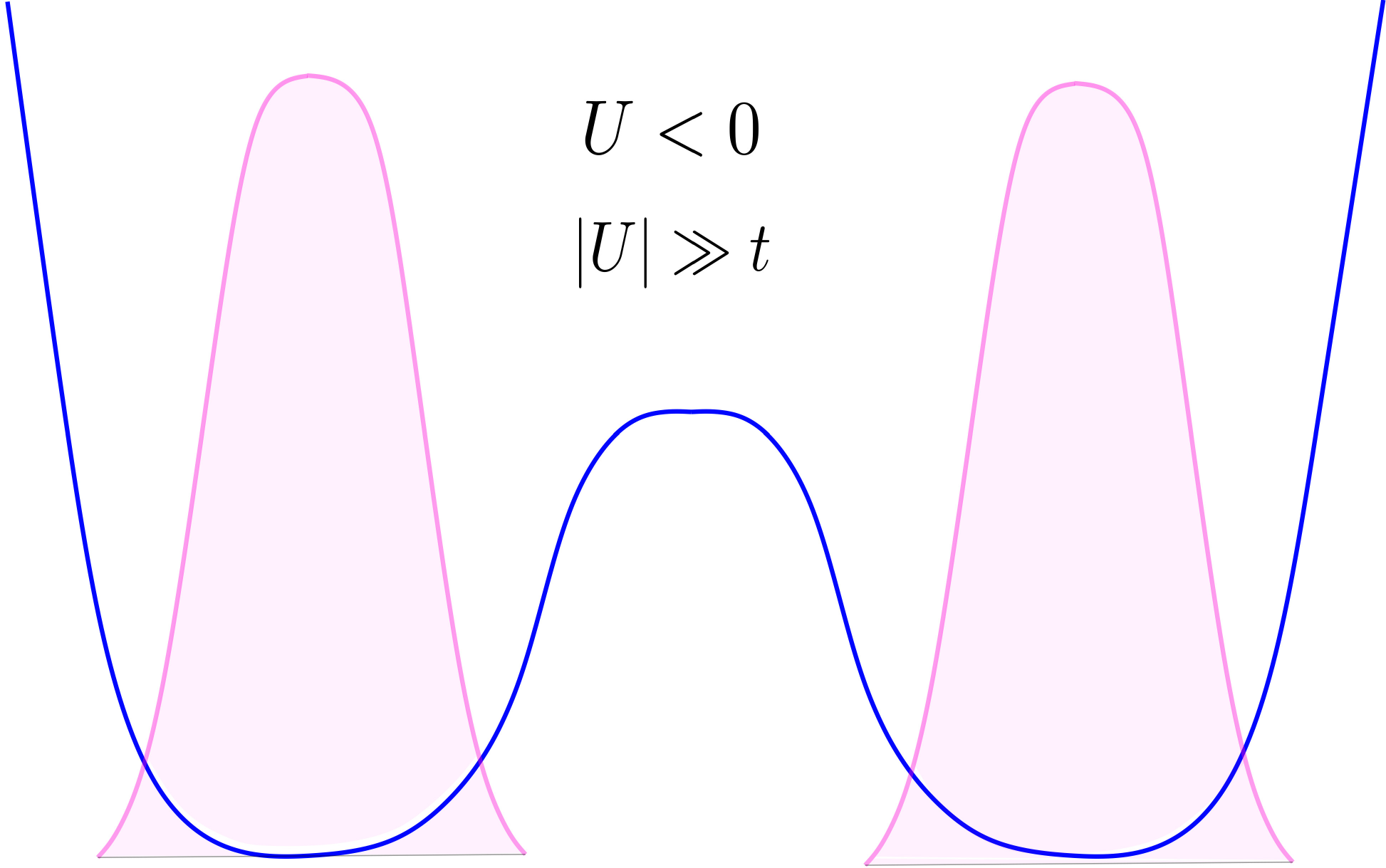}
        \caption{We pictorially represent the ground states corresponding to the three regimes of interest. From left to right: the coherent state, the Fock state and the cat state. The opacity in the figures above represents how localized the state is. The coherent state is a completely delocalized state with a nonvanishing projection onto all of the basis number states, the Fock state corresponds to two localized places at which the state exhibits a peak and the cat state corresponds to a delocalized state corresponding to a superposition of all the particles being in one of the wells or the other.} 
        \label{Fig:Ground_States}
    \end{center}
\end{figure}

Prior to discussing the ground states, it is convenient to introduce the concept of the single-particle reduced density matrix $\rho_{\textsc{sp}}$. Such density matrix is defined as 
\begin{equation}
    \left( \rho_{\textsc{sp}} \right)_{ij} = \tr \left( \rho a^{\dagger}_{i} a_j  \right),
\end{equation}
where $\rho$ represents the density matrix of the system. According to Leggett's nomenclature~\cite{Leggett2006}, a condensate will be fragmented if its single-particle density matrix has more than one macroscopic eigenvalues, i.e., more than one eigenstates with eigenvalue of order $\order{N}$. Intuitively, this would correspond to a situation in which condensation occurs simultaneously in different single-particle states. Armed with this tool, let us now analyze the ground state in the three limiting cases: the free case, the repulsive interacting case and the attractive interacting case. The corresponding states are pictorially represented in Fig.~\ref{Fig:Ground_States}. 

\textbf{Free condensate:} The free case can be explicitly solved via an ordinary Bogolyubov transformation, described by the following $SO(2)$ rotation within the space of operators $\{ a_1^{\dagger}, a_2^{\dagger} \}$: 
\begin{equation}
    b_{1,2}^{\dagger} = \left(a_1^{\dagger} \pm a_2^{\dagger} \right)/\sqrt{2}.
\end{equation}
Thus, we find that for the single-particle subspace, the eigenstates of the Hamiltonian are the symmetric and antisymmetric states with energies $-t$ and $t$, respectively. Thus, the ground state will be given by the fully symmetric state
\begin{equation}
    \ket{C} = \frac{1}{\sqrt{2^N} N!} \left( a_1^{\dagger} + a_2^{\dagger}\right)^N \ket{0},
    \label{Eq:CoherentKet}
\end{equation}
because they do not interact among themselves. $\ket{0}$ represents the Fock vacuum, i.e., the state annihilated by $a_1$ and $a_2$. The $C$ stands for the coherent state, since it corresponds to $N$ bosons coherently delocalized among the two wells with equal probability of finding them in one of the wells. The single particle density matrix for this state reads
\begin{equation}
    \rho_{\textsc{sp}} (C) = \frac{N}{2} \left( \begin{matrix} 1&1 \\ 1&1 \end{matrix} \right).
\end{equation}
This state has a single macroscopic eigenvalue $N$. According to Leggett's classification~\cite{Leggett2006}, it corresponds to a non-fragmented condensate since we just have one macroscopic eigenvalue of the single-particle density matrix. Physically, it corresponds to a condensation on the single-particle state which is approximately a superposition of the two Gaussians peaked around the center of each well in position space, describing the ground state of each well. Since the ground state $\ket{C}$ is a linear combination of number states $\ket{n_1,n_2} = a_1 ^{\dagger n_1} a_2 ^{\dagger n_2} \ket{0}/ \sqrt{n_1 ! n_2 !}$, the number of particles in each well has enormous fluctuations. We can compute them by writing the coherent state $\ket{C}$ in the number basis. For even $N$ (we consider such case for simplicity), we directly obtain from Eq.~\eqref{Eq:CoherentKet} 
\begin{equation}
    \ket{C} = \sum_{\ell = -N/2}^{N/2} \Psi_{\ell}^{(0)} \ket{\ell},
\end{equation}
where $\ket{\ell} \equiv \ket{\frac{N}{2} + \ell, \frac{N}{2} - \ell}$ and 
\begin{equation}
    \Psi_{\ell}^{(0)} = \left( \frac{N!}{2^N \left( \frac{N}{2} + \ell \right)! \left( \frac{N}{2} - \ell \right)!} \right)^{1/2} \approx \frac{e^{-\ell^2/N}}{(\pi N /2)^{1/4}}. 
\label{Eq:Gaussiandistrib}
\end{equation}
The fluctuations in the number of particles on each well are given by~\cite{Mueller2006}
\begin{equation}
    \expval{\Delta n_i ^2}_C = \expval{(n_i - \expval{n_i}_C)^2}_C = N/4,
\end{equation}
with $i=1,2$. 

Let us move on to discuss the strongly interacting case, i.e., \mbox{$ t=0$}, \mbox{$U \neq 0$}. For this purpose, taking into account that the total number of particles is conserved \mbox{$n_1 + n_2 = N$}, we can rewrite the Hamiltonian as
\begin{equation}
    H = \frac{U}{4} \left[ (n_1 - n_2)^2 + N^2 - 2N \right],
\end{equation}
which parametrizes the interaction by the difference in the number of particles on each well. 

\textbf{Strong repulsive interactions:} We will begin with the repulsive interactions \mbox{$(U>0)$}. Such case clearly favors the minimum difference in the number of particles between the two wells. Thus, for even $N$, the ground state for the repulsive interactions is given by
\begin{equation}
\ket{F} = \frac{a_1^{\dagger N/2} a_2^{\dagger N/2}}{(N/2)!} \ket{0},
\label{Eq:Fock_state}
\end{equation}
where the $F$ stands for the Fock state. Its single particle density matrix is
\begin{equation}
\rho_{\textsc{sp}} (F) = \frac{N}{2} \left( \begin{matrix} 1&0 \\ 0&1 \end{matrix} \right),
\label{Eq:MatrixFock}
\end{equation}
which corresponds to a fragmented condensate since it has a macroscopic eigenvalue $N/2$ with multiplicity $2$. It corresponds to two uncorrelated condensates (notice that the state is a product state) having half of the particles on each of the wells. Such a result for the ground state in this limit is quite intuitive since having one more particle than the half of them on one of the two wells would imply paying a penalty in energy. The fluctuations in the number of particles on each well vanish identically
\begin{equation}
\expval{(\Delta n_i)^2}_F=0.
\end{equation}
Thus, in this case, we have a fragmented condensate with a well-defined number of particles on each well. In that sense, from the point of view of Bose-Einstein condensation, it can be regarded as an independent condensation on each of the wells. 

\textbf{Strong attractive interactions:} For attractive interactions $U<0$, the situation is the opposite: the favored states are those containing a huge difference in the number of particles on each well. This means that the ground state is the subspace spanned by the vectors  $\{ \ket{N,0}, \ket{0,N} \}$. However, if we begin with the coherent state $\ket{C}$, the ground state of the free theory, and we turn on adiabatically the attractive interactions, the state that we will reach once $t$ becomes negligible will be a concrete superposition of both vectors (we will come back to this point later). Actually, it corresponds to a Schr\"{o}dinger-cat like state
\begin{equation}
\ket{\textrm{cat}} = \frac{1}{\sqrt{2}} \left( \ket{N,0} + \ket{0,N} \right).
\label{Eq:Cat_state}
\end{equation}
It is also fragmented since its single-particle density matrix has also two eigenvalues 
\begin{equation}
\rho_{\textsc{sp}} (\textrm{cat}) = \frac{N}{2} \left( \begin{matrix} 1&0 \\ 0&1 \end{matrix} \right) ,
\end{equation}
being equal to $\rho_{\textsc{sp}} (F)$, the one that we had for the Fock state. Particle number fluctuations on each of the wells allow us to distinguish between both states though: in this case they do not vanish but
\begin{equation}
\expval{(\Delta n_i)^2}_{\textrm{cat}} = N^2/4.
\end{equation}
Thus, we have seen that although the Fock and cat states are fragmented according to the standard definition of Leggett (their one-particle density matrices have two macroscopic eigenvalues), they physically correspond to very different states. While the Fock state has a definite number of particles on each well, the cat state has an indefinite number of particles on each well, it corresponds to having all the bosons completely delocalized among them. 

Until now, we have presented the three states that will be relevant for our purposes. Summarizing, the coherent state $\ket{C}$, is the ground state of the free theory $t \neq 0, U =0$; the Fock state $\ket{F}$,  is the ground state of the strongly repulsive interacting theory \mbox{$t=0, U >0$}; and the cat state $\ket{\textrm{cat}}$, is the ground state of the strongly attractive interacting theory $t=0, U<0$. The coherent state corresponds to a non-fragmented condensate, whereas the Fock and cat states correspond to fragmented ones. However, while the Fock state exhibits no fluctuations in the number of particles in each well, the coherent and cat states display large fluctuations in particle number across the wells.

\textbf{Intermediate interactions:} We analyze what happens as we turn on the interactions adiabatically, i.e., we begin with the free Hamiltonian and assume we slowly turn on the interactions in such a way that the ground state of the free theory accommodates to the ground state of the interacting theory. Although numerically it is possible to obtain explicitly the wave function of the ground state for the interacting theory, we will take advantage of the ansatz introduced in~\cite{Mueller2006} and use their families of vectors as explicit and analytic states closely approximating the ground states. 

These ans\"{a}tze for the interacting theory will be based on the ground state wave function for the noninteracting theory, the $U=0$ case of the Hamiltonian~\eqref{Eq:BoseHubbard}. The starting point will be to write the ground state in terms of the number basis
\begin{equation}
\ket{\Psi} = \sum_{\ell=-N/2}^{N/2} \Psi_{\ell} \ket{\ell},
\end{equation}
and write the Schr\"{o}dinger equation for this system $ H \ket{\Psi} = E \ket{\Psi}$, which gives the following expression 
\begin{equation}
    E \Psi_{\ell} = - t_{\ell + 1} \Psi_{\ell + 1} - t_{\ell} \Psi_{\ell -1} + U \ell^2 \Psi_{\ell},
    \label{Eq:Tightbinding}
\end{equation}
with 
\begin{equation}
t_{\ell} = t \sqrt{(N/2 + \ell)(N/2 - \ell +1 )}. 
\end{equation}
The problem is equivalent then to a one-dimensional tight-binding model in a harmonic potential with \mbox{nonuniform} tunneling matrix elements \cite{Mueller2006}. The nonuniformity is such that wave functions $\Psi_{\ell}$ with large amplitudes near $\ell \sim 0$ always have less energy than the ones that spread around different values of $\ell$. Actually, in the free case, the wave function was a narrow Gaussian centered at $\ell = 0$. 

With this expression for the eigenvalue problem, let us start by considering the repulsive case $(U>0)$. These interactions make the coherent state~\eqref{Eq:Gaussiandistrib} squeeze into an even narrower distribution. We can introduce the following ansatz to capture this feature~\cite{Mueller2006}
\begin{equation}
    \Psi_{\ell}(\sigma) = \frac{e^{-\ell^2/\sigma^2}}{(\pi \sigma^2 / 2)^{1/4}}.
    \label{Eq:Familystates}
\end{equation}
As $\sigma^2$ varies from $ \sqrt{N}$ to small values, the initial coherent state $\ket{C}$ starts looking much more like the Fock state. Actually we can obtain a relation between $\sigma$ and the value of $U$ by taking a continuum limit on~\eqref{Eq:Tightbinding}, see~\cite{Mueller2006} for details. In this limit, the equation reduces to that of a harmonic oscillator potential and we can obtain the value of $\sigma (U)$ since in that case the problem is exactly solvable: $\sigma^{-2} = (2/N)(1+U N/t)^{1/2}$. The single particle density matrix for these states reads
\begin{equation}
    \rho_{\textsc{sp}} = \frac{N}{2} \left( \begin{matrix} 1&e^{-1/(2 \sigma^2)} \\ e^{-1/(2 \sigma^2)}&1 \end{matrix} \right).
\label{Eq:Matrixfocktocoherent}
\end{equation}
It has eigenvalues $ N (1 \pm e^{-1/\sigma^2}) /2$ and the fluctuation in the number of particles on each well is $\expval{(\Delta n_i)^2}=\sigma^2 /2$. As the gaussian width increases, the number of particles on each well vary from $(N,0)$ to $(N/2,N/2)$ and the number fluctuations $\expval{(\Delta n_i)^2}$ varies from $N$ to $0$. 

Let us discuss now the case of attractive interactions $U<0$. When the interaction is attractive, states having a huge difference in the number of particles on each well, or, in other words, a huge amount of particles on a single well, are favored. Thus, the effect of slowly turning on an attractive interaction is to split the Gaussian of the noninteracting state~\eqref{Eq:Gaussiandistrib} into a symmetric distribution with two peaks, a process which is captured by the family of states 
\begin{equation}
    \Psi_{\ell}(a) = K \left( e^{-(\ell-a)^2/2 \sigma'^2} +  e^{-(\ell+a)^2/2 \sigma'^2} \right),
\end{equation}
being $2a$ is the separation between the peaks, $\sigma'$ their width and $K$ is a normalization factor. As $a$ varies from $0$ to $N/2$ and $\sigma'$ reduces at the same time from $1 / \sqrt{N}$ to 0, the coherent state of the free theory evolves to the cat state (which is reached in the $U/t \rightarrow -\infty$ limit). 

The question now arises as to whether perturbations on top of the state can be understood as a form of acoustic excitations that, in some sense, probe an effective superposition of spacetimes. For this interpretation to hold, it is crucial that these small perturbations, do not fully destabilize the system, at least on timescales shorter than the lifetime of the BEC in its condensed phase. Recall that the condensed state itself is metastable, with a finite lifetime. In the next sections, we will evaluate the validity of this perspective through a stability analysis of the states and investigate the extent to which these states can indeed be interpreted as superpositions of spacetimes.

\subsection{Stability of the ground states}
\label{Subsection:Stability}

We observed that, although the one-particle density matrices of the Fock and cat states were identical, they described fundamentally different states: while the cat state exhibited enormous number fluctuations, the Fock state had zero number fluctuations. This implies that higher-order correlation functions are required to accurately characterize and distinguish these fragmented states. Furthermore, the significant number fluctuations in the cat state already indicate something about the system's stability: a delocalized system of this kind is evidently prone to instability under local perturbations that hit independently each of the wells. To illustrate this point, let us consider a generic interaction modeled by the Hamiltonian
\begin{equation}
    H' = \epsilon \left( c_{1}^{\dagger} a_1 + c_1 a_1^{\dagger} \right).
    \label{Eq:Perturbation}
\end{equation} 
Here $ \epsilon$ is a small parameter controlling the perturbation, $a_1$ is the usual annihilation operator for bosons in the first well and $c_{1}^{\dagger}$ denotes the creation operator for a generic excited state of the first well. Consequently, this Hamiltonian represents a local perturbation in the first well. Under this small perturbation, the Fock state~\eqref{Eq:Fock_state} is robust, as it transitions simply into another Fock state, $\ket{F'}$, under the action of this Hamiltonian operator
\begin{equation}
\ket{F'} = H' \ket{F} = K' c_{1}^{\dagger} a_1^{\dagger (N/2-1)} a_2^{\dagger N/2} \ket{0},
\end{equation}
with $K'$ an irrelevant normalization constant. This feature illustrates the robustness of Fock states under standard local perturbations. However, the cat state does not display such robustness. Actually, the cat state tends to collapse to a localized condensate once we take them into account
\begin{equation}
H' \ket{\textrm{cat}} = K'' c_{1}^{\dagger} a_1^{\dagger (N-1)} \ket{0},
\end{equation}
with $K''$ another irrelevant normalization constant, since this state corresponds to a localized state on the first well. It corresponds to a number state with \mbox{$N-1$} particles on the ground state and one particle in the excited state described by the operator $c_{1}^{\dagger}$. 

The conclusion we extract from this analysis is that the system tends to avoid being in delocalized macroscopic superpositions, since they are unstable under local perturbations that hit the two wells independently. We have just considered interactions of the type~\eqref{Eq:Perturbation} which will be contained in a generic condensed matter Hamiltonian modeling a realistic BEC system. This can be seen by noting that, in general, we can work in the second quantized formalism in which the fundamental object we consider is the field $\psi(x)$,
whose expansion in a concrete basis of one-particle states reads
\begin{equation}
\psi = \sum_i \left( a_i f_i + a_i^{\dagger} f_i^{*} \right). 
\end{equation}
The interaction Hamiltonian will generically be of the form, 
\begin{equation}
    H_{\textrm{int}} = \int \dd^3 x h(\psi ^{\dagger} (x), \psi(x)),
\end{equation}
where $h$ represents the Hamiltonian density.

The previous simplification used to illustrate the instability, effectively assumed the presence of only the mode $f_1$, representing an excited state of the first well. However, in practice, there exists an entire tower of excited states above it, which can also be regarded as perturbations of the form~\eqref{Eq:Perturbation}. This makes the situation even worse, as the system has numerous states into which it can destabilize. 

\subsection{Attempts to generate a superposition of two effective spacetimes}
\label{Subsection:Superposing_Spacetimes}

After describing the three main states that can appear in a Bose-Hubbard model and analyzed their stability properties, we now pose the question of whether any of them can be interpreted as an effective superposition of two geometries. The idea is that by manipulating the potential structure within the wells, one can engineer configurations such that, if the condensation occurred in a non-fragmented manner (within one well exclusively), they would correspond to distinct geometries. If we introduce such small-scale structure and, on the scale of the wells, design the system to transition into a fragmented state like those described in the previous section, can the resulting state be understood as a superposition of two effective spacetimes? Without a mechanism to propagate signals and probe these superpositions, i.e., without an operational way of describing them, this question remains somewhat ambiguous.

In fact, the crucial property that characterizes analogue systems is that the collective excitations of the system follow the causal structure fixed by the background, which can be effectively described as a geometry. In the BEC, collective excitations manifest as sound waves, corresponding to the compression and expansion of the fluid along the direction of wave propagation. A fundamental prerequisite for the emergence of these sound waves is the presence of an effective repulsion between the constituents of the BEC. Additionally, it is essential that the wavelengths of these excitations are significantly larger than the healing length of the condensate. We recall that the healing length can be understood as the scale over which inhomogeneities in the atomic density of the condensate are smoothed out, rendering them imperceptible to the propagating waves. 

The Fock state, being a product state and hence simply describing an independent condensation on each of the two wells, cannot be interpreted by any means as an effective superposition of two spacetimes. In fact, it simply describes the propagation of sound waves independently on each of the two wells. Thus, we only have two natural candidates for describing superpositions of spacetimes: the coherent and the cat states. 

The coherent state can be taken as representing a system in a macroscopically delocalized state. However, we have seen that this state appears at the cost of eliminating the local interactions between the composing bosons. Having negligible local interactions, this state lacks the possibility of producing a rich causal physics. For instance, for $U=0$ the state develops no acoustic excitations and it does not really serve as an analogue model of gravity. However, given our analysis of the previous section in which we see that there is a smooth interpolation among the states with $U=0$ and $U > 0$, we can turn on a small repulsion without breaking the qualitative features of the state and allowing sound-like excitations. However, this delocalized state would have a wave-function with a profile varying too fast spatially to allow for the hydrodynamic approximation to be accurate if $U$ is not big enough. So we see that this coherent state does not seem suitable to describe superpositions of spacetimes either.

We are left with the cat state as the only one that we can identify, at least momentarily, with a superposition of two different spacetimes. Actually, it seems to be also the naive state that one would also build in a quantum gravity theory as a putative superposition of spacetimes: a linear combination of two states that separately give rise to smooth spacetimes in each of them. However, it does not give rise to any kind of sound-like excitations since it is in a regime in which no repulsive interactions are present among the atoms. Hence, it seems that it is not possible that any quantum sound emerges. However, one can think of first building the state under consideration, either by adiabatically switching on the attractive interactions ($U<0$), beginning with the $U=0$ case, to build the cat state. This switch-on of the attractive interactions can be engineered through Feshbach resonances~\cite{Chin2010}. Once the system settles in the appropriate state, one could turn the interaction towards the repulsive regime in such a way that sound-like excitations may emerge.

Hence, if we construct a condensate in the cat state, i.e., a would-be ``superposition of spacetimes'' in the analogue system and then turn on repulsive interactions once the system is settled in the state, what happens with the potential acoustic perturbations propagating on top of such a spacetime? Is there an effective spacetime resulting from this superposition? The idea is that there should be a limitation in describing the acoustic perturbations propagating on top of such a superposition as acoustic perturbations on a single different spacetime. Otherwise, this would mean that the superposition of analogue spacetimes would be also a spacetime, contrary to what one would expect to happen even with gravitational systems. Indeed, such limitation appears in the form of a huge instability of the system under consideration. As we discussed in Subsection~\ref{Subsection:Stability}, the cat state tends to decay to a Fock-like state under small perturbations. The propagation of sound-like excitations itself would be enough to destabilize such superposition. This translates into a decay of the putative superposition of spacetimes into a mixture of two spacetimes, i.e., two effective spacetimes which behave independently from each other. In that sense, this end state would be similar to a Fock like state. 

Actually, this impossibility recalls Penrose's ideas on the possibility of superposing spacetimes~\cite{Penrose2014,Penrose1992,Penrose1996}. On the one hand, Penrose argues that the superposition of spacetimes is an ill-posed concept in gravity, since it leads to a causal structure which is not well-defined. Since already in gravity there are hints that these superpositions are ill-posed, we could have expected in advance a fundamental limitation to find them in analogues. On the other hand, Penrose relates this ill-definiteness of the causality of superposed spacetimes with the mechanisms of quantum state reduction. We will come back to Penrose's idea in Section~\ref{Section:Penrose_Idea}, where we will contrast it with our findings in the analogue setup. 

\subsection{Same localization, different causality}
\label{Subsection:Same_Localization}

We will now discuss the viability of generating a superposition of spacetimes, but now we will consider a different situation. Instead of superposing two spacetimes with similar causal properties but based around different locations, we will now consider superposing two spacetimes at the same location with different causal properties. Roughly speaking, instead of attempting to superpose two geometries that in GR would be generated by two lumps of matter spatially translated, we want to superpose two lumps of matter with different shapes and densities located around the same region so that we have a blurred causal structure for the analogue. We will analyze this possibility in general terms.

As the simplest situation, one could think of a homogeneous BEC at rest but with a superposition of sound velocities. We recall that the sound velocity in a condensate, $c_s^2=g n_c /m$, is controlled by the coupling constant $g$ (from the $\abs{\psi}^4$ term of the Gross-Pitaevskii equation), the effective mass of the bosons $m$ and their number density $n_c$. The numbers $g$ and $m$ are parameters of the system which are in principle not subject to quantum rules. The masses are renormalized due to the interactions and, in principle, it is even possible to achieve negative masses as mentioned in the previous chapter~\cite{Khamehchi2017}. On the other hand, the coupling constant can be controlled by an external magnetic field using Feshbach resonances~\cite{Chin2010}. Then, we can ask, whether we can engineer this magnetic field to be in a quantum superposition. The magnetic field is a macroscopic entity, so what we are doing with this inquiry is just translating the problem of generating a macroscopic superposition from one place to another. Instead, it is typically assumed that the values of the parameter of a system are phenomenological values to which one could in principle associate infinitely precise values. As the causality of the system depends directly on these values one could think that they imprint a classical flavor into the notion of causality. Hence, this tuning of magnetic fields does not seem to be a way of achieving such a superposition. 

The other quantity entering the definition of the sound velocity is the density of condensed particles $n_c$. This quantity can have fluctuations if we deal with a system allowing flows of particles (grand-canonical ensembles), but in a condensed matter system the number of particles is completely fixed: the creation of new particles is forbidden by a huge energy gap. If one were hypothetically able to generate a BEC in an ultrarelativistic regime, the possibility of having a superposition of states with different number of particles would exist. But again, would it be possible to have a stable superposition of two mutually noninteracting gases each having a different number of particles? The possibility does not sound realistic, the very notion of macroscopic (thermodynamic) behavior would try to produce states with a peaked distribution for the number of particles. Actually, the standard ensemble in terms of which condensed matter systems are described is the Grand Canonical Ensemble, which allows not only for energy fluxes but also particle density fluxes between the system and the environment. In such scenarios, thermal homogeneity is needed to reach thermodynamical equilibrium although it is not sufficient. Chemical homogeneity is also required, which means having equal chemical potentials for the environment and the system. If the number of particles is not highly peaked, it is impossible to reach a stationary state. Hence, it seems that this is also a no-go pathway, since an almost stationary situation in which the condensate is equilibrated is required for the analogue picture to be sensible at all. 
       
The only element of the causality in a BEC remaining is the flow velocity of the fluid. The flow velocity that appears in the effective analogue gravity metric in a BEC corresponds to a macroscopic occupation of a particular phase structure for the mono-particle wave functions. Once again, the question is whether it is possible to have a stable superposition of condensates with different phases or not. The situation would be parallel to our discussion of the stability properties of the cat and Fock states, respectively. Depending on whether we perform a superposition of the two states with all the particles on each state or a more democratic Fock-like state in which half of the particles are on each of the phase states, we would produce an unstable state or a stable state, respectively. The additional difference is the external potential which, for the formation of standard condensates, always favors one of the two phase states. Thus, although under the absence of external potentials it would be possible to generate a stable Fock-like state, in a real system the particles would always tend to be projected onto one of the two fixed phase states. 

Thus, this analysis suggests also the inability to perform superpositions of different causalities, i.e. to have a BEC sitting on a single location but giving rise to different causal structures for the propagation of sound-like waves. Nonetheless, further study is required to fully understand and characterize whether there is a universal constraint on the possibility of simulating such superposition of analogue geometries in other analogue systems. Our belief is that a generic limitation should appear in any analogue system, although we do not have a general argument as the one described in the previous chapter for chronological pathologies. 

\section{Penrose's ideas on superposing spacetimes}
\label{Section:Penrose_Idea}

The results that we have found in our analogue system strongly resonate with Penrose's ideas about the \emph{instabilities} that one finds when superposing spacetimes~\cite{Penrose1989,Penrose1992,Penrose1996,Penrose2014}. This has not been a mainstream research focus, although there has been a steady flux of works trying to see whether the standard formalism and interpretation of quantum mechanics, which one could associate with von Neumann's formalization~\cite{vonneumann1955}, could fail in some regime~\cite{Wigner1962,Bialyanicki1976,Pearle1976,Ellis1983,Pearle1989,Ghirardi1986,Ghirardi1990,Bassi2013,Feldmann2012,Bassi2016}. Within those attempts, there have been many similar in spirit to Penrose's, i.e., attempts trying to trace these modifications of quantum mechanics back to a gravitational origin~\cite{Karolyhazy1966,Komar1969,Karolyhazy1974,Diosi1986,Coleman1988,Diosi1989,Gisin1989,Penrose1989,Ghirardi1990b,Penrose1992,Percival1995,Pearle1996,Penrose1996,Frenkel2002,Hu2003,Giulini2011,Adler2014,Hu2014,Penrose2014,Sharma2014,Bera2015,Singh2015,Donadi2020}. Here we are going to focus our comparison with Penrose's specific proposal as it is presented in the closest language to our analogue model. An overview of other ideas can be found in Appendix~\ref{AppA}. 

Penrose argues that superposing spacetimes is an ill-defined concept because the resulting state does not incorporate a clear definition of time evolution. In turn, he argues that this uncertainty in the definition of time evolution should lead to an uncertainty in the definition of energy.  Furthermore, this should entail the decay of such states into mixed states where correlations among the superposed spacetimes are lost. His ideas are not at the level of a precise dynamical theory and hence we do not attempt to make a precise analogy here: we simply discuss its similarities and differences with our analogue toy model. 

To put it more explicitly, consider that we have some theory of quantum gravity. A mild assumption is that there should exist states which represent an almost classical stationary spacetime sourced by an equivalently almost classical stationary distribution of matter. We represent these states as $\ket{\psi,G_{\psi}}$, where $\psi$ is associated with a particular material configuration and $G_\psi$ represents the gravitational field generated by such matter content. Following the superposition principle of quantum mechanics, within the set of possible states one has to include some of the form
\begin{equation}
    \alpha \ket{\psi,G_{\psi}} + \beta \ket{\phi,G_{\phi}},
    \label{superpgravmatt}
\end{equation}
with the two terms independently representing almost classical configurations. We take the complex coefficients $\alpha,\beta$ to satisfy $\abs{\alpha}^2 + \abs{\beta}^2=1$, in order to keep the state normalized. The key observation is that, even if each of the states in this combination is stationary, it is not clear whether (\ref{superpgravmatt}) is a stationary state or not. In stationary globally hyperbolic spacetimes, we have a well defined notion of stationary state since we have a global timelike Killing vector $K$ that generates time translations. It is natural to use this classical notion in order to define stationarity in the quantum scenario, since gravity should be well described (at least approximately) by a classical spacetime. Thus, stationary states $\ket{S_E}$ are those eigenstates of the Killing vector operator $\hat{K}$. These states turn out to be also states of well defined energy, 
\begin{equation}
    -i \hat{K} \ket{S_E} = E \ket{S_E},
\end{equation}
where $\ket{S_E}$ represents a generic stationary state of the form $\ket{\psi,G_{\psi}}$ with energy $E$. The key observation is that the state (\ref{superpgravmatt}) is a superposition of states of the form $\alpha \ket{S_E} + \beta \ket{S'_{\tilde{E}}}$, both stationary states with respect to their own Killing vector field $K, K'$. The problem is that we have, in some sense, a superposition of spacetimes and the concept of time translation in this superposed spacetime is an ill-defined concept. To talk about stationary states, we would need to be able to identify the corresponding time translation generators. However, Penrose argues that there is a fundamental difficulty with it. This ambiguity introduces an associated uncertainty in the concept of energy for the superposed state. According to the standard application of Heisenberg uncertainty principle, as used to estimate the lifetime of unstable particles, an uncertainty in the energy, which we denote as $\varepsilon$, would be related to the lifetime of the superposed state via the expression $\tau \sim \varepsilon^{-1}$, assuming we could estimate $\varepsilon$. Thus, the superposition of almost classical spacetimes is an ill-posed concept since it is unstable.

In fact, even for Newton-Cartan spacetimes, where time coordinates can be identified, establishing a canonical correspondence between the various spatial sections of one spacetime and another, there is no canonical way to identify individual points within those spatial sections. It is this lack of definite pointwise identification between spatial sections that leads to problems when trying to define the notion of time-translation for the superposition of gravitational fields. 

\begin{figure}
    \begin{center}
        \includegraphics[width = 0.22 \textwidth]{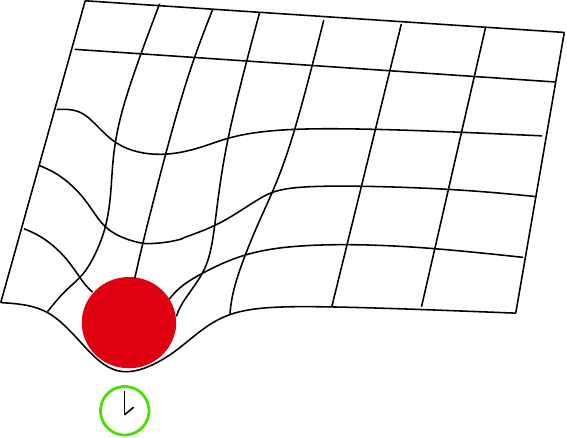}
        \includegraphics[width = 0.20 \textwidth]{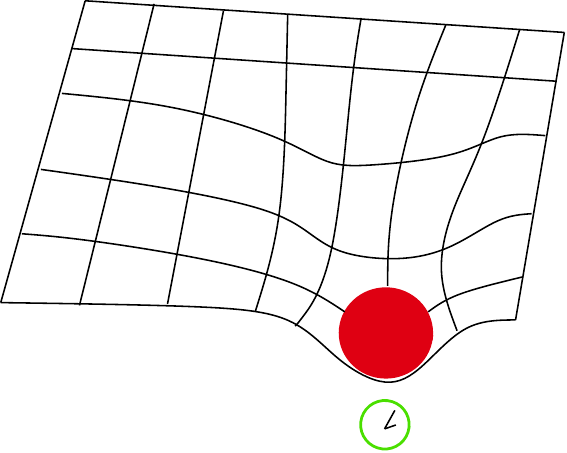}
        \caption{We pictorially represent the two configurations that we want to superpose. They correspond to the gravitational field associated with a certain lump of matter (a star, for instance), which is located in two different places for each of the states.}
        \label{Fig:Penrose_Picture}
    \end{center}
\end{figure}

Let us consider the case in which we have a superposition of configurations based on different locations, for example, two spacetimes corresponding to the gravitational field generated by a lump of matter, such as a star, located at two different positions. If the star is not too compact, the effective spacetimes can be well-described within the Newtonian limit. For this setup, which is depicted in Fig.~\ref{Fig:Penrose_Picture}, Penrose is able to quantify the incompatibility of the superposition. He quantifies $\varepsilon$ as the gravitational self-energy of the difference between the mass distributions of each of the two locations of the lump of mass: 
\begin{equation}
    \varepsilon = - 4 \pi G \int \dd^3 x \dd^3 y \frac{\left[ \rho(x) - \rho'(x) \right] \left[ \rho(y) - \rho'(y) \right]}{ \abs{x-y}},
\end{equation}
where $\rho, \rho'$ are the mass densities of the two lumps. This quantity would control the lifetime $\tau$ of the superposition for these simple examples of cat-like states for the matter distribution. 

For a careful analysis of the physical impact of gravitationally induced decoherence and its comparison with standard environmental decoherence, see~\cite{Howl2019}.  Moreover, recently it has been argued that natural models with $\tau = \eta  \hbar /\varepsilon$ and $\eta$ of order 1 are falsified experimentally\footnote{We restore here $\hbar$ to match the formulas in the references.}: the relaxation times these models imply appear to be much shorter than the ones inferred in the experiment~\cite{Donadi2020}. However, it is important to recall that the indirect results of this experiment crucially rely on assuming a random diffusive behavior for the density matrix of the system. For us it is not clear to what extent this needs to be a compulsory assumption.

The mechanism causing decoherence in our model, although similar in spirit to Penrose's proposal, differs from it in several respects. In our toy model the instability under local perturbations that the cat state suffers comes from the huge fluctuations in the particle number density that the system displays. Recalling that the speed of sound at a point in the condensate, $c_s$, is proportional to the local number density of particles $n_c$, huge fluctuations in the number of particles would induce huge fluctuations of the underlying causal structure for the propagation of sound waves in the effective geometry. Such fluctuations arise because the cat state is highly delocalized. However, Penrose argues that the source of instability for the superpositions of spacetimes arises due to a purely geometric reason: it arises from the inability to make sense of the notion of time evolution for such states. The absence of a well-defined notion of time, i.e., the absence of a well-defined parameter with respect to which we can talk about evolution when considering these kind of superpositions, makes impossible to regard the system as having some kind of causal structure. Although some fuzziness of the causal structure when entering the quantum regime is to be expected, attempting to superpose two radically different causalities would drive us completely into this regime where the causal structure itself dilutes and becomes a meaningless concept. Thus, the reason for the decay of a delocalized superposition seems to be the huge instability these states develop due to being far from having a well-defined causal structure.

One difference between our model and Penrose's ideas is that, strictly speaking, in our formalism the loss of coherence is only effective as the underlying system is an $N$-body quantum mechanical system. On the contrary, in Penrose's description, the gravitational decoherence appears as something fundamental. In a sense, our proposal seems to be closer to the environmental decoherence proposal than Penrose's. However, we rather think that our analysis highlights that the line separating an effective decoherence from a fundamental decoherence is blurred. Our toy model suggests that even taking a purely quantum theory as the high-energy theory, any theory constructed by macroscopic observers would contain quantum elements but also some elements that should not be treated quantum mechanically: there are elements in the effective theory that are collective excitations of already quantized degrees of freedom and they must not be quantized again. The difference with standard environmental decoherence is that the environmental degrees of freedom in this case are not so controllable from the effective theory point of view.

It is also worth remarking that somehow, all mechanisms that try to account for the reduction of state in quantum mechanics need to introduce parameters controlling how macroscopic a system is. Standard environmentally induced decoherence generically uses the number of particles in the system $N$ as a measure of how macroscopic the system is. Penrose's model measures how complex and how macroscopic a system is in terms of the gravitational field it can create. In general terms, we can expect that for ordinary macroscopic matter both measures display similar behaviors regarding the decoherence time-scales, but conceptual differences appear. If one were able to monitor the whole set of particles composing a system, it would be possible in principle to distinguish among both mechanisms~\cite{Howl2019}.  

However, a second important difference of our model with respect to Penrose's idea is that in BECs we can make sense of the superposition of two Minkoswki spacetimes with different speeds of sound. In Penrose's geometrical language, these spacetimes are one and the same due to the diffeomorphism invariance. If causality were an emergent property, much like the sound speed of collective excitations, then it would have additional properties beyond those associated with the effective geometrical description. Our model suggests that the inability of superposing different causalities is a robust idea that could survive in different approaches to quantum gravity. Although in a completely different setup, similar resonant conclusions are reached in Euclidean Dynamical Triangulations and Causal Dynamical Triangulations. In the former, one sums over arbitrary Euclidean manifolds and the result is dominated by configurations that do not resemble at all a smooth and almost flat (on suitable scales) spacetime that we live in. On the other hand, in the latter one constrains \emph{ab initio} the geometries one sums over including only those that lead to a well-defined causal structure. In this case, it is possible to find almost macroscopic spacetimes~\cite{Loll2019}.

As a final remark, let us consider how our model relates to the fundamental challenges of quantum mechanics, which were Penrose's original motivation for his ideas. A fully satisfactory interpretation of quantum mechanics must address two key issues. First, the formalism does not inherently distinguish between microscopic and macroscopic systems, implying that the principle of superposition should, in principle, apply universally. Second, it must explain why measurements yield definite outcomes despite the intrinsic statistical nature of quantum mechanics. Without invoking the Born rule, the unitary evolution of quantum mechanics neither defines what constitutes a measurement nor accounts for the source of nondeterminism. As a caveat, it is worth noting that hidden-variable theories, such as Bohmian mechanics, offer a resolution to these issues—albeit at the cost of introducing explicit nonlocality into the formulation~\cite{Bohm1952,Bohm1995}.

Our model, like Penrose's proposal, directly accounts for the first of these problems: how a superposition ends up becoming a mixture (a state lacking quantum coherences). For the second problem, it does not offer a complete solution since it would require making an explicit model of measurement. Actually, this is a long-standing problem that has accumulated a vast literature~\cite{Schlosshauer2005}. Nonetheless, we feel that our analogue toy model adds some interesting new perspectives to this second conundrum.

The philosophy underlying the analogue gravity program and in a more complete sense, emergent gravity scenarios is that everything that we perceive is constituted by low-energy collective excitations of some fundamental microscopic degrees of freedom. For instance, the spacetime itself and the quantum fields describing the Standard Model of Particle Physics would be part of such collective excitations. In that sense, although the microscopic theory may not display a Lorentzian causality, the effective low-energy regime does. Within this logic, observers would also be ``emergent" in the sense of being composed of excitations of these effective fields. The notion of observer that we are using is that of a conscious being, understood not literally as a human being but as any primitive form of consciousness or even a recording device. In general any sufficiently complex system relying on causality does the job. In fact a crucial prerequisite for consciousness to appear is causality itself\footnote{There even exist works suggesting that our consciousness developed because of our progressive ability to simulate progressions of hypothetical events organized as causes and effects in our minds~\cite{Jordan2012}.}. 

Now, recall that the main conclusion that we have drawn up from our analysis is that causality is a special macroscopic property in the sense that it is not amenable to be in quantum superposition. Then, an immediate consequence of this tight relation between causality and consciousness seems to be that different states of consciousness will also tend to rapidly decohere. 

Our construction in this paper takes standard nonrelativistic quantum mechanics, call it NRQM1, as the microscopic fundamental theory with no special commitment to any of its interpretations. Then, inner observers living in the system could make another nonrelativistic quantum theory, call it NRQM2, to describe the phenomena they perceive whenever the regimes they are probing are nonrelativistic. NRQM2 should of course become a Lorentz-invariant quantum field theory at high energies from the internal perspective but still low energies from the perspective of laboratory observers, that would still use NRQM1. Now, consider for example that NRQM1 as a high-energy theory happens to provide only practical decoherence and nothing similar to a collapse to one of the reduced states. Even in this case the description of NRQM2 will tend to incorporate in its rules a collapse postulate. This will be so because conscious beings will never have fuzzy experiences associated with being in cat-like states, i.e., they will never perceive macroscopic quantum coherences between states with very different causalities. If on the contrary the formulation of NRQM1 incorporated a mechanism for the selection of one and only one alternative under individual acts of measurement (by incorporating, for example, a stochastic rule \emph{\`a la} Ghirardi-Rimini-Weber~\cite{Ghirardi1986}), the NRQM2 would have the same form as before. Thus, the phenomenology of NRQM2 is independent of the measurement theory taken in NRQM1. In this way our example illustrates how difficult it is to learn things about a microscopic fundamental theory if we only have access to the collective excitations belonging to its low-energy sector.  

This is closely related to Wigner's discussion on the role of consciousness in our physical description~\cite{Wigner1962}. Essentially, he advocated that sufficiently macroscopic or complex processes (not just those involving conscious beings as complex as human beings) will in practice decohere, marking a separation between the classical and quantum realms. Notice that here, complexity and macroscopic are used in a nontechnical sense referring to an imprecise characterization of such magnitudes. The main difference of our approach is that it highlights the decoherent character of causality as a more primitive notion than consciousness.

\section{Conclusions and discussion}
\label{Section:Discussion}

This chapter has explored whether a meaningful notion of superpositions of spacetimes can arise in quantum analogue gravity models. Our starting point was a BEC in a double-well potential, used as a toy model to simulate the superposition of two effective spacetimes localized at different positions. This model exhibits three distinct phases: the coherent phase, when interactions are negligible; the fragmented uncorrelated phase, characterized by the dominance of repulsive interactions and represented by the Fock state; and the fragmented correlated phase, where attractive interactions dominate, giving rise to the so-called cat state if the interactions are switched on adiabatically.

In principle, the coherent phase cannot support causal signaling propagating on top of it due to the lack of repulsive interactions. Even when repulsive interactions are introduced to induce signaling, the resulting state does not resemble what one would expect from a superposition of spacetimes. On the other hand, the Fock state lacks any correlation between the two wells and it is therefore interpreted as two independent effective spacetimes, each confined to one of the wells, rather than a quantum superposition. These two distinct spacetimes are stable under local perturbations, posing no challenges either from a condensed matter perspective or from the analogue gravity interpretation point of view.

By contrast, the cat state could be naively interpreted as an analogue of the superposition of spacetimes, at least formally. As a linear combination of two states, each of which independently gives rise to an effective spacetime, the cat state represents the intuitive candidate for what one might imagine as a superposition of spacetimes. However, we have encountered two significant obstacles in attempting to give it an operational interpretation. While it may be possible to formally construct such spacetimes, the analogy becomes tenuous if there is no signaling that can be interpreted as probing the superposition of causalities. One major issue is the inherent instability of the state under local perturbations. Even small disturbances in one of the wells tend to destabilize the configuration, driving it toward a Fock-like state, a state that no longer resembles any kind of superposition. This leads to substantial limitations in realizing and interpreting such superpositions in a meaningful way.

Finally, we have considered, on general grounds, the possibility of superposing two spacetimes centered around the same location but exhibiting distinct causal structures, such as differing speeds of sound. While these spacetimes would be equivalent from a GR perspective, related through a diffeomorphism, in the analogue framework they become distinguishable because the underlying Galilean structure prevents their identification as the same spacetime. The presence of a preferred reference frame allows us to differentiate the two configurations. However, we observed that generating such a superposition of causalities inevitably requires engineering the superposition of a macroscopic quantity, such as the number of particles in the condensed phase or the magnetic field controlling the bosons' self-interaction properties. This, in essence, shifts the challenge: the problem of creating a superposition of spacetimes becomes the problem of producing a macroscopic superposition of another physical quantity, such as the magnetic field or particle number. In this sense, and \emph{mutatis mutandis}, we merely reframe the difficulty rather than solving it. Thus, we conclude that this second attempt to generate a superposition of causal structures does not appear to be a feasible either. 

While our second attempt to superpose effective causal structures revealed general limitations that discourage the pursuit of specific toy models, seemingly doomed to fail from the outset, our first attempt, using a BEC in a double-well potential, initially appeared promising as a suitable analogue. The instability that we found strongly parallels the phenomenology advocated by Penrose with his ideas on the ``gravitization'' of quantum mechanics. Penrose heuristically argues that superpositions of spacetimes should rapidly decay into mixtures of spacetimes, with the decay rate proportional to a measure of the system's macroscopicity. In Penrose's framework, this measure corresponds to the strength of the gravitational field generated by the system. In our model, it is represented by the number of condensed particles and the ratio between the hopping parameter and the interactions $(t/U)$. 

Penrose's argument rests on the idea that a superposition of spacetimes is inherently ill-defined, lacking a coherent causal structure, even for superpositions of stationary causal geometries. Similarly, in our model, fluctuations in the particle number between the wells quantify the instability of the BEC under local perturbations. Since the sound speed depends on the local number density of bosons, large fluctuations in density correspond to significant indefiniteness in the effective causal structure. In this sense, our analogue model highlights the challenges of pushing the causal structure of analogue systems too far.

The results of this chapter also connect with those of the previous one, where we analyzed the possibility of chronological pathologies arising in the emergent causalities of gravitational analogues. Specifically, we uncovered a mechanism of chronology protection rooted in the system inheriting the ``safe'' causality from the Galilean background structure. This protection manifests through the system's dynamics, which actively avoid chronologically pathological situations by breaking down the analogue description before such regimes are reached. A similar observation arises here: the system resists configurations that would lead to ill-behaved emergent causalities, as its dynamics inherently forbid such states. Although the phenomenology differs when the system is pushed toward these two distinct causally pathological scenarios, the shared conclusion is that the background causal structure enforces stability and prevents these pathologies.

The main takeaway from these two chapters is that the presence of background structures in analogue gravity systems inherently prevents certain pathological situations that, in GR (or in naive quantum mechanical GR setups involving superpositions of spacetimes), must be somehow removed by hand or through additional arguments. This observation suggests that background structures could play a meaningful role in constructing theories beyond GR that integrate quantum features. As discussed in the Introduction, the criteria for discarding such theories often seem to be historical, rooted in the success of GR as a background-independent theory, and aesthetic, typically justified in terms of simplicity. However, addressing pathological behaviors in GR often requires imposing restrictions on the energy-momentum tensor and, by extension, the system's dynamics. This approach is problematic, as no known energy condition is universally valid in physically realistic scenarios like gravitational collapse, where vacuum polarization effects can be unbounded. These analyses unveil the potential value of background structures in addressing such challenges by building dynamical theories of gravity beyond GR which incorporate them.

The next two chapters build upon this key insight, taking it as their starting point. Specifically, we assume that background structures may not only be useful but perhaps even necessary for developing gravitational theories beyond GR that incorporate features of the quantum realm.


\chapter{Gravity as a theory of interacting gravitons: the role of background structures}
\label{Ch3:SelfCoupling}

\fancyhead[LE,RO]{\thepage}
\fancyhead[LO,RE]{Gravity as a theory of interacting gravitons: the role of background structures}


In the previous chapters, we have conducted analyses within the framework of analogue gravity. As discussed, analogue gravity reproduces the kinematic properties of gravitational fields, but not the dynamical features of GR. Nevertheless, even without being limited by the GR dynamics, we found that situations considered unphysical in GR, such as chronological pathologies, or scenarios where the gravitational field is quantum and involves a superposition of spacetimes, are, in fact, prohibited by the dynamics of the system. The common thread between these two scenarios is the absence of a well-defined causality. In fact, our analyses have unveiled that the background structure of the analogue system plays a crucial role in preventing these extreme causality scenarios.

When we transition to setups where the gravitational field is dynamical, we realize that the role of background structures has, to some extent, been overlooked in the literature. This chapter is dedicated to revisiting the construction of gravitational theories from the perspective of self-consistent interacting theories of massless spin-2 excitations, i.e., gravitons, with a particular focus on the role of background structures. From a particle physics point of view, the gravitational interaction must be described by a massless spin-2 excitation, as it can be inferred from the universal coupling property of gravity and its long range~\cite{Feynman1996,Weinberg1965}. While this approach differs from the original conception of GR by Einstein, which was based on geometric arguments, it proves to be much more convenient for the purpose of building emergent theories of gravity. 

In fact, a common belief is that the only self-consistent nonlinear completion of a linear theory of a massless spin-2 particle is GR. Thus, from an emergent gravity perspective, a naive interpretation of this results would suggest that any scenario where an emergent Lorentz invariance arises and massless spin-2 excitations appear in the form of a some sort of collective excitations, would inevitably lead to GR, a completely
background independent theory. Our findings in this chapter are that this common belief is actually incorrect and we illustrate it with explicit counterexamples. 

First of all, we find that there are ambiguities in the process of building a theory of interacting spin-2 particles through a bootstrapping procedure, and as such, some decisions need to be taken in the construction at every step. In fact, depending on the specific choices that are made, one ends up with a nonlinear theory or another. We put some examples that explicitly illustrate this point: how different nonlinear completions can be obtained for the same linear theory depending on the choices made in the bootstrapping procedure. Second, the classic analyses were somehow contaminated by the hidden assumption that the resulting theory needed to be background independent. In a sense, they were contaminated by the expectation to find GR as an output of the procedure. In fact, the only analyses that are done allowing for the presence of background structures in the resulting theory are incorrect since they use some results that are only valid if the resulting theory turns out to be background independent~\cite{Barcelo2014}. 

The bootstrapping of other theories of gravity, for instance higher order theories or metric-affine theories, is an even more obscure problem, since there have been carried different analyses and it is not clear to what extent they are equivalent~\cite{Deser2017,Ortin2017}. Given that it is straightforward to analyze them with the tools that we have developed, we present also a careful analysis of the problem and clarify some misconceptions present in the literature. This chapter is largely based on article~\cite{Delhom2022b}, together with some of the introductory material presented in~\cite{Garcia-Moreno2023}, both articles published during this thesis. 

For the general considerations of some structural aspects of gravitational theories discussed in this chapter and the following ones, we will work in $D+1$ spacetime dimensions unless otherwise stated.

\section{Linear spin-2 theories}
\label{Sec:LinearSpin2}

Elementary particles are associated with unitary irreducible representations of the proper orthocronous Poincar\'e group. Such representations can be classified according to the mass $m$ of the particle and the quantum numbers of its little group~\cite{Wigner1939,Weinberg1995,Bekaert2006}. For massive particles \mbox{$m \neq 0$}, we can always choose a reference frame such that they are at rest and their momentum reads $P^{\mu} = (m, 0 , ..., 0)$. The little group corresponds to the transformations that leave invariant this choice of momentum: the group $SO(D)$ of \mbox{$D$-dimensional} rotations. For the special case $D=3$, the label coming from the little group is related to the angular momentum $j$, cataloging the projective representations of $SO(3)$. For a particle with angular momentum $j$, we have $2j +1$ states labeled by the polarization $s$, which runs from $s = -j$ to $s = +j$ in steps of one unit. For massless particles, $m=0$, the best we can do is to choose a reference frame such that the momentum reads $P^{\mu} = (E, 0 , ... , 0 ,E)$, being $E$ the energy of the particle. Now, the little group corresponds to the group $ISO(D-1)$, i.e., the $(D-1)$-dimensional Euclidean group. Again, for the special case $D=3$, the corresponding representations are labeled with the angular momentum $j$ but just the states with polarizations $s = \pm j$ survive. Thus, in the following, we will refer to a massless particle with angular momentum $j=2$ as a graviton, which in $D=3$ will only have two polarizations.

In general, a field representing a massless integer-spin-$j$ particle would be tentatively described by a tensor with $j$ indices. However, such tensors contain significantly more components than the physical polarizations of the massless particle they are meant to describe. For example, a massless spin-1 particle has $D-1$ physical polarizations, whereas a vector field $A^{\mu}$ has $D+1$ components. Similarly, a spin-2 field carries $(D+1)(D-2)/2$ polarizations, while a symmetric tensor $h^{\mu \nu}$ has \mbox{$(D+1)(D+2)/2$} independent components.

If we define the graviton field as a symmetric tensor $h^{\mu \nu}$, we must ensure that only its physical polarizations correspond to actual degrees of freedom. However, it is impossible to construct a tensor with nontrivial projection exclusively onto the subspace of physical states while maintaining Lorentz invariance, as Lorentz transformations do not preserve such subspace. Thus, we find a clash between massless particles and their Lorentz invariant representations as tensors. The resolution to this issue lies in the introduction of gauge symmetry.

\subsection{The origin of gauge symmetry}
\label{Subsec:Poincaré}

Let us begin with the spin-1 field propagating on top of flat spacetime as an illustrative example. If we have a vector field $A^{\mu}$, its $D+1$ components may lead to the propagation of $D+1$ degrees of freedom. However, from the group theory perspective, we know that to describe the propagation of massless particles, we need to ensure that it propagates only $D-1$ degrees of freedom~\cite{Wigner1939,Weinberg1995}. Let us take a configuration that represents a plane wave
\begin{align}
    & A_{\mu} (x) = \epsilon_{\mu} (p) e^{i  p \cdot x}, \\
    & p^2 = 0,
\end{align}
with $p \cdot x := p_\mu x^\mu$ and $\epsilon_{\mu} (p)$ denoting a vector that encodes the polarization. First of all, notice the $A^{\mu}$ field decomposes into $1 \oplus D$ as irreducible representations of the Poincaré group. Actually, the trivial representation corresponds to the scalar encoded in $A^{\mu}$, namely the projection in  the direction of $p^\mu$, i.e., $p_{\mu} \epsilon^{\mu}(p)$ or in coordinate space $\partial_{\mu} A^{\mu}(x)$. We can remove that scalar degree of freedom by simply imposing a constraint $p_{\mu} \epsilon^{\mu}(p)=0$ which, equivalently, in coordinate space is $\partial_{\mu} A^{\mu} = 0$, the so-called Lorenz condition. However, we still need to eliminate one of the remaining states to describe a massless particle. The first thing to notice is that there are no more Lorentz invariant constraints that we can impose on $A^{\mu}$ in order to remove degrees of freedom. If we intend to preserve Lorentz invariance explicitly, we cannot impose a non-Lorentz invariant constraint like the Coulomb gauge for example. It is this absence of additional potential Lorentz-invariant constraints what leads to the introduction of gauge symmetry. Notice that if we impose the constraint $\epsilon \cdot p = 0$, any shift on the polarization vector of the form
\begin{equation}
    \epsilon_{\mu}  \rightarrow \epsilon'_{\mu}  = \epsilon_{\mu} + \alpha (p) p_{\mu}
    \label{eq:transfeps}
\end{equation}
will lead to a new vector that still verifies the constraint due to the massless dispersion relation
\begin{equation}
     \epsilon \cdot p = 0  \rightarrow \epsilon' \cdot p = 0, \qquad p^2  = 0.
\end{equation}
Thus, it is natural to impose that configurations related by a transformation of the form \eqref{eq:transfeps}, are physically equivalent, i.e., they belong to the same gauge orbit. We can write down this equivalence in coordinate space and it leads to the standard form of gauge symmetry:
\begin{align}
    & A_{\mu} \rightarrow A_{\mu} + \partial_{\mu} \alpha (x) , \nonumber \\
    & \Box \alpha = 0. 
    \label{Eq:Gauge_Spin1}
\end{align}
Up to this point, we have seen that in order to describe a massless spin-1 particle through a vector field $A_{\mu}$, we need to ensure that it contains a dispersion relation of the form $p^2 = 0$, that the vector field is divergenceless $\partial_{\mu} A^{\mu} = 0$ and it displays the gauge symmetry in Eq.~\eqref{Eq:Gauge_Spin1}.

Let us now repeat the analysis with a tensor $h^{\mu \nu}$ aiming to describe a massless spin-2 particle. The first thing that we notice is that the tensor $h^{\mu \nu}$ contains $  (D+1)(D+2) /2 $ potential degrees of freedom, but we want it to propagate only the \mbox{$  (D+1) \left( D -2 \right) /2 $} degrees of freedom associated with a massless spin-2 particle~\cite{Wigner1939,Weinberg1995}. Again, take $h_{\mu \nu}$ to be a plane wave:
\begin{align}
    & h_{\mu \nu} = \epsilon_{\mu \nu} (p) e^{i  p \cdot x}, \\
    & p^2 = 0. 
\end{align}
The first thing to notice is that also $h_{\mu \nu}$ decomposes nontrivially into irreducible representations of the Poincaré group. For instance, it contains a symmetric traceless representation, a vector representation and two scalars, i.e., we can explicitly write the decomposition as \mbox{$ \frac{1}{2} (D+1)(D+2) = 1 \oplus 1 \oplus D \oplus \frac{1}{2}(D+2)(D-1)$}. To be more precise, the first scalar encoded in $h_{\mu \nu}$ is of course the trace $h = h^{\mu \nu} \eta_{\mu \nu}$; then we have the $(D+1)$-vector $A^{\mu} = \partial_{\nu} h^{\nu \mu}$, which can be broken into the scalar $ \partial_{\mu} A^{\mu} = \partial_{\mu}\partial_{\nu} h^{\mu \nu} =:\ddh$ and the corresponding divergenceless vector, as in the previous section. In addition, we have the remaining traceless symmetric tensor $h_{\mu \nu}$ which contains more components than the ones a massless particle contains. We can remove the two scalars and the vector component by imposing suitable conditions. These conditions are that the tensor is traceless $h = 0$, which removes one of the scalars; and the Lorentz transversality condition $p_{\mu} \epsilon^{\mu \nu} = 0$, that removes simultaneously the other scalar $\ddh = 0$, and the vector. This is the best that we can do by using constraints that preserve Lorentz invariance. To remove the remaining components of $h_{\mu \nu}$ in a manifestly Lorentz invariant way, we need to introduce gauge symmetries again. Given that we are imposing the constraints $\epsilon^{\mu \nu} \eta_{\mu \nu} = 0$ and $p_{\mu} \epsilon^{\mu \nu} = 0$, any transformation on the tensor $\epsilon_{\mu \nu}$ of the form
\begin{equation}
    \epsilon_{\mu \nu} \rightarrow \epsilon'_{\mu \nu} = \epsilon_{\mu \nu} + \xi_{\mu} (p) p_{\nu} + \xi_{\nu} (p) p_{\mu}, \qquad \alpha \cdot p = 0,
\end{equation}
will preserve the constraints:
\begin{equation}
    p^{\mu} \epsilon_{\mu \nu} = \epsilon_{\mu \nu}\eta^{\mu \nu} = 0 \rightarrow p^{\mu} \epsilon^\prime_{\mu \nu} = \epsilon^\prime_{\mu \nu}\eta^{\mu \nu} = 0, \qquad p^2 = 0.
\end{equation}
Hence, it is natural to impose that configurations related by a transformation of this form are physically equivalent. In coordinate space, these transformations take the form:
\begin{align}
    & h_{\mu \nu} \rightarrow h_{\mu \nu} + \partial_{\mu} \xi_{\nu} + \partial_{\nu} \xi_{\mu}, \nonumber \\
    & \partial_{\mu} \xi^{\mu} =  0,\quad \Box \xi^{\mu}=0.  
    \label{Eq:Gauge_Spin2}
\end{align}
Thus, for describing a massless spin-2 particle through a tensor field $h_{\mu \nu}$ in a minimal way, we need to ensure that it contains a dispersion relation of the form $p^2 = 0$, the tensor needs to be traceless $h = 0$ and divergenceless $\partial_{\mu} h^{\mu \nu} = 0$, and realizes the gauge symmetry as in Eq.~\eqref{Eq:Gauge_Spin2}. 

\subsection{Non-minimal realizations: enlarging the gauge symmetries}
\label{Subsec:Enlarging}

In practice, it is cumbersome to work with constrained fields (traceless, transverse, etc.). Thus, for practical purposes, working with the previous minimal constructions is not useful. The idea to avoid it is that we can eliminate the constraints that we are imposing at the expense of enlarging the gauge symmetry.

Let us illustrate this with the spin-1 field. Here there is only one way of enlarging the gauge symmetry in such a way that we only have the $A^{\mu}$ field as a configuration variable and we still propagate only the $D-1$ desired degrees of freedom associated with the massless particle. Once we relax the condition $\partial_{\mu} A^{\mu} = 0$, we can consider more general transformations, not only those that preserve the constraint. This means that we no longer need to impose any constraint on $\alpha$, the gauge parameter of the transformation, and we can perform arbitrary transformations of the form
\begin{align}
    A_{\mu} \rightarrow A_{\mu} + \partial_{\mu} \alpha.
\end{align}
Now, it is easy to write down a Lagrangian for such a theory. We need to only write down the most general quadratic Lagrangian displaying this gauge symmetry and such that it gives rise to a massless dispersion relation. This is of course the Maxwell Lagrangian: 
\begin{equation}
    \mathcal{L}_{\text{Maxwell}} = - \frac{1}{4} F_{\mu \nu} F^{\mu \nu}\,,
\end{equation}
where $F_{\mu \nu} := \partial_{\mu} A_{\nu} - \partial_{\nu} A_{\mu}$.

The situation is different for the massless spin-2 particle. Here, there are two ways in which one can enlarge the amount of gauge symmetry until one works with an unconstrained field $h_{\mu \nu}$. It is instructive to enlarge the gauge symmetry in two steps. First of all, we relax the constraint $\partial_{\mu} h^{\mu \nu} = 0$. With this, we notice that we can now do more general gauge transformations, since we can relax the condition that the vector generating them obeys a wave equation, as in the previous case. The result is a transformations of the form
\begin{align}
    h_{\mu \nu} \to  h'_{\mu \nu}=h_{\mu \nu} + \partial_{\mu} \xi^\sT_{\nu} + \partial^\sT_{\nu} \xi_{\mu}, 
\end{align}
where $\xi^\sT_{\mu}$ is an arbitrary transverse vector field $\eta^{\mu \nu} \partial_{\mu} \xi^\sT_{\nu}  = 0$, in order to preserve the traceless condition on $h_{\mu \nu}$. It is possible to write a Lagrangian that gives rise to a linear dispersion relation and implements this symmetry as long as we keep the constraint that the tensor is traceless: 
\begin{align}
    \mathcal{L}_{\text{TDiff}} = - \frac{1}{2} \partial_{\mu} h_{\nu \rho} \partial^{\mu} h^{\nu \rho} + \partial_{\mu} h^{\mu \rho} \partial_{\nu} h^{\nu}{}_{\rho}. 
\end{align}
Hereinafter, we drop boundary terms depending on gauge parameters. Finally, we come to the point of relaxing the constraint $h = 0$. This can be achieved by two different enlargements of the gauge transformations. First of all, we can relax the constraint that the vector generating the transformation is divergenceless $\partial_{\mu} \xi^{\sT \mu} = 0$, since we do not need to ensure that the trace of $h$ is preserved anymore. To put it explicitly, this means that we work with a tensor field $h_{\mu \nu}$ with the following gauge symmetry
\begin{align}
    h_{\mu \nu} \rightarrow h'_{\mu \nu} = h_{\mu \nu} + \partial_{\mu} \xi_{\nu} + \partial_{\nu} \xi_{\mu}
\end{align}
for arbitrary $\xi_{\mu}$. These transformations enlarge the set of linearly realized transverse diffeomorphisms to the whole set of linearly realized diffeomorphisms. The most general Lagrangian that one can write down displaying this gauge symmetry and giving rise to a quadratic dispersion relation is of course the Fierz-Pauli Lagrangian~\cite{Fierz1939}
\begin{align}
    \mathcal{L}_{\text{FP}} = - \frac{1}{2} \partial_{\rho} h_{\mu \nu} \partial^{\rho} h^{\mu \nu} + \partial_{\mu} h^{\mu}{}_{\nu} \partial_{\rho} h ^{\rho \nu} - \partial_{\mu} h^{\mu \nu} \partial_{\nu} h + \frac{1}{2} \partial_{\mu} h \partial^{\mu} h.
    \label{Eq:FP_Lagrangian}
\end{align}
However, there is another way of relaxing the constraint $h = 0$. We can keep the transverse diffeomorphisms and introduce the linearly realized Weyl transformations on the tensor $h_{\mu \nu}$. Explicitly, this means that we consider the following transformations
\begin{align}
    h_{\mu \nu} \rightarrow h'_{\mu \nu} = h_{\mu \nu}+ \partial_{\mu} \xi^\sT_{\nu} + \partial_{\nu} \xi^\sT_{\mu} + \varphi \eta_{\mu \nu}, 
\end{align}
where the $ \varphi$ is an arbitrary scalar function and $\xi^\sT_{\mu}$ is transverse (divergenceless). Imposing that this gauge group, the set of linear Weyl transformations and transverse diffeomorphisms, is realized at the Lagrangian level together with the massless condition, leads, up to a global constant, to the so-called Weyl Transverse-Diffeomorphism (WTDiff) Lagrangian~\cite{Alvarez2006}
\begin{align}
    \mathcal{L}_{\text{WTDiff}} =  - \frac{1}{2} \partial_{\rho} h_{\mu \nu} \partial^{\rho} h^{\mu \nu} + \partial_{\mu} h^{\mu}{}_{\nu} \partial_{\rho} h^{\rho \mu} - \frac{2}{D+1} \partial_{\mu}h^{\mu \nu} \partial_{\nu} h + \frac{D+3}{2(D+1)^2} \partial_{\mu} h \partial^{\mu} h .
    \label{Eq:WTDiff_Lagrangian}
\end{align}
As a summary, the most general action for a massless spin-2 field $h^{\mu \nu}$ that one can consider is
\begin{align}
    S = \int \dd^{D+1}x \sum_{n=1}^{4} M^{\alpha \tau}_{(n)   \beta  \gamma \rho \sigma}  \partial_{\alpha} h^{\beta \gamma} \partial_{\tau} h^{\rho \sigma},
    \label{masslessspin2}
\end{align}
up to boundary terms, and the tensors $M^{\alpha \tau}_{(n) \beta \gamma \rho \sigma}$ depend on three real parameters \linebreak \mbox{$\vartheta_1, \vartheta_2, \vartheta_3 \in \mathbb{R}$} and can be expressed as
\begin{align}
    & M^{\alpha \tau}_{(1)  \beta \gamma \rho \sigma} = - \frac{1}{4} \eta^{\alpha \tau} \eta_{\beta (\rho} \eta_{\sigma) \gamma}, \nonumber \\
    & M^{\alpha \tau}_{(2)   \beta \gamma \rho \sigma} =  \frac{1}{4} ( 1 + \vartheta_1) \left( \eta_{\beta (\rho} \delta^{\alpha}_{\sigma)} \delta^{\tau}_{\ \gamma} + \eta_{\gamma  (\rho} \delta^{\alpha}_{\sigma)} \delta^{\tau}_{\ \beta } \right), \nonumber \\
    & M^{\alpha \tau}_{(3)   \beta \gamma \rho \sigma} = - \frac{1}{4}( 1 + \vartheta_2) \left( \delta^{\alpha}_{\ (\beta} \delta^{\tau}_{\ \gamma)} \eta_{\rho \sigma} +  \delta^{\tau}_{\ (\rho} \delta^{\alpha}_{\ \sigma)} \eta_{\beta \gamma}  \right), \nonumber \\
    & M^{\alpha \tau}_{(4)   \beta \gamma \rho \sigma} = \frac{1}{4} (1 + \vartheta_3) \eta^{\alpha \tau} \eta_{\beta \gamma} \eta_{\rho \sigma }.
    \label{Eq:Tensors}
\end{align}
The maximum amount of gauge symmetry is found just for two sets of parameters~\cite{Alvarez2006}. On the one hand, for $\vartheta_1 = \vartheta_2 = \vartheta_3 = 0$, we recover Fierz-Pauli theory~\cite{Fierz1939}. As it is well known, such theory is invariant under linearized diffeomorphisms generated by arbitrary vector fields $\xi^{\mu}$ as 
\begin{equation}
    h_{\mu \nu} \to h^{\prime}_{\mu \nu} = h_{\mu \nu} +  \partial_{\mu} \xi_{\nu} + \partial_{\nu} \xi^{\mu}.
\end{equation}
On the other hand, for $\vartheta_1 = 0, \ \vartheta_2 = -1 /2, \  \vartheta_3 = -5/8$, we have the so-called WTDiff theory. It is invariant under linearized Weyl scalings and linearized Transverse Diffeomorphisms (TDiffs). They are generated by arbitrary scalar fields $\varphi$ and transverse vector fields $\xi^\sT_{\mu}$, i.e., those obeying $ \partial^{\rho} \xi^{\sT}_{\rho} = 0$, as  
\begin{equation}
    h_{\mu \nu} \to h^{\prime}_{\mu \nu} = h_{\mu \nu} + \varphi \eta_{\mu \nu}  + \partial_{\mu} \xi^\sT_{\nu}+  \partial_{\nu} \xi^\sT_{\mu}.
\end{equation}
The group of linearly realized Weyl transformations and transverse diffeomorphisms was argued in~\cite{Alvarez2016b} to have the structure of a semidirect product \mbox{$G = \textrm{Weyl} \rtimes \textrm{TDiff}$.}

\section{Coupling gravity: Conserved currents and bootstrapping}
\label{Sec:Bootstrapping}

This section is devoted to explaining the bootstrapping procedure used to construct nonlinear extensions of linear theories, with a particular focus on the spin-2 theories discussed in the previous section. The idea is to add the nonlinear interacting terms order by order in the theory (first cubic terms, then the quartic ones, etc.). When no gauge symmetries are present, the process is straightforward since there are no constraints to satisfy. However, in the presence of gauge symmetries, special care must be taken to avoid introducing additional degrees of freedom with the interactions. We examine in detail the ambiguities inherent to this procedure and the distinctive features of gravitational theories.

\subsection{Coupling to conserved currents}

Consider a linear theory with a given number of degrees of freedom. Our goal is to construct a nonlinear theory that retains the same number of degrees of freedom while reducing to the original linear theory in the weak field limit. In some cases, this may be trivial. However, when the linear theory exhibits gauge symmetry, the possible self-couplings that preserve the degrees of freedom are highly constrained. Typically, one seeks an iterative approach in which, starting from a quadratic action, self-couplings consistent with the gauge symmetry are introduced order by order.

Since the equations are divergenceless as a consequence of gauge symmetry, the nonlinear extension is achieved by coupling the theory to a conserved current. This means, that we need to add a conserved current on the right-hand side of the equations. This current must be selected from the set of conserved currents in the free theory. This process leads to a self-interacting nonlinear theory that satisfies a deformed version of the original linear gauge symmetry. 

Practically, this procedure is implemented by identifying currents that satisfy conservation order by order. Introducing a self-coupling at a given order typically generates nontrivial corrections to the current at the next order, necessitating additional higher-order self-couplings. The process terminates if, at some order, the self-coupling no longer produces higher-order corrections to the current. Otherwise, the procedure extends to infinite order, and for it to be successful, one must establish a general relation between the self-couplings at a given order and the currents at the next order.

To be more specific, assume that we have a set of fields which we denote as $\Phi^{I}$, being $I$ a multi-index including possible internal indices or spacetime indices. For an action that is quadratic on the fields, we will find linear equations of motion of the form
\begin{align}
    \mathcal{D}_{I J} \Phi^J = 0,
\end{align}
where $\mathcal{D}_{I J}$ is a given differential operator. Assume that we identify a conserved current\footnote{It might be the case that this includes several conserved currents.} $j^{I}$ which we want to add as a source, so that it couples as
\begin{align}
    \mathcal{D}_{I J} \Phi^{J} = \lambda  j_{I},
\end{align}
where the coupling $\lambda$ will be useful as a bookkeeping dimensionless parameter for the iterative procedure. Notice that the current will be a function of the fields $j^{A}$. In order to derive this coupling from a variational principle, we would need to add a term to the action which would schematically be of the form
\begin{align}
    \Delta S \sim  \lambda \int \dd^{D+1} x\  j_I \Phi^I.
\end{align}
However, since $j_{I}$ generally depends nontrivially on $\Phi^{I}$, the term $\Delta S$ will also contribute to the right-hand side of the field equations, resulting in a modified source term $j^{I} + \Delta j^{I}$. Consequently, if we seek an action that correctly reproduces the updated equations, an additional term of the next order must be included in the action. This process repeats iteratively.

A sufficient condition for the iteration to terminate after a finite number of steps $N$ is that the correction term $\Delta j_{I}$ at order $\lambda^N$ does not contain derivatives of the fields, ensuring that no further correction is required at the next order. Indeed, from the structure of Noether currents, it follows that if $\Delta j_{}$ at order $N$ lacks derivatives of the fields, it will not contribute further to these currents. This condition is already satisfied at order $N=2$ when applying this procedure to the gauge-invariant spin-1 field~\cite{Deser1970}. However, in other cases, such as spin-2 field theory working in second order formalism, the recursive process requires an infinite number of steps~\cite{Butcher2009}. The inclusion of additional matter fields follows naturally within this framework: the iterative procedure remains virtually unchanged, with the only difference being that the initial seed is now a conserved current incorporating the additional fields used as sources.

\subsection{Conserved currents}
\label{SubSec:ConservedCurrent}
To apply the bootstrapping procedure, we must first characterize the set of conserved currents in the free theory. According to Noether's theorem, identifying these conserved currents is equivalent to identifying the symmetries of the theory. However, the specific form of a conserved current associated with a given symmetry is not unique. In fact, any conserved current can be expressed as follows:
\begin{align}
    J^{\mu} = W^{\mu} + S^{\mu},
\end{align}
where $S^{\mu} = \partial_{\nu} N^{[\nu \mu]}$ is a superpotential, i.e., the divergence of an antisymmetric tensor $N^{\mu \nu}$, that is identically, i.e., not only on-shell conserved. Once the charge is computed on-shell as an integral of $J^0$, the second term always results in a boundary contribution evaluated at the spatial boundary of spacetime, as dictated by Gauss's theorem. In general, appropriate boundary conditions supplementing the equations of motion ensure sufficient fall-off behavior, causing this contribution to vanish~\cite{Julia1998}. Moreover, we can modify the current by adding terms that are identically divergence-free, as these can always be expressed as the divergence of a certain superpotential without affecting the conservation of the current.

If the symmetry is a gauge symmetry, it implies that $W^{\mu}$ vanishes on-shell ($W^{\mu} \mid_{\mathcal{S}} = 0$, where $\mathcal{S}$ represents the space of solutions of the theory). Consequently, the associated current does not lead to a nonzero charge. On the other hand, if the symmetry is a physical symmetry, $W^{\mu}$ remains nonzero on-shell, allowing us to define a conserved charge associated with it for a given solution. In this framework, a system is considered to be integrable if the number of conserved charges matches the number of degrees of freedom since they can be used to parametrize the space of solutions.

At this point, it is convenient to note that, in the bootstrapping procedure sketched above, the quantity entering the right-hand side of the equations of motion is actually the current and not the charge. As such, we may wonder whether different choices of the current, for instance, adding a different superpotential to the same current, affect the result of the bootstrapping procedure. This question is at the heart of the ambiguities of the procedure and it is not totally clear in the literature.   

\subsection{Bootstrapping gravitational theories}

Let us now explore the peculiarities and subtleties of the gravitational case. We consider a linear theory of a graviton field $h^{\mu \nu}$. As discussed in Sec.~
\ref{Sec:LinearSpin2}, there are only two possible maximal sets of gauge symmetries that can eliminate the unphysical degrees of freedom associated with a symmetric rank-2 tensor, leaving only the two standard gravitational polarizations. These are either linearized diffeomorphisms, which lead to the Fierz-Pauli action, or the combination of Weyl transformations and Transverse Diffeomorphisms (the WTDiff group). In both cases, the equations of motion are schematically of the form
\begin{align}
    \mathcal{D}_{\mu \nu \rho \sigma} h^{\rho \sigma } = 0,
    \label{eq:schematiclinearized}
\end{align}
with $\mathcal{D}_{\mu \nu \rho \sigma}$ a suitable differential operator. These equations are divergenceless and symmetric in the indices $\mu$ and $\nu$. Additionally, for the Weyl-transverse theory, the equations are also traceless. Therefore, we must couple them to an object that is symmetric and divergenceless on-shell, and traceless in the case of WTDiff. The only natural candidate is the energy-momentum tensor, as it is the only object with two indices that is symmetric and conserved on-shell for a Lorentz invariant theory. In the case of WTDiff, the coupling would need to be to the traceless part of the energy-momentum tensor and the trace of such object would need to be constant. 

Hence we need to add to the right hand-side of Eq.~\eqref{eq:schematiclinearized} a term which would correspond to the ``gravitational energy-momentum tensor''. Such a tensor cannot be both gauge invariant and conserved at the same time, since there is no local definition of energy-momentum for the gravitational field in a diffeomorphism-invariant theory. Therefore, either we construct a non-gauge-invariant tensor, or we construct a nonconserved quantity~\cite{Weinberg1980b,Barcelo2005,Barcelo2021b}. We will refer to the gravitational energy-momentum as the conserved tensor computed following any of the procedures outlined below, in Subsection~\ref{Subsec:ComputingEMT}. Adding the energy-momentum tensor of additional matter fields is straightforward. 

Once this gravitational energy-momentum tensor has been added to the right-hand side of Eq.~\eqref{eq:schematiclinearized}, following the bootstrapping procedure, we need to add a piece to the original action in order to derive such a term from an action principle. The main difference with respect to the spin-1 case is that the iterative process is infinite. Thus, finding a consistent nonlinear extension of a linear theory requires finding a general recursion formula. This is a complicated task, if not impossible at all. However, given a nonlinear theory, it is a much simpler task to do the reverse exercise, seeing if it can be reconstructed from its linearization through a bootstrapping mechanism. Specially if the resulting theory is diffeomorphism invariant, following the approach of~\cite{Butcher2009}. One just needs to perform a series expansion on top of flat spacetime, and see if the terms at a given order can be obtained from the ones at a previous order.

\subsection{Computing the energy-momentum tensor}
\label{Subsec:ComputingEMT}
Once we have determined that the energy-momentum tensor is the conserved current to which the graviton field couples, we must specify how to compute it. Generally, there are two approaches to computing the energy-momentum tensor: the canonical approach and the Einstein-Hilbert prescription. The canonical approach defines it as Noether currents associated with the symmetries of flat spacetime, while the Einstein-Hilbert-like prescription derives it through a variational method. Both methods involve ambiguities, which manifest as identically conserved terms that can be added to the currents, i.e., the divergences of superpotentials, as discussed in Section~\ref{SubSec:ConservedCurrent}.
\paragraph*{\textbf{Canonical currents:}}
Let us begin with the canonical definition of the energy-momentum tensor. Consider a field theory described by a Lagrangian of the type $\mathcal{L}=\mathcal{L}(\Phi^I,\partial_\mu\Phi^I)$ which is invariant under the Poincar\'e group. Then, the Noether currents associated with translations, namely the energy-momentum tensor, is given by
\begin{align}
    T_\text{can\,}{}^\mu{}_\nu &:= \frac{\partial \mathcal{L}}{\partial \partial_\mu \Phi^I}\partial_\nu\Phi^I - \mathcal{L} \delta^\mu{}_{\nu}\,.\label{Eq:CanonicalTmunu}
\end{align}
As an explicit example, consider the massless free scalar field, described by
\begin{align}
    S[\Phi] = - \frac{1}{2} \int \dd^{D+1} x\ \eta^{\mu \nu} \partial_{\mu } \Phi \partial_{\nu} \Phi.
    \label{Eq:Scalar_Field_action}
\end{align}
A straightforward computation leads to the following currents
\begin{align}
    T_\text{can}{}_{\mu \nu} = -  \partial_{\mu} \Phi \partial_{\nu} \Phi + \frac{1}{2} \eta_{\mu \nu} \partial_{\rho } \Phi \partial^\rho \Phi,
    \label{Eq:Scalar_Field_EMT}
\end{align}
The physical significance of Noether currents arises from their conservation on-shell, provided the system respects the corresponding global symmetry. However, as specified above, we can always add identically conserved terms that do not spoil the conservation while preserving the same conserved charges. For instance, for the energy-momentum tensor current we can always add a term of the type $\Delta T^{\mu \nu} =  \partial_{\rho} \chi^{[ \rho \mu] \nu}$, where $\chi$ is a three-index tensor built from the fields and their derivatives. To illustrate this, let us explicitly show an example that we will use later: for the case of the massless scalar field in Eq.~\eqref{Eq:Scalar_Field_action}, consider a term of the form
\begin{align}
    \Delta T_{\mu \nu} = \alpha \left( \partial_{\mu} \partial_{\nu} \Phi - \eta_{\mu \nu} \partial^2 \Phi \right),
    \label{Eq:Example_SPT}
\end{align}
where the associated superpotential is
\begin{align}
    \chi^{\rho \mu \nu} = 2  \alpha \partial^{[\rho} \Phi \eta^{\mu] \nu}
    \label{Eq:Example_SPT2}
\end{align}
and $\alpha$ is an arbitrary constant. In some cases, the canonical energy-momentum tensor \eqref{Eq:CanonicalTmunu} is not symmetric, and it is possible to find an appropriate superpotential term to make it symmetric, as famously done through Belinfante's procedure~\cite{Belinfante1940,Ortin2017}. 

\paragraph*{\textbf{Hilbert's prescription:}}

Let us now move on to the other method to compute the energy-momentum, known as Hilbert's prescription. Let us begin by examining theories that can be extended to arbitrary geometries using only the metric field. Note that in theories with spinor fields, the appropriate field variable would be the vielbein, instead, and we will come back to this point later. In Hilbert's prescription, the starting point is a certain action defined in Minkoswki spacetime, and the prescription is implemented in three steps:
\begin{enumerate}
\item Firstly, we extend the action to a curved spacetime by following a minimal-like coupling procedure. If we are only interested in deriving the energy-momentum tensor, it is enough to promote the Minkowski metric to a general one $g_{\mu\nu}$, as well as $\partial_\mu$ to covariant derivatives with respect to the Levi-Civita connection of $g_{\mu\nu}$.

\item The resulting matter action should be understood as a functional of the metric and the matter fields, $S_{\text{M}}[\boldsymbol{g}, \Phi]$. The second step is to vary such an action with respect to the metric.

\item Finally, we evaluate the geometrical fields in the resulting expressions by setting a Minkowski spacetime, which leads to the definition
\end{enumerate}
\begin{align}
    T_{\text{H}}{}_{\mu \nu} &:= \frac{-2}{\sqrt{-g}} \frac{\delta S_{\text{M}}[\boldsymbol{g}, \Phi]}{\delta g^{\mu \nu}} \bigg\rvert_{ g^{\mu \nu} = \eta^{\mu \nu}}. \label{Eq:defTH}
\end{align}
Note that, by construction, the tensors obtained from this procedure are always symmetric under the exchange of their two indices
\begin{equation}
    T_{\text{H}}{}_{\mu \nu}=T_{\text{H}}{}_{(\mu \nu)}\,.
\end{equation}
To relate this procedure more closely to Noether's method for deriving canonical currents, it is important to note that there is a counterpart to the ambiguities encountered in that procedure. Specifically, the freedom to add an identically divergenceless term to the canonical currents now manifests as the possibility of introducing non-minimal couplings in the action, those that vanish identically in flat spacetime (for example, those involving the curvature tensors). To illustrate this more clearly, let us revisit the case of the scalar field theory in Eq.~\eqref{Eq:Scalar_Field_action}. The action in curved spacetime is:
\begin{align}
    S_{\text{M}}[\boldsymbol{g},\Phi] =  - \frac{1}{2} \int \dd^{D+1} x\,\sqrt{-g}\  g^{\mu \nu} \partial_{\mu} \Phi \partial_{\nu} \Phi,
\end{align}
and it is not difficult to check that the corresponding Hilbert energy-momentum tensor coincides with the result in Eq.~\eqref{Eq:Scalar_Field_EMT}. However, one can add to the scalar field action a non-minimal coupling of the type
\begin{align}
    S_{\text{nm}}[\boldsymbol{g}, \Phi] = - \frac{\alpha}{2} \int \dd^{D+1} x\, \sqrt{-g}\ \Phi R(\boldsymbol{g}),
\end{align}
which yields a correction to the Hilbert energy-momentum tensor with exactly the same form as the one arising from the superpotential in Eq.~\eqref{Eq:Example_SPT2} in the canonical prescription. Thus, non-minimal couplings lead to identically conserved pieces in the energy-momentum tensor.

\subsection{Gravity as interacting gravitons: a historical note}
\label{Subsec:Literature}

To contextualize the contributions of this thesis to the understanding of gravity as a theory of interacting gravitons, we find it useful to provide a historical overview of the key results in the literature now that the general framework is presented.

To the best of our knowledge, the first reformulation of GR as an interacting field theory of a massless spin-2 field on Minkowski spacetime was worked out by Rosen~\cite{Rosen1940a,Rosen1940b}, who showed that GR can be rewritten as a field theory on top of Minkowski spacetime. This field carries the degrees of freedom corresponding to the self-interacting massless spin-2 representation of the Poincaré Group, i.e., the graviton~\cite{Wigner1939,Fierz1939,Weinberg1995}.

Later on, Gupta~\cite{Gupta1954} suggested that the structure of GR is such that it would be the only possible nonlinear extension of Fierz-Pauli, suggesting that Rosen's construction of the interacting theory of a massless spin-2 field was unique. There were many works addressing this problem and the uniqueness of the construction (see, e.g.,~\cite{Kraichnan1955} and \cite{Huggins1962}). Among them, we wish to highlight Feynman's contributions \cite{Feynman1996}, where he showed that the unique nonlinear self-consistent theory that one could build starting from the linear massless spin-2 field theory is one in which, order by order, the field couples to its own energy-momentum tensor. This approach is today called a \textit{bootstrapping} procedure. The approach taken in this line of work is that, after adding up the infinite series of self-consistent interactions, one would end up with a nonlinear theory which ideally should be uniquely GR. Later, Deser showed that GR is a solution to the consistency problem by implementing a similar idea in the first order formalism \cite{Deser1970} (see also~\cite{Ortin2015} for a clear presentation of this approach), although uniqueness still slipped away.

In a different line, Wald proposed another approach based on a different guiding principle, which led to similar conclusions \cite{Wald1986,Wald1986b}. To be more specific, instead of focusing on the self-coupling to its own energy-momentum tensor, Wald insisted only on preserving the linear Bianchi identities associated with the gauge invariance of the linearized theory. His consistency conditions were that the divergencelessness of the lowest-order equations needs to be enough to ensure the divergencelessness of the higher order terms entering in the right-hand side, to avoid the appearance of extra restrictions. For spin-1 fields, he showed that this was enough to fully characterize the set of consistent nonlinear theories (Yang-Mills theories). However, in the case of Fierz-Pauli theory, he showed the existence of two families of theories: one corresponding to diffeomorphism-invariant theories and another one which was not diffeomorphism-invariant. In any case, he gave plausibility arguments for discarding the non-diffeomorphism-invariant theories: he argued that they would become inconsistent once one includes matter in the picture~\cite{Wald1986}. A related approach to that of Wald is the one taken by Ogievetsky and Polubarinov~\cite{Ogievetsky1965}, reaching also similar conclusions.

In more recent times, the work of Padmanabhan~\cite{Padmanabhan2008} questioning the uniqueness of Deser's construction attracted some interest back into the problem~\cite{Butcher2009,Deser2010}, and the issue of the uniqueness of the construction does not seem to be fully closed. In fact we will show that in dimensions greater than four, GR is not the only possible nonlinear completion to Fierz-Pauli. In addition to it, the analyses in the literature were affected by the preconceived expectation that the resulting theory would be background independent.

Furthermore, Deser's analysis involves using an auxiliary field which makes the analysis subtle. Indeed, Deser's analysis in the first order formalism includes an extra field in the linear theory that is not actually bootstrapped but used so that the procedure works straightforwardly. This extra auxiliary field ends up being the connection. Whereas this turns out to be mild for bootstrapping Fierz-Pauli to find GR, as GR is equivalent in the Palatini and the metric formulations, for higher derivative theories and metric-affine ones it leads to subtleties that have somehow been overlooked~\cite{Butcher2009,Deser2017,Ortin2017}. In addition to this, even for GR, the proof of Deser is only of existence and not of uniqueness of the construction. 

\section{Bootstrapping metric theories of gravity}
\label{Sec:Bootstrapping_Metric}

Once the idea of the bootstrapping mechanism is clear, along with the ambiguities inherent to it, we can take a step further to address some key issues and misconceptions regarding the bootstrapping of Fierz-Pauli and WTDiff. First of all, we exemplify the nonuniqueness of the construction showing how different superpotentials added to the current may yield to different final theories. The derivation of the main formulas that we use in this section and the next one are presented in Appendix~\ref{AppB}. In particular, we show that: i) Different nonlinear theories can arise from Fierz-Pauli at the nonlinear level for dimensions greater than four, ii) the bootstrapping of WTDiff can be done in such a way that one ends up finding a background-dependent theory: UG. 

\subsection{Nonuniqueness of the construction}

As emphasized above, the absence of a unique conserved energy-momentum tensor means that different choices of superpotentials can actually lead to different resulting theories. To illustrate explicitly how two different nonlinear theories can arise from the same linear theory, it is convenient to perform a reverse engineering exercise, following the idea's presented in~\cite{Butcher2009}. 

Consider a nonlinear diffeomorphism and background independent theory depending only on the metric $g_{\mu \nu}$ as dynamical variable. Assume that the theory is described by the following action principle $S[\boldsymbol{g}]$. For the metric, we can perform an expansion of the form
\begin{align}
    g^{\mu\nu} = \bar{g}^{\mu\nu} + \lambda h^{\mu\nu},
\end{align}
where we are choosing $\bar{g}^{\mu\nu}$ to be a \textit{generic} solution to the vacuum field equations, i.e.,
\begin{align}
    \frac{\delta S [\boldsymbol{g}]}{\delta g^{\mu\nu}} \bigg\rvert_{g^{\mu\nu} = \bar{g}^{\mu\nu}} = 0,
    \label{Eq:Background}
\end{align}
and $h^{\mu\nu}$ represents the deviation with respect to the nondynamical background $\bar{g}^{\mu\nu}$. The parameter $\lambda$ could be absorbed in $h^{\mu \nu}$ but we will use it as a bookkeeping device. Explicitly we have that the expansion is given by
\begin{align}
    S[\boldsymbol{g}] = \sum_{n = 0}^{\infty} \lambda^n S^{(n)} [\boldsymbol{\bar{g}}, \boldsymbol{h}], 
    \label{Eq:FullAction}
\end{align}
with the partial actions reading
\begin{align}
    S^{(n)} [\boldsymbol{\bar{g}}, \boldsymbol{h}] = \frac{1}{n!} \frac{\dd^n}{\dd \lambda^n} S [\boldsymbol{\bar{g}} + \lambda \boldsymbol{h}] \bigg\rvert_{\lambda  = 0}.
    \label{Eq:Partial_Actions}
\end{align}
Note that the $n$-th partial action contains $n$ powers of the $h^{\mu \nu}$ field. Given a field configuration for $\bar{g}^{\mu\nu}$ and $h^{\mu\nu}$, the derivatives with respect to $\lambda$ are just total derivatives for a function of a real variable. However, they can be written as functional derivatives with respect to the background metric by means of the relation
\begin{align}
    \frac{\dd}{\dd \lambda} S[\boldsymbol{\bar{g}} + \lambda \boldsymbol{h}] = \int \dd^{D+1} x \ h^{\mu\nu} (x) \frac{\delta}{\delta \bar{g}^{\mu\nu} (x) } S[\boldsymbol{\bar{g}} + \lambda \boldsymbol{h}].
    \label{eq:LambdaDerivativeAsFunctionalDerivative}
\end{align}
To perform variations with respect to $\boldsymbol{\bar{g}}$, it is important that we consider it to be an arbitrarily curved metric in the expressions that involve its functional derivative. Eq.~\eqref{eq:LambdaDerivativeAsFunctionalDerivative} holds up to surface integrals that arise when integrating by parts. From now on, we neglect them as they contribute neither to the equations of motion nor to the energy-momentum tensor.

By applying repeated differentiation, we can express $n$-th $\lambda$-derivatives as
\begin{align}
    \frac{\dd^n}{\dd \lambda^n} S[\boldsymbol{\bar{g}} + \lambda \boldsymbol{h}] = \left[ \int \dd^{D+1} x\ h^{\mu\nu}(x) \frac{\delta}{\delta \bar{g}^{\mu\nu} (x) }\right]^n S[\boldsymbol{\bar{g}} + \lambda \boldsymbol{h}].
    \label{Eq:Partial_Derivatives}
\end{align}
Therefore, we can combine Eqs.~\eqref{Eq:Partial_Actions} and~\eqref{Eq:Partial_Derivatives} to express all the higher-order partial actions $S^{(n)}$, for $n>2$, in terms of derivatives of $S^{(2)}[\bar{g}, h]$ as
\begin{align}
    S^{(n)}[\boldsymbol{\bar{g}}, \boldsymbol{h}] =  \frac{2}{n!} \left[ \int \dd^{D+1} x\ h^{\mu\nu}(x) \frac{\delta}{\delta \bar{g}^{\mu\nu} (x) }\right]^{n-2} S^{(2)}[\boldsymbol{\bar{g}}, \boldsymbol{h}].
    \label{Eq:SnfromS2}
\end{align}
More details on the derivation of this equation can be found in Appendix~\ref{App:General_Fields_genformula}.

Note that, from the bottom-up approach, i.e., starting from the linear theory in Minkowski spacetime, we do not know the full functional form of $S^{(2)}[\boldsymbol{\bar{g}}, \boldsymbol{h}]$ for any arbitrary background metric $\bar{g}^{\mu\nu}$, but only $S^{(2)}[\eta, h]$ in flat spacetime. To use the generating formula \eqref{Eq:SnfromS2}, we need to promote  $\eta^{\mu\nu}\to\bar{g}^{\mu\nu}$ with the appropriate non-minimal couplings. 

From Eq.~\eqref{Eq:SnfromS2} it is possible to show that metric perturbations couple order by order to their own energy-momentum tensor. Specifically, at each order in the expansion $S^{(n)}$, the perturbation $h^{\mu\nu}$ couples to the energy-momentum tensor derived from the preceding term $S^{(n-1)}$. To explicitly verify this, we note that, to the lowest order in the expansion, the equations of motion for the metric perturbation are:
\begin{align}
    \lambda^2  \frac{1}{\sqrt{-\bar{g}}} \frac{\delta S^{(2)} [\boldsymbol{\bar{g}}, \boldsymbol{h}]}{\delta h^{\mu\nu}} = \order{\lambda^3}.
    \label{Eq:Lowest_Order_Bootstrapping}
\end{align}
In the case that $S[\boldsymbol{g}]$ is the Einstein-Hilbert action, this would simply yield to the Fierz-Pauli equation with non-minimal coupling terms that arise when we linearize the action around arbitrary backgrounds. We will come back to this example later. Now, we want to check that $h^{\mu\nu}$ couples to its own energy-momentum tensor. For such purpose, we can apply Hilbert's prescription, according to which the energy-momentum tensor associated with $S^{(2)}[\boldsymbol{\bar{g}}, \boldsymbol{h}]$ is
\begin{align}
    t^{(2)}_{\mu\nu} := - \frac{\lambda^2}{\sqrt{-\bar{g}}} \frac{\delta S^{(2)} [\boldsymbol{\bar{g}}, \boldsymbol{h}]}{\delta \bar{g}^{\mu\nu}}.
\end{align}
Strictly speaking, it would only correspond to the energy-momentum tensor that we have defined in Section~\ref{Sec:Bootstrapping} once we evaluate on $\bar{g}^{\mu\nu} = \eta^{\mu\nu}$ and up to a factor $2$, which we will absorb in the energy-momentum tensor definition for convenience. To build all the partial actions, we need to maintain the background metric $\bar{g}^{\mu\nu}$ arbitrary, i.e., we need to know the partial actions $S^{(n)}[\boldsymbol{\bar{g}},\boldsymbol{h}]$ in an open neighborhood of the flat spacetime metric $\eta^{\mu\nu}$. Now, we want $t^{(2)}_{\mu\nu}$ to appear to the next order on the right-hand side of Eq.~\eqref{Eq:Lowest_Order_Bootstrapping}, i.e., we want the terms $\order{\lambda^3}$ to be precisely $\lambda t^{(2)}_{\mu\nu}$. Furthermore, we need this term to be derivable from an action upon variation of $h^{\mu\nu}$, i.e., we need a partial action $S^{(3)} [\boldsymbol{\bar{g}}, \boldsymbol{h}]$ such that
\begin{align}
     \frac{\lambda^3}{\sqrt{-\bar{g}}} \frac{\delta S^{(3)} [\boldsymbol{\bar{g}}, \boldsymbol{h}]}{\delta h^{\mu\nu}} = \lambda t^{(2)}_{\mu\nu},
\end{align}
in such a way that $\lambda t_{\mu\nu}^{(2)}$ acts as source for the quadratic field equations of $h^{\mu\nu}$. Doing this to all orders would ensure that the field $h^{\mu\nu}$ couples to its own energy-momentum tensor order by order. This would be fulfilled trivially if we manage to show that the following equation holds at all orders in the perturbative expansion
\begin{align}
    \frac{\lambda^n}{\sqrt{- \bar{g}}}\frac{\delta S^{(n)} [\boldsymbol{\bar{g}},\boldsymbol{h}]}{\delta h^{\mu\nu}} = \lambda t^{(n-1)}_{\mu\nu}.
    \label{Eq:Bootstrap_Eq_EMT}
\end{align}
Given the definition of the energy-momentum tensor through Hilbert's prescription, we have that the $n$-th order partial action contributes to the energy-momentum tensor as
\begin{align}
    t^{(n)}_{\mu\nu} := - \frac{\lambda^{n}}{\sqrt{-\bar{g}}} \frac{\delta S^{(n)} [\boldsymbol{\bar{g}}, \boldsymbol{h}]}{\delta \bar{g}^{\mu\nu}},
    \label{Eq:Gravitational_EMTensor}
\end{align}
and then we have that Eq.~\eqref{Eq:Bootstrap_Eq_EMT} can be rewritten as the key expression
\begin{align}
    \frac{\delta S^{(n)} [\boldsymbol{\bar{g}},\boldsymbol{h}]}{\delta h^{\mu\nu}} = \frac{\delta S^{(n-1)} [\boldsymbol{\bar{g}},\boldsymbol{h}]}{\delta \bar{g}^{\mu\nu}}.
    \label{Eq:Key_Bootstrap}
\end{align}
That this holds for arbitrary background independent metric theories is proved in Appendix~\ref{App:General_Fields_funidentity}, see Eq.~\eqref{Eq:Key_Bootstrap_Gen}. Furthermore, since the action is diffeomorphism-invariant, the discussion in Subsection~\ref{Subsec:ComputingEMT} guarantees that the right-hand side is precisely what we have been calling the energy-momentum tensor. 

\paragraph*{\textbf{Example: General Relativity.}}

Let us now work out a simple example by applying this procedure to the linearized version of GR. The Einstein-Hilbert action is given by
\begin{align}
    S_{\text{EH}}[\boldsymbol{g}] = \frac{1}{2 \kappa_D}\ \int \dd^{D+1} x \sqrt{-g}\, R (\boldsymbol{g}),
\end{align}
where $R(\boldsymbol{g})$ represents the Ricci scalar tensor of the metric $g_{\mu\nu}$ and $\kappa_D$ is the corresponding Einstein constant of the chosen dimension. Now, we would need to expand the action in terms of a general metric $\bar{g}^{\mu \nu}$ and its perturbations on top of it, $h^{\mu \nu}$. This would result in a structure for the action which would be
\begin{align}
    S_{\text{EH}}[\boldsymbol{\bar{g}} + \lambda \boldsymbol{h}] = \frac{1}{4 \kappa_D} \int \dd^{D+1} x \sqrt{- \bar{g}} &\sum_{n=2}^{\infty} \Big[  M_{(n)}{}^{\alpha_1 \alpha_2}{}_{\mu_1 \nu_1 \ldots \mu_n \nu_n } \bar{\nabla}_{\alpha_1 } h^{\mu_1 \nu_1} \bar{\nabla}_{\alpha_2}  h^{\mu_2 \nu_2}  h^{\mu_{3} \nu_{3} } \ldots h^{\mu_{n} \nu_{n}} \nonumber \\
    &\qquad  + H_{(n) \mu_1 \nu_1 \ldots \mu_n \nu_n} h^{\mu_1 \nu_1 } \ldots h^{\mu_n \nu_n}\Big],
    \label{Eq:SEH_MH}
\end{align}
where the tensors $M_{(n)}$ and $H_{(n)}$ are built only from curvature invariants, their covariant derivatives, the background metric tensor $\bar{g}_{\mu\nu}$, its inverse $\bar{g}^{\mu\nu}$, and the Kronecker $\delta^{\mu}{}_{\nu}$ tensor. This action clearly has the structure of Eq.~\eqref{Eq:Partial_Derivatives}. As we have discussed and show explicitly in Appendix~\ref{App:General_Fields_genformula}, only the first term in this series is required to reconstruct the whole action, for which we have:
\begin{align}
    M_{(2)}{}^{\alpha \beta}{}_{\mu\nu \rho \lambda} &= -\frac{1}{2} \left[ \bar{g}^{\alpha\beta} \bar{g}_{ \mu (\rho} \bar{g}_{\lambda) \nu} - \bar{g}^{\alpha\beta}\bar{g}_{\mu\nu} \bar{g}_{\rho\lambda} - 2 \delta^{\alpha}{}_{(\rho} \bar{g}_{\lambda) (\mu} \delta^{\beta}{}_{\nu)} + \delta^{\alpha}{}_{(\rho} \delta^{\beta}{}_{\lambda)} \bar{g}_{\mu \nu} + \delta^{\beta}{}_{(\mu} \delta^{\alpha}{}_{\nu)} \bar{g}_{\rho\lambda} \right]\,, \nonumber \\
    H_{(2) \mu \nu \rho \lambda}  &=  \frac{1}{2} R \left( \boldsymbol{\bar{g}} \right) \left( \bar{g}_{\mu \rho} \bar{g}_{\lambda \nu} + \frac{1}{2} \bar{g}_{\mu \nu} \bar{g}_{\rho \lambda} \right) - R_{ \mu \nu} \left( \boldsymbol{\bar{g}} \right) \bar{g}_{\rho\lambda}
\end{align}
(note that $M_{(2)}{}^{\alpha \beta}{}_{\mu\nu \rho \lambda}$ yields the Fierz-Pauli term in an arbitrary coordinate system). Although $H_{(n) \mu_1 \nu_1 \ldots \mu_n \nu_n}$ is identically zero when evaluated on solutions of the vacuum equations of motion, i.e., Ricci-flat backgrounds, it is needed for the self-coupling procedure to yield a coupling between $h^{\mu\nu}$ and its own energy-momentum tensor: the variation with respect to those terms gives a nonvanishing contribution. This is an instance of the arbitrariness that appears when writing the flat spacetime action on an arbitrarily curved spacetime, and, as such, it leads to an identically conserved current. However, we emphasize again that they are crucial to reconstruct the action~\eqref{Eq:FullAction} from the quadratic action in Eq.~\eqref{Eq:SnfromS2}. 

Let us now discuss the problem raised by Padmanabhan in~\cite{Padmanabhan2008} regarding the statement that the bootstrapping of the Fierz-Pauli action actually yields GR. In that reference, he argued that if one takes the Fierz-Pauli action in flat spacetime and replaces the flat metric and its Levi-Civita compatible derivatives with those of an arbitrary background, one is not able to reconstruct the whole Einstein-Hilbert action. This was also mentioned in~\cite{Butcher2009} and interpreted as saying that it is not possible to recover GR from the linear Fierz-Pauli theory through a bootstrapping procedure. However, we think that this conclusion is reached due to them having a very strict definition of the energy-momentum tensor. Such definition does not allow the addition of superpotentials that do not contribute to the Noether charges in flat spacetime and, therefore, should be mild from the point of view of the Noether current. 

In order for the bootstrapping to work as expected, the addition of the correct superpotential, or accordingly the correct non-minimal coupling, proves to be crucial. Indeed, if we apply the bootstrapping procedure to the Fierz-Pauli action, we only recover GR if we add the non-minimal coupling $H_{(2)}$, as well as the rest of the non-minimal couplings arising at each order in the process, $H_{(n)}$. Otherwise, we are not granted to find a diffeomorphism-invariant theory which couples to its own energy-momentum.

\paragraph*{\textbf{Nonuniqueness of the construction: Lovelock theories.}}

However, it is convenient to note that the choice of $H_{(2)}$, and more in general, all the $H_{(n)}$ terms, that lead to a consistent nonlinear theory are not unique. In fact, for the purpose of illustrating the nonuniqueness of the construction it is convenient to think of spacetime dimensions greater than four. In such spacetime dimensions, there is a celebrated theorem by Lovelock~\cite{Lovelock1971}, which characterizes all the theories that give rise second order equations. In $D+1$ spacetime dimensions, the most general Lagrangian is the following\footnote{Notice that we are ignoring the cosmological constant since flat spacetime is not a solution to the vacuum equations in the presence of a cosmological constant.}
\begin{align}
    \mathcal{L} = \sqrt{-g} \sum_{n=1}^{N} a_n \delta^{\mu_1}{}_{[\alpha_1}  \delta^{\nu_1}{}_{\beta_1} \cdots \delta^{\mu_n}{}_{\alpha_n} \delta^{\nu_n}{}_{\beta_n]} \prod_{r=1}^{n} R^{\alpha_r \beta_r}{}_{\mu_r \nu_r} \left( \boldsymbol{g} \right),
\end{align}
where we have $N = (D-2)/2$ for even spacetime dimensions and $N = (D-1)/2$, and $a_n$ represent the coupling constants of the theory. All the terms that are only included in greater dimensions can be ignored in lower dimensions since they lead to boundary terms, i.e., they are topological in lower dimensions. For instance, the term for $n=2$ in the sum is simply the Gauss-Bonnet term, i.e. \mbox{$ \mathcal{L}_{GB}:= \sqrt{-g}\left( R^2 - 4R_{\mu \nu} R^{\mu \nu} + R^{\alpha \beta \mu \nu} R_{\alpha \beta \mu \nu} \right)$}, which is a topological invariant in $D=3$~\cite{Nakahara2003}. Given that the equations following from such Lagrangian are second order and diffeomorphism invariant, their linearization on top of flat spacetime leads to Fierz-Pauli theory. In fact, if we linearize them on top of an arbitrary background we find that they only differ from the Einstein-Hilbert action in the $H_{(2)}$ terms, that are different to the Einstein-Hilbert counterpart, and different among the different theories. 

At this point, it is clear that any Lovelock theory can be obtained from the Fierz-Pauli action at quadratic order, together with the identically conserved terms in the energy-momentum tensor (the $H_{(n)}$ terms) derived from the nonlinear theory through reverse engineering \emph{\`a la} Butcher~\cite{Butcher2009}. In fact, it is straightforward that any Lovelock theory will obey the recursion relations derived above, namely Eqs.~\eqref{Eq:FullAction} and~\eqref{Eq:SnfromS2}, since the only requirement imposed for their derivation is that the original action is diffeomorphism invariant. 

Thus, we conclude that any Lovelock theory, including GR, could be obtained from Fierz-Pauli depending on the superpotentials added, i.e., the energy-momentum tensor used in the construction. This example clearly illustrates the nonuniqueness of the construction, even if we assume \emph{ab initio} that the resulting theory should be diffeomorphism invariant and expressible solely in terms of curvature scalars built from a metric tensor $g_{\mu \nu}$, at least in dimensions greater than four.

\paragraph*{\textbf{Strongly coupled theories.}}

One can wonder whether this analysis is exhaustive, in the sense that whether all the diffeomorphism invariant theories that can be obtained from Fierz-Pauli through the bootstrapping mechanism are Lovelock theories. Although they constitute highly pathological examples, there is the possibility of having nonlinear theories that contain higher order derivatives, and as such, they propagate more degrees of freedom than the two of Fierz-Pauli, but still their linearization on top of flat spacetime degenerates and leads to Fierz-Pauli theory. These theories have the property that the flat spacetime is a strongly coupled background~\cite{Delhom2022}. A strongly coupled background in a theory means that such background is a solution to the equations of motion, but the perturbations on top of it propagate less degrees of freedom than those that the full theory has. This means that from a space of solutions point of view, the solution is pathological since it constitutes an isolated point. In fact, in most of the cases, strongly coupled backgrounds act as separatrixes in the space of solutions, separating different behaviors of the solutions. In any case, these backgrounds tend to be unstable since the dynamics tries to push the system away: any generic perturbation tends to excite the additional degrees of freedom and makes the system flows away from such background.

Although a blind application of the reverse engineering technique that we have used could be used to argue that these theories can be obtained through a bootstrapping mechanism, the whole scheme would be rather inconsistent. In that sense, the full nonlinear theory, which would propagate more degrees of freedom than the Fierz-Pauli theory taken as starting point, would have been obtained from a perturbative expansion on top of a singular background. The very first step in which we generalize the Fierz-Pauli action to that specific curved background action obtained from the nonlinear theory would be introducing the additional degrees of freedom that the nonlinear theory has. In that sense, although the radius of convergence of the expansion in $h^{\mu \nu}$ is generally unknown, it is clear that for such strongly coupled backgrounds it becomes zero since the expansion is singular.

An example of a theory in which the flat spacetime displays this strong coupling property is Einstenian cubic gravity~\cite{Bueno2016}. The quadratic order in perturbation theory reduces to the Fierz-Pauli action on top of maximally symmetric spacetimes, but in generic backgrounds the theory propagates more degrees of freedom. In fact, the analyses reported in~\cite{BeltranJimenez2020,BeltranJimenez2023} pinpoint the instability of those strongly coupled backgrounds, as expected on general grounds. 

Thus, we conclude that aside from Lovelock theories, we expect that no other diffeomorphism invariant theory in which the flat spacetime is not a singular solution, namely that it is not strongly coupled, can be obtained through a consistent self-coupling of the graviton degrees of freedom from the Fierz-Pauli action. The next section is devoted to discussing precisely what happens if the resulting theories are not diffeomorphism invariant.

\subsection{Role of background structures} 

Let us now move on and consider the possibility that the theory resulting from the bootstrapping retains the background structure present at the outset. In other words, that the flat spacetime metric $\eta_{\mu \nu}$ does not fully decouple in the final theory. Although Eq.~\eqref{Eq:SnfromS2} might suggest that for any $S^{(2)}[\boldsymbol{\bar{g}},\boldsymbol{h}]$ we can generate a whole series of partial actions $S^{(n)}[\boldsymbol{\bar{g}},\boldsymbol{h}]$, a theory generated in this way does not couple to its own energy-momentum tensor. In fact, for the proof of Eq.~\eqref{Eq:Key_Bootstrap}, it was crucial that the $S^{(2)}[\boldsymbol{\bar{g}},\boldsymbol{h}]$ action arose as the quadratic order of an expansion for a background independent action $S[\boldsymbol{g}]$. Arbitrary choices of $S^{(2)}[\boldsymbol{\bar{g}},\boldsymbol{h}]$ would not result in a finally diffeomorphism invariant theory. For instance, this would happen for the actions considered in~\cite{Padmanabhan2008,Barcelo2014}.

\paragraph*{\textbf{Arbitrary choices of $S^{(2)}[\boldsymbol{\bar{g}},\boldsymbol{h}]$ do not bootstrap.}}

An explicit example that illustrates this point is the following action $S_{\text{ex}}^{(2)}[\boldsymbol{\bar{g}},\boldsymbol{h}]$
\begin{align}
    S^{(2)}_{\text{ex}}[\boldsymbol{\bar{g}},\boldsymbol{h}] = \int \dd^{D+1} x \sqrt{-\bar{g}}  \bar{g}_{\mu \nu} \bar{g}_{\rho \sigma} h^{\mu \nu} h^{\rho \sigma}.
\end{align}
We could naively define $S_{\text{ex}}^{(3)}[\boldsymbol{\bar{g}},\boldsymbol{h}]$ using Eq.~\eqref{Eq:SnfromS2}
\begin{align}
    S_{\text{ex}}^{(3)}[\boldsymbol{\bar{g}},\boldsymbol{h}] = \frac{2}{3!} \int \dd^{D+1}  x h^{\mu \nu} \frac{\delta}{\delta \bar{g}^{\mu \nu}} S_{\text{ex}}^{(2)}[\boldsymbol{\bar{g}},\boldsymbol{h}], 
\end{align}
and given that 
\begin{align}
    \frac{\delta}{\delta \bar{g}^{\mu \nu}} S_{\text{ex}}^{(2)}[\boldsymbol{\bar{g}},\boldsymbol{h}] = - \sqrt{-\bar{g}} \left( + \frac{1}{2} \bar{g}_{\mu \nu} h^2 + 2 h_{\mu \nu} h \right),
    \label{Eq:ExampleS2g}
\end{align}
where we are using the notation $h:= \bar{g}_{\mu \nu} h^{\mu \nu}$, we find
\begin{align}
    S_{\text{ex}}^{(3)}[\boldsymbol{\bar{g}},\boldsymbol{h}] = -  \int \dd^{D+1}  x \sqrt{-\bar{g}} \left[  \frac{1}{6}h^3 + \frac{2}{3} h^{\mu \nu} h_{\mu \nu} h  \right].
\end{align}
Now we can perform the variation with respect to $h^{\mu \nu}$, and we find:
\begin{align}
    \frac{\delta }{\delta h^{\mu \nu}} S^{(3)}_{\text{ex}}[\boldsymbol{\bar{g}},\boldsymbol{h}] = - \sqrt{ - \bar{g}} \left[ \frac{1}{2} h^2 \bar{g}_{\mu \nu} + \frac{2}{3} \bar{g}_{\mu \nu} \bar{g}_{\alpha \tau} \bar{g}_{\beta \sigma} h^{\alpha \beta} h^{\tau \sigma} + \frac{4}{3} h h^{\rho \sigma} \bar{g}_{\rho \mu} \bar{g}_{\sigma \nu} \right],
\end{align}
which clearly disagrees with Eq.~\eqref{Eq:ExampleS2g}:
\begin{align}
    \frac{\delta S_{\text{ex}}^{(3)} [\boldsymbol{\bar{g}},\boldsymbol{h}]}{\delta h^{\mu\nu}} \neq \frac{\delta S_{\text{ex}}^{(2)} [\boldsymbol{\bar{g}},\boldsymbol{h}]}{\delta \bar{g}^{\mu\nu}}.
\end{align}
Thus, we see that a fundamental requisite for our derivation of Eq.~\eqref{Eq:Key_Bootstrap} to hold is that the partial actions arise from the expansion of a theory that only depends on the combination $\boldsymbol{\bar{g}} + \lambda \boldsymbol{h}$. If it is not the case and we simply generate the partial actions using Eq.~\eqref{Eq:SnfromS2} from an arbitrarily chosen $S^{(2)}_{\text{ex}}[\boldsymbol{\bar{g}},\boldsymbol{h}]$, the probability that it self-couples to its own energy-momentum tensor is really low. This clearly happens for the example that we have chosen, and it is clear that such an action cannot arise from the linearization of any background independent theory (it does not contain any derivatives of the graviton field). In that sense, the approach that we are taking here is only convenient to see if a nonlinear theory can be reconstructed from this bootstrapping mechanism, although it is not so successful from the point of view of generating a nonlinear theory from a linear one. 

To highlight the challenges of dealing with theories that incorporate background structures, consider a fully nonlinear theory derived from Fierz-Pauli or WTDiff in such a way that the background (flat) metric cannot simply be absorbed through a field redefinition. In this case, we would obtain a nonlinear theory of the form $S_{BD}[\boldsymbol{\eta},\boldsymbol{h}]$. It is convenient to reinterpret this as a bimetric theory by redefining\footnote{In general, there is no guarantee that $\boldsymbol{g}$ is a Lorentzian metric, as it may not even possess a well-defined signature.} $\boldsymbol{h} \to \boldsymbol{g} - \boldsymbol{\eta}$, leading to an action of the form $S_{\text{BD}} [ \boldsymbol{\eta}, \boldsymbol{g}]$. Performing the same expansion on top of an arbitrary background we can find a set of partial actions
\begin{align}
    S_{\text{BD}} [\boldsymbol{\eta}, \boldsymbol{\bar{g}} + \boldsymbol{h}] = \sum_{n=0}^{\infty} S^{(n)}_{\text{BD}} [\boldsymbol{\eta}, \boldsymbol{\bar{g}},  \boldsymbol{h}] ,
\end{align}
which will still obey the recursion relation
\begin{align}
    \frac{\delta S_{\text{BD}}^{(n)} [\boldsymbol{\eta}, \boldsymbol{\bar{g}},  \boldsymbol{h}]}{\delta h^{\mu\nu}} = \frac{\delta S_{\text{BD}}^{(n-1)} [\boldsymbol{\eta}, \boldsymbol{\bar{g}},  \boldsymbol{h}]}{\delta \bar{g}^{\mu\nu}}.
\end{align}
However, it is no longer possible to interpret the right-hand side as the energy-momentum tensor, given the presence of the two metrics $\boldsymbol{\eta}, \boldsymbol{\bar{g}}$. Although it does not look impossible to find bimetric theories that can be consistently obtained from Fierz-Pauli or WTDiff procedure through a bootstrap procedure we have not been able to find any of them. However, we have been able to show that WTDiff can be completed into a nonlinear theory which inherits the background volume form of flat spacetime. 

\paragraph*{\textbf{UG from bootstrapping WTDiff.}}

UG is closely related to GR but differs in certain aspects, particularly in the treatment of the cosmological constant. In the next chapter, we will provide a systematic comparison between GR, its generalizations, and their UG counterparts. For now, we briefly outline the simplest formulation of UG to illustrate how it arises as a consistent self-coupling of WTDiff.

The formulation of the unimodular version of any general relativistic action $S_{\text{GR}}[\boldsymbol{g}]$ requires to introduce a privileged background volume form $\boldsymbol{\omega} = \omega (x)~\dd x^0 \wedge ... \wedge \dd x^{D}$. Then, we introduce an auxiliary metric $\tilde{g}_{\mu \nu}$ defined in terms of our dynamical metric $g_{\mu \nu}$ and the scalar density $\omega$ as 
\begin{equation}
    \tilde{g}_{\mu \nu} (\boldsymbol{g}) = g_{\mu \nu} \left( \frac{\omega^2}{\abs{g}} \right)^{\frac{1}{D+1}}.
    \label{Eq:Auxiliary_Metric}
\end{equation}
With this, we can write an action for the dynamical variable $g_{\mu \nu}$, taking advantage of this auxiliary metric as
\begin{equation}
    S_{\text{UG}}[\boldsymbol{g}; \boldsymbol{\omega}] = S_{\text{EH}} [\boldsymbol{\tilde{g}(g)}].
\end{equation}
One can work out the equations of motion for this theory and immediately find out that the equations of motion are the traceless part of the equations of motion of the general relativistic counterpart~\cite{Carballo-Rubio2022}. Explicitly, we have the following
\begin{equation}
   \frac{\delta S_{\text{UG}}[\boldsymbol{g}; \boldsymbol{\omega}]}{\delta g^{\mu \nu}} = E_{\mu \nu} [\boldsymbol{\tilde{g}(g)}] - \frac{1}{D+1} \tilde{g}_{\mu \nu} \tilde{g}^{\alpha \beta} E_{\alpha \beta} [\boldsymbol{\tilde{g}(g)}],
\end{equation}
where
\begin{align}
E_{\mu \nu} [\boldsymbol{g}] := \dfrac{\delta S_{\text{GR}}[\boldsymbol{g}]}{\delta g^{\mu \nu}}.
\end{align}
The gauge group of these theories is not the full group of diffeomorphisms. Instead, they are invariant under transverse diffeomorphisms and Weyl scalings. Explicitly, these transformations are:
\begin{align}
    & \delta_{\xi} g_{\mu \nu} = 2 \nabla_{( \mu} \xi_{\nu)}, \qquad \nabla_{\mu} \xi^{\mu} = - \frac{1}{2} \xi^{\sigma} \partial_{\sigma} \log \left( \frac{\omega^2}{\abs{g}} \right), \\
    & \delta_{\varphi} g_{\mu \nu} =  \varphi g_{\mu \nu}.
\end{align}
If one includes matter or additional fields, they need to be coupled through the metric $\tilde{g}_{\mu \nu} (\boldsymbol{g})$. The equations that one obtains in this way $S_{\text{M}} [\boldsymbol{g}, \Phi]$ are again the traceless equations of motion of the general relativistic equations of motion, i.e.,
\begin{align}
    E_{\mu \nu} \left[\boldsymbol{\tilde{g} (g)} \right] - \frac{1}{D+1} \tilde{g}_{\mu \nu} \tilde{g}^{\alpha \beta} E_{\alpha \beta} \left[ \boldsymbol{\tilde{g}(g)}\right] = T_{\mu \nu} \left[\boldsymbol{\tilde{g} (g)} \right] - \frac{1}{D+1} \tilde{g}_{\mu \nu} \tilde{g}^{\alpha \beta} T_{\alpha \beta }\left[\boldsymbol{\tilde{g} (g)} \right] .
    \label{Eq:Unimodular_Matter_Eqs}
\end{align}
Here the energy-momentum tensor is computed through Hilbert's prescription, i.e., given $S_{\text{M}} [\boldsymbol{g},\Phi]$, the energy-momentum tensor is directly computed as
\begin{align}
    T_{\mu \nu} [\boldsymbol{g}]= \frac{-2}{\sqrt{-g}}  \frac{\delta S_{\text{M}} [\boldsymbol{g}, \Phi]}{\delta g^{\mu \nu}}.
\end{align}
Taking the divergence on both sides in Eq.~\eqref{Eq:Unimodular_Matter_Eqs} and performing a trivial integration, one can find that a cosmological constant appears as an integration constant (more on this in the next chapter), and the equations can be rewritten as exactly their general relativistic counterpart. We will come back to this point in the next chapter.

Let us now briefly discuss the effect of using this alternative symmetry principle to build theories and their bootstrapping. In particular, we will again invoke a reverse engineering exercise~\cite{Butcher2009}. We perform the expansion of the metric again in a background plus a perturbation of the form $g^{\mu \nu} = \bar{g}^{\mu \nu} + \lambda h^{\mu \nu}$, where the background is taken to be a solution to the equations of motion. In particular, if we particularize for $\bar{g}^{\mu \nu} = \eta^{\mu \nu}$ and we take $\boldsymbol{\omega}$ to be the background volume form of flat spacetime, $\omega = \sqrt{-\eta}$, we find that the quadratic term for $h$ is precisely the WTDiff theory presented in Section~\ref{Sec:LinearSpin2}.

For these theories, the recursion relation that we found is still valid even in the presence of the background form
\begin{equation}
    \frac{\delta S^{(n-1)} [\boldsymbol{\omega}, \boldsymbol{\bar{g}}, \boldsymbol{h}]}{\delta \bar{g}^{\mu \nu} } = \frac{\delta S^{(n)} [\boldsymbol{\omega}, \boldsymbol{\bar{g}}, \boldsymbol{h}]}{\delta h^{\mu \nu}}.
\label{Eq:Bootstrap_Unimodular}
\end{equation}
We now take into account that metric $g_{\mu \nu}$ always enters the action through the auxiliary metric $\tilde{g}_{\mu \nu}$ and we have that
\begin{equation}
    \frac{ \delta S^{(n)}[\boldsymbol{\omega}, \boldsymbol{\tilde{g}(\bar{g})}, \boldsymbol{h}]}{\delta \bar{g}^{\mu \nu}} = \frac{\delta S^{(n)} [\boldsymbol{\omega}, \boldsymbol{\tilde{g}(\bar{g})}, \boldsymbol{h}]}{\delta \tilde{g}^{\mu \nu}} - \frac{1}{D+1} \tilde{g}_{\mu \nu} \tilde{g}^{\alpha \beta } \frac{\delta S^{(n)} [\boldsymbol{\omega}, \boldsymbol{\tilde{g}(\bar{g})}, \boldsymbol{h}]}{\delta \tilde{g}^{\alpha \beta}}.
\end{equation}
We can identify the energy-momentum tensor from the definition above, at order $n-1$ in $\lambda$, to find the following expression
\begin{equation}
    \frac{\delta S^{(n)} [ \boldsymbol{\omega}, \boldsymbol{\bar{g}}, \boldsymbol{h}]}{\delta h^{\mu \nu}}  = -  \sqrt{-\bar{g}} \left( t^{(n-1)}_{\mu \nu}   - \frac{1}{D+1} \bar{g}_{\mu \nu} \bar{g}^{\alpha \beta} t^{(n-1)}_{\alpha \beta} \right).
\end{equation}
This expression simply indicates that in this case, $h^{\mu \nu}$ couples to the traceless part of its own energy-momentum tensor rather than the full energy-momentum tensor. 

This analysis shows that it is possible to construct self-consistent theories in which the background structure still appears in the formulation of the nonlinear theory. We have found an explicit example: UG. It is a nonlinear completion of WTDiff obtained via a bootstrapping procedure also, which retains part of the flat spacetime metric structure: the volume form associated with it. 

As discussed above, the possibility of genuinely bimetric theories, where the flat spacetime metric still appears in the nonlinear theory, remains an open question, as we have not yet identified any. However, we see no fundamental obstruction to their existence. In fact, it seems plausible that the theories introduced by Wald in~\cite{Wald1986,Wald1986b} could also emerge from the bootstrapping approach outlined here, though verifying this would require additional analytical tools. 

\section{Extension to metric-affine theories}

In this section, we extend the bootstrapping mechanism to theories with more than the two degrees of freedom associated with the graviton. Specifically, we explore the bootstrapping of higher-order theories of gravity from their linearizations. We study theories containing more than two time derivatives and in particular, metric-affine theories, where the connection is treated as an independent dynamical field rather than being constrained to be the Levi-Civita connection \emph{ab initio}. 

In addition to the intrinsic interest in understanding the self-coupling of metric-affine theories, this analysis also clarifies apparent discrepancies in the literature regarding the bootstrapping of higher-order theories of gravity~\cite{Deser2017,Ortin2017}. The key issue revolves around the formalism employed: Deser's analysis is conducted in the first-order formalism, often called the Palatini formalism, where the metric and connection are treated as independent fields. While this approach, yields to the same equations of motion as GR, since they enforce the Levi-Civita connection\footnote{Strictly speaking, the equations exhibit a projective symmetry, allowing for a family of connections related by projective transformations. However, this symmetry is a gauge symmetry, and it only manifests as a reparametrization of geodesics when considering their behavior under these transformations\mbox{~\cite{Bernal2016,JimenezCano2021}}.}, the equivalence does not hold for higher-order theories of gravity. By framing the discussion within the broader context of bootstrapping metric-affine theories of gravity, these discrepancies become entirely transparent.

\subsection{Metric-affine theories}

Metric-affine theories of gravity are theories in which the spacetime manifold is taken to be a triad $(M,\boldsymbol{g},\boldsymbol{\Gamma})$, being $M$ the manifold, $\boldsymbol{g}$ the metric and $\boldsymbol{\Gamma}$ the connection which is not uniquely determined in terms of the metric $\boldsymbol{g}$, i.e., it is not the Levi-Civita connection compatible with it. To distinguish between the Levi-Civita connection of $g_{\mu \nu}$ and the general connection, in this section we will use the symbols ($\boldsymbol{\LCg}$, $\CDg$, $\Rg$, ...) to represent the former, and the symbols ($\boldsymbol{\Gamma}$, $\nabla$, $R$) to represent the latter. Furthermore, it is convenient to introduce the torsion and the nonmetricity tensors, respectively, as follows:
\begin{align}
   \mathcal{T}_{\mu \nu}{}^\rho &:= \Gamma_{\mu\nu}{}^\rho -  \Gamma_{\nu\mu}{}^\rho\,,\\
   \mathcal{Q}_{\mu \nu \rho} &:= -\nabla_\mu g_{\nu\rho}\,.
\end{align}
With this in mind it can be shown that the general connection can be always expressed as the Levi-Civita connection plus a tensorial deviation that we split into two contributions~\cite{JimenezCano2021}:
\begin{equation}
    \Gamma_{\mu\nu}{}^\rho = \LCg_{\mu\nu}{}^\rho + K_{\mu\nu}{}^\rho + L_{\mu\nu}{}^\rho\,.
\end{equation}
In this decomposition we have introduced
\begin{align}
   K_{\mu\nu}{}^\rho &:= \frac{1}{2}g^{\rho\sigma}(\mathcal{T}_{\mu \nu \sigma}+\mathcal{T}_{\sigma\mu \nu }-\mathcal{T}_{\nu \sigma\mu })\,,\label{Eq:defContorsion}\\
   L_{\mu\nu}{}^\rho &:= \frac{1}{2}g^{\rho\sigma}(\mathcal{Q}_{\mu \nu \sigma}+\mathcal{Q}_{\nu \sigma\mu }-\mathcal{Q}_{\sigma\mu \nu })\,,
\end{align}
which are called contorsion and disformation tensors, respectively. They have the symmetry properties:
\begin{equation}
    K_{\mu\nu\rho}=-K_{\mu\rho\nu}\,,\qquad L_{\mu\nu}{}^\rho=L_{\nu\mu}{}^\rho\,.
\end{equation}
When working in an arbitrary orthogonal frame, i.e., with a vielbein $e^\mu{}_a$ in terms of which we have
\begin{align}
    g_{\mu \nu} e^\mu{}_A e^\nu{}_B= \eta_{AB},
\end{align}
the components of the connection $\nabla$ become:
\begin{equation}
    \omega_{\mu A}{}^{B} :=  \Gamma_{\mu\nu}{}^\rho e^\nu{}_A e_\rho{}^B - e^\nu{}_A\partial_\mu e_\nu{}^B.
\end{equation}
The decomposition into contorsion and disformation still holds in this formulation:
\begin{equation}
    \omega_{\mu A}{}^B=\mathring{\omega}_{\mu A}{}^B+K_{\mu A}{}^B+L_{\mu A}{}^B,
    \label{Eq:generalconnection2}
\end{equation}
where $\mathring{\omega}_{\mu A}{}^B$ is the contribution of the Levi-Civita connection, and
\begin{equation}
    K_{\mu A}{}^B:= e^\nu{}_A e_\rho{}^B K_{\mu \nu}{}^\rho \,,\qquad L_{\mu A}{}^B:=e^\nu{}_A e_\rho{}^B L_{\mu \nu}{}^\rho.
\end{equation}
A generic metric-affine theory is naturally formulated in terms of an action that depends on both the connection and the metric. Since the connection can be fully expressed in terms of the contorsion and disformation tensors, along with the Levi-Civita connection, it is convenient to treat these tensors as the fundamental independent variables representing the dynamical connection. Consequently, any metric-affine action takes the general form $S_{\text{MAG}}[\boldsymbol{g},\boldsymbol{K},\boldsymbol{L},\Phi]$, where $\boldsymbol{K}$ and $\boldsymbol{L}$ denote the contorsion and disformation tensors, respectively, and we have included any additional matter fields $\Phi$.

The idea now is to perform an expansion on top of a background solution:
\begin{align}
     g^{\mu\nu} &= \bar{g}^{\mu\nu} + \lambda h^{\mu\nu}, \\
     K_{\mu\nu}{}^\rho &= \bar{K}_{\mu\nu}{}^\rho + \lambda k_{\mu\nu}{}^\rho ,\\
     L_{\mu\nu}{}^\rho &= \bar{L}_{\mu\nu}{}^\rho + \lambda \ell_{\mu\nu}{}^\rho ,\\
     \Phi^I &= \bar{\Phi}^{I} + \lambda \phi^I, 
\end{align}
where $\{\bar{g}^{\mu\nu},\,\bar{K}_{\mu\nu}{}^\rho, \bar{L}_{\mu\nu}{}^\rho, \,\bar{\Phi}^{I}\}$ represents an arbitrary solution of the equations of motion. We are interested in the flat spacetime background, i.e., $\bar{g}^{\mu\nu} = \eta^{\mu \nu}$ and $\bar{K}_{\mu\nu}{}^\rho = \bar{L}_{\mu\nu}{}^\rho =  \bar{\Phi}^{I} = 0$), but again, in order to recover the correct superpotentials for the bootstrapping it is required to consider the expansion in an arbitrary background. In fact, we can perform the expansion in partial actions as before
\begin{equation}
    S_{\text{MAG}}[\boldsymbol{\bar{g}} + \lambda \boldsymbol{h}, \boldsymbol{\bar{K}}+\lambda \boldsymbol{k}, \boldsymbol{\bar{L}}+\lambda \boldsymbol{\ell}, \bar{\Phi}+ \lambda \phi] = \sum_{n=0}^{\infty} \lambda^n S^{(n)}_{\text{MAG}}[\boldsymbol{\bar{g}}, \boldsymbol{h}, \boldsymbol{\bar{K}}, \boldsymbol{k}, \boldsymbol{\bar{L}}, \boldsymbol{\ell}, \bar{\Phi}, \phi],
\end{equation}
and the general recursive formula \eqref{Eq:Key_Bootstrap_Gen} shows that the source of the equations of motion for $n$-th order perturbations is the variation with respect to the background of $(n-1)$-th order action. In particular, we have:
\begin{align}
    & \frac{\delta S^{(n)}_{\text{MAG}}[\boldsymbol{\bar{g}}, \boldsymbol{h}, \boldsymbol{\bar{K}}, \boldsymbol{k}, \boldsymbol{\bar{L}}, \boldsymbol{\ell}, \bar{\Phi}, \phi]}{\delta h^{\mu\nu}} =\frac{\delta S^{(n-1)}_{\text{MAG}}[\boldsymbol{\bar{g}}, \boldsymbol{h}, \boldsymbol{\bar{K}}, \boldsymbol{k}, \boldsymbol{\bar{L}}, \boldsymbol{\ell}, \bar{\Phi}, \phi]}{\delta \bar{g}^{\mu\nu}}, \label{Eq:recursiveh} \\
    & \frac{\delta S^{(n)}_{\text{MAG}}[\boldsymbol{\bar{g}}, \boldsymbol{h}, \boldsymbol{\bar{K}}, \boldsymbol{k}, \boldsymbol{\bar{L}}, \boldsymbol{\ell}, \bar{\Phi}, \phi]}{\delta k_{\mu\nu}{}^{\rho}} =\frac{\delta S^{(n-1)}_{\text{MAG}}[\boldsymbol{\bar{g}}, \boldsymbol{h}, \boldsymbol{\bar{K}}, \boldsymbol{k}, \boldsymbol{\bar{L}}, \boldsymbol{\ell}, \bar{\Phi}, \phi]}{\delta \bar{K}_{\mu\nu}{}^{\rho}}, \label{Eq:recursivek} \\
    & \frac{\delta S^{(n)}_{\text{MAG}}[\boldsymbol{\bar{g}}, \boldsymbol{h}, \boldsymbol{\bar{K}}, \boldsymbol{k}, \boldsymbol{\bar{L}}, \boldsymbol{\ell}, \bar{\Phi}, \phi]}{\delta \ell_{\mu\nu}{}^{\rho}} =\frac{\delta S^{(n-1)}_{\text{MAG}}[\boldsymbol{\bar{g}}, \boldsymbol{h}, \boldsymbol{\bar{K}}, \boldsymbol{k}, \boldsymbol{\bar{L}}, \boldsymbol{\ell}, \bar{\Phi}, \phi]}{\delta \bar{L}_{\mu\nu}{}^{\rho}}.
    \label{Eq:recursivel}
\end{align}
The first of these equations is identical to the one obtained when considering only metric theories of gravity in Section~\ref{Sec:Bootstrapping_Metric}. Given that the remaining equations follow a similar structure, it is natural to explore whether the perturbations of the contorsion and disformation tensors can be interpreted in a similar way. Specifically, we aim to determine whether the fully nonlinear metric-affine theory, taken as the starting point, can be understood as a consistent self-coupling of its degrees of freedom to a well-defined conserved current with a clear physical meaning.

In other words, we seek to determine whether the dynamics of the geometric degrees of freedom encoded in the connection emerge from the bootstrapping of the theory's linearization, just as the dynamics of the metric can be understood as the consistent self-coupling of the metric perturbation to its own energy-momentum tensor. To address this, we now analyze each equation in detail.
\begin{itemize}
    \item \textbf{Energy-momentum tensor:} The right-hand side of Eq.~\eqref{Eq:recursiveh} can be understood as the energy-momentum tensor computed through Hilbert's prescription, as the discussion from Section~\ref{Subsec:ComputingEMT} illustrates. 

    \item \textbf{Spin-density current:} The right-hand side of Eq.~\eqref{Eq:recursivek} corresponds to the spin-density current of the field theory, computed through Hilbert's prescription. In addition to the energy-momentum tensor, a field theory which is Poincar\'e invariant displays an additional Noether current which is the angular momentum tensor, associated with the invariance under Lorentz transformations. Take a Lorentz invariant theory described by the following Lagrangian $\mathcal{L}=\mathcal{L}(\Phi^I,\partial_\mu\Phi^I)$. Then, the conserved current reads:
    \begin{align}
        J_\text{can\,}{}^{\mu\nu\lambda} &:= T_\text{can\,}{}^{\mu [\lambda}x^{\nu]} + S_\text{can\,}{}^{\mu\nu\lambda}\,,\label{Eq:CanonicalAngularMom}
    \end{align}
    where $T_\text{can\,}{}^\mu{}_\nu$ is the canonical energy-momentum tensor defined above in Eq.~\eqref{Eq:CanonicalTmunu} and we have introduced the canonical spin-density tensor as
    \begin{equation}
        S_\text{can\,}{}^{\mu\nu\lambda} := \frac{1}{2} \sum_I\frac{\partial \mathcal{L}}{\partial \partial_\mu\Phi^I} (\Lambda^{\nu\lambda}_{{\Phi}^I}) \Phi^I,
    \end{equation}
    with $\Lambda^{\nu\lambda}_{{\Phi}^I}$ the generator of the Lorentz algebra that correspond to the representation under which the fields $\Phi^I$ transform. Notice that the spin-density current is not conserved on its own, since the conserved quantity is the full angular momentum tensor $J_\text{can\,}{}^{\mu\nu\lambda}$, but it clearly admits an interpretation as the contribution from the internal spin of the fields to the angular momentum. To compute it through a straightforward generalization of Hilbert's prescription, one simply needs to promote the flat spacetime metric to a curved metric $g_{\mu \nu}$, and the derivatives $\partial_{\mu}$ to covariant derivatives with respect to a torsional (but metric-compatible) connection. Instead of the torsion tensor, it is convenient to parametrize the action in terms of the contorsion tensor $K$ since from this action $S_{\text{M}}[\boldsymbol{g},\boldsymbol{K},\Phi]$, we can directly obtain the spin-density current as:
    \begin{align}
    S_{\text{H}}{}^{\mu \nu \lambda} \eta_{\lambda\rho} &:= \frac{1}{\sqrt{-g}} \frac{\delta S_{\text{M}}[\boldsymbol{g},\boldsymbol{K},\Phi]}{\delta K_{\mu\nu}{}^{\rho}} \bigg\rvert_{ g^{\mu \nu} = \eta^{\mu \nu}, K_{\mu\nu}{}^{\rho} = 0}, \label{Eq:defSH}
    \end{align}
    which is precisely the right-hand side of Eq.~\eqref{Eq:recursivek}. From this, we automatically conclude that the dynamics of the torsion degrees of freedom of a metric-affine theory can be understood as the consistent self-coupling of their linearization to the spin-density current order by order. The nonindependent conservation of the spin-density current is not a problem since the equations for the torsion are not divergenceless. Actually, it is precisely the on-shell conservation of the full angular momentum that guarantees their consistency.
    
    \item \textbf{Dilation-shear current:}  If we blindly apply the bootstrapping procedure defining the currents only in terms of the Hilbert prescription, the master equation for the perturbations still applies and the disformation perturbations couple order by order to the dilation-shear current which is defined as the variation of the action with respect to the disformation tensor~\cite{Hehl1995}, i.e., the right-hand side of Eq.~\eqref{Eq:recursivel}. Everything is completely parallel in that sense to the already studied case of torsion. However, the dilation-shear current cannot be computed as a Noether current in the canonical approach, it can only be computed through Hilbert’s prescription. In fact, to be able to obtain such a current from the canonical approach, we would need the theory to be invariant under the whole ${\rm GL}(D+1,\mathbb{R})$, and the general theories that we consider are only Poincaré invariant. The reason for this, is that a general connection which is not metric-compatible is such that the connection becomes a ${\rm GL}(D+1,\mathbb{R})$-connection instead of a Lorentz one. Thus, although the recursion relation~\eqref{Eq:recursivel} still holds, its physical meaning is not so clear. 
\end{itemize}
%
\subsection{Including fermions.}
So far, our discussion has been limited to bosonic fields, as fermionic matter, i.e., spinor fields, does not couple directly to the metric but instead requires the introduction of a frame field (vielbein). In other words, while bosonic matter fields necessitate non-minimal couplings to interact with torsion, fermions naturally couple to torsion through the standard minimal coupling prescription. Here, we introduce a slight modification of the bootstrapping procedure to demonstrate how it applies in the vielbein formalism.

For that purpose, let us consider an action that depends on the vielbein instead of the metric tensor $W[\boldsymbol{e},\Phi]$. The set $\Phi$ collectively denotes the full set of matter fields. Let us expand for perturbations on top of a background in the form
\begin{align}
    & e^{\mu}{}_{A} = \bar{e}^{\mu}{}_{A} + \Tilde{\lambda} \epsilon^{\mu}{}_{A}, \\
    & \Phi^I = \bar{\Phi}^I + \Tilde{\lambda} \phi^I,
\end{align}
which translates into an expansion of the action as
\begin{align}
    W[\boldsymbol{e}, \Phi] = \sum_{n=0}^{\infty} \tilde{\lambda} ^n W^{(n)} [\boldsymbol{\bar{e}}, \boldsymbol{\epsilon}, \bar{\Phi},\phi].
\end{align}
From our analysis in Appendix \ref{App:General_Fields_funidentity}, it is clear that we can find the recursive relation
\begin{align}
    \frac{\delta W^{(n)} [\boldsymbol{\bar{e}}, \boldsymbol{\epsilon}, \bar{\Phi},\phi] }{\delta \epsilon^{\mu}{}_{A}}  = \frac{\delta W^{(n-1)} [\boldsymbol{\bar{e}}, \boldsymbol{\epsilon}, \bar{\Phi},\phi] }{\delta \bar{e}^{\mu}{}_{A} },
\end{align}
where the right hand side is understood as the energy-momentum tensor at order $n-1$ in the vielbein formulation. To be more precise, we introduce the following notation for such partial energy-momentum tensors
\begin{equation}
    \mathfrak{t}^{(n)}{}_{\mu }{}^A := \frac{\tilde{\lambda}^n}{\abs{\bar{e}}}\frac{\delta W^{(n)} [\boldsymbol{\bar{e}}, \boldsymbol{\epsilon}, \bar{\Phi},\phi] }{\delta \bar{e}^{\mu}{}_{A} },
\end{equation}
in complete analogy with Eq.~\eqref{Eq:Gravitational_EMTensor}. We thus see that for every theory in which the gravitational field is described by a vielbein, gravitational perturbations couple to their own energy-momentum tensor. 

It is convenient to note though, that for theories described only in terms of the metric, i.e., theories of the form $S[\boldsymbol{g},\Phi]$, the expansion that we are performing here is different from the one that we performed in the metric tensor perturbation.  In other words, if we express such an action in terms of vielbein fields rather than the metric tensor, we would have
\begin{align}
     W[\boldsymbol{e}, \Phi] := S[\boldsymbol{g(e)},\Phi],
\end{align}
where we are emphasizing that the action $W$ depends on the vielbein only through the metric. However, the expansions in metric and vielbein perturbations are not directly related, as the vielbein itself is expressed as a series in $\lambda$, the parameter characterizing the metric expansion. To highlight this difference, we have used a different parameter, $\tilde{\lambda}$, for the vielbein expansion.

More explicitly, the relationship between the partial energy-momentum tensors $\mathfrak{t}^{(n)}{}_{\mu}{}^{a}$ and the $t^{(n)}_{\mu \nu}$ from previous sections is not straightforward. Identifying the two series would require systematically expanding vielbein perturbations in terms of $h^{\mu \nu}$, i.e., performing a $\lambda$ series for it, which does not appear feasible in a systematic manner to all orders. Therefore, establishing a direct connection between the \textit{partial} energy-momentum tensors computed in the metric formulation with those obtained in the vielbein formulation is not straightforward. However, at the full nonlinear order, both approaches must agree, as a direct consequence of the identity \eqref{Eq:Key_Bootstrap_Gen}, which we derive in Appendix~\ref{App:General_Fields_funidentity}.

To explicitly illustrate this point, it is convenient to consider a simple example. Consider a scalar of the form
\begin{align}
    \mathcal{S} = \mathcal{M}_{\mu \nu} g^{\mu \nu} ,
\end{align}
where $\mathcal{M}_{\mu \nu}$ is assumed to be a tensor which is independent of the metric. We now want to perform an expansion in $\lambda$, i.e., in the $\boldsymbol{g}+\lambda \boldsymbol{h}$ decomposition as:
\begin{align}
    \mathcal{S} = \mathcal{M}_{\mu \nu} \bar{g}^{\mu \nu} + \lambda \mathcal{M}_{\mu \nu} h^{\mu \nu}, 
\end{align}
and in $\tilde{\lambda}$, i.e., in the $\boldsymbol{\bar{e}} + \tilde{\lambda} \boldsymbol{\epsilon}$ decomposition:
\begin{align}
    \mathcal{S} = \mathcal{M}_{\mu \nu} \bar{e}^{\mu}{}_{A} \bar{e}^{\nu}{}_{B}\eta^{A B} + 2 \tilde{\lambda} \mathcal{M}_{\mu \nu} \bar{e}^{\mu}{}_{A} \epsilon^{\nu}{}_{B}\eta^{AB} + \tilde{\lambda}^2 \mathcal{M}_{\mu \nu} \epsilon^{\mu}{}_{A} \epsilon^{\nu}{}_{B}\eta^{AB}\,.
\end{align}
This example clearly shows that the two expansions are different: while the metric expansion includes only up to $\order{\lambda^2}$ terms, the vielbein expansion extends up to $\mathcal{O} ( \tilde{\lambda}^3 )$. To reconcile the two, one would need to express the full nonlinear vielbein as a series in $\lambda$ to all orders, a challenging task. 

Thus, in conclusion, we have found that vielbein perturbations consistently bootstrap by coupling to their own energy-momentum tensor. However, the connection between the expansions in the metric and vielbein formalisms is not straightforward.

If instead of only the vielbein, we consider also the connection as an independent variable, i.e., a torsionful connection, we have an action $S_{\text{F}} [e, K, \Phi ]$ that can be expanded as
\begin{equation}
    S_{\text{F}}[\boldsymbol{\bar{e}} + \tilde{\lambda} \boldsymbol{\epsilon}, \boldsymbol{\bar{K}}+\tilde{\lambda} \boldsymbol{k}, \tilde{\lambda} \phi] = \sum_{n=0}^{\infty} \lambda^n S^{(n)}_{\text{F}}[\boldsymbol{\bar{e}} , \boldsymbol{\epsilon}, \boldsymbol{\bar{K}}, \boldsymbol{k}, \bar{\Phi}=0, \phi],    
\end{equation}
with
\begin{align}
    &S^{(n)}_{\text{F}}[\boldsymbol{\bar{e}} , \boldsymbol{\epsilon}, \boldsymbol{\bar{K}}, \boldsymbol{k}, \bar{\Phi}=0,\phi]  =  \nonumber\\
    & \left[\int \dd^{D+1} x\ \left(\epsilon^{\mu}{}_{A} \frac{\delta }{\delta \bar{e}^{\mu}{}_{A} (x)} + k_{\mu AB}\frac{\delta }{\delta \bar{K}_{\mu AB} (x) }\right) \right]^{n-2} S^{(2)}_{\text{F}}[\boldsymbol{\bar{e}} , \boldsymbol{\epsilon}, \boldsymbol{\bar{K}}, \boldsymbol{k}, \bar{\Phi}=0,\phi] \bigg\rvert_{\tilde{\lambda} = 0}.
\end{align}
Analogous to the partial energy-momentum tensors, one can introduce the $n$-th order spin-density current, which has only the contribution coming from $S_{\text{F}}$ and is given by:
\begin{equation}
    \mathfrak{s}^{(n)}{}^{\mu A B} = \mathfrak{s}^{(n)}_{\text{F}}{}^{\mu A B} := \frac{\tilde{\lambda}^n}{|\bar{e}|}\frac{\delta S^{(n)}_{\text{F}}[\boldsymbol{\bar{e}} , \boldsymbol{\epsilon}, \boldsymbol{\bar{K}}, \boldsymbol{k}, \bar{\Phi}, \phi]}{\delta \bar{K}_{\mu A B}}.
\end{equation}
Again, we see that the distortion tensor couples order by order to the spin-density current. 

For the nonmetricity, the analysis is completely parallel to the one carried on above. The disformation tensor couples order by order to the dilation-shear current, although we again lack an interpretation in terms of canonical current for the shear since the connection $\omega_{\mu A}{}^B$ becomes a ${\rm GL}(D+1,\mathbb{R})$ connection. In fact, this issue becomes specially problematic for spinors, since they do not have well definite transformation properties under ${\rm GL}(D+1,\mathbb{R})$, but only under the Lorentz subgroup. Indeed, this is a consequence of ${\rm GL}(D+1,\mathbb{R})$ not admitting finite spinor representations, but only infinite-dimensional ones~\cite{Neeman1978}. 

In summary, we have analyzed the bootstrapping process for metric-affine theories of gravity by decomposing the connection into its torsion and nonmetricity components. To facilitate the analysis, we worked with the contorsion and disformation tensors, demonstrating that they satisfy recursive relations alongside the metric: at each order $n+1$, the perturbation is sourced by the variation of the action at order $n$ with respect to the corresponding background field. We identified the energy-momentum tensor, the spin-density current, and the dilation-shear current as the variations with respect to the background metric, background contorsion, and background disformation tensor, respectively.

For the energy-momentum tensor, the interpretation is straightforward, as metric perturbations couple directly to it. Similarly, we found that contorsion perturbations couple to the spin-density current, which, despite not being independently conserved, has a well-defined physical interpretation. However, interpreting the recurrence relation for disformation perturbations is more challenging, as the dilation-shear current lacks a clear canonical meaning.

\subsection{Bootstrapping higher order theories of gravity}

The results from Section~\ref{Sec:Bootstrapping_Metric} already made it evident that any metric theory of gravity can be reconstructed through the bootstrapping procedure from its linearization. Specifically, we have shown that the theory couples to its own energy-momentum tensor. The reverse-engineering approach introduced in~\cite{Butcher2009} naturally identifies the correct energy-momentum tensor by determining the necessary superpotentials to obtain the nonlinear theory. Similarly, \cite{Ortin2017} reaches the same conclusion. However, Deser argues in~\cite{Deser2017} that an additional constraint must be imposed for this result to hold.

To clarify Deser’s approach, it is important to recall that he works within the first-order formalism, where the metric (or metric perturbations) and the connection are treated as independent variables. While this formalism is equivalent to the standard second-order metric formulation in the case of GR, the equivalence breaks down for higher-derivative theories of gravity. The constraint Deser imposes to recover the nonlinear theory is that the connection must be the Levi-Civita connection compatible with the metric. By doing so, he effectively eliminates any potential degrees of freedom associated with the independent connection that would propagate otherwise. From this perspective, it is unsurprising that when treating the connection as an independent variable, he needs to constrain it to be Levi-Civita in order to recover metric-based higher-derivative theories of gravity.

In contrast, if one adopts the Palatini formalism for the nonlinear theory and deals with a metric-affine theory, it becomes necessary to analyze the bootstrapping of the affine connection (or, equivalently, the torsion and nonmetricity) since the resulting nonlinear theory has a different dynamical structure. A careful examination, such as the one carried out here, reveals that the theory still couples to well-defined currents: the metric couples to the energy-momentum tensor and the torsion couples to the spin-density current. Although the nonmetricity couples to the shear-dilation current we argued that such current does not have a straightforward interpretation from the canonical point of view.

\section{Conclusions and discussion}

This section summarizes the chapter's key results and their relevance to the emergent gravity framework underlying this thesis. It is divided into two parts: the first focuses on the chapter's technical findings, while the second explores their implications for the emergent paradigm.

\subsection{General results of the chapter}

One of the main novel results of this chapter is the demonstration of the nonuniqueness of GR as a self-consistent extension of Fierz-Pauli theory. We have shown that, in arbitrary spacetime dimensions, all Lovelock theories can emerge from Fierz-Pauli through the bootstrapping procedure. As a consequence of showing it, this analysis has also clarified the inherent ambiguities in the bootstrapping process. Since the energy-momentum tensor, like any conserved current, is not unique, allowing for the addition of identically conserved terms, different choices can, in principle, lead to distinct nonlinear theories. Our study explicitly confirmed this possibility through concrete examples.

As a byproduct of this analysis, and given the growing interest in the topic in the past years, we have explored the reconstruction of higher-order theories of gravity from their linear counterparts. To this end, we extended the bootstrapping procedure to the broader metric-affine framework for the first time. This extension served two purposes. First, it permits the analysis of a more general class of theories by incorporating additional geometric fields, such as the connection in the bootstrapping procedure. These fields couple to different conserved currents: for instance, the spin-density current in the case of torsion. Whereas the nonmetricity tensor obeys recursion relations that are close in spirit to the bootstrapping equations, it is hard to interpret them as the coupling to a conserved current. This is because the would-be conserved current would need to be associated with some transformations that, in general, are not symmetries of all Poincar\'e invariant theories.  

Second, some previous analyses of the bootstrapping of higher order theories of gravity in the literature were conducted within the first-order formalism, where the metric and connection were treated as independent variables. In these approaches, reconstructing the nonlinear theory from its linear counterpart required imposing a constraint, namely, the compatibility of the connection with the metric. This constraint is natural, as the equivalence between the Palatini and second-order formalisms holds only for GR, not for higher-derivative extensions. In our framework, this distinction becomes particularly clear: whether we choose to bootstrap the connection independently of the metric directly affects the resulting theory, leading to different possible nonlinear extensions. In particular, one can obtain metric-affine theories of gravity if one allows the potential degrees of freedom encoded in the connection to propagate. 

Finally, and particularly relevant for the emergent gravity framework, we have examined the role of background structures. We have explicitly demonstrated that the bootstrapping procedure can lead to background-dependent theories, as seen in the case of UG, which emerges as the nonlinear completion of its linear counterpart, WTDiff. Furthermore, we have shown that UG can be consistently reconstructed as a self-interacting theory of gravitons, where these gravitons couple universally to the traceless part of the energy-momentum tensor.

In this case, the starting point of the bootstrapping process is a free theory of gravitons propagating on flat spacetime. The resulting nonlinear theory from this consistent self-coupling procedure includes only the volume form associated with the flat metric. While we have not been able to identify any theories where the full flat spacetime metric appears in the nonlinear regime, we have argued that it is reasonable to expect that some theories reported in the literature might arise through this mechanism. Whether this is the case, and whether other theories exist in which the flat spacetime metric remains present in the final construction, remains an open question worthy of further investigation.

\subsection{Interplay with the emergent gravity paradigm}

The key takeaway for the emergent gravity paradigm from this chapter is that the only consistent extension of a massless spin-2 theory is not necessarily a diffeomorphism-invariant theory, and in particular, not uniquely GR. We have demonstrated that self-consistent nonlinear extensions of the linear theory can exist where the flat spacetime metric does not entirely disappear from the formulation, challenging the notion that GR is the sole possible outcome of such a process. 

The explicit example we identified is UG, which retains the background volume form of the flat spacetime metric. While UG was introduced almost as early as GR, and various analyses of its properties and generalizations exist in the literature, a comprehensive and systematic comparison with GR was missing. Therefore, the next chapter is dedicated to providing a detailed comparison between GR and UG, along with their respective generalizations. 

An immediate consequence of this result is that the presence of a spin-2 excitation in a system does not necessarily imply that their interactions must be described by GR at leading order and for the energy regime where the theory exhibits the emergent Lorentz invariance. For example, spin-2 excitations have been reported in certain condensed matter systems~\cite{Xu2006,Gu2006,Xu2006b,Gu2009,Xu2010}, and our analysis leaves open the possibility that their nonlinear interactions may differ from those of GR.

Our analysis also has broader implications for other emergent frameworks. For example, for string theory. It raises the question of whether the low-energy limit of string theory must necessarily be GR. A common argument asserting that it must be the case given that GR is the only consistent theory of interacting gravitons is incorrect, as our results demonstrate. In fact, in the next chapter, we will also explore the relationship between UG and string theory, illustrating that both GR and UG could, in principle, describe the low-energy sector of string theory.


\chapter{Unimodular gravity and WTDiff invariance}
\label{Ch4:UG}

\fancyhead[LE,RO]{\thepage}
\fancyhead[LO,RE]{Unimodular gravity and WTDiff invariance}


The previous section explored self-consistent extensions of free massless spin-2 theories, revealing that, contrary to popular belief, GR is not the only consistent theory of interacting gravitons in flat spacetime. Specifically, we found that the two linear theories with maximal gauge symmetry, namely Fierz-Pauli and WTDiff, admit two distinct nonlinear completions: GR and UG. Whereas GR is completely background independent, UG displays a background structure, the spacetime volume form, that is fixed \emph{ab initio}. Furthermore, although both theories share the same equations of motion, they differ fundamentally in their treatment of the cosmological constant, as we will discuss. In GR, the cosmological constant appears as a coupling constant in the action, whereas in UG, it emerges as an integration constant.

The first work on UG dates back to Einstein, who wrote his traceless equations for a metric with fixed determinant (in suitable coordinates) $\abs{g} = 1$ for the first time in~\cite{Einstein1919}, with the first translation of this article appearing in~\cite{Einstein1952}. Already there, Einstein pointed out the benefit of having the cosmological constant as an integration constant. However, the condition $\abs{g} = 1$ reduces the gauge group to only the subset of transverse diffeomorphisms, which correspond to those preserving the volume form associated with the metric determinant. It is possible to consider a direct generalization of this theory to make everything coordinate independent, by introducing a fixed background volume form and working with an unconstrained metric. To the best of our knowledge, this was pointed out for the first time in~\cite{Anderson1971}, see also \cite{Unruh1989} for further developments. This makes the theory invariant also under Weyl rescalings of the unconstrained metric $g_{\mu \nu}$. Actually, it makes just the conformal structure of the metric fluctuate. We will refer as WTDiff-invariant theories or UG theories to theories in which this volume form is written explicitly and display this combination of gauge symmetries: Weyl rescalings of the metric and transverse diffeomorphisms (WTDiff transformations). 

This chapter is dedicated to systematically address the question of the equivalence between Diff-invariant theories and WTDiff-invariant theories. For that purpose, we have looked at several regimes and situations in which both theories might look differently, with the aim of being as exhaustive as possible. Whereas some of these regimes and situations were already studied in the literature, we have found convenient to critically review them here and we present them in a unified way. Other sections include completely new material. We will work in $D+1$ spacetime dimensions only when we are dealing with generic features of the theories, in particular those required for analyzing the embedding of UG in string theory. The chapter is based on the articles~\cite{Carballo-Rubio2022,Garay2023,Garcia-Moreno2023} that have been published during this thesis. Appendix~\ref{AppC} provides complementary analyses and review material, along with additional technical details

\section{Classical linear theory}
\label{Sec:ClassicalL}

In Section~\ref{Sec:LinearSpin2}, we discussed the representations of a free massless spin-2 field, i.e., a graviton field that we denote as $h^{\mu \nu}$. We focused on theories that maximize the gauge symmetry for such fields, as this allows working with unconstrained variables. Two theories meet this criterion: Fierz-Pauli and WTDiff. However, we did not previously explore their equations of motion or analyze their dynamics in detail. This section fills that gap by examining their equations of motion and demonstrating that, despite propagating the same number of local degrees of freedom, they are not equivalent due to differences in their space of solutions.
\subsection{Same local degrees of freedom, different number of global degrees of freedom}
\label{Subsec:GlobalDof}

Both Fierz-Pauli and WTDiff propagate the same number of local degrees of freedom. However, as we will now discuss, they differ in the number of global degrees of freedom, which leads to a key distinction between the two.
\paragraph*{\textbf{Fierz-Pauli case.}}
Let us first consider the Fierz-Pauli theory. The corresponding equations of motion are
\begin{align}
    0=\eqFP_{\mu \nu} := \Box h_{\mu \nu} - 2 \partial_{(\mu | } \partial_{\rho} h^{\rho}{}_{| \nu)} + \partial_{\mu} \partial_{\nu} h + \eta_{\mu \nu} \ddh - \eta_{\mu \nu} \Box h ,
    \label{Eq:Fierz_Pauli_Eqs}
\end{align}
which fulfill
\begin{align}
    \eta^{\mu \nu}\eqFP_{\mu \nu}=(D-1)\Big[\ddh-\Box h\Big]\,,\qquad\qquad \partial^\nu\eqFP_{\mu \nu}=0\,\,,
\end{align}
where we recall that $ \ddh := \partial_{\mu} \partial_{\nu} h^{\mu \nu}$. We can now perform a gauge fixing to fulfill simultaneously the conditions $h = 0$ and $\partial_{\mu} h^{\mu \nu} = 0$. Notice that given a value for $\partial_{\mu} h^{\mu \nu}$ and $ h$, we can always perform a diffeomorphism $h_{\mu \nu} \to h^{\prime}_{\mu \nu}$:
\begin{align}
    \partial_{\mu} h^{\prime \mu \nu} &= \partial_{\mu} h^{\mu \nu} + \Box \xi^{\nu} + \partial^{\nu} \partial_{\mu} \xi^{\mu}\,, \\
    h' &= h + 2 \partial_{\mu} \xi^{\mu}\,,
\end{align}
such that $\partial_{\mu} h^{\prime \mu \nu} = h' = 0$. This leads to the following system of equations for $\xi^\mu$ that always admits a solution:
\begin{align}
    & \partial_{\mu} \xi^{\mu} = -\frac{1}{2} h, \\
    & \Box \xi^{\mu} = \frac{1}{2} \partial^{\mu} h - \partial_{\nu} h^{\nu \mu}.
\end{align}
Hence, we can always make a gauge fixing such that the equations~\eqref{Eq:Fierz_Pauli_Eqs} reduce to a sourceless wave equation for a transverse traceless tensor:
\begin{align}
    & \Box h_{\mu \nu} = 0, \\
    & \partial_{\mu} h^{\mu \nu} = h = 0. 
\end{align}

\paragraph*{\textbf{WTDiff case.}}
Following~\cite{Bonifacio2015}, we can try to do the same for the WTDiff theory \eqref{Eq:WTDiff_Lagrangian}. In this case, the dynamical equations are\footnote{
    Due to the Weyl symmetry, this equation is actually independent of the trace of $h_{\mu \nu}$. Indeed, if we decompose $h_{\mu \nu}=\hat{h}_{\mu \nu}+\frac{1}{D+1}\eta_{\mu \nu}h$ (where by construction $\eta^{\mu \nu}\hat{h}_{\mu \nu}=0$), we find the identity:
    \[
        \eqWTDiff_{\mu \nu} = \Box \hat{h}_{\mu \nu} - 2 \partial_{(\mu|} \partial_{\rho} \hat{h}^{\rho}{}_{|\nu)}  + \frac{2}{D+1} \eta_{\mu \nu} (\partial^2\!\cdot\!\hat{h}) \,.
    \]
}
\begin{equation}
    0 = \eqWTDiff_{\mu \nu} := \Box h_{\mu \nu} - 2 \partial_{(\mu|} \partial_{\rho} h^{\rho}{}_{|\nu)} + \frac{2}{D+1} \partial_{\mu} \partial_{\nu} h + \frac{2}{D+1} \eta_{\mu \nu} \ddh - \frac{D+3}{(D+1)^2} \eta_{\mu \nu} \Box h,
    \label{Eq:WTDiff_Eqs}
\end{equation}
which fulfills
\begin{equation}
    \eta^{\mu \nu}\eqWTDiff_{\mu \nu}=0\,,\qquad\qquad \partial^\nu\eqWTDiff_{\mu \nu}=-\frac{D-1}{D+1}\partial_\mu\Big[\ddh-\frac{1}{D+1}\Box h\Big] = 0\,.
\end{equation}
From the second one we can derive the general result that, on-shell, the quantity in the square bracket must be constant, i.e.,
\begin{equation}
    \ddh-\frac{1}{D+1}\Box h = c\,,\label{eq:conddivWTDiff}
\end{equation}
where $c$ is an integration constant. 

Now one can try to find a gauge leading to $\Box h_{\mu \nu} = J_{\mu \nu}$ such that the field $h_{\mu \nu}$ is traceless and transverse and such that the source $J_{\mu \nu}$ is independent of $h_{\mu \nu}$. However, it is not difficult to check that, due to the presence of $c$, one cannot achieve all the conditions simultaneously. Let us show this in more detail. First, we perform a Weyl transformation to reach $h = 0$. It is always possible to do it since under a Weyl transformation $h$ transforms as $h \to h^{\prime}$
\begin{align}
    h' = h + (D+1) \chi, 
\end{align}
and taking $\chi = - h/(D+1)$, we reach $h'=0$. The condition \eqref{eq:conddivWTDiff} implies then that, in this gauge, $\ddh = c$. This can be integrated to obtain an expression for the divergence of $h^{\mu \nu}$
\begin{align}
    \partial_{\mu} h^{\mu \nu} = \frac{c}{D+1} x^{\nu} + a^{\sT \nu}, 
\end{align}
where $a^\sT_{\nu}$ is any arbitrary divergenceless vector field $\partial^{\nu} a^\sT_{\nu} = 0$. The vector $a^\sT_{\nu}$ can be simply removed through a transverse diffeomorphism since we have
\begin{align}
    \partial_{\mu } h^{\prime \mu \nu} = \partial_{\mu} h^{\mu \nu} + \Box \xi^{\sT \nu}, 
\end{align}
so the condition $\partial_{\mu} h^{\prime \mu \nu} = \frac{c}{D+1} x^{\nu}$, leads to an equation for $\xi^{\sT \nu}$ that always admits a solution
\begin{align}
    \Box \xi^{\sT}_{\nu} = - a^\sT_{\nu}. 
\end{align}
Hence, in this gauge, we end up again with a sourceless wave equation for $h_{\mu \nu}$, 
\begin{align}
    \Box h_{\mu \nu} = 0,\label{eq:boxhfixedWTDiff}
\end{align}
with
\begin{align}
    \partial_{\mu} h^{\mu \nu} = \frac{c}{D+1} x^{\nu}\,,\qquad\qquad h=0\,.
\end{align}
Note that the transversality condition is violated for nonvanishing $c$. Another way to look at this result is making a field redefinition $h_{\mu \nu} \to H_{\mu \nu}$ such that
\begin{align}
    h_{\mu \nu} =  H_{\mu \nu} + \frac{1}{2} \frac{c}{D+1} \eta_{\mu \nu} x^2.
\end{align}
This new tensor $H_{\mu \nu}$ is transverse, although it is not traceless anymore:
\begin{equation}
    \partial_{\mu} H^{\mu \nu} = 0, \qquad  H_\nu{}^\nu = - \frac{1}{2} c x^2. 
\end{equation}
The equation of motion~\eqref{eq:boxhfixedWTDiff} in terms of $H_{\mu \nu}$ contains a constant source term proportional to $c$:
\begin{align}
    \Box H_{\mu \nu} = - c \eta_{\mu \nu}\,.
    \label{Eq:Dhceta}
\end{align}
It is noteworthy that Eq.~\eqref{Eq:Dhceta} matches the lowest-order contribution of a cosmological constant $c$ to the Fierz-Pauli equation, as derived from the linearization of the Einstein-Hilbert action with a cosmological constant term. Later, we will see that in the full nonlinear theory, this integration constant $c$ is promoted to the full cosmological constant in Einstein’s equations. Since $c$ is arbitrary, it serves as an additional degree of freedom in WTDiff, which must be specified as part of the initial conditions.

Thus far, we have observed that while both theories propagate the same number of local degrees of freedom, they are not equivalent due to the extra parameter $c$ required to define the initial conditions in WTDiff. In this sense, they cannot be considered different formulations of the same theory. We explore this distinction further in the next section.

\subsection{Stueckelberg-ing}
\label{Subsec:Linear_Stueck}
It is always possible to extend the gauge symmetries of a theory by introducing additional fields, often referred to as Stueckelberg fields, which increase the number of configuration variables at the expense of enlarging the set of gauge transformations of the theory too. A natural question to ask is whether there exists a more general ``parent" theory that includes additional fields beyond the tensor $h^{\mu \nu}$, and remains invariant under both arbitrary diffeomorphisms and Weyl transformations acting linearly on $h^{\mu \nu}$, such that Fierz-Pauli and WTDiff arise as particular gauge fixings. Although our discussion on the global mode of WTDiff suggests that such a theory should not exist, explicitly analyzing this possibility provides valuable insight into the gauge structure of the theory. 

\paragraph*{\textbf{Fierz-Pauli case.}}
Let us begin with Fierz-Pauli. The ``parent'' theory that we seek is a theory in which the action is invariant under linearized Weyl transformations on the tensor $h_{\mu \nu}$ and linearized diffeomorphisms. The way to achieve this using Stueckelberg's approach is to introduce an additional field $\varphi$ in the action by making the replacement \mbox{$h_{\mu \nu} \rightarrow h_{\mu \nu} + \varphi \eta_{\mu \nu}$}. In this way, the Weyl symmetry is realized in the $\varphi$ field as a shift symmetry
\begin{align}
    h_{\mu \nu} &\to h'_{\mu \nu} = h_{\mu \nu} + \chi \eta_{\mu \nu}\,, \\
     \varphi &\to \varphi'=\varphi - \chi\,. 
\end{align}
After introducing the field $\varphi$, we are led to the following Lagrangian: 
\begin{equation}
     \mathcal{L}_{\text{FP-St}} = \mathcal{L}_{\text{FP}} -  (D-1) \partial_{\mu}h^{\mu \nu} \partial_{\nu} \varphi + (D-1) \partial_{\mu}h \partial^{\mu} \varphi + \frac{1}{2} (D)(D-1) \partial_{\mu} \varphi \partial^{\mu} \varphi\,, 
    \label{Eq:Stueckelberg_GR}
\end{equation}
whose equations of motion are
\begin{align}
    [\text{EoM }\varphi]&\qquad  \Box \varphi - \frac{1}{(D)(D-1)} \eta^{\mu \nu}\eqFP_{\mu \nu} = 0\,, \\
   [\text{EoM } h^{\mu \nu}] &\qquad \eqFP_{\mu \nu} + (D-1)( \partial_{\mu} \partial_{\nu} \varphi - \eta_{\mu \nu}\Box \varphi )= 0\,. \label{eq:StFP}
\end{align}
One important thing to notice is that the equation of $\varphi$ is the trace of the second one, so it is redundant. Precisely because the equation of $\varphi$ is redundant, the unitary gauge $\varphi = 0$ is a legal one. Here by legal we mean that fixing the gauge in the action and fixing it directly in the equations leads to equivalent equations of motion. We give a more elaborate analysis of legal and illegal gauge fixings together with some simple examples in Appendix~\ref{Sec:Illegal_gauge}.

The question now is whether one can find a legal gauge fixing that leads to the WTDiff dynamics. 
In principle, fixing the gauge at the level of the action to be \mbox{$\varphi = - h /(D+1) $} leads to the WTDiff Lagrangian~\eqref{Eq:WTDiff_Lagrangian}, and one could naively conclude that then the Lagrangian in Eq.~\eqref{Eq:Stueckelberg_GR} is the desired ``parent'' Lagrangian. However, it turns out to be an illegal gauge fixing, since it does not lead to the WTDiff equations of motion when performed at the level of the equations.

First, to show that it is always possible to reach this gauge, we recall that under a generic gauge transformation the fields change as:
\begin{align}
    h_{\mu \nu}&\to h'_{\mu \nu} = h_{\mu \nu} + 2 \partial_{(\mu} \xi_{\nu)} + \chi \eta_{\mu \nu}, \noindent\\
    \varphi&\to \varphi' = \varphi - \chi \,.\label{eq:totalsym-FPS}
\end{align}
We want to see if for generic $\varphi$ and $h_{\mu \nu}$, we can find $\xi^{\mu}$ and $\chi$ such that we end up with $ \varphi' = - h'/(D+1)$. The latter condition fixes the divergence of the vector $\xi_{\mu}$:
\begin{align}
    \partial_{\mu} \xi^{\mu} = - \frac{1}{2} h - \frac{D+1}{2} \varphi.
\end{align}
A trivial example of such a vector field would be:
\begin{align}
    \xi^{\mu} =  \delta^{\mu}{}_0 \int^{t}_{t_0} \dd t'  \left( - \frac{1}{2} h (t',x^i) - \frac{D+1}{2} \varphi(t',x^i) \right).
    \label{Eq:Diff_Transf}
\end{align}
Let us now show that this gauge fixing is illegal. In this gauge, the only independent equation of motion, Eq.~\eqref{eq:StFP}, reduces to:
\begin{equation}
    \eqWTDiff_{\mu \nu}+\frac{D-1}{D+1}\eta_{\mu \nu}\big[ \ddh - \frac{1}{D+1} \Box h \big]=0\,. \label{eq:StFPFix}
\end{equation}
Since $\eqWTDiff_{\mu \nu}$ is traceless and the second term is pure trace, this equation is fulfilled if and only if both terms vanish independently. The first one leads to the WTDiff equations and this implies \eqref{eq:conddivWTDiff} with the appearance of the arbitrary integration constant $c$. However, the second term enforces $c=0$. This proves that the gauge fixing \mbox{$\varphi = - h/ (D+1)$} is illegal since the equations resulting from fixing the gauge at the level of equations do not reproduce the dynamics of the WTDiff Lagrangian, which admits an arbitrary integration constant $c$.

\paragraph*{\textbf{WTDiff case.}}
We can carry out a similar analysis for WTDiff by introducing a Stueckelberg field to extend the gauge symmetries from Weyl and transverse diffeomorphisms to Weyl and full diffeomorphisms. This comes at the cost of adding an extra Stueckelberg field. Notably, any arbitrary diffeomorphism can always be locally decomposed into a transverse (divergenceless) component, $\xi_{\mu}^\sT$, and a longitudinal (irrotational) component, $\sigma$, as follows
\begin{equation}
    \xi_{\mu} = \xi_{\mu}^\sT + \partial_{\mu} \sigma \qquad (\partial^{\mu} \xi_{\mu}^\sT = 0)\,.
\end{equation}
When we expand such a transformation acting on $h_{\mu \nu}$ we have
\begin{align}
    h_{\mu \nu} \rightarrow h'_{\mu \nu} = h_{\mu \nu} + \partial_{\mu} \xi^\sT_{\nu} + \partial_{\nu} \xi^\sT_{\mu} + 2 \partial_{\mu} \partial_{\mu} \sigma.\label{eq:htransfxi}
\end{align}
In order to make the WTDiff Lagrangian presented in Eq.~\eqref{Eq:WTDiff_Lagrangian} invariant also under longitudinal diffeomorphisms, we need to introduce a Stueckelberg field making the replacement
\begin{align}
    h_{\mu \nu} \to h_{\mu \nu} + 2 \partial_{\mu} \partial_{\nu} \varphi.
    \label{eq:stuckWTDiff}
\end{align}
The resulting Lagrangian, after some integrations by parts, is:
\begin{align}
    & \mathcal{L}_\text{WTDiff-St} =  \mathcal{L}_\text{WTDiff}  \nonumber \\
    & + 2 \frac{(D-1)}{D+1} \partial_{\mu} h^{\mu \nu} \partial_{\nu} \Box \varphi - 2 \frac{D-1}{(D+1)^2} \Box \partial^{\mu} \varphi \partial_{\mu} h + 2 \frac{(D)(D-1)}{(D+1)^2} \Box \partial_{\mu} \varphi  \Box \partial^{\mu} \Box \varphi\,,
    \label{Eq:WTDiff-St_lagrangian}
\end{align}
and its equations of motion are:
\begin{align}
    [\text{EoM }\varphi]&\qquad  \partial^\mu\partial^\nu\eqWTDiff_{\mu \nu} - \frac{2(D)(D-1)}{(D+1)^2}\Box^3\varphi=0\,, \\
   [\text{EoM } h^{\mu \nu}] &\qquad \eqWTDiff_{\mu \nu} - \frac{2(D-1)}{D+1}\left[  \partial_{\mu}\partial_{\nu} \Box\varphi -\frac{1}{D+1}\eta_{\mu \nu}\Box^2\varphi\right] = 0\,.\label{eq:StWTDiff}
\end{align}
Observe that, again, the first of the equations can be obtained by taking $\partial^\mu\partial^\nu$ in the second one so it is redundant. Thus, we can focus only on the second equation. As a consequence of the substitution~\eqref{eq:stuckWTDiff}, the simultaneous transformations~\eqref{eq:htransfxi} and
\begin{align}
    \varphi  \to \varphi' = \varphi - \sigma,
\end{align}
become a symmetry of the theory. Therefore, the Lagrangian in Eq.~\eqref{Eq:WTDiff-St_lagrangian} exhibits the following symmetry:
\begin{align}
    h_{\mu \nu}&\to h'_{\mu \nu} = h_{\mu \nu} +\partial_{\mu} \xi^\sT_{\nu} + \partial_{\nu} \xi^\sT_{\mu} + 2 \partial_{\mu} \partial_{\nu} \sigma + \chi \eta_{\mu \nu}, \noindent\\
    \varphi&\to \varphi' = \varphi - \sigma \,.\label{eq:totalsym-WTDiffS}
\end{align}
Notice that this is not the same Lagrangian that we found in Eq.~\eqref{eq:totalsym-FPS} by Stueckelberging Fierz-Pauli to make the theory invariant under Weyl transformations. This shows explicitly two different ways of realizing the gauge group of Weyl transformations and full diffeomorphisms using only a scalar field $\varphi$ and a tensor $h_{\mu \nu}$. We will now delve into the inequivalence between both of them, even if they display the same gauge symmetries. 

Fixing a gauge where $\Box \varphi = - \frac{1}{2} h$ at the level of the action results in the Fierz-Pauli Lagrangian. This appears contradictory to the existence of a global degree of freedom, as imposing this gauge condition would seemingly eliminate the degree of freedom associated with the constant $c$. To understand this point, we first note that this gauge can always be attained through an appropriate Weyl transformation, given that the trace $h$ and the field $\varphi$ transform as
\begin{align}
    & h \rightarrow h' = h +  (D+1) \chi , \\
    & \varphi \rightarrow \varphi' = \varphi ,
\end{align}
where $\chi$ is the gauge parameter of the Weyl transformation. By imposing $\Box \varphi' = - \frac{1}{2} h'$, we find that we can always choose $\chi$ that does the job:
\begin{align}
    \chi = - \frac{1}{D+1} \left( h + 2 \Box \varphi \right). 
\end{align} 
To demonstrate that this is an illegal gauge fixing, we observe that the only independent equation, Eq.~\eqref{eq:StWTDiff}, reduces to the following after imposing the gauge condition \mbox{$\Box \varphi = - \frac{1}{2} h$:}
\begin{align}
    &\Box \left[ \ddh - \Box h \right]=0\,, \label{eq:FPx1}\\
    &\eqFP_{\mu \nu} - \frac{D-1}{D+1} \eta_{\mu \nu} \left[ \ddh -\Box h \right] = 0 \,.
\end{align}
By taking the divergence of the second one, and recalling that the $\eqFP_{\mu \nu}$ part is divergenceless, we find that the combination $\left[ \ddh -\Box h \right]$ must be constant. Therefore, Eq.~\eqref{eq:FPx1} is redundant and the system of equations becomes:
\begin{equation}
    \eqFP_{\mu \nu} =  c\eta_{\mu \nu} \,,
    \label{Eq:fixing_constant}
\end{equation}
for some constant $c$. The equations that follow from fixing the gauge at the level of the action do not contain an integration constant, since the Lagrangian reduces to Fierz-Pauli. However, the equations obtained by performing the gauge fixing at the level of the equations of motion do contain such a constant~\eqref{Eq:fixing_constant}, and hence we conclude that the gauge fixing is illegal. 

One might think that this inconsistency comes from the fact that we are comparing WTDiff with Fierz-Pauli without a cosmological constant and that we should add it. However the key point is that WTDiff contains intrinsically an \emph{arbitrary} integration constant, whereas Fierz-Pauli would simply contain an additional coupling constant. 

In summary, at the linear level analyzed so far, we conclude that there is no ``parent'' theory from which both Fierz-Pauli and WTDiff emerge as legitimate gauge fixings. The key reason is that, despite propagating the same number of local degrees of freedom, WTDiff includes an additional global degree of freedom, which can be interpreted as the linear counterpart of the cosmological constant. 

\section{Classical nonlinear theory}
\label{Sec:ClassicalNL}

\subsection{Einstein-Hilbert action: Diff or WTDiff principle}
\label{Subsec:Einstein_Hilbert}
GR can be obtained from a variational principle in which the action is diffeomorphism invariant. Diffeomorphism invariance dictates that the unique volume form that we can use to integrate is the one associated with the metric tensor, which in a chart reduces to an integration with respect to the weight $\dd^{D+1}x \sqrt{-g}$. Now, we need to choose diffeomorphism invariant objects to build the action. In principle, any curvature scalar could serve for this purpose. The Ricci scalar $R(\boldsymbol{g})$ with a possible cosmological constant term is the lowest-dimensional operator that we can add to the action and leads to second order equations of motion for the metric. This defines the Einstein-Hilbert action, which is given by the following expression~\cite{Wald1984}
\begin{equation}
S_{\text{EH}}[\boldsymbol{g}] = \frac{1}{2 \kappa_D^2} \int \dd^{D+1} x \sqrt{-g} \left[-2 \Lambda + R (\boldsymbol{g}) \right],
\label{einstein_hilbert_diff}
\end{equation}
where $\Lambda$ is the cosmological constant. We can add another piece to the action representing additional matter fields, $S_{\textrm{matter}}[\Phi,\boldsymbol{g}]$, where $\Phi$ collectively represents all of the matter fields present in the theory. Diffeomorphism invariance means that the action is invariant under the following transformations generated by a vector field $\xi^{\mu}$ (we lower and raise indices with the metric $g_{\mu \nu}$ and its inverse $g^{\mu \nu}$, respectively)
\begin{equation}
    g_{\mu \nu} \rightarrow g_{\mu \nu} + \delta_{\xi} g_{\mu \nu}, \qquad \delta_{\xi} g_{\mu \nu} = 2 \nabla_{( \mu} \xi_{\nu)}. 
    \label{gdiffstransform}
\end{equation}
The equations of motion correspond to the Einstein equations
\begin{equation}
R_{\mu \nu} (\boldsymbol{g})- \frac{1}{2}  g_{\mu \nu} R(\boldsymbol{g}) + \Lambda g_{\mu \nu} = \kappa_D^2 T_{\mu \nu} (\boldsymbol{g}),
\end{equation}
where 
\begin{align}
T_{\mu \nu} \left( \boldsymbol{g} \right) = \frac{-2}{\sqrt{|g|}} \frac{\delta S_{\textrm{matter}}}{\delta g^{\mu \nu}}
\label{Eq:EM_Tensor_Definition}
\end{align}
is the energy-momentum tensor as defined in the previous chapter. 

Let us now write down a theory in which the guiding principle is not that of Diff invariance but that of WTDiff invariance. First, we note that we need to introduce a privileged background volume form which we represent as
\begin{align}
    \boldsymbol{\omega} = \frac{1}{(D+1)!} \omega(x) \dd x^0 \wedge ... \wedge \dd x^D.
\end{align}
As discussed in Appendix~\ref{App:Diffs}, the fact that it constitutes a background structure implies that it is nondynamical, i.e., it is fixed \emph{a priori}, and it does not transform under \emph{active diffeomorphisms}. Using this volume form, we can construct the following Weyl-invariant auxiliary metric
\begin{equation}
    \tilde{g}_{\mu \nu} = g_{\mu \nu} \left( \frac{\omega^2}{|g|} \right)^{\frac{1}{D+1}}.
    \label{auxiliary_metric}
\end{equation}
Any curvature scalar constructed from the auxiliary metric $\tilde{g}_{\mu \nu}$ will automatically be invariant under Weyl transformations, which act on the metric as
\begin{align}
    g_{\mu \nu} \to e^{\phi} g_{\mu \nu},
    \label{Eq:finite_weyl}
\end{align}
since the auxiliary metric $\boldsymbol{\tilde{g}}$ is already Weyl invariant. The infinitesimal version of the transformation is
\begin{equation}
    \delta_{\phi} g_{\mu \nu} =  \phi g_{\mu \nu}. 
    \label{infinitesimal_Weyl_transformation}
\end{equation}
Furthermore, curvature scalars of $\boldsymbol{\tilde{g}}$ are invariant under Weyl transformations and transverse diffeomorphism, which are those that preserve the background volume form $\boldsymbol{\omega}$. Mathematically, a diffeomorphism generated by a vector $\xi^{\mu}$ is a transverse diffeomorphism if the volume form $\boldsymbol{\omega}$ has zero Lie derivative along $\xi^{\mu}$
\begin{equation}
    \mathcal{L}_{\xi} \boldsymbol \omega = 0 \quad\rightarrow \quad\tilde{\nabla}_{\mu} \xi^{\mu} = 0, 
\end{equation}
where $\tilde{\nabla}$ represents the covariant derivative compatible with the auxiliary metric $\tilde{g}_{\mu \nu}$. Transverse diffeomorphisms act on $\boldsymbol{g}$ as
\begin{equation}
    \delta_{\xi} g_{\mu \nu} = 2 \nabla_{( \mu} \xi_{\nu)},
\end{equation}
with $\xi_{\nu} = g_{\nu \mu} \xi^{\mu}$. The transverse condition, which is the zero-divergence condition on the vector $\xi^{\mu}$ with respect to the covariant derivative compatible with the auxiliary metric $\tilde{g}_{\mu \nu}$, can equivalently be written as 
\begin{equation}
    \tilde{\nabla}_{\mu} \xi^{\mu}  = 0\quad \rightarrow\quad \nabla_{\mu} \xi^{\mu} = - \frac{1}{2} \xi^{\sigma} \partial_{\sigma} \log \left( \frac{\omega^2}{\abs{g}} \right),
    \label{transverse_diff_divergence}
\end{equation}
in terms of the covariant derivative compatible with the dynamical metric $g_{\mu \nu}$. This can be understood as a consequence of the covariant derivative $\tilde{\nabla}$ being compatible with an integrable Weyl connection~\cite{Salim1996,Romero2012,Yuan2013,Barcelo2017}. Notice that these transformations can be equivalently expressed as transformations acting on the auxiliary metric $\tilde{g}_{\mu \nu}$ as follows: 
\begin{equation}
    \delta_{\xi} \tilde{g}_{\mu \nu} = 2 \tilde{\nabla}_{(\mu} \tilde{\xi}_{\nu)}, \qquad \tilde{\nabla}_{\mu} \xi^{\mu} = 0, 
    \label{variation2}
\end{equation}
where $\tilde{\xi}_{\mu} = \tilde{g}_{\mu \nu} \xi^{\nu}$ and $\tilde{\xi}_{\mu} = \left( \omega^2 / \abs{g} \right)^{1/(D+1)} \xi_{\mu}$. 

Furthermore, writing everything in terms of the auxiliary metric ensures that just the conformal structure of spacetime fluctuates, with the volume form of the geometry to which we couple matter fields and particles fixed \emph{a priori}. With this, the lowest order operator that we can write down is the Ricci scalar in terms of $\tilde{\boldsymbol{g}}$. The cosmological constant operator is no longer a dynamical term in the action since it is independent of $\boldsymbol{g}$, and hence it simply corresponds to adding a constant to the action. With this, we can now write down the action
\begin{equation}
    \tilde{S}_{\text{EH}}[\boldsymbol{g}; \boldsymbol{\omega}] = S_{\text{EH}}[\boldsymbol{\tilde{g}}] = \frac{1}{2 \kappa_D^2} \int \dd ^{D+1} x\,\omega  R ( \boldsymbol{\tilde{g}} ) ,
\label{einstein_hilbert_wtdiff}
\end{equation}
where, although we use the object $\boldsymbol{\tilde{g}}$ as the metric with respect to which we compute the scalar of curvature and we build the matter action, we consider the field $\boldsymbol{g}$ as the dynamical field. Furthermore, the action of matter that we would add now would be the same we had before, except for replacing $\boldsymbol{g}$ with the auxiliary metric $\boldsymbol{\tilde{g}}$:  $ \tilde{S}_{\textrm{matter}} (\Phi, \boldsymbol{g}) = S_{\textrm{matter}} (\Phi, \boldsymbol{\tilde{g}})$. 

This formulation ensures that the variation with respect to $g_{\mu \nu}$ always leads to traceless equations since
\begin{equation}
    \delta \tilde{g}^{\mu \nu} = \delta g^{\mu \nu} - \frac{1}{D+1} g^{\mu \nu} g_{\alpha \beta} \delta g^{\alpha \beta},
\end{equation}
as it can be derived from the relation between $\tilde{g}_{\mu \nu}$ and $g_{\mu \nu}$ in Eq.~\eqref{auxiliary_metric}. Notice that $g_{\mu \nu} g^{\alpha \beta} = \tilde{g}_{\mu \nu} \tilde{g}^{\alpha \beta}$ since the conformal factor that is multiplying is canceled by the one that appears in the denominator. With this, the variation of the Ricci-scalar part of the action reads
\begin{equation}
    \delta \tilde{S}_{\text{EH}} [\boldsymbol{g}] = \int \dd ^{D+1} x\, \omega \left(R_{\mu \nu} (\boldsymbol{\tilde{g}}) - \frac{1}{D+1} R (\boldsymbol{\tilde{g}}) \tilde{g}_{\mu \nu } \right) \delta g^{\mu \nu}.
\end{equation}
The variation of the matter action, on the other hand, gives the traceless part of the energy-momentum tensor automatically, since the metric that enters the matter action is $\tilde{g}_{\mu \nu}$. Combining both parts we find the following equations of motion:
\begin{align}
    R_{\mu \nu} (\boldsymbol{\tilde{g}}) - \frac{1}{D+1} R (\boldsymbol{\tilde{g}}) \tilde{g}_{\mu \nu } = \kappa_D^2 \left( T_{\mu \nu} (\boldsymbol{\tilde{g}}) - \frac{1}{D+1} T (\boldsymbol{\tilde{g}}) \tilde{g}_{\mu \nu} \right).
    \label{traceless_einstein}
\end{align}
Taking the covariant derivative, assuming the conservation of the energy-momentum tensor $\tilde{\nabla}_{\mu} T^{\mu \nu} (\boldsymbol{\tilde{g}}) = 0$~\cite{Ellis2010,Ellis2013}, and applying Bianchi identities \mbox{$\tilde{\nabla}_{\mu} \left[ R^{\mu \nu} (\boldsymbol{\tilde{g}})- R (\boldsymbol{\tilde{g}})  \tilde{g}^{\mu \nu} / 2 \right] = 0$}, we can find
\begin{equation}
    \tilde{\nabla}_{\mu} R (\boldsymbol{\tilde{g}}) = -  \kappa_D^2 \tilde{\nabla}_{\mu} T (\boldsymbol{\tilde{g}}),
\end{equation}
i.e., the Ricci scalar is related to the trace of the energy-momentum tensor
\begin{equation}
R (\boldsymbol{\tilde{g}}) +  \kappa_D^2 T (\boldsymbol{\tilde{g}}) = 4\Lambda ,
\end{equation}
with $\Lambda$ being an arbitrary integration constant. Plugging it on Eq.~\eqref{traceless_einstein} we find
\begin{equation}
    R_{\mu \nu} (\boldsymbol{\tilde{g}}) - \frac{1}{2} R (\boldsymbol{\tilde{g}}) \tilde{g}_{\mu \nu } + \Lambda \tilde{g}_{\mu \nu}= \kappa_D^2 T_{\mu \nu} (\boldsymbol{\tilde{g}}),
\end{equation}
which are equivalent to Einstein equations upon performing the gauge fixing $\omega = \sqrt{|g|}$. To understand why this condition is a gauge fixing, we just notice from Eq.~\eqref{Eq:finite_weyl}  that we can always perform a Weyl transformation that fixes the determinant of the metric to be any desired function. 

We emphasize that Weyl invariance does not automatically imply conformal invariance. Whereas Weyl invariance here refers to the invariance of the action under conformal rescalings of the metric only (no action on matter fields whatsoever), conformal invariance implies also a suitable compensating rescaling of the fields involved in the construction~\cite{Birrell1982}. 

Notice that the cosmological constant here enters as an integration constant, not as a constant in the Lagrangian. This is the main difference from WTDiff-invariant theories and Diff-invariant ones, the implications of which will be discussed extensively later. In conclusion, we have that the space of solutions in UG is the space of solutions of GR for all the possible values of the cosmological constant, i.e., it contains one additional degree of freedom, as in the linear theory.

\subsection{Common parent theory?}
\label{Subsec:Parent}

Although the analysis of the linear theories above already suggests that a parent theory from which both UG and GR emerge as gauge fixings cannot exist, it is still insightful to examine it in detail. Such an analysis not only clarifies the gauge structure of both theories but also highlights the role of background structures in UG.

\subsubsection{Making GR gauge invariant under Weyl-rescalings of the metric}
\label{Subsec:GR_Weyl}

Let us begin with the GR action with an arbitrary cosmological constant~\eqref{einstein_hilbert_diff}. We want this action to be invariant under Weyl rescalings of the metric, this means, transformations that act on the metric as 
\begin{align}
    g_{\mu \nu} \rightarrow e^{2 \phi (x)} g_{\mu \nu}. 
    \label{Eq:Weyl_Resc}
\end{align}
For such purpose, we introduce a Stueckelberg field $\pi(x)$ in the action as
\begin{align}
    S_{\text{GR-} \Lambda \text{-St}}  [\boldsymbol{g},\pi] = S_{\text{GR-}\Lambda} [ e^{2 \pi} \boldsymbol{g} ]. \label{eq:GRStdef}
\end{align}
In order for this action to be invariant under Weyl rescalings of the metric we need that the $\pi$ field transforms through a shift
\begin{align}
    \pi \rightarrow \pi - \phi,
\end{align}
which in conjunction with Eq.~\eqref{Eq:Weyl_Resc} allows us to deduce that the product $e^{2 \pi} g_{\mu \nu} $ is invariant. Hence, the action $S_{\text{GR-} \Lambda \text{-St}}[\boldsymbol{g},\pi]$ is Weyl invariant. We can use the transformation properties of the Ricci scalar under these rescalings of the metric to write down the action $S_{\text{GR-} \Lambda \text{-St}}[\boldsymbol{g},\pi]$ explicitly in terms of the Stueckelberg field $\pi$ and the metric $\boldsymbol{g}$. Under a transformation $\boldsymbol{g} \rightarrow e^{2\phi} \boldsymbol{g}$ the Ricci scalar transforms as (Eq. D.9 from~\cite{Wald1984} with  $\Omega = e^{\phi}$)
\begin{align}
    R\left( e^{2\phi} \boldsymbol{g} \right) = e^{-2\phi} \Big( R\left( \boldsymbol{g} \right) - 2 D g^{\mu \nu}\nabla_\mu \nabla_\nu \phi - D (D-1) g^{\mu \nu} \nabla_\mu \phi \nabla_\nu  \phi \Big), 
\end{align}
where $\Box := g^{\mu \nu} \nabla_\mu \nabla_\nu$ and $ (\partial \phi )^2 := g^{\mu \nu} \nabla_\mu \phi \nabla_\nu \phi$. The change in the determinant is a simple exponential factor in $\pi$: $\abs{g} \rightarrow e^{2 (D+1) \pi } \abs{g}$ and we obtain the following action after integration by parts
\begin{align}
    S_{\Lambda\text{GR-St}} [\boldsymbol{g},\pi] = \frac{1}{2 \kappa_D^2} \int \dd^{D+1}x\, \sqrt{\abs{g}} e^{ (D-1) \pi} \left[ - 2 \Lambda e^{2 \pi} + R (\boldsymbol{g})+ D (D-1) ( \partial \pi)^2 \right]. 
    \label{Eq:Stueckelberg}
\end{align}
We now want to fix a gauge in which the action is still invariant under transverse diffeomorphisms and Weyl transformations, but no longer under longitudinal diffeomorphisms, to see if it corresponds to the UG version of the original action. To achieve this, we can write down the metric as follows: 
\begin{align}
    g_{\mu \nu} = \abs{g}^{\frac{1}{D+1}} \frac{g_{\mu \nu}}{\abs{g}^{\frac{1}{D+1}}} = \abs{g}^{\frac{1}{D+1}} g'_{\mu \nu},
\end{align}
where we have introduced the auxiliary field $g'_{\mu \nu} := g_{\mu \nu} / \abs{g}^{\frac{1}{D+1}}$ which is automatically invariant under WTDiff transformations, though not under longitudinal diffeomorphisms. To end up in an action which is only Weyl and TDiff invariant, we fix the longitudinal diffeomorphism by imposing 
\begin{align}
    e^{2 \pi } \abs{g}^{\frac{1}{D+1}}  = \omega^{\frac{2}{D+1}},
    \label{eq:gaugefix}
\end{align}
where $\omega$ is arbitrary but nondynamical. Consistency of Eq.~\eqref{eq:gaugefix} requires $\omega$ to be a scalar density with the same weight as $\sqrt{- g}$. We can always reach this condition through a longitudinal diffeomorphism since the value of the factor $e^{2 \pi} \abs{g}^{\frac{1}{D+1}}$, let us call it $F(x)^{\frac{2}{D+1}}=e^{2 \pi} \abs{g}^{\frac{1}{D+1}}$, changes as
\begin{align}
    F^{\frac{2}{D+1}}\quad  \rightarrow \quad \big(F\abs{J}\big)^{\frac{2}{D+1}}  , 
\end{align}
where we have used that $e^{2 \pi}$ is a scalar and that $\abs{g}$ picks a factor $\abs{J}$ which is the determinant of the Jacobian $J$. Thus, to reach the gauge \eqref{eq:gaugefix}, we just make a transformation with 
\begin{align}
    \abs{J}=\frac{\omega}{F}= \frac{\omega}{ \sqrt{\abs{g}}\, e^{(D+1) \pi}}.
\end{align}
If we fix this gauge at the level of the action, and introduce 
\begin{align}
    \tilde{g}_{\mu \nu}  := \omega^{\frac{2}{D+1}} g'_{\mu \nu}  =  \left( \frac{\abs{g}}{\omega^2} \right)^{-\frac{1}{D+1}}g_{\mu \nu}  , 
\end{align}
we get
\begin{align}       
    S_{\text{GR-}\Lambda\text{-St}} [\boldsymbol{g}|_{\eqref{eq:gaugefix}},\pi] = S_{\text{GR-}\Lambda} [\boldsymbol{\tilde{g}}]=    
    \frac{1}{2 \kappa_D^2} \int \dd^{D+1}x\, \omega R\left(\boldsymbol{\tilde{g}}\right)-\frac{\Lambda}{ \kappa^2} \int \dd^{D+1}x\, \omega  \,, \label{eq:SLGRStfixed}
\end{align}
where in the first equality we used \eqref{eq:GRStdef} and in the second one we took into account that $| \tilde{g}| = \omega^2$. In Eq.~\eqref{eq:SLGRStfixed} we see that, as a result of the gauge fixing, we obtain the UG action presented in Eq.~\eqref{einstein_hilbert_wtdiff} plus a nondynamical constant contribution.

Fixing this gauge at the level of the action leads to UG, but this gauge fixing turns out to be illegal. To see it, we can compute the equations of motion from~\eqref{Eq:Stueckelberg}. By varying with respect to $\pi$ and $g^{\mu \nu}$, we respectively get (after recasting the equations a bit):
\begin{align}
    0&= (D-1) R \left( e^{2 \pi} \boldsymbol{g} \right) -2 (D+1) \Lambda \, , \label{eq:eqpiGRS}\\
    0&= R_{\mu \nu}\left(\boldsymbol{g}\right) - \frac{1}{2} R \left( \boldsymbol{g} \right)  g_{\mu \nu} + \Lambda e^{2 \pi}g_{\mu \nu}- (D-1) \nabla_{\mu} \nabla_{\nu} \pi + (D-1) g_{\mu \nu} \Box \pi \nonumber \\
    &\qquad + (D-1) \partial_{\mu} \pi \partial_{\nu} \pi + \frac{1}{2} (D-1)(D-2) g_{\mu \nu} (\partial \pi)^2.\label{eq:eqgGRS}
\end{align}
Notice that the first one implies  the following constraint between the Ricci scalar $R \left( e^{2 \pi} \boldsymbol{g} \right)$ and the cosmological constant
\begin{align}
    R \left( e^{2 \pi} \boldsymbol{g} \right) = \frac{2(D+1)}{D-1}\Lambda\, . 
\end{align}
Fixing the unitary gauge $\pi = 0$ in Eqs.~\eqref{eq:eqpiGRS}-\eqref{eq:eqgGRS} leads to Einstein equations with cosmological constant $\Lambda$ and its trace, meaning that they are not independent. In particular, this equality also holds with the gauge fixing $ e^{ (D+1) \pi} = \omega\abs{g}^{-1/2}$:
\begin{align}
    R ( \boldsymbol{\tilde{g}} ) = \frac{2(D+1)}{D-1}\Lambda\, . 
\end{align}
At the level of the action, the situation is different, since this gauge fixing leads to UG whose Ricci scalar is not fixed by $\Lambda$. Hence, we conclude that the gauge fixing is illegal, i.e., it cannot be performed at the level of the action.

\subsubsection{Making UG gauge invariant under longitudinal diffeomorphisms}
\label{Subsec:UG_LDiffs}

Let us now consider UG and let us make it invariant under the full set of diffeomorphism. For such purpose, we will introduce the Stueckelberg fields $Y^{\mu}(x)$ as follows:
\begin{align}
    & S_{\text{UG-St}}[\boldsymbol{g},Y; \boldsymbol{\omega}]  =  \int  {\dd}^{D+1} x \omega (x) \\
    & \times R \left[ \left( \frac{\omega(x)^2}{\abs{g(Y(x))}}\right)^{\frac{1}{D+1}} \abs{\det(\frac{\partial Y^\mu (x)}{\partial x^\nu})}^{-2/(D+1)}  g_{\rho \sigma} (Y(x)) \frac{\partial Y^\rho (x)}{\partial x^\mu} \frac{\partial Y^\sigma (x)}{\partial x^\nu} \right].
\end{align}
This can be obtained from the UG action \eqref{einstein_hilbert_wtdiff} by making the following replacement everywhere:
\begin{align}
    g_{\mu \nu}(x) \rightarrow G_{\mu \nu} (Y(x)) =  \frac{\partial Y^\rho (x)}{\partial x^\mu} \frac{\partial Y^\sigma (x)}{\partial x^\nu} g_{\rho \sigma} (Y(x)). 
\end{align}
Here, $G_{\mu \nu}$ should not be mistaken for the Einstein tensor. The fields $Y^\mu$ should be invertible functions. The theory enjoys a symmetry which is: $\mathrm{WTDiff} \times \mathrm{Diff}'$. Here, $\text{WTDiff}$ is the symmetry that UG already displays, realized in such a way that it leaves the $Y^{\mu}$ untouched, and the diffeomorphism group $ \mathrm{Diff}'$ comes from the introduction of the Stueckelberg fields $Y^\mu$. They are realized as:
\begin{align}
    g_{\mu \nu} (x) &\xrightarrow{\mathrm{Diff}'}  \frac{\partial f^\rho}{\partial x^\mu} \frac{\partial f^\sigma}{\partial x^\mu} g_{\rho \sigma} (f(x)), \\
    \omega(x) &\xrightarrow{\mathrm{Diff}'} \omega(x), \\
    Y^\mu(x) &\xrightarrow{\mathrm{Diff}'} (f^{-1})^\mu ( Y(x) ), \label{eq:transfY}
\end{align}
where the transformations of $Y^{\mu}$ have been chosen to leave $G_{\mu \nu}(Y(x))$ invariant
\begin{align}
    G_{\mu \nu} (Y(x)) &\xrightarrow{\mathrm{Diff}'} G_{\mu \nu} (Y(x)). 
\end{align}
In order to recover the original UG action, one just has to take the gauge fixing \mbox{$Y^\mu(x)= x^\mu$}. This can always be achieved since the fields $Y^{\mu}$ are invertible, by just performing the diffeomorphism for which $f^\mu = Y^\mu$, so that in this gauge we find
\begin{equation}
    G_{\mu \nu}(Y(x))\Big|_{Y^\mu(x)=x^\mu} = g_{\mu \nu}(x).
\label{eq:Gtogfix}
\end{equation}
Let us now derive the equations of motion. It is convenient to introduce the notation
\begin{equation}
    \tilde{G}_{\mu \nu}(Y(x)):=\left( \frac{\omega(x)}{\abs{G(Y(x))}}\right)^{\frac{1}{D+1}}G_{\mu \nu}(Y(x)),\qquad G:=\det(G_{\mu \nu}),
\end{equation}
where we notice that $Y$ does not appear in the dependence of $\omega$ on $x$. Now, we realize that the dynamical fields $g^{\mu \nu}$ and $Y^\mu$ only appear in the action through the combination $\tilde{G}^{\mu \nu}$, so we have
\begin{align}
    \delta S_{\text{UG-St}} = \int {\dd}^{D+1} x \, \omega(x) \,\left[ R_{\mu \nu} ( \boldsymbol{\tilde{G}} )- \frac{1}{D+1} R ( \boldsymbol{\tilde{G}} ) \tilde{G}_{\mu \nu}  \right]_{Y(x)} \delta G^{\mu \nu}(Y(x)), \label{eq:prevarUGsT}
\end{align}
where we took into account that varying with respect to $\tilde{G}^{\mu \nu}(Y(x))$ is equivalent to varying with respect to the traceless part of $G^{\mu \nu}(Y(x))$ since $\tilde{G}^{\mu \nu}(Y(x))$ has fixed determinant. In Eq.~\eqref{eq:prevarUGsT}, the subscript indicates that the object in the square bracket is evaluated in $Y(x)$. Now we can compute the variation of $G^{\mu \nu}$ coming from both $Y^\mu$ and $g^{\mu \nu}$. It is convenient to introduce
\begin{equation}
    M^\mu{}_\nu := \partial_\nu Y^{\mu},\qquad N^\mu{}_\nu := \partial_\nu (Y^{-1})^{\mu},
\end{equation}
where $\partial_\mu$ just means partial with respect to the $\mu$-th slot. These matrices are the inverse of each other in the following sense $M^\mu{}_\nu (x)N^\nu{}_\rho (Y(x))=N^\mu{}_\nu (Y(x)) M^\nu{}_\rho (x) =\delta^\mu{}_\rho$. It is a consequence of the chain rule, which also implies that
\begin{equation}
    \delta N^\mu{}_\sigma (Y(x)) = - N^\mu{}_\rho (Y(x))\, N^\tau{}_\sigma(Y(x))\, \delta M^\rho{}_\tau (x) \,.\label{eq:MMprop}
\end{equation}
Then: 
\begin{align}
    \delta G^{\mu \nu}(Y(x)) &= \delta \left[ g^{\sigma \rho} N^\mu{}_\sigma N^\nu{}_\rho \right]_{Y(x)}\nonumber\\
    &= \left[\delta g^{\sigma \rho} N^\mu{}_\sigma N^\nu{}_\rho + 2 g^{\sigma \rho} N^\mu{}_\sigma \delta N^\nu{}_\rho\right]_{Y(x)}\nonumber\\
    &= \left[\delta g^{\sigma \rho} N^\mu{}_\sigma N^\nu{}_\rho\right]_{Y(x)} - 2 \left[g^{\sigma \rho} N^\mu{}_\sigma N^\nu{}_\tau N^\delta{}_\rho \right]_{Y(x)} \delta M^\tau{}_\delta (x), \label{eq:varG}
\end{align}
where in the last step we made use of Eq.~\eqref{eq:MMprop}. The first piece of Eq.~\eqref{eq:varG}, which correspond to the variations with respect to $g^{\mu \nu}$, leads to:
\begin{align}
    \left[N^\mu{}_\sigma N^\nu{}_\rho \left(R_{\mu \nu} ( \boldsymbol{\tilde{G}} ) - \frac{1}{D+1} R (\boldsymbol{\tilde{G}} ) \tilde{G}_{\mu \nu}  \right))\right]_{Y(x)}=0, 
\end{align}
which can simply be written as the traceless Einstein equations by multiplying by appropriate factors of $M^\mu{}_\nu(x)$:
\begin{align}
    \left[R_{\mu \nu} ( \boldsymbol{\tilde{G}} ) - \frac{1}{D+1} R (\boldsymbol{\tilde{G}} ) \tilde{G}_{\mu \nu} \right]_{Y(x)}=0, \label{eq:UGStgeq}
\end{align}
On the other hand, the second term in Eq.~\eqref{eq:varG} corresponds to the equations of motion of $Y^\mu$. It contains $\delta M^\tau{}_\delta(x) = \partial_\delta \delta Y^\tau (x)$, can be integrated by parts to obtain:
\begin{align}
    \partial_{\rho}\left\{ \omega(x) \left[  2 g^{\sigma \delta}  N^\rho{}_\delta N^\mu{}_\sigma N^\nu{}_\tau \left(  R_{\mu \nu} ( \boldsymbol{\tilde{G}} ) - \frac{1}{D+1} R (\boldsymbol{\tilde{G}} ) \tilde{G}_{\mu \nu}   \right) \right]_{Y(x)}\right\}=0 . \label{eq:UGStYeq}
\end{align}
This equation is redundant because it is an immediate consequence of Eq.~\eqref{eq:UGStgeq}.

We can now try to fix a gauge in which we reach GR. For that purpose, we would make a Weyl transformation \eqref{Eq:Weyl_Resc} in which the following condition holds
\begin{align}
    \left( \frac{\omega(x)}{\abs{g(Y(x))}}\right)^{\frac{1}{D+1}} \abs{\det(M^\mu{}_\nu)}^{-2/(D+1)}=1\quad \Leftrightarrow\quad 
    \omega(x) = \sqrt{\abs{g(Y(x))}}\abs{\det(M^\mu{}_\nu)} .
\end{align}
In this gauge, the action reduces to:
\begin{align}
    \int  {\dd}^{D+1} x \ \abs{\det(\frac{\partial Y^\mu(x)}{\partial x^\nu})} \sqrt{\abs{g(Y(x))}} \ R \left(  g_{\sigma \rho} (Y(x)) \frac{\partial Y^\sigma (x)}{\partial x^\mu} \frac{\partial Y^\rho(x)}{\partial x^\nu} \right),
\end{align}
and this is the GR action after applying an active diffeomorphism $x^\mu \to Y^\mu(x)$. Of course, this gauge transformation is illegal, in the sense that fixing a gauge and computing the equations of motion does not commute. However, when imposed directly at the level of the equations of motion, Eq.~\eqref{eq:UGStgeq} simply reduces to the traceless part of Einstein's equations evaluated at $x^\mu$, while Eq.~\eqref{eq:UGStYeq} is automatically satisfied as a consequence of the former.

\subsection{Higher-derivative generalizations}
\label{Subsec:Higher_Derivatives}

In this section we will prove that higher-derivative generalizations of the Einstein-Hilbert action with Diff invariance~\eqref{einstein_hilbert_diff} and WTDiff invariance~\eqref{einstein_hilbert_wtdiff}, are classically equivalent in everything but the cosmological constant, which always reenters the equations of motion as an integration constant. 

\paragraph*{\textbf{Diff-invariant theories.}} We start with Diff-invariant generalizations of GR: consider a general theory with higher curvature terms in the action, like $f(R)$ theories, for instance. The only requirement for the action is to be invariant under diffeomorphisms. We will then have 
\begin{equation}
    S_{\text{Diff}} [\boldsymbol{g}]= \int \dd^{D+1}x \sqrt{-g} \big[f(R_{\alpha \beta \gamma}{}^{\delta} (\boldsymbol{g}) , \boldsymbol{g})-2\Lambda\big],
\label{action_diff_higher}
\end{equation}
where the function $f$ is a scalar of curvature built from the metric tensor and its derivatives, through the Riemann tensor. The action can also contain derivatives of the curvature tensor, i.e., more precisely we would need to write down a term of the form \mbox{$f\left(\boldsymbol{g}, R_{\alpha \beta \gamma}{}^{\delta} (\boldsymbol{g}), \nabla_{\tau} R_{\alpha \beta \gamma}{}^{\delta} (\boldsymbol{g}), \ldots \right) $} although we will skip the derivatives to avoid a cumbersome notation. The variation of such an action will contain three pieces
\begin{equation}
    \delta S_{\text{Diff}} [\boldsymbol{g}] = \int \dd^{D+1} x \sqrt{-g} \Big[ E_{\mu \nu} (\boldsymbol{g}) - \frac{1}{2} g_{\mu \nu} f \left( R_{\alpha \beta \gamma}{}^{\delta} (\boldsymbol{g}) , \boldsymbol{g} \right) +\Lambda g_{\mu\nu}\Big] \delta g^{\mu \nu} ,
\end{equation} 
where the first term corresponds to the variation of the $f(R_{\alpha \beta \gamma}{}^{\delta} (\boldsymbol{g}) ,\boldsymbol{g})$ piece of the action, while the last two terms correspond to the variation of the determinant of the metric. Explicitly, the equations of motion for this theory are
\begin{equation}
     E_{\mu \nu} (\boldsymbol{g}) - \frac{1}{2} g_{\mu \nu} f(R_{\alpha \beta \gamma}{}^{\delta} (\boldsymbol{g}) , \boldsymbol{g} ) +\Lambda g_{\mu \nu} = 0. 
\end{equation}
Now we will use the fact that the theory is invariant under diffeomorphisms. Let us begin with the case $\Lambda = 0$, since of course the action is diffeomorphism invariant independently of the value of $\Lambda$. As such, if we take a variation of the form $\delta g^{\mu \nu} = 2 \nabla^{(\mu} \xi^{\nu)}$ such that it vanishes on the boundary, we will have $\delta S = 0$. This means that we have
\begin{align}
    0 =  &  
    \int \dd^{D+1} x \sqrt{-g} \Big[ E_{\mu \nu} (\boldsymbol{g}) - \frac{1}{2} g_{\mu \nu} f(R_{\alpha \beta \gamma}{}^{\delta} (\boldsymbol{g}) , \boldsymbol{g} ) \Big] \nabla^{(\mu} \xi^{\nu)}  \\
    = & - 
    \int \dd^{D+1} x \sqrt{-g} \xi^{\nu} \nabla^{\mu} \Big[ E_{\mu \nu} (\boldsymbol{g})  - \frac{1}{2} g_{\mu \nu} f(R_{\alpha \beta \gamma}{}^{\delta} (\boldsymbol{g}) , \boldsymbol{g}) \Big],
\end{align} 
where we have integrated by parts and discarded boundary terms again. The vector fields $\xi^{\mu}$ are unconstrained, and hence this means that the vanishing of the previous expression implies that we have the identity
\begin{align}
    \nabla_{\mu} \Big[ E^{\mu \nu} (\boldsymbol{g}) - \frac{1}{2} g^{\mu \nu} f(R_{\alpha \beta \gamma}{}^{\delta} (\boldsymbol{g}) , \boldsymbol{g} ) \Big] = 0,
    \label{Eq:Generalized_Bianchi_Id}
\end{align}
which is a generalization of the contracted Bianchi identities for higher-derivative theories. Despite the derivation we have made from a variational principle, notice that Eq.~\eqref{Eq:Generalized_Bianchi_Id} is simply a geometrical identity. We have merely used the variation with respect to transformations generated by the vector fields $\xi^{\mu}$ as a tool to uncover it. The same happens in GR with the identity $\nabla_{\mu} \big[R^{\mu \nu} (\boldsymbol{g}) - g^{\mu \nu} R (\boldsymbol{g}) /2 \big] =  0$, i.e., the vanishing divergence of the Einstein tensor is simply a geometric identity but it can also be found as a consequence of the diffeomorphism invariance of the Einstein-Hilbert action in Eq.~\eqref{einstein_hilbert_diff}.

Now, we can repeat the same argument but with a nonvanishing cosmological constant $\Lambda \neq 0$. This means that instead of Eq.~\eqref{Eq:Generalized_Bianchi_Id}, we would reach
\begin{align}
    \nabla_{\mu} \Big[ E^{\mu \nu} (\boldsymbol{g}) - \frac{1}{2} g^{\mu \nu} f(R_{\alpha \beta \gamma}{}^{\delta} (\boldsymbol{g}) , \boldsymbol{g} ) + \Lambda g_{\mu \nu} \Big] = 0.
    \label{Eq:Generalized_Bianchi_Id_CC}
\end{align}
If we now use the identity that we have already derived, i.e., Eq.~\eqref{Eq:Generalized_Bianchi_Id}, the expression~\eqref{Eq:Generalized_Bianchi_Id_CC} reduces to the identity $\nabla^{\nu} \Lambda = 0$, confirming that the cosmological constant must indeed be a true constant.

\paragraph*{\textbf{WTDiff-invariant theories.}}

Let us now discuss the same version of the action but with a fixed volume background, i.e., the closest action to~\eqref{action_diff_higher} obeying a WTDiff-invariance principle. For this WTDiff-invariant theory, the action would read
\begin{equation}
    S_{\text{WTDiff}} [\boldsymbol{g}] = S_{\text{Diff}} [\boldsymbol{\tilde{g}}] = \int \dd^{D+1} x\, \omega f \left[ R_{\alpha \beta \gamma}{}^{\delta}  ( \boldsymbol{\tilde{g}}), \boldsymbol{\tilde{g}} \right], 
\label{generic_higher_derivative_wtdiff}
\end{equation}
which is the same function we had for the GR generalization, but instead of writing the curvature scalars terms of the metric $g_{\mu \nu}$, we do it in terms of the auxiliary metric $\tilde{g}_{\mu \nu}$ introduced in Eq.~\eqref{auxiliary_metric}. Again, there is no cosmological constant in the action because it would be a nondynamical variable. The equations of motion are now computed as follows
\begin{equation}
    \delta S_{\text{WTDiff}}  [\boldsymbol{g}]  = \int \dd^{D+1} x\, \omega E_{\mu \nu} ( \boldsymbol{\tilde{g}} ) \delta \tilde{g}^{\mu \nu},
\end{equation}
which can be rewritten in terms of the metric $\boldsymbol{g}$ using
\begin{equation}
\delta \tilde{g}^{\mu \nu} = \delta g^{\mu \nu} - \frac{1}{D+1} g^{\mu \nu} g_{\alpha \beta} \delta g^{\alpha \beta}.
\end{equation}
Again, we can use $g_{\mu \nu} g^{\alpha \beta} = \tilde{g}_{\mu \nu} \tilde{g}^{\alpha \beta}$ to express the variation of the action as
\begin{equation}
    \delta S_{\text{WTDiff}} [\boldsymbol{g}] = \int \dd^{D+1} x\, \omega \left( E_{\mu \nu} (\boldsymbol{\tilde{g}}) - \frac{1}{D+1} E (\boldsymbol{\tilde{g}} ) \tilde{g}_{\mu \nu}   \right) \delta g^{\mu \nu}, 
\end{equation}
where $ E (\boldsymbol{\tilde{g}} ) = \tilde{g}^{\mu \nu} E_{\mu \nu} (\boldsymbol{\tilde{g}}) $. Thus, we have the equations of motion
\begin{equation}
    E_{\mu \nu} ( \boldsymbol{\tilde{g}} ) - \frac{1}{D+1} E ( \boldsymbol{\tilde{g}} ) \tilde{g}_{\mu \nu} = 0. 
\end{equation}
Taking the covariant derivative $\tilde{\nabla}_{\mu}$ and using the geometrical identity~\eqref{Eq:Generalized_Bianchi_Id}, we find that 
\begin{equation}
    \tilde{\nabla}_{\mu} E ( \boldsymbol{\tilde{g}} ) = 0.
\end{equation}
Thus, $E(\boldsymbol{\tilde{g}})$ needs to be a constant. Choosing it adequately to have the same normalization we had for the GR generalization, we reach the following equations of motion
\begin{equation}
    E_{\mu \nu} (\boldsymbol{\tilde{g}}) - \frac{1}{2} \tilde{g}_{\mu \nu} f(R_{\alpha \beta \gamma}{}^{\delta} (\boldsymbol{\tilde{g}}) , \boldsymbol{\tilde{g}} ) + \Lambda \tilde{g}_{\mu \nu} = 0,
\end{equation}
which are equivalent to the ones obtained in higher-derivative Diff-invariant generalizations of GR upon fixing the gauge $\sqrt{|g|} = \omega$.

\paragraph*{\textbf{Matter fields.}}

The generalization to include an energy-momentum tensor is almost straightforward. If we assume the same matter content is included in both theories and it is described by an energy-momentum tensor $T_{\mu \nu}$, we have the following equations for the Diff-invariant theory 
\begin{equation}
   E_{\mu \nu} (\boldsymbol{g})- \frac{1}{2} g_{\mu \nu} f(R_{\alpha \beta \gamma}{}^{\delta} (\boldsymbol{g}) , \boldsymbol{g} ) +  \Lambda g_{\mu \nu} = \kappa_D^2 T_{\mu \nu}  (\boldsymbol{g}).
\label{Diff_higher_order_generalizations}
\end{equation}
Following the nomenclature of the previous chapter, given an action for the matter sector which we add to the gravitational part, $S_{\text{M}}(\boldsymbol{g},\Phi)$, the energy-momentum tensor is given by
\begin{align}
    T_{\mu \nu}  (\boldsymbol{g}) = - \frac{2}{\sqrt{-g}} \frac{\delta S_{\text{M}}(\boldsymbol{g},\Phi) }{\delta g^{\mu \nu}},
\end{align}
although we recall that not every energy-momentum tensor needs to be obtained from a variational principle. For the WTDiff version of such theory, we would couple the theory to the traceless part of the energy-momentum tensor. If we have an action principle, for the theory, this follows immediately from the replacement $\boldsymbol{g} \to \boldsymbol{\tilde{g}}$ in the action, i.e., $S_{\text{M}}(\boldsymbol{\tilde{g}}, \Phi)$ since we find
\begin{align}
    \frac{\delta S_{\text{M}}(\boldsymbol{g},\Phi) }{\delta g^{\mu \nu}} = \left[ \frac{\delta S_{\text{M}}(\boldsymbol{g},\Phi) }{\delta \tilde{g}^{\mu \nu}} - \frac{1}{D+1} \tilde{g}_{\mu \nu} \tilde{g}^{\alpha \beta} \frac{\delta S_{\text{M}}(\boldsymbol{g},\Phi) }{\delta \tilde{g}^{\alpha \beta}} \right],
\end{align}
which after expressing in terms of the energy-momentum tensor leads to the following equations of motion
\begin{equation}
E_{\mu \nu} ( \boldsymbol{\tilde{g}} ) - \frac{1}{D+1} \tilde{g}_{\mu \nu} E ( \boldsymbol{\tilde{g}})  = \kappa_D^2 \left( T_{\mu \nu} ( \boldsymbol{\tilde{g}})- \frac{1}{D+1} T ( \boldsymbol{\tilde{g}} ) \tilde{g}_{\mu \nu} \right).
\end{equation}
Taking the derivative and performing the same manipulations introduced above, one reaches the equation 
\begin{equation}
    E_{\mu \nu} (\boldsymbol{\tilde{g}}) - \frac{1}{2} \tilde{g}_{\mu \nu} f(R_{\alpha \beta \gamma}{}^{\delta} (\boldsymbol{\tilde{g}}) , \boldsymbol{\tilde{g}} ) + \Lambda \tilde{g}_{\mu \nu} = \kappa_D^2 T_{\mu \nu} (\boldsymbol{\tilde{g}}).
\end{equation}
This set of equations agree with those for a Diff-invariant theory~\eqref{Diff_higher_order_generalizations}, upon fixing the gauge $\sqrt{|g|} = \omega$. Again, $\Lambda$ is an integration constant or a parameter from the action, depending on whether we are on the WTDiff-invariant or the Diff-invariant version. 

As a final comment, let us emphasize that working with a general dynamical connection that is independent of the metric (this means, one does not include the Levi-Civita one), as it is done in the metric-affine formalism, can be easily incorporated in our discussion. For all practical purposes, the general affine connection, which can be expressed in terms of the torsion and nonmetricity tensors as in the previous chapter, would simply be treated as additional matter fields. In that sense, no differences would appear in either approach as long as one does not violate the conservation of the energy-momentum tensor. 

\subsection{Non-minimal couplings. Einstein and Jordan frames}
\label{Subsec:Nonminimal_couplings}

Up to this point, we have implicitly assumed that all couplings to gravity follow the minimal-coupling prescription. In this section, we will examine the potential inequivalence, at the classical level, of theories with non-minimal couplings formulated under Diff and WTDiff principles. We will begin by reviewing the well-known results for non-minimal couplings in Diff-invariant theories before extending the discussion to WTDiff-invariant cases. For simplicity, we will focus on the Einstein-Hilbert action, considering only the Ricci scalar as the curvature invariant in the equations of motion. However, our analysis can be directly extrapolated to arbitrary non-minimal couplings.

\paragraph*{\textbf{Non-minimal couplings in Diff-invariant theories.}}

Non-minimal couplings in Diff-invariant theories can be mapped to an equivalent description in which the theory displays just minimal couplings~\cite{Faraoni1998}. These transformations are associated with conformal rescalings of the metric variable and we refer to this operation as a change of frame. We will first discuss the mathematical grounds on which these transformations are performed and then move on to their physical interpretation. 
 
To begin with, let us consider the following action for a scalar field $\phi$ coupled to the metric $g^{\jordanf}_{\mu \nu}$
\begin{equation}
    S^{\jordanf} [\boldsymbol{g}^{\jordanf},\phi] = \frac{1}{2 \kappa_D^2} \int \dd^{D+1} x \sqrt{|g^{\jordanf}|} \left[ F^{\jordanf} (\phi) R (\boldsymbol{g}^{\jordanf}) + G^{\jordanf} (\phi) \partial_{\mu} \phi \partial^{\mu} \phi + V^{\jordanf}  ( \phi ) \right],
    \label{diff_action_jordan}
\end{equation}
where the superscript (J) means that we are working in the so-called Jordan frame. Performing a conformal transformation of the form
\begin{equation}
    g^{\einsteinf}_{\mu \nu} = \left[ F(\phi) \right]^\frac{2}{D-1} g^{\jordanf}_{\mu \nu},
\end{equation}
one finds the action 
\begin{equation}
    S^{\einsteinf} [\boldsymbol{g}^{\einsteinf}, \phi] = \frac{1}{2 \kappa_D^2} \int \dd^{D+1} x \sqrt{|g^{\einsteinf}|} \left[ R (\boldsymbol{g}^{\einsteinf}) + G^{\einsteinf} (\phi) \partial_{\mu} \phi \partial^{\mu} \phi + V^{\einsteinf}  ( \phi ) \right],
    \label{diff_action_einstein}
\end{equation}
where the superscript (E) means that we have moved to the so-called Einstein frame~\cite{Faraoni1998}. The functions $G^{\einsteinf} (\phi),V^{\einsteinf} (\phi)$ differ from the ones in the Jordan frame, $G^{\einsteinf} (\phi) \neq G^{\jordanf} (\phi)$ and $V^{\einsteinf} (\phi) \neq V^{\jordanf} (\phi)$. The explicit relation is irrelevant for our discussion and it can be worked out using the standard properties of the Ricci scalar under conformal transformations. 

There has been ongoing debate over the years regarding which of the two frames is the ``physical" one, meaning on which of both frames do massive timelike particles propagate on geodesic trajectories~\cite{Faraoni1998}. Here, we simply point out that the equivalence between the two descriptions implies that if particles move along geodesics in one frame, they will generally not do so in another. In other words, a metric theory of gravity is defined not only by its equations of motion but also by how local probes couple to the underlying geometry~\cite{Sotiriou2007,Will2014}. The same equations of motion with different couplings to local probes would represent different theories. In that sense, one needs to understand that the equivalence among frames requires also the mapping of the probe dynamics between frames. Hence, specifying a physical theory for gravity coupled to a matter scalar field requires more information than a simple frame: it requires also the dynamics of the probes that will propagate on the corresponding geometry. 

In order to discuss the existence of a theory obeying a WTDiff principle with the same equations of motion in a suitable gauge, let us prove a useful identity which is the analogue to Eq.~\eqref{Eq:Generalized_Bianchi_Id}. Following the same procedure used above, let us introduce the equations of motion associated with the action from Eq.~\eqref{diff_action_jordan},
\begin{equation}
    K_{\mu \nu} (\boldsymbol{g}^{\jordanf}, \phi) = 0, \quad \textrm{with} \quad K_{\mu \nu} (\boldsymbol{g}^{\jordanf} , \phi) := \frac{\delta S}{\delta g^{\jordanf \mu \nu}} .
    \label{non_minimal_jordan_eqs}
\end{equation}
Using the diffeomorphism invariance of the action under transformations \mbox{$\delta g^{{\jordanf} \mu \nu} = 2 \nabla^{\jordanf( \mu} \xi^{\nu)}$} and performing an integration by parts we find
\begin{equation}
    0 = 2 \int \dd^{D+1}x \sqrt{|g|} K_{\mu \nu} (\boldsymbol{g}^{\jordanf}, \phi)\nabla^{ \jordanf ( \mu} \xi^{\nu)} = - 2 \int \dd^{D+1}x\, \xi^{\nu} \nabla^{{\jordanf} \mu} K_{\mu \nu} (\boldsymbol{g}^{\jordanf}, \phi). 
\end{equation}
Thus, we have a generalization of Bianchi identities for a theory with non-minimal couplings
\begin{equation}
    \nabla^{\jordanf}_{\mu} K^{\mu \nu} (\boldsymbol{g}^{\jordanf}, \phi) = 0.
    \label{Bianchi_identity_jordan}
\end{equation}
This relation will be used later.

\paragraph*{\textbf{Non-minimal couplings in WTDiff-invariant theories.}}

For WTDiff theories, we cannot perform frame changes since they are gauge transformations. However, as we have discussed above, frame changes by themselves are not enough to specify a theory, since one needs to specify the frame in which particles propagate in timelike geodesics. In the WTDiff version of these theories, this choice of frame manifests simply as a choice of action. Hence, no superscript associated with frames will be used for these considerations.

First, observe that for a generic Diff-invariant theory in the Einstein frame as considered in Eq.~\eqref{diff_action_einstein}, which is equivalent to those introduced in Eq.~\eqref{einstein_hilbert_diff} with a suitable matter content, there exists a corresponding WTDiff-invariant theory within the family described by Eq.~\eqref{einstein_hilbert_wtdiff}.

Furthermore, we note that the equations of motion of a given Diff-invariant theory in the Jordan frame, such as those in Eq.~\eqref{diff_action_jordan}, possibly including an additional cosmological constant term, can also be derived from a suitably constructed WTDiff-invariant action. Specifically, for the action~\eqref{diff_action_jordan}, the corresponding WTDiff-invariant theory is described by the following action:
\begin{equation}
    \tilde{S}^{\jordanf} [\boldsymbol{g},\phi] = \frac{1}{2 \kappa_D^2} \int \dd^{D+1} x\, \omega \left[ F^{\jordanf} (\phi) R (\boldsymbol{\tilde{g}}) + G^{\jordanf} (\phi) \partial_{\mu} \phi \partial^{\mu} \phi + V^{\jordanf}  ( \phi ) \right].
    \label{wtdiff_action_jordan} 
\end{equation}
The variation of this action gives the following equations of motion
\begin{equation}
    K_{\mu \nu} (\boldsymbol{\tilde{g}},\phi) - \frac{1}{D+1} \tilde{g}_{\mu \nu} K ( \boldsymbol{\tilde{g}},\phi )  = 0.
\end{equation}
Here, $K_{\mu \nu}( \boldsymbol{\tilde{g}},\phi)$ has the same functional dependence on $\boldsymbol{\tilde{g}}$ and its derivatives as $K_{\mu \nu}( \boldsymbol{g},\phi)$ has on $\boldsymbol{g}$. We have also introduced its trace, defined as $K ( \boldsymbol{\tilde{g}},\phi) = \tilde{g}^{\mu \nu} K_{\mu \nu} (\boldsymbol{\tilde{g}},\phi)$. Taking the covariant derivative $\tilde{\nabla}_{\mu}$ and using the identity~\eqref{Bianchi_identity_jordan}, we arrive at the following equations after a straightforward integration of the resulting expression:
\begin{equation}
    K_{\mu \nu} ( \boldsymbol{\tilde{g}},\phi)+ \Lambda \tilde{g}_{\mu \nu} =0. 
\end{equation}
Fixing the gauge in which $\tilde{g}_{\mu \nu} = g_{\mu \nu}$, we reach the equations we had for the Diff-invariant theory, those from Eq.~\eqref{non_minimal_jordan_eqs}. Hence, we have proved that for a generic theory with non-minimal couplings in a Diff-invariant theory, we have equivalent descriptions in a theory obeying a WTDiff-invariance principle. This is independent of whether test particles follow geodesics on either the Jordan frame or the Einstein frame, although we emphasize again that they correspond to different physical theories. The same logic would hold for other kind of theories with other forms of non-minimal couplings.

\subsection{Nonconserved energy-momentum tensor?}
\label{Subsec:Non_Conserv}

Up to this point, we have assumed that the energy-momentum tensor of the matter sector is a conserved quantity. However, unlike in GR, UG does not require energy-momentum conservation for the consistency of its equations of motion. To clarify this distinction, let us first analyze the GR case.\footnote{Notice that the following arguments can be straightforwardly extended to consider higher derivative generalizations of the equations too.} Take Einstein equations with an arbitrary energy-momentum tensor:
\begin{align}
    R^{\mu \nu} - \frac{1}{2} g^{\mu \nu} R = \kappa_D^2 T^{\mu \nu}.
\end{align}
Taking the divergence of this equation, the left-hand side vanishes identically off-shell due to the Bianchi identities. Consequently, for consistency, the energy-momentum tensor must also be divergenceless on-shell. In many cases, particularly for matter sectors derived from a diffeomorphism-invariant action principle, this conservation follows directly from the equations of motion. Consider, for instance, a matter sector described by an action $S_{\text{matter}} \left[ \boldsymbol{g}, \Phi \right]$. If we perform an on-shell diffeomorphism transformation, this means that the equations of motion are imposed, and only the metric transforms as $\delta g^{\mu \nu} = 2 \nabla^{(\mu} \xi^{\nu)}$ we have
\begin{align}
    0  & = 2 \int \dd^{D+1} x \sqrt{-g} \frac{\delta S_{\text{matter}}}{\delta g^{\mu \nu}} \nabla^{(\mu} \xi^{\nu)} \\
    & = - \int \dd^{D+1} x \sqrt{-g} \xi^{\nu} \nabla^{\mu} \left[ \frac{1}{\sqrt{-g}} \frac{\delta S_{\text{matter}}}{\delta g^{\mu \nu}} \right],
\end{align}
where we have passed from the first line to the second line through an integration by parts and discarding boundary terms since the variation vanishes there. If we introduce now the definition of the energy-momentum tensor from Eq.~\eqref{Eq:EM_Tensor_Definition}, we are led to the conclusion that it has to be conserved on-shell $\nabla_{\mu} T^{\mu \nu} = 0$. 

We can illustrate this with an explicit example which is the energy-momentum tensor of a free real scalar field supplemented with the Klein-Gordon equation of motion. The energy-momentum tensor is
\begin{align}
    T_{\mu \nu} = \nabla_{\mu} \phi \nabla_{\nu} \phi - \frac{1}{2}g_{\mu \nu} \nabla_{\alpha} \phi \nabla^{\alpha} \phi, 
\end{align}
we can take the divergence to find after a cancellation among some of the terms that appear:
\begin{align}
    \nabla^{\mu} T_{\mu \nu} =  \nabla^{\mu} \nabla_{\mu} \phi \nabla_{\nu} \phi,
\end{align}
which is automatically zero on the space of solutions since \mbox{$\nabla_{\mu} \nabla^{\mu} \phi = 0$.} 

However, there are examples of energy-momentum tensors that are not derived from an action principle, or, at least, it is not easy to do it. For instance, the energy-momentum tensor of a perfect fluid is given by
\begin{align}
    T^{\mu \nu} = \left( \rho + p \right)u^{\mu} u^{\nu} + p g^{\mu \nu}.
\end{align}
In principle, the conservation is not ensured \emph{a priori}. Furthermore, the equations of motion of the fluid on the other hand are unspecified. What is usually done is to derive the equations of motion of the fluid precisely by imposing the conservation of the energy-momentum tensor. This leads to the following equations of motion:
\begin{align}
    & \left( \nabla_{\mu} \rho \right) u^{\mu} + \left( \rho + p \right) \nabla_{\mu} u^{\mu} = 0, \\
    & \left( \rho + p \right) u^{\nu} \nabla_{\nu} u^{\mu} + \left( g^{\mu \nu} + u^{ \mu} u^{\nu} \right) \nabla_{\nu} p = 0,
\end{align}
with $p$ the pressure, $\rho$ the density, and $u^{\mu}$ the velocity of the fluid. Once we are within the UG framework, demanding conservation of the energy-momentum tensor is not mandatory anymore, since we can simply allow for its nonconservation. The main problem is that, in general situations, it is not easy to find well-motivated nonconserved energy-momentum tensors. An example where such a nonconservation is motivated from a microscopic principle is~\cite{Perez2017}. In that paper, using some heuristic arguments, the authors propose a concrete functional form for an effective-field theory-like expansion of the particular nonconserved energy-momentum tensor divergence current $J_{\nu} = \nabla^{\mu} T_{\mu \nu}$. They make the key assumption that the one form $J_{\nu}$ is exact, i.e., $\nabla_{[\mu} J_{\nu]} = 0$. In this way, it is (at least locally) an exact form that can be written as $J_{\mu} = \nabla_{\mu} Q$. Then, they are able to obtain some effective equations with a cosmological constant where the cosmological constant is the sum of an arbitrary constant and an integral of the current $J_{\mu}$. 

To the best of our knowledge, there has been no systematic investigation into the phenomenology of different types of nonconserved energy-momentum tensors. Even for the model described above, without the key assumption that the $\boldsymbol{J}$ form is exact, making further progress is not straightforward.

\section{Semiclassical regime}
\label{Sec:Semiclassical}

So far, we have analyzed the theory solely at the classical level by deriving the equations of motion from the classical action. A step toward a full theory of quantum gravity involves studying the behavior of quantum fields in curved spacetime~\cite{Birrell1982,Wald1995}. In this framework, the metric is treated as a fixed, generically curved background, which does not evolve in response to the backreaction of quantum fields. Even with that limitation, this approach already reveals a wealth of new phenomena such as Hawking radiation or cosmological particle production.

The semiclassical regime precisely constitutes the regime in which the backreaction of those quantum fields on the geometry is taken into account. For that purpose, one would include the contribution of the quantum fields to the energy-momentum tensor on the right-hand side of Einstein equations. The main problem that one faces is that the energy-momentum tensor, even for free fields, contains at least the product of the fields and/or its derivatives evaluated at the same spacetime point. In QFT, the classical fields $\Phi(x)$ are promoted to operator valued distributions $ \widehat{\Phi}(x)$, making the expression inherently divergent and requiring renormalization for a sensible result. While, in principle, a similar renormalization procedure could be applied to interacting theories by systematically redefining composite operators involving the product of more than two fields and/or their derivatives, the standard practice is to focus on free fields. 

After renormalization, we have that the energy-momentum tensor is promoted to a well-defined operator valued distribution $\widehat{T}^{\text{R}}_{\mu \nu} (x)$, and we can model the backreaction on the geometry by inserting its expectation value in the suitable state under consideration $ \langle \widehat{T}^{\text{R}}_{\mu \nu} (x) \rangle_{\Omega}$.\footnote{From now on we will skip the dependence on the spacetime point $x$.} Choosing a state $\Omega$ is quite a complicated task in arbitrarily curved spacetimes. In fact, different choices of $\Omega$ lead to completely different QFT algebras and observables~\cite{Wald1995}.

In that sense, whereas in flat spacetime we have a preferred vacuum which is the only one that is Poincar\'e invariant, in curved spacetimes there is no univocal criterion for selecting among the possible vacuum states. Only in highly symmetric geometries, e.g., a static spacetime, or horizonless spacetimes with flat regions or asymptotically flat regions, it is possible to invoke physical arguments to single out a preferred vacuum. Although the renormalized energy-momentum tensor  itself ends up depending only on the spacetime point $x$, its dependence on the metric $\boldsymbol{g}$, might be nonlocal. Furthermore, it depends on a bunch of renormalized coupling constants that might arise from the renormalization procedure and would need to be fixed experimentally. 

\subsection{Local and nonlocal terms}
To see this explicitly, we can work in the path integral formalism, although any approach would give rise to the same result. In particular, we will compute the effective action $\Gamma[\boldsymbol{g}]$ which is defined as
\begin{align}
    e^{i \Gamma_{\Omega}[ \boldsymbol{g} ]} = \int \mathcal{D} \Phi_{\Omega} e^{i S[\Phi, \boldsymbol{g}] + i S_{\text{bare}} [\boldsymbol{g}]},
    \label{integration_out_effective_action}
\end{align}
where we have introduced $S_{\text{bare}} [\boldsymbol{g}]$ representing any potential counterterms that might be needed to have a finite effective action. From the effective action, it is straightforward to obtain the energy-momentum tensor as
\begin{align}
    \frac{\delta \Gamma_{\Omega} [ \boldsymbol{g} ] }{\delta g^{\mu \nu}} = \langle \widehat{T}^{\text{R}}_{\mu \nu} \rangle_{\Omega}. 
\end{align}

We have explicitly introduced the subscript $\Omega$ in both the integration measure of the path integral and the effective action to emphasize their dependence on the choice of vacuum state for the fields. In particular, computing the path integral requires making sense of the inverse of the propagator. For instance, in the case of a scalar field, this involves inverting an operator such as $\left( \Box - m^2 - \xi R (\boldsymbol{g}) \right)$ where $m^2$ and $\xi$ are parameters of the theory. This process is equivalent to selecting a Green function for the corresponding Klein-Gordon equation. In fact, it is directly analogous to what occurs in canonical quantization, where different choices of quantization modes lead to different Green functions~\cite{Barcelo2011}.

The UV divergences arise as local terms in curvature and they are state independent, so they can be removed through the renormalization procedure by conveniently choosing the local counterterms. Those terms are schematically of the following form in four spacetime dimensions
\begin{equation}
    \Gamma_{\text{local}} (\boldsymbol{g})  = \int \dd^4 x \sqrt{-g} \big[ - \lambda_0 + \lambda_1 R (\boldsymbol{g})+ \lambda_2 R^2(\boldsymbol{g}) + \lambda_3 C_{\mu \nu \rho \sigma} (\boldsymbol{g})C^{\mu \nu \rho \sigma} (\boldsymbol{g}) \big],
    \label{sakharov_effective_action}
\end{equation}
where we have introduced the Weyl tensor $C^{\mu \nu \rho \sigma} (\boldsymbol{g})$ in the previous expression following the conventions of~\cite{Misner1974}. We omit the $R_{\mu \nu} R^{\mu \nu}$ operator since in four spacetime dimensions the Gauss-Bonnet term is a topological invariant, and it allows to eliminate one of the quadratic operators in curvature. These constants can be separated in three pieces: a divergent piece, $\lambda_i ^{\textrm{div}}$, arising from the divergent part of the loop computation; a finite part, $\lambda_i ^{\textrm{finite}}$, arising from the finite parts of the loop computation; and another part coming from the bare action for the gravitational field $S_{\textrm{bare}} (\boldsymbol{g})$
\begin{equation}
    \lambda_i =   \lambda_i^{\textrm{div}} + \lambda_i ^{\textrm{finite}} + \lambda_i ^{\textrm{bare}}.
    \label{one_loop_coupling_constants}
\end{equation}
The infinite part of the bare coupling constants needs to be chosen in such a way that they contain a divergent piece which exactly cancels out the infinities in $\lambda_i^{\textrm{div}}$. In other words, at one loop we have:
\begin{align}
    \lambda_i^{\textrm{bare}} = - \lambda_i ^{\textrm{div}} +  \tilde{\lambda}_i^{\textrm{bare}},
\end{align}
where $\tilde{\lambda}^{\textrm{bare}}_i$ represents the finite part of the coupling constants in the bare action. The resulting coupling is actually the sum of the finite pieces
\begin{align}
    \lambda_i^{\text{obs}} = \lambda_i^{\textrm{div}} + \tilde{\lambda}_i^{\textrm{bare}}, 
\end{align}
and it would need to be fixed somehow experimentally or by some theoretical arguments. In standard QFT, these couplings can be fixed by measuring, for instance, some scattering amplitudes at a given energy scale and then all the amplitudes at every energy scale (for which the theory is perturbatively defined) are settled. For example, in Quantum Electrodynamics this simply corresponds to measuring the fine structure constant by measuring a given amplitude at a given energy scale, and then the scattering of an arbitrary number of photons, electrons and positrons is fully determined for arbitrary energy scales, at least well-below the Landau pole. The main difference in semiclassical gravity regarding the determination of those couplings is that they can only be measured through their backreaction on the metric, and it is not easy to find a way to measure them independently, and then analyze in detail their implications for semiclassical gravity. 

On the other hand, we have the state-dependent part of the action which is encoded somehow in the nonlocal terms, although it is not easy to perform a sharp splitting between these nonlocal terms and the finite pieces from the local terms. In that sense, the nonlocal piece of the action is often ignored as being less relevant for the purposes of finding a renormalized action and, specially, for massive fields. However, those terms are the ones responsible for the huge backreaction effects in extreme causal scenarios, such as spacetimes on the verge of developing chronological horizons, the ones responsible for the Hawking radiation, etc. This is the case even for massive fields, since these kind of extreme causal situations are such that the almost infinite redshift, ensures that all the fields behave as if they were massless. In that sense, these terms are crucial for understanding how the classical gravitational phenomena change when the backreaction of quantum fields is taken into account.

At the end of the day, it is the combination of both of the local and nonlocal pieces that appear in the renormalized energy-momentum tensor entering the semiclassical Einstein equations as:
\begin{align}
    R_{\mu \nu} \left( \boldsymbol{g} \right) - \frac{1}{2} R \left( \boldsymbol{g} \right) g_{\mu \nu} = 8 \pi \kappa_D^2 \left[ T^{\text{C}}_{\mu \nu} + \langle \widehat{T}^{\text{R}}_{\mu \nu}\rangle_{\Omega} \right].
\end{align}
Notice that all the local higher order curvature terms appearing in Eq.~\eqref{integration_out_effective_action} required to remove the UV divergences are included in $\langle \widehat{T}^{\text{R}}_{\mu \nu}\rangle_{\Omega}$. This means that we are regarding them as part of the renormalized energy-momentum tensor, instead of modifications of GR require to renormalize the fields. For practical purposes, it is irrelevant whether these local terms are included on the left-hand side or the right-hand side of the equation.

Now, we can repeat the analysis with a WTDiff invariant theory. Consider that instead of the fields propagating on top of a geometry, we assume that they propagate on top of a conformal structure like the one that a generic WTDiff-invariant theory displays. Hence, we would replace everywhere in the action the metric $\boldsymbol{g}$ by the auxiliary metric $\boldsymbol{\tilde{g}}$ from Eq.~\eqref{auxiliary_metric}. Furthermore, we would replace the metric volume form $\sqrt{|g|}$ with the privileged background volume form $\boldsymbol{\omega}$. With this, all the results obtained above would be obtained here but with $\boldsymbol{g}$ replaced by $\boldsymbol{\tilde{g}}$, except for the cosmological constant term, that disappears~\cite{Carballo-Rubio2015}. To put it explicitly, we get the following local terms for a WTDiff-invariant theory 
\begin{equation}
    \tilde{\Gamma}^{(1)}_{\textrm{eff}} (\boldsymbol{\tilde{g}})  = \int \dd^4x \,\omega \left( \lambda_1 R (\boldsymbol{\tilde{g}}) + \lambda_2 \tilde{R}^2(\boldsymbol{\tilde{g}}) + \lambda_3 C_{\mu \nu \rho \sigma}(\boldsymbol{\tilde{g}}) C^{\mu \nu \rho \sigma} (\boldsymbol{\tilde{g}}) \right),
    \label{one_loop_effective_action_wtdiff}
\end{equation}
and the same nonlocal terms, upon the substitution $\boldsymbol{g} \to \boldsymbol{\tilde{g}}$. In that sense, from the perspective of semiclassical gravity, it seems that the difference between a Diff and a WTDiff principle (as long as we follow the same prescription for computing the renormalized energy-momentum tensor) is again restricted to the different behavior of the cosmological constant. The trace anomaly, while it still appears in the renormalized energy-momentum tensor, has no dynamical consequences since the metric couples only to the traceless part in the approach that we are taking. The trace would only appear upon using the conservation equation.

For diffeomorphism-invariant theories, under a set of well-motivated assumptions, including the requirement that the energy-momentum tensor be divergence-free, Wald demonstrated~\cite{Wald1977,Wald1978,Wald1995} that any two renormalized energy-momentum tensors satisfying these conditions can differ only by local curvature terms. This reflects the fact that ambiguities in the renormalization procedure are limited to the coefficients of the higher-curvature terms appearing in the action~\eqref{sakharov_effective_action}. Remarkably, in WTDiff-invariant theories, more general scenarios become possible. As discussed in Section~\ref{Subsec:Non_Conserv}, conservation of the energy-momentum tensor is no longer required, allowing for a broader class of viable tensors. However, identifying a well-motivated criterion for selecting a non-conserved energy-momentum tensor remains a significant challenge.

\subsection{Potential Weyl anomaly}
\label{Subsec:Weyl}

There is a point that is worth mentioning here concerning the possibility that the Weyl symmetry becomes anomalous at the quantum level. Anomalies arise when a theory exhibits two symmetries at the classical level but preserving both during quantization is impossible. In such cases, one symmetry must be sacrificed in favor of the other. A well-known example is the chiral anomaly, which involves a trade-off between the conservation of vector and axial currents for massless fermions. If we employ a regularization scheme that preserves one of these currents, the other necessarily becomes anomalous. The standard approach ensures gauge invariance by preserving the vector current while allowing the axial current to be anomalous. 

In the gravitational context, it is well-known that Weyl symmetry becomes anomalous and gives rise to the so-called conformal anomaly~\cite{Deser1976}. In this case, the clash between symmetries is precisely between Diff invariance and Weyl transformations. The insistence on using a Diff invariant quantization scheme, automatically renders the Weyl symmetry anomalous. In particular, the trade-off happens between longitudinal diffeomorphisms and Weyl transformations. However, quantizing a theory which is only invariant under transverse diffeomorphisms preserving them does not compromise Weyl symmetry. 

To demonstrate this, we follow the arguments in~\cite{Carballo-Rubio2015} and work in the path integral formalism.\footnote{As always, to give a proper definition of the path integral one needs to work in Euclidean time and then rotate back to the Lorentzian signature to get sensible results.} Anomalies in this framework occur when a symmetry that is preserved at the classical level is broken at the quantum level due to the noninvariance of the path-integral measure. In other words, while the classical action remains symmetric, the measure itself may transform nontrivially. To determine whether the measure is truly invariant, Fujikawa’s method \cite{Fujikawa1979,Fujikawa1980,Weinberg1996} provides a systematic approach. 

The typical procedure to define such measure is to endow the space of fields with an inner product $(\cdot, \cdot)$. For a scalar field, it can be straightforwardly done as follows
\begin{equation}
    (\phi_1, \phi_2) = \int \dd \mu (x) \phi_1 (x) \phi_2(x),
\end{equation}
with an appropriate integration measure $\dd \mu (x)$. In the case of Diff-invariant theories, the natural choice is the standard Riemannian measure: $\dd \mu (x) = \dd^{D+1}x \sqrt{g (x)}$.  For WTDiff theories, however, it corresponds to: $\dd \tilde{\mu} (x) =  \dd^{D+1}x \sqrt{\tilde{g} (x)} = \dd^{D+1}x \omega (x)$, which incorporates the preferred background volume form. With this in place, we can introduce an orthonormal basis $\{\phi_n \}_n$, such that a generic field in this basis would be decomposed as
\begin{equation}
    \phi (x) =  \sum_n c_n \phi_n(x), \qquad  c_n = \int \dd \mu (x) \phi_n(x) \phi(x),
\end{equation}
and define the measure of integration that enters the path integral as\footnote{Notice that we omit the choice of state $\Omega$ in the measure for simplicity in the notation.}
\begin{equation}
    \mathcal{D} \phi = \prod_{n} \frac{ \dd c_n}{\sqrt{2 \pi}}.
\end{equation}
Let us start by examining the case of diffeomorphism invariance. By construction, the measure remains invariant under both longitudinal and transverse diffeomorphisms. However, it does not retain the invariance under Weyl transformations. Fujikawa's method makes this explicit by computing the Jacobian associated with the transformation of the measure under Weyl transformations. Specifically, under an infinitesimal Weyl transformation, where the metric varies as $\delta g_{\mu \nu} = \varphi g_{\mu \nu} $, the coefficients $c_n$ change as
\begin{equation}
    \delta c_n = \int \dd \mu (x) \varphi^2(x) \phi_n (x) \phi (x), 
\end{equation}
since the $ \dd \mu$ changes as
\begin{equation}
    \delta \dd \mu (x) = \dd \mu (x) \varphi^2 (x) .
\end{equation}
The Jacobian $J$ associated with this integral is given by 
\begin{equation}
    \log J = \lim_{N \rightarrow \infty} \sum_{n=1}^{\infty} \int \dd \mu(x) \varphi (x) \phi_n (x) \phi_n(x),
\end{equation}
where a proper regularization needs to be considered. After such regularization, we find a finite Jacobian which is the manifestation of the conformal anomaly~\cite{Deser1976}. 

The situation in WTDiff is different because the measure of integration $\dd \tilde{\mu}(x)$ now is just invariant under transverse diffeomorphisms. This allows it to be invariant under the conformal transformations, since now we simply have that 
\begin{equation}
    \delta \dd \tilde{\mu} (x) = \dd^{4} x\, \delta \omega = 0.
\end{equation}
Since the volume form remains unchanged under Weyl transformations, we conclude that scalar fields do not exhibit Weyl anomalies. The cases of spin-$\frac{1}{2}$ and spin-$1$ fields can be analyzed in a similar manner, and no anomalies are found either~\cite{Carballo-Rubio2015}.

\subsection{The cosmological constant and the WTDiff principle}
\label{Subsec:CC}

The cosmological constant problem has been a major focus of theoretical physics research~\cite{Weinberg1988,Martin2012,Burgess2013,Padilla2015}. In Diff-invariant theories, it is associated with the coefficient of the operator $\sqrt{-g}$ in the effective gravitational action, and as such it is one of the coupling constants of the theory. The cosmological constant problem is often described as the vast discrepancy between its observed value and the theoretical prediction based on the vacuum energy of Standard Model quantum fields. However, determining the specific value of the cosmological constant falls outside the scope of the effective field theory framework~\cite{Polchinski1992,Manohar2018}. In this approach, the values of coupling constants cannot be predicted but are instead fixed by experimental measurements.

The most compelling approach to sharply formulating the cosmological constant problem involves its connection to the concept of technical naturalness. A coupling is considered not technically natural if \emph{generic} extensions of the theory, either UV-complete theories or Effective Field Theories (EFTs) with a higher cutoff, lead to a large discrepancy (by orders of magnitude) in its predicted value. In Burgess's terminology~\cite{Burgess2013,Burgess2020}, a coupling is technically natural if there is an answer to the following two questions: \mbox{i) Why} is it small compared to other physical scales? ii) Why does it remain small after integrating out heavy degrees of freedom to derive an effective field theory valid at those scales? A specific way to ensure technical naturalness, often called 't Hooft naturalness~\cite{tHooft1979}, is through the presence of an approximate symmetry that protects the coupling from acquiring larger values.

We believe that the notion of naturalness itself is quite problematic though. First, there are formulations of quantum field theory in which scale hierarchies are not inherently troublesome, and technically unnatural values of coupling constants do not appear to present any fundamental issue~\cite{Mooij2021,Mooij2021b,Mooij2024,Ageeva2024}. Moreover, the idea of \emph{generic} completions of a theory is somewhat vague, as it requires assigning a probability distribution to the coupling constants of the high-energy completion, an arbitrary choice with no clear justification~\cite{Hossenfelder2018}. Additionally, since most quantum field theories in spacetime dimensions higher than four are only well-defined perturbatively,\footnote{Only highly symmetric theories such as supersymmetric theories admit fully satisfactory nonperturbative formulations.} the concept of \emph{generic} couplings is biased toward those that allow for a perturbative expansion and limits the possible completions of the theory that are considered. Finally, it is worth noting that many of the so-called ``successes" of the naturalness principle in predicting new physics are actually \emph{postdictions}~\cite{Dijkstra2019}, e.g., the electron mass or the pion mass difference. Even when apparently unnatural values of a coupling constants~\cite{Arkani-Hamed2021} can be reconciled with naturalness~\cite{Craig2021}, the explanation often requires an overly intricate analysis, for example, invoking a very complicated symmetry. In such cases, applying Occam’s razor may favor abandoning the notion of naturalness altogether, rather than insisting on its preservation as a guiding principle for model-building.

Regardless of these concerns, let us set them aside and adopt the most conventional perspective on the cosmological constant problem. In Diff-invariant theories, the second of the questions that Burgess puts forward to determine whether a coupling constant is technically natural lacks a satisfactory answer. Specifically, the cosmological constant term in Eq.~\eqref{sakharov_effective_action} is expected to be of the order of the cutoff scale, $\lambda_0 \sim E^4$~\cite{Polchinski1992}. Here, $E$ represents the energy scale at which new degrees of freedom emerge, completing the low energy effective field theory which includes the cosmological constant operator. The highest energy scale that has been probed until now with colliders is of around $E_{\text{LHC}} \sim 13 ~\text{TeV}$~\cite{PDG2024}, providing a lower bound for the cutoff of at least $\sim 10^{13} ~ \text{eV}$.\footnote{Note that the cutoff could be slightly lower if precision measurements at lower energy scales reveal new physics. However, for the argument that is expelled here, such a difference in order of magnitude is not relevant.} Even under the most conservative assumptions about new physics, one would expect new phenomena to appear at the Planck scale $E_{\text{P}} \sim 10^{19} ~\text{GeV}$, where perturbative quantum gravity becomes strongly coupled. This settles an upper bound for the scale of new physics of approximately $10^{28} ~ \text{eV}$. 

This value is drastically different from the cosmological constant inferred from observations. However, cosmological measurements always involve extensive modeling (see Chapter 7 of~\cite{Ellis2012} for a thorough discussion), making it challenging to definitively rule out alternative cosmological scenarios. Within the standard cosmological model, the so-called $\Lambda$CDM model, the universe is assumed to be homogeneous and isotropic on large scales, allowing it to be described by a Friedmann-Lema\^{i}tre-Robertson-Walker (FLRW) metric sourced by various perfect fluid components: dark energy, dark matter, baryonic matter, and radiation. This approximation is generally valid for scales $ L \gtrsim 100 ~ \text{Mpc} $, while at smaller scales, inhomogeneities and anisotropies become noticeable: galaxies typically span a few kpc and cluster into structures of a few Mpc. In this framework, the dark energy component is identified with the cosmological constant, which can be determined from global fits, such as Planck’s analysis of CMB anisotropies~\cite{Planck2018}, or from local measurements of the universe’s expansion through the so-called ``Local Distance Ladder''~\cite{Riess2021}. These observations consistently suggest a vacuum energy density of approximately $\rho_V \sim (2 ~ \text{meV})^4$. This value would correspond to a natural cutoff scale of $E \sim 10^{-3} ~ \text{eV}$, many orders of magnitude below the expected fundamental scales, thus presenting a clear violation of the naturalness principle.

The situation with a WTDiff principle is different. Although the definition of naturalness is introduced by Burgess only for parameters defining the theory, UG makes the cosmological constant entering Einstein equations to be automatically insensitive to the UV completion and hence it can easily accommodate that it takes small values~\cite{Carballo-Rubio2015,Barcelo2018}. To see this, notice that the cosmological constant term in Einstein equations enters merely as an integration constant. The question is whether such integration constant is altered by radiative corrections. It was shown in~\cite{Carballo-Rubio2015} that the fixed background volume form $\boldsymbol{\omega}$ protects the cosmological constant from receiving any quantum corrections, hence being completely stable since it is decoupled from the fluctuations of the quantum fields. Apart from this, all the other operators generated by radiative corrections with a Diff principle, like the Riemann-squared terms, appear in WTDiff with the same coefficients, except for the fact that they appear written in terms of the auxiliary metric $\boldsymbol{\tilde{g}}$ from Eq.~\eqref{auxiliary_metric} instead of the dynamical metric $\boldsymbol{g}$. 

At the semiclassical level, the WTDiff principle appears more favorable than the Diff principle for model building if one prioritizes technical naturalness and adopts the cosmological constant value inferred from $\Lambda$CDM. However, it is important to recognize that this determination is entirely dependent on the FLRW framework, which already exhibits tensions between different parameter estimations~\cite{Abdalla2022}. Furthermore, alternative models exist that can account for some cosmological observations without requiring a cosmological constant. For instance, inhomogeneous models provide an explanation for local universe measurements without invoking a cosmological constant, see for instance Chapter 15 of~\cite{Ellis2012}. While these models remain too simplistic to fully accommodate all cosmological observations~\cite{Bull2011}, they serve as a proof of principle that viable alternatives to the standard paradigm may exist. Given these considerations, our take is that relying on the cosmological constant value inferred from $\Lambda$CDM, along with the somewhat problematic notion of naturalness, as a guiding principle in the search for new physics should be approached with caution.

In any case, if one adopts standard naturalness as a guiding principle alongside the cosmological constant value inferred from $\Lambda$CDM observations, an approach commonly taken in the literature, then the WTDiff principle would be strictly preferable to the Diff principle. In this framework, the cosmological constant can naturally take any possible value, including values significantly smaller than the Planck or electroweak scales. 

As a final comment, it is worth recalling one often-overlooked issue: the potential evolution of the cosmological constant within the $\Lambda$CDM model. Specifically, phase transitions in the early universe could induce changes in the vacuum energy~\cite{Weinberg1988}. For instance, the strong interacting sector undergoes a phase transition at a temperature of \mbox{$ T \sim 1 ~ \text{GeV}$}, involving a jump in the vacuum energy, while the electroweak sector experiences a crossover transition at temperatures of $T \sim  10^3~\text{GeV}$~\cite{Kajantie1996,Kajantie1997}, which results in a smooth variation of the vacuum energy. Assuming energy-momentum tensor conservation, vacuum energy behaves the same way in both GR and UG, meaning the WTDiff principle does not provide additional insight into this aspect.

\subsection{Sakharov's induced gravity}
\label{Subsec:Sakharov}

In a seminal work, Sakharov~\cite{Sakharov1967} proposed that gravitational dynamics could emerge purely from the backreaction of quantum fields propagating on a nondynamical background. In this approach, the gravitational field acts as a passive spectator, with its dynamics induced entirely by the quantum excitations on top of it. A comprehensive review of this idea can be found in~\cite{Visser2002}.

In essence, Sakharov’s proposal suggests that even if the bare coupling constants initially vanish, i.e., $\lambda_i^{\text{bare}} = 0$, they will acquire finite values due to quantum corrections when evaluated at a finite cutoff scale $\mu$~\cite{Sakharov1967}. Following this original idea, several modifications have been proposed. Notably, the approach by Frolov and Fursaev (see, e.g.,~\cite{Frolov1998}) is particularly interesting because, despite being somewhat restrictive in terms of matter content, it allows for explicit and cutoff-independent computations. In such models, the coupling constants become calculable quantities. 

If instead of a full metric structure we consider only a conformal structure, as in a WTDiff theory, a similar mechanism would induce its dynamics through the background structure, making Sakharov’s proposal equally relevant. Ultimately, when it comes to the induced dynamics \emph{\`a la} Sakharov, the only notable difference between Diff and WTDiff principles appears in the treatment of the cosmological constant term again.

\section{Perturbative quantum theory}
\label{Sec:Perturbative}

Up to this point, we have treated the geometry in Diff-invariant theories, and the conformal structure in WTDiff-invariant theories purely as classical fields. In this section, we turn our attention to their perturbative quantization. Specifically, we decompose the dynamical metric $g^{\mu \nu}$ in a background $\bar{g}^{\mu \nu}$ and a perturbation $h^{\mu \nu}$, which represents the graviton field. Our analysis will be conducted in a flat spacetime background, with a focus on computing scattering amplitudes for asymptotic graviton states. Throughout this section, whenever we refer to the metric $g^{\mu \nu}$, it should be understood as an implicit sum of the background metric $\bar{g}^{\mu \nu}$ and the perturbation $h^{\mu \nu}$. 

\subsection{Quantization of linearized Diff and WTDiff theories}
\label{Subsec:Quantization}

Following reference~\cite{deBrito2021}, we will demonstrate that a suitable quantization scheme exists in which both, Diff-invariant and WTDiff-invariant theories, formally reduce to the same type of path integral computation. This would establish their perturbative equivalence at the quantum level. A plausible argument supporting this equivalence was proposed in~\cite{Padilla2014}, though it did not constitute a formal proof. Further evidence of agreement between Diff-invariant and WTDiff-invariant theories in the value of their scattering amplitudes was presented in~\cite{Gonzalez-Martin2017,Gonzalez-Martin2018}. However, other studies have suggested potential differences between the two theories at the quantum level~\cite{Bufalo2015,Upadhyay2015} or in the presence of non-minimal couplings~\cite{Herrero-Valea2020}. Our perspective is that, since a quantization scheme exists in which both theories yield identical results for scattering amplitudes, any perceived differences likely arise from employing different quantization schemes that are not directly comparable.

Let us begin the discussion with an arbitrary Diff-invariant theory. Let us consider a generic diffeomorphism invariant action  $S_{\textrm{Diff}} [\boldsymbol{g}]$  for a metric field $\boldsymbol{g}$. Formally, we are interested in computing the object
\begin{equation}
    Z_{\textrm{Diff}} = \int \frac{\mathcal{D} h_{\mu \nu}}{V_{\textrm{Diff}}} e^{-S_{\textrm{Diff}}[\boldsymbol{g}]},
    \label{perturbative_path_integral}
\end{equation}
where we have introduced the integration measure  $\mathcal{D}h_{\mu \nu}$ which we assume to be Diff-invariant so that no gauge anomaly is present. Furthermore,  we are dividing by the volume of the gauge group $V_{\textrm{Diff}}$. As usual it should be understood as a formal expression requiring regularization given that such volume is infinite. 

Let us apply Faddeev-Popov procedure~\cite{Faddeev1967} to perform a partial gauge fixing of this theory. The gauge symmetries of this theory are transformations generated by a vector field $\xi^{\mu}$ acting on the metric as in Eq.~\eqref{gdiffstransform}. We can decompose every vector field $\xi^\mu$ as a sum of a transverse (divergenceless) part $\xi_\textrm{T}^\mu$ and a longitudinal (irrotational) one that can be written locally as $\nabla^{\mu} \varphi$ 
\begin{align}
    \xi^{\mu} = \xi^{\mu}_{\textrm{T}} + \nabla^{\mu} \varphi, \qquad \nabla_{\mu} \xi^{\mu}_{\textrm{T}} = 0.
    \label{perturbative_gauge_transf_diff}
\end{align}
That every longitudinal diffeomorphism can be written locally as $\nabla^{\mu} \varphi$ is immediate. Global topological obstructions to this might appear~\cite{Nakahara2003}, although they are not relevant at the perturbative quantum level. 

The subgroup of transverse diffeomorphisms is generated by transverse vectors $  \xi^{\mu}_{\textrm{T}}$. We want to make a gauge fixing that breaks the Diff group down to the TDiff group, in order to compare with WTDiff, where we will also perform a gauge fixing of the WTDiff group down to the TDiff group. Following the standard Faddeev-Popov procedure, we will insert a trivial factor of $1$ in the path integral as follows:
\begin{equation}
    1 = \Delta_{F} (\boldsymbol{g}) \int \mathcal{D} \varphi\, \delta \left[ F \left( \boldsymbol{g}^{\varphi} \right) \right],
    \label{faddeev_popov}
\end{equation}
where $F(\boldsymbol{g}) = 0$ is a particular gauge fixing functional and $\boldsymbol{g}^{\varphi}$ represents the finite transformation associated with a purely longitudinal generator $ \nabla^{\mu} \varphi$, whose infinitesimal action is
\begin{align}
    \delta_\varphi g_{\mu\nu} = 2\nabla_\mu\nabla_\nu\varphi.
\end{align}
Introducing this into the path integral~\eqref{perturbative_path_integral} we find
\begin{equation}
    Z_{\textrm{Diff}} = \int \frac{\mathcal{D} h_{\mu \nu}}{V_{\textrm{Diff}}} \left\{ \Delta_{F} (\boldsymbol{g}) \int \mathcal{D} \varphi\, \delta \left[ F \left( \boldsymbol{g}^{\varphi} \right) \right] \right\} e^{-S_{\textrm{Diff}}[\boldsymbol{g}]}.
\end{equation}
We assume that both the action $S_{\textrm{Diff}}[\boldsymbol{g}]$ and the measure are Diff invariant, and hence, we can perform a change of variables $\boldsymbol{g} \rightarrow \boldsymbol{g}^{\varphi}$. To avoid a cumbersome notation, we rename $\boldsymbol{g}^{\varphi}$ and write $\boldsymbol{g}$ instead. Hence the path integral reads 
\begin{equation}
    Z_{\textrm{Diff}} = \int \frac{\mathcal{D} \varphi \mathcal{D} h_{\mu \nu}}{V_{\textrm{Diff}}}  \Delta_{F} (\boldsymbol{g})   \delta \left[ F \left( \boldsymbol{g} \right) \right]e^{-S_{\textrm{Diff}}[\boldsymbol{g}]}.    
    \label{partial_gauge_fix}
\end{equation}
Once we have reached this point, we would like to factor the volume of the Diff group into a piece which is the volume of the TDiff group and another piece. Since the whole set of diffeomorphisms is generated by arbitrary vector fields $\xi^{\mu}$, the $V_{\textrm{Diff}}$ should be written as
\begin{align}
    V_{\textrm{Diff}} = \int \mathcal{D} \xi^{\mu}, 
\end{align}
whereas the volume of the TDiff group should be computed by plugging a $\delta$-functional that ensures that we just integrate over divergenceless vectors
\begin{equation}
    V_{\textrm{TDiff}} = \int \mathcal{D} \xi^{\mu} \delta \left[ \nabla_{\mu} \xi^{\mu} \right]. 
\end{equation}
Both of them can be related~\cite{Percacci2017,Ardon2017} and obey
\begin{equation}
    V_{\textrm{Diff}} =V_{\textrm{TDiff}} \det ( - \nabla^2 )  \int \mathcal{D} \varphi.
\end{equation} 
We immediately see that the integrals over purely longitudinal diffeomorphisms formally cancel in Eq.~\eqref{partial_gauge_fix}. Furthermore, to fix the gauge on purely longitudinal diffeomorphisms we can choose the following gauge fixing
\begin{equation}
    F(\boldsymbol{g}) = \abs{g} - \omega^2,
    \label{gauge_fixing1}
\end{equation}
where $\omega^2$ represents a fixed scalar density. The Faddeev-Popov determinant can be computed as 
\begin{equation}
    \Delta_F (\boldsymbol{g}) = \det    (- \nabla^2 ) .
\end{equation}
Collecting everything and plugging it into Eq.~\eqref{partial_gauge_fix}, we reach the following expression for the path integral
\begin{equation}
    Z_{\textrm{Diff}} = \int \frac{\mathcal{D} h_{\mu \nu}}{V_{\textrm{TDiff}}}  \delta (\abs{g} - \omega^2 ) e^{-S_{\textrm{Diff}}[\boldsymbol{g}]}.     
\end{equation}

Let us now perform an analogue gauge fixing for a theory obeying a WTDiff principle. The starting point now is a WTDiff action $S_{\textrm{WTDiff}} [\boldsymbol{g}]$ in which the dynamical metric is $\boldsymbol{g}$. We study formally the WTDiff action which is parallel to the Diff-invariant theory $S_{\textrm{Diff}} [\boldsymbol{g}]$, namely we consider that $S_{\textrm{WTDiff}} [\boldsymbol{g}] = S_{\textrm{Diff}} [\boldsymbol{\tilde{g}}]$, where $\boldsymbol{\tilde{g}}$ is the auxiliary metric introduced in Eq.~\eqref{auxiliary_metric}. Now the path integral reads 
\begin{equation}
    Z_{\textrm{WTDiff}} = \int \frac{\mathcal{D} h_{\mu \nu}}{V_{\textrm{WTDiff}}} e^{-S_{\textrm{WTDiff}}[\boldsymbol{g}]},
    \label{perturbative_path_integral_WTDiff}
\end{equation}
where we have introduced the integration measure   $\mathcal{D}h_{\mu \nu}$ which in this case we assume to be WTDiff invariant. Instead of $V_{\textrm{Diff}}$, now the factor that appears dividing is the (again infinite) volume of the WTDiff group $V_{\textrm{WTDiff}}$. 

We will now apply also the Faddeev-Popov procedure to fix the gauge symmetry associated with Weyl transformations. The gauge symmetries of the theory are now transformations acting on the metric as 
\begin{align}
    & g_{\mu \nu} \rightarrow g_{\mu \nu} + 2 \nabla_{(\mu} \xi_{\text{T}, \nu)} + \frac{1}{2} \varphi g_{\mu \nu}, \qquad
    \tilde{\nabla}_{\mu} \xi^{\mu}_{\textrm{T}} = 0,
    \label{perturbative_gauge_transf_wtdiff}    
\end{align}
where $\tilde{\nabla}$ is the covariant derivative compatible with the auxiliary metric $\tilde{g}_{\mu \nu}$. 

We now want to perform a partial gauge fixing of the Weyl symmetry in order to prove the formal equivalence between both path integrals. Again, we will do it via Faddeev-Popov procedure by introducing a gauge fixing of the form~\eqref{faddeev_popov}. Let, $\boldsymbol{g}^{\varphi}$ represent the transformed metric under a Weyl rescaling generated by $\varphi$. By the same arguments exposed above, we find 
\begin{equation}
    Z_{\textrm{WTDiff}} = \int \frac{\mathcal{D} \varphi \mathcal{D} h_{\mu \nu}}{V_{\textrm{WTDiff}}}  \Delta_{F} (\boldsymbol{g})   \delta \left[ F \left( \boldsymbol{g} \right) \right]e^{-S_{\textrm{WTDiff}}[\boldsymbol{g}]}.    
    \label{partial_gauge_fix2}
\end{equation}
For WTDiff we have that the structure of the group allows for the factorization
\begin{align}
    V_{\textrm{WTDiff}} = V_{\textrm{TDiff}} \int \mathcal{D} \varphi.
\end{align}
Furthermore, it is possible to choose the same gauge fixing function as we did before
\begin{equation}
    F(\boldsymbol{g}) = \abs{g} - \omega^2,
\end{equation}
with $\omega$ representing the background volume form. Now, the Faddeev-Popov determinant does not contain any differential operator and hence can be absorbed in the measure of integration. Putting everything together, we find the following for the WTDiff path integral
\begin{equation}
    Z_{\textrm{WTDiff}} = \int \frac{\mathcal{D} h_{\mu \nu}}{V_{\textrm{TDiff}}}   \delta ( \abs{g} - \omega^2 )e^{-S_{\textrm{WTDiff}}[\boldsymbol{g}]}.    
\end{equation}
Notice that the auxiliary metric~\eqref{auxiliary_metric} is such that we have the following identity
\begin{equation}
    \delta ( \abs{g} - \omega^2 )e^{-S_{\textrm{WTDiff}}[\boldsymbol{g}]} =  \delta ( \abs{g} - \omega^2 ) e^{-S_{\textrm{Diff}}[\boldsymbol{g}]},  
\end{equation}
since $\delta ( \abs{g} - \omega^2 ) $ ensures that the auxiliary metric $\tilde{g}_{\mu \nu}$ collapses to the dynamical metric $g_{\mu \nu}$, and we have chosen $S_{\textrm{WTDiff}}[\boldsymbol{g}] = S_{\textrm{Diff}}[\boldsymbol{\tilde{g}}]$. Thus, the two theories are formally equivalent at the quantum level under perturbative quantization. In this way, any scattering amplitude that we compute, for which the cosmological constant does not play a role, gives the same result in both theories. More formal analysis, for instance the Becchi-Rouet-Stora-Tyutin (BRST) quantization of the theories~\cite{Kugo2021,Kugo2022b,Kugo2022}, are consistent with this picture. 

\subsection{Recursion relations: constructibility of WTDiff-trees}
\label{Subsec:Constructibility}

The proof in the previous section ensures that the perturbative computation of scattering amplitudes should yield identical results for both theories. An explicit calculation in~\cite{Alvarez2016} confirmed this for the four- and five-point Maximal-Helicity-Violating (MHV) amplitudes, which are more tractable. Despite the propagator and interaction vertices differing significantly from those of Diff-invariant GR due to the chosen gauge fixing, full agreement was found with GR results for these amplitudes. However, this Feynman diagram-based approach becomes increasingly impractical for scattering amplitudes involving more particles.

On-shell methods have proven to be an invaluable tool for understanding quantum field theories~\cite{Elvang2015}. Particularly relevant are recursion relations for scattering amplitudes, which allow amplitudes involving $n$ particles to be expressed in terms of those involving less particles. The first such relations, known as Britto-Cachazzo-Feng-Witten (BCFW) recursion relations~\cite{Britto2004,Britto2005}, were established for Yang-Mills theories and later extended to gravity~\cite{Benincasa2007,ArkaniHamed2008}. It was shown that general relativistic tree-level amplitudes are constructible in the sense that all the tree level amplitudes are fully determined from the three particles scattering amplitude~\cite{Benincasa2007b,Cheung2015}. However, standard proofs of constructibility rely on technical assumptions, such as the propagator's momentum-space structure. As noted in~\cite{Alvarez2016}, UG's off-shell propagator does not allow for the standard BCFW shift, preventing the use of the same techniques employed in GR to prove the constructibility of UG tree-level amplitudes.

To put it explicitly, the idea is that, first of all, the three point scattering amplitudes involving massless particles are completely fixed by the requirements of Lorentz invariance and locality. In particular, we have the following for scattering amplitudes of massless particles of spin $2$:
\begin{align}
    A \left( 1^{-},2^{-},3^{+} \right) = M_P \left(  \frac{\expval{12}^3}{\expval{13}\expval{23}} \right)^{2} , \quad A \left( 1^{+},2^{+},3^{-}\right) = M_P \left(  \left[ 1 2 \right] \left[ 2 3 \right] \left[ 1 3 \right] \right)^{2},
\end{align}
to the lowest order, i.e., the amplitudes arising from $\sim \sqrt{-g}R$ terms in an action. See Appendix~\ref{App:Scattering} for further details. As discussed in the appendix, amplitudes involving equal helicities for the three particles originate from higher-dimensional operators of the form $\sim \sqrt{-g} R^3$, where $R^3$ refers generically to cubic terms in the curvature tensor. Therefore, if we focus only on the Einstein-Hilbert term in the action, these amplitudes are identically zero. 

\begin{figure}
\begin{center}
\includegraphics[width=0.65 \textwidth]{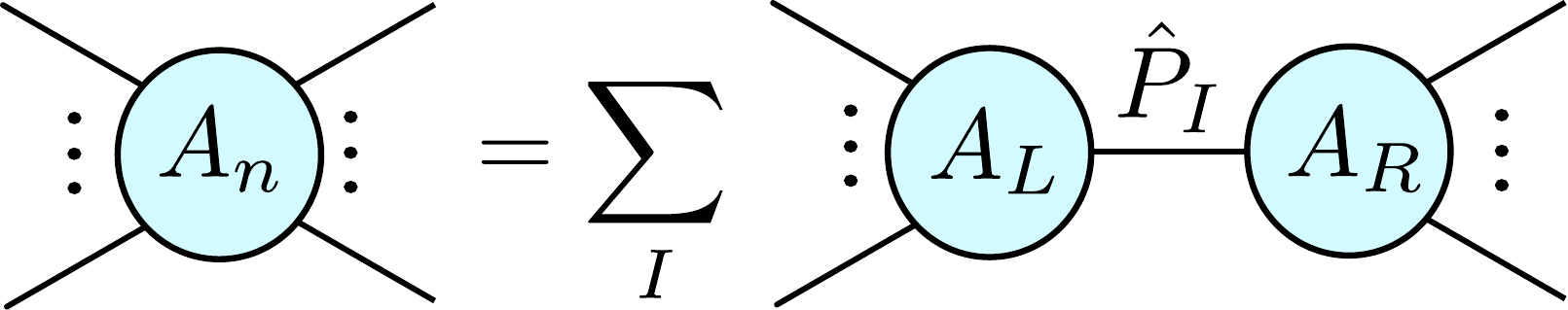}
\caption{Pictorial representation of how the scattering amplitude involving $n$ external states can be decomposed in terms of amplitudes involving less states.}
\label{Fig:Amplitudes}
\end{center}
\end{figure}

Now, a scattering amplitude for $n$ of massless particles is a function of the momenta and helicities of the particles exclusively:
\begin{align}
    A (p_1^{h_1}, \ldots , p_i^{h_i},  \ldots, p_n^{h_n}),
\end{align}
where the momenta obey $p_i^2 = 0$ and momentum conservation holds $\sum_{i=1}^n p_i^{\mu} = 0$. To find the recursion relations, we shift some of the momenta introducing a set of vectors $\{ r_i^{\mu}\}$ (some of them can vanish) as
\begin{align}
    p_i^{\mu} \rightarrow \hat{p}_i^{\mu} = p_i^{\mu} + z r_i^{\mu}, \quad z \in \mathbb{C}, 
\end{align}
with $\sum_i^n r_i^{\mu} = 0$; $r_i \cdot r_j = 0$ and $r_i \cdot p_j  = 0$ for any pair $(i,j)$ of index. This automatically implies that the amplitude evaluated on the shifted momenta remains a valid amplitude. Furthermore, we can write down for any subset $\{ p_i \}_{i \in I}$ containing between $2$ and $n-2$ of the momenta the sum $P_I^{\mu} = \sum_{i\in I} p_i^{\mu}$. Its shifted counterpart is $\hat{P}_I^{\mu}$ and it obeys
\begin{align}
    \hat{P}_I^{2} = P_I^2 + z 2 P_I \cdot R_I,
\end{align}
with $R_I = \sum_{i \in I} r_i$, which can equivalently be rewritten as
\begin{align}
    \hat{P}_I^{2} = - \frac{P_I^2}{z_I} \left(z - z_I \right), \quad \text{with} \quad z_I = - \frac{P_I^2}{2 P_I \cdot R_I}.
\end{align}
Now, at tree-level, the function $\hat{A }_n (z)$ can contain, at most, single poles located away from the origin $z=0$. This is a consequence of locality, i.e., the amplitudes can be derived from some local Lagrangian, according to which the poles can only arise from propagators. Thus, the only poles that we get are of the form $1 / \hat{P}_I^2$ at $z = z_I$. Furthermore, the different poles arising from the combinations of the different momenta, are always located at different values of $z$. If we now consider the function $\hat{A}_n (z) /z$ and consider its integral along contour enclosing only the pole at $z = 0$, we have that the result is precisely $A_n = \hat{A }_n (0)$ due to Cauchy's theorem. We can deform the contour to surround all the other poles together with a contour enclosing all of the poles, for which we can safely consider the $\abs{z} \to \infty$ limit. Applying again Cauchy's theorem we have:
\begin{align}
    A_n = - \sum_{z_I} \text{Res}_{z = z_I} \frac{\hat{A}_n(z)}{z} + B_n, 
\end{align}
with $B_n$ the integral from a contour enclosing all the poles, and it can be found as $\order{z^0}$ term in the $\abs{z} \to \infty$ expansion of $A_n$. The residues at $z_I$ are such that the propagator $\abs{z} \to \infty$ goes on-shell, and locality ensures that the amplitude factorizes in two on-shell terms 
\begin{align}
    \sum_{z_I} \text{Res}_{z = z_I} \frac{\hat{A}_n(z)}{z} = - \hat{A}_L(z_I) \frac{1}{P_I^2} \hat{A}_R (z_I),
\end{align}
where the $\hat{A}_L(z_I)$ and $\hat{A}_R(z_I)$ are called subamplitudes and simply correspond to amplitudes involving less than $n$ particles evaluated at $z = z_I$. Notice that the denominator contains the unshifted momenta $P_I^2$. The sum over all the residues simply translates into a sum over all the possible factorization channels 
\begin{align}
    - \sum_{z_I} \text{Res}_{z = z_I} \frac{\hat{A}_n(z)}{z} = \sum_{\text{channels I}}  \hat{A}_L(z_I) \frac{1}{P_I^2} \hat{A}_R (z_I),
\end{align}
which we recall that depend on the choice of shifts, i.e., the $\{ r_i^{\mu} \}$. The idea now is that if one shows that the $B_n$ terms vanish, or equivalently, that $\hat{A}_n(z) \to 0$ for $\abs{z} \to \infty$, the amplitude itself can be fully expressed as
\begin{align}
    A_n =  \sum_{\text{channels I}}  \hat{A}_L(z_I) \frac{1}{P_I^2} \hat{A}_R (z_I). 
\end{align}
This relates the $n$-point amplitude with amplitudes involving at most $n-1$ particles. Thus, if we are able to find a clever choice of shifts for which we can show that $B_n = 0$ for all the amplitudes, we can determine all the $n$-point amplitudes for $n>3$ from the $n = 3$ amplitude. A pictorial representation can be found in Fig.~\ref{Fig:Amplitudes}.

The problem with UG is that the off-shell propagator contains a piece~\cite{Alvarez2016}
\begin{align}
    \frac{p_{\mu} p_{\nu} p_{\rho} p_{\sigma}}{p^6}, 
\end{align}
which under the conventional BCFW shift diverges and thus cannot be used since the $B_n$ terms are not controlled. This issue was resolved in~\cite{Carballo-Rubio2019}. Instead of using a traditional BCFW shift, the authors followed the approach in~\cite{Elvang2016,Laddha2017}, first proving that UG obeys soft theorems and then introducing a suitable shift that exploits this structure to establish constructibility. Since UG's $n$-particle scattering amplitudes can be recursively expressed in terms of $n-1$-particle amplitudes, and the three-, four-, and five-particle amplitudes match those of GR, it follows that UG and GR have identical tree-level amplitudes.

Beyond tree level, recursion relations for GR in vacuum at arbitrary loop order are unlikely to exist due to its nonrenormalizability beyond two loops~\cite{Goroff1985,vandeVen1991}. However, pure GR is known to be one-loop finite~\cite{tHooft1974}, and while a general framework for expressing one-loop amplitudes in terms of tree-level ones is still lacking, some specific one-loop amplitudes have been successfully written in this form~\cite{Brandhuber2007}.

The results of this section are consistent with those of the previous one, as the perturbative quantization of both UG and GR was shown to be equivalent. In fact, the one-loop effective action for WTDiff was explicitly computed in~\cite{Ardon2017,Percacci2017}, and the result was found to be consistent with its counterpart in a Diff-invariant theory.

\subsection{Asymptotic safety program and WTDiff principle}
\label{Subsec:Asymptotic_Safety}

Asymptotic Safety is a program seeking to formulate a consistent theory of quantum gravity within the framework of perturbative quantum field theory. Originally proposed by Weinberg~\cite{Weinberg1980}, the key idea is that the running coupling constants approach finite, scale-invariant values at high energies. While the space of coupling constants is, in principle, infinite-dimensional, it is expected that the critical surface is finite-dimensional, meaning that the theory can be effectively described by a finite number of parameters. In this section, we only focus on the interplay between the Asymptotic Safety program and the WTDiff principle.

The key tool used nowadays in Asymptotic Safety is the so-called Wetterich equation~\cite{Wetterich1993}, that was introduced for gravity in~\cite{Reuter1996}. It is an equation describing the evolution of the effective average action, which we call $\Gamma_k$, that incorporates quantum fluctuations above the energy scale $k$ as we vary such scale
\begin{equation}
    k \frac{\partial}{\partial k} \Gamma_{k} = \frac{1}{2} \tr \left[ \left( \frac{1}{\frac{\delta^2 \Gamma_k}{\delta \Phi^i \delta \Phi^i} + R_k } \right) k \frac{\partial}{\partial k} R_k \right],
    \label{Wetterich_equation}
\end{equation}
where $R_k$ is a cutoff function chosen in such a way that suppresses the excitations below the scale $k$~\cite{Percacci2017b}, and $\Phi^i$ collectively represents the fields entering the description. Here, the trace represents a sum over all momenta (or spacetime points if we are working in position space), as well as a sum over fields and their components. We notice that the standard practice in Asymptotic Safety is to work in Euclidean signature. However, whether a Wick rotation can be performed to recover a Lorentzian theory remains uncertain, as there is no guarantee that the deformation of the integration contour will avoid singularities\cite{Bonanno2020}. Nevertheless, for our purpose of comparing Diff and WTDiff theories, this issue does not introduce any additional complications to the ones that are present in standard Diff-invariant Asymptotic Safety analyses. 

In practical terms, the first step in the Asymptotic Safety program is to choose a truncation for the basis of operators entering in $\Gamma_k$. This involves selecting a finite set of operators $\mathcal{O}_i$, namely 
\begin{equation}
    \Gamma_k = \sum_{i = 1}^M g_i (k) \mathcal{O}_i.
\end{equation}
Then, given an initial set of data at $k = k_0$, Eq.~\eqref{Wetterich_equation} has a well-posed initial-value problem for evolving the coupling constants $g_i(k)$ from $k_0$ to an arbitrary scale $k$. As $k\rightarrow \infty$, one should hit a fixed point living on the critical surface to have a UV complete fixed point.

From that ultraviolet fixed point at $k = \infty$, that is already present within the truncation studied and to the level of approximation (loop expansion) computed, one can obtain the full quantum effective action
\begin{equation}
    \Gamma_{\textrm{eff}} = \lim_{k \rightarrow 0} \Gamma_k, 
\end{equation}
from which the low energy phenomenology can be extracted. 

The main novelty that a WTDiff-invariance principle offers with respect to a Diff-invariance one in Asymptotic Safety has already been explored in previous sections: the cosmological constant does not receive quantum corrections. Apart from this generic feature of WTDiff theories, several truncations have been studied for UG within the framework of Asymptotic Safety, and no observable differences have been found with respect to their Diff-invariant versions~\cite{Eichhorn2013,Eichhorn2015} even when matter is included~\cite{deBrito2019,deBrito2020}. 

In fact, the difference that was originally observed in the Renormalization Group (RG) flows for UG~\cite{Eichhorn2013} seems to do with the field parametrization that was used. In particular, instead of the standard splitting $g_{\mu \nu} = \bar{g}_{\mu \nu} + h_{\mu \nu}$, that is conventionally used, the clever parametrization
\begin{align}
    g_{\mu \nu} = \bar{g}_{\mu \rho} \left( e^{X} \right)^{\rho}_{\ \nu}, 
\end{align}
is introduced, where $X^{\rho}_{\ \nu} = \bar{g}^{\rho \sigma} h_{\sigma \nu}$ is traceless. This parametrization, which is particularly convenient for UG theories, leads to different RG flow behaviors. However, it is important to note that RG flows are not directly observable unless they correspond to essential couplings: those that govern the running of gauge-invariant quantities~\cite{Herrero-Valea2020}. Given the quantization scheme established in the previous section, which ensures agreement between both theories, and the lack of any reported differences in the running of essential couplings, we conclude that any observed discrepancies stem from the use of different quantization schemes.

\section{Nonperturbative path integral formulation: Euclidean quantum gravity}
\label{Sec:Euclidean}

The Euclidean path integral for gravity mirrors the standard procedure in local quantum field theory, where path integrals are typically defined in Euclidean space, with Lorentzian results obtained through analytic continuation. Given that quantum gravity is expected to accommodate topological fluctuations~\cite{Wheeler1964,Hawking1979}, it is natural to consider a sum over different topologies. Moreover, Euclidean black-hole saddle-point configurations exhibit nontrivial topology~\cite{Hartle1976}, and their inclusion is essential for correctly reproducing the Bekenstein-Hawking formula for black-hole entropy. Consequently, the most straightforward formulation of Euclidean quantum gravity would involve computing the following object:
\begin{equation}
    Z = \sum_{\textrm{topologies}} \int \mathcal{D} g \mathcal{D} \phi_{\textrm{matter}} e^{-S_\text{E}[\boldsymbol{g},\Phi]},
    \label{Euclidean_gravity}
\end{equation}
where one needs to sum over inequivalent geometries and topologies. We need to supplement this with suitable boundary conditions which are crucial. For example,  to  compute  the  partition  function  of a  canonical  ensemble  of  spaces  that  asymptote  a  certain  three-geometry,  we  fix  the  time coordinate  of  the  $(3+1)$-decomposition to be a periodic variable with period $\beta$, the inverse of the temperature~\cite{Gibbons1978}. 

However, we must point out that there is no known way of mathematically making sense of the integral in Eq.~\eqref{Euclidean_gravity}, except for some two-dimensional models~\cite{Saad2019}. Furthermore, in those models, the path integral seems to give rise to an average over an ensemble of theories instead of a single theory path integral. Thus, strictly speaking, it cannot be an \textit{ab initio} definition of quantum gravity. The best we can do to make sense of the previous expression is to do a perturbative expansion around saddle points of the action that obey the prescribed boundary conditions and add the relevant ones. This means that we can write
\begin{equation}
    Z \approx \sum_{\textrm{saddle points}} e^{-S_\text{E} [\boldsymbol{g}_{\textrm{sad} }, \Phi_{\textrm{sad}}] +  S^{(1)} + \cdots},
\end{equation}
where $S^{(1)}$ represents the one-loop contribution and the ellipsis represent higher-order contributions. From an interpretative standpoint, it remains unclear whether summing over all inequivalent Euclidean geometries is truly relevant for quantum gravity, which is inherently formulated in Lorentzian spacetime. Beyond this fundamental issue, there is also the challenge of summing over inequivalent topologies. Since four-manifolds are not classifiable, making precise sense of such a summation is, at best, highly nontrivial~\cite{Hawking1979}. The WTDiff formulation of Euclidean gravity does not provide any new insights into these longstanding issues. However, where it may offer a valuable perspective is in addressing the conformal factor problem, which we will review in the next subsection. That said, we will later demonstrate that an analogous issue arises in the Euclidean quantum gravity version of WTDiff, linked to longitudinal diffeomorphisms. Consequently, from the standpoint of Euclidean quantum gravity, the Diff and WTDiff principles can be considered on equal footing.

\subsection{The conformal factor problem in the Diff-invariant theory}
\label{Subsec:Conformal_Factor}

There is a severe problem in Euclidean quantum gravity related to the fact that the Euclidean Einstein-Hilbert action is not bounded from below. To illustrate this, we can consider a conformal transformation of the metric appearing in the Einstein-Hilbert action $g_{\mu \nu} \rightarrow \Omega^2 g_{\mu \nu}$,\footnote{Note that, based on the terminology used above, this would correspond to a Weyl transformation. However, in this section, we adopt the standard nomenclature and refer to it as a conformal transformation.} in $3+1$ spacetime dimensions to be specific, under which it transforms as follows 
\begin{equation}
    S_{\text{EH}}[\boldsymbol{g}]= - \frac{1}{2\kappa_D^2} \int \dd ^{4}x \sqrt{g}\, \Omega^2 R (\boldsymbol{g})- \frac{3}{ 2 \kappa^2} \int \dd ^{4} x \sqrt{g}\, g^{\mu \nu} \nabla_{\mu} \Omega \nabla_{\nu} \Omega. 
\end{equation}
By choosing a conformal factor that oscillates rapidly with a given frequency, the action can be made arbitrarily large and negative. As a result, saddle points of the action acquire negative modes. The presence of these negative modes leads to purely imaginary contributions to the energy, indicating an instability of the classical vacuum due to tunneling effects. This suggests that an infinite number of negative modes may exist around every classical saddle point. 

A major challenge in addressing this issue lies in the gauge symmetry, which must be properly fixed through suitable constraints to account for the physical degrees of freedom of the theory. Ideally, these gauge constraints would eliminate the negative modes associated with the conformal sector. However, since the constraints are imposed in Lorentzian signature, it is unclear how they translate to Euclidean space. The interpretation of these potential negative modes around classical saddle points in Euclidean signature remains an open question, complicating the understanding of this formulation of quantum gravity. 

The most widely used approach to address this problem is the Gibbons-Hawking-Perry (GHP) prescription. This method involves rotating the contour of integration so that the conformal factor of the metric takes on imaginary values~\cite{Gibbons1978}. At first glance, this prescription may appear to be an \emph{ad hoc} adjustment, introduced solely to enforce the convergence of the path integral without any physical justification. However, in~\cite{Gross1982}, this approach was applied to compute the free energy of a gas of free gravitons using path-integral methods. The expected result reproduced the Stefan-Boltzmann law for particles with two physical polarizations (the two graviton polarizations), as expected on physical grounds. Furthermore, when the same prescription was applied to a black hole background, it was found to describe a bounce process in which hot flat space tunnels to nucleate a black hole, highlighting the instability of flat space under gravitational perturbations. These physically sensible results provide strong evidence in favor of the GHP prescription.

Nevertheless, a key unresolved question is whether this prescription extends consistently to arbitrary backgrounds. A more refined analysis of the analytic continuation of the trace and traceless modes was carried out in~\cite{Mazur1989} at the one-loop level, showing that this treatment is equivalent to the GHP prescription, at least for GR at one loop. Further developments were presented in~\cite{Mottola1995}, where a geometrically motivated approach to defining the path integral at one loop was proposed. This method avoids the introduction of the standard Faddeev-Popov ghosts typically required for gauge fixing and provides an independent justification for the GHP prescription.

\subsection{The longitudinal diffeomorphism problem of WTDiff}
\label{Subsec:Longitudinal_Diff}

For WTDiff theories, the conformal factor problem is not a problem itself since these conformal transformations are gauge transformations. The potential issue is that longitudinal diffeomorphisms, which are not gauge transformations, also generate negative modes. Let us consider the Euclidean version of WTDiff in $3+1$ dimensions to be specific
\begin{equation}
S_\text{E}[\boldsymbol{g}] = - \frac{1}{2 \kappa_D^2}\int \dd^{4} x\, \omega (x) R (\boldsymbol{\tilde{g}}),
\end{equation}
where we recall that the Ricci scalar constructed out of the auxiliary metric reads 
\begin{equation}
\tilde{g}_{\mu \nu} = g_{\mu \nu} \left( \frac{\omega^2}{\abs{g}} \right)^{\frac{1}{4}}.
\end{equation}
Let us perform an infinitesimal transformation characterized by a vector field $\xi^{\mu}$ whose action in the auxiliary metric can be written as 
\begin{equation}
    \delta_{\xi} \tilde{g}_{\mu \nu} = 2 \tilde{\nabla}_{(\mu} \tilde{\xi}_{\nu)} + \frac{1}{2} \left[ - \frac{1}{2} \tilde{g}_{\mu \nu} \xi^{\sigma} \partial_{\sigma} \log \left( \frac{\omega^2}{g} \right) - \tilde{g}_{\mu \nu} \nabla_{\alpha} \xi^{\alpha}\right].
\end{equation}
The choice $ \nabla_{\alpha} \xi^{\alpha} = - \frac{1}{2} \xi^{\sigma} \partial_{\sigma} \log \left( \omega^2 / g \right)$ amounts to transverse diffeomorphisms as introduced above in Eq.~\eqref{transverse_diff_divergence}. Had we chosen the divergence of $\nabla_{\alpha} \xi^{\alpha} = - \frac{1}{2} \xi^{\sigma} \partial_{\sigma} \log \left( \omega^2 / g \right) - f$, for some $f$, the variation of the action would be of the form
\begin{equation}
    \delta_{\xi} S \propto \int \dd ^4 x\, \omega f(x) R (\boldsymbol{\tilde{g}}).
\end{equation}
Thus, we arrive at the same conclusion as in GR: the mode associated with diffeomorphisms generated by a vector $\xi^{\mu}$ containing a longitudinal part can become arbitrarily negative, rendering the action unbounded from below. This demonstrates that the conformal factor problem in GR directly translates into the longitudinal diffeomorphism problem in WTDiff, as previously noted in Appendix B of~\cite{Blas2008}. At least in its naive formulation, Euclidean WTDiff does not provide any new insights into the conformal factor problem, which simply manifests as the longitudinal diffeomorphism mode problem. Even if an appropriate prescription were found to handle this mode, all the other fundamental challenges of Euclidean gravity would persist. These include the problem of summing over topologies and the difficulty of assigning physical meaning to the results, specifically, the issue of analytically continuing to Lorentzian signature to extract physically relevant conclusions.

\section{Embedding WTDiff in string theory}
\label{Sec:Strings}

It is generally assumed that the low-energy EFT describing the massless states of strings is governed by a Diff-invariant principle. While it is true that one can construct an effective action that respects Diff invariance, this does not imply that such an action is unique. In this section, we argue that adopting the standard Diff-invariant general relativistic action as the effective theory is a choice rather than a necessity. One could just as well formulate an alternative effective action that is WTDiff-invariant. Two main arguments are typically put forward to support the claim that the low-energy dynamics of string theory is best described by a Diff-invariant theory. 

The first of these arguments is that it is possible to compute the string scattering amplitudes involving asymptotic graviton states with a low-energy effective description obeying a Diff-invariance principle~\cite{Green1987a,Green1987b}. On the one hand, one computes these scattering amplitudes involving gravitons via correlation functions on the worldsheet in string perturbation theory. Then, one shows that such scattering amplitudes can be computed at low energies equivalently within GR.

The second of these arguments, is that in order to avoid gauge anomalies in the worldsheet when the string propagates on an arbitrary background, one needs to ensure that the so-called $\beta$-functionals vanish~\cite{Polchinski1998a,Polchinski1998b}. They are computed order by order in the string tension $\alpha'$ assuming that the background fields vary much slower than the string scale. Those $\beta$-functionals can be equivalently derived as the equations of motion associated with GR to the lowest-order in the $\alpha'$ expansion, and higher-order Diff-invariant theories at higher orders in the $\alpha'$ expansion, with a suitable matter content associated with the other massless excitations of the theory. We critically examine these arguments and explore whether a WTDiff-invariant formulation can be an equally viable alternative. We will focus on bosonic string theory for simplicity. In the case of superstrings, incorporating worldsheet fermions does not introduce significant changes to the bosonic sector of the theory, which includes graviton excitations and thus the gravitational field, except for the change in the critical dimension from $D=26$ to $D=10$. 

 \paragraph*{\textbf{Notation in this section.}}

In this section, latin indices from the beginning of the alphabet $(a,b, \ldots )$ are used to represent tensor indices in the worldsheet. The target space metric is represented as $\mathcal{G}_{\mu \nu}$, with the Riemann and Ricci tensor and scalar represented as $\mathcal{R}_{\mu \nu \rho}^{\ \ \ \lambda}$, $\mathcal{R}_{\mu \nu}$ and $\mathcal{R}$, respectively. This is done to avoid a cumbersome notation.

\subsection{Graviton scattering from string theory}
\label{Subsec:Graviton_Scattering}

This section contains a review of the quantization of strings in a flat background as well as the computation of string scattering amplitudes for gravitons from string theory. This is well-known material that can be found in any textbook~\cite{Polchinski1998a,Green1987a} and review here for the sake of completeness.

The starting point of our discussion of perturbative string theory is the action describing relativistic strings propagating in flat spacetime. For relativistic free particles it is natural to consider the action to be the proper time of the particle trajectory, i.e., the embedding of the world-line in the target space. In the same way, for strings it is natural to consider the area swept out by the worldsheet. For that purpose, let us introduce a coordinate system in the worldsheet, a pair $\sigma^a$ ($a = 0,1$) which corresponds to the time coordinate $\sigma^0 \in (-\infty,\infty)$ and a spatial coordinate $\sigma^1$. Furthermore, we will restrict our attention to closed strings (those giving rise to graviton excitations) for which we periodically identify the spacelike coordinate $\sigma^1 \sim \sigma^1 + 2 \pi$. 

We endow the $(D+1)$-   dimensional flat spacetime with coordinates $X^{\mu}$, and we look for an action corresponding to the area density swept by the string expressed in terms of derivatives of the embedding $X^{\mu} (\tau, \sigma)$. The induced metric on the worldsheet is given by 
\begin{align}
    h_{a b} = \eta_{\mu \nu} \partial_{a} X^{\mu} \partial_{b} X^{\nu},
\end{align}
and the area can be straightforwardly expressed in terms of it as
\begin{align}
    S_{NG} [X] = - \frac{1}{2 \pi \alpha'} \int \dd ^2 \sigma \sqrt{-h},
\end{align}
which is the so-called Nambu-Goto action for relativistic strings. The constant $\alpha'$ represents the string tension, i.e., the energy density per unit length. Although this action is perfectly reasonable classically, it poses challenges for quantization. The issue arises because it is not quadratic in its variables, as the square root appears explicitly in the action. To circumvent this problem, one can use the Polyakov action, which is given by 
\begin{align}
    S_P[X,\boldsymbol{\gamma}] = -\frac{1}{4 \pi \alpha'} \int \dd^2 \sigma \sqrt{-\gamma} \gamma^{a b} \partial_{a } X^{\mu} \partial_{b} X^{\nu} \eta_{\mu \nu}. 
\end{align}
In this action, an additional configuration variable, $\boldsymbol{\gamma}$, is introduced as a metric on the worldsheet. This formulation makes the action explicitly quadratic in the $X^{\mu}$ variables facilitating the path-integral quantization of the theory. In fact, following standard conventions, we define a two-dimensional energy-momentum tensor as the variation of the Polyakov action with respect to the worldsheet metric $\gamma_{ab}$ as: 
\begin{align}
    T_{a b} = -  \frac{1}{\sqrt{-\gamma}} \frac{\delta S_P}{\delta \gamma^{ab}} = \frac{1}{4 \pi \alpha'} \left[ \partial_{a} X^{\mu} \partial_b X^{\nu} - \frac{1}{2} \gamma_{a b} \gamma^{a d}  \partial_{c} X^{\mu }\partial_{d} X^{\nu} \right]\eta_{\mu \nu}.
\end{align}
The Polyakov action does not contain any derivatives of the worldsheet metric $\gamma_{a b}$, meaning that the equations of motion for $\gamma_{ab}$ act as a constraint, enforcing $T_{a b} = 0$. Consequently, $\gamma_{ab}$ is not a dynamical variable in the usual sense. This constraint allows us to express $\gamma_{ab}$ in terms of the embedding coordinates $X^{\mu}$. Substituting this solution back into the Polyakov action recovers the original Nambu-Goto action, demonstrating their equivalence. It is worth pausing at this point and discussing the continuous symmetries of the theory:
\begin{itemize}
    \item Poincar\'e invariance. This is a global symmetry on the worldsheet 
    \begin{align}
        & X^{\mu} \rightarrow \Lambda^{\mu}_{\ \nu} X^{\nu} + c^{\mu}.
    \end{align}
    
    \item Diffeomorphism invariance in the worldsheet \mbox{$\sigma^a \rightarrow \Tilde{\sigma}^a (\sigma) $}. Whereas the $X^{\mu}$ fields transform as worldsheet scalars, $\gamma_{ab}$ transforms as a two-index covariant tensor: 
    \begin{align}
        & X^{\mu} (\sigma) \rightarrow X^{\mu} (\tilde{\sigma}) = X^{\mu} (\sigma) , \\
        & \gamma_{a b} (\sigma) \rightarrow \tilde{\gamma}_{ab} (\tilde{\sigma}) = \frac{\partial \sigma^c}{ \partial \tilde{\sigma}^a} \frac{\partial \sigma^d}{ \partial \tilde{\sigma}^b} \gamma_{c d} (\sigma).
    \end{align}
    \item Weyl invariance of the worldsheet metric $\gamma_{ab}$. This transformation leaves invariant the $X^{\mu}$ coordinates and the metric gets a local rescaling 
    \begin{align}
        & X^{\mu} (\sigma) \rightarrow X^{\mu} (\sigma) , \\
        & \gamma_{a b} (\sigma) \rightarrow e^{2 \phi (\sigma) } \gamma_{a b} (\sigma).
    \end{align}
\end{itemize}
We can distinguish now between oriented, those that have a well defined transformation law under parity transformations (i.e., $\sigma^1 \rightarrow 2 \pi - \sigma^1$), and unoriented strings, which do not. 

Not all the symmetries that we have introduced are straightforwardly preserved through the process of quantization. The Weyl symmetry is anomalous and, since it is a gauge symmetry, we must insist on preserving it at the quantum level to remove unphysical states. As discussed in Sec~\ref{Sec:Semiclassical} above, anomalies arise as a clash of preserving two different symmetries at the quantum level. In Diff- and Weyl-invariant theories, the clash occurs between longitudinal diffeomorphisms and Weyl transformations. Therefore, it is expected that, at least, some consistency conditions may arise in order to preserve both sets of transformations at the quantum level.

Let us perform the quantization of the theory through a path-integral procedure and analyze the spectrum of the theory. For that, we define the generating functional of correlation functions following the usual Faddeev-Popov procedure. Rotating to Euclidean space to make the path integral sensible, we can write down the generating functional as
\begin{equation}
    Z = \frac{1}{V(\text{gauge})} \int \mathcal{D} \boldsymbol{\gamma} \mathcal{D} X e^{-S_P [X,\boldsymbol{\gamma}]},
\end{equation}
where $V(\text{gauge})$ represents the volume of the gauge group and it is introduced to avoid counting more than once configurations related through a gauge transformation. As usual, we introduce a Faddeev-Popov determinant $\Delta_{FP} [\boldsymbol{\gamma}]$ to take this volume into account. The integral over the gauge orbits cancels with the volume of the gauge group, and we reach the following expression for the generating functional
\begin{align}
    Z[\boldsymbol{\gamma}]  =  \int  \mathcal{D} X \Delta_{FP} [\boldsymbol{\gamma} ]  e^{-S_P [X,\boldsymbol{\gamma}]}. 
\end{align}
Choosing a convenient normalization for the action, we can rewrite the Faddeev-Popov determinant as 
\begin{align}
    \Delta_{FP} [\boldsymbol{\gamma}] = \int \mathcal{D} b \mathcal{D} c e^{- S_{\text{g}} [b,c]},
\end{align}
where $b$ and $c$ are ghost Grassmann variables that anticommute and 
\begin{align}
    S_{\text{g}} = \frac{1}{2 \pi} \int \dd^2 \sigma \sqrt{\gamma} b_{a b} \nabla^a c^b.
\end{align}
At this point, we have reduced the evaluation of the path integral for the bosonic string theory to the evaluation of the path integral: 
\begin{align}
    Z = \int \mathcal{D}b \mathcal{D}c \mathcal{D}X e^{- S_{P}[\boldsymbol{\gamma},X] - S_{\text{g}} [\boldsymbol{\gamma}, b, c]}, 
\end{align}
which is the Conformal Field Theory (CFT) of $D+1$ scalar fields (the $X^{\mu}$) and the $bc$-ghost system~\cite{Polchinski1998a,Nakayama2013}. Weyl invariance requires that the trace of the energy-momentum tensor vanishes. In the specific case of two-dimensional theory under consideration, this trace is entirely determined by the central charge and the intrinsic curvature of the worldsheet
\begin{align}
    \expval{T^a_{\ a}} = - \frac{c}{12} R \left( \boldsymbol{\gamma} \right).
\end{align}
To preserve Weyl invariance, we need that the central charge vanishes. This is precisely the consistency condition that we anticipated. The system of the $X^{\mu}$-scalars and the $bc$-ghost system is linear, and hence the total central charge is the sum of the central charges of the two systems independently:
\begin{align}
    c = c_{g} + c_{X}.
\end{align}
The $bc$-ghost system~\cite{Polchinski1998a} has a central charge $c_{g} = - 26$ while each scalar field gives a contribution of $1$ to the central charge $c_{X} = D + 1 $. Ensuring Weyl-invariance means that we need the spacetime dimension to be $26$. This is the well-known way in which the critical dimension of bosonic string theory emerges. 

We now move on to analyze the spectrum of the theory. To show the presence of a tachyon, a dilaton, and a graviton, we can skip a detailed BRST analysis and focus only on the states generated by $X$-fields which are the ``physical fields''. We use the so-called state-operator map for CFTs~\cite{Qualls2015,DiFrancesco1997}, in which states are replaced by operator insertions that generate them by acting in a neighborhood of the vacuum. For this purpose, it is first easier to use complex coordinates $\sigma \rightarrow (z,\bar{z})$ on the worldsheet. Furthermore, we now need the operators to be gauge invariant. Diffeomorphism invariance can be ensured by integrating local operators $\mathcal{O}(z, \bar{z})$ over the worldsheet as
\begin{align}
    V = \int \dd^2z \mathcal{O} (z, \bar{z}), 
\end{align}
with $V$ standing for vertex operators. Weyl invariance is ensured by choosing the operators $\mathcal{O}$ to transform adequately under Weyl rescalings, i.e., having a suitable weight. The measure of integration, $\dd^2z$ has a conformal weight $(-1,-1)$ under such rescalings. Hence, $\mathcal{O}$ needs to be a primary operator of the CFT with weight $(+1,+1)$ to compensate it. 

The kind of operators that give rise to the lowest energy states of the string are $e^{i p \cdot X}$ and $P_{\mu \nu} \partial X^{\mu} \partial X^{\nu} e^{i p \cdot X}$, with $p$ a given momentum that we endow the string with and $P_{\mu \nu}$ the polarization tensor~\cite{Green1987a,Polchinski1998a}. The operator $e^{i p \cdot X}$ gives rise to the tachyon, since we need to impose that $ - p^2 = - 4/\alpha' < 0$, for the operator to be Weyl invariant. The operator $P_{\mu \nu} \partial X^{\mu} \partial X^{\nu} e^{i p \cdot X}$ corresponds to the dilaton (pure trace part of $P_{\mu \nu}$) and the symmetric part of $P_{\mu \nu}$ gives rise to the graviton, since $p^2 = 0$ (massless condition) and $p^{\mu} P_{\mu \nu} = 0$ (transverse condition) needs to be imposed to ensure the Weyl invariance. The antisymmetric part does not give rise to any states for unoriented strings. 

Although we have identified the massless states corresponding to the dilaton and the graviton, the Polyakov action per se does not give rise to interactions among them. In fact, we can add the Einstein-Hilbert term to the Polyakov action, which respects all the symmetries of the theory and is purely topological in two dimensions:
\begin{align}
    S_{\text{int}} = \frac{\lambda}{4 \pi} \int \dd^2 \sigma \sqrt{\gamma} R(\boldsymbol{\gamma}) = 2 \lambda (1-g), 
\label{Eq:Int_Action}
\end{align}
where $g$ represents the genus of the worldsheet and $\lambda$ is a coupling constant which we assume to be small in order to do perturbation theory. Hence, if we add this term to the string action, we get
\begin{align}
    Z = \sum_{\text{topologies}} \int \mathcal{D} X \mathcal{D} \boldsymbol{\gamma} e^{-S_P -  S_{\text{int}}} = \sum_{g=0}^{\infty} e^{-2\lambda (1-g)} \int \mathcal{D} X \mathcal{D} \boldsymbol{\gamma} e^{-S_P}.
\end{align}
If we introduce the string coupling constant as $g_s := e^{\lambda}$, this gives a good expansion as long as $g_s \ll 1$. The whole series is known to be a divergent series~\cite{Gross1988}, as the perturbative series in QFT typically are~\cite{Dyson1952}. In addition to this problem, there is a harder problem which is the finiteness of \emph{each} of the terms in the series, i.e., the path integral over the different geometries. For a fixed topology, the path integral with the Polyakov action requires to compute a sum over the moduli of conformally inequivalent surfaces. In general, for higher loop orders, i.e., nontrivial topologies, this requires to perform an integral over a moduli space that is not obviously convergent. A fully satisfying proof is still lacking, although some results in the literature suggest its finiteness~\cite{Mandelstam1991}  

Now we arrive at the task of computing some observables. In string theory, the primary observable is the string $S$-matrix. This involves taking an ``in" state from the free string spectrum and computing the probability amplitude for generating another ``out" state of free string spectrum when the interactions~\eqref{Eq:Int_Action} are taken into account. Such states are generated by introducing their corresponding vertex operators.

For our purposes of analyzing how GR or UG might emerge from string theory, we are interested in computing the scattering amplitude involving $m$ gravitons with momenta $\boldsymbol{p}_i$ and polarization tensors $\boldsymbol{e}_i$, which we represent as $\mathcal{A}^{(m)} (p_1,e_1; p_2,e_2; \ldots; p_m, e_m)$. This is computed as a suitable path integral for the Polyakov action $S_P$ that schematically reads~\cite{Green1987a,Polchinski1998a}
\begin{equation}
    \mathcal{A}^{(m)} (1^{h_1},2^{h_2},...,m^{h_m}) = \frac{1}{g_s^2} \frac{1}{V_\textrm{gauge}} \int \mathcal{D} X \mathcal{D} \boldsymbol{g} \,e^{- S_\text{P} [X,\boldsymbol{g}]} \prod_{i=1}^{m} V_{i} (p_i, h_i),
\end{equation}
where $V_i$ represents the vertex operator associated with a graviton insertion with a given spin and momentum. To begin with, we particularize the amplitude for three gravitons and we find 
\begin{equation}
    \mathcal{A} (p_1,e_1; p_2, e_2; p_3, e_3) = i \frac{g_s (\alpha^{\prime})^6}{2}(2 \pi)^{26} \delta^{26} \left( p_1 + p_2 + p_3 \right) e_{1 \mu \nu} e_{2 \alpha \beta} e_{3 \gamma \delta} T^{\mu \alpha \gamma} T^{\nu \beta \delta},
\end{equation} 
where
\begin{equation}
    T^{\mu \alpha \gamma} = p_{23}^{\mu} \eta^{\alpha \gamma} + p_{31}^{\alpha} \eta^{ \gamma \mu } + p_{12}^{\gamma} \eta^{\mu \alpha} + \frac{\alpha'}{8} p^{\mu}_{23} p^{\alpha}_{31} p^{\gamma}_{12},
\end{equation}
and $p_{ij}^{\mu} := p^{\mu}_i - p^{\mu}_{j}$. The terms of order $ \order{\alpha'}$ in $T^{\mu \alpha \gamma}$ contribute as $\order{p^4}$ to the amplitude. If we focus only on the lowest-order terms $\order{p^2}$, this amplitude is equivalent to those computed at tree level from the Einstein-Hilbert action, upon the identification $\kappa_{25} = g_s ( \alpha')^6$. The same agreement is found for amplitudes involving an arbitrary number of gravitons: if we neglect the higher-order contributions from the string amplitude, they agree with those computed from the Einstein-Hilbert action~\cite{Polchinski1998a,Green1987a}, with the same identification of $\kappa_{25}$ and the string constants $g_s, \alpha'$.  

As it has been already discussed~\cite{Alvarez2016,Carballo-Rubio2019}, the tree-level scattering amplitudes of gravitons computed in GR and UG are identical. Hence, from the point of view of scattering amplitudes, string theory does not point toward GR in a univocal way: both UG and GR are equivalent from a low-energy effective field theory point of view. 

\subsection{Strings in general backgrounds}
\label{Sec:Background_String_Theory}

Up to now, we have only considered strings propagating in flat spacetime. However, the string spectrum includes excitations that interact among themselves, potentially generating a nontrivial background. Just as a laser can be described as a coherent state of photons, a coherent state of gravitons is expected to resemble a curved background, requiring an appropriate description for a string propagating in such a setting. The same reasoning applies to the dilaton field. Moreover, at low energies, only the massless excitations contribute significantly. Consequently, we can construct the most general renormalizable action incorporating these fields, which takes the form of the following nonlinear $\sigma$-model  
\begin{align}
    & S[X, \boldsymbol{\gamma}] = S_P[X,\boldsymbol{\gamma}] + S_D[X,\boldsymbol{\gamma}] \nonumber \\
    & = -\frac{1}{4 \pi \alpha'} \int \dd^2 \sigma \sqrt{-\gamma} \left[ \gamma^{a b} \mathcal{G}_{\mu \nu} (X) \partial_{a } X^{\mu} \partial_{b} X^{\nu} + \alpha' R \left( \boldsymbol{\gamma} \right) \Phi (X) \right],
\label{Eq:Action_String}
\end{align}
where $\mathcal{G}_{\mu \nu} (X)$ represents a metric (graviton excitations), $\Phi(X)$ represents the dilaton background field, and $R[\boldsymbol{\gamma}]$ represents the Ricci-scalar of the two-dimensional metric. The last term explicitly breaks Weyl invariance in the worldsheet and it is of higher dimension than the Weyl-invariant terms, that is why it does not contain any dimensionful constant in front of it. 

There are two missing terms that still give rise to a renormalizable theory. The first of these terms is the coupling to the Kalb-Ramond field. However, if we focus on unoriented strings, we can skip it since the divergences of the rest of the terms do not require this term to be renormalized. In case we deal with oriented strings, this term gives a contribution to the conformal anomaly~\cite{Polchinski1998a}.

The additional term that we can add to the action corresponds to a coupling to the background tachyon field $T(X)$
\begin{align}
    S_T = \frac{1}{4\pi} \int \dd^2 \sigma \sqrt{- \gamma} T \left( X \right).
\end{align}
In principle this term is needed to cancel some of the quadratic divergences arising from vacuum to vacuum diagrams. However, if we use a renormalization scheme such that those divergences are absent (for example, dimensional regularization), we can safely skip those terms. Furthermore, it is worth mentioning that supersymmetry in the worldsheet ensures that those quadratic divergences are absent in superstrings due to the characteristic cancellation among fermionic and bosonic degrees of freedom, with independence of the renormalization scheme. 

\subsubsection{Determination of the Weyl anomaly}
\label{Subsec:Betas}

For string theories, Weyl invariance is potentially anomalous, as discussed above, due to the trade-off between Weyl and full Diff-invariance symmetry. By using a regularization scheme that maintains diffeomorphism invariance, Weyl symmetry may become anomalous. To avoid this, the nonlinear sigma model, i.e., the background fields, must be carefully chosen. In particle physics terms, this is analogous to canceling potential gauge anomalies. 

A parallel can be drawn with the Standard Model, where the chiral nature of gauge interactions means that an arbitrary matter content could introduce anomalies. However, the actual matter content is such that these anomalies cancel out. Similarly, in the previous section, we ensured Weyl invariance by fixing the target space dimension to 26; otherwise, the symmetry would have been anomalous. Here, we expect additional constraints on the background fields in the nonlinear sigma model. Specifically, constraints that the metric $G_{\mu \nu} (X)$ and the dilaton $\Phi(X)$ fields must obey. 

We want now to write down the most general form that the Weyl anomaly can display. Following D'Hoker~\cite{Deligne1999}, it is possible to show that the structure of the anomaly for unoriented strings in a curved background needs to be of the form
\begin{align}
    \expval{T_{a}^{\ a}} = \beta_{\mu \nu}^{\mathcal{G}}(X) \partial_{a} X^{\mu} \partial_{b} X^{\nu} \gamma^{ab}+ \beta^{\Phi} \left( X \right) R \left( \boldsymbol{\gamma} \right)  + \beta^V_{\mu}(X) g^{a b} D^{\ast}_{a} \partial_{b} X^{\mu},
\label{Eq:Trace_Anomaly}
\end{align}
where $D^{\ast}_a$ represents the covariant derivative on the product space of the cotangent space of the worldsheet and the tangent space of the target space, and it can be explicitly written down as  
\begin{align}
    D^{\ast}_a \partial_b X^{\mu} = \partial_a \partial_b  X^{\mu} - \Gamma_{ab}{}^{c} \partial_c X^{\mu}+ \Gamma_{\nu \rho}{}^{\mu} \partial_{a} X^{\nu} \partial_b X^{\rho}, 
\end{align}
where $\Gamma_{ab}{}^{c}$ are the Christoffel symbols of the metric $\gamma_{a b}$ and $\Gamma_{\nu \rho}{}^{\mu}$ represent the Christoffel symbols of the metric $\mathcal{G}_{\mu \nu}$. The last term in the Weyl anomaly, $\beta^V$ can be removed through a transformation on the $X^{\mu}$ fields. This leaves only two independent $\beta$ functionals: $\beta^{\mathcal{G}}$ and $\beta^{\Phi}$\footnote{For oriented strings there would be another $\beta$-functional associated with the Kalb-Ramond field.}.

Hence we need to determine the $\beta$ functionals obtained from the action~\eqref{Eq:Action_String}. They are computed through an $\alpha'$ expansion, which assumes that the background fields $\mathcal{G}_{\mu \nu} (X)$ and $\Phi(X)$ vary smoothly with respect to the scale $\alpha'$. It is standard practice to perform computations using the background field formalism, where the fields $X^{\mu}$ are decomposed into a background component $X_0^{\mu}$ and quantum fluctuations $Y^{\mu}$ 
\begin{equation}
    X^{\mu} \left( \sigma \right) = X_0^{\mu} (\sigma) + Y^{\mu} \left( \sigma \right),
\end{equation}
where the integration is now performed with respect to $Y^{\mu}$ instead of $X^{\mu}$. We define the effective action $\Gamma [ X_0,g ] $ following~\cite{Callan1989} as 
\begin{align}
    e^{- \Gamma [ X_0,g ] } = \int \mathcal{D} Y e^{- \left[ S (X_0,Y)- S(X_0) - \int \dd^2 \sigma Y^{\mu} (\sigma) \frac{\delta S}{\delta X_0^{\mu}} \right] },
\end{align}
which is the generating functional of the Feynman diagrams relevant for the computation of the $\beta$-functionals. 

At this point, it is better to pause and mention a crucial step in the computations. The coordinate difference does not transform in a covariant way under changes of coordinates. Hence, in order to obtain results that are manifestly covariant, it is better to do the computation in manifestly covariant variables too. This can be done as follows. Imagine that the coordinates $X^{\mu}_0$ correspond to a given point $p_0$ and the coordinates \mbox{$X^{\mu} = X^{\mu}_0 + Y^{\mu}$} to a point $p$. If both points are close enough, there exists only one geodesic with respect to $\mathcal{G}_{\mu \nu}$ connecting both of them. Hence, we can replace the coordinate difference $Y^{\mu}$ which characterizes the point $p$, by the tangent vector $t^{\mu}$ of that geodesic at the point $p_0$, which transforms covariantly under changes of coordinates.

In fact, we can use the tangent vector $t^{\mu}$ to perform a covariant Taylor expansion around $X_0^{\mu}$ for any tensor defined on the target manifold. Explicitly, any tensor $T_{\mu_1 \hdots \mu_n} (X)$ can be expanded as
\begin{align}
    T_{\mu_1 \hdots \mu_n} (X_0 + t) = \sum_{k=0}^{\infty} T^{(k)}_{\mu_1 ... \mu_n \nu_1 ... \nu_k} (X_0) t^{\nu_1 } \hdots t^{\nu_k},
\end{align}
where each of the terms $T^{(k)}_{\mu_1 ... \mu_n \nu_1 ... \nu_k} $ is a combination of covariant derivatives of the tensor $T_{\mu_1 \hdots \mu_n} $ and contractions with curvature tensors evaluated at $X_0$. This expansion can be carried out using the normal coordinate expansion. However, we emphasize that it remains valid in any coordinate system since it is expressed in a manifestly covariant form. We are interested in the expansion of $\mathcal{G}_{\mu \nu}(X)$, $\Phi(X)$ and $\partial_a \left( X_0^{\mu} + Y^{\mu} \right)$. These expansions can be computed (see~\cite{Callan1989} for details) and we find
\begin{align}
    \mathcal{G}_{\mu \nu} (X) & = \mathcal{G}_{\mu \nu} (X_0) + \frac{1}{3} \mathcal{R}_{\mu \rho \sigma \nu} (X_0) t^{\rho} t^{\sigma } + ..., \\
    \Phi(X) & = \Phi(X_0) + \nabla_\mu \Phi(X_0) t^{\mu} + \frac{1}{2} \nabla_\mu \nabla_\nu \Phi(X_0) t^{\mu} t^{\nu} + ..., \\
    \partial_a \left( X^{\mu}_0 + Y^\mu\right) & = \partial_a X^{\mu}_0 + \nabla_a t^{\mu} + \frac{1}{3} \mathcal{R}^{\mu}_{\ \nu \rho \sigma} (X_0) \partial_a X^{\sigma}_0 t^{\nu} t^{\rho} + ..., 
\end{align}
where $\mathcal{R}^{\mu}_{\ \nu \rho \sigma}$ represents the Riemann tensor associated with $\mathcal{G}_{\mu \nu}$. The quadratic term for the quantum fields $t^{\mu}$ contains an arbitrary metric in front of it, $\mathcal{G}_{\mu \nu} (X_0) \nabla_a t^{\mu} \nabla_b t^{\nu}$, that we would need to invert to find the propagator. The way to deal with this problem and obtain a simple propagator is to introduce a vielbein $e^{A}_{\ \mu} (X_0)$ which fulfills
\begin{align}
    e^{A}_{\ \mu} (X_0) e^{B}_{\ \nu} (X_0) \eta_{A B} = \mathcal{G}_{\mu \nu} (X_0),
\end{align}
with $\eta_{AB}$ a Lorentzian metric. In this way, we can rewrite all the vector expressions in the nonholonomic basis $e^{A}_{\ \mu}$ and get a trivial propagator for the $t^{A}= e^{A}_{\ \mu} t^{\mu}$ fields. This comes with a subtlety, because now the derivatives $\nabla_a$ involve the spin-connection of the spacetime $\omega_{\mu}^{\ AB}$; for example, 
\begin{align}
    \nabla_a t^A = \partial_a t^A + \omega_{\mu}^{\ AB} \partial_a X^{\mu}_0 t^{C} \eta_{BC}.
\end{align}
Obtaining a trivial propagator requires breaking the $SO(D,1)$ invariance of the theory. However, since we are working within a manifestly gauge-covariant formalism, we can be certain that the diagrammatic expansion will always include contributions that restore explicit gauge covariance at intermediate steps. Summarizing all the information gathered so far, we have arrived at the following expansion for the Polyakov term in the action:
\begin{align}
    S_{P} & = S_P[X_0] + \frac{1}{2\pi \alpha'} \int \dd^2 \sigma \sqrt{\gamma} \gamma^{ab} \mathcal{G}_{\mu \nu} (X_0) \partial_a X_0^{\mu} \nabla_b t^{\nu} \nonumber\\
    & + \frac{1}{4 \pi \alpha'} \int \dd^2 \sigma \sqrt{\gamma} \gamma^{ab} \left[ \eta_{A B} \nabla_a t^A \nabla_b t^B \right] \nonumber \\
    & + \frac{1}{3 \pi \alpha'} \int \dd^2 \sigma \sqrt{\gamma} \gamma^{ab} \mathcal{R}_{\mu A B C} (X_0) \partial_a X_0^{\mu} t^A t^B \nabla_{b} t^C \nonumber \\
    & + \frac{1}{12 \pi \alpha'} \int \dd^2 \sigma \sqrt{\gamma} \gamma^{ab} \mathcal{R}_{A B C D}  (X_0) t^{B} t^{C} \nabla_{a} t^{A}  \nabla_{b} t^{D} + \ldots ,
\end{align}
and for the dilaton part we have the trivial structure: 
\begin{align}
    S_D [X_0 + t] = & S_D[X_0] - \frac{1}{8 \pi} \int \dd^2 \sigma \sqrt{\gamma}  \nabla_A  \Phi(X_0) t^A \nonumber \\
    & - \frac{1}{16 \pi} \int \dd^2 \sigma \sqrt{\gamma}  \nabla_A \nabla_B \Phi(X_0) t^A t^B + \ldots \ . 
\end{align}
We can safely impose the equations of motion for the classical fields and drop the linear terms, since it is tantamount to a legitimate field redefinition.

Now we can determine the $\beta$-functionals of the trace anomaly~\eqref{Eq:Trace_Anomaly} from the effective action. The computation requires an additional loop order for the dilaton field, as its dimensionality is different from other operators. The computation is rather lengthy and hence we do not reproduce it here~\cite{Callan1989}. We simply write down the result as 
\begin{align}
    & \beta^{\mathcal{G}}_{\mu \nu} =  \mathcal{R}_{\mu \nu} \left( \boldsymbol{\mathcal{G}} \right) - \nabla_{\mu} \nabla_{\nu} \Phi + \order{\alpha'}, \\ 
    & \beta^{\Phi} = \frac{D-26}{6} + \alpha' \left[ - \mathcal{R} \left( \boldsymbol{\mathcal{G}} \right)  + 2 \nabla^2 \Phi +(\nabla \Phi)^2 \right] +  \order{\alpha^{\prime 2}}.
\end{align}
If we are dealing with a flat worldsheet, the vanishing of $\beta^G$ is enough to ensure Weyl invariance at the quantum level, as long as we are working in $ D = 26$ dimensions, the critical dimension (see Eq.~\eqref{Eq:Trace_Anomaly}. Hence, in principle, we expect that the same applies to nonflat worldsheets, i.e., that the condition $\beta^{\Phi} = 0$ is not independent of $\beta^\mathcal{G} = 0$. Actually, we have a nontrivial constraint coming from the Bianchi identity
\begin{align}
    & \nabla^{\mu} \left( \mathcal{R}_{\mu \nu} \left( \boldsymbol{\mathcal{G}} \right) - \frac{1}{2} \mathcal{R} \left( \boldsymbol{\mathcal{G}} \right) \mathcal{G}_{\mu \nu} \right) = 0.
\end{align}
This ensures that we have 
\begin{align}
    \nabla^{\mu} \beta^\mathcal{G}_{\mu \nu} =  \nabla_{\nu} \beta^{\Phi} = 0, 
\end{align}
whenever $\beta^{\mathcal{G}}_{\mu \nu} = 0$, as can be seen by direct calculation order by order. This implies that $\beta^{\Phi}$ is a constant as long as $\beta^\mathcal{G} =  0$. By continuity, this automatically implies that $\beta^{\Phi} = 0$ for $D=26$~\cite{Callan1989}. From now on we will restrict ourselves to work in $D=26$ and make a comment on strings on noncritical dimension later.

\subsubsection{Including string-loop corrections}
\label{Subsec:String_Loops}

At this point, we have only focused on the zeroth-order in the $g_s$-expansion. Although it is clear that string loops should modify the results, it is not completely clear how those corrections must be included. One of the most accepted proposals is the Fischler-Susskind approach~\cite{Fischler1986,Fischler1986b,Fischler1987}. The idea behind such mechanism is that string loop divergences can be absorbed through a renormalization of the background fields in the nonlinear sigma model. Let us illustrate this explicitly for unoriented closed bosonic strings. For the purpose of this section, it is simpler to work with a sharp cut-off as regularization scheme. 

Divergences in string loops arise when summing over conformally inequivalent surfaces of fixed topology (genus). For a given nontrivial topology, this sum corresponds to an integral over the finite-dimensional Teichm\"uller space~\cite{Polchinski1998a,Green1987a}. These integrals develop divergences when handles shrink to zero size, which are analogous to those from inserting a local operator on a worldsheet of trivial genus. For example, in flat spacetime, the divergence of torus topology can be removed by inserting the operator $\frac{\log \Lambda}{2 \pi} \gamma^{ab} \eta_{\mu \nu}  \partial_a X^{\mu}\partial_b X^{\nu}$ with an appropriate coefficient, where $\Lambda$ acts as a cutoff in Teichm\"uller space. 

If we extend our analysis to a curved background with a nontrivial profile for the dilaton field, we need to introduce the metric $G_{\mu \nu}$ accordingly and introduce a relative factor $e^{-\Phi}$ to account for the dependence of the path integral on the surface topology. The asymptotic value of the dilaton field, $\lambda = \expval{\Phi}$, is related to the string coupling constant by the exponential relation $g_s = e^{\lambda}$, as can be seen by comparing the actions in Eq.\eqref{Eq:Action_String} and Eq.\eqref{Eq:Int_Action}. Specifically, for torus topology, the resulting divergences take the following form:
\begin{align}
    \delta S^{\text{loop}} = \frac{\log \Lambda}{2 \pi} \int \dd^2 \sigma \sqrt{-\gamma} \gamma^{ab} e^{-\Phi} \mathcal{G}_{\mu \nu} (X) \partial_a X^{\mu}\partial_b X^{\nu}.
\end{align}
The factor $e^{-\Phi}$ ensures that, when evaluated on a worldsheet with trivial topology, it correctly captures the divergences present at the torus level. If the dilaton field has a nontrivial background profile $\Phi(X)$ rather than just a constant zero mode $\lambda$, replacing $\Phi$ with $\Phi(X)$ is expected to introduce a leading-order correction in the $\alpha'$-expansion. This correction modifies the $\beta$-functionals, in particular the one associated with the metric, by introducing an additional term $\delta \beta^{\mathcal{G}}_{\mu \nu}$ to the original $\beta^{\mathcal{G}}_{\mu \nu}$. From this point forward, we will refer to $\beta$-functionals modified due to string loop corrections as $\tilde{\beta}$:
\begin{align}
    \tilde{\beta}^{\mathcal{G}}_{\mu \nu} = \beta^{\mathcal{G}}_{\mu \nu} + \delta \beta^{\mathcal{G}}_{\mu \nu}.
\end{align}
Specifically, the contribution from the string loops looks like a cosmological constant term, i.e.,
\begin{align}
    \delta \beta^{\mathcal{G}}_{\mu \nu} =  C e^{- \Phi} \mathcal{G}_{\mu \nu} ,
\end{align}
where $C$ is an arbitrary constant that arises in the renormalization procedure. Similarly, an additional contribution to the dilaton, denoted as $\delta \beta ^{\Phi}$, will also emerge, though its explicit evaluation is challenging. Instead, it is more practical to determine it using a consistency argument~\cite{Fischler1986,Fischler1986b,Fischler1987}. As discussed earlier, the vanishing of the modified \mbox{$\tilde{\beta}^{\Phi}$-functional} due to string loop corrections is not independent of the vanishing of $\tilde{\beta}^{\mathcal{G}}_{\mu \nu}$. In the computation from the previous section, we saw that the constancy of $\beta^{\Phi}$ directly followed from the vanishing of the other functionals. Assuming that this consistency condition still holds when string loops are included, we can derive an equation for the $\tilde{\beta}^{\Phi}$-functional. 

Taking the divergence of the $\tilde{\beta}^{\mathcal{G}}_{\mu \nu}$ and simplifying it through Bianchi identities and using also the vanishing of $\tilde{\beta}^{\mathcal{G}}_{\mu \nu}$ itself, we find: 
\begin{align}
    \nabla^{\mu} \tilde{\beta}^{\mathcal{G}}_{\mu \nu} =  \nabla_{\nu} \left( \frac{1}{2} \mathcal{R} \left( \boldsymbol{\mathcal{G}} \right) - \nabla^2 \Phi - \frac{1}{2} \left( \nabla \Phi \right)^2 \right).
\end{align}
This leads us to the following $\tilde{\beta}^{\Phi}$ functional for the dilaton field:
\begin{align}
    \tilde{\beta^{\Phi}}=  \alpha' \left[ -  \mathcal{R} \left( \boldsymbol{\mathcal{G}} \right) + 2 \nabla^2 \Phi + \frac{1}{2} \left( \nabla \Phi \right)^2 \right]. 
\end{align}
Since we know this term is a constant, we can safely set it to zero. However, if we were dealing with strings in a noncritical dimension, an additional factor of $D-26/6$ factor should be included arising from the $bc$-ghost system contribution to the Weyl-anomaly at the string tree level. Notice that we have introduced $\alpha'$ as a dimensionful parameter. 

\subsubsection{EFTs for the massless sector of the theory}
\label{Subsec:EFTs}

At this stage, it is useful to pause and summarize our progress so far. We began by analyzing the $\alpha'$-expansion of the sigma model that describes string propagation in arbitrary backgrounds. We then determined the lowest-order $\beta$-functionals associated with the Weyl anomaly. From there, we addressed the inclusion of string-loop corrections, which must necessarily modify the constraints on the background fields. To incorporate these corrections, we observed that divergences from string loops can be absorbed through a renormalization of the background fields $\mathcal{G}_{\mu \nu}$ and $\Phi$. As a result, we have now derived a set of equations that these background fields must satisfy to ensure the consistent propagation of strings. 

In fact, these equations closely resemble the equations of motion of a given field theory for $\mathcal{G}_{\mu \nu} (X)$ and $\Phi(X)$: 
\begin{align}
    & \Tilde{\beta}^{\mathcal{G}}_{\mu \nu} = \mathcal{R}_{\mu \nu} \left( \boldsymbol{\mathcal{G}} \right) - \nabla_{\mu} \nabla_{\nu} \Phi +  C e^{-\Phi} \mathcal{G}_{\mu \nu}  + \order{\alpha'}, \\
    & \Tilde{\beta}^{\Phi} = \frac{D-26}{6} + \alpha' \left[ - \mathcal{R} \left( \boldsymbol{\mathcal{G}} \right) + 2 \nabla^2 \Phi +(\nabla \Phi)^2 \right] +  \order{\alpha^{\prime 2}}.
\end{align}
Setting $C=0$ corresponds to omitting the string loop corrections. 

The natural question now is whether an action exists whose equations of motion imply the vanishing of these functionals. In addition, such an action should correctly account for the scattering amplitudes involving excitations of the string. There are two effective actions that fulfill these criteria corresponding to a GR-like EFT, i.e., Diff invariant, and a UG-like EFT, i.e. WTDiff invariant. The GR-like EFT would correspond to
\begin{align}
    S_{\text{eff}}^{\text{GR}} = & \frac{1}{2 \kappa_D^2} \int \dd^{D+1} X \sqrt{-\mathcal{G}} e^{ \Phi}  \nonumber \\
    & \times \left(  - \frac{(D-26)}{6 \alpha'}  - 2 C e^{-\Phi} + \mathcal{R} \left( \boldsymbol{\mathcal{G}} \right) + (\nabla \Phi)^2 \right) + \order{\alpha'}.
    \label{Eq:Eff_Action_GR}
\end{align}
From this action principle it is straightforward to obtain the $\beta$-functionals as 
\begin{align}
    & \tilde{\beta}^{\Phi} = - 2 \kappa_D^2 \frac{ e^{-\Phi}}{\sqrt{-\mathcal{G}}} \frac{\delta S^{\text{GR}}_{\text{eff}} }{\delta \Phi} , \\
    & \tilde{\beta}^{\mathcal{G}}_{\mu \nu} = 2 \kappa_D^2 \frac{e^{- \Phi }}{\sqrt{-\mathcal{G}}} \left( \frac{\delta S^{\text{GR}}_{\text{eff}} }{\delta \mathcal{G}^{\mu \nu}} + \frac{1}{2} \mathcal{G}_{\mu \nu} \frac{\delta S^{\text{GR}}_{\text{eff}}}{\delta \Phi } \right).
\end{align}
Furthermore, it is possible to perform a field redefinition to map this action to the Einstein Frame~\cite{Green1987a}. 

However, given our discussion from Section~\ref{Sec:ClassicalNL}, we know that it is also possible to write down an action principle which reproduces the same equations of motion that Eq.~\eqref{Eq:Eff_Action_GR} displays, with the cosmological constant $C$ entering as an integration constant instead of a coupling constant. To be concrete, we can write down the following action principle: 
\begin{align}
    S_{\text{eff}}^{\text{UG}} = \frac{1}{2 \kappa_D^2} \int \dd^{D+1} X \omega e^{ \Phi} \left(  - \frac{(D-26)}{6 \alpha'}   + R ( \boldsymbol{\tilde{G}} ) + ( \tilde{\nabla} \Phi)^2 \right) + \order{\alpha'}, 
    \label{Eq:Eff_Action_UG}
\end{align}
where we are introducing the fixed background volume form $\boldsymbol{\omega}$ and the auxiliary metric $\boldsymbol{\tilde{\mathcal{G}}}$ whose components are:
\begin{align}
    \tilde{\mathcal{G}}_{\mu \nu} = \left( \frac{\omega^2}{\abs{\mathcal{G}}} \right) ^{\frac{1}{D+1}}\mathcal{G}_{\mu \nu}.
\end{align}
If we compute the variation with respect to $\mathcal{G}_{\mu \nu}$ we obtain the traceless version of the equations obtained from Eq.~\eqref{Eq:Eff_Action_GR}. Explicitly, if we define 
\begin{align}
     \frac{\delta S_{\text{eff}}^{\text{GR}}}{\delta \mathcal{G}_{\mu \nu}} =  K_{\mu \nu} \left( \boldsymbol{\mathcal{G}} \right) - \frac{1}{2}  K \left( \boldsymbol{\mathcal{G}} \right) \mathcal{G}_{\mu \nu},
\end{align}
for the variation of $S_{\text{eff}}^{\text{UG}}$ we obtain the following: 
\begin{align}
    \frac{\delta S_{\text{eff}}^{\text{UG}}}{\delta \mathcal{G}^{\mu \nu}} = K_{\mu \nu} ( \boldsymbol{\tilde{\mathcal{G}}} ) - \frac{1}{D+1} K ( \boldsymbol{\tilde{\mathcal{G}}} )  \tilde{\mathcal{G}}_{\mu \nu} = 0,
\end{align} 
with $K ( \boldsymbol{\tilde{\mathcal{G}}} ) = \tilde{\mathcal{G}}^{\mu \nu} K_{\mu \nu} ( \boldsymbol{\tilde{\mathcal{G}}} ) $. Upon taking the divergence and using the generalized Bianchi identities for the corresponding tensor $K$ entering the equations we find: 
\begin{align}
    E_{\mu \nu} ( \boldsymbol{\tilde{\mathcal{G}}} ) = K_{\mu \nu} ( \boldsymbol{\tilde{\mathcal{G}}} ) - \frac{1}{2} K ( \boldsymbol{\tilde{\mathcal{G}}} ) \Tilde{\mathcal{G}}_{\mu \nu} + C \tilde{\mathcal{G}}_{\mu \nu} = 0, 
\end{align}
where we notice that  
\begin{align}
    E_{\mu \nu} \left( \boldsymbol{\mathcal{G}} \right) =  \frac{\delta S_{\text{eff}}^{\text{GR}}}{\delta \mathcal{G}^{\mu \nu}}. 
\end{align}
Again, a suitable combination of these equations with the equation obtained from the equation of motion for $\Phi$ we find: 
\begin{align}
    & \tilde{\beta}^{\Phi} = - 2 \kappa_D^2 \frac{ e^{-\Phi}}{\omega} \frac{\delta S^{\text{UG}}_{\text{eff}} }{\delta \Phi} , \\
    & \tilde{\beta}^{\mathcal{G}}_{\mu \nu} = 2 \kappa_D^2 \frac{e^{- \Phi }}{\omega} \left( E_{\mu \nu} (\boldsymbol{\tilde{\mathcal{G}}}) + \frac{1}{2} \tilde{\mathcal{G}}_{\mu \nu} \frac{\delta S_{\text{eff}}}{\delta \Phi } \right),
\end{align}
confirming our claim that a WTDiff invariant action~\eqref{Eq:Eff_Action_UG} reproduces the $\beta$-functionals. Notice that this effective action does not only reproduce the $\beta$-functionals but it also reproduces all of the scattering amplitudes involving massless excitations of the string. Thus, from our perspective, neither action is preferred over the other as an effective field theory for describing the massless modes of the string.

\section{Conclusions and discussion}
\label{Sec:ConclusionsUG}

This chapter has provided a detailed comparison of gravitational theories based on Diff invariance and WTDiff invariance. Broadly speaking, the key distinction between the two lies in the behavior of the cosmological constant, though this difference manifests in various ways depending on the regime considered. We now summarize the results of the chapter, discuss some open questions, and examine the connection between WTDiff-invariant theories and the emergent gravity framework.

\subsection{Summary of the comparison of the theories}

\paragraph*{\textbf{Classical theories.}}

At the classical level, the primary distinction between the two theories lies in the treatment of the cosmological constant. In Diff-invariant theories, the cosmological constant is a fixed coupling constant set at the level of the action. In contrast, in WTDiff-invariant theories, it emerges as an additional global degree of freedom. Consequently, the solution space of WTDiff-invariant theories corresponds to the solution space of their corresponding Diff-invariant counterpart for all possible values of the cosmological constant, regardless of the inclusion of higher-derivative terms or non-minimal couplings. This holds as long as the couplings are constructed to mirror those in Diff-invariant theories, that is, ensuring that the energy-momentum tensor remains conserved in the equations of motion and the couplings preserve the WTDiff symmetry.

\paragraph*{\textbf{Semiclassical theories.}}

At the semiclassical level, we have found that as long as the matter content couples through the Weyl-invariant metric $\boldsymbol{\tilde{g}}$ metric, constructed from the dynamical metric $\boldsymbol{g}$ and the volume form $\boldsymbol{\omega}$, the backreaction of quantum fields enters the semiclassical equations in the same way as in Diff-invariant theories. A key aspect of this result is the absence of a Weyl anomaly. While this may seem counterintuitive at first, it arises because the quantization procedure no longer needs to preserve longitudinal diffeomorphisms, which are the transformations that inherently conflict with Weyl invariance. Since enforcing longitudinal diffeomorphisms is not required, there is no fundamental obstruction preventing Weyl symmetry from being realized non-anomalously at the quantum level.

The only distinction, once again, lies in the treatment of the cosmological constant. In Diff-invariant theories, the cosmological constant appears as the coupling constant of an operator in the action and is therefore subject to renormalization. In contrast, in WTDiff-invariant theories, it remains a global degree of freedom even at the semiclassical level. As a result, it is effectively decoupled from radiative corrections. 

This has direct consequences for the conventional formulation of the cosmological constant problem. We have attempted to present the issue in its sharpest possible form, namely, in terms of technical (un)naturalness. Our analysis shows that the standard arguments used to highlight this problem do not apply to WTDiff, precisely because the cosmological constant remains decoupled from radiative corrections. However, we have also revisited the rationale behind using naturalness as a guiding principle in model building and argued that one should approach it with caution. Moreover, in the specific case of the cosmological constant, we have emphasized its strong dependence on cosmological observations and, in particular, on accepting the standard $\Lambda$CDM model.

\paragraph*{\textbf{Perturbative quantization.}}

Regarding the perturbative quantization of both theories, we have reviewed the existence of a quantization scheme in which they yield identical results. Consequently, we have concluded that, at least at tree level, both theories can be quantized in an equivalent manner concerning local observable computations. The natural observables in this context are on-shell scattering amplitudes. As a complementary analysis, we have examined on-shell methods to demonstrate that the scattering amplitudes in both theories coincide and can be fully determined from the three-point function alongside the principles of locality and unitarity. At the loop level (in particular two loops and higher), since neither theory is renormalizable, they must be treated as effective field theories (EFTs). Nevertheless, the existence of a common quantization scheme implies that their predictions as EFTs remain consistent also.

\paragraph*{\textbf{Euclidean path integral.}}

The nonperturbative path integral quantization of gravity presents significant challenges for several reasons. Most of the difficulties encountered in the Diff-invariant formulation remain essentially unchanged in the WTDiff-invariant approach. The key distinction lies in the conformal factor problem of Diff-invariant theories: the mode associated with conformal transformations of the metric is unbounded from below, rendering the path integral ill-defined. In the WTDiff formulation, this issue arises \emph{mutatis mutandis}, as the mode associated with longitudinal diffeomorphisms exhibits the same pathological behavior.

\paragraph*{\textbf{Interplay with string theory.}}

Finally, we have explored the relationship between gravitational theories with a WTDiff invariance principle and string theory. It is often stated that the low-energy limit of string theory is described by GR coupled to matter fields representing the massless excitations of the string spectrum, along with higher-order corrections arising from the expansion in the string tension parameter, $\alpha'$. However, our findings suggest that this claim is made too hastily. Three common arguments are used to support this claim:

\begin{enumerate}
    \item The theory's spectrum contains a graviton, and GR is the only consistent theory of interacting gravitons at low energies.
    
    \item Scattering amplitudes computed in string perturbation theory match those derived from GR.

    \item The requirement that the Weyl anomaly cancels in arbitrary background spacetimes enforces that these backgrounds must satisfy Einstein's equations, coupled to additional matter fields, at leading order in $\alpha'$. 
\end{enumerate}

We have shown in the previous chapter that the first argument is invalid, as UG can also emerge as a self-consistent theory of interacting gravitons. The second argument is similarly flawed, since the scattering amplitudes of GR and UG are equivalent. Finally, the third argument also does not hold upon closer examination: determining whether the background gravitational field in string theory corresponds to a fully dynamical metric (as in GR) or a conformal structure with a fixed volume form (as in UG) is ultimately a choice. Thus, at this level, we conclude that it remains a matter of choice whether the low-energy description of string theory is formulated using a Diff-invariant or WTDiff-invariant principle.
   
\subsection{Open questions}

Given our discussion, it is quite clear that locally both theories are equivalent. However, given that UG, and in general WTDiff invariant theories, display an additional global degree of freedom, it is a legitimate question whether such degree of freedom is somehow directly measurable or not. In particular, it is interesting that the canonically conjugated variable to the cosmological constant is a time variable that foliates the spacetime. In that sense, it would be interesting to investigate if one is able to directly probe somehow such global degree of freedom, and not only indirectly through metric measurements. 

It also is worth remarking that although the WTDiff-invariant theories presented here are somehow equivalent to Diff-invariant theories, aside from the key difference of the cosmological constant, alternative formulations of the same theory may naturally suggest different extensions. For example, the UG equations motivated the exploration of nonconserved energy-momentum tensors, an idea that was not traditionally considered viable within GR. In this spirit, an intriguing direction for future research would be to investigate whether the global degree of freedom could acquire a nontrivial dynamics. In the current framework, this degree of freedom remains trivial in the sense that the cosmological constant is completely fixed by initial conditions and it does not change with time. 

In general, giving rise to ``dynamical'' cosmological constants, i.e., a theory that allows for an approximate description in some regime resembling a time-dependent cosmological constant, is challenging, as typical Diff-invariant modified gravity models often suffer from instabilities~\cite{Delhom2022}. However, some of these issues might be avoided if the cosmological constant arises as a global degree of freedom. It would be particularly interesting to explore whether certain modifications or extensions of standard WTDiff-invariant theories could accommodate a dynamical cosmological constant, which by being a global degree of freedom, exhibits an effective Newtonian-like dynamics and is coupled to the local degrees of freedom. In particular, those setups might be useful for analyzing cosmological models beyond the standard $\Lambda$CDM model, in which the cosmological constant is nondynamical.

\subsection{Emergent gravity framework interplay}

Emergent gravity approaches, as defined in the introduction, are scattered throughout the literature, often employing distinct frameworks, methodologies, and tools. The analysis conducted here within the WTDiff-invariance principle provides new insights into emergent gravity theories and lays the groundwork for deeper investigations of its interplay with emergent approaches. 

String theory, the most well-known emergent gravity approach, has the advantage of being well-defined enough to permit detailed analyses like the one conducted here. We revisited the key arguments used in perturbative string theory to demonstrate that GR necessarily emerges as its low-energy limit. This follows from the assumption that the background on which strings propagate is modeled as a metric, with graviton degrees of freedom arising as fluctuations of this metric. However, we argue that one could just as well adopt a WTDiff-invariance principle, introducing a background volume form and treating the fluctuating conformal structure as the fundamental dynamical field instead. This serves as a proof of principle that the emergence of GR in such frameworks requires careful consideration, as it may simply reflect an implicit choice to model gravity using Diff invariance.

Another example in which a WTDiff principle and UG naturally arise in the context of emergent gravity is Jacobson's derivation of Einstein’s equations~\cite{Jacobson1995}. Notably, the equations Jacobson initially derives are the traceless version of Einstein’s equations. It is only by assuming the conservation of the energy-momentum tensor that he can manipulate them, much like we have done with UG equations, to recover the full Einstein equations, with an integration constant appearing as the cosmological constant. In this sense, WTDiff-invariant theories fit nicely within the thermodynamically-inspired emergent gravity approaches \emph{\`a la Jacobson}.  

Finally, it is worth discussing the connection between UG and the emergent gravity approach in the spirit of Volovik. According to Volovik, the ground state of quantum many-body systems can replicate properties of the vacuum of quantum field theories. In particular, the system's thermodynamic behavior ensures that, in isolation, the vacuum energy vanishes. However, in out-of-equilibrium situations, the system may exhibit a nonzero cosmological constant, which then evolves toward equilibrium, eventually reaching a vanishing value~\cite{Volovik2006,Volovik2009}.

In this context, models where the cosmological constant behaves as a global degree of freedom, potentially realized through extensions of WTDiff-invariant theories, could provide a useful framework for describing such dynamics. For instance, if this global degree of freedom follows a Newtonian-like equation as proposed above, an emergent theory inspired by Volovik’s ideas would naturally predict that the zero value of the cosmological constant corresponds to a minimum in the potential. As a direction for future research, it would be interesting to explore models grounded in Volovik's reasoning that can accommodate this type of phenomenology.



\part{Horizonless ultracompact objects}
\label{pt2}

\chapter{Almost no-hair theorems for black holes within and beyond General Relativity}
\label{Ch5:Nohair}

\fancyhead[LE,RO]{\thepage}
\fancyhead[LO,RE]{Almost no-hair theorems for black holes within and beyond general relativity}

This second part of the thesis explores ultracompact objects beyond GR. Here, we broadly define an ultracompact object as any object that closely resembles a black hole observationally and can replicate most of its properties. The general framework of emergent approaches, along with the analyses conducted in the first part of the thesis, suggests that singularities and extreme causal scenarios should be absent. While temporarily apparent horizons may form and disappear, genuine event horizons, which are eternal by definition, are not expected to emerge. 

No-hair theorems, which are theorems that constrain the form of the metric of black hole spacetimes, are one of the key results in GR~\cite{Chrusciel2012}. These theorems are typically formulated under idealized assumptions, involving a mixture of local (regularity of the horizon) and global aspects (everywhere vacuum spacetime and asymptotic flatness). Even within the framework of GR, this limits their applicability to astrophysical scenarios of interest such as binary black holes and accreting systems. Furthermore, it is not possible to apply any of these results as they stand to any kind of horizonless object. 

In this chapter we push the no-hair theorems beyond by considering two specific extensions although we work in the simplified setup of staticity and axisymmetry. First of all, we extend no-hair theorems to situations in which the black hole is in an astrophysical background, i.e., we consider that the black hole is immersed in an external gravitational field. The only previous results in this direction are due to G\"urlebeck~\cite{Gurlebeck2015}, who showed that the Weyl multipole moments that characterize a spacetime close to infinity can be expressed as a sum of the contributions from black hole regions and regions with a nonvanishing energy-momentum tensor. The contributions from black holes are those associated with the Schwarzschild black holes. That is, a distorted black hole does not modify the contribution to the multipole structure of the external gravitational field in which it is embedded.

However, no specific statement of the fact that there exists only one geometry compatible with such environment is presented. We fill this gap in the literature and explicitly show that only a one-parameter family of black-hole geometries is compatible with a given external gravitational field, where the parameter can be identified with the mass of the corresponding black hole. In other words, we show that the specific shape of the horizon, i.e., its intrinsic geometry, is fully determined by the gravitational background. 

Second, we extend the no-hair theorems to horizonless objects that are close to forming a horizon. To quantify this, we assume that Einstein's equations hold up to a surface, defined by a finite maximum redshift, and perform an expansion in such redshift. By imposing the physical condition that curvatures remain bounded, we constrain the possible metric structures, showing that deviations from the standard black hole geometry must vanish as the system approaches the black hole limit. This allows us to determine how much the metric can differ from its general relativistic black hole counterpart, depending on the maximum curvature and redshift permitted for the object. This chapter is based on the article~\cite{Barcelo2024} that has been published during this thesis.

\section{Revisiting Israel's theorem: local vs global aspects}
\label{Sec:IsraelRev}

The general no-hair theorem applies to stationary, asymptotically flat spacetimes with an event horizon that satisfy Einstein’s equations, along with certain additional technical assumptions~\cite{Chrusciel2012}\footnote{Some of these assumptions, such as analyticity, are particularly difficult to relax. It remains unclear whether this is merely a mathematical challenge or if there are deeper physical obstructions that prevent it.}. The proof of the general theorem begins by separating the analysis into two cases: static and nonstatic spacetimes. Static spacetimes are the subset of stationary spacetimes in which there exists a spacelike hypersurface orthogonal to the Killing vector associated with stationarity. In this thesis, we have focused on static black holes as a first step toward the general analysis. Toward the end of this chapter, we will briefly discuss possible extensions of this work to more general scenarios, which remain subjects of ongoing and future research.

The theorem that constrains the metric for a black hole static spacetime was proved by Israel~\cite{Israel1967}. We review here the content of the theorem with the aim of disentangling local and global aspects of the theorem, which are essential to formulate the extensions presented afterwards. The starting point of the discussion is a general static metric which can be written in the following form~\cite{Israel1967}:
\begin{align}
    \dd s^2 = - \mathscr{V}^2(x) \dd t^2 + g_{ij} (x) \dd x^i \dd x^j,
\end{align}
where $\mathscr{V}(x)$ represents a potential function, and $g_{ij}$ is an arbitrary static 3-dimensional metric. This parametrization can be used to describe any horizonless configuration, with $\mathscr{V}\neq 0$, or in the case of having horizons, i.e., surfaces at which $\mathscr{V} \to 0^+$, the geometry exterior to them. From a pure geometrical, i.e., kinematical perspective, it is interesting to recall that a geometry can have horizons with any shape and topology. Hawking's theorem on the topology of black holes in four spacetime dimensions makes the additional dynamical assumption that the matter content at the horizon must satisfy the dominant energy condition~\cite{Hawking1973}. In fact, this theorem shows that the topology of a horizon must be spherical or exceptionally toroidal. In this chapter, we exclusively focus on spherical horizons, and leave the discussion of toroidal horizons for the next section. 

Now, the vacuum Einstein equations for the 3-dimensional metric can be written in the following form:
\begin{align}
    & {}^{3}R = 0, \\
    & {}^{3}R_{ij} + \mathscr{V}^{-1} \nabla^{(g)}_i \nabla^{(g)}_j \mathscr{V} = 0, 
\end{align}
where ${}^{3}R_{ij}$ and ${}^{3}R$ represent the Ricci tensor and the Ricci scalar of the spatial metric $g_{ij}$ and $\nabla^{(g)}_i$ is the covariant derivative constructed with the Levi-Civita connection of $g_{ij}$. Combining both equations, we can replace the equation associated with the vanishing of the Ricci scalar ${}^{3}R$ with a Laplace equation for $\mathscr{V}$,
\begin{align}
    \nabla^2_{(g)} \mathscr{V} = 0. 
\end{align}
Israel's theorem~\cite{Israel1967} imposes the following conditions: 
\begin{itemize}
    \item Take $\Sigma$ to be any $t=\text{constant}$ spatial hypersurface that is regular, empty, noncompact, and asymptotically flat, namely that there exists a set of coordinates (that are asymptotically identified with the coordinates $x^i$) in which:
    \begin{align*}
        & g_{ij} = \delta_{ij} + \order{r^{-1}}, \qquad \partial_{k} g_{ij} = \order{r^{-2}}, \\
        & \mathscr{V} = 1 - m/r + \eta (r), \qquad m = \text{constant}, \\
        & \eta = \order{r^{-2}}, \qquad \partial_i \eta = \order{r^{-3}}, \qquad \partial_i \partial_j \eta = \order{r^{-4}}, 
    \end{align*}
    when $ r = \left( \delta_{ij} x^i x^j \right)^{1/2} \rightarrow \infty $.
    \item The equipotential surfaces $\mathscr{V} = \text{constant} >0,$ $t = \text{constant}$ are regular, simply connected, closed 2-spheres. 
    \item The Kretschmann scalar $\mathcal{K} = R_{\mu \nu \rho \sigma} R^{\mu \nu \rho \sigma}$ is bounded everywhere on $\Sigma$. 
    \item If $ \mathscr{V} $ has a vanishing lower bound on $\Sigma$, the two-surfaces  $\mathscr{V}=c \in \mathbb{R}$ approach a limit as $c \rightarrow 0^{+}$, corresponding to a closed regular 2-space of finite area.  
\end{itemize}
Israel showed that the only static metric that satisfies all of them is the Schwarzschild metric, which also turns out to be spherically symmetric. The core of the proof is a sort of shooting method in which the equations above are integrated from the asymptotic $\mathscr{V} = 1$ surface to the $ \mathscr{V} = 0^{+}$ surface. Then, Israel shows that the Schwarzschild solution is the only one satisfying all the assumptions and resulting in a regular geometry at $\mathscr{V}=0^{+}$. In order to implement this procedure, the first step is to introduce a set of adapted coordinates on the spatial hypersurfaces, which correspond to the coordinate $\mathscr{V}$ and a set of angular-like coordinates $(\theta^1,\theta^2)$ on the $\mathscr{V}=\text{constant}$ surfaces. 

The first hypothesis sets the problem, in the sense that one considers a vacuum spacetime and ensures that the spacelike slices are noncompact and have an asymptotic infinity. Furthermore, the asymptotically flat behavior can be seen as providing the initial data for the inward integration. The second assumption, in combination with the emptiness assumption, is required to ensure that there is only ``one single spheroidal central object'' acting as a source of the gravitational field. Relaxing these assumptions (vacuum and single central spheroid) is one of the aims of the chapter. 

Finally, the last two assumptions ensure that the geometry is regular at the surface $\mathscr{V} = 0^{+}$, i.e., that the spacetime displays a regular Killing horizon (which in this case coincides with the event horizon). In this way, the shooting procedure singles out the Schwarzschild geometry, since any deviation from the Schwarzschild metric would lead to a singular geometry (a singular Kretschmann scalar) on the event horizon. Thus, one concludes the uniqueness of the solution. As a side note, it is puzzling to realize that the theorem specifically avoids the analysis of the strange case in which a putative horizon of infinite area forms. We will discuss such case below, when we introduce the Curzon metric.

From this result, we could draw  too quickly the conclusion that any distortion of an event horizon away from spherical symmetry leads to a singular behavior. This misunderstanding may be reinforced due to the reported vanishing of the Love numbers of black holes (see~\cite{Binnington2009,Poisson2020} for an introduction and~\cite{Damour2009,Damour2009b,Kol2011} for the computation for static black holes) being often interpreted as the impossibility of deforming an event horizon. However, in this paper we will clearly show that we can separate the deformations of an event horizon from spherical symmetry in terms of the ``agent'' that generates the deformation of the horizon: whether it is an intrinsic deformation (due to the internal structure of the object itself), or a deformation due to an external gravitational field. Whereas the former kind of deformations does indeed lead to a singular geometry at $\mathscr{V} = 0^{+}$, the latter gives rise to a perfectly smooth event horizon that departs from spherical symmetry. 

Actually, let us show that the vacuum Einstein equations are compatible with any geometry induced at the horizon, as long as it is topologically a 2-sphere. We can show this through a constructive procedure, following the approach in~\cite{Frolov1985}. If an analytic static metric exists in an open neighborhood outside a regular topologically-spherical horizon, then one can conveniently choose adapted coordinates $(t,\mathscr{V},\vec{\theta})$ so that the metric reads:
\begin{align}
    \dd s^2 = -\mathscr{V}^2 \dd t^2 + \frac{\dd \mathscr{V}^2}{\kappa^2 (\mathscr{V}, \vec{\theta})} + h_{ab} (\mathscr{V}, \vec{\theta}) \dd \theta^a \dd \theta^b,
\end{align}
with the horizon located at $\mathscr{V} \rightarrow 0^{+}$. In this expression, $\kappa$ represents the surface gravity and $h_{a b}$ the induced metric on the horizon. Then, it is possible to write the Einstein vacuum equations in a neighborhood of $\mathscr{V} = 0$ for $\kappa$ and $h_{ab}$, and solve them for a series expansion of the form:
\begin{align}
    \kappa = \sum_{n = 0}^{\infty} \kappa^{(n)} ( \vec{\theta} ) \mathscr{V}^n, \qquad h_{ab} = \sum_{n =0 }^{\infty} h_{ab}^{(n)} (  \vec{\theta} ) \mathscr{V}^n.
\end{align}
One can show that the consistency of this procedure requires $\kappa^{(0)}$ to be a constant (in agreement with the zeroth law of black hole dynamics~\cite{Bardeen1973}), which corresponds to a choice of units. Apart from this condition, the Cauchy-Kovalevskaya theorem~\cite{Evans2010} ensures that there always exists a unique and analytical local solution for every analytic tensor $h_{ab}^{(0)}(\vec{\theta})$. Hence, it is possible to have a local black hole with any kind of horizon shape. Then, one can continue integrating outwards starting from one of these arbitrarily shaped horizons. However, Israel's theorem tells us that only the spherically symmetric shape leads to an asymptotically flat spacetime when we integrate outwards using the Einstein vacuum equations. We will show this explicitly in the next section, in which we restrict our attention to the case of static and axisymmetric solutions, where it is possible to provide sharper and clearer results.
 
Let us make two final remarks before discussing static and axisymmetric solutions in the next section. First, we expect that even if there is matter in the exterior region, there is an empty region around the black hole horizon without any matter. To be more precise, any kind of matter, for instance an accretion disk, would be expected to occur at radii larger than the innermost stable circular orbit. Hence, it is reasonable to assume that there exists a neighborhood of the horizon that is devoid of matter. Second, it is convenient to rewrite the line element for a static metric in a form which is closer to the line element that we will use in the static and axisymmetric case:
\begin{align}
    \dd s^2 = - e^{2U} \dd t^2 + e^{-2U} \gamma_{ij} \dd x^i \dd x^j, 
    \label{Eq:gamma_metric}
\end{align}
where $\gamma_{ij}$ is an Euclidean 3-dimensional metric. This form is related to the previous line element through the mapping $e^{2U} = \mathscr{V}^2$ and $g_{ij} = e^{-2 U } \gamma_{ij}$. The Einstein equations in vacuum for this form of the metric can be written as
\begin{align}
    &  R_{ij} \left( \gamma \right) + 2 \nabla^{(\gamma)}_{i} U \nabla^{(\gamma)}_j U = 0, \\
    & \nabla^2_{(\gamma)} U = 0,
\end{align}
where $R_{ij} \left( \gamma \right)$ represents the Ricci tensor of the metric $\gamma_{ij}$ and $\nabla_{(\gamma)}^2$ represents the Laplacian with respect to the metric $\gamma$ itself. A putative horizon in these coordinates would be located at $U \rightarrow - \infty$.   

\section{Static and axisymmetric geometries}
\label{Sec:NoHair}

Let us restrict the discussion to static and axisymmetric geometries since the simplicity of this setup allows for an exhaustive and explicit analysis. Furthermore, we will also assume that the topology of the horizon is a sphere. A discussion of nonspherical topologies will be provided in the next chapter. These metrics can be written in the Weyl form, which is well adapted for the symmetries
\begin{align}
    \dd s^2 = - e^{2 U} \dd t^2 + e^{-2 U} \left[ e^{2V} \left( \dd r^2 + \dd z^2 \right) +r^2 \dd \varphi^2 \right],
\label{Eq:Metric}
\end{align}
where $U = U(r,z)$ and $V=V(r,z)$ are functions that depend only on $r$ and $z$. The vector generating the time translations is $\bm{k} = \partial_t$ and the vector generating the axisymmetric transformations is $\bm{m} = \partial_{\varphi}$. This is a particular form of the generic static metric presented in Eq.~\eqref{Eq:gamma_metric} with the metric $\gamma_{ij}$ expressed in coordinates adapted to staticity and axisymmetry. More explicitly we have:
\begin{align}
    \gamma_{ij} \dd x^i \dd x^j = e^{2 V} \left( \dd r^2 + \dd z^2 \right) + r^2 \dd \varphi^2.
\end{align}
This form can always be achieved by means of coordinate transformations since the Lie algebra spanned by these two vectors is trivial, $[\bm{k},\bm{m}] = 0$~\cite{Carter1970}. 

In the presence of matter ($T^{\mu}_{\ \nu} \neq 0$) Einstein equations are given by:
\begin{align}
8 \pi  T^t_{\ t} & = e^{2 U - 2V} \left[ -  2 \nabla^2 U + \nabla^2 V - \frac{1}{r} \partial_r V + \left( \partial_r U \right)^2 + \left( \partial_r U \right)^2 \right], \nonumber \\
8 \pi T^{r}_{\ r} & = e^{2U - 2V}  \left[- \left( \partial_r U \right)^2 + \frac{1}{r} \partial_r V +\left( \partial_z U \right)^2  \right], \nonumber \\
8 \pi T^{r}_{\ z} & = e^{2U - 2V} \left[ -2 \partial_{r} U \partial_{z} U + \frac{1}{r} \partial_{z} V \right], \nonumber \\
8 \pi T^{z}_{\ z} & = e^{2U - 2V} \left[ \left( \partial_r U \right)^2 - \frac{1}{r} \partial_r V -\left( \partial_z U \right)^2  \right], \nonumber \\
8 \pi  T^{\varphi}_{\ \varphi} & = e^{2 U- 2 V} \left[ \nabla^2 V - \frac{1}{r} \partial_r V + \left( \partial_r U \right)^2 + \left( \partial_z U \right)^2 \right].
\label{Eq:Einsteineqs}
\end{align}
In these equations $\nabla$ represents the covariant derivative associated with a fiduciary Euclidean 3-space, expressed in cylindrical coordinates $(r,z,\varphi)$. As we will see, the appearance of this fiduciary Euclidean 3-space is a formal advantage that permits a complete control of the static axisymmetric case as opposed to the general static case. 

We can subtract the first equation from the last one and manipulate it to find:
\begin{align}
    \nabla^2 U = 4 \pi e^{2V- 2U} \left( - T^{t}_{\ t} + T^{\varphi}_{\ \varphi}  \right),
    \label{Eq:NewtonAnalogue}
\end{align}
which is similar to an effective Poisson equation in a flat 3-space, although the source depends nonlinearly on $U$ and $V$. In empty regions of spacetime we simply have  
$\nabla^2 U=0$; in this case, we will call finding $U(r,z)$ the \emph{associated Laplacian problem} (ALP).

In Newtonian gravity, given some sources, one only needs to solve Poisson equation on a flat 3D space to obtain the gravitational potential $U$. Any source is compatible with the equation, for example, two masses at rest at a fixed distance. Poisson equation by itself does not discern whether a configuration of sources that interact only gravitationally is consistent or not with staticity. However, GR is a constrained system. One cannot give arbitrary conditions but only those compatible with the constraints. In this specific case, this general issue appears in the solutions of the equation for $V$. As we will see, even if $U$ is perfectly regular it might happen that the associated $V$ exhibits irregularities. These irregularities inform us about some inconsistency that the selected source might have. 

In particular, in the absence of matter, the function $V$ must vanish on the $z$-axis, i.e., $V(0,z)=0$, to avoid the presence of conical singularities. This ensures that the length of a proper orbit of the axial Killing vector $\bm m$ that passes through a point $q$ located at a proper distance $d$ from the axis (fixed points of $\bm m$) is $2 \pi d$ at leading order (see~\cite{Mars1992} for a rigorous demonstration).

\begin{figure}
\begin{center}
\includegraphics[width=0.25 \textwidth]{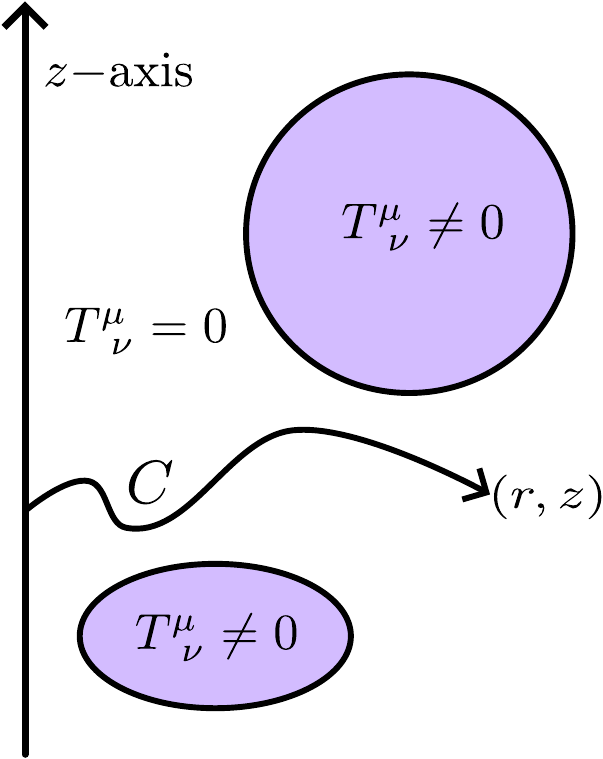}
\caption{Schematic representation of the problem. The $z$-axis is depicted as a black solid line. We illustrate the presence of matter in some compact regions that are colored. Notice that those regions would correspond to topological tori once we consider the $\varphi$ direction too. Furthermore, we depict a generic curve $C$ that begins in the $z$-axis and ends up in the point of coordinates $(r,z)$. The integral of $\boldsymbol{\omega}$ along $C$ would give us the function $V(r,z)$.}
\label{Fig:Sketch}
\end{center}
\end{figure}

To delve into this problem, we first realize that, in regions devoid of matter, the second and third equations in~\eqref{Eq:Einsteineqs} can be rewritten as
\begin{align}
    & \partial_r V = r \left[ (\partial_r U)^2 - ( \partial_z U)^2 \right], \label{Eq:DrV}\\
    & \partial_z V = 2 r \partial_r U \partial_z U, \label{Eq:DzV}\\
    & \nabla^2 U = 0, 
\end{align}
where the last equation is precisely the ALP for $U$. It is convenient now to introduce the one-form (in the $rz$ plane)
\begin{align}
    \boldsymbol{\omega} := r \left[ (\partial_r U)^2 - ( \partial_z U)^2  \right] \dd r + 2 r \partial_r U \partial_z U \dd z.
\end{align}
This form is defined wherever the coordinates are valid, even if $\nabla^2 U\neq0$. If we take the exterior derivative of this form, we find that it is proportional to the Laplacian of $U$, namely we find 
\begin{align}
    \dd \boldsymbol{\omega} = -2 r \left( \nabla^2 U \right) \left(\partial_z U \right) \dd r \wedge \dd z.
\end{align}
Thus, for every simply connected open set of the spacetime which is in vacuum, we can write $\boldsymbol \omega$ as the differential of the scalar function $V$. Thus, in vacuum open sets, once the ALP for $U$ is solved, $V$ can be determined simply by performing an integral of $\boldsymbol\omega$, which  is defined in terms of $U$ through Eqs.~\eqref{Eq:DrV}-\eqref{Eq:DzV}, as
\begin{align}
   V (r,z) = V(r_0,z_0) +  \int_{C} \boldsymbol{\omega},
\end{align}
with $C$ being any curve that ends in the point with coordinates $(r,z)$ and $V(r_0,z_0)$ represents a boundary condition for the function $V$ that is required to solve the full problem. Given the regularity condition that we have to impose on the $z$-axis, we choose $r_0=0$ so that $V(0,z_0) = 0$. The curve $C$ is any curve that starts at the $z$-axis and avoids the regions in which there is some matter content. In fact, we will be always focusing on the vacuum regions. See Fig.~\ref{Fig:Sketch} for a pictorial sketch of the setup. 

The simplest curve that one can take is precisely a curve that is orthogonal to the $z$-axis, which leads to the following expression for $V(r,z)$:
\begin{align}
    V(r,z) = \int^{r}_{0} \dd r' r' \left[ \left(\partial_{r'} U(r',z)\right)^2 - \left(\partial_z U(r',z) \right)^2 \right].
    \label{Eq:V_cuadrature}
\end{align}
In addition, asymptotic flatness requires that $V(r,z)$ vanishes also at infinity. This cannot be ensured automatically through an additional boundary condition, since the equations for $V(r,z)$ are of first order and we can only impose one boundary condition. Since we have decided that $V(0,z)=0$, the vanishing at infinity would require that $U(r,z)$ satisfies the following nonlocal constraint
\begin{align}
    \int_0^{\infty} \dd r r \left[ \left(\partial_{r} U(r,z)\right)^2 - \left(\partial_z U(r,z) \right)^2 \right]= 0. 
    \label{eq:equilibriumzeroinfty}
\end{align}
By inspecting some examples, such as a spherically symmetric star surrounded by a ring (a Saturn-like configuration), we realize that whenever the matter content linked to the ALP is not in equilibrium, this integral does not vanish. For example, for a ring located in the equatorial plane, which divides the star in two equal half-spheres, the integral straightforwardly vanishes. When the ring is not on the equatorial plane, the integral gives a nonvanishing result. We will illustrate this explicitly in Subsection~\ref{Subsec:OutsideDefs} with a black hole solution surrounded by a matter ring. 

Let us  now  establish a firm  relation between regularity at the axis $V(0,z)=0$, asymptotic flatness $V(\infty,z)=0$, the condition in Eq.~\eqref{eq:equilibriumzeroinfty}, and the notion that the effective matter content in the ALP is in equilibrium. With this aim, let us first prove that Eq.~\eqref{eq:equilibriumzeroinfty} is equivalent to
\begin{equation}
\oint_\gamma\boldsymbol\omega=0
\label{eq:equilibriumcond}   
\end{equation}
for any closed curve $\gamma$ fully contained in the vacuum region.
\begin{figure}
\begin{center}
\includegraphics[width=0.75 \textwidth]{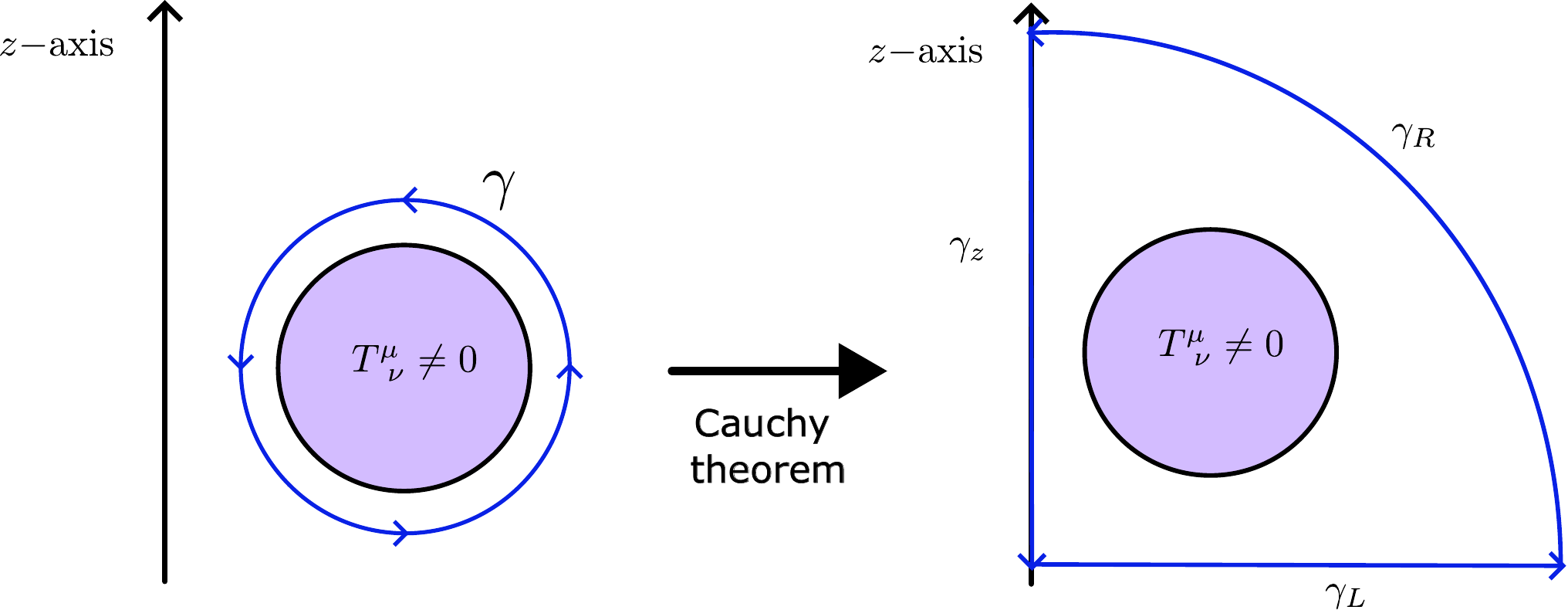}
\caption{Schematic representation of the argument to show the equivalence between Eq.~\eqref{eq:equilibriumzeroinfty} and Eq.~\eqref{eq:equilibriumcond}. The $\gamma$ contour can be deformed into the $\gamma_z+\gamma_L+\gamma_R$ due to Cauchy theorem. The integral along $\gamma_z$ is shown to vanish at the axis whereas the integral along $\gamma_L$ which is precisely the integral in Eq.~\eqref{eq:equilibriumzeroinfty}.}
\label{Fig:Sketch2}
\end{center}
\end{figure} 
A closed contour $\gamma$ as the one depicted in the left of Fig.~\ref{Fig:Sketch2} can be deformed to the contour $\gamma_z+\gamma_L+\gamma_R$ in the right without changing the value of the integral (Cauchy's theorem):
\begin{align}
\oint_\gamma\boldsymbol\omega=\int_{\gamma_L}\boldsymbol\omega+\int_{\gamma_R}\boldsymbol\omega+\int_{\gamma_z}\boldsymbol\omega.
\end{align}
In the limit $R\to\infty$, the first term in the right hand side is simply the integral in Eq.~\eqref{eq:equilibriumzeroinfty}. The second term can be obtained from a series expansion in $1/R$ (multipole expansion) of $U$, whose leading order is $1/R$. Therefore, 
\begin{align}
    \int_{\gamma_{R}} \boldsymbol{\omega} \sim-\frac{1}{R^2}+  \order{\frac{1}{R^4}},
\end{align}
which vanishes in the limit $R \rightarrow \infty$. Finally, the last integral is equal to $V(0,z_0)-V(0,R)$; Eq. \eqref{Eq:DzV} implies that $V$ is constant at the axis and therefore this term also vanishes. Consequently, $\oint_\gamma\boldsymbol\omega=0$ if and only if Eq.~\eqref{eq:equilibriumzeroinfty} holds as we wanted to show.

Now we can use Stoke's theorem to rewrite the condition $\oint_\gamma\boldsymbol\omega=0$  as a surface integral over the surface $S$ enclosed by $\gamma$ (in the $rz$-plane)
\begin{align}
   \oint_\gamma\boldsymbol\omega = \int_{S} \dd z \dd r r \partial_z U\nabla^2 U=0.
    \label{Eq:Equilibrium_Condition}
\end{align}
In the absence of nonempty regions, it vanishes identically because $\nabla^2 U=0$. However, if the closed curve $\gamma$ encloses a surface $S$ with some matter content, it is not straightforward that it vanishes since the integrand is nonvanishing. In the Newtonian limit, this condition translates into the vanishing of the projection of the net force on the $z$-axis $ F_{z}=-\partial_z U $, which is a natural equilibrium condition. For this reason, we will call Eq.~\eqref{Eq:Equilibrium_Condition} the \emph{equilibrium condition.} 

The nonvanishing of the integral in Eq.~\eqref{Eq:Equilibrium_Condition}, i.e. a violation of the equilibrium condition is translated into the fact that asymptotic flatness and regularity of the axis cannot hold simultaneously. This means that there has to be some matter either at the axis or at infinity that supports the configuration. It would be precisely that matter that is responsible for the staticity of a system which, otherwise, would not in equilibrium. In that sense, the nonvanishing of $V$ at the $z$-axis would correspond to a cosmic-string-like behavior on the axis of the energy-momentum tensor and the nonvanishing of $V$ at infinity would correspond to some matter content located at infinity.

\subsection{Putative black-hole spacetimes}
\label{Subsec:Putative}

For static metrics, an event horizon is a Killing horizon~\cite{Hawking1973} and hence the redshift function $e^{U}$ should vanish. This means that the function $U$ for which we need to solve an ALP, needs to go to $-\infty$ at any putative horizon. Thus, the horizon itself would not be covered by these coordinates. Furthermore, from the ALP point of view, the regions at which the function $U$ goes to $-\infty$ must correspond to those in which there is some kind of singular ($\delta$-like distributional) source. However, we highlight that, generically, a regular horizon would be in vacuum (the actual energy-momentum tensor would vanish there), and these sources are fictitious sources that arise as an artifact of the ALP. Here, we focus exclusively on smooth horizons, specifically at least $\mathcal{C}^2$, to ensure the Einstein equations remain well-defined.

We focus on asymptotically flat spacetimes with matter (either corresponding to real matter or to distributional matter that represents the location of a horizon) restricted to live in compact regions of the spacetime. In such setup, the potential sources that might lead to a blow-up of $U$ are necessarily objects of codimension 3 or 2 for the ALP, i.e., point-like contributions or rod-like contributions. Other extended objects of codimension 1 or 0 cannot generate divergences in $U$ if they are located in a compact region. 

Given the axisymmetric setup that we are considering, this translates into the fact that all the potential black hole sources need to be point-like and rod-like distributions located in the symmetry axis and any number of infinitely thin rings around the symmetry axis. Any solution of the ALP involving a divergence in $U$ associated with these elements is in principle a potential black hole candidate. At this stage, we will call them ``putative black holes'' since it is not clear whether these divergences can be identified with regular or singular horizons until further analysis. 

As the ALP is a linear equation, we can always separate the effects of the different putative black holes present in the system. Thus, let us discuss first the elementary cases in which there is just one of these singular behaviors, namely a point-like distribution at the axis, a ring-like distribution around the symmetry axis, or a constant density rod-like distribution sitting on the axis, and there is no additional matter content anywhere. This is tantamount to considering the putative black holes to be isolated.

\paragraph*{\textbf{Curzon's putative black holes.}}

Let us first analyze a source term for the ALP which is a point-like distribution. The Curzon metric (and its higher multipole generalizations) correspond to a source term for the ALP which is precisely a point-like distribution. This can be a simple delta source (it can also be a more singular multipole profile) located at a point on the axis. 

For the purpose of describing these metrics it is more convenient to change coordinates in the 3D space. Instead of using coordinates $(r,z,\varphi)$, we can use spherical coordinates $(R,\theta,\varphi)$. Thus we have $R = \sqrt{r^2 + z^2}$ and 
\begin{align}
    & r = R \sin \theta, \label{Eq:rSpherical}\\
    & z = R \cos \theta \label{Eq:zSpherical}.
\end{align}
Then, in these coordinates the metric in Eq.~\eqref{Eq:Metric} becomes:
\begin{align}
    ds^2 = - e^{2 U} dt^2 +e^{-2U} \left[ e^{2V} \left( dR^2 + R^2 d \theta ^2 \right) + R^2 \sin^2 \theta d \varphi^2 \right].
\label{Eq:Line-Element-v2}
\end{align}
Taking a monopolar solution for the ALP located at the origin, i.e., a source of the form $\rho(\boldsymbol{x}) = M\delta(\boldsymbol{x})$, yields
\begin{align}
    & U(R) = - \frac{M}{R}, \\
    & V(R, \theta) = - \frac{M^2 \sin^2 \theta}{R^2}, 
\end{align}
which is the so-called Curzon metric~\cite{Curzon1925}. We focus on the positive mass case here, since the negative mass case represents a naked singularity, with no horizon, and thus it is not of direct interest for us. A more detailed analysis of the properties of the Curzon metric, including the negative mass case, is presented in Appendix~\ref{App:Curzon}. 

\begin{figure}[H]
\begin{center}
\includegraphics[width=0.45 \textwidth]{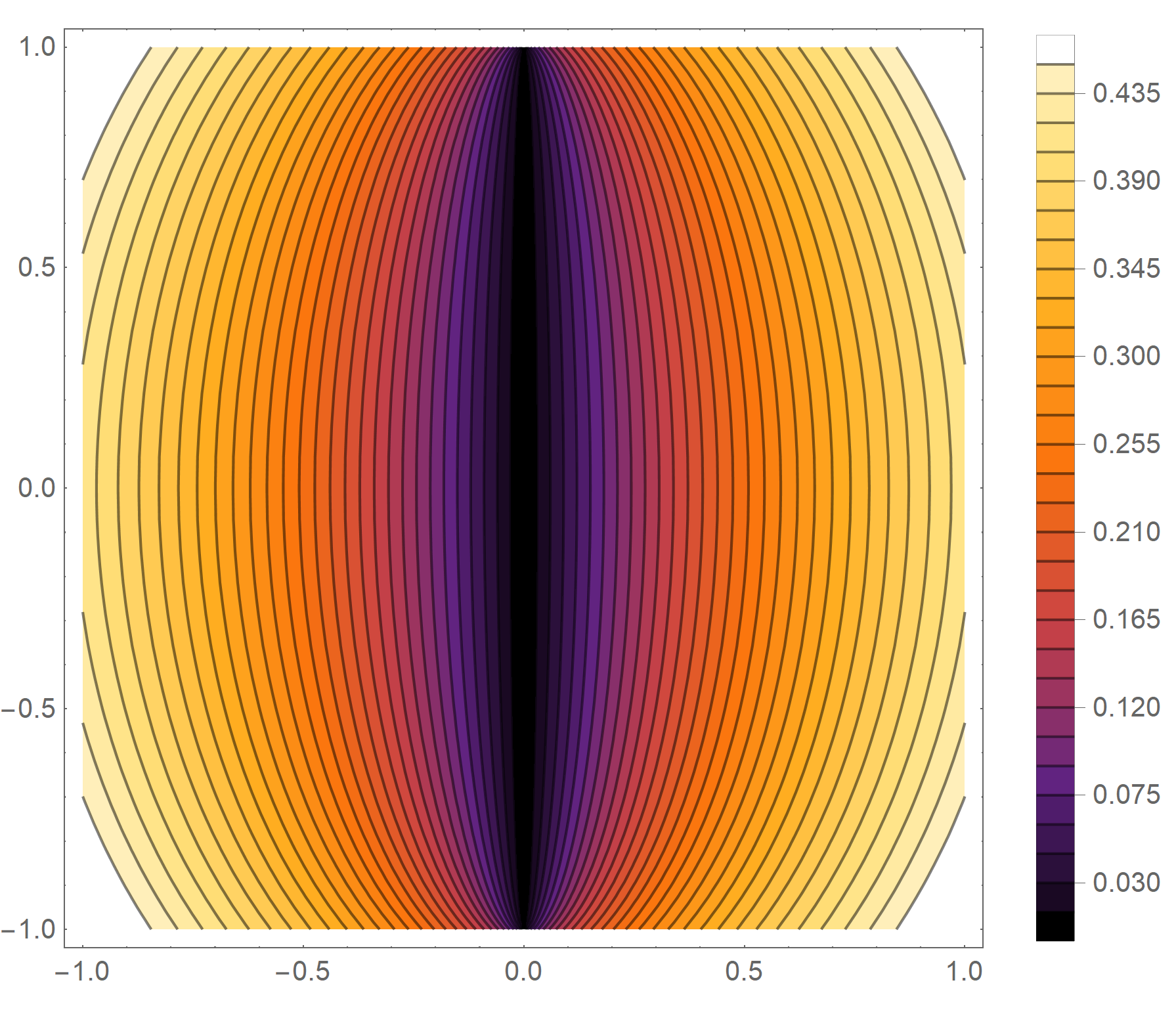}
\includegraphics[width=0.45 \textwidth]{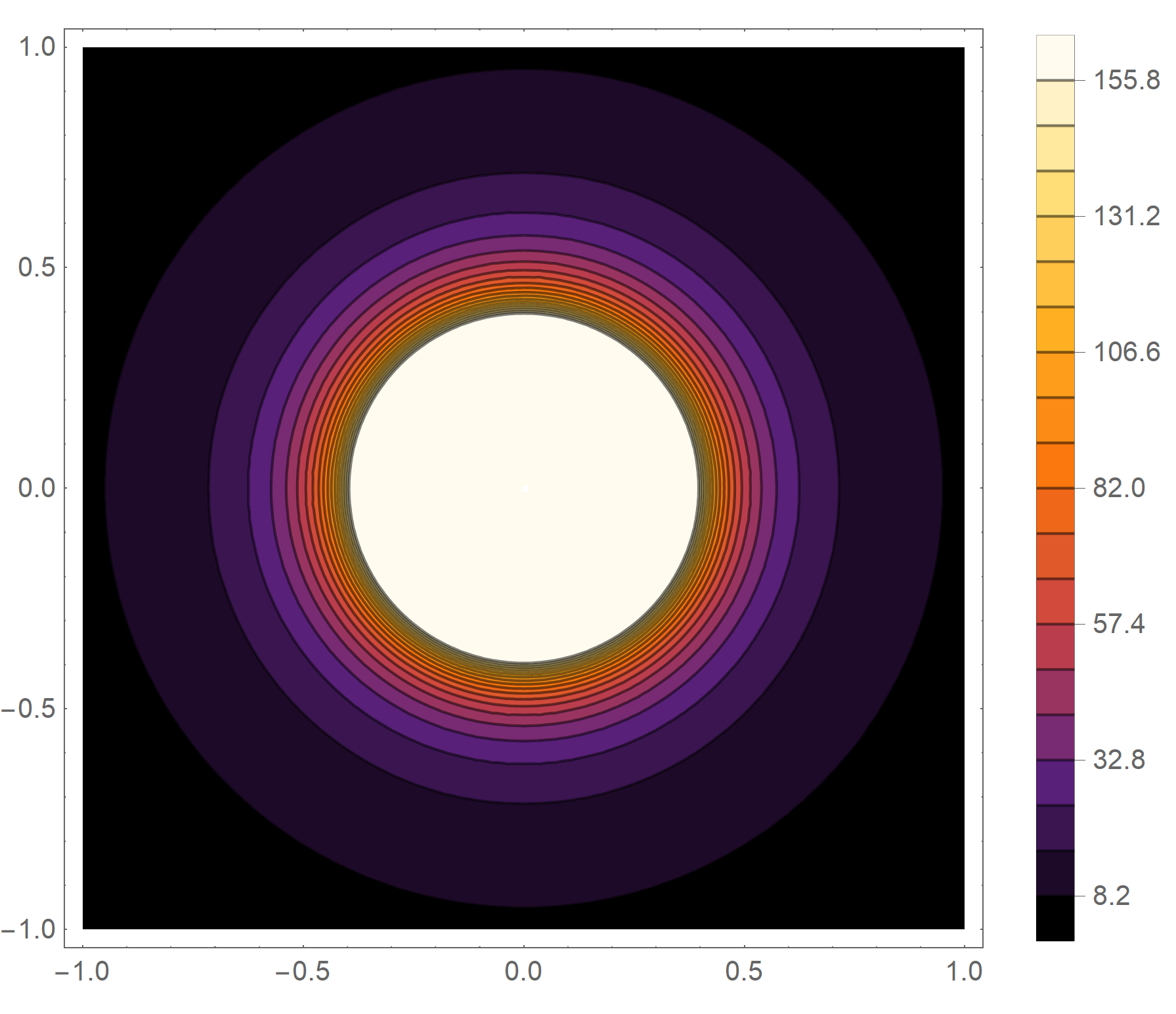}
\includegraphics[width=0.45 \textwidth]{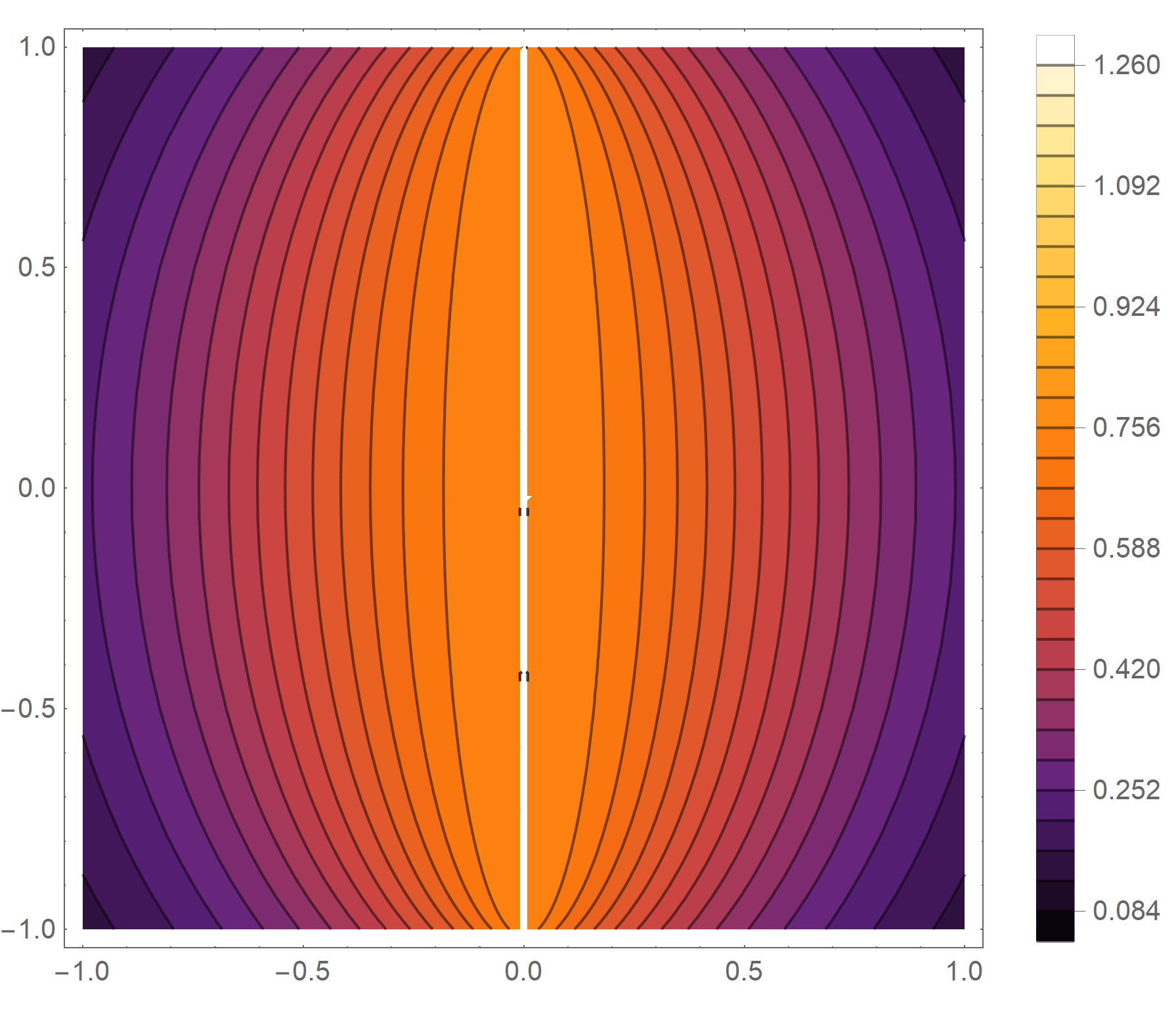}
\includegraphics[width=0.45 \textwidth]{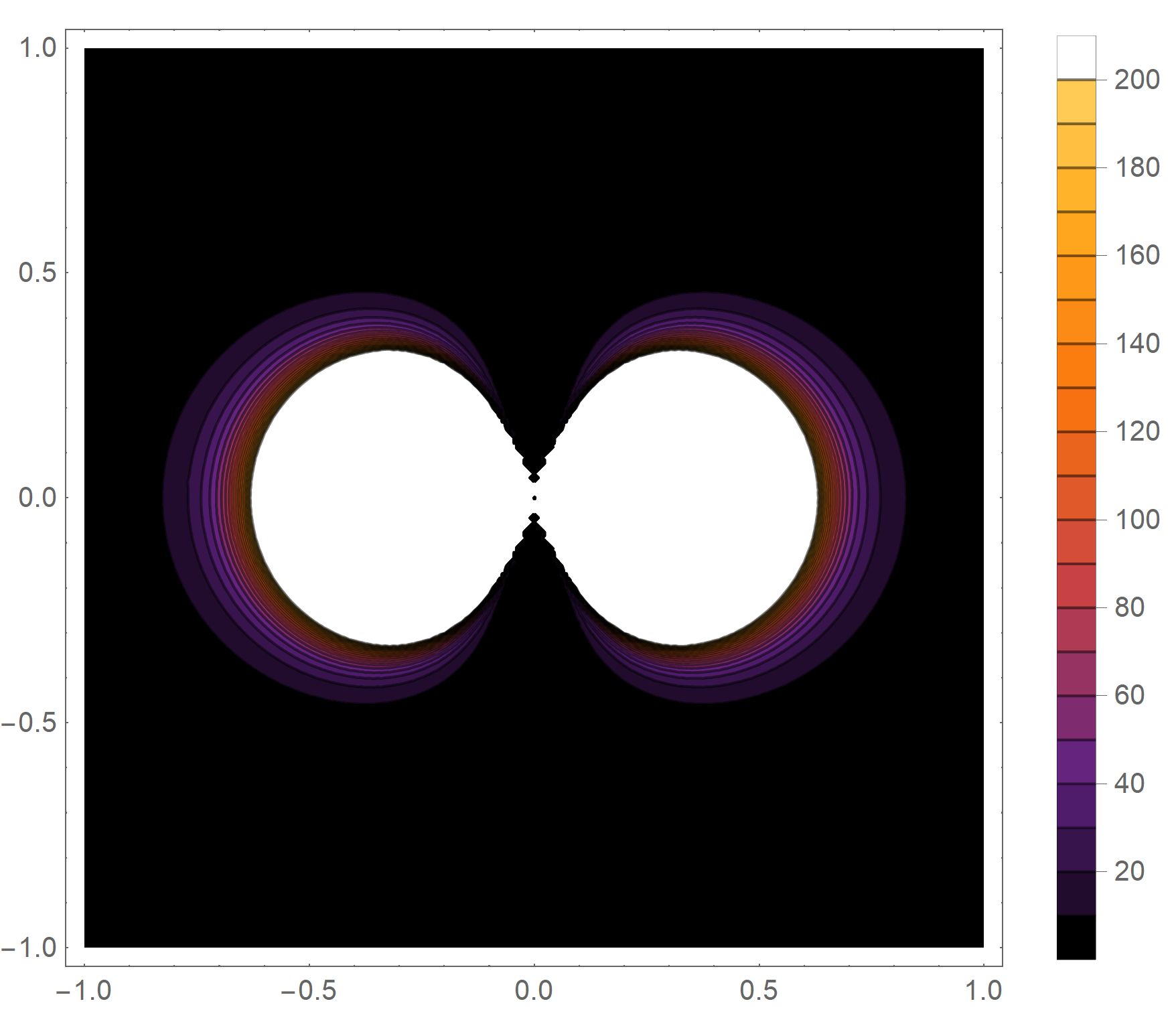}
\caption{The upper panel shows constant redshift surfaces on a plane of fixed $\varphi$ for the Schwarzschild solution (left) and the Curzon solution (right). The lower panel displays the Kretschmann scalar for the same configurations: Schwarzschild on the left and Curzon on the right. We have taken $M = 1$ for both of them.}
\label{Fig:Curvatures_Redshift}
\end{center}
\end{figure}

Notice that, although the redshift function does not depend on $\theta$, the whole metric is not spherically symmetric due to the functional dependence of $V$ on the angle $\theta$. This percolates to the Kretschmann scalar which also depends on $\theta$:
\begin{align}
    \mathcal{K} = e^{2 M \left[M \sin ^2(\theta )-2 R\right]/{R^2}}
    \times {\mathcal{R}^{-4}(R,\theta;M)},
\end{align}
where $\mathcal{R}^4(R,\theta;M)$ is a function constructed from $M,R$ and $ \theta$, with dimensions of length to the fourth power:
\begin{align}
    \mathcal{R}^4(R,\theta;M)= \frac{R^{10}}{8 M^2 \left[2 M^3 \sin ^2(\theta ) (M-3 R)-3 R^2 \left[M^2 (\cos (2 \theta )-3)+4 M R-2 R^2\right] \right]}.
\label{Eq:MathcalR}
\end{align}
It clearly displays a directional behavior: for any value of $\theta \not \in \{0, \pi\}$, the Kretschmann scalar diverges as we approach $R \rightarrow 0$ since it has the behavior
\begin{equation}
    \mathcal{K}_{\theta \not\in\{0, \pi\}} \sim e^{N M^2/R^2 + \cdots} \times \mathcal{R}^ {-4}(R,\theta;M),
\end{equation}
where $N$ is a positive number and we are omitting the subdominant term in the argument of the exponential. For $\theta \in\{0, \pi\}$, the Kretschmann has a different behavior and actually vanishes as we approach $R \rightarrow 0$ since in that case it exhibits a behavior 
\begin{equation}
    \mathcal{K}_{\theta \in\{0, \pi\}} \sim e^{-4 M/R} \times \mathcal{R}^{-4}(R,\theta;M).
\end{equation}
In both cases, the exponential behavior dominates either to make the Kretschmann scalar blow up or vanish. Figure~\ref{Fig:Curvatures_Redshift} displays the equipotential surfaces and the surfaces of constant Kretschmann scalar, alongside their Schwarzschild counterparts. 
Another way of illustrating the nonspherically symmetric nature of the Curzon metric is by considering the induced metric on surfaces of constant redshift for the spatial sector of the metric:
\begin{align}
    \dd s_{U=\text{constant}}^2 = e^{-2U} \frac{M^2}{U^2}\left[ e^{-2 U^2 \sin^2 \theta} \dd \theta^2 + \sin^2 \theta \dd \varphi^2 \right],
\end{align}
which is clearly not spherically symmetric since its scalar curvature explicitly depends on~$\theta$.

\begin{figure}[H]
\begin{center}
\includegraphics[width=0.75 \textwidth]{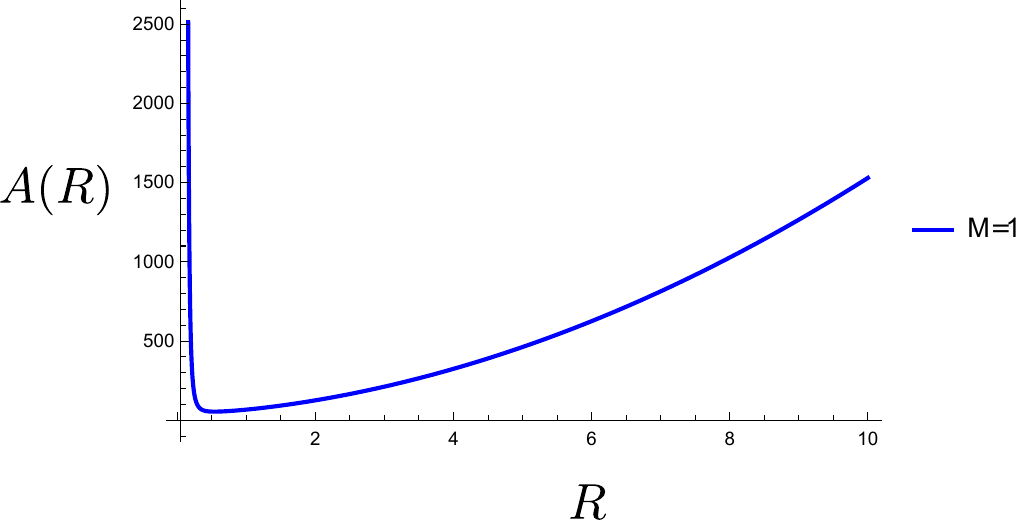}
\caption{We are depicting here the area function as a function of $R$ for $M=1$. The area decreases as we decrease $R$ until we reach a minimum $R \simeq 0.538905$ and it increases as we decrease the value of $R$ from that value on, see Appendix~\ref{App:Curzon} for further details.}
\label{Fig:AreaSingle}
\end{center}
\end{figure}

The Curzon metric exhibits another noteworthy feature. The area of the equipotential surfaces (constant $R$ surfaces) goes to infinity as $R \to 0$, see Fig.~\ref{Fig:AreaSingle} for a representation of the area of constant $t$ and $R$ surfaces as a function of the radius $R$. Its explicit analytic expression can be found in Appendix~\ref{App:Curzon} although it is not specially illuminating. This automatically excludes them from Israel's theorem scope, due to the last assumption of the theorem. Furthermore, it also shows that the putative horizon at $R \rightarrow 0$ is not regular in a different way. In fact, the directional behavior of the Kretschmann scalar, combined with the infinite extent of the constant $R$ region, suggests that the object, which appears point-like in these coordinates, actually represents an extended configuration.
In fact, the behavior of the geodesics in the Curzon metric was studied in~\cite{Morgan1973,Scott1985,Scott1985b}. Null geodesics approach the coordinate singularity $R=0$ in such a way that they enter parallel to the $z$-axis. Given that the surfaces $R = \text{constant}$ have an area that increases as we approach $R=0$, the surface $R=0$ is not really a point, but a whole plane of infinite extension. The geometry can be regarded as the end state of a very nonspherical anisotropic collapse, which tends to compress all the matter into an infinitely diluted plane such that the matter content is moved away to infinity where a ring singularity is formed. Given that the plane is infinitely diluted, the curvature that is found by entering through the disk vanishes in an infinitely differentiable way, through the archetypal $e^{-M/R}$ behavior of nonanalytic $C^{\infty}$ functions. In fact, this is what generically null geodesics find, since they enter towards the origin through the $z$-axis. 

We can attach a flat spacetime to this solution, corresponding to what light rays crossing $R= 0$ would encounter, which we recall represents a whole surface. If we interpret this as the final stage of a collapse scenario, this would be the only extension needed, similar to how black hole interiors are extended. However, for an eternal configuration, we must also extend the solution backward in time, continuing outgoing geodesics into the past and attaching another flat spacetime region. Additional extensions are possible depending on the behavior of other geodesics. Moreover, since the metric is not analytic, the extension is not unique, there exist infinitely many $C^{\infty}$ extensions of the metric across $R=0$.

We can also consider the behavior of other singular pure multipole sources which are similar to the pure monopole behavior. The functions $U$ and $V$ in the metric for an $\ell$-order multipole configuration are \cite{Stephani2003}
\begin{align}
    & U^{(\ell)} = - \frac{M^{(\ell)}}{R^{\ell + 1}} P_{\ell} \left( \cos \theta \right), \nonumber\\ 
    & V^{(\ell)} = - \frac{(\ell + 1 ) \left[ M^{( \ell )} \right]^2}{2 R^{2 \ell + 2}} \left[ P_{\ell}^2 \left( \cos \theta \right) - P_{\ell+1}^2 \left( \cos \theta \right) \right].\label{Eq:U_Multipoles}
\end{align}
Here, $P_{\ell} \left( \cos \theta \right)$ represents the Legendre polynomial of degree $\ell$. 

The Kretschmann scalar displays a behavior which is qualitatively similar to the monopole one as we approach $R\rightarrow 0$, although with a more convoluted directional structure. Instead of having only two directions along which it vanishes, apparently there are a set of directions $\theta \in \{ \theta_i \}_{i \in \mathcal{J}_{\ell}}$, with $\mathcal{J}_{\ell}$ a finite set for every $\ell$ along which the Kretschmann goes to zero. However, it is not easy to determine them in full generality and for every value of $\ell$. For any $\theta \neq \theta_i$, the leading behavior of the Kretschmann is 
\begin{align}
    \mathcal{K} \sim e^{N \frac{ \left[ M^{(\ell)} \right]^2}{R^{2 \ell + 2 }} + ... }\times \mathcal{R}^{-4}_{\ell} \left( R, \theta ; M^{(\ell)}\right), 
    \label{Eq:Kretschmann_Divergences}
\end{align}
where again $N$ is a given number and $\mathcal{R}^{-4}_{\ell} \left( R, \theta ; M^{(\ell)}\right)$ is a different polynomial that depends on $\ell$. For our purposes, it is enough to realize that for all the values of $\ell$, there exist directions for which the Kretschmann scalar becomes divergent as $R \rightarrow 0$. 

Exactly the same logic applies to the case in which we consider the superposition of different multipole configurations or even superpositions of the multipole configurations and the Schwarzschild solution, since the Kretschmann will still contain these divergences that we have just discussed. 

All in all, this leads to the conclusion that the Curzon metric and its higher multipole versions do not constitute proper black hole solutions since curvature singularities are found when approaching the vanishing redshift along some directions. They seem to represent another kind of objects. For a further discussion of this issue, specially focused in the Curzon (monopolar) metric, see Appendix~\ref{App:Curzon}.

\paragraph*{\textbf{Ring singularities as putative black holes.}}

Consider the solution of the ALP  associated with an infinitely thin ring of matter located at $z=0$ and $r=r_{\rm ring}$, i.e., a distributional density  of the form $\rho(\boldsymbol{x}) = M_{\text{ring}}  \delta (z) \delta( r -r_{\rm ring} )/2 \pi$. This solution is often called the Bach-Weyl ring and is reviewed in detail in Appendix~\ref{App:Rings}. Axisymmetry implies that the ring has constant linear density along its circumference. There are several pathological features of the geometry associated with this solution. 

First of all, as we approach the ring, the equipotential surfaces are toroids with smaller and smaller radii.

This implies that they do not satisfy the hypotheses of Israel's theorem. It is easy to see that the area of these equipotential toroids goes to zero as they approach the source, since the vector generating axisymmetry $\boldsymbol{m}$ degenerates. The norm of $\boldsymbol{m}$ is given by $r e^{U}$, and $r$ remains bounded whereas $e^{U}$ vanishes as $r \rightarrow r_{\rm ring}$. Thus, the area of the toroids also vanishes since it is proportional to $2 \pi$ times the norm of $\boldsymbol{m}$ times the integral in the additional direction. This, by itself, tells us that they cannot approach a proper black hole horizon of finite well-defined area, as it was noticed by Geroch and Hartle~\cite{Geroch1982}. 
\begin{figure}
\begin{center}
\includegraphics[width=0.35 \textwidth]{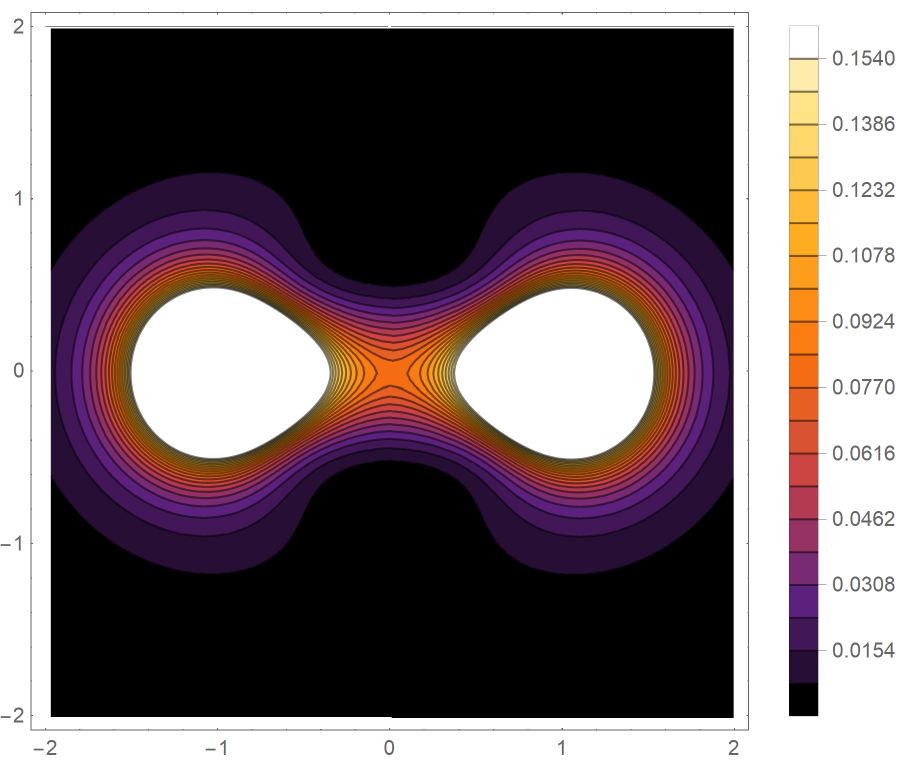}
\includegraphics[width=0.45 \textwidth]{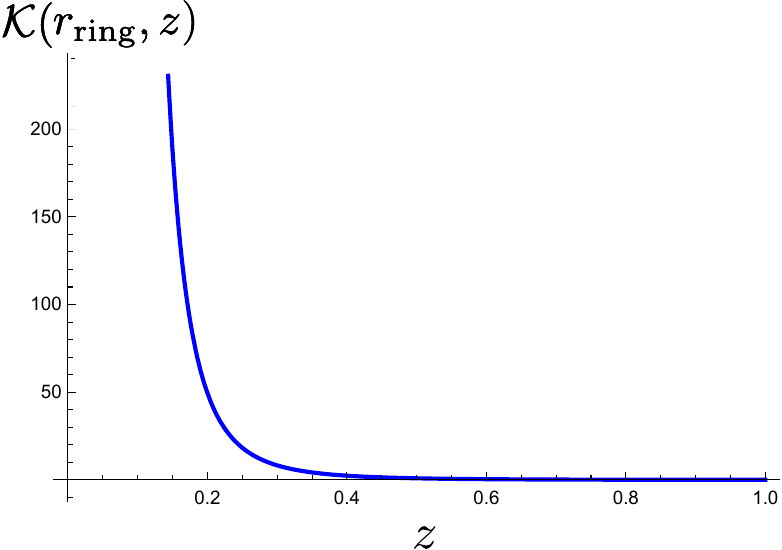}
\caption{The left panel shows the Kretschmann scalar on a constant-$\varphi$ plane for parameter values \mbox{$M_{\text{ring}} = 0.1$} and \mbox{$r_{\text{ring}} = 1$}, with the vertical axis corresponding to the $z$-coordinate and the horizontal axis to the transverse direction. The right panel plots the Kretschmann scalar as a function of $z$ along a vertical line orthogonal to the ring, clearly illustrating its divergence as $z \rightarrow 0$.}
\label{Fig:RingKretschmann}
\end{center}
\end{figure} 

An additional pathology of these geometries that we have analyzed is the behavior of the Kretschmann scalar. The Kretschmann scalar diverges at the ring itself as it is illustrated in Fig.~\ref{Fig:RingKretschmann}, leading to a singular geometry. It is actually a naked singularity, similar to the Schwarzschild solution with negative mass. Thus, even though they are excluded from Israel's analysis by the assumption on the topology of constant redshift surfaces, we see that they cannot even represent toroidal black holes with regular event horizons. This behavior is generic of any singularity that is not located at the $z$-axis. Thus, we now focus on singularities located exclusively in the $z$-axis.
 
\paragraph*{\textbf{Schwarzschild black hole.}}

The remaining type of source in the ALP to which we may associate with an isolated black hole is that of a constant density rod located in the symmetry axis. Taking it to lie between $z = - M$ and $ z = + M$, we have that the functions $U$ and $V$ for this source read:
\begin{align}
    & U_S (r,z) = \frac{1}{2} \log \left( \frac{R_{+} + R_{-} - 2M}{R_{+} + R_{-} + 2M} \right),
    & V_S(r,z) = \frac{1}{2} \log \left( \frac{(R_{+} + R_{-})^2 - 4 M^2 }{4 R_{+} R_{-}} \right), \label{Eq:Schwarzschild}
\end{align}
where we have introduced the following shortcut in the notation
\begin{align*}
    R_{\pm}^2 = r^2 + (z \pm M)^2.
\end{align*}
As is well known, this is precisely the Schwarzschild metric written in Weyl coordinates~\cite{Stephani2003}. The leading order behavior of $U$ and $V$ as $r \rightarrow 0$ for \mbox{$z \in (-M,M)$} is $\log (r)$. Thus, we automatically have the behavior required for a putative black hole, namely,
\begin{align}
    \lim_{\substack{r \to 0 \\ z \in (-M,M)}} U(r,z)= -\infty  .
\end{align}
It has two crucial properties that might appear as surprising from the point of view of the Weyl coordinates, whereas they are well-known features of the Schwarzschild geometry.
On the one hand, the area of the equipotential surfaces, which are ellipsoids in the $(r,z,\varphi)$ coordinates, tends to a finite constant $A= 16 \pi M^2$ when these surfaces approach the rod. On the other hand, the Kretschmann scalar is constant in the equipotential surfaces and also goes to a finite value when the rod is approached. Therefore, the rod itself represents a regular geometrical boundary, i.e., a proper horizon through which one can extend the geometry. We can see in this way that the source of the ALP problem does not correspond to any kind of matter content from the GR point of view in this case, as the Schwarzschild solution can be extended beyond the horizon as a vacuum solution~\cite{Hawking1973}. In other words, the constant density rod is an artifact of the ALP since the Schwarzschild solution has vanishing Einstein tensor everywhere.  

\subsection{Deforming a black hole from the inside}
\label{Subsec:InsideDefs}

The previous discussion has shown that from the initial list of simple putative black holes, only the Schwarzschild geometry represents a proper black hole. By deforming a Schwarzschild black hole from the inside we mean here to eventually perturb the Schwarzschild source term in the ALP problem, maintaining the condition of an empty exterior. Starting from the Schwarzschild solution we can try to see whether there are additional sources apart from the uniform rod leading to a regular horizon. It is clear that we cannot add a distributional point-like structure inside a rod. This will result in a Curzon-like contribution to the geometry which leads to irregular horizons. Thus, the only remaining possibility is to modify the linear density of the rod to make it inhomogeneous, i.e., to consider a nonconstant function $\lambda(z)$ for the linear density of the rod. 

This situation is carefully analyzed in Appendix B of~\cite{Geroch1982}. Any deformation from the inside of the Schwarzschild spacetime needs to be encoded in the presence of inhomogeneities along the initially homogeneous rod, i.e., a function $\lambda(z)$. It translates into a behavior \mbox{$ U = f(z) \log r + \text{o} \left( \log r \right)$} as we approach the horizon $r \to 0$ with a fixed $z\in (-M,M)$. Here $f(z)=2\lambda(z)$. This is easy to understand, as when one approaches a segment closely enough, the local linear density dictates the leading behavior of the potential.

When a black hole has a singular horizon, a first hint comes from a calculation of its surface gravity
\begin{align}
    \kappa^2 = \lim_{r\rightarrow 0} e^{4U - 2V} \left[ \left( \partial_r U \right)^2 + \left( \partial_z U \right)^2 \right].
\end{align}
It needs to be bounded to have a regular geometry. Now, boundedness of the surface gravity requires that the density of the rod is homogeneous. Any kind of inhomogeneous behavior leads to a divergent surface gravity. To show it, we consider the leading behavior for $U = f(z) \log r $, as $r \rightarrow 0$ for $z \in (-M,M)$, so that the radial equation for $V$ becomes:
\begin{align}
    \partial_r V = r \left[ \left(  \frac{f(z)}{r} \right)^2 - \left( f'(z) \log r\right)^2 \right].
\end{align}
Since the second term is of the form $\sim r \log r$, it gives a subdominant contribution as we approach $r \rightarrow 0$ and the leading order behavior of $V$ is similar to that of $U$, namely $V = f^2(z) \log r + \text{subdominant}$. If we plug this behavior in the surface gravity and the Kretschmann scalar we find:
\begin{align}
     \kappa = & f(z) r^{f(z) - 1} + \text{finite terms}, \label{Eq:Inhomogeneous_Kappa}\\
    \mathcal{K} = &  \frac{12 (f(z)-1)^2 f(z)^2}{r^4} + \nonumber \\
    & \frac{8 f'(z)^2\left[3 f(z)^2 \log ^2(r) -3 f(z) \log ^2(r) +\log ^2(r) +2 \log (r) +2 \right]}{r^2} \nonumber \\
    & + \text{finite terms} \label{Eq:Inhomogeneous_Kretschmann}. 
\end{align}
We see that if $f(z)$ is different from a constant function with a value either $0$ or $1$ we find a singular geometry at $r\to 0$. A possible interpretation is the following. The constant density rod in the ALP represents an auxiliary source and when analyzing the black hole geometry we realize that it does not map to any kind of matter content (the Einstein tensor identically vanishes on the horizon). The inhomogeneous profiles do not correspond to such auxiliary source and do indeed represent a real matter content that manifests in a nonvanishing Einstein tensor. Given that it is static matter on top of an event horizon, it gives rise to a singular behavior which is what we see in Eqs.~\eqref{Eq:Inhomogeneous_Kappa} and~\eqref{Eq:Inhomogeneous_Kretschmann}. 

Therefore, we conclude that there is no other distributional source in the ALP aside from the homogeneous rod leading to a proper black hole geometry in vacuum, i.e., with a regular horizon. In fact, $z$-dependent profiles for the density actually lead to naked singularities. 

\subsection{Deforming a black hole from the outside}
\label{Subsec:OutsideDefs}

Let us consider now more general black holes that are locally in vacuum but do not necessarily extend the vacuum region all the way up to infinity. Geroch and Hartle~\cite{Geroch1982}, showed that every black hole solution with an event horizon whose topology is a two-sphere is such that the function $U(r,z)$ takes the same values at both extremes of the horizon, i.e., $U(0,M) = U(0,-M)$. Furthermore, they also showed that $U(r,z)$ should have the same distributional character as its Schwarzschild counterpart $U_S$ on the event horizon, namely that of a infinitely thin homogeneous rod located on the $z$ axis between $-M$ and $+M$. Equivalently, this means that every local black hole solution can be represented as 
\begin{align}
    U(r,z) = U_S(r,z) + \Delta U (r,z),
\end{align}
with $\Delta U$ any solution to Laplace equation that is analytic and regular, even at the horizon. Such a solution represents a deformed black hole due to an external gravitational field, with the function $V$ being obtained by direct integration. These are all the local black holes in vacuum.\footnote{Although they were found previously~\cite{Israel1964,Lawrence1966,Israel1973}, an exhaustive and systematic analysis was presented by Geroch and Hartle~\cite{Geroch1982}, including the case of black hole horizons that have toroidal topology.} However, as we discussed at the beginning of the section, finding a local solution for $U$ is not the whole story, since we still need to ensure that $V$ has a nonsingular behavior everywhere. 

The horizon in these coordinates is located in the $z$-axis, between $-M$ and $M$. Hence, the $z$-axis is disconnected and made of two semi-axis pieces, $  (-\infty,-M) \cup (M,\infty) $. Thus, even if we set $V(0,z)$ to vanish on one of the semi-axis, it is not immediately clear that it will also vanish on the other. The condition $U(0,M) = U(0,-M)$ guarantees that if $V(0,z)$ vanishes on, for example, the positive semi-axis, it will also vanish on the negative one. Let us see why. Since we expressed $U = U_S + \Delta U$, we can similarly decompose $V = V_S + \Delta V$ where both $U_S$ and $V_S$ satisfy the vacuum equations. However, $\Delta V$ must satisfy the following equation:
\begin{align}
    \partial_z \Delta V = 2 r \left( \partial_r U_S \partial_z \Delta U + \partial_r \Delta U \partial_z U_S + \partial_r \Delta U \partial_z \Delta U \right). 
\end{align}
Given that $U_S \sim \log r$ as $r \rightarrow 0$, performing an integration parallel to the $z$ axis near $r = 0$, leads to the conclusion that
\begin{align}
    \Delta V(0,M) - \Delta V(0,-M) \sim \Delta U(0,M) - \Delta U(0,-M).
\end{align}
Hence, we can ensure that $V$ vanishes on both pieces of the axis as long as $U(0,M) = U(0,-M)$, where this expression should be understood as a limit. This ensures that we can make the function $V$ regular on the whole axis taking it as the integral in Eq.~\eqref{Eq:V_cuadrature}. 

The final step is to check asymptotic flatness, which requires that the system is in equilibrium. If the ALP of the black hole surrounded by some matter content is in equilibrium, it leads to a solution that is regular and can be extended as a vacuum solution all the way up to infinity, with the matter located only in compact regions.

\begin{figure}
\begin{center}
\includegraphics[width=0.25 \textwidth]{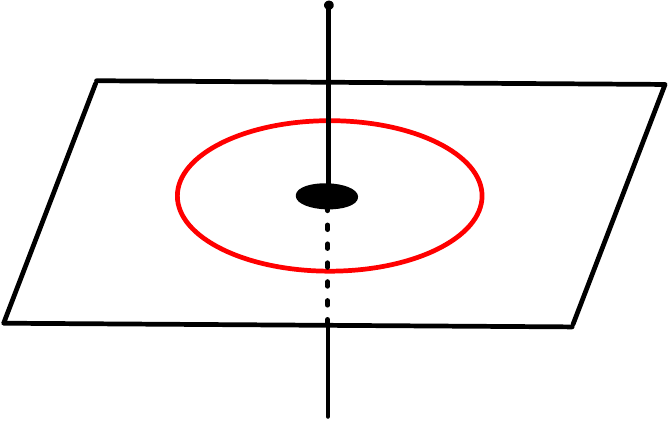}
\includegraphics[width=0.25 \textwidth]{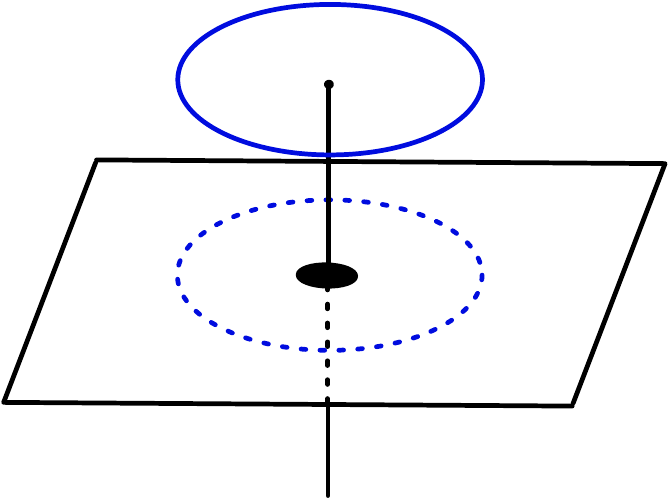}
\caption{The shadowed black region represents the black hole, and the blue ring can be located either in the same plane as the black hole, or in a plane that is not the one of the black hole. In the former case we find that the integral vanishes, whereas in the latter case we find a finite result.}
\label{Fig:Saturns}
\end{center}
\end{figure}

An illustrative example that we can explicitly construct is precisely a black hole deformed by the presence of an external ring of matter. For details about how the functions $U_{\text{ring}}$ and $V_{\text{ring}}$ are obtained for the so-called Bach-Weyl ring, see Appendix~\ref{App:Rings}:
\begin{align}
    U = U_S + U_{\text{ring}}. 
\end{align}
The ring can have a finite thickness, but for computational simplicity, it is convenient to consider an infinitely thin ring. In this case, we have to ignore the singular behavior at the ring itself, as it is an artifact of the approximation. If the ring lies in a plane that symmetrically intersects the black hole, the resulting geometry remains perfectly regular, and the integral~\eqref{Eq:Equilibrium_Condition} vanishes since the system is in equilibrium along the $z$-direction, i.e., $\partial_zU=0$. The ring does not contract due to its internal pressures, which can be directly derived from Einstein equations. However, if the ring is positioned on a different plane, the equilibrium condition \eqref{Eq:Equilibrium_Condition} is not satisfied, making it impossible to satisfy both regularity at the axis and asymptotic flatness simultaneously.\footnote{Although now the ALP contains an homogeneous rod at the axis and is not pure vacuum, the same argument still holds if we slightly deform the contour $\gamma_z$ in Fig.~\ref{Fig:Sketch2} to avoid the distributional matter content representing the horizon.} See Fig.~\ref{Fig:Saturns} for a pictorial representation.

\begin{figure}
\begin{center}
\includegraphics[width=0.75 \textwidth]{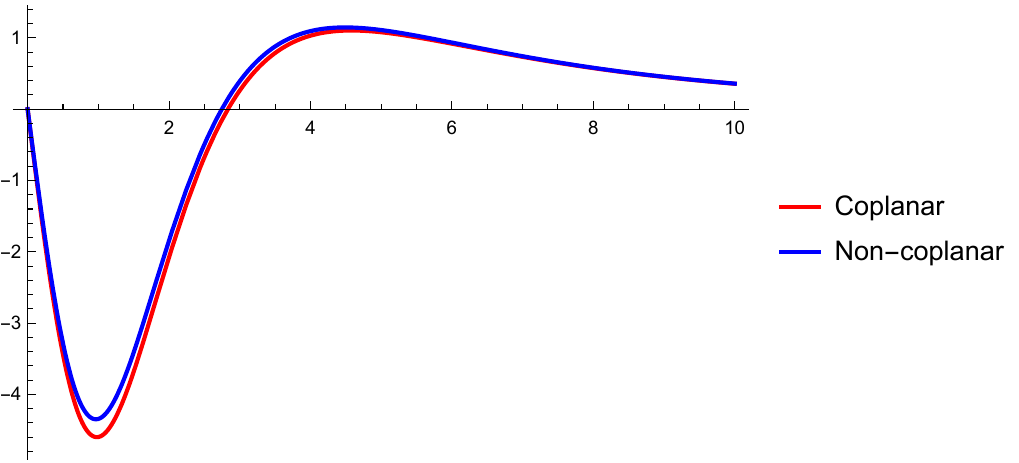}
\caption{Here we represent the integrand of Eq.~\eqref{Eq:Equilibrium_Condition} as a function of $r$ for some specific choices of the parameters, namely we are depicting the value for $z = 3$ and we are taking the radius and mass of the ring to be $M_{\text{ring}} = 1, r_{\text{ring}}= 1$ and the black hole mass $M= 1 $. The red curve corresponds to the integrand when the ring and the black hole are on the same plane, whereas the blue one corresponds to them lying on different planes, namely five units displaced in the z-axis. \emph{A priori}, there is no apparent symmetry reason to expect a cancellation of the integral.}
\label{Fig:Saturns_Integrands}
\end{center}
\end{figure}

We have not been able to perform the integral~\eqref{Eq:Equilibrium_Condition} analytically in either case, and it would be quite unlikely to conclude that the integral vanishes simply by inspecting the integrand if we did not have the equilibrium interpretation, see Fig.~\ref{Fig:Saturns_Integrands}. 

More in general, we can take any kind of matter content in equilibrium surrounding the black hole and consider a deformation which is given directly by solving the ALP with the suitable boundary conditions that correspond to the matter content included. In fact, the boundary conditions associated with the matter content can be replaced with an effective density $\rho_{\text{eff}} (\boldsymbol{x})$ that we can integrate with the flat space Green function $G(\boldsymbol{x},\boldsymbol{x'}) = 1/4 \pi \abs{\boldsymbol{x} - \boldsymbol{x'}}$, as $U(r,z)$ obeys the Laplace equation in vacuum regions, to find the contribution to the $U$ function given by the matter content:
\begin{align}
    \Delta U (\boldsymbol{x})= \int \dd ^3 x' \frac{\rho_{\text{eff}} ( \boldsymbol{x'})}{\abs{\boldsymbol{x}-\boldsymbol{x'}}}.
\end{align}
We find that as long as there is no density in the horizon, we are avoiding the coincidence limit in the Green function and no singular behavior arises.

\section{Black holes in astrophysical environments}
\label{Sec:NoHair_External}

After an extensive discussion of the potential deformations of black holes in the restricted case of static and axisymmetric configurations, we are now in a position to formulate our no-hair result for black holes in astrophysical environments. 

To the best of our knowledge, the only previous results in the literature that go in this direction are those of G\"urlebeck~\cite{Gurlebeck2012,Gurlebeck2015}, which also apply in the static and axisymmetric case. G\"urlebeck showed that the Weyl multipoles of a spacetime can be expressed as a sum of different contributions. These contributions take the form of integrals that are localized in compact regions around coordinate singularities (such as horizons or defects) and regions containing some matter content. Furthermore, G\"urlebeck showed that the contribution from black-hole-like singularities is always the same as the one from Schwarzschild isolated black holes. Here we are going to present the no-hair theorem in a different way which constrains the form of the metric itself, and is closer to the classic no-hair results.

\noindent
\textbf{Theorem:} Given a regular (in the sense precised below) gravitational environment, there exists only one static, axisymmetric and asymptotically flat geometry containing a black hole which is nonsingular. Moreover, the horizon of the black hole will depart from spherical symmetry (the shape when the environment is trivial) in a unique and specific manner that is completely dictated by the environment.

As a first step, let us characterize what we mean by an external gravitational environment in the restricted setup of staticity and axisymmetry that we are considering. 

\begin{figure}
\begin{center}
\includegraphics[width=0.35 \textwidth]{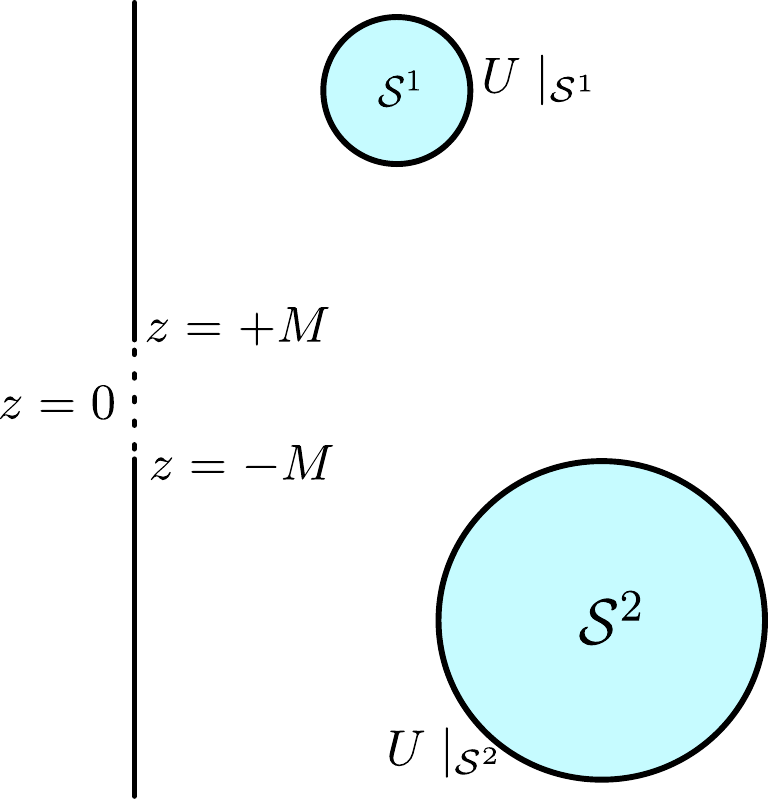}
\caption{Schematic representation of the setup. The $z$-axis is depicted as a black solid line, with a dashed segment between $z=-M$ and $z=M$ where the event horizon is located. Finally, we have illustrated some compact surfaces enclosing matter content (they would correspond to topological torus once we consider the $\varphi$ direction too) and we replace the potential matter content inside with a boundary condition at the surface of the object.}
\label{Fig:BH_screens}
\end{center}
\end{figure}

Consider a homogeneous rod source for the ALP located at $r=0$ between $z = - M$ and $z = M$ in the coordinate system that we are using. Furthermore, consider a set of compact and $\mathcal{C}^{\infty}$ toroidal two-surfaces\footnote{While it may be possible to consider less regular surfaces, we assume them to be arbitrarily smooth for simplicity in our analysis.} $S^i$ enclosing the external regions containing matter, see Fig.~\ref{Fig:BH_screens} for a pictorial representation. Inside these surfaces the vacuum equations are not satisfied and a careful analysis of the matter content would be required to determine the metric inside. However, to analyze their influence outside, we can replace the matter content by the value of the function $U$ induced on those surfaces. We restrict ourselves here to situations in which the functions $U$ at these surface are at least $\mathcal{C}^2$.\footnote{Although it must be possible to relax this condition, this is enough for our purposes.} Notice that we are not only demanding that $U$ restricted to the surface is $\mathcal{C}^2$ but also across the surface, which automatically excludes the presence of any distributional matter content, i.e., thin shells. Then, the values of $U$ at the different toroidal surfaces act as Dirichlet boundary conditions.\footnote{This is similar to the analysis of systems like the electromagnetic field between mirrors, where a perfectly reflecting boundary condition is imposed as a way to model in a simple way the interaction between the field and the matter content in the boundary. In this way, we can focus only on the determination of the field from its equations of motion supplemented with this boundary condition.}

As we have shown in the previous section, the central black hole can only be represented by a homogeneous rod source with a single free parameter: its length. Such length corresponds to the mass of the black hole $M$. Knowing this internal source, the value of the function $U$ at the toroidal two-surfaces, and the asymptotic condition that $U \to 0$ at infinity to ensure asymptotic flatness, we can solve the ALP equation for all vacuum regions in spacetime. In fact, we have a well-posed elliptic problem with Dirichlet boundary conditions, i.e., we know the harmonic function $U$ at the horizon, at the surfaces $S^i$, and we know that it goes to zero at infinity as a polynomial in $1/r$. Hence, the solution to the Laplace equation exists and is unique outside the surfaces by virtue of the classic theorems~\cite{Gilbarg2001}.

The final ingredient that we need to require is consistency, i.e., that the equilibrium condition~\eqref{Eq:Equilibrium_Condition} is satisfied everywhere. This is required so that the system does not display any kind of singular (for instance, cosmic-string-like) behavior, and we are really representing a black hole configuration in a gravitational environment. The specific form of the function $\Delta U=U-U_S$ determines the specific departure from spherical symmetry of the distorted black hole. 

Putting everything together, we have shown that:
\begin{enumerate}
    \item Given a gravitational environment, the horizon of a black hole departs from spherical symmetry in a unique and specific manner dictated by the environment. Any deviation from this natural shape will result in an irregularity of the geometry.
    \item Any geometry for a topologically spherical horizon is, in principle, possible; one would just need to determine the required gravitational environment to support it.
    \item Deformations of the horizon generated by external matter content are mild in terms of curvature, in the sense that they give rise to bounded curvature contributions at the horizon.
    \item In any case, for a solution with black holes and external matter to be regular everywhere, we need the equilibrium condition \eqref{Eq:Equilibrium_Condition} in the ALP. 
\end{enumerate}

\section{Almost-no-hair results for horizonless ultracompact objects}
\label{Sec:NoHair_ECOs}

For static and axisymmetric black holes, we have found a uniqueness result under complete generality, without requiring some of the conditions in Israel's theorem. In particular, we did not require the constraint on the topology of the constant redshift surfaces or the finiteness of the area of the minimum redshift surface. Now, let us extend our analysis to horizonless ultracompact objects. By this, we refer to material bodies whose first equipotential surface outside matter has an extremely high but finite redshift, characterized by a parameter $e^{2U_{\rm surf}} = \epsilon$, where $0 < \epsilon \ll 1$. In other words, we assume that the vacuum Einstein equations are satisfied up to a surface with redshift value $\epsilon$. 

Since no-hair theorems rely on the presence of a horizon, one could think that in the absence of a horizon, any ultracompact object would be possible. This would include, for example, compact but finite Curzon objects, thin rings, and spheroidal but not spherically-symmetric objects. However, a new form of no-hair results can be obtained by requiring that the Kretschmann scalar does not acquire trans-Planckian values~\cite{Barcelo2019}, i.e., ${\cal K}< {\cal K}_P=1/\ell_P^4$ with $\ell_P$ representing the Planck length. More generally, this condition aligns with the limiting curvature hypothesis, which postulates that all curvature invariants should remain bounded~\cite{Markov1982,Markov1987,Frolov1988,Frolov2021,Frolov2022}. This curvature constraint effectively replaces the standard condition of a regular horizon in traditional no-hair theorems. It imposes stringent limitations on the permissible geometries, specially in terms of their deviation from spherical symmetry in the static case. 

The primary physical motivation for this assumption is that GR itself becomes unreliable at extreme curvature scales. Imposing a curvature constraint not only rules out nonspherical geometries but also excludes Planck-scale spherically symmetric configurations, such as Planck-sized Schwarzschild black holes. This stands in contrast to Israel’s black hole uniqueness theorem, which remains agnostic to the specific curvature value as long as it remains finite. In general, GR tends to avoid configurations with extremely high curvatures, except in cases where gravity becomes so strong that singularities inevitably form. Therefore, it is reasonable to expect that any viable extension of GR would naturally favor configurations with lower curvatures if such alternatives are accessible dynamically. 
\subsection{Curvature-induced bounds}
As previously discussed, in the static and axisymmetric case, only three possible sources in the ALP can lead to infinite redshifts. Let us now examine the constraints that the curvature bound imposes on an object that, in the infinite redshift limit, approaches each of these three GR solutions.

\paragraph*{\textbf{Curvature-induced bounds for ultracompact lenticular objects.}}

Let us consider an object whose geometry in the external region is well-approximated by the (not spherically symmetric) Curzon metric. At least, it should correspond to the leading order as the minimum redshift surface is approached. For a fixed value of $\epsilon$ representing the minimum redshift attained by the configuration in the external region, we directly have a relation between the Weyl monopole $M^{(0)}$, the value of the coordinate $R$ at that surface,~$R_0$, and~$\epsilon$:
\begin{align}
    \epsilon = \exp \left( -\frac{2M^{(0)}}{R_0} \right).
\end{align}
To have $\epsilon \ll 1$, the ratio $M^{(0)}/R_0$ must be large. From this equation, we can obtain an expression for the monopole in terms of $R_0$ and the redshift:
\begin{align}
    M^{(0)}=\frac{1}{2} R_0 \abs{\log \epsilon}.
    \label{Eq:LogSeparation}
\end{align}
Now, the maximum value of the Kretschmann scalar in the Weyl sphere of radius $R_0$ is
\begin{align}
    \mathcal{K}_{\rm max} \sim \frac{1}{{\cal R}_0^4} \exp\left( \frac{ 2 [M^{(0)}]^2 }{R_0^2} \right),
\end{align}
where we have neglected subdominant terms for small $R_0$. The limitation on the maximum value of the Kretschmann amounts to an inequality of the form  
\begin{align}
    \frac{{\cal R}_0}{\ell_P} \gtrsim \exp\left( \frac{[M^{(0)}]^2 }{2R_0^2} \right) = \exp\left( \frac{1}{8} |\log\epsilon|^2 \right).
\end{align}
For very small $\epsilon$ we find that 
\begin{align}
\frac{R_0}{\ell_P} \gtrsim \frac{1}{\sqrt{2}} \abs{\log \epsilon}^{3/2} \exp\left( \frac{1}{8} |\log\epsilon|^2 \right),
\label{Eq:ParametricalSeparation}
\end{align}
after substituting the specific function $\mathcal{R}_0 \left( R, \pi/2; M^{0}\right)$ from Eq.~\eqref{Eq:MathcalR}. We are taking $\theta = \pi/2$ which is the direction in which the Kretschmann grows faster as we approach $R \rightarrow 0$.

In conclusion, an object whose metric outside a finite redshift region given by $\epsilon$ is the Curzon metric, must be extremely large $R_0 \gg \ell_P$ with an even larger $M^{(0)}$ to avoid generating trans-Planckian curvatures. In that sense, it must satisfy the following separation of scales
\begin{align}
    M^{(0)} > R_0 \gg \ell_P.
\end{align}
While the separation between $M^{(0)}$ and $R_0$ depends logarithmically on $\epsilon$ and therefore does not need to be exceptionally large [see Eq.~\eqref{Eq:LogSeparation}], the separation between $R_0$ and $\ell_P$ must [see Eq.~\eqref{Eq:ParametricalSeparation}]. 

\begin{figure}[H]
\begin{center}
\includegraphics[width=1 \textwidth]{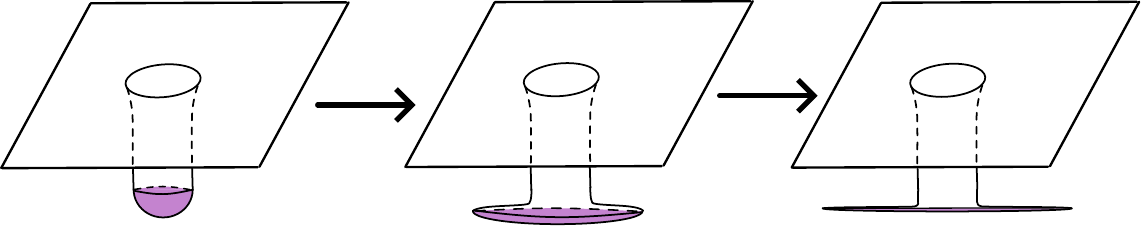}
\caption{Pictorial representation of gravitational collapse into a Curzon-type object. The colored region represents the material content in the inner region. Rather than generating an increasingly deep gravitational potential well, the collapse compresses the matter into a plane and stretches it along that plane.}
\label{Fig:Lenticular_Collapse}
\end{center}
\end{figure}

These configurations are not spherically symmetric, even if the redshift function only depends on $R$. As discussed earlier, this can be interpreted as the collapse of a matter cloud, which gradually forms a large, thin, lenticular object, as illustrated in Fig.~\ref{Fig:Lenticular_Collapse}. This represents a highly anisotropic collapse, where matter is compressed in one direction while spreading out in the other two. Our analysis indicates that the resulting object cannot be excessively flattened, as this would lead to trans-Planckian curvatures at the rim of the lenticular shape.
An equivalent analysis can be applied to any ultracompact object that is well-described in the external region by a different Weyl multipole expansion. The key difference is that equipotential surfaces are no longer spherical in Weyl coordinates, leading to objects with more complex shapes. However, we expect these differences to primarily affect the quantitative bounds, while the qualitative behavior of Curzon-like objects remains representative of this kind of objects. 

\paragraph*{\textbf{Curvature-induced bounds for ultracompact rings.}}

We can repeat the analysis for objects that are well-approximated in the external region by a Bach-Weyl ring geometry. We have found a quite convoluted expression for the Kretschmann scalar using the software \textit{Mathematica} with the help of the xAct package~\cite{xAct}, which is not very illuminating. Thus, we have decided to perform an analysis based on purely dimensional grounds and inspecting the leading orders of the divergences in the potential and Kretschmann scalar. Qualitatively speaking the analysis is very similar to the previous case.

Close to the ring, the equipotential surfaces are tori. For a given value of the mass $M$, the minimum value of the redshift $\epsilon$ is acquired for a torus of given minor radius $d$ located around the ring. Such radius can be understood as the thickness of the ultracompact ring under consideration. We have a relation between the mass of the ring $M$, such radius $d$ and the redshift $\epsilon$:
\begin{align}
    \epsilon \sim \exp(-M/d) \quad\Longrightarrow\quad M \sim d   \abs{ \log \epsilon}.
\end{align}
We can obtain an expression for $M$ in terms of $\epsilon$ and $d$. Now, the bound on the curvature implies
\begin{align}
    {\cal K} \sim e^{N(M/d)^n}{\cal R}^{-4} < \frac{1}{\ell_P^4}. 
\end{align}
The function $\mathcal{R}$ is a function of the coordinates, since the value of the scalar depends on which direction we approach the torus. Its specific functional form is quite convoluted and we omit it here. Inserting the previous expression for $M$ as a function of $d$ and $\epsilon$ we will obtain a condition of the form
\begin{align}
    d > \ell_P F (\abs{\log \epsilon}),
\end{align}
with $F$ a growing function of its argument $\abs{\log \epsilon}$ which tends to infinity as $\epsilon \to 0$. Notice that its specific functional form is determined by the function $\mathcal{R}$. At the end, we find again a separation of scales $M > d \gg \ell_P$, meaning that the ring cannot have a Planck sized radius. 

\paragraph*{\textbf{Curvature-induced bounds for ultracompact spheroids.}}

As a final case, let us consider the most astrophysically relevant scenario: ultracompact, stellar-like objects that closely resemble general relativistic black holes. To effectively mimic black holes, these objects must possess a deep gravitational well, making them nearly indistinguishable through observations. It is constructive to first consider a spherically symmetric configuration which, in virtue of Birkhoff's theorem~\cite{Wald1984}, will necessarily correspond to the Schwarzschild solution. In Weyl coordinates, spherically symmetric surfaces take the form of ellipsoids. To avoid trans-Planckian curvatures at the maximum redshift surface~$\epsilon$, we require the object to be macroscopic, meaning that $M \gg M_P$.

Consider now an ultracompact object such that the innermost equipotential surface takes on a very small value, $\epsilon\ll 1$, outside the region where Einstein equations do not hold. Such surface will generically be distorted with respect to the ellipsoids that describe spherically symmetric configurations in Weyl coordinates.\footnote{Notice that the surface from which vacuum Einstein equations do not hold does not necessarily need to be a constant redshift surface. We are taking the innermost equipotential surface from the region where they are satisfied.} To describe the deviations of a configuration from spherical symmetry in the static case it is convenient to resort to the Geroch multipoles~\cite{Geroch1970}. In the axisymmetric case under consideration, the multipoles reduce to a sequence parametrized by a single number $\{ M_{\rm G}^{(\ell)} \}_{\ell=0,1,\ldots}$, with the monopolar contribution representing the mass of the object $M_{\rm G}^{(0)}=M$. 

In principle, Geroch and Weyl multipoles are defined in a different way. However, there exists a subtle relation between both of them. Such relation is not invertible, namely it only allows to express the Geroch multipoles in terms of the Weyl multipoles of a spacetime. This suggests that the same expansion in terms of Geroch multipoles might admit several expansions in terms of Weyl multipoles. To be more precise about this relation, far from a matter source, a Geroch multipole of order $\ell$ can be decomposed as a sum of Weyl multipoles [defined in Eq.~\eqref{Eq:U_Multipoles}] as:
\begin{align}
    M_{\rm G}^{ (\ell)} = \sum_{n=1}^{\ell} \sum_{k_{1}...k_{n}} c_{k_{1}...k_{n}} M^{(k_{1})} \times \cdots \times M^{(k_{n})}; \qquad \sum_{i=1}^{n} k_{i} = \ell,
\end{align}
where the second equality is a constraint over the sum $\sum_{k_{1}...k_{n}}$, and $c_{k_{1}...k_{n}}$ are coefficients, for which there is not an explicit closed expression, but which can be found at any finite order through the algorithm presented in~\cite{Fodor1989}. 

A note of caution is in order at this point, since the matching of both expansions is only well-suited in the asymptotic region. As an illustrative example, consider the Geroch pure monopole which corresponds the Schwarzschild metric. In the Weyl multipole expansion, the Schwarzschild solution is an infinite series of even multipoles located at the origin given by
\begin{align}
    M^{(2\ell)} = - \frac{1}{\ell +1 } M^{2 \ell + 1} \quad \ell \in \mathbb{Z}^{+} \cup \{ 0\},
\end{align}
with all the odd multipoles identically vanishing $ M^{(2\ell + 1 )} = 0 \quad \ell \in \mathbb{Z}^{+} \cup \{ 0\}$. See Appendix~\ref{App:Multipolar} for a derivation. Furthermore, in the Appendix it is also shown that the radius of convergence of the Weyl multipole series is finite, it only converges for $R > M$, where $R$ is the spherical radial coordinate introduced above Eqs.~\eqref{Eq:rSpherical}-\eqref{Eq:zSpherical}. In that sense, the Weyl multipole expansion centered at the origin is not well-suited for describing near horizon properties, which lets us conclude that it is not well-suited for describing spherically symmetric configurations either.

In the absence of horizons, the same gravitational field can arise from infinitely many compactly supported matter distributions. However, when we consider the limit in which a horizon develops, $\epsilon \to 0$, any deviation from the Schwarzschild metric leads to huge curvatures in the metric that become divergent in the $\epsilon = 0$ limit. These divergences can fall into two families, depending on whether we unveil a behavior $f(z) \log(r)$ of the redshift function or a $1/r^{\ell+1}$ behavior as $\epsilon \to 0$. In the second case, the geometry would be dominated by the Weyl multipoles, so the configuration would be qualitatively like the Curzon-like objects discussed above and the bounds found there would apply to them. 

Thus, we focus now exclusively on ultracompact objects whose leading order behavior for $U$ is $f(z)\log r$ as $r \rightarrow 0$ for $z \in (-M,M)$. For the purpose of this analysis, it is convenient to introduce some constraints on the function $f(z)$:
\begin{align}
    0 = & \int_{-M}^{M} \dd z \left[ f(z) - 1 \right] \label{Eq:FixedMass} , \\ 
    0 = & \int_{-M}^{M} \dd z z \left[ f(z) - 1 \right]. \label{Eq:CenterMass}
\end{align}
They can be easily understood by recalling that $\lambda(z) = f(z)/2$, where $\lambda(z)$ is the density of the rod, as Eq.~\eqref{Eq:FixedMass} ensures that the total mass of the system is fixed to be $M$ and Eq.~\eqref{Eq:CenterMass} ensures that the center of mass of the system is located at the origin of coordinates. Now we analyze what happens when we approach the minimum redshift surface $\epsilon \to 0$. Such minimum redshift surface will correspond, in general, to some sort of geometrically deformed ellipsoid defined by the equation:
\begin{align}
    \epsilon \sim \exp \left[ 2 f(z) \log \left( \frac{r}{M} \right)   \right]. 
    \label{Eq:RedshiftSpheroids}
\end{align}
The leading order behavior of the Kretschmann scalar for this configuration is given by:
\begin{align}
   \mathcal{K} \sim  \frac{12[f(z)-1]^2 f(z)^2}{r_e^4}
   \label{Eq:Inhomogeneous_Kretschmann_leading}. 
\end{align}
Using Eq.~\eqref{Eq:RedshiftSpheroids}, the curvature bound leads to the constraint
\begin{align}
   \frac{1}{\ell_P^4} \gtrsim \frac{12[f(z)-1]^2 f(z)^2}{r^4}, 
\end{align}
which we can equivalently express as 
\begin{align}
   \frac{r}{\ell_P} \gtrsim \left\{12[f(z)-1]^2 f(z)^2\right\}^{1/4}.
   \label{Eq:size-condition}
\end{align}
For $f(z)=1$, which corresponds to the Schwarzschild geometry, the right-hand side identically vanishes. Thus, one would need to look at the subleading terms. All the divergent terms vanish as it can be seen from Eq.~\eqref{Eq:Inhomogeneous_Kretschmann}, and the only constraint that arises to avoid trans-Planckian curvatures is that the object itself should remain macroscopic, namely that $M \gg \ell_P$. On the other hand, when $f(z) \neq 1$, the bigger the departure of $f(z)$ from unity, the larger would have to be the ultracompact object if one does not want to generate trans-Planckian curvatures. Notice that Eq.~\eqref{Eq:FixedMass}, which fixes the monopole, automatically excludes the case in which $f(z) = 0$ everywhere. Actually, $f(z)$ can become zero at some points, although for instance, if it is a simple zero, the subleading term in Eq.~\eqref{Eq:Inhomogeneous_Kretschmann} would still impose constraints on $f'(z)$. This remains true for some values of $z$ even in the case in which $f(z) = 0$ is zero on an interval $I \subset (-M,M)$, which physically would represent two disconnected objects separated by a vacuum region in between. 

There is an alternative way to state the previous result. Let us consider that both the mass of the object~$M$ and the minimum redshift~$\epsilon$ are fixed. Then, after some manipulations of Eq.~\eqref{Eq:size-condition} in combination with Eq.~\eqref{Eq:RedshiftSpheroids}, we are led to the following constraint,
\begin{align}
    \log \left( \frac{M}{\ell_P} \epsilon \right) \gtrsim f (z) \times \log \left[ \abs{ f(z) \left( f(z) - 1 \right) } \right]
    \label{Eq:multipole-condition}. 
\end{align}
If we now keep $M$ fixed and make $\epsilon \to 0$, we clearly see that the $\epsilon$ term dominates on the left-hand side. As such, the object inside the square brackets on the right-hand side should approach $0$ faster than $\epsilon$. This means that, as the object becomes more and more compact, the deviation from spherical symmetry has to die off to avoid trans-Planckian curvatures.

Finally, it is worth remarking that from every $f(z)$, one can obtain its associated Geroch multipoles $M_{\rm G}^{ (\ell)}$. The specific functional relation can be however quite convoluted and we do not present it here in detail, we will just sketch for the sake of completeness how it could be established. First of all, we notice that the Weyl form of the Schwarzschild metric in Eq.~\eqref{Eq:Schwarzschild} can be transformed into the standard Schwarzschild coordinates $(u,\theta)$ by performing the transformation
\begin{align}
    r=\sqrt{u(u-2M)} \sin \theta, \qquad z=(u-M)\cos \theta. 
\end{align}
Applying this transformation blindly to a generic Weyl metric~\eqref{Eq:Line-Element-v2} we obtain
\begin{align}
    \dd s^2 = & - e^{2 U} \dd t^2  \nonumber \\
    & + e^{-2 U} \bigg[ e^{2V} \left( (1 + \frac{M^2 \sin^2 \theta }{u^2-2Mu}) \dd u^2 + \left( u^2-2Mu + M^2 \sin^2 \theta \right) \dd \theta^2 \right) 
    \nonumber \\
    &  + u(u-2M) \sin^2 \theta \dd  \varphi^2 \bigg],
    \label{SchwCoor}
\end{align}
where the functions $U$ and $V$ need to be expressed in terms of $u$ and $\theta$. The potential $U$ associated with an inhomogeneous rod of density $\lambda(z)$ is given by
\begin{align}
    U(r,z) = - \int_{-M}^{M} \dd z' \frac{\lambda(z')}{\sqrt{r^2 + (z'-z)^2}}.
\end{align}
Let us recall that, for a fixed $z \in (-M,M)$ in the limit $r \to 0$, we have 
\begin{align}
    U(r,z) \simeq  2 \lambda(z) \log r = f(z) \log r.
    \label{Eq:Uleading}
\end{align}
As discussed at the beginning of the chapter, from $U$ one can directly obtain $V$ through quadratures, although in general there is no closed expression for $V$-functions. Because of the previous conditions on the center of mass, we can be sure that the Schwarzschild coordinates that we are using are Thorne's ``Asymptotically Cartesian and Mass-Centered" (ACMC) coordinates~\cite{Thorne1980}. Then, from the form of the metric one could extract Thorne's multipoles, and in turn Geroch's multipoles~\cite{Gursel1983}.

Furthermore, we expect that in the limiting $\epsilon = 0$ case, there is a one-to-one correspondence between the families of functions $f(z)$ and the Geroch multipoles. Deviations from the Schwarzschild monopolar contribution that preserve the center of mass are described by the $\ell >1$ Geroch multipoles. It is then clear that Geroch multipoles have to die off as compactness becomes very small to avoid the generation of trans-Planckian curvatures, in the same way that the function $f(z)$ needs to approach $1$.

\subsection{Interplay with previous results in the literature}

Let us now mention two previous results in the same direction and briefly compare them with our findings. In the analysis presented in~\cite{Barcelo2019}, the static case is studied without additional assumptions. There, it is shown that, under the same conditions imposed here, the Geroch multipole structure of the configuration parametrically approaches that of the Schwarzschild solution as the redshift $\epsilon$ tends to zero. 
The paper was restricted to a perturbative analysis of the possible deformations around the Schwarzschild solution that have a smooth limit to it in the $\epsilon \to 0$ limit. This leaves the set of nonperturbative solutions that do not have such a smooth limit outside the analysis, which in the case under consideration here are easily analyzed. In fact, we have been able to completely constrain the structure of different ultracompact objects under the assumption that the curvatures do not become trans-Planckian and the additional assumption of axisymmetry.

Something similar happens with the analysis in~\cite{Raposo2019}. They study linear perturbations on top of a patch of the Schwarzschild metric. Specifically, they consider the Schwarzschild solution with mass $M$ from a surface of proper radius slightly larger than the would-be event horizon, i.e., the region between $r = 2M (1 + \epsilon)$ and $r \to \infty$. There it is shown that the boundedness of curvature invariants leads to a parametric suppression of the Geroch multipole deviations from the Schwarzschild solution. They find that all Geroch-Hansen mass multipoles vanish polynomially in $\epsilon$ if they are ``nonspin induced", meaning that they vanish as the first current multipole of the leading-order perturbation goes to zero. Since we restrict ourselves to the static case, these ``nonspin induced" multipoles correspond to the ones that we have analyzed here. Thus, in the intersecting domain of validity of these analyses, their conclusions and ours seem to be equivalent. However, again, their results are intrinsically perturbative and only apply to objects that parametrically approach the Schwarzschild solution as $\epsilon \to 0$. 

As a final comment, it is important to highlight that, for an ultracompact object with a small but fixed $\epsilon$, distinguishing whether deviations from spherical symmetry arise from internal structure or external gravitational influences is not straightforward, as both can produce finite distortions. To resolve this, one would need to examine the surrounding gravitational field and determine whether the object's shape exhibits distortions beyond those expected from external effects alone. What we have shown here is that any internal distortion must be minimal for high redshifts $\epsilon \ll 1$, and the more compact the object, the smaller these deviations should become.

\section{Conclusions and future work}
\label{Sec:ConclusionsNH}

This chapter has been devoted to extensions of the no-hair theorems within the context of a static and axisymmetric spacetimes. First, we noted that this case allows one to show that the only vacuum and asymptotically flat black hole solution with a regular horizon is Schwarzschild spacetime. When compared with Israel's theorem~\cite{Israel1967} for static geometries, we clearly see that the axisymmetry restriction allows to get rid of two premises of the theorem: the condition on the topology of the equipotential surfaces and the requirement that the horizon area be finite. The extensions of the theorems presented here fall in two categories: a result for black holes for which the external region is not everywhere vacuum, and a result for objects that display a maximum redshift surface, under the assumption that the Kretschmann scalar does not become trans-Planckian.

\subsection{Black hole deformations: gravitational environments}

Regarding the analysis of black holes in gravitational environments, our results demonstrate that in a static and axisymmetric scenario, there exists at most a one-parameter family of horizon shapes that can coexist with an external gravitational field (or a given matter configuration). This parameter represents an overall size for the black hole i.e. it represents its mass. For the environment to be consistent we have shown that it is necessary that it exhibits a gravitational equilibrium condition, as worked out in detail in Section~\ref{Sec:NoHair}. Therefore, we can say that from a pure geometric (or kinematic) point of view one can have any shape for the black hole horizon. All the uniqueness results come from the dynamical requirements imposed by the vacuum Einstein equations, along with the boundary conditions of the entire gravitating system, one of which is asymptotic flatness.

To the best of our knowledge, the only prior research in this direction is by G\"urlebeck \cite{Gurlebeck2012,Gurlebeck2015}, who demonstrated that the Weyl multipoles for a static and axisymmetric configuration can be expressed as a sum of surface integrals surrounding both the matter and black hole regions. In their interpretation, since the contribution to the multipole structure from the black hole, i.e., the ``hair'', is captured by a single parameter, they conclude that this constitutes a version of the no-hair theorem for black holes in external gravitational fields. However, we have sharpened the analysis and formulated a theorem in a manner closer to more traditional uniqueness results. 

At this point, we find it necessary to clarify a common misconception about the deformability of black holes, leveraging the results of G\"urlebeck. Love numbers, well established in Newtonian gravity, and with some working definitions in general relativistic contexts~\cite{Binnington2009,Poisson2020}, are known to vanish for static black holes~\cite{Damour2009,Damour2009b,Kol2011}. For a static object with a specific multipole structure deformed by an external gravitational field, Love numbers quantify the extent to which the object's contribution to the multipole structure deviates in response to the strength of the external field. As we have shown, this interpretation holds, at least in the static and axisymmetric case under consideration here. However, the vanishing of Love numbers must not be confused with a nondeformability of the cross-sectional geometry of event horizons. Indeed, as the solutions presented here demonstrate, it is evident that the intrinsic geometry of the horizon cross section is deformed due to the presence of external matter. 

Regarding potential extensions of results here, the first obvious extension that would be interesting to consider is the extension to the general static but nonaxisymmetric case. The main problem that one faces is that, in that case, Einstein equations do not reduce to a linear equation and equations that can be solved through quadratures. Although the redshift function still obeys a Laplacian equation, it is with respect to the curved metric on the constant $t$-slices. Thus, a careful analysis of which kind of sources on an arbitrarily curved geometry lead to infinite redshifts would be required, to characterize the presence of event horizons. Furthermore, the equilibrium conditions which will also arise would need to be characterized, and we expect that they do not reduce to only one condition as in the axisymmetric case. 

The second extension that would be desirable to perform is to relaxing the assumption of staticity and dealing with the stationary and axisymmetric case. In fact, this second case seems much more manageable. In that case, the Einstein vacuum equations simplify to the well-known Ernst equation, an integrable partial differential equation, alongside additional functions that can be determined directly through quadratures~\cite{Stephani2003}. This setup closely parallels the static and axisymmetric case, although the analysis of Ernst equation would require a dedicated treatment. The expectation is that the outcome of this analysis is an understanding of how external matter sources do not modify the content of no-hair theorems, still allowing only a biparametric family of geometries compatible with the existence of a vacuum horizon in GR, while modifying the event horizon's geometry with respect to its Kerr counterpart. This analysis could also illuminate the ongoing discussion around the vanishing Love numbers for Kerr black holes, see~\cite{Charalambous2021,LeTiec2020,Poisson2020,Chia2020} and references therein.

\subsection{Ultracompact objects}

Our explanation of why black holes cannot be distorted from the inside is somewhat technical, so here we offer a more intuitive, physical interpretation. A lump of matter can resist its own gravitational pull by generating internal positive pressures that counteract it. When an object is not very compact, this internal pressure largely determines its shape. For example, objects such as asteroids with very low compactness can take on very irregular, nonspherical shapes. However, as an object’s compactness increases, gravity becomes a dominant factor in shaping it, and nonrotating objects tend to become more spherical. This spherical shape minimizes the object’s gravitational self-energy. 

When the compactness of an object surpasses the black-hole threshold, i.e., when an event horizon forms, any structure capable of resisting gravitational collapse would require violating energy conditions, either by creating negative pressures or negative energy densities. For instance, one could construct spherically symmetric black holes that only differ from the Schwarzschild black hole in an inner core located in the black hole region~\cite{Frolov1989}. In contrast, any nonspherically symmetric regular black hole would require external matter to sustain the deviation from spherical symmetry. This can be understood as follows: consider a geometry with a nonvanishing $T_{\mu \nu}$ in an inner core that is not spherically symmetric but depends continuously on a parameter representing its compactness. As the configuration becomes progressively more compact, there will be a point where a horizon forms. However, our analysis indicates that such a horizon would inevitably leave some matter outside. If the horizon were to enclose all the matter content, it would necessarily result in a spherically symmetric horizon.

In the static and axisymmetric case considered here, we have been able to constrain all the possible ultracompact structures, including deformations of spherically symmetric ultracompact stars by requiring that the Kretschmann scalar does not reach trans-Planckian values. For deviations from the Schwarzschild metric described by an inhomogeneity function in the direction of the symmetry axis, we find that these inhomogeneities, which are equivalent to deviations from spherical symmetry, have to die off as the structure becomes more and more compact. 

Regarding the extensions of this analysis to static nonaxisymmetric and the stationary and axisymmetric case, the main problem lies on characterizing the horizon in terms of suitable variables. In the general static case the equations are such that the Laplacian equation satisfied by the redshift function is with respect to a generically curved metric induced on the constant-$t$ slices. Thus, a classification of all the possible distributional sources that give rise to an infinite redshift surface is subtle. Regarding the stationary and axisymmetric case, the potential entering Ernst equation is not straightforwardly related with the metric, and, equivalently, one would need to characterize the solutions that display an event horizon. This last analysis is highly interesting, since, as mentioned above, in Ref.~\cite{Raposo2019}, it is claimed that the rotationally induced multipoles can deviate from the Kerr ones in a slower way as the object approaches the black hole limit. 


\chapter{Toroidal black holes in four spacetime dimensions}
\label{Ch6:ToroidalBHs}

\fancyhead[LE,RO]{\thepage}
\fancyhead[LO,RE]{Toroidal black holes in four spacetime dimensions}


From a purely geometric (kinematic) perspective, black holes in four-dimensional spacetimes could have event horizons with arbitrary topologies. It is only when energy conditions are imposed that the horizon's shape is constrained to be spherical~\cite{Hawking1973}. While in the case of toroidal black holes, this exotic matter could live far from the horizon, for higher genus this matter must permeate the horizon itself. Although such configurations are unlikely to be relevant in astrophysical contexts, since energy condition violations are expected only in extreme causal scenarios or when large curvatures appear, not in macroscopic settings, it is still interesting to analyze them from a theoretical point of view. Studying these exotic horizons provides valuable insights into the structural aspects of GR and helps develop intuition about the implications of energy condition violations.

This chapter builds on the tools developed in the previous one. In the restricted framework of staticity and axisymmetry, requiring that the axial Killing field does not degenerate at the horizon limits the possible horizon topologies to spherical and toroidal~\cite{Geroch1982}. While the properties of local vacuum black holes in this setup (those with an open neighborhood around the horizon free of matter) were previously characterized~\cite{Geroch1982} (see also~\cite{Xanthopoulos1983}), complete solutions, describing the entire exterior region of such toroidal black holes without singularities, have not been reported yet. Building on earlier work by Thorne on static and axisymmetric configurations~\cite{Thorne1975}, Peters~\cite{Peters1979} attempted to construct an explicit example. However, his analysis showed that, for that line element, black holes with regular event horizons inevitably give rise to naked singularities in the exterior region.

The tools developed in the previous chapter enable us to present, to the best of our knowledge, the first families of toroidal black holes in four-dimensional spacetimes that remain nonsingular in the external region. These solutions are simple and allow for a thorough analysis of violations of the energy conditions. This chapter is largely based on the article~\cite{Barcelo2025} submitted to publication during this thesis, and it slightly diverges from the main themes of this thesis, as it explores exotic horizon topologies that we do not expect to occur in realistic astrophysical scenarios. The project grew out of a series of informal conversations among colleagues (and friends) about gravitational phenomena, yet we believe that the analysis presented here offers meaningful insights. 

In this chapter, it is important to distinguish between limits taken from the right and from the left and the conventional notation $\lim_{x \to a^{\pm}} f(x)$ is potentially confusing. To this end, we denote the right-hand limit as $\lim_{x \searrow a}$ and the left-hand limit as $\lim_{x \nearrow a}$. When dealing with functions defined on an interval, e.g., a function $f(x)$ with \mbox{$x \in [a,b]$}, we define the limit of its derivative at the boundaries as $f'(a) := \lim_{x \searrow a} f'(x)$ and $f'(b) := \lim_{x \nearrow b} f'(x)$. The same notation applies to higher-order derivatives and to partial derivatives.

\section{Overview of previous literature}
\label{Sec:Literature}

In this chapter we will be working in the same setup that we were considering in the previous one. This means, that we are taking a static and axisymmetric line element
\begin{align}
    \dd s^2 = - e^{2 U} \dd t^2 + e^{-2 U} \left[ e^{2V} \left( \dd r^2 + \dd z^2 \right) +r^2 \dd \varphi^2 \right],
    \label{Eq:WeylEl}
\end{align}
where $U = U(r,z)$ and $V=V(r,z)$ are functions that depend only on $r$ and $z$. The vector generating time translations is $\bm{k} = \partial_t$ and the vector generating axial transformations is $\bm{m} = \partial_{\varphi}$. We recall that in vacuum, the equations of motion reduce to a Laplace equation for $U$ and two equations for $V$, that can be obtained through quadratures:
\begin{align}
    & \nabla^2 U = 0, \label{Eq:Laplace}\\
    & \partial_r V = r \left[ (\partial_r U)^2 - ( \partial_z U)^2 \right], \label{Eq:drV}\\
    & \partial_z V = 2 r \partial_r U \partial_z U. \label{Eq:dzV}
\end{align}
Our aim here is to analyze black holes with exotic topologies in this setup. The rest of the section is dedicated to reviewing previous results in the literature concerning toroidal black hole solutions.

\subsection{Topology of event horizons}
\label{Subsec:Topology_Horizons}

\begin{figure} 
\begin{center}
\includegraphics[width=0.3 \textwidth]{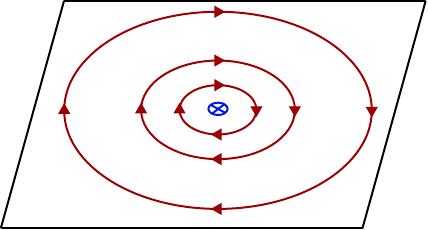}
\caption{The index of the axial Killing vector field $\bm m$ is always $+1$ since its orbits that can be mapped to a circle $S^1$, and it generates a rigid rotation around the zero of the vector, which is fixed by the action of the isometry group.}
\label{Fig:IndexVector}
\end{center}
\end{figure}

Every cross section $\mathcal{S}$ of the horizons considered here is a compact, orientable (since it is a boundary) surface with a positive definite metric, and an axial Killing vector field $\bm m$. According to the Poincar\'e-Hopf theorem~\cite{Milnor1997}, the sum of the indices of the zeros of $\bm m$ should be equal to the Euler number of the cross section $\chi (\mathcal{S})$:
\begin{align}
    \sum_{i} \text{index}_i (\bm m) = \chi (\mathcal{S}) = 2  - 2g(\mathcal{S}),
\end{align}
with $g(\mathcal{S})$ being the genus of the surface $\mathcal{S}$. The vector field $\bm m$ can have at most isolated zeros. Given that $\bm m$ is the Killing vector generating axial symmetry, we have that the index of those zeros is always $+1$, as they need to behave locally around the zero as rotations, since the flow of $\bm m$ leaves the zero as a fixed point (see Fig.~\ref{Fig:IndexVector}).
Hence, $2 - 2g \geq 0$ and $g \leq 1$, so it can only take the values $g = 0$, corresponding to a spherical horizon where the Killing vector has two zeros at the two poles of the sphere; and $g = 1$, corresponding to a toroidal horizon corresponding to the vector having no zeros. 
\begin{figure}
\begin{center}
\includegraphics[width=0.3 \textwidth]{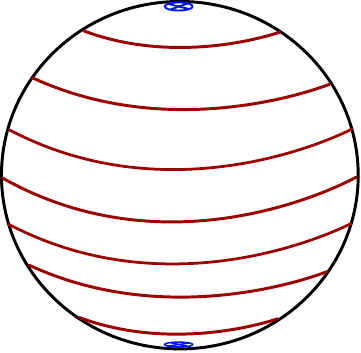}
\includegraphics[width=0.3 \textwidth]{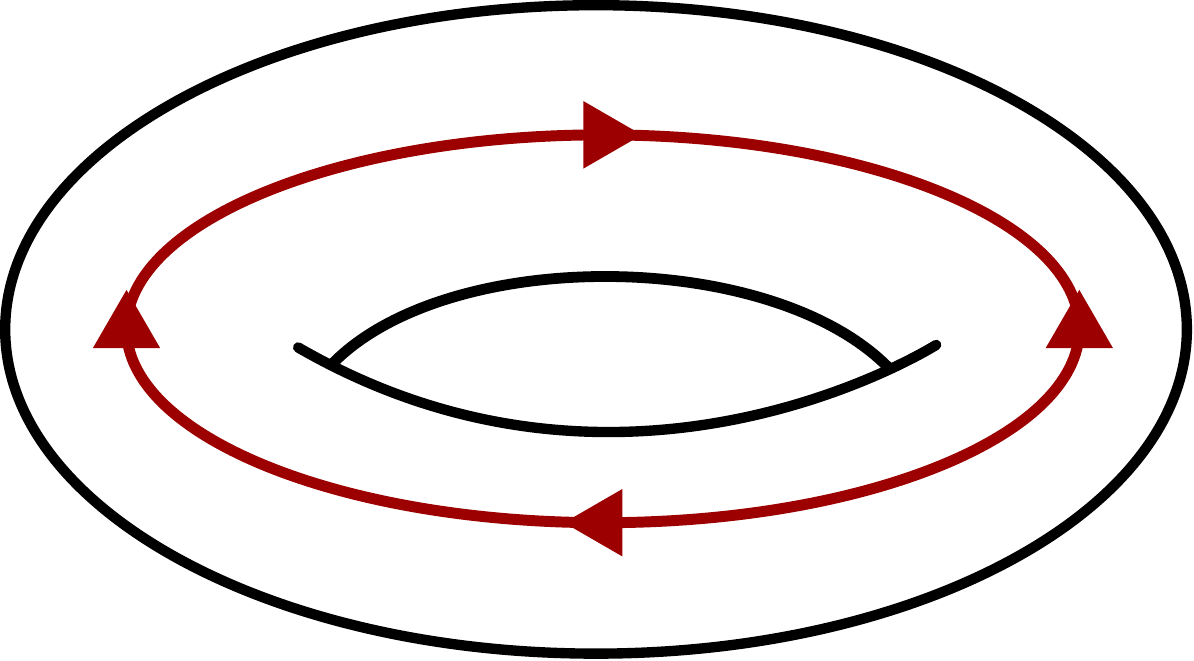}
\caption{Here we depict the two possible configurations for the cross sections of the horizon: a two-sphere with the two poles at which the vector vanishes and the toroidal configuration.}
\label{Fig:Topologies}
\end{center}
\end{figure} 

\subsection{Local toroidal black holes}
\label{Subsec:Toroidal}

Building on the results of the previous chapter, one may ask under which conditions a black hole horizon could possess a toroidal cross-section. Specifically, one can wonder whether there exists a type of matter content that can distort the black hole horizon to the extent that its topology becomes a torus, rather than a sphere. As established by Hawking’s theorem (see Prop. 9.3.2 of \cite{Hawking1973}), such scenarios necessarily involve violations of the energy conditions in the exterior region. Additionally, Galloway’s theorem, which relaxes some global assumptions, provides further insight into this possibility~\cite{Galloway1993}. For further developments, see also~\cite{Chrusciel1994,Jacobson1994}.

However, for toroidal horizons, violations of the energy conditions do not need to occur directly on the black hole horizon itself. In fact, all the toroidal black holes whose horizons are in vacuum were characterized by Geroch and Hartle~\cite{Geroch1982}, see also~\cite{Xanthopoulos1983} for a complementary analysis.

The construction parallels that of black holes with spherical horizons discussed in Chapter~\ref{Ch5:Nohair}, but now we take as reference metric the vacuum solution characterized by the functions $U_T = \log (r /2M)$ and $V_T = \log (r/M)$. It corresponds to the solution of the ALP sourced by an infinite rod located at $r=0$. To be more precise, we are actually solving again the Poisson equation with a distributional source, but as we will see, this source is actually ``fictitious'', meaning it is not associated with any real energy-momentum tensor. Plugging these functions in the line element~\eqref{Eq:WeylEl} we get
\begin{equation}
    \dd s^2= - \left(\frac{r}{2M}\right)^2 \dd t^2 + 4\left[ \dd r^2 +   \dd z^2  + M ^2 \dd \varphi^2 \right].
    \label{Eq:GerochHartleBH}
\end{equation}
As the solution does not depend on $z$, we can construct a compactified version of it by just introducing the following identifications on the Euclidean three-space described in cylindrical coordinates by $(r,z,\varphi)$: $(r,z=M,\varphi) \sim (r,z = -M, \varphi + \alpha)$, with $M$ and $\alpha$ any real numbers; and \mbox{$(r,z,\varphi) \sim (r,z,\varphi + 2 \pi)$}. $\alpha$ is a twisting angle and can be safely taken to be zero. Furthermore, the geometry is flat, as can be checked by directly computing the Riemann tensor that identically vanishes. Actually, it simply corresponds to the flat spacetime metric described in Rindler coordinates~\cite{Wald1984}. 

There are two important features of the geometries under construction that are worth emphasizing. First, regarding the asymptotic region: if we take the line element in Eq.~\eqref{Eq:GerochHartleBH} to describe the spacetime all the way to spatial infinity ($r \to \infty$), the periodic identification of the $z$-coordinate implies that the resulting geometry is not asymptotically flat. Instead, it describes an infinitely extended toroidal tube. However, our goal is not to construct such infinite toroidal black hole tubes, but to obtain asymptotically flat toroidal black holes. Therefore, the line element in Eq.~\eqref{Eq:WeylEl} should be interpreted as valid only in a finite patch of the exterior region.

Second, one needs to ensure that the the geometry is smooth and can be extended across the horizon, which in these coordinates is approached by taking $r \to 0^{+}$. For the reference metric with $U = U_T$ and $V=V_T$, we notice that it can be analytically continued by performing the change of coordinates: 
\begin{align}
    & R^2 - T^2 = 4 r^2, 
    & T/R = \tanh \left( \frac{t}{4M} \right),
\end{align}
so that the metric is explicitly flat
\begin{align}
    \dd s^2 = - \dd T^2 + \dd R^2 + 4 \dd z^2 + 4M^2 \dd \varphi^2. 
\end{align}
In this way, we can smoothly extend the metric across the horizon located at $r = 0^{+}$, and we see that it is locally in vacuum. A similar extension can be performed for any distortion from this reference metric, which amounts to replacing $ U = U_T$ and $V = V_T $ by $ U = U_T + \hat{U} $, where $ \hat{U} $ is any solution to the Laplace equation that is analytic at $ r = 0 $, and the $ V $ function obtained by quadratures from Eqs.~\eqref{Eq:drV}-\eqref{Eq:dzV}. The analyticity of $ \hat{U} $ at $r = 0$ guarantees that the extension can be carried out.

The hatted quantities would represent deformations from the reference metric induced by external matter sources, supported wherever the Laplacian of $\hat{U}$ is nonzero, i.e., where $\nabla^2 \hat{U} \neq 0$~\cite{Geroch1982,Barcelo2024}. In fact, we must encounter that some of the matter outside the horizon must necessarily violate energy conditions~\cite{Hawking1971}. This matter must either source $\hat{U}$ in the region described by the line element in Eq.~\eqref{Eq:WeylEl}, be located in another coordinate patch of the external region, or both.

\subsection{Locally vacuum toroidal black holes: previous attempts}
\label{Subsec:Peters}

An attempt to construct an explicit toroidal black hole solution in 4D was done by Peters~\cite{Peters1979}, following up on a previous solution to the Einstein vacuum equations presented by Thorne~\cite{Thorne1975}. However, Peters found out that a regular event horizon for that specific solution requires the presence of some singularities in the external region. Let us discuss it in detail. 

We start by describing Thorne's solution. The idea is to consider the Einstein vacuum equations \eqref{Eq:Laplace}-\eqref{Eq:dzV} for a static and axisymmetric metric expressed in Weyl coordinates~\eqref{Eq:WeylEl}. In this case, the potential $U$ is taken to be ``originated'' in a line segment singularity, which we will see in a moment that this is generically promoted to a real spacetime naked singularity, lying between $z = -a$ and $z = + a$ in the $z$-axis, i.e., $r=0$. In addition, the potential is constrained to have $\partial_z U=0$ at two disks located at $z= \pm a$ and having radius $b$. Thorne shows that far away from the constraining disks, for $\sqrt{r^2 + z^2 } \gg b $, the potential is almost spherical since $U = - M / r + \order{r^{-2}} $ and $V = \order{r^{-2}}$, and hence the metric is asymptotically flat with mass $M$.

From the point of view of the solution to the Laplace equation, it corresponds to the solution which obeys the following set of conditions: 
\begin{enumerate}
    \item The solution behaves as $U  \sim (M/a) \log (r) $ as $r \rightarrow 0$ with $z \in (-a,a)$.
    \item $\partial_z U \to 0$ as $z \rightarrow \pm a$ with $r \in (0,b)$. 
    \item $\nabla^2 U = 0$ everywhere for $r >0$. 
    \item $U \sim - M / \sqrt{r^2 + z^2}$ as $\sqrt{r^2+z^2} \rightarrow \infty$.
    \item $V$ obeys Eqs~\eqref{Eq:drV}-\eqref{Eq:dzV} everywhere and vanishes as $\sqrt{r^2+z^2} \rightarrow \infty$. 
\end{enumerate}
To endow the source with a toroidal topology, an additional ingredient must be introduced, specifically, the identification of the outer faces of the upper and lower disks in the three-dimensional Euclidean space described by the coordinates $(r,z,\varphi)$, i.e.,
\begin{align}
    \lim_{\epsilon \searrow 0} (r,z = a + \epsilon , \varphi) = \lim_{\epsilon \searrow 0} (r,z = - a - \epsilon , \varphi), \quad r \in [0,b],
\end{align}
and the inner faces of the upper and lower disks too
\begin{align}
    \lim_{\epsilon \searrow 0} (r,z = a - \epsilon , \varphi) = \lim_{\epsilon \searrow 0} (r,z = - a + \epsilon , \varphi), \quad r \in [0,b].
\end{align}
This automatically guarantees that the singularity possesses a toroidal topology. The solution depends on three free parameters: $(M, a, b)$. In general, it is not possible to obtain a closed-form analytic expression for the function $U$, and consequently not for $V$ either. However, useful approximations exist in the limiting cases $a/b \ll 1$ and $b/a \ll 1$. For explicit expressions in these cases, see~\cite{Thorne1975,Peters1979}.

At this point, once the metric is characterized, one can realize that there are two potential spacetime singularities associated with it: one at $r \rightarrow 0$ for \mbox{$z \in (-a,a)$} and another one which is precisely lying at the edge of the disks (the common edge, given the periodic identification), i.e., $r \to b$ and $z = \pm a$. Thorne (see Sec.~IIF of~\cite{Thorne1975}) identified that removing the latter requires imposing a constraint among the three parameters $(M,a,b)$. Regarding the other potential singularity at $r \rightarrow 0$, for $z \in (-a,a)$, we have that the asymptotic form of the metric as $r \rightarrow 0$ is:
\begin{align}
    \dd s^2 = -  r^{2M/a} \dd t^2 + r^{(2M/a) \times (M/a - 1)} (\dd r^2 + \dd z^2) + r^{-2(M/a-1)} \dd  \varphi^2 + (\text{finite terms}).
\end{align}
In fact, for $M/a = 1$, Peters~\cite{Peters1979} showed that the leading-order behavior of the metric corresponds exactly to flat spacetime in Rindler coordinates, with the $z$-coordinate periodically identified, that is, the same line element we discussed earlier for $U = U_T$ and $V = V_T$. Furthermore, he demonstrated that the subleading terms in the solution still yield regular curvature invariants as $r \to 0$, which permits a smooth extension of the geometry into the interior region. These terms correspond to the contribution from analytic functions $\hat{U}$ and $\hat{V}$ described earlier. However, as we previously noted, such contributions must originate from some form of matter or singular structure in the exterior region. In this particular case, they stem from the singular behavior at the edges of the disks.

\begin{figure}
\begin{center}
\includegraphics[scale=.8]{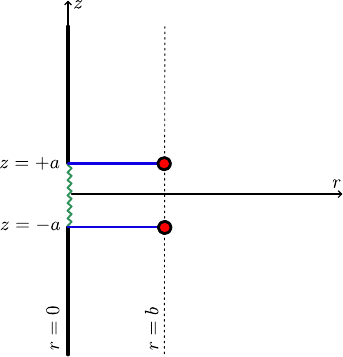}
\caption{Pictorial representation of Thorne toroidal solution. The singularity at $r = 0$ and $z \in (-a,a)$ is depicted as a zigzag green line, and the disks that are identified are depicted in blue. The red dots at the edge of the disks represent the putative singularity that might arise depending on the values of the parameters.}
\label{Fig:Thorne_Solution}
\end{center}
\end{figure}

Peters further showed in Section II of~\cite{Peters1979} that removing the singularity at the edges of the disks requires $M/a >2$, which is clearly incompatible with the $M/a = 1$ requirement of a smooth horizon at $r = 0$. A pictorial representation of the solution for arbitrary values of the parameters can be found in Fig.~\ref{Fig:Thorne_Solution}.
Thus, within this solution, it is not possible to simultaneously remove the singularity at $r \to 0$ and the external singularities located at the edges of the disks. In fact, this could have been anticipated based on Hawking's theorem, given that there is no matter content in the exterior region.

\section{Building toroidal black holes}
\label{Sec:Construction}
\begin{figure}
\begin{center}
\includegraphics[scale=1.]{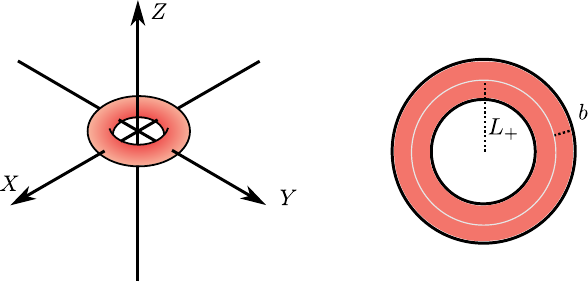}
\caption{Representation of the outermost region $\Omega_1$ for a fixed time $T$ and the limiting hypersurface $\Sext$ (in red).}
\label{Fig:Toro}
\end{center}
\end{figure}

In this section, we provide our construction of toroidal black holes. The geometries that we consider here describe the transition from a flat spacetime region to a region in which we have a toroidal black hole, the latter being locally described by the line element in Eq.~\eqref{Eq:WeylEl} and the functions in Eqs.~\eqref{Eq:drV}-\eqref{Eq:dzV}. For that purpose, we distinguish three regions of spacetime that we call $\Omega_1$, $\Omega_2$ and $\Omega_3$.
\begin{figure}
    \centering
    \includegraphics[scale=.8]{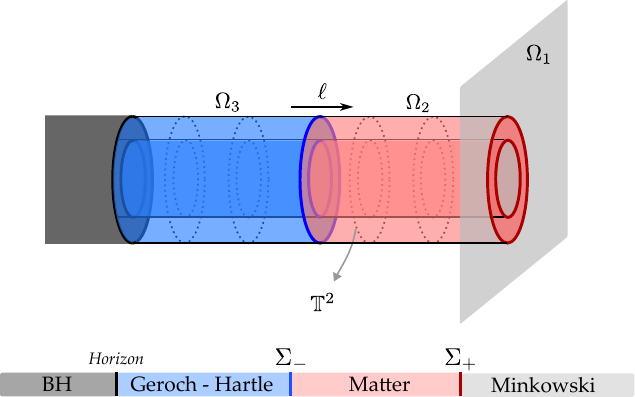}
    \caption{Schematic representation of the toroidal black hole that we construct.}
    \label{Fig:Om1Om2Om3}
\end{figure}
The regions $\Omega_2$ and $\Omega_3$ are delimited by the values of a certain coordinate $\ell$, $\ell\in(\Rint, \Rext)$ and $\ell\in (0, \Rint)$, respectively, whereas $\Omega_1$ represents an outermost flat spacetime region with a torus removed. Notice that, by construction, we are taking $0<\Rint<\Rext$. In addition, there will be two special matching hypersurfaces, $\Sext$ and $\Sint$, at $\ell= \Rext$ and $\ell=\Rint$ joining, respectively, $\Omega_1$ and $\Omega_2$, and $\Omega_2$ and $\Omega_3$. The technical details of the matching between these regions using Israel's junction conditions~\cite{Israel1966} are provided in Appendix~\ref{AppF}. See Fig.~\ref{Fig:Om1Om2Om3} for a pictorial representation.
\paragraph*{\textbf{Exterior flat region $\Omega_1$}.}
The first region $\Omega_1$ is a flat spacetime region described by the set of coordinates $(T,X,Y,Z)$ in terms of which the metric looks like
\begin{align}
    \dd s^2_1 = -\dd T^2 + \dd X^2 + \dd Y^2 + \dd Z^2\,.
    \label{Eq:Region1}
\end{align}
This describes the exterior region of the black hole, which extends up to a surface $\Sext$ with toroidal topology. In the chosen coordinate system, this surface is defined by the following equation:
\begin{align}
    \left( \sqrt{X^2 + Y^2} - \Rext \right)^2 + Z^2 =  b^2,
    \label{Eq:TorusEq}
\end{align}
where $\Rext$ is the radial distance from the origin to the central circumference of the tube, and the positive parameter $b$ is the radius of the tube. See Fig.~\ref{Fig:Toro}. Note that the coordinates $(X,Y,Z)$ are restricted to values satisfying $\left( \sqrt{X^2 + Y^2} - \Rext \right)^2 + Z^2 >  b^2$.
%
\paragraph*{\textbf{Interior toroidal region $\Omega_3$.}}
We also consider a region $\Omega_3$ which is described by the Geroch-Hartle line element: 
\begin{align}
\label{Eq:gOm3}
    \dd s^2_3 = - \frac{\ell^2}{M^2} \dd t^2 + \dd \ell^2 +  \frac{M^2}{\pi^2}  \dd z^2 +  M^2 \dd \varphi^2,
\end{align}
where $z \sim z +2 \pi$ and $\varphi \sim \varphi + 2 \pi$. This metric describes the same geometry as \eqref{Eq:GerochHartleBH}. Indeed Eq.~\eqref{Eq:gOm3} can be obtained from Eq.~\eqref{Eq:GerochHartleBH} by performing the rescalings of the coordinates:
\begin{equation}
    z\to \frac{M}{\pi}z\,, \qquad t\to 2t\,,\qquad r \to \ell/2\,,
\end{equation}
followed by $M\to M/2$. Here, the coordinate $\ell$ ranges over the interval $(0, \Rint]$. The value $\ell = \Rint$ corresponds to the hypersurface $\Sint$, which, at any fixed time $t$, has the geometry of a flat torus, since its two-dimensional intrinsic curvature vanishes.

\begin{figure}
\begin{center}
\includegraphics[scale=1.]{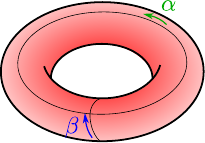}
\caption{Geometrical meaning of the coordinates $\alpha$ and $\beta$.}
\label{fig:alphabeta}
\end{center}
\end{figure}

\paragraph*{\textbf{Interpolating region $\Omega_2$.}}
Finally, we consider the region $\Omega_2$ that will interpolate between the flat torus $\Sint$ and the curved torus $\Sext$, satisfying Israel junction conditions at both surfaces. For this region, we can take a line element of the form: 
\begin{equation}\label{eq:region2General}
    \dd s^2_2 = - H(\ell) \dd \tau^2 + \dd \ell^2 + \mathcal{F}(\ell, \beta) \dd \alpha^2 + b^2 \dd  \beta^2\,,
\end{equation}
with the angular coordinates periodically identified $\alpha \sim \alpha + 2 \pi$ and $\beta \sim \beta + 2 \pi$, and the functions $H(\ell)$ and $\mathcal{F}(\ell, \beta)$ should satisfy the conditions:
\begin{equation}\label{eq:calFH_interpcond}
    H(\Rint)=H(\Rext)=1\,,\quad \mathcal{F}(\Rint, \beta) = \frac{M^2}{\pi^2}\,,\quad \mathcal{F} (\Rext, \beta) =  \left(\Rext + b \cos \left(\beta\right) \right)^2.
\end{equation}
A specifically simple choice of interpolating function is 
\begin{equation}\label{eq:FG}
    \mathcal{F}(\ell, \beta) =  F(\ell) \left( \Rext+ b \cos (\beta) \right)^2 + G(\ell)\,,
\end{equation}
which makes the line element to be of the form:
\begin{equation}\label{eq:gregion2}
    \dd s^2_2 = - H(\ell) \dd \tau^2 + \dd \ell^2 + \left[ F(\ell) \left( \Rext+ b \cos (\beta) \right)^2 + G(\ell) \right] \dd \alpha^2 + b^2 \dd  \beta^2.
\end{equation}
Notice that this region is spatially foliated by 2-dimensional tori, for each value of $\tau$ and $\ell$ we have a torus. They are represented as the dotted lines in Fig.~\ref{Fig:Om1Om2Om3}. In particular, $\beta$ runs around the loop of the torus and $\alpha$ is the angle of revolution of the torus, which corresponds to an isometric direction, i.e., $\partial_\alpha$ is a Killing vector of the metric~\eqref{eq:gregion2}. We have depicted both angles in Fig.~\ref{fig:alphabeta} for clarity. Moreover, the interpolating functions $F(\ell)$ and $G(\ell)$ should satisfy:
\begin{equation}\label{eq:FGH_interpcond}
    F(\Rint) = 0\,,\quad F(\Rext) = 1\,,\quad G(\Rint) = \frac{M^2}{\pi^2}\,,\quad G(\Rext) = 0\,.
\end{equation}
This region is confined between $\ell = \Rext$, which describes the surface $\Sext$ and $\ell=\Rint$ corresponding to $\Sint$. The simplest example that we can think of is the following choice of functions:
\begin{align}\label{eq:exampleFGH}
    F(\ell) = \dfrac{\ell^2-\Rint^2}{\Rext^2-\Rint^2}\,,\qquad G(\ell) = \dfrac{M^2}{\pi^2} \dfrac{\Rext^2-\ell^2}{\Rext^2-\Rint^2}\,,\qquad H(\ell)=1\,.
\end{align}
%

\subsection{Simplest attempt}
\label{Subsec:Try1}

At first sight, one might wonder why three distinct regions are required, and why it would not be sufficient to consider a single toroidal shell directly connecting $\Omega_1$ and $\Omega_3$. The primary issue with such a construction is that it would result in a discontinuity in the metric, since the induced metric on the shell would necessarily inherit a curved geometry from the exterior and a flat one from the interior.

\begin{figure}
\begin{center}
\includegraphics[scale=0.75]{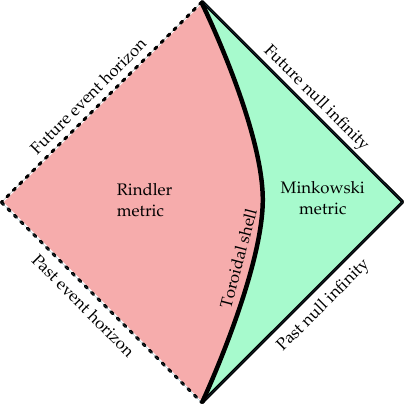}
\caption{Pictorial representation of the simplest setup to match an exterior flat spacetime and an interior Rindler metric with the periodic identification of one of the transverse coordinates. The matching would need to occur on a surface with the topology of a two torus. However, given that the intrinsic metric that the torus inherits from the inside and the outside is different (whereas the one from the inside is flat, the one from the exterior is curved), it is not possible to match them and satisfy Israel junction conditions.}
\label{Fig:Toroidal_Shell_Try}
\end{center}
\end{figure}

Explicitly, imagine that we want to match an external flat region with an interior region described by the line element~\eqref{Eq:WeylEl}. This means that we take the spacetime to be of the form~\eqref{Eq:WeylEl} for $r \in (0,R)$, and match it with a Minkowski spacetime from which we have removed the interior of a torus at every time, see a pictorial representation in Fig.~\ref{Fig:Toroidal_Shell_Try}. Although both surfaces display the same topology, it is not possible to match them smoothly within GR. The first of the Israel junction conditions requires that the metric is continuous across the matching surface in order to have a well-defined distributional curvature. The continuity of the metric implies that the two induced metrics (one from each side) need to describe the same intrinsic geometry. In the matching scenario considered in our model, the torus inherits a flat geometry from the interior region, i.e., it possesses no intrinsic curvature, whereas the induced metric derived from the exterior region is curved.

One might wonder whether it could be possible to choose a different smooth embedding of the torus $\Sext$ in $\mathbb{R}^3$ allowing for a flat induced metric from $\Omega_1$. However, it is not possible to construct a $C^{2}$-embedding of this kind. To see this, let us proceed by \emph{reductio ad absurdum}. Suppose there exists such an embedding of the torus in flat space with identically vanishing Gauss curvature. Given that it is a compact surface, it would be possible to enclose it with a sphere, which has positive Gaussian curvature at every point. Now, suppose that we reduce the radius of the sphere until it touches the torus in some contact point(s). The curvatures would need to agree at that point and hence the curvature of the torus would need to be zero and positive at the same time, reaching a contradiction and demonstrating the impossibility to embed a flat torus in flat space with at least $C^2$-differentiability. This forbids this simple construction and requires us to perform some more subtle matching between the external and internal region. Specifically, we need to include an interpolating four-dimensional region in between.

It is still interesting to note that if one relaxes the $C^2$ regularity and only demands $C^1$ regularity, this proof breaks down (since the Gauss curvature is no longer defined for the torus). In fact, it was shown by the seminal theorems of Nash and Kuiper~\cite{Nash1954,Kuiper1955,Kuiper1955b} that $C^1$-embeddings of flat torus into the flat spacetime actually exist. To make sense of this physically, one would need to interpret the discontinuity on the metric.\footnote{Notice however that $C^2$-embeddings of the flat torus are possible in more than 3 spatial dimensions, e.g., the Clifford torus in $\mathbb{R}^4$.}

\subsection{Two shell model}
\label{Subsec:Try2}

The regions $\Omega_1$ and $\Omega_3$ are flat and hence are automatically vacuum solutions. The region $\Omega_2$ displays a nontrivial matter content, in addition to the distributional matter content at the two matching surfaces. In fact, the distributional matter content $S_{ij}$ at the hypersurfaces $\Sint$ and $\Sext$ can be directly obtained from Israel junction conditions as described in detail in Appendix~\ref{AppF}.

\paragraph*{\textbf{Energy-momentum tensor in region $\Omega_2$.}}
In this section we compute the energy-momentum tensor supporting the nonflat geometry in the region $\Omega_2$. The only nonvanishing components for the energy-momentum tensor are the following:
\begin{align}
    8 \pi T^{\tau}{}_{\tau}& = -\frac{1}{4b^2}\left[\left(\frac{\partial_\beta \mathcal{F}}{\mathcal{F}}\right)^2 + b^2 \left(\frac{\partial_\ell \mathcal{F}}{\mathcal{F}}\right)^2 - 2 \left(  
    \frac{\partial^2_\beta \mathcal{F}}{\mathcal{F}} + b^2 \frac{\partial^2_\ell \mathcal{F}}{\mathcal{F}} \right) 
    \right]\,, \\[2ex]
    8 \pi T^{\ell}{}_{\ell} & = -\frac{1}{4b^2}\left[\left(\frac{\partial_\beta \mathcal{F}}{\mathcal{F}}\right)^2 - 2   
    \frac{\partial^2_\beta \mathcal{F}}{\mathcal{F}} - b^2 \frac{H'}{H}\frac{\partial_\ell \mathcal{F}}{\mathcal{F}} \right]
    \,,\\[2ex]
    8 \pi T^{\ell}{}_{\beta} & = \frac{1}{4}\left[\frac{\partial_\beta \mathcal{F}\partial_\ell \mathcal{F}}{\mathcal{F}^2} - 2 \frac{\partial_\beta\partial_\ell \mathcal{F}}{\mathcal{F}} \right] \qquad = 8 \pi b^2 T^{\beta}{}_{\ell}\,, \\[2ex]
    8 \pi T^{\alpha}{}_{\alpha} & 
    = \frac{1}{4}\left[2\frac{H'' }{H }-\left(\frac{H'}{H}\right)^2\right]\,,\\[2ex] 
    8 \pi T^{\beta}{}_{\beta} & 
    = \frac{1}{4}\left[ 2
    \frac{\partial^2_\ell \mathcal{F}}{\mathcal{F}}
    -\left(\frac{\partial_\ell \mathcal{F}}{\mathcal{F}}\right)^2
    +2\frac{H''}{H}-\left(\frac{H'}{H}\right)^2
    +\frac{H'}{H}\frac{\partial_\ell \mathcal{F}}{\mathcal{F}}\right]\,,
\end{align}
where we have omitted the coordinate dependencies of the functions $\mathcal{F},H$ and their derivatives.

The energy-momentum tensor that we found for the region $\Omega_2$ has the form
\begin{equation}
    \begin{pmatrix}
        T^{\tau}{}_{\tau} & 0 & 0 & 0 \\
        0 & T^{\ell}{}_{\ell} & 0 & T^{\ell}{}_{\beta} \\
        0 & 0 & T^{\alpha}{}_{\alpha} & 0 \\
        0 & T^{\beta}{}_{\ell} & 0 & T^{\beta}{}_{\beta}
    \end{pmatrix}\,,
\end{equation}
which has a nontrivial timelike eigenvector and therefore belongs to the Segré-Pleba\'nski class [111,1]~\cite{Stephani2003}, or, equivalently, type I in the Hawking-Ellis nomenclature~\cite{Hawking1973}.

\paragraph*{\textbf{Distributional energy-momentum tensor in $\Sext$.}}
The induced metric on the surface $\Sext$ is given by:
\begin{align}
    \dd s_+^2 = - \dd \Text^2 +\left(\Rext + b \cos \bext \right)^2 \dd \aext^2 + b^2 \dd \bext^2\,. 
\end{align}
The nontrivial components of the distributional energy-momentum tensor are given by (see App.~\ref{Sec:Sext})
\begin{align}
    & 8 \pi S^+_{\Text\Text} = -\frac{1}{b}- \frac{\cos(\bext)}{\Rext + b \cos (\bext)} + \frac{\partial_\ell \mathcal{F}(\Rext,\bext)}{2(\Rext + b \cos (\bext))^2}, \label{Eq:S+tautau} \\
    & 8 \pi S^+_{\aext\aext} =  (\Rext + b \cos (\bext))^2\left(\frac{1}{b}-\frac{1}{2}H'(\Rext)\right), \label{Eq:S+alpha alpha} \\
    & 8 \pi S^+_{\bext\bext} = \left[\frac{\cos(\bext)}{\Rext + b \cos (\bext)}-\frac{1}{2}H'(\Rext)-\frac{\partial_\ell \mathcal{F}(\Rext,\bext)}{2(\Rext + b \cos (\bext))^2}\right]b^2\, . \label{Eq:S+betabeta}    
\end{align}

\paragraph*{\textbf{Distributional energy-momentum tensor in $\Sint$}.}
Regarding the shell at $\Sint$, we have that the induced metric reads
\begin{align}
       \dd s_-^2 = -\dd  \Tint^2 + \frac{b^2}{\pi^2} \dd  \aint^2 + b^2 \dd  \bint^2,
\end{align}
where the identification $M = b$ is required to ensure the continuity of the metric. The nontrivial components of the distributional energy-momentum tensor are given by (see App.~\ref{Sec:Sint}):
\begin{align}
    8 \pi S^-_{\Tint\Tint} & = -\frac{\pi^2}{2b^2}\partial_\ell\mathcal{F}(\Rint,\bint) , \\
    8 \pi S^-_{\aint\aint} & = \frac{b^2}{\pi^2}\left[\frac{1}{2}H'(\Rint) -\frac{1}{\Rint}\right] , \\
    S^-_{\bint\bint} & =\pi^2 S^-_{\aint\aint} - b^2 S^-_{\Tint\Tint} . \label{Eq:Sbetabeta} 
\end{align}

\subsection{Only one shell model}
\label{Subsec:Try_one_shell}
By carefully selecting the functions $\mathcal{F}(\ell, \beta)$ and $H(\ell)$, we can match the two regions without requiring a thin-shell at $\ell = \Rint$. In fact, we just need:
\begin{equation}
    \lim_{\ell\searrow \Rint} \partial_\ell\mathcal{F}(\ell,\beta) = 0\,,\qquad \lim_{\ell\searrow \Rint} H'(\ell) = \frac{2}{\Rint}\,,\label{eq:no-shell}
\end{equation}
so that both $S^-_{\Tint\Tint}$ and $S^-_{\aint\aint}$ are zero, and consequently $S^-_{\bint\bint}$ is also zero, as it can be seen from Eq.~\eqref{Eq:Sbetabeta}. 

A simple choice of functions that satisfy the interpolation conditions~\eqref{eq:calFH_interpcond}, as well as the conditions~\eqref{eq:no-shell} required to avoid the presence of an internal shell at $\ell = \Rint$ is given by the function $\mathcal{F}(\ell, \beta)$ defined in Eq.~\eqref{eq:FG}, with the following specific choices for $F(\ell)$, $G(\ell)$ and $H(\ell)$:
 \begin{align}
     F(\ell) &= \left(\frac{\ell-\Rint}{\Rext-\Rint}\right)^2\,,\nonumber\\
     G(\ell) &= \dfrac{M^2}{\pi^2} \frac{(\Rext-\ell)(\ell+\Rext-2\Rint)}{(\Rext-\Rint)^2}\,,\nonumber\\
     H(\ell)&=\frac{-2\ell^2+(\Rext+\Rint)(2 \ell-\Rint)}{\Rint(\Rext-\Rint)}\,,
\end{align}
all of which are strictly positive in the range $\ell\in(\Rint,\Rext)$. In particular, $F(\ell)$ and $G(\ell)$ satisfy~\eqref{eq:FGH_interpcond}.

It is not possible to do the same with the external shell for this class of geometries in which the line element for the interpolating region is given by Eq.~\eqref{eq:region2General}. The absence of external shell requires the vanishing of the right hand-side of Eqs.~\eqref{Eq:S+tautau}--\eqref{Eq:S+betabeta}. From Eq.~\eqref{Eq:S+alpha alpha} we obtain the condition
\begin{align}
    \frac{1}{b} - \frac{1}{2} H'( \Rext) = 0. 
    \label{Eq:NoShell1}
\end{align}
If we now add Eqs.~\eqref{Eq:S+tautau} and Eq.~\eqref{Eq:S+betabeta} times $1/b^2$, we find
\begin{align}
    -\frac{1}{b} - \frac{1}{2} H' \left( \Rext \right) = 0,
\end{align}
which clearly is in contradiction with Eq.~\eqref{Eq:NoShell1}. Although it may be possible to avoid the external thin shell considering a broader family of interpolating regions, we prefer to restrict our consideration to this family of functions for the sake of simplicity.

\section{Energy conditions violations}
\label{Sec:EC}

As discussed above, classic theorems ensure that violations of energy conditions need to occur in the external region to have a toroidal topology~\cite{Hawking1973}. On a heuristic basis, we expect energy condition violations since: i) the flat spatial torus of the Geroch-Hartle line element is converted into a curved one so that it can match the flat exterior, ii)~the interpolating region between the horizon and infinity, which is geometrically a tube, is transformed into a flat region with spheres whose area increases without a bound as we approach infinity. In both processes, geodesic congruences diverge, implying violations of the energy conditions. For certain choices of interpolating functions, the two processes occur simultaneously in the same region, making them difficult to disentangle. This is for instance what would happen if we had a model without any thin shell. The next two subsections are dedicated to study of the violations of energy conditions due to these two processes separately as a warm-up exercise to gain intuition.

\subsection{Developing some intuition: infinite toroidal tube}
\label{SubSec:InfinteTorus}

To understand where and how does the geometry violate the energy conditions due to the curved metric of the torus, it is instructive to look first at a simpler geometry. Consider the spacetime $\mathbb{R}^2\times \mathbb{T}^2$ with metric
\begin{align}
    \dd s^2 = - \dd  \tau^2 + \dd \ell^2 + \left[ \Rext + b \cos(\beta) \right]^2  \dd \alpha^2 + b^2 \dd\beta^2 \,.
\end{align}
Its spatial sector for fixed $\tau$ is an infinite tube of toroidal sections that inherits a curved metric for each value of $\ell$. We are taking the torus to be nondegenerate, i.e., we consider $\Rext> b$ again. 

The energy-momentum tensor supporting this geometry (assuming Einstein equations) violates the Null Energy Condition (NEC), and consequently, all other pointwise energy conditions as well: the Weak Energy Condition (WEC), the Dominant Energy Condition (DEC) and the Strong Energy Condition (SEC)~\cite{Kontou2020}. To prove it, it is sufficient to take the null vector field $k^\mu := (\partial_\tau)^\mu + b^{-1}(\partial_\beta)^\mu$ and contract it with the Einstein tensor to find
\begin{equation}
    G_{\mu \nu} k^\mu k^\nu =\frac{1}{b} \frac{\cos(\beta)}{\Rext + b\cos(\beta)}\,,
\end{equation}
which becomes negative for $\beta \in (\pi/2, 3\pi/2)$.

As an intuitive way of understanding why energy conditions are violated, we can consider a congruence of geodesic observers at constant $\alpha=\alpha_0$ and $\ell=\ell_0$, i.e., timelike curves going around the $\beta$ direction. For instance, we can take the parametrization
\begin{equation}\label{Eq:geodesicszeta}
    (\tau, \ell_0, \alpha_0, \beta(\tau)) \qquad \text{with}\qquad \beta(\tau)= \frac{A}{b} \tau\quad \text{for}\quad A\in(0,1),
\end{equation}
whose corresponding normalized tangent vector is given by:
\begin{equation}
    \zeta^\mu = \frac{1}{\sqrt{1-A^2}}\left((\partial_\tau)^\mu + \frac{A}{b}(\partial_\beta)^\mu\right)\,.
\end{equation}
This congruence is twist-free and the expansion is given by:
\begin{equation}
    \theta(\tau) = - \frac{A}{\sqrt{1-A^2}} \frac{\sin(\beta(\tau))}{\Rext + b \cos(\beta(\tau))}\,.
\end{equation}
The derivative with respect to the parameter is:
\begin{equation}
    \frac{\dd \theta}{\dd \tau}  = - \frac{A^2}{\sqrt{1-A^2}} \frac{b+\Rext\cos(\beta(\tau))}{\Big[\Rext + b \cos(\beta(\tau))\Big]^2}\,,
\end{equation}
whose sign depends on the sign of the function $b+\Rext\cos(\beta(\tau))$ which is changing since $\Rext> b$. This indicates that different observers of this congruence become closer or separate, and hence violations of the SEC occur, depending on the value of $\beta$. 
\begin{figure}[H]
    \centering
    \includegraphics[scale=1]{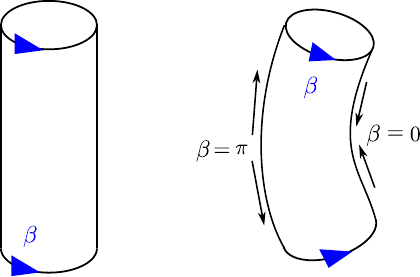}
    \caption{Pictorial representation of the geometrical operation (stretching-compressing) required to go from a flat to a curved torus.}
    \label{fig:flatcurvedtorusdeform}
\end{figure}
%

\subsection{Developing intuition: external thin shell}
\label{Subsec:ExternalThinShell}

\begin{figure}[H]
    \centering
    \includegraphics[scale=1]{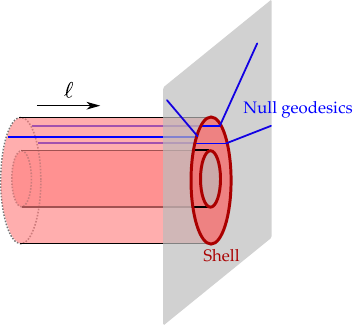}
    \caption{Pictorial representation of the light rays diverging when they cross the external shell and reach the asymptotic flat region.}
    \label{Fig:NEC}
\end{figure}

In addition to the energy-condition violations stemming from the curvature of the toroidal spatial slices, further violations are expected from the matching between the toroidal tube and the asymptotically flat region. While one might attempt a smooth transition, the most straightforward approach pursued here involves introducing a distributional energy-momentum tensor associated with a thin shell. It is easy to see that the energy-momentum tensor associated with this shell must violate the NEC, and hence, all other pointwise energy conditions as well. Null rays traveling parallel to the tube and approaching the shell will enter the flat region, where they begin to diverge, that is, they move apart from one another. A pictorial representation of the diverging rays can be found in Fig.~\ref{Fig:NEC}. 

To isolate the violations of energy conditions arising from this matching, we can work out a simpler example before considering the energy-condition violations occurring in our full model. Consider taking the infinite torus from the previous section, but now cutting it at $\ell = \Rext$, i.e., we take the line element
\begin{align}
    \dd s^2 = - \dd  \tau^2 + \dd \ell^2 + \left[ \Rext + b \cos(\beta) \right]^2  \dd \alpha^2 + b^2 \dd\beta^2 \,,
    \label{Eq:InfiniteTubeCut}
\end{align}
for $\ell \leq \Rext$. We glue it at $\ell = \Rext$ where the induced metric reads
\begin{align}
    \dd s^2 = - \dd \tau^2 +  \left[ \Rext + b \cos(\beta) \right]^2  \dd \alpha^2 + b^2 \dd\beta^2 \,, 
\end{align}
with the external region of the flat spacetime from which we have removed a torus at each time, i.e., we take the line element
\begin{align}
    \dd s^2_1 = -\dd T^2 + \dd X^2 + \dd Y^2 + \dd Z^2, 
\end{align}
for the region
\begin{align}
    \left( \sqrt{X^2 + Y^2} - \Rext \right)^2 + Z^2 \geq b^2.
\end{align}
The equality holds at the matching surface. In fact, the Israel junction conditions have already been worked out for the matching between these geometries as the line element in Eq.~\eqref{Eq:InfiniteTubeCut} simply corresponds to the one in Eq.~\eqref{eq:region2General} upon the identification $H(\ell) = 1$ and $\mathcal{F}(\ell, \beta) = \left( \Rext + b \cos \left( \beta \right) \right)^2 $. Thus, by plugging this specific choice of functions in Eqs.~\eqref{Eq:S+tautau}-\eqref{Eq:S+betabeta} we find 
\begin{align}
    & 8 \pi S^+_{\Text\Text} = -\frac{1}{b}- \frac{\cos(\bext)}{\Rext + b \cos (\bext)} ,\\
    & 8 \pi S^+_{\aext\aext} =  \frac{\left(\Rext + b \cos (\bext)\right)^2}{b}, \\
    & 8 \pi S^+_{\bext\bext} = \frac{b^2 \cos(\bext)}{\Rext + b \cos (\bext)}\, . 
\end{align}
It corresponds to the energy-momentum tensor of a fluid on the surface\footnote{It corresponds to the type I in Hawking-Ellis classification~\cite{Hawking1973}.} with anisotropic pressures and density that are given by
\begin{align}
     \sigma & = - \frac{1}{b} - \frac{\cos \beta }{\Rext+ b\cos \beta} , \\
    p_{\alpha} & = \frac{1}{b}, \\ 
    p_{\beta} & = \frac{\cos \beta }{\Rext+ b\cos \beta}.
\end{align}
It is immediate to conclude that this energy-momentum tensor violates the NEC. To see this, we just recall that for the NEC we would need to have $\sigma + p_i \geq 0$ for $i = \alpha,\beta$ and we have:
\begin{align}
    & \sigma + p_{\alpha} = - \frac{\cos \beta }{\Rext+ b\cos \beta}, \\
    & \sigma + p_{\beta}  = - \frac{1}{b}.
\end{align}
We have $\sigma + p_{\alpha}<0$ everywhere and $\sigma + p_{\beta} <0$ for $\beta \in [0, \pi/2) \cup (3\pi/2,0]$. Thus, we automatically conclude that the NEC is violated. 

\subsection{General analysis near the end of the interpolating region}
\label{Subsec:Finite}

The proof for the finite interpolating toroidal region is essentially an adaptation of the argument presented in the previous subsection to this specific setup. Since the solution near $\Rext$ closely resembles that of the infinite torus, we also expect violations of energy conditions.

To study potential violations of the WEC and the NEC\footnote{We recall that violations of the NEC imply also violations of the WEC.}, we introduce the one-parameter family of vector fields:
\begin{equation}
    (A_\vpar)^\mu := H(\ell)^{-1/2} (\partial_\tau)^\mu + \vpar b^{-1} (\partial_\beta)^\mu\,,
\end{equation}
with $\vartheta \in [0,1]$, and define the corresponding scalar
\begin{equation}\label{eq:scalarEC}
        \mathcal{Z}_\vpar := 8\pi T_{\mu \nu}\, (A_\vpar)^\mu (A_\vpar)^\nu\,,
\end{equation}
This family of vector fields interpolates between two situations:
\begin{itemize}
    \item Case $\vpar = 0$. We are considering free falling observers in orbits of $\partial_{\tau}$ 
    \begin{equation}\label{eq:observertau}
        u^\mu := (A_0)^\mu = H(\ell)^{-1/2} (\partial_\tau)^\mu \,.
    \end{equation}
    Then the scalar \eqref{eq:scalarEC} is proportional to the energy density $T_{\mu \nu}u^\mu u^\nu$ measured by the observers:
    \begin{equation}
        \mathcal{Z}_0 = 8\pi T_{\mu \nu}u^\mu u^\nu = \frac{1}{H(\ell)} G_{\tau\tau} \,.
    \end{equation}
    Wherever $\mathcal{Z}<0$, the WEC is violated. We highlight, though, that $\mathcal{Z}_0\geq 0$ does not imply that the WEC is satisfied, since $\mathcal{Z}_0$ only represents the energy density measured by the particular observer $u^\mu$, whereas the WEC requires the energy density measured by \emph{every} observer to be non-negative.
    
    \item Case $\vpar = 1$. This corresponds to a null vector field
    \begin{equation}
        k^\mu := (A_1)^\mu = H(\ell)^{-1/2} (\partial_\tau)^\mu +  b^{-1} (\partial_\beta)^\mu \,,
    \end{equation}
    and the scalar $\mathcal{Z}_1$ can be used to explore violations of the NEC. As in the previous case, if $\mathcal{Z}_1<0$ the NEC is violated. Again, the case $\mathcal{Z}_1\geq 0$ does not allow us to conclude whether the NEC is violated or not, since we are focusing only on a specific null vector, $k^\mu$.
\end{itemize} 
We can compute $\mathcal{Z}_\vpar$ for the metric \eqref{eq:region2General} and we find:
\begin{align}\label{eq:Z_calFH}
     \mathcal{Z}_\vpar &= \frac{1}{4b^2}\left[\left(\frac{\partial_\beta \mathcal{F}}{\mathcal{F}}\right)^2  - 2   
    \frac{\partial^2_\beta \mathcal{F}}{\mathcal{F}} 
    \right]
    +\frac{1-\vpar^2}{4}\left[ \left(\frac{\partial_\ell \mathcal{F}}{\mathcal{F}}\right)^2 - 2  \frac{\partial^2_\ell \mathcal{F}}{\mathcal{F}}  
    \right] \nonumber\\
    &\qquad +
    \frac{\vpar^2}{4}\left[2\frac{H''}{H}-\left(\frac{H'}{H}\right)^2
    +\frac{H'}{H}\frac{\partial_\ell \mathcal{F}}{\mathcal{F}}\right]
    \,,
\end{align}
from which the limiting cases $\vartheta =0,1$ can be easily read.

Now, we notice that for any metric of the form~\eqref{eq:region2General} such that the functions $H(\ell)$ and $\mathcal{F} (\ell, \beta)$ are at least $C^2$ and satisfy the conditions~\eqref{eq:calFH_interpcond} involving the matching at $\Sext$, we have the following expansion for $\mathcal{Z}_\vpar$
\begin{align}
        \mathcal{Z}_\vpar & =   
        - \frac{1}{b(\Rext-b)} +\frac{1-\vpar^2}{4}\left[\left(\frac{\partial_\ell \mathcal{F}(\Rext,\pi)}{(\Rext-b)^2}\right)^2 -2 \frac{\partial^2_\ell \mathcal{F}(\Rext,\pi)}{(\Rext-b)^2}\right] \nonumber\\
        &\qquad +
        \frac{\vpar^2}{4}\left[2H''(\Rext)- H'(\Rext)^2
        +\frac{H'(\Rext) \partial_\ell \mathcal{F}(\Rext,\pi)}{(\Rext-b)^2}\right] + \ldots, \label{eq:lemma_Z}
\end{align}
where the dots represent subleading terms of order 1 in $\beta-\pi$ and/or order 1 in $\ell-\Rext$.

If we impose the additional conditions
\begin{align}
    \lim_{\ell \nearrow \Rext}\partial_\ell\mathcal{F}(\Rext, \beta) = \lim_{\ell \nearrow \Rext}\partial^2_\ell \mathcal{F}(\Rext, \beta)=0\qquad\forall \beta\in[0,2\pi),
    \label{Eq:Condition1}
\end{align}
on the partial derivatives with respect to $\ell$, Eq.~\eqref{eq:lemma_Z} reduces to 
\begin{equation}\label{eq:asymp_FG}
    \mathcal{Z}_\vpar \sim   - \frac{1}{b(\Rext-b)} + \frac{\vpar^2}{4}(2H''(\Rext)-H'(\Rext)^2) \qquad  (\ell\to \Rext, \beta\to \pi)\,.
\end{equation}
In the case $\vpar=0$ we get
\begin{equation}\label{eq:auxthm}
    \mathcal{Z}_0 \sim   - \frac{1}{b(\Rext-b)}\qquad  (\ell\to \Rext, \beta\to \pi)\,,
\end{equation}
which is negative since we are assuming $\Rext>b$. This indicates a violation of the WEC, as anticipated based on the intuition developed from the infinite torus example discussed above. If we take $\vpar = 1$ and instead and further assume
\begin{equation}\label{eq:limRext_H}
    H'(\Rext) = H''(\Rext) = 0\,,
\end{equation}
we get exactly the same leading order term found in Eq.~\eqref{eq:auxthm}, so the NEC is also violated in a neighborhood of $\ell=\Rext$ and $\beta=\pi$.

Finally, if we take the function $\mathcal{F} (\ell, \beta)$ as in Eq.~\eqref{eq:FG} so that the line element is given by Eq.~\eqref{eq:gregion2}, and impose the condition~\eqref{Eq:Condition1}, which for this choice of function can be achieved enforcing
\begin{align}
    F'(\Rext)= F'' (\Rext)= G'(\Rext)= G''(\Rext)=0,
    \label{Eq:Conditions2}
\end{align}
we find violations of the WEC close to $\ell=\Rext$ and $\beta=\pi$. Additionally, if we impose the condition
\begin{align}
    H'(\Rext) = H''(\Rext) = 0,
    \label{Eq:Conditions3}
\end{align}
the NEC is also violated, as discussed for the general $\mathcal{F}(\ell,\beta)$ function. 

\subsection{Some specific models}
\label{Subsec:specific}

The analysis in the previous subsection has been eminently local near the external shell at $\ell=\Rext$. More detailed features of the spacetimes satisfying the results above, such as the way in which the negative energy density is distributed for $\ell\in(\Rint,\Rext)$, require specifying the functions appearing in the metric. 

For simplicity, we now focus on the lowest-degree polynomial ansatz of the type~\eqref{eq:FG} that satisfies the interpolation conditions~\eqref{eq:FGH_interpcond}, the no-internal-shell condition~\eqref{eq:no-shell} and conditions~\eqref{Eq:Conditions2}-\eqref{Eq:Conditions3}. It can be obtained through Hermite interpolation~\cite{Davis1963} and the result reads:
 \begin{align}
     F(\ell) &= \frac{(\ell-\Rint)^2\big[3\ell^2-2\ell(4\Rext-\Rint)+6\Rext^2-4\Rext\Rint+\Rint^2\big]}{(\Rext-\Rint)^4}\,,\nonumber\\
     G(\ell) &= \dfrac{m^2}{\pi^2} \frac{(\Rext-\ell)^3(3\ell+\Rext-4\Rint)}{(\Rext-\Rint)^4}\,,\nonumber\\
     H(\ell)&=\frac{2\ell(\Rext-\ell)^3+2\ell\Rint(\ell^2-3\ell\Rext+3\Rext^2)-\Rint(\Rext^3+3\Rext^2\Rint-3\Rext\Rint^2+\Rint^3)}{\Rint(\Rext-\Rint)^3}\,,\label{eq:greatFGH}
\end{align}
which are strictly positive for the interval $\ell\in(\Rint,\Rext)$.

\begin{figure}[H]
    \centering
    \includegraphics[scale=0.49]{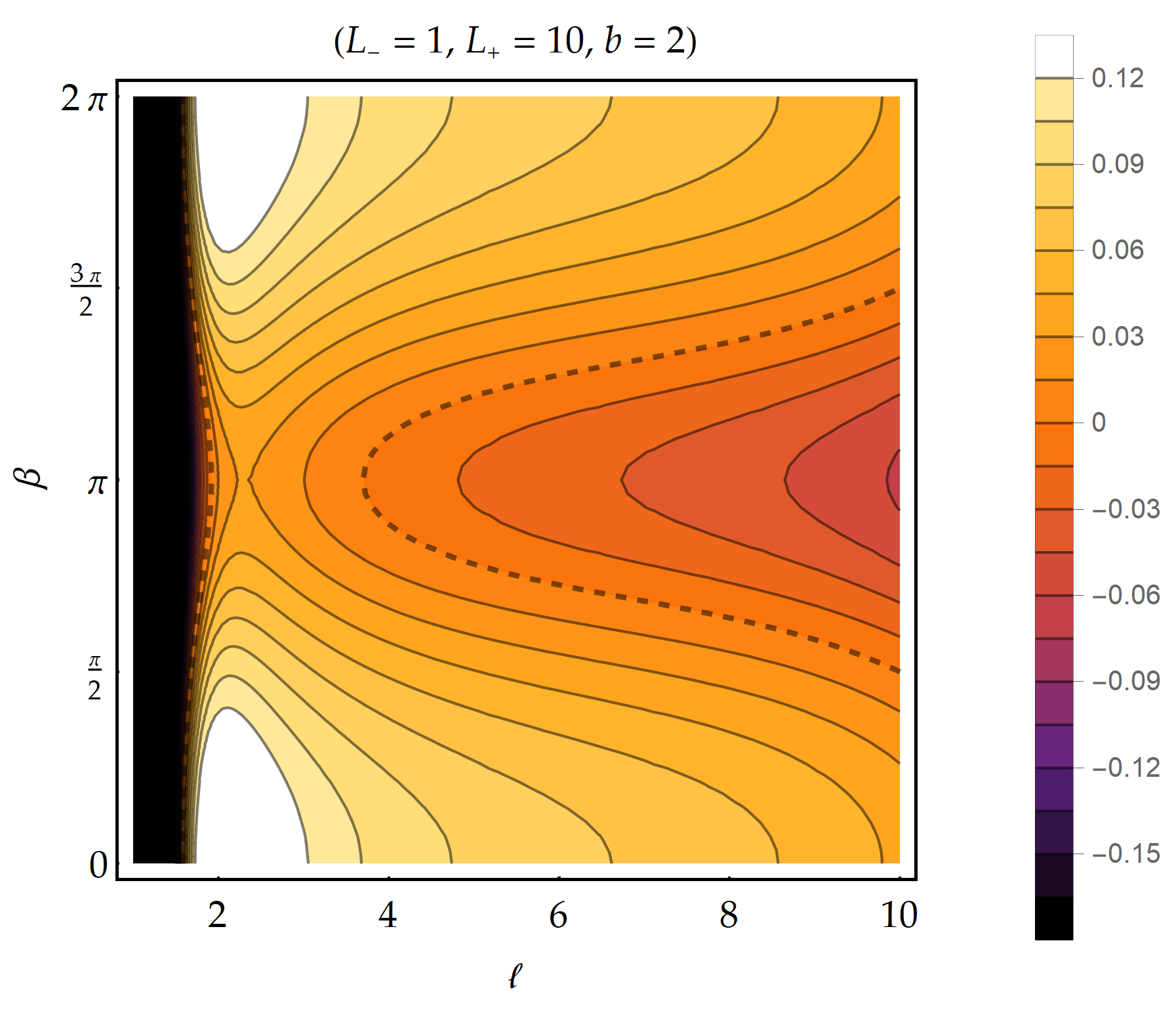}
    \includegraphics[scale=0.49]{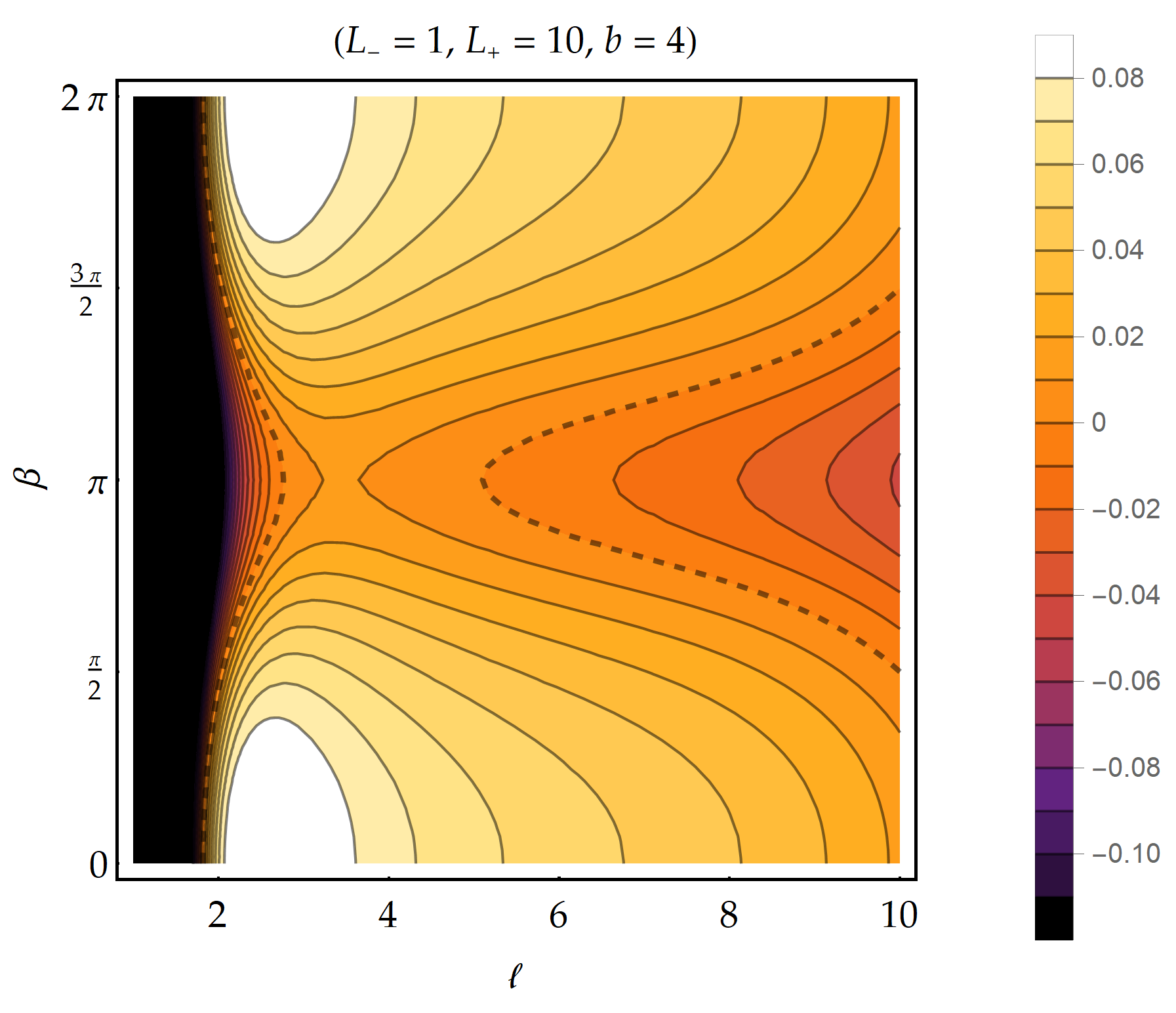}
    \includegraphics[scale=0.49]{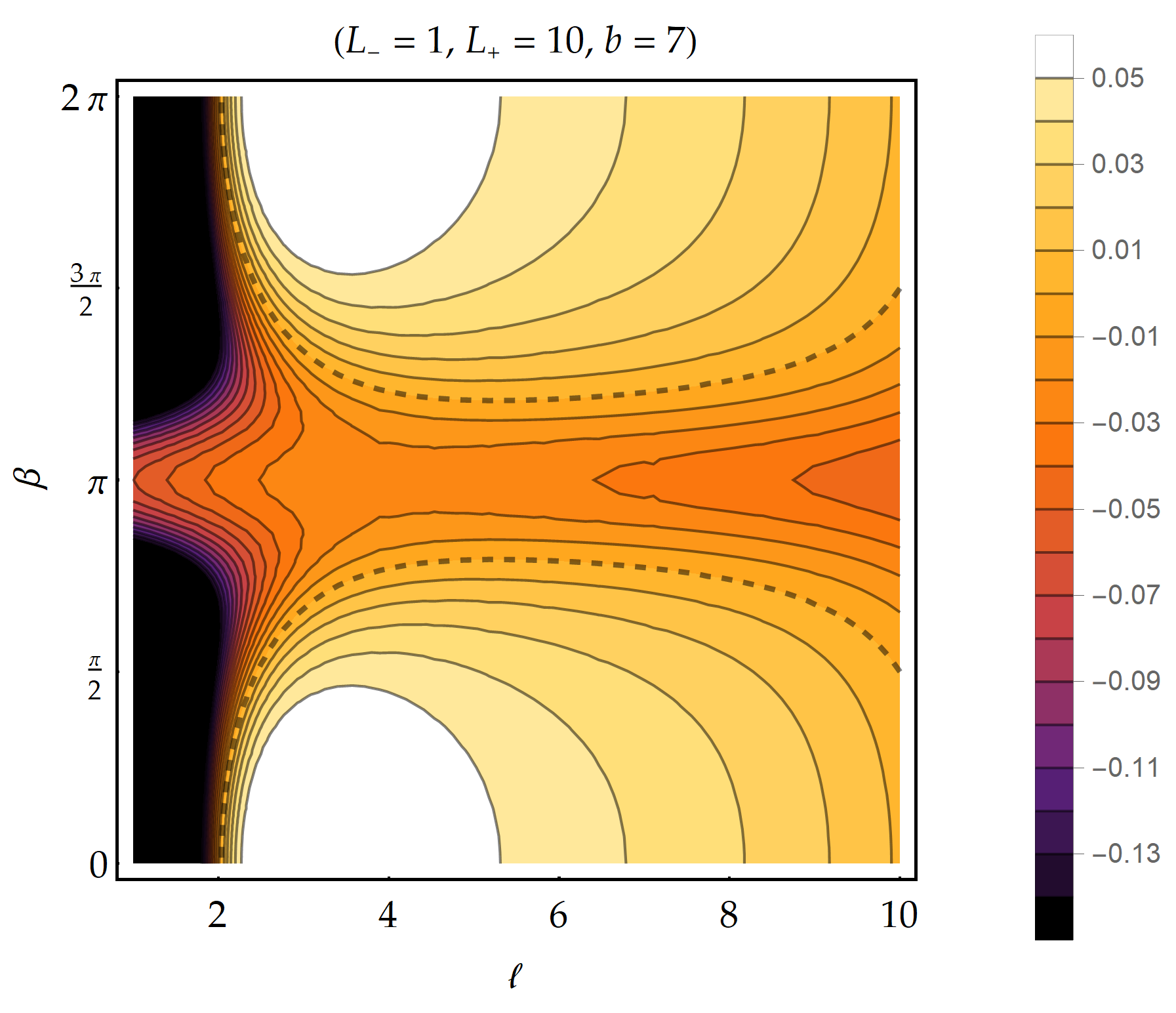}
    \includegraphics[scale=0.49]{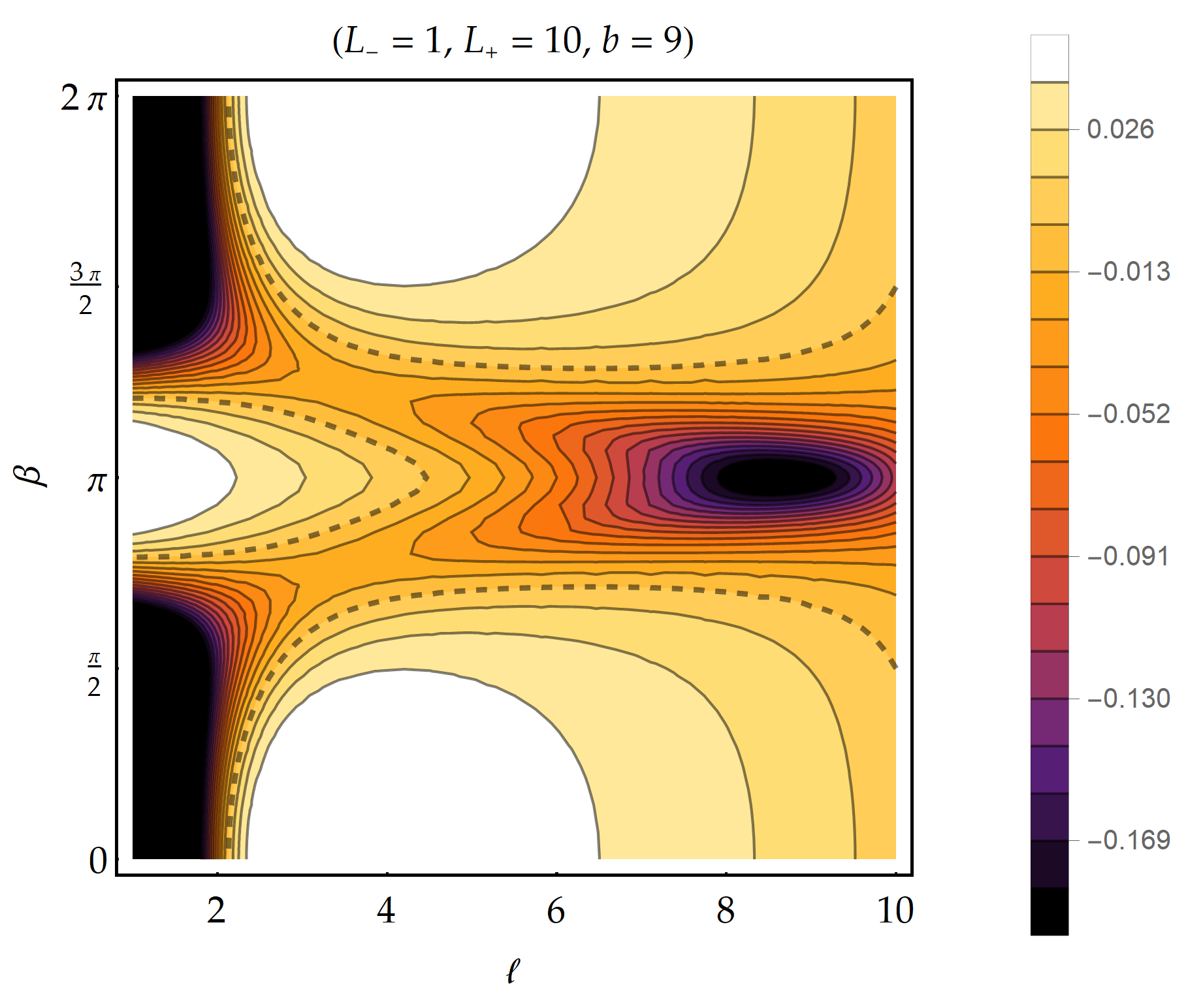}
    \caption{Contour plot for the values of $\mathcal{Z}_0(\beta, \ell)$ within the interpolating region $\Omega_2$ for the ansatz \eqref{eq:greatFGH} and for different values of $b$. The dashed line represents the limit between positive and negative energy densities ($\mathcal{Z}_0=0$). For the plots, we took $\Rint=1$ and $\Rext=10$.}
    \label{fig:Z0FH0}
\end{figure}

In the following, we analyze violations of the energy conditions throughout the entire $\Omega_2$ region for this specific choice. In Fig.~\ref{fig:Z0FH0}, we represent the function $\mathcal{Z}_0$ (essentially, the energy density seen by observers in the orbits of $u^\mu$) as a function of the coordinates $\ell$ and $\beta$ of the interpolating region for a particular set of parameters. The contour plot clearly shows that the negative energy region enclosed by the surface of vanishing energy density (the dashed line) will be reached by any observer sufficiently close to $\ell = \Rext$ and $\beta = \pi$. 

\begin{figure}[H]
    \centering
    \includegraphics[scale=0.49]{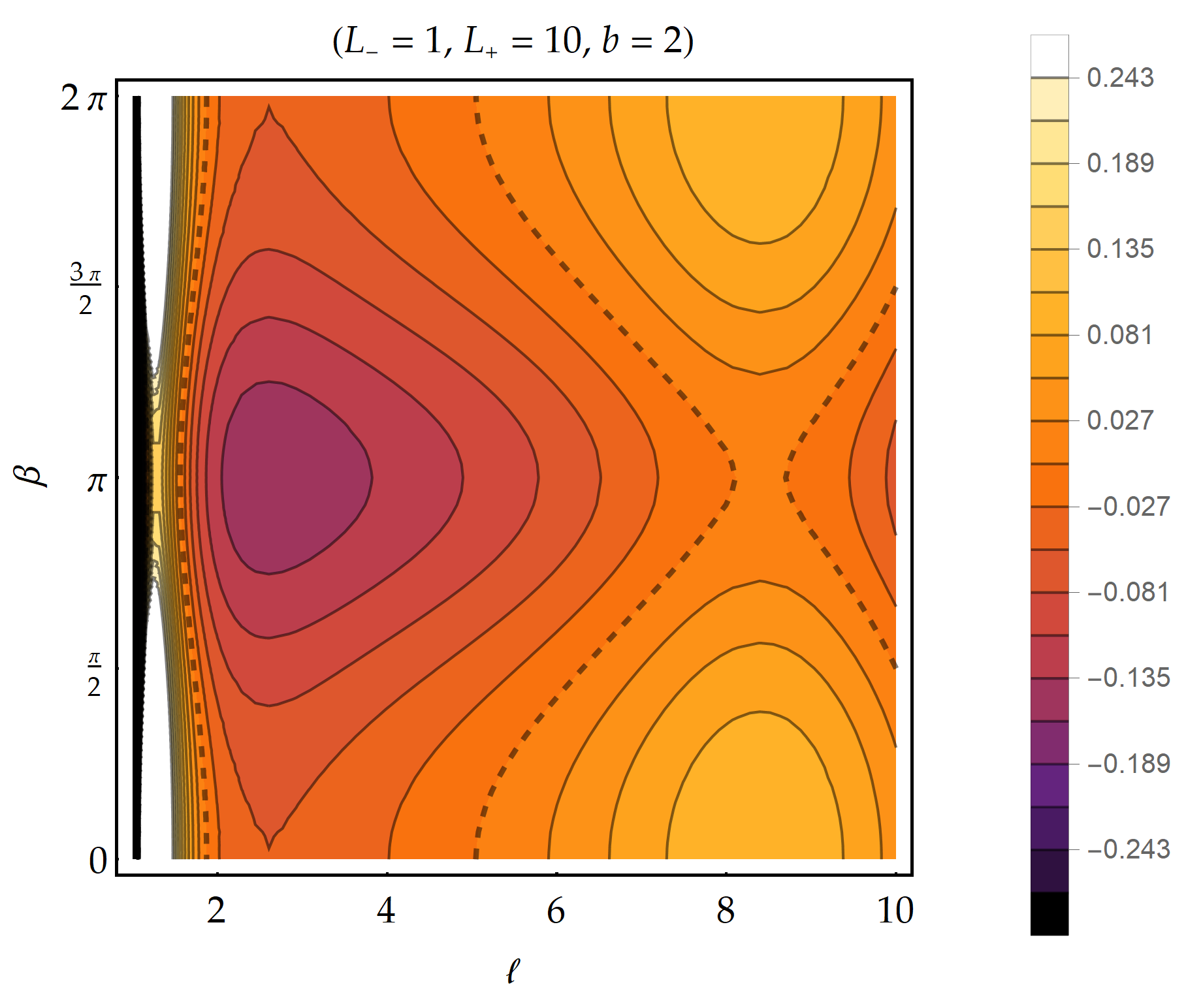}
    \includegraphics[scale=0.49]{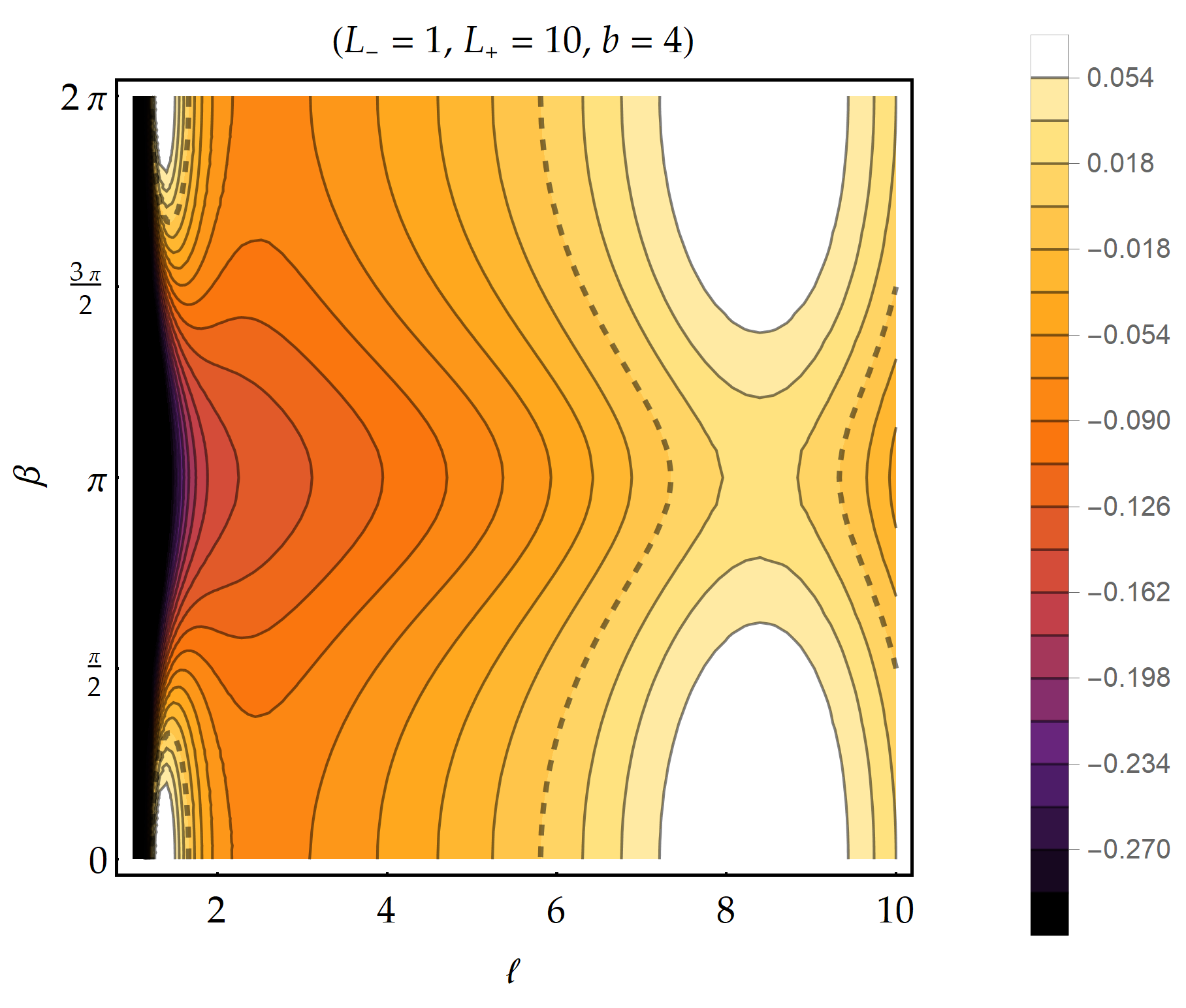}
    \includegraphics[scale=0.49]{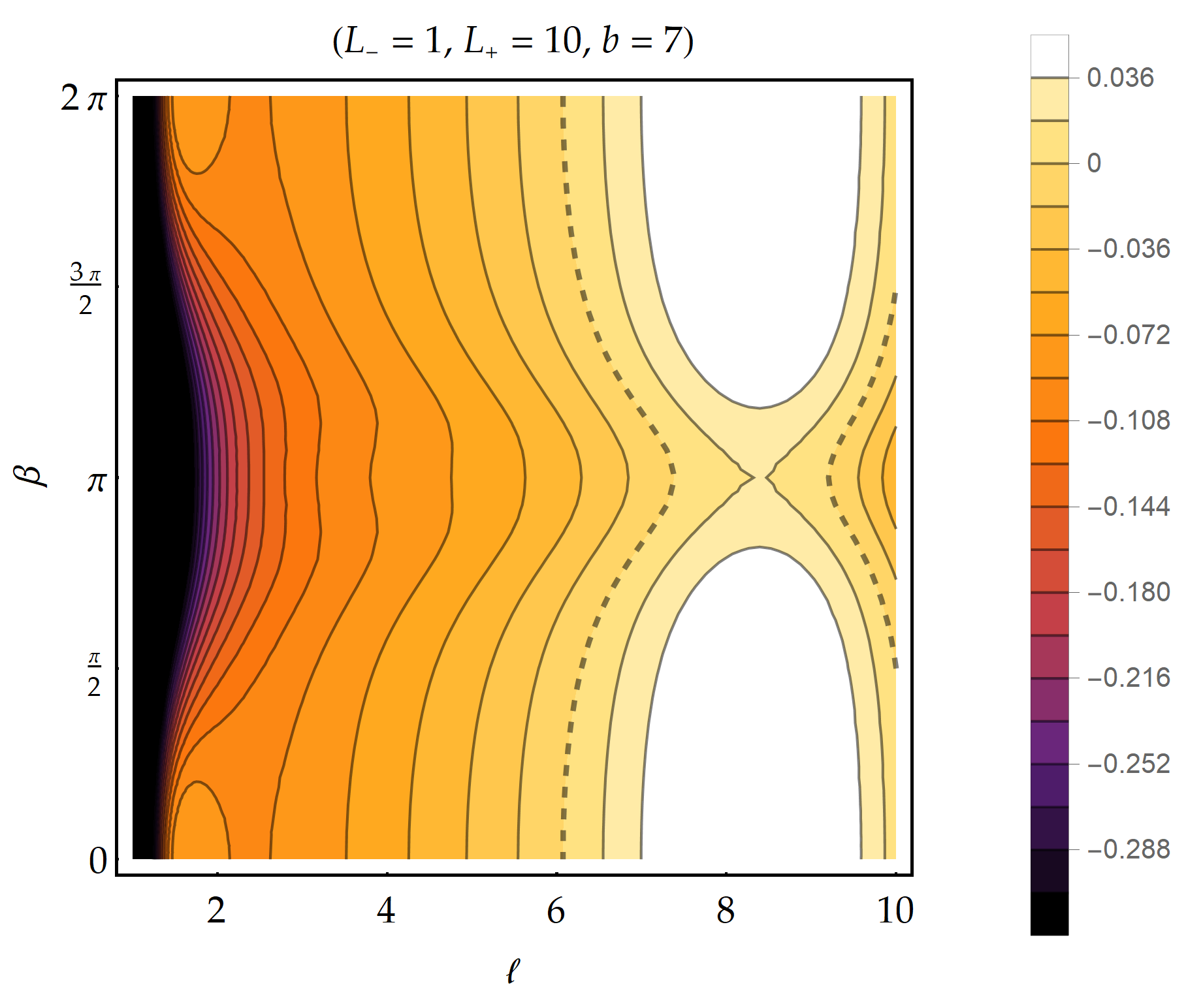}
    \includegraphics[scale=0.49]{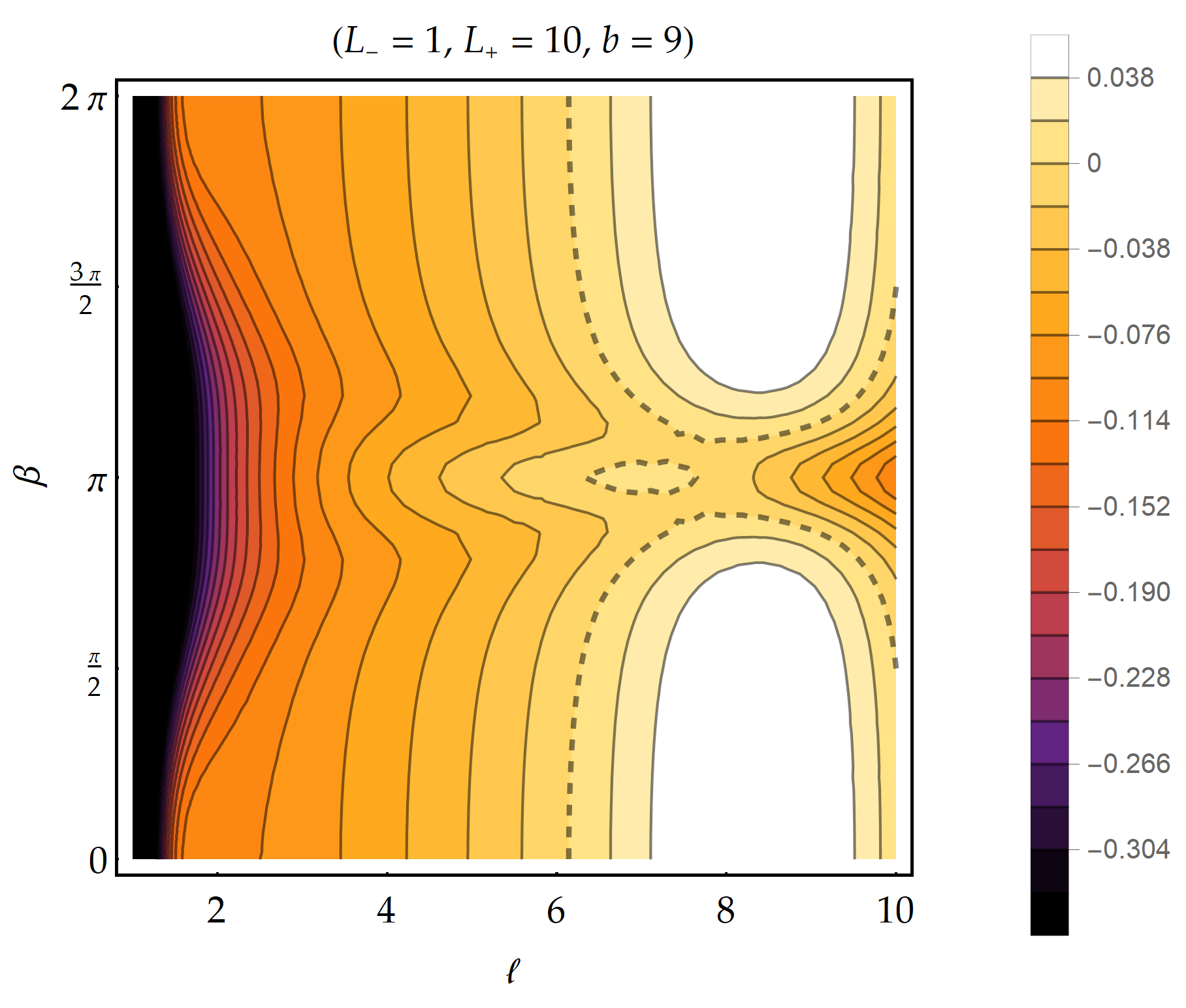}
    \caption{Contour plot for the values of $\mathcal{Z}_1(\beta, \ell)$ within the interpolating region $\Omega_2$ for the ansatz \eqref{eq:greatFGH} and for different values of $b$. The dashed line represents the limit between positive and negative values ($\mathcal{Z}_1=0$). For the plots, we took $\Rint=1$ and $\Rext=10$.}
    \label{fig:Z1FH0}
\end{figure}

We also present, in Fig.~\ref{fig:Z1FH0}, the scalar $\mathcal{Z}_1$ as a function of $\ell$ and $\beta$ for the same particular set of parameters. The dark regions represent those in which the NEC is violated.
\section{Conclusions}
\label{Sec:ConclusionTorus}

In this chapter, we have presented a comprehensive overview of static and axisymmetric toroidal black holes in four dimensions. First, we have revisited the key assumptions underlying the classic theorems that forbid their existence, with a special emphasis on the role of energy condition violations. We have also reviewed previous attempts to build toroidal black holes in 4D and how and why such attempts failed in providing a nonsingular external region. 

We then introduced the simplest construction of a toroidal black hole, realized by matching three distinct regions: an external flat spacetime, an inner region described by the simplest locally vacuum toroidal black hole (a Rindler line element with periodically identified transverse directions), and an intermediate region containing a nontrivial matter distribution that connects the two. We analyzed the matter content in this intermediate region, which consists of an anisotropic fluid and two thin shells: one located at the interface with the inner region, and the other at the interface with the outer region.

While a specific choice of the interpolating matter profile can eliminate the internal shell, the external shell remains unavoidable within the family of geometries that we consider. Although it is possible to consider a broader family of geometries for the interpolating regions such that the external thin-shell is not present, we have focused on this particular geometry for its easier interpretation. Additionally, we explained why a more straightforward construction in which one directly attempts to match the inner toroidal black hole with the external Minkowski spacetime via a single shell does not work: the induced metric on the shell would be curved on one side and flat on the other, thereby violating the first of Israel’s junction conditions, which requires continuity of the first fundamental form across the shell.

We have explicitly studied the violations of energy conditions and confirmed that the DEC is violated, in agreement with the implications of Hawking's theorem. In particular, we have demonstrated the violation of the WEC and NEC both near and on the external shell, as approached from the interpolating region. Additionally, we have examined the properties of the fluid in the intermediate region for specific choices of the interpolating functions. 

There are several promising directions for future research building on these results. First, the assumption of a flat external region is a significant simplification, and it would be interesting to replace it with a spacetime that is only asymptotically flat, such as the Schwarzschild or Curzon metric. Additionally, the analysis could be extended to the more general case of stationary and axisymmetric configurations. Given the integrability of the Ernst equation, addressing this problem in such a setup seems feasible. However, it would require a comprehensive characterization of local stationary and axisymmetric vacuum black holes, analogous to the Geroch-Hartle analysis for the static case.

Finally, the results discussed here are restricted to asymptotically flat scenarios, and it remains an open question whether energy condition violations are still necessary to support these configurations in alternative contexts, such as black holes embedded in cosmological spacetimes.


\chapter{Compactness bounds for spherically symmetric objects}
\label{Ch7:Buchdahl}

\fancyhead[LE,RO]{\thepage}
\fancyhead[LO,RE]{Compactness bounds for spherically symmetric objects}


If one were to start from the premise that the final state of any gravitational phenomenon must be horizonless, a natural question arises: what, then, are astrophysical black holes, and what is their true geometry? In fact, verifying whether the dark and compact objects that we observe in the universe strictly correspond to general relativistic black holes is still an open question that is attracting a large amount of interest and effort~\cite{Collmar1998,Abramowicz2002}. 

In astrophysics, it has long been assumed that any object with sufficiently large gravitational mass confined within a small enough region has been considered to be a black hole. This belief has been supported by some classic results in GR putting constraints to the mass and compactness that stellar-like configurations can sustain. On one side, Chandrasekhar’s classic limit provides an upper bound on the mass a compact object can have before the degeneracy pressure of fermions is no longer sufficient to prevent gravitational collapse~\cite{Chandrasekhar1931a,Chandrasekhar1931b}\footnote{Apparently, this limit was first found by Edmund Stoner~\cite{Nauenberg2008}.}. On the other, the Buchdahl and Bondi bounds constrain the compactness of the objects i.e., the quotient between the total mass $M$ and the radius $R$ of spherically symmetric objects: $C(R):=2M/R$.

First, Buchdahl proved that the compactness is bounded as $C(R)<8/9$. The theorem makes no strong hypothesis regarding the equation of state of matter beyond being barotropic. Its two central assumptions are just the isotropy of the fluid, namely that pressures in the tangential and radial directions are equal, and that the energy density is a positive, (outwards) monotonically decreasing function of the radius. 

Even when the density profile is not required to be a monotonically decreasing function of the radius, it is still possible to establish upper bounds on compactness, provided the density remains non-negative. Bondi~\cite{Bondi1964} found for this situation a general upper bound of $C(R) < 12 \sqrt{2} - 16 \approx 0.971$, which is less restrictive than the Buchdahl bound \mbox{$C(R)<8/9 \approx 0.889$} but still well-below the black hole compactness $C(R)=1$. Bondi then proposed a model to illustrate how this limiting compactness could be attained, although he described them in a vague way. Bondi~\cite{Bondi1964} also demonstrated that if negative energy densities are permitted, the compactness can approach the black hole limit as close as desired. Again, the specific model used to illustrate this result is based on a thin-shell construction that is not fully characterized.

This chapter provides a careful and dedicated study of both theorems, highlighting the main assumptions behind them. In the case of Buchdahl's theorem, we explore the consequences and the physical viability of relaxing each individual assumption. In particular, we are interested in finding simple and realistic physical models that allow us to understand how relaxing these assumptions allows one to bypass the limit. We restrict ourselves to static and spherically symmetric configurations assuming for simplicity a Schwarzschild exterior, i.e., vacuum, staticity and spherical symmetry. 

First, we explore the role of the monotonicity and positivity of the energy density by introducing a stellar model composed of two thick layers of constant energy density: an internal core and an external crust, with the condition that the density is always larger for the crust than for the core. For this model, we analyze two very different situations: one in which we constrain the core density to be positive, and another in which we allow it to take negative values. We show that, in the first situation, we are able to approach numerically Bondi's bound $C(R) <  12 \sqrt{2} - 16$ as much as desired. With negative solutions, we build solutions as close to the black hole limit as desired, at least as long as we do not impose any energy condition on the fluid. We compare our models to Bondi's, and in fact make a careful analysis of the specific Models that he introduces clarifying their physical meaning.

As a separate situation, we test the relevance of relaxing the isotropy of the pressure. For this, we present a toy model consisting of a thin shell matching an internal Minkowskian core with an external Schwarzschild solution. We show that the shell can be placed as close to its Schwarzschild radius as desired, making the configuration arbitrarily close to the black hole compactness. However, imposing energy conditions sets compactness bounds in the anisotropic case. We revisit and compare our findings with similar results in the literature presenting bounds for anisotropic configurations.  

\section{Classic results: Buchdahl and Bondi}
\label{Sec:Classic_Results}

\subsection{Buchdahl's limit}
\label{SubSec:Buchdahl_Limit}

We begin by reviewing the Buchdahl limit, which establishes an upper bound on the compactness of isotropic stars in hydrostatic equilibrium with outwards monotonically decreasing densities in GR~\cite{Buchdahl1959}. More importantly, we will systematically outline the underlying assumptions.

Consider the following line element that represents static and spherically-symmetric spacetimes: 
\begin{align}
    \dd s^2 = - f(r) \dd t^2 + h(r) \dd r^2 + r^2 \dd \Omega^2_2.
    \label{Eq:LineElement2}
\end{align}
For some purposes it is useful to use an alternative set of functions:
\begin{align}
    f(r) = e^{2 \Phi(r)}, ~~~~~~~~ h(r)= \frac{1}{1-\frac{2m(r)}{r}},
    \label{Eq:LineElement1}
\end{align}
where $\Phi(r)$ is the redshift function and $m(r)$ the Misner-Sharp mass~\cite{Misner1964,Hernandez1966}. We want to study solutions representing a star sustained by the energy-momentum tensor of a perfect fluid:
\begin{align}
    T_{\mu \nu} = \rho u_\mu u_\nu + p \left( g_{\mu \nu} + u_\mu u_\nu \right),
    \label{Eq:SET}
\end{align}
where $u^\mu$ is the vector field representing the velocity of the fluid, $\rho$ is the density of the fluid, and $p$ is the pressure. Staticity requires that the vector $u^\mu$ is aligned with the timelike Killing vector associated with the staticity of the spacetime, $u^\mu = e^{-\Phi} \delta^\mu{}_{t}$. Furthermore, we will restrict our considerations to barotropic fluids, i.e., fluids admitting an equation of state of the form $\rho=\rho(p)$. We focus on solutions that describe a perfect fluid sphere for $r<R$,  matched to vacuum for $r>R$, where both the energy density and pressure vanish, i.e., $\rho = p = 0$. By virtue of Birkhoff's theorem~\cite{Wald1984}, the exterior spacetime must be described by the Schwarzschild metric, as it is static and spherically symmetric. Therefore, we have:
\begin{align}\label{Eq:SchwMetric}
    & \Phi(r) = \frac{1}{2} \log \left( 1 -  \frac{2M}{r} \right), \\ 
    & m(r) = M, 
\end{align}
with $M \in \mathbb{R}^{+}$ for $r>R$. 

If we plug the ansatz from Eqs.~\eqref{Eq:LineElement2} and~\eqref{Eq:LineElement1} into Einstein equations, we are led to the following equation that determines $\Phi(r) $:
\begin{align}
    \frac{ \dd \Phi(r)}{\dd r} = \frac{m(r) + 4 \pi r^3 p}{r \left[r-2m(r)\right]},
    \label{Eq:Redshift}
\end{align}
with $m(r)$ defined in terms of an integral of the energy density given in Eq.~\eqref{Eq:SET},
\begin{align}
    m(r) = 4 \pi \int^{r}_0 \dd r' r^{\prime 2} \rho(r').
    \label{Eq:Mass}
\end{align}
Combining Eq.~\eqref{Eq:Redshift} with the conservation of the energy-momentum tensor, which is:
\begin{align}
    \frac{\dd p(r) }{\dd r}=-\left(\rho+p\right) \frac{\dd \Phi(r)}{\dd r},
    \label{Eq:Cons}
\end{align}
we arrive at the Tolman-Oppenheimer-Volkoff (TOV) equation, which is given by
\begin{align}
    \frac{\dd p}{\dd r} = - (p+\rho) \frac{m(r) + 4 \pi r^3 p}{r \left[r-2m(r)\right]}.
    \label{Eq:TOV}
\end{align}
Alternatively, this equation can be rewritten using a local compactness function defined in terms of the Misner-Sharp mass function as \mbox{$C(r) = 2 m(r)/r$}:
\begin{align}
    \frac{\dd p}{\dd r} = - \frac{(p+\rho)}{2r} ~ \frac{\left[ C(r) + 8 \pi r^2 p \right]}{\left[1-C(r)\right]}.
    \label{Eq:P-TOV}
\end{align}
Expressions~(\ref{Eq:Mass}--\ref{Eq:TOV}), together with the equation of state $\rho(p)$, form a closed system of differential equations to be solved for $\Phi,m,\rho$, and $p$. In order to find a specific solution, we take a central density $\rho(0) = \rho_c$ which, by virtue of the equation of state determines a central pressure $p (0) = p_c$. This allows to integrate the TOV equation together with the equation of state and Eq.~\eqref{Eq:Mass} for $m(r)$. Finally, it is possible to determine the redshift by integrating Eq.~\eqref{Eq:Redshift} [or, equivalently, Eq.~\eqref{Eq:Cons}] and matching the redshift of the surface $r=R$ with its Schwarzschild counterpart, given in Eq.~\eqref{Eq:SchwMetric}. 

Buchdahl's limit sets an upper bound to the total mass $M$ that a star with a given radius $R$ can have. To understand what we need in order to surpass Buchdahl's compactness limit, we will express its underlying assumptions in its most conservative version. This is best done by stating the result in the form of a theorem.

\medskip

\noindent \textbf{Buchdahl's theorem:} Consider the problem of finding the solution to the TOV equations~(\ref{Eq:Mass}--\ref{Eq:TOV}) assuming a perfect fluid with an equation of state that fulfills the following properties: 
\begin{enumerate}
    \item $f(r)$ is at least a piecewise $\mathcal{C}^2$ function, while $h(r)$ is at least piecewise $\mathcal{C}^1$. In particular, both $f'$ and $h$, where the prime represents differentiation with respect to $r$, are continuous at $r=R$. We integrate the equations from the center $r=0$ with given pressures and densities $\{\rho_c, p_c\}$ towards the surface $r=R$, defined by the condition $p(R)=0$. From that point on ($r>R$) we match an exterior (vacuum) Schwarzschild solution with $\rho = p =0$, independently of the value of $\rho \left( R^{-}\right)$. 
    
    \item The density is a monotonically decreasing function. The continuity and monotonicity assumptions along with the boundary conditions require that \mbox{$\rho(r) \geq 0$}. 
\end{enumerate}
For any such solution, the compactness satisfies the inequality $C(R) \leq 8/9$.

\medskip

\textbf{Proof:} Demanding staticity automatically enforces $2M/R \leq 1$, since otherwise the $rr$ component of the metric changes its sign and becomes timelike. In fact, the lower compactness limit can be found integrating Einstein equations with the assumptions $\rho(r) \geq 0 $, $\dd \rho/ \dd r \leq 0$ and requiring regularity of the solutions. Using the parametrization of the metric given by Eq.~\eqref{Eq:LineElement2}, Einstein equations read
\begin{align}
    G^{t}_{\ t} & = (rh^2)^{-1} h' + r^{-2} \left( 1 - h^{-1} \right) = 8 \pi \rho, \label{Eq:00component}\\
    G^{r}_{\ r} & = (rfh)^{-1} f' - r^{-2} \left( 1 - h^{-1} \right) = 8 \pi p, \\
    G^{\theta}_{\ \theta} & = \frac{1}{2} (fh)^{-1/2} \frac{\dd}{\dd r} \left[ (fh)^{-1/2} f' \right] + \frac{1}{2} (rfh)^{-1} f' - \frac{1}{2} (rh^2)^{-1} h = 8 \pi p. 
\end{align}
Eq.~\eqref{Eq:00component}, associated with the $tt$ component, is precisely Eq.~\eqref{Eq:Mass} in the $\Phi(r), m(r)$ parametrization of the line-element. Moreover, because we have assumed isotropic pressure, we can equate $G_{rr}$ and $G_{\theta  \theta}$. Using Eq.~\eqref{Eq:LineElement1} and after some manipulations, we find the following identity:
\begin{align}\label{Eq: equality Gs isotropic}
    \frac{\dd}{\dd r} \left[ r^{-1} h^{-1/2} \frac{\dd f^{1/2}}{\dd r} \right] = (fh)^{1/2} \frac{\dd }{\dd r} \left( \frac{m(r)}{r^3} \right). 
\end{align}
We now use the second assumption: that $\rho$ is monotonically decreasing. If $\rho(r)$ is monotonically decreasing, this means that the average density $\bar{\rho} (r)$ that we introduce as
\begin{align}
    \bar{\rho} (r) = \frac{3 m(r)}{4 \pi r^3},
\end{align}
is also monotonically decreasing. To see this, we compute its derivative and use Eq.~\eqref{Eq:Mass} to find
\begin{align}
    \frac{\dd \bar{\rho}}{\dd r} (r) = \frac{3}{r} \left[ \rho(r) - \frac{3}{r^3} \int^{r}_0 \dd r' r^{\prime 2}  \rho (r') \right].
\end{align}
Noting that $\rho(r)$ is positive and it is monotonically decreasing, we have
\begin{align}
    \int^{r}_0 \dd r' r^{\prime 2}  \rho (r') \geq \rho(r) \int^{r}_0 \dd r' r^{ \prime 2} = \frac{\rho(r)}{3} r^3, 
\end{align}
since $\rho(r)$ is the minimum of the function in the interval $[0,r]$, with the equality holding only if the density is constant. Putting everything together, we have that
\begin{align}
    \frac{\dd \bar{\rho}}{\dd r} (r) \leq  \frac{3}{r} \left[ \rho(r) - \rho(r) \right] = 0.
\end{align}
Thus, the average density is also a monotonically decreasing function. This implies that the right-hand side of Eq.~\eqref{Eq: equality Gs isotropic} needs to be nonpositive, since it is proportional to the derivative of the average density. Of course, this implies that the left-hand side needs to be nonpositive as well, hence
\begin{align}
    \frac{\dd }{\dd r} \left[ r^{-1} h^{-1/2} \frac{\dd f^{1/2}}{\dd r} \right] \leq 0. 
\end{align}
Integrating this inequality from the surface of the star located at a radius $R$ to some smaller radius $r$, we find the inequality
\begin{align}
    \frac{1}{rh^{1/2}(r)} \frac{\dd f^{1/2}}{\dd r} & \geq \frac{1}{Rh^{1/2}(R)} \frac{\dd f^{1/2}}{\dd r}(R) \nonumber \\
    & = \left. \frac{(1-2M/R)^{1/2}}{R} \frac{\dd}{\dd r} \left( 1 - \frac{2M}{r} \right)^{1/2} \right\rvert_{r = R} = \frac{M}{R^3},
    \label{Eq:Matching}
\end{align}
where we are using the continuity of $f'$ and $h$ to match their values at the surface $r=R$ with their known Schwarzschild counterparts. We now multiply the equation by $r h^{1/2}$ and integrate inwards from the surface $r=R$ to the center at $r= 0$. We use the value of $f$ at the surface and the expression of $h$ in terms of $m(r)$ to find
\begin{align}
    f^{1/2} (0) \leq \left( 1 - 2M/R \right)^{1/2} - \frac{M}{R^3} \int^R_0 \dd r \frac{r}{ \sqrt{ 1 - \frac{2 m(r)}{r}}}. 
    \label{Eq:fbound}
\end{align}
The monotonicity condition on $\rho(r)$ implies that $m(r)$ can be, in the best scenario, as small as the value it would have for a uniform density function, that is,
\begin{align}
    m(r) \geq Mr^3/R^3. 
    \label{Eq:}
\end{align}
Thus, the best upper bound that we can provide for $f^{1/2}(0)$ from Eq.~\eqref{Eq:fbound} corresponds to the case when $m(r) = Mr^3 /R^3$, i.e., the constant density scenario yields the optimal bound. Performing the integral for this $m(r)$ we find
\begin{align}
    f^{1/2} (0) \leq \frac{3}{2} \left( 1 - 2 M/R \right)^{1/2} - \frac{1}{2}.
\end{align}
But for regular stellar configurations $f$ must be positive, so we are left with the bound
\begin{align}
    (1 - 2M/R)^{1/2} > 1/3,
\end{align}
and hence
\begin{align}
    C(R)=\frac{2M}{R} < 8 /9, 
\end{align}
which is the so-called Buchdahl limit. 

\emph{Observation 1:} Notice that, by the TOV equation, it is interchangeable to assume $\{\rho \geq 0, \dd \rho/\dd r \leq 0 \}$ and $\{p\geq 0, \dd p/\dd \rho \geq 0\}$. In that sense, imposing that the pressure and density are positive, together with the causality condition $\dd p/\dd \rho \geq 0$, ensure that the density is monotonically decreasing toward the surface of the star. 

\emph{Observation 2:} Notice that we are starting with a perfect fluid and hence the distribution of pressures is isotropic. As we illustrate below, the anisotropic case in which pressures in the angular directions are allowed to grow without bounds does not lead to any limit on the compactness of the object. By forcing the fluid to satisfy energy conditions, though, it becomes possible to find some compactness bounds as well~\cite{Lindblom1984,Ivanov2002,Barraco2003,Boehmer2006,Andreasson2007,Urbano2018} (we will revisit this in Section~\ref{Sec:Anisotropy}).

\emph{Observation 3:} Notice that the proof also tells us which density profile saturates the bound: the uniform density profile. Actually, instead of putting an upper bound to the mass to radius ratio, we can think of it as setting a limit to the maximum mass that a sphere of a given uniform density can display. The properties of this solution are discussed in detail in Appendix~\ref{App:FluidSpheres}. One interesting feature of these constant density solutions is that, for compact enough configurations, the dominant energy condition is violated. This is the case for values of the compactness $2M/R > 6/8$, where we have that $p_c > \rho_0$. This will become relevant later on when we analyze the anisotropic toy model in Section~\ref{Sec:Anisotropy}.

\subsection{Bondi's Limit}
\label{SubSec:Bondi_Limit}

This section reviews Bondi's analysis of compactness bounds for isotropic stars~\cite{Bondi1964}, see also~\cite{Karageorgis2007} for an ulterior analysis closing some technical details in Bondi's proof. The framework of Bondi’s analysis is the one presented at the beginning of Section~\ref{SubSec:Buchdahl_Limit}. However, its key novelty lies in relaxing the assumption that the density profile must be monotonically decreasing.

Bondi's analysis begins by introducing the convenient variables $u(r) : = m(r)/r$ and \mbox{$v(r) := 4 \pi r^2 p (r)$}. In terms of these variables, the system reduces to:
\begin{equation}\label{Eq:uveqs}
    \frac{1}{r}  \frac{\dd r}{\dd u} = \frac{(1-2u) \frac{\dd v}{\dd u} + u + v}{2v - (u^2 + 6uv+v^2)},\quad  4 \pi r^2 \rho = u \frac{\dd v/\dd u-\beta}{\dd v/\dd u-\alpha}, 
    \end{equation}
    with 
    \begin{equation}
    \alpha = -\frac{u+v}{1-2u}, \qquad \beta = - \frac{v}{u} \frac{2 - 5u - v}{1 - 2u }.
\end{equation} 
Written in the $(u,v)$ variables, stellar solutions become trajectories in a $(u,v)$ diagram. From those trajectories, it is possible to obtain the physical variables immediately from Eqs~\eqref{Eq:uveqs}, together with the equation for the redshift $\Phi$
\begin{align}
    \frac{1}{2} r \frac{\dd \Phi}{\dd r} = \frac{u+v}{1-2u}.
\end{align}
Constant compactness curves, those described by $ \dd u/\dd r = 0$, correspond in these variables to the hyperbolas
\begin{align}
    H \equiv 2v - (u^2 + 6uv+v^2)=0 . 
\end{align}
Also, the integrals of $\dd v/\dd u = \alpha$ are important and correspond to curves along which $\dd r/\dd u = 0$, i.e., there is a change in the compactness but the radius of the spheres is kept constant. They are described by the family of parabolas parametrized by $A$:
\begin{align}\label{Eq:PACurves}
    P_A \equiv v = \sqrt{A (1-2u)} - 1 + u. 
\end{align}
Bondi identifies these parabolas with solutions describing thin shells although he does not characterize them by analyzing their energy-momentum tensor.

\begin{figure}
    \centering
    \includegraphics[width=0.5\linewidth]{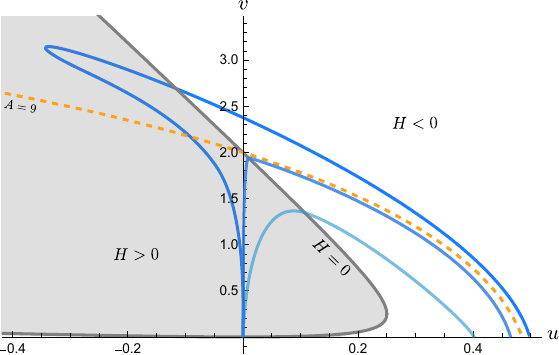}
    \caption{ $(u,v)$ diagram of the kind used in Bondi's proof~\cite{Bondi1964}. Regular stellar solutions with non-negative densities lie in the first quadrant, i.e., $u,v>0$. They describe curves that start from $(0,0)$ inside the $H>0$ region and increase until they cross $H=0$. After that, moving within the $H<0$ region they decrease, always with a more negative slope than $P_{A}$ curves, until crossing $v=0$. These solutions are always bound to take values below the dashed orange $P_{\rm A}$-curve with $A=9$, i.e., a compactness $C(R)<C_{\rm B}$. Solutions with negative densities $u<0$ but $v>0$, start from $(0,0)$ and can take unbounded negative $u$ values. Then, they have a turning point at which they begin moving to greater values of $u$, cross $H=0$, and keep decreasing until crossing the $v=0$ axis. These solutions can have any compactness $C(R)<1$.}
    \label{Fig:uvDiagram}
\end{figure}

Now we need to consider how do solutions describing a stellar interior behave in these $(u,v)$ variables. If we restrict to non-negative values of the density, this means that $u$ needs to be non-negative too. Furthermore, Bondi also shows~\cite{Bondi1964} that non-negative densities enforce non-negative pressures through the hydrostatic equilibrium equations. Thus, an interior with non-negative densities is confined to lie within the first quadrant of the $(u,v)$ diagram in Fig~\ref{Fig:uvDiagram}. In this figure we have drawn, in light blue, two curves representing stars with non-negative densities.

In the $(u,v)$ diagram, a generic solution describing a stellar interior with finite central pressures needs to be such that it starts from the origin $(0,0)$ and moves within region $H>0$, intersecting $P_{A}$ curves of increasing $A$ until it hits the hyperbola $H=0$ with exactly the same slope as the $P_{A}$ that is intersecting it at the crossing point (to ensure that Eqs.~\eqref{Eq:uveqs} are finite). Then, the solution enters the region $H<0$ and acquires negative slope, intersecting $P_{A}$ curves with decreasing $A$, eventually reaching $v=0$ at a given value of $u$ which fixes the compactness of the star.

Now, we attempt to construct the most compact solution possible within the positive $(u,v)$ region. Restricting ourselves to $u \geq 0$, the highest point at which it is possible to cross $H = 0$ is $v = 2$. Once in the $H<0$ region, the curve that optimizes the compactness is the $A=9$ parabola, which intersect $v=0$ at $u = 6 \sqrt{2} -8$. Since a regular solution must lie below both these curves, this proves the existence of an upper bound for the compactness of stars with non-negative densities, namely 
\begin{equation}\label{Eq:BondiBound}
    C(R)<C_{\rm B}\equiv 12 \sqrt{2} - 16.
\end{equation}

Allowing solutions to take values in the $u<0, v>0$ region, it is possible to obtain curves that grow in $v$ towards negative $u$ values and which are then matched with curves arbitrarily close to $P_{A}$ parabolas with $A$ values as large as desired. Since the $P_{A}$ curves cross $v = 0$ at the positive $u$ values
\begin{equation}
    u_{+} = 1-A+\sqrt{A\left(A-1\right)},
\end{equation}
we see that
\begin{equation}
    \lim_{A\to\infty}u_{+}=\frac{1}{2}-\frac{1}{8A}+\order{\frac{1}{A^{2}}},
\end{equation}
hence we have the bound $C(R)<1$ and the black hole limit can be approached as much as desired. 
The darkest blue line in Fig.~\ref{Fig:uvDiagram} represents a generic configuration with a negative energy core and a compactness beyond Buchdahl and Bondi bounds.

After presenting the general framework, Bondi proposed two specific stellar models, Model I and Model II. These models serve, respectively, to illustrate how to saturate the Bondi bound and how to construct configurations that approach the black hole compactness as much as desired with negative density cores. However, it is not straightforward to extract the physical meaning of these models in a clear way. While it is straightforward to see that, by the way they are constructed, they exhibit distributional thin-shell layers, such layers should not exhibit tangential pressures as this would be against one of the hypothesis of the set-up, namely the isotropy of the pressures. 

Model I is described as a vanishing energy core with nonvanishing pressure connecting through a thin envelope to an external Schwarzschild geometry. The vanishing energy core is represented by the curve $ u = 0$ and $v \in [0, 2 - \epsilon]$, $\epsilon >0$, where
\begin{align}
    \rho = 0, \qquad p = \frac{p_c}{1+2\pi p_c r^2}, \qquad 2 \pi p_c r^2 = \frac{v}{2-v}, \qquad 0 \leq r \leq \sqrt{\frac{2 - \epsilon}{2 \pi p_c \epsilon}},
\end{align}
with $p_c$ being the value of the pressure at the center $r=0$. The envelope is represented by $v = ( 3 - \epsilon) (1 - 2u)^{1/2} - (1-u)$ and $u \in [0, u_s]$, with 
\begin{align}\label{Eq:uBound}
    u_s = \left( 3 - \epsilon \right) (8 -6 \epsilon + \epsilon^2)^{\frac{1}{2}} - \left( 8 - 6 \epsilon + \epsilon^2 \right) = 6 \sqrt{2} - 8 - \order{\epsilon}.
\end{align}
As the envelope is strictly following one of the parabolic curves, it represents from the start an infinitely-thin limit. This is so because along such curve $r$ does not vary and the energy density becomes infinite, see Eqs.~\eqref{Eq:uveqs}. By making $\epsilon$ arbitrarily small, the model can approach Bondi's compactness limit as closely as desired.

To better understand it, let us perform a detailed thin-shell analysis following Israel formalism~\cite{Israel1966}. Take a spherical shell with radius $R$, such that its proper area is $4 \pi R^2$. Let the spacetime be described by external Schwarzschild patch that we denote with a $+$ sign and the internal patch that we denote with a $-$ sign. Generically, the metric reads:
\begin{align}
    & \dd s^2_{+} = - f_{+}(r_{+}) \dd t_{+}^2 + h_{+}(r_{+}) \dd r_{+}^2 + r_{+}^2 \dd \Omega_2 ^2, \\
    & \dd s^2_{-} = - f_{-}(r_{-}) \dd t_{-}^2 + h_{-}(r_{-}) \dd r_{-}^2 + r_{-}^2 \dd \Omega_2 ^2.
    \label{Eq:thinshellbulks}
\end{align}
The external patch corresponds to $f_{+}(r_{+}) = 1 - 2M/r_{+}$ and \mbox{$h_{+}(r_{+}) = 1/f(r_+)$}, the Schwarzschild metric.  The internal geometry has zero mass function and as such \mbox{$h(r_{-})=1$}. On the other hand, the redshift function can be obtained through a straightforward integration to find
\begin{equation}
    f(r_{-})=(1 + 2\pi p_c r_{-}^2)^2.    
\end{equation}
We locate the thin shell at $r_{+} = r_{-} = R$. The matching is worked out in detail in Appendix~\ref{App:AnisotropicShell}, and we find that the shell displays a distributional energy-momentum tensor with the following densities and pressures:
\begin{align}
    & \rho = \sigma \delta (r - R), \label{Eq:RhoDist} \\
    & p_r = 0, \label{Eq:PrDist} \\
    & p_t = \tilde{p}_t \delta(r -R), \label{Eq:PtDist}
\end{align}
with $\sigma$ and $\tilde{p}_t$ given by
\begin{align}
     \sigma & = \frac{1}{4 \pi R} \left(1 - \sqrt{1-\frac{2M}{R}} \right), \nonumber \\
     \tilde{p}_t & = \frac{1}{8  \pi R } \left(\sqrt{1-\frac{2M}{R}} -1 \right) +\frac{1}{16  \pi R} \left(\frac{2M}{R} \frac{1}{\sqrt{1-\frac{2M}{R}}} - \frac{8\pi p_c R^2}{(1 + 2 \pi p_c R^2)}\right).
\end{align}
Now, for every value of the compactness there is always a specific value of $p_c$ such that the tangential pressure vanishes, i.e., $\tilde{p}_t=0$, which corresponds to
\begin{align}
 p_c= -\frac{R \left(\sqrt{1-\frac{2 M}{R}}-1\right)+M}{2 \pi  R^2 \left(R \left(3 \sqrt{1-\frac{2 M}{R}}-1\right)+M\right)}.
\end{align}
In this scenario, the shell is held in place by the pressure from the matter content of the inner region, despite the shell itself being pressureless. This situation is peculiar because it involves an interior composed of unusual matter that exerts pressure without possessing any energy density. We discover that the compactness cannot be increased to the black hole limit, as there is a specific value at which the solution becomes singular for $r=0$, as $p_c \to \infty$. Increasing the compactness from that value only moves this singularity to a finite nonzero radius. The threshold value for $p_c$ corresponds to the following compactness $C(R)$:
\begin{align}
    C(R) = 12 \sqrt{2} - 16.
\end{align}
This is precisely Bondi's bound \mbox{$C(R) = C_B$}. 

In the following sections, we introduce clear and simple physical models that demonstrate how the Buchdahl and Bondi limits can be arbitrarily approached or even exceeded when some of the assumptions are relaxed.

\section{Bilayered stars with nonmonotonically decreasing density profiles}
\label{Sec:Bilayer}
\begin{figure}
    \centering
    \includegraphics[width= 0.35\textwidth, height= 0.35\textwidth]{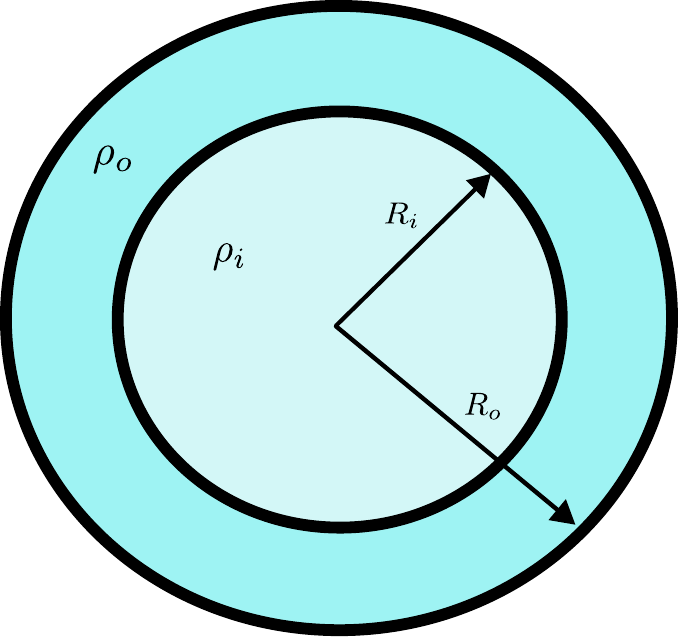}
    \caption{Pictorial representation of the two-layered toy model that we are considering in this section. The density of the outer layer is bigger than that of the inner layer, $\rho_i < \rho_o$.}
    \label{Fig:Bilayer}
\end{figure}

In this section we study a toy model displaying a very simple outward increasing density profile. As a consequence, it violates one of the key assumptions behind Buchdahl theorem. Specifically, we consider a star whose internal structure displays two constant-density layers: 
\begin{align}
    \rho(r) = \begin{cases}
      \rho_i  & r < R_i \\
      \rho_o & R_i < r < R \\
      0 & r> R
    \end{cases},
\end{align}
with $\rho_i < \rho_o$ and $R_i < R$. As indicated in previous sections, the exterior metric ($r > R$) is Schwarzschild. The setup is pictorially represented in Fig.~\ref{Fig:Bilayer}.
We impose that the Misner-Sharp mass function is continuous everywhere. Physically, this means that all the mass of the system is entirely provided by the two layers, i.e., there are not additional thin-shell contributions. This implies that we have
\begin{align}
    m(r) = \begin{cases}
      \frac{4}{3} \pi r^{3} \rho_{i}  & r < R_i \\
      \frac{4}{3} \pi R_i^3 \rho_{i} + \frac{4}{3} \pi \rho_{o} (r^3 - R_i^3) & R_i < r <R \\
      M & r> R
    \end{cases},
    \label{Eq:MisnerSharpBi}
\end{align}
together with the constraint:
\begin{align}
   M =  \frac{4}{3} \pi R_i^3 \rho_{i} + \frac{4}{3} \pi \rho_{o} (R^3 - R_i^3).
   \label{Eq:Constraint}
\end{align}
As mentioned before, $M$, which represents the total mass of the spacetime, cannot surpass the black hole limit: $M<R/2$. The inner radius $R_i$ can, at most, range between the center of the star and the total radius
\begin{align}
    0 < R_i < R,
\end{align}
since the star would become single-layered in both limits. The outer density $\rho_o$ is allowed to vary within the range
\begin{align}
    \rho_c <\rho_{o}<\infty,
\end{align}
with $\rho_{c}=3M/4\pi R^3$ being the critical density, which corresponds to the limiting case in which $\rho_i = \rho_o$, i.e., the star has constant density and it cannot surpass the Buchdahl limit.

This model automatically satisfies Israel junction conditions at $R_{i}$ since, by virtue of Eq.~\eqref{Eq:Cons}, the discrete jump in $\rho$ between the inner and outer layers translates into a compensatory discontinuity in $p'$ that guarantees the continuity of $\Phi'$. Consequently, the metric components $g_{tt}$ and $g_{rr}$ are piecewise $\mathcal{C}^2$ and $\mathcal{C}^{1}$ functions, respectively. This discontinuous behavior in the $tt$ component of the energy-momentum tensor is of the same kind as the one that occurs at the surface of the star and, therefore, does not carry along the introduction of distributional sources of any kind.

The structure of this bilayered fluid sphere is specified by the set of five parameters $\{\rho_{i},R_{i},\rho_{o},R,M\}$. However, not all of the parameters are independent. First of all, since the TOV equation we are solving does not incorporate any additional length scales, we can set one of the parameters as a reference unit for the numerical problem. We find it convenient to choose $R$ as such parameter: for all plots, we set $R = 1$.  Moreover, the constraint given in Eq.~\eqref{Eq:Constraint} reduces the number of independent parameters by one. We choose to express the parameter $\rho_i$ in terms of the remaining parameters
\begin{align}
    \rho_{i}=  \frac{3 M}{4\pi R_{i}^3} +\rho_{o}\left(1 - \frac{R^3}{R_{i}^3} \right).
    \label{Eq:RhoI}
\end{align}
Thus, we take the independent parameters to be \mbox{$\{ M, R_i, \rho_o \}$} and $R$ that provides the units.

Furthermore, we integrate the equations from the surface of the star towards its center. Although this is entirely equivalent to outward integrations from a regular center, our strategy allows to select the compactness of the fluid sphere beforehand, thus proving to be more efficient to search for compactness bounds from the numerical point of view.

Solutions for the functions $p$ and $\Phi$ at the inner layer are straightforward to obtain by integrating the TOV equation (see Appendix~\ref{App:FluidSpheres}), and fixing the corresponding integration constants in terms of the parameters of the outer layer. 
Defining the constants $\Phi_{\text{i}}$ and $p_{i}$ as
\begin{equation}
\Phi_{i}= \Phi(R_i),\quad p_{i}= p(R_i),
\end{equation}
we obtain
\begin{equation}
  \Phi(r)=\Phi_{i}\log\left[3\left(p_{i}+\rho_{i}\right)-\left(3p_{i}+\rho_{i}\right)\sqrt{\frac{3-8\pi r^2\rho_{i}}{3-8\pi R_{i}^2\rho_{i}}}~\right]-\Phi_{i}\log\left(2\rho_{i}\right), \label{Eq:InnerPhi}
\end{equation}
and
\begin{equation}
    p(r)=-\rho+e^{-\Phi(r)+ \Phi_i}\left[p_{i}+\rho_{i}\right],  \label{Eq:InnerP}
\end{equation}
for $r < R_i$, i.e., inside the inner layer.
For the outer layer, instead, the solution can be written as combinations of elliptic functions of cubic roots~\cite{Wyman1939}, resulting in lengthy and non-illuminating expressions that we omit here but which were obtained and evaluated numerically using the software \textit{Mathematica.} The query about the maximum compactness allowed by these bilayered toy models reduces to an inquiry about whether there exists a maximum possible value for the total mass $M$ (always below the black hole limit, or $M < R/2$), beyond which there is not any solution in the family preserving a finite and regular pressure function everywhere inside (and equivalently for the redshift function). 

We distinguish two situations: the case in which the inner density $\rho_i$ is strictly positive, which we explore in detail in Subsection~\ref{Subsec:PositiveBilayer}, with the limiting case $\rho_i = 0$ presented separately in Subsection~\ref{Subsec:ZeroBilayer}; and the case in which the density of the inner layer is negative, which we discuss in Subsection~\ref{Subsec:NegativeBilayer}. For positive inner densities, the Misner-Sharp mass function remains positive throughout the configuration, and we demonstrate that the maximum compactness bound found by Bondi, Eq.~\eqref{Eq:BondiBound}, can be approached arbitrarily. For negative inner densities, there are regions with negative Misner-Sharp mass in the interior, allowing for configurations that reach arbitrarily high compactness, all the way up to the black hole limit, in full agreement with Bondi's conclusions. Unlike Bondi's Model I and II, the bilayer model presented here gives rise to a nondistributional matter content throughout.

\subsection{Bilayered stars with positive core densities}
\label{Subsec:PositiveBilayer}

Let us begin analyzing the case in which we have $\rho_i > 0$. The positivity of the right hand side of Eq.~\eqref{Eq:RhoI} imposes a lower bound on $R_{i}$. Explicitly, we have that
\begin{align}
    \left(1-\rho_{c}/\rho_{o}\right)^{1/3} R < R_{i} < R.
    \label{Eq:RadI}
\end{align}
We demonstrate that there is an upper bound to the compactness that these stars can achieve, beyond which no regular solutions exist. To establish this, we fix the parameters $R_{i}$ and $\rho_{o}$, and we determine numerically the value of $M_{\infty}(R_{i}, \rho_{o})$ that leads to a solution where the central pressure becomes infinite. This solution defines the maximum mass that a sphere with outer density $\rho_{o}$ and inner radius $R_{i}$ can sustain. Spheres with masses below this maximum will have finite pressures throughout, whereas for larger masses, an ``infinite pressure surface'' emerges at some finite radius of the star $r > 0$.
\begin{figure}
    \centering
    \includegraphics[width=0.6 \linewidth]{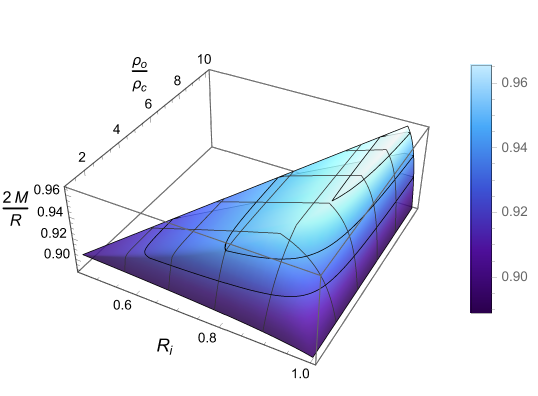}
    \caption{Three dimensional representation of the surface in the $\{R_{i},\rho_{o},M\}$ space corresponding to solutions with an infinite central pressure and  $\rho_{i}>0$. Every point on this surface lies above the standard Buchdahl limit  $2M/R=8/9$, which is reached in the $\rho_{o}/\rho_{c}\to1,~R_{i}/R\to1$ limits. Lighter colors denote larger values of $2M/R$, and the black contour lines correspond to the \mbox{$2M/R=\{0.92,0.94,0.96\}$} planes. 
    This surface can be extended further in the $\rho_{o}/\rho_{c}\gg 1$ direction where we find an asymptotic approach to the maximum compactness value of $2M/R\approx 0.9706$.}
    \label{Fig:BilayeredRhoP}
\end{figure}

\begin{figure}
    \centering
    \includegraphics[width=0.5\linewidth]{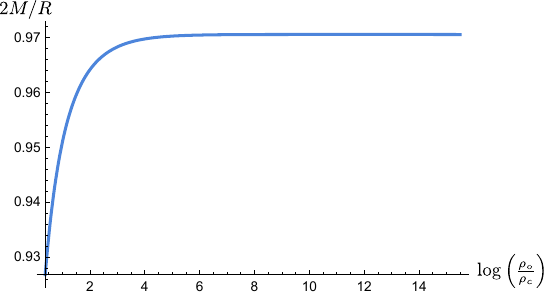}
    \caption{Maximum compactness limit as a function of the quotient between the energy density of the outer layer and the critical density for stars with $\rho_{i}>0$.  For $\rho_{o}/\rho_{c}\to1$, the compactness limit tends to the Buchdahl limit whereas for $\rho_{o}/\rho_{c}\to\infty$ it approaches the Bondi limit.}
\label{Fig:MaxCP}
\end{figure}

\begin{figure}
    \centering
    \includegraphics[width=0.6\linewidth]{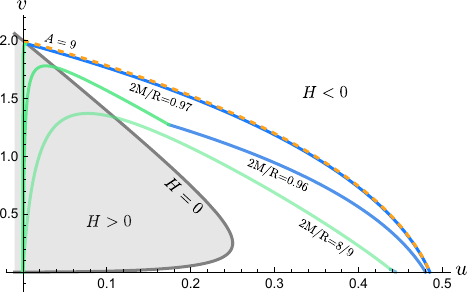} 
    \caption{$(u,v)$ diagram of bilayered stars with non-negative energy densities. 
    We have represented the $v(u)$ curves from the inner layer in green and those from the outer layer in blue. As the compactness approaches the Bondi compactness, the bilayered model acquires characteristics very similar to Bondi's Model I, namely an inner layer that lies on the vertical axis and an outer layer that approaches the $P_{A}$-curve with $A=9$.}
    \label{Fig:UVPos}
\end{figure}

By varying the values of $\rho_o$ and $R_i$, we can map out a surface in the three-dimensional parameter space $\{R_{i}, \rho_{o}, M\}$ that precisely represents the value of $M_{\infty}(R_i, \rho_o)$ for each pair $\left(R_i,\rho_o\right)$. This surface is shown in Fig.~\ref{Fig:BilayeredRhoP}. The vertical axis represents the maximum mass allowed for a given $R_i$ and $\rho_o$. All points on this surface lie between two constant $2M/R$ planes: one above Buchdahl's limit ($2M/R = 8/9$), which it intersects in the limits $\rho_{o}/\rho_{c} \to 1$ and $R_{i}/R \to 1$, and the other below the black hole limit ($2M/R = 1$), which it neither intersects nor approaches asymptotically. Extending Fig.~\ref{Fig:BilayeredRhoP} in the direction of $\rho_{o}/\rho_{c} \gg 1$, we find that $C(R)$ saturates at a maximum value which approximately is
%
\begin{align}\label{Eq:CompLimRhoP}
    C_{\text{max}} = 2\sup_{ R_{i},\rho_o } \frac{M_{\infty} (R_i,\rho_o)}{R} \approx 0.9706.
\end{align}
To the level of our numerical precision it agrees with Bondi's bound~\eqref{Eq:BondiBound}: we do not find a gap between this uppermost bound~\eqref{Eq:CompLimRhoP} and Bondi's bound. Fig.~\ref{Fig:MaxCP} shows how this maximum compactness limit scales with the quotient between the outer and critical densities. Finally, in Fig.~\ref{Fig:UVPos} we have plotted two solutions belonging to this family.


\subsection{Bilayered stars with vanishing core densities}
\label{Subsec:ZeroBilayer}
A particularly illustrative case is when the density of the inner core is zero, a feature shared with Model I presented by Bondi. We treat it separately because its simplicity allows to find the compactness limit in an easier way. Moreover, we can provide numerical evidence for the absence of any finite gap between the maximum compactness attained by this bilayered model and Bondi's bound. In the present situation, the Misner-Sharp mass vanishes everywhere inside the core, and, for a fixed value of the density $\rho_o$ and the mass $M$, Eq.~\eqref{Eq:RhoI} enforces the radius $R_i$ to be precisely
\begin{equation}\label{Eq:Rirho0}
    R_{i}=\left(1-\frac{\rho_{c}}{\rho_{o}}\right)^{1/3}R,
\end{equation}
where we recall that the expression of $\rho_c$ in terms of $\{M,R\}$ is $\rho_c = 3M/4 \pi R^3$. This constraint also restricts the pressure to be positive, inwards-increasing monotonic function. This is because, starting from a positive, inwards-growing pressure, the numerator of the TOV equation~\eqref{Eq:TOV} is only allowed to change sign if $m(r)<0$. Actually, we can explicitly integrate the TOV equation in the inner layer and find the following pressure profile
\begin{equation}
    p=\frac{1}{2\pi r^2+K},
\label{Eq:PressureZero}
\end{equation}
where $K$ is an arbitrary integration constant whose value is obtained from the matching to the outer layer, i.e., 
\begin{equation}\label{Eq:KappaValue}
    K=\frac{1}{p_{i}}-2\pi R_{i}^2=\frac{1}{p_{i}}-2\pi R^2\left(1-\frac{\rho_{c}}{\rho_{o}}\right)^{2/3}.
\end{equation}
In Subsection~\ref{Subsec:PositiveBilayer} we noticed that the regular bilayered star with positive density that exhibits the largest possible compactness $C_{\text{max}}$ would be obtained in the limit of a thin and infinitely dense shell surrounding an interior of (nearly) vanishing mass. In the present case where the inner core has exactly zero mass, the  maximum compactness is attained by the solution~\eqref{Eq:PressureZero} with $K=0$, precisely the one that makes the pressure profile diverge exactly at $r=0$. To find this compactness bound, we  calculate how $p_{i}$ scales in the $\rho_{o}/\rho_{c}\to\infty$ limit (the limit in which the outer shell is arbitrarily thin) for different values of the compactness. 

In our numerical explorations, we see that $p_i$ always approaches a constant positive value, which we denote as $P_{i}^{~(\text{max})}$, as we vary $\rho_o$ while keeping the mass fixed for $\rho_{o}\gg \rho_{c}$.  By calculating $P_{i}^{~(\text{max})}$ for a range of compactness values between the Buchdahl and black hole limits, we find that, if the density of the outer layer is made very large, the value $K=0$ is found very close to the Bondi limit, defined in Eq.~\eqref{Eq:BondiBound}. In particular, we find the scaling
\begin{equation}\label{Eq:LimitScaling}
    C(R)\approx C_{\rm{B}}-\frac{\alpha}{R^{2}\rho_{o}}, 
\end{equation}
where $\alpha$ is a dimensionless positive constant. The Bondi bound is only reached in the singular limit $\rho_{o}\to\infty$, which would correspond to putting all the mass of the star inside a thin shell since it implies $R_{i}\to R$ in virtue of Eq.~\eqref{Eq:Rirho0}. Although the solution with $K = 0$ is, strictly speaking, singular, we can find solutions whose compactness approaches Bondi's bound as much as desired through a physical model that is purely isotropic and can be appropriately matched with a Schwarzschild exterior.

\paragraph*{\textbf{Comparison with Bondi's Model I.}}
In Fig.~\ref{Fig:UVPos}, we have plotted two configurations in the $(u,v)$ variables from our family of solutions with vanishing core densities. The portion of the curve with a negative slope, where $v$ decreases as $u$ increases, corresponds to the thick shell region. As the external shell becomes thinner, the compactness approaches Bondi's bound more closely. Regardless of the shell thickness, these curves never coincide with the parabolic curves defined by a fixed parameter $A$. Instead, they consistently move downward, intersecting parabolas with different values of $A$, which is the crucial difference with respect to Bondi's Model I.

In fact, Bondi's Model I simply corresponds to the distributional limit of our models which, in that sense, are much more appealing from a physical point of view. They are always smooth and well-defined geometries, without the need to invoke any distributional limit. In fact, for a given compactness $C_1(R)$, which is below Bondi's limit but lying between it and Buchdahl's, i.e., $C_{\rm Buchdahl}<C_1(R)<C_{\rm Bondi}$, there are always bilayered models with vanishing density cores and sufficiently thin crusts that lead to a completely regular configuration.

\subsection{Bilayered stars with negative densities in the core}
\label{Subsec:NegativeBilayer}

For a non-negative, outward-increasing density profile, the compactness can surpass the Buchdahl limit, but it is constrained by a new bound established by Bondi. We have built a toy model that can approach this bound as closely as desired. Bondi further argued that allowing negative inner densities removes this restriction, enabling configurations that approach the compactness limit of a black hole. In this subsection, we present an explicit example within the bilayered model, demonstrating that for negative values of $\rho_i$, objects can be constructed arbitrarily close to this limit. 

In the case examined previously, we found that the solution which maximizes the compactness looks (approximately) like a star with all its positive mass stored in a very narrow outer layer. The inner layer, on the contrary, occupies the whole bulk of the star and does not contribute to the mass.  Such arrangement of the two layers is optimal in the sense that it maximizes the mass of the configuration. Instead, allowing for a negative density core enables the construction of stars for which the configuration that minimizes the central pressure does not require a very narrow and highly dense outer layer. Examples of these solutions are found by selecting the position of $R_{i}$ in such a way that the pressure remains almost constant inside the core. In the remainder of the section, we show that the compactness of these solutions can be arbitrarily close to the black hole limit.

Assume a bilayered model with an internal layer of negative density $\rho_{i}$. In order to proceed with the proof, we need to take a step back and consider for a moment the uniform-density, single-layer model ($R_{i}=0$). By inspection of Eqs.~\eqref{Eq:MisnerSharpBi} and~\eqref{Eq:Constraint}, we see that, for $\rho_{o}>\rho_{c}$, the Misner-Sharp mass $m(r)$ takes negative values in the interval $0\leq r<\left(1-\rho_{c}/\rho_{o}\right)^{1/3} R$. Extrapolated to $r=0$, this would point to the existence of a negative delta contribution to the density at the core $r=0$, meaning that the extrapolation of this configuration all the way to the center of the star is not regular. 

Here, we only use this irregular configuration to generate the outer layer of the total geometry. In these configurations it is easy to see that the width of the negative-mass region grows with $\rho_{o}$. Now, by the TOV equation~\eqref{Eq:TOV}, at the surface $r=R$, pressure always increases inwards:
\begin{equation}
    p' = - \frac{M\rho_{o}}{\left(R-2M\right)R}<0.
\end{equation}
Moreover, a negative Misner-Sharp mass can produce a maximum in the pressure at some $r=r_{\text{max}}$, which, by Eq.~\eqref{Eq:TOV}, corresponds to \mbox{$m(r_{\text{max}})=-4\pi r_{\text{max}}^3 p(r_{\text{max}})$}. Values of $\rho_o$ bigger than a critical value $\rho_{\text{sep}}$ (whose specific value can be found numerically) ensure that it is possible to generate configurations with pressure profiles that are everywhere finite. One starts from $p(R)=0$ at the surface, then reaches a maximum value at $r=r_{\text{max}}$, to finally decrease until the center $r=0$. In the $r\to0$ limit, we obtain the following behavior:
\begin{equation}
        p\simeq -\rho + k \sqrt{r},\quad k>0.
\end{equation}  
As mentioned above, if only one layer is allowed, these configurations display a curvature singularity at $r=0$. However, if we introduce an inner layer with $\rho_{i}<0$, to account physically for the negative Misner–Sharp mass, then there exists a range of matching radii $R_{-} \leq R_{i} \leq R_{+}$ that readily ensures $p' \geq 0$ throughout the core, with the pressure vanishing exactly at $r=0$. This results in a finite and well-defined pressure profile everywhere.

The two boundaries of this interval $R_{-}$ and $R_{+}$ correspond to the choices of internal radii that make $p_i = - \rho_i/3$ and $p_i = - \rho_i$ respectively. Regular solutions can also be found for $R_{i}\gtrsim R_{+}$ and for $R_{i}\lesssim R_{-}$ although we do not focus on them here since our aim is only to show that it is possible to find solutions as close to the black hole compactness as desired. For such purpose, it is enough to consider such interval. Inside the core, the solutions for the  Misner-Sharp mass, redshift, and pressure are given by Eqs.~(\ref{Eq:MisnerSharpBi}--\ref{Eq:InnerP}). 

\begin{figure}
    \centering\includegraphics[width=0.6\textwidth]{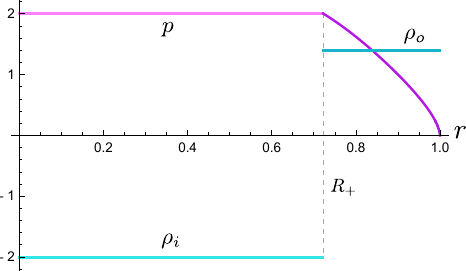}
    \caption{Typical pressure and density profiles for stars displaying a constant-pressure inner layer such that $p+\rho_{i}=0$. The star shown in this figure has $2M/R=0.99,~R^{2}\rho_{o}=1.4$ and $R_{i}=R_{+}\approx0.72R$. In the limit where the outer layer is infinitesimally thin, these solutions reduce to the Anti-de Sitter (AdS) star model (see Section~\ref{Sec:AdSStars} below).
    It is always possible to find stars belonging to this family for any compactness $2M/R<1$.}
    \label{Fig:RP}
\end{figure}

\begin{figure}
    \centering\includegraphics[width=0.6\textwidth]{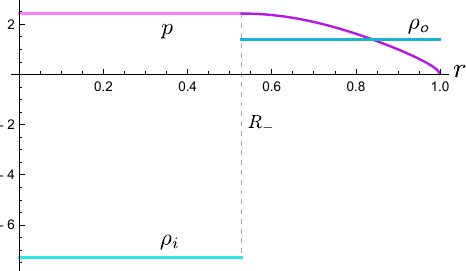}
    \caption{Typical pressure and density profiles for stars displaying a constant-pressure inner layer such that $3p+\rho_{i}=0$. The star shown in this figure has $2M/R=0.99,~R^{2}\rho_{o}=1.4$ and \mbox{$R_{i}=R_{-}\approx0.53R$}. In the limit where the outer layer is infinitesimally thin, these solutions reduce to the Einstein static star model (see Section~\ref{Sec:AdSStars} below).
    It is always possible to find stars belonging to this family for any compactness $2M/R<1$.}
    \label{Fig:RM}
\end{figure}

On the one hand, for $R_i = R_{+}$, the inner core has constant (positive) pressure in a way that guarantees $p+\rho_{i}=0$ everywhere inside the core. The redshift function adopts the simple form
\begin{equation}\label{Eq:Phi_i}
    \Phi(r)=\Phi_{i}\log\left(\sqrt{\frac{3-8\pi r^2\rho_{i}}{3-8\pi R_{i}^2\rho_{i}}}\right).
\end{equation}
It is straightforward to check that this solution is regular in the range $0\leq r<R_{i}$, since the only divergences in~\eqref{Eq:Phi_i} either come from
\begin{equation}
    8\pi r^2 \rho_{i}-3=0,\quad \text{or}\quad 8\pi R_{\text{i}}^2 \rho_{i}-3=0,
\end{equation}
which do not have real roots if $\rho_{i}<0$. On the other hand, the choice $R_i = R_{-}$ leads to an inner core that is characterized by a constant redshift function $\Phi(r)=\Phi_{i}$ and which satisfies the condition $3p+\rho_i=0$. Figs.~\ref{Fig:RP} and~\ref{Fig:RM} show plots of these two solutions with constant-pressure inner cores. These special values will be further studied in Section~\ref{Sec:AdSStars}. 

For any choice of $R_i$ in between those two values, namely $R_{-}<R_{i}<R_{+}$, we obtain regular solutions with $p'>0$ inside the core. To see this, first we note that for any value of $\rho_o$ greater than a given value $\rho_{\text{sep}}$ the pressure is bounded. Recall that the density of the inner layer is fixed in terms of the other parameters through Eq.~\eqref{Eq:RhoI}. In the $R_{i}\to0$ limit, it diverges towards negative infinity as
\begin{equation}
    \rho_{i}\propto \left(\rho_{c}-\rho_{o}\right)\frac{R^3}{R_{i}^{3}},
\end{equation}
provided that  $\rho_{o}\geq\rho_{\text{sep}}> \rho_{c}$. Since we have chosen $\rho_{o}$  such that pressure in the outer layer remains bounded regardless of the radius $R_{i}$ of the inner layer, and since by moving $R_{i}$ we can construct inner regions with  $\rho_{i}\in(-\infty,0)$, there is always an interval $R_{i}\in[R_{-},R_{+}]$ for which $p_{i} /\rho_{i}\in\left[-1,-1/3\right]$. Within this interval of the parameter space, the interior solution for the redshift, given by Eq.~\eqref{Eq:InnerPhi}, is regular, and approaches the solution of Eq.~\eqref{Eq:Phi_i} as $R_{i}\to R_{-}$, and the constant solution $\Phi(r)=\Phi_{i}$ as $R_{i}\to R_{+}$. Finally, since  $\rho_{\text{sep}}$ exists for every $2M/R<1$, we conclude that, by violating the assumption $\rho_{i}\geq0$, we can obtain a new family of solutions describing stars with constant-pressure inner layers for which there is no upper compactness bound. However, we also emphasize for values of $R_i$ outside the interval $[R_{-},R_{+}]$, it is also possible to find arbitrarily compact solutions although the pressure is no longer constant in the inside layer.
\begin{figure}
    \centering
    \includegraphics[width=0.7\linewidth]{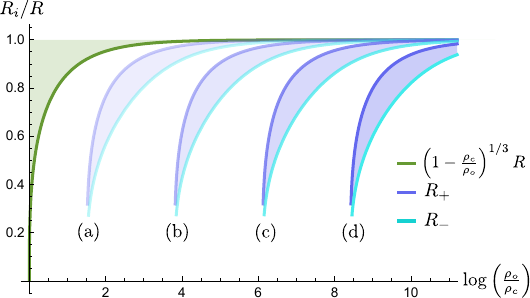}
    \caption{Regions of regular solutions for different compactness values. The shaded green area represents the parameter space for stars with $\rho_i>0$, with its boundary indicated by the green curve. Within this region we find the maximum compactness value $C_{\text{max}}\approx 0.9706 $. 
    The shaded blue regions (a), (b), (c) and (d) correspond to regions of the space of parameters (bounded by the curves $R_{+}$ and $R_{-}$) for which we find regular solutions with $2M/R=\left\{0.998,0.9998,0.99998,0.999998\right\}$, respectively. These solutions are necessarily in the $\rho_{i}<0$ region and can be found for any compactness value $2M/R<1$.}
    \label{fig:enter-label}
\end{figure}

As a summary, the main consequence of not constraining the sign of $\rho_i$ is that we can explore a broader region of the space of parameters $ \{ R_i, \rho_o, M\}$. This is depicted in Fig.~\ref{fig:enter-label} where, for some fixed mass value, only solutions inside the green region are allowed to have $\rho_i>0$. The green curve $R_{i}=\left(1-\rho_{c}/\rho_{o}\right)^{1/3}R$, where $\rho_{i}=0$, acts as the boundary of this region. Within the green portion of the diagram we find the maximum compactness bound~\eqref{Eq:CompLimRhoP}. Once we move below said curve, thus allowing for solutions with $\rho_{i}<0$, it is always possible to find perfectly regular stars whose compactness is \textit{arbitrarily close} to the black hole compactness. In particular, we have identified a set of constant pressure stars in the inner layer that fulfills this. 

\begin{figure}
    \centering
    \includegraphics[width=0.6\linewidth]{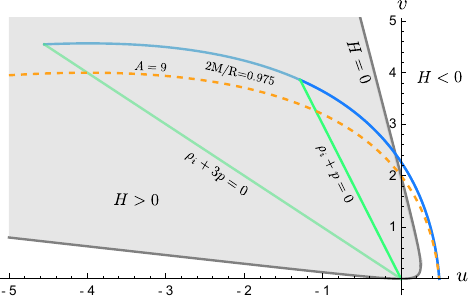}
    \caption{$(u,v)$ diagram of bilayered stars with negative energy densities. 
    The $v(u)$ curves from the inner layer appear in green and those from the outer layer in blue. We have plotted two example solutions satisfying $\rho_{i}+p=0$ and $\rho_{i}+3p=0$, respectively, which surpass Bondi's limit. The compactness of these solutions can be arbitrarily close to the black hole limit.}
    \label{Fig:UVNeg}
\end{figure}

These solutions have a complementary interpretation in terms of the $(u,v)$ variables used in Subsection~\ref{SubSec:Bondi_Limit}. The $v(u)$ curve describing the constant-density core solutions is given by
\begin{equation}\label{Eq:vuNegrho}
    v=-\frac{u\left(v_{0}+3\sqrt{3-6u}\right)}{v_{0}+\sqrt{3-6u}},
\end{equation}
where $v_{0}$ is an integration constant related to the central pressure via
\begin{equation}
    v_{0}=-\frac{3\sqrt{3}\left(\rho_{i}+p(0)\right)}{\rho_{i}+3p(0)}
\end{equation}
The constant-pressure curves $\rho+p_{i}=0$ and $\rho+3p_{i}=0$ are just the straight lines $v=-3u$ and $v=-u$ in the $(u,v)$ diagram from Fig.~\ref{Fig:uvDiagram}. These are obtained by evaluating~\eqref{Eq:vuNegrho} in the limits $v_{0}\to0$ and $v_{0}\to\pm\infty$, respectively. Since these $v$ lines always intersect the $P_{A}$ parabolas~\eqref{Eq:PACurves} at some $u<0$ for any $A>1$, it is possible to match them to an outer constant-density layer so that the compactness approaches the black hole compactness $u=1/2$ $(A\to\infty)$ as much as desired. Fig. \ref{Fig:UVNeg} shows the $v(u)$ curves for the two models with constant-pressure inner layers that we presented above.

\section{Anisotropy} 
\label{Sec:Anisotropy}

Another key assumption in Buchdahl's theorem is that the matter supporting the geometry is a perfect, i.e., isotropic fluid. However, it is natural to consider scenarios where the supporting matter exhibits anisotropic pressures. For completeness, we explore this possibility in the present section. In general, the energy-momentum tensor can be written as:
\begin{align}
    T_{\mu \nu} =  \rho T_{\mu} T_{\nu} + p_r R_\mu R_\nu + p_t \left( \Theta_\mu \Theta_\nu + \Phi_\mu \Phi_\nu \right),
    \label{Eq:GenericFluid}
\end{align}
where $\{ T_\mu, R_\mu, \Theta_\mu, \Phi_\mu \}$ represents a tetrad adapted to the symmetry of the problem, and we have introduced the density $\rho$, the radial pressure $p_r$, and the tangential pressure $p_t$. Notice that the two tangential pressures need to be equal due to spherical symmetry. 

\begin{figure}
\begin{center}
\includegraphics[width=0.4 \textwidth]{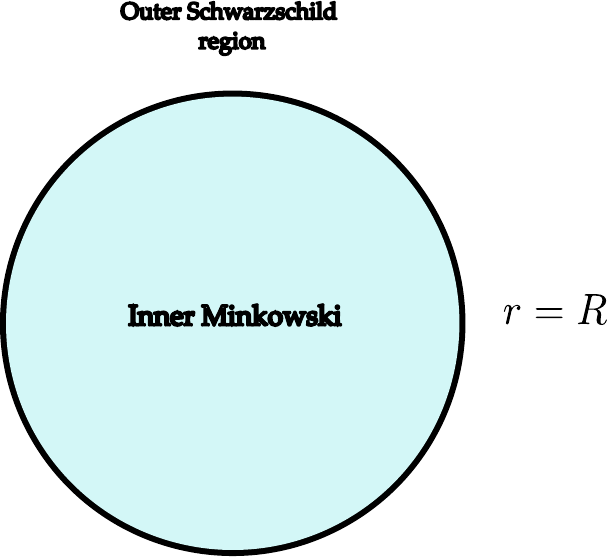}
\caption{Here we represent pictorially the matching of the inner Minkowskian core with the external Schwarzschild region through a spherically symmetric thin shell.}
\label{Fig:Shell}
\end{center}
\end{figure}

Buchdahl's theorem assumes a perfect fluid meaning that $p_r = p_t$. However, once we deviate from isotropy, there is no bound to the compactness that we can have. Let us illustrate this point with a very simple example of a shell with anisotropic pressures such that it can be made as compact as desired up to the black hole limit. The setup that we are considering is pictorially depicted in Fig.~\ref{Fig:Shell}.

Let us specify the construction in detail. Take a spherical shell with radius $R$, such that its proper area is $4 \pi R^2$. Let the spacetime outside the shell be Schwarzschild with mass $M$. Inside the shell, we take the spacetime to be flat, i.e., the Schwarzschild solution with $M =0$. Let us denote the two patches with a $+$ sign for the Schwarzschild patch outside and a $-$ sign for the flat patch inside. The metric is now given by the two line elements in Eq.~\eqref{Eq:thinshellbulks} with $f_{+}(r_{+}) = 1 - 2M/r_{+}$, $h_{+}(r_{+})=1/f_{+}(r_{+})$ and $f_{-}(r_{-}) = 1$, $h_{-}(r_{-})=1$. The shell is located at $r_{+} = r_{-} = R$.

The matching is worked out in detail in Appendix~\ref{App:AnisotropicShell} using Israel's junction conditions, and we find that the shell displays a distributional energy-momentum tensor as the one in Eqs.~\eqref{Eq:RhoDist}-\eqref{Eq:PtDist} with $\sigma$ and $\tilde{p}_t$ given by
\begin{align}
     \sigma & = \frac{1}{4 \pi R} \left( 1 - \sqrt{f_{+}(R)} \right), \nonumber \\
     \tilde{p}_t & = \frac{1}{16  \pi R \sqrt{f_{+}(R)}} \left(- 2\sqrt{f_{+}(R)} + 2 f_{+}(R) + f_{+}'(R) R  \right).
    \label{Eq:Densities}
\end{align}
Given that $\tilde{p}_t $ blows up at $R= 2M$, and that it goes to zero faster than $\sigma$ as $R \to \infty$, there is a crossover at a given $R = R_0$ for which $\sigma = \tilde{p}_t$. This means that we can expect violations of energy conditions. In fact, the DEC requires
\begin{align}
    \rho \geq p_t. 
\end{align}
Given that the two of them have the same distributional behavior, for our particular example this would imply
\begin{align}
    \sigma \geq \tilde{p}_t.
\end{align}
If we plug in the explicit values of $\sigma$ and $\tilde p_t$ given in Eq.~\eqref{Eq:Densities} and impose the DEC, we find the bound:

\begin{align}
    C= \frac{2M}{R} \leq \frac{24}{25} = 0.96.
\end{align}
This bound is less stringent than the Buchdahl bound for isotropic situations, $C<8/9 \sim 0.888$ ~\cite{Buchdahl1959}. But it is still more stringent than the bound that follows from just demanding the positivity of the $g_{rr}$ component of the metric, which corresponds to the black hole limit $C<1$. 

In~\cite{Ivanov2002}, bounds were derived for the maximum redshift that the surfaces of anisotropic stars can have under several assumptions, namely:
\begin{align}
    s = \frac{1}{(1-2M/R)^{1/2}}-1 \leq s_{\text{max}},
\end{align}
where the value of the $s_{\text{max}}$ depends on the assumptions invoked. In terms of the compactness, we have that the bounds are of the form:
\begin{align}
    \frac{2M}{R} \leq \frac{s_{\text{max}}(s_{\text{max}}+2)}{(1+s_{\text{max}})^2}.
\end{align}
In~\cite{Ivanov2002}, the bounds are derived assuming that the radial pressure is strictly positive (for stability reasons). It is also assumed that $\rho$ is positive and a decreasing function, although the possibility of a constant density for a finite interval is contemplated. Assuming the DEC is obeyed, the upper bound for the redshift found is:
\begin{align}
    s_{\text{max}}^{\text{DEC}} = 5.421 \rightarrow \frac{2M}{R} \leq 0.974,
\end{align}
whereas imposing the condition that $\rho \geq 2 p_t$ leads to a more stringent bound:\footnote{We believe that although Ivanov refers to this condition as the SEC, this is actually incorrect. If the condition $p_t \geq p_r$ is satisfied and the density and pressures are positive as they discuss, the SEC is actually much more restrictive.}
\begin{align}
    s_{\text{max}}^{\text{SEC}} = 3.842 \rightarrow \frac{2M}{R} \leq 0.958.
\end{align}
We can compare these bounds with the shell model that we are considering. In fact, the shell model begins to violate the DEC whenever the compactness is greater than $2M/R = 0.96$, which still obeys the bound $2M/R \leq 0.974$ found by Ivanov~\cite{Ivanov2002}. 

Further results that put bounds on the compactness of the objects with similar assumptions are the results reported in~\cite{Barraco2003,Boehmer2006,Andreasson2007,Urbano2018}. In~\cite{Barraco2003}, under roughly speaking the same assumptions as Ivanov's, it is shown that anisotropic models fall into two classes according to whether $p_r > p_t$ or $p_t >p _r$. In the case $p_r > p_t$, they manage to show that the $p_r$ function is strictly greater than the one in a fiduciary isotropic model with the same function $m(r)$. This means that these stars cannot be super-Buchdahl; actually, isotropic stars obeying DEC need to be less compact than Buchdahl~\cite{Barraco2002}. Regarding the case $p_r < p_t$, they show that the radial pressure is always bigger than the one in the fiduciary isotropic model with the same Misner-Sharp mass profile. Imposing the DEC, they derive bounds that are consistent with Ivanov's. In~\cite{Boehmer2006}, under some additional assumptions about the surface density behavior motivated on physical grounds, slightly sharper bounds are proved. Finally, in~\cite{Lindblom1984,Urbano2018} it is shown that imposing a causality condition, namely that the speed of sound is smaller than the speed of light, there is a maximum compactness that can be reached, which turns out to be below Buchdahl's limit, namely $2M/R \lesssim 0.768$. 

In any case, the simple example exhibited in this section shows that in the absence of energy conditions, anisotropic configurations can be made as compact as desired. Energy conditions set a limit to the compactness, but it is nevertheless far from Buchdahl's, leaving the door open to having ultracompact objects that are close to forming event horizons. In fact, such configurations can even display photon rings, since for that it is only required that $2M/R > 2/3 \sim 0.666$. 

\section{AdS Stars and Einstein Static stars} 
\label{Sec:AdSStars}

Among the bilayered models with a negative energy density at the core, we identified two with particularly notable characteristics, whose pressure profiles are shown in Figs.~\ref{Fig:RP}-\ref{Fig:RM}. Motivated by these solutions, we can consider a thin-shell limit of these models with compactness close to but below the black hole limit ($8/9<2M/R\sim 1$):
\begin{enumerate}
\item In the core they have a negative density that makes the pressure exactly constant, such that \mbox{$p(r)=p_i = -\rho_i$}. In the limit in which the outer thick layer becomes infinitesimally thin (distributional), we shall call these stellar configurations \textit{AdS stars}.

\item  In the core they have a negative density that makes the pressure exactly constant, such that \mbox{$p(r)=p_i=-\rho_i/3$}. In the limit in which the outer thick layer becomes infinitesimally thin (distributional), we shall call these stellar configurations \textit{Einstein static stars}. 
\end{enumerate}
AdS stars are constituted by an anti-de Sitter interior glued through a thin shell to a Schwarzschild exterior. The geometry can be built through a cut and paste procedure with the help of Israel junction conditions, as the thin-shell model of the previous section. The interesting thing here is that the qualitative behavior of this geometry constructed \emph{ad hoc} naturally represents a limiting case of the exact regular bilayered model. The only difference is that in such case, the transition from the inside core and the outside region is smooth, with no shell distribution behavior.

The setup is the same as the one depicted in Fig.~\eqref{Fig:Shell}, but considering the core to be Ads, instead of flat spacetime. The metric is now given by the two line elements in Eq.~\eqref{Eq:thinshellbulks}) with $f_{+}(r_{+}) = 1 - 2M/r_{+}$, $h_{+}(r_{+})=1/f_{+}(r_{+})$ and $f_{-}(r_{-}) = 1 + \abs{\lambda} r^2$, $h_{-}(r_{-})=1/f_{-}(r_{-})$. The real parameter $\lambda$ corresponds to a negative cosmological constant. Again, we locate the shell at $r_{+} = r_{-} = R$. Using the equations from Appendix~\ref{App:AnisotropicShell}, we find that the shell displays a distributional energy-momentum tensor as the one in Eqs.~\eqref{Eq:RhoDist}-\eqref{Eq:PtDist} with $\sigma$ and $\tilde{p}_t$ given by
\begin{align}
     \sigma & = \frac{1}{4 \pi R} \left(\sqrt{1+|\lambda|R^2} - \sqrt{1-\frac{2M}{R}} \right), \nonumber \\
     \tilde{p}_t & = -\frac{1}{8  \pi R } \frac{d}{dR} \left(
     R\sqrt{1+|\lambda|R^2} - R \sqrt{1-\frac{2M}{R}} \right).
\end{align}
In the crust of the bilayered model, the pressure increases from zero to $p_i$. In the distributional limit, it becomes a discontinuous jump at the thin-shell. The closer the compactness is to $1$, the larger this jump in pressure becomes. Although our aim here is to generate configurations surpassing the Buchdahl limit, it is remarkable that the structure of AdS stars is precisely the backbone of the semiclassical stellar solutions found previously in~\cite{Arrechea2021}. The only energy-momentum tensor supporting the stellar configurations reported there is that of a perfect fluid of constant density, together with the self-consistent semiclassical renormalized energy-momentum tensor of a massless minimally coupled scalar field computed in the Polyakov approximation.

In this self-consistent semiclassical solutions, there is always a positive energy density coming from the fluid which, in the core, is overcome by a negative semiclassical contribution from the renormalized energy-momentum tensor of the quantum field. In the limit in which the crust is very thin, these stars can be approximated by AdS stars. In that sense, one can consider AdS stars as a simplified proxy for these semiclassical stars.

It is important to highlight that these AdS stars are different from the AdS black shells of~\cite{Danielsson2017}. The biggest difference comes from the compactness. In the AdS black shell construction, the shell is located around the Buchdahl radius $R=9M/4$. In our case, the shell can be located arbitrarily close to the black hole limit.

Einstein stars are those in which the core has constant positive pressure satisfying \mbox{$\rho_i = -3 p_i$}. We have called them Einstein stars because of their similarities with Einstein static cosmological model~\cite{Einstein1917}. In that model, a 3-sphere filled with a pressureless dust, representing the matter content of the universe, is counterbalanced by the presence of a cosmological constant. As its cosmological acceleration is zero, we have that \mbox{$\rho_{\rm total}+ 3p_{\rm total}=0$}. In the cosmological case we have that 
\begin{align}
    \rho_{\rm dust}+ \rho_\Lambda +3p_\Lambda = \rho_{\rm dust} - \Lambda/(4 \pi)=0,
\end{align}
with $\Lambda$ the cosmological constant. In the case of our semiclassical stars we have 
\begin{equation}
    \rho_{\rm class} + \rho_{\rm semiclass} + 3 p_{\rm semiclass}=0,
\end{equation}
with $\rho_{\rm semiclass}$ being the only negative term. Geometrically speaking, these stars have a constant nonzero redshift core.  

In that sense, these models could serve as simple templates of black hole mimickers inspired by semiclassical effects but easier to handle analytically. One central element responsible for both the features of their ringdown~\cite{Cardoso2019} and their associated shadows~\cite{Ayzenberg2023} is the redshift, and the consequent time delay effect suffered by light rays crossing their interiors. While the AdS star has a redshift function that decreases towards the center, the one associated to the Einstein static star is fully constant. This would result in radically different crossing times for null rays and, in turn, potentially different observational signatures. 

\section{Discussion}
\label{Sec:Discussion}

The relaxation of one or both of the main assumptions of Buchdahl's theorem, namely, the outward decreasing monotonicity of the density profile, and the isotropy of the energy-momentum tensor, allows the existence of super-Buchdahl configurations. In this chapter we have developed different models capturing each of these possibilities. On the one hand, we have discussed how, for nondecreasing (yet everywhere non-negative) density profiles, a new limit to the compactness arises. This compactness limit was first found by Bondi~\cite{Bondi1964}.

On the other hand, we have constructed a geometrically smooth toy model that can approach this limit as closely as desired. Furthermore, we demonstrate that if negative densities are permitted, thereby violating the WEC, our model allows for objects with compactness arbitrarily close to that of a black hole, aligning with Bondi’s findings. 

Additionally, we also introduced a toy model of an anisotropic shell, demonstrating that, in the absence of additional constraints (such as energy conditions), anisotropic pressures can allow for objects to approach the black hole compactness limit arbitrarily closely. However, when energy conditions are imposed, the Buchdahl limit can still be violated, but new, stricter upper bounds on compactness emerge.

\subsection{Energy conditions and causality}

Once we allow violations of energy conditions, we need to be careful, since we are opening the door to a plethora of pathological behavior. For instance, violations of the DEC are often interpreted as leading to acausal behavior. In fact, the DEC can be understood as representing that the speed of the energy flow of the matter field is always subluminal. This is demonstrated by a theorem from Hawking~\cite{Hawking1970,Hawking1973}, which states that if the energy-momentum tensor satisfies the DEC and vanishes on a portion of a Cauchy slice, it must also vanish throughout the domain of influence of that region. This can be understood as implying that signals constrained by the DEC cannot propagate faster than light~\cite{Wong2010}.

However, that the shell or the constant-density star do not obey the DEC does not necessarily mean that its perturbations are acausal. In fact, perturbations with respect to a configuration that violates the DEC are not guaranteed to violate the DEC as well. To put it explicitly, given our configuration with background values of the density and pressure $\bar{\rho}, \bar{p}$ that violate the DEC, i.e., $\abs{\bar{\rho}} \leq \abs{\bar{p}}$, the perturbations $\delta \rho, \delta p$ do not necessarily obey $ \abs{\delta p} \leq \abs{\delta \rho}$. Although it might be useful to understand this in terms of a background and a perturbation energy-momentum tensor, such splitting always involves an intrinsic arbitrariness. This makes it difficult to precisely determine how a background configuration might violate the DEC without leading to pathological behavior in perturbations that otherwise respect it.

Furthermore, the superluminal behavior that is tied to violations of the DEC may not necessarily be problematic, as there are examples of situations in which apparently superluminal behaviors do not necessarily entail a pathology~\cite{Babichev2007,Geroch2010,Barcelo2022}. We believe that the DEC violation in this case would point to an instability of the background for generic perturbations, which in the shell case that we analyze is expected since the object would tend to collapse. The theorems in which the DEC is used, namely, to prove many of the results concerning no-hair theorems, black hole mechanics, and to ensure the well-posedness of the initial value problem in its general formulation~\cite{Curiel2014}, do not seem to contradict this interpretation.

When we allow nonmonotonically decreasing density profiles, we may expect that if the matter content is that of an ordinary fluid, it will display an unstable behavior, but no energy conditions need to be violated, \textit{a priori}. If we go further and allow negative energy densities, causally pathological behavior might arise due to the violation of the WEC. However, if such behavior of the effective densities and pressures is motivated by an underlying healthy and causal physical mechanism, e.g., semiclassical physics, the pathological behavior might not arise. Indeed, this behavior is expected if one assumes that the relationship between density and pressure perturbations mirrors that of the background quantities. However, this assumption does not necessarily hold, and, in the case of semiclassical gravity, it is definitively violated.


\setcounter{secnumdepth}{-1}
\chapter{Conclusions}
\label{ChF:Conclusions}

\fancyhead[LE,RO]{\thepage}
\fancyhead[LO,RE]{Conclusions}


\section*{Summary of the results of the thesis}

This section is devoted to summarizing the main results of the thesis. We also outline possible directions for further research that naturally emerge from this work.

\subsection*{Lessons from analogue gravity}

General Relativity without any constraints on the energy-momentum tensor, such as those imposed by energy conditions, leads to an unreasonable theoretical framework. In such a setting, causal pathologies abound, undermining the predictability of the theory and allowing for problematic spacetime geometries. The imposition of energy conditions eliminates many of these issues, ruling out, for instance, the existence of closed timelike curves. A conceptual twist, however, emerges at the semiclassical level, where violations of energy conditions naturally occur. In this context, it becomes unclear how the geometry evolves under semiclassical gravity. These concerns motivated Hawking to propose his well-known Chronology Protection Conjecture~\cite{Hawking1991}, which posits that spacetimes with pathological causal features, such as those admitting closed timelike curves, should be excluded from physical theories.

Although most physicists agree that spacetimes with pathological causal structures should be excluded, the precise mechanism that enforces this exclusion remains unclear. In Chapter~\ref{Ch1:ChronologyAnalogue}, we explored this issue from a different angle by employing the framework of analogue gravity, where the causality encoded in the effective metric is emergent, while the underlying Galilean (and ultimately Lorentzian) causal structure of the laboratory remains intact. Within this context, our analysis was purely kinematical, as the emergent geometries do not satisfy any deformed version of the Einstein equations. Rather than providing a sharp analogy for gravitational dynamics, our analysis offers conceptual insights into the potential role of background structures, particularly those determining causality, in preventing the formation of causally pathological geometries.

In the analogue gravity framework, we have shown that even though energy conditions no longer constrain the form of the emergent metric, it remains impossible to construct time machines or causally ill-behaved configurations. This limitation arises from the underlying causal structure of the laboratory system, which enforces causal consistency, albeit in a way that may not be evident to observers that are only able to probe the emergent geometry. For example, the construction of a closed timelike curve within the analogue metric would require an infinite amount of work, effectively forbidding its realization.

If the geometry of General Relativity were in fact emergent, underpinned by a more fundamental causal structure in nature, one could anticipate a similar phenomenology to that observed in analogue gravity systems. However, establishing a direct connection between the absence of closed timelike curves and the existence of this deeper causal structure is far from straightforward. Even if such a structure were ultimately responsible for forbidding them, making the link explicit would require a way to probe or access it.

Some analogue systems possess a quantum substratum, meaning that the system itself has an inherently quantum nature. This opens the door, at least theoretically, to exploring more extreme limits of causality. In particular, one can ask what happens when excitations propagate on top of a quantum superposition of two states, each of which individually gives rise to a different effective geometry. This is precisely the question addressed in Chapter~\ref{Ch2:AnalogueSuperp}.

The superposition of two emergent geometries results in a blurred notion of causality that proves to be unstable: any small perturbation tends to disrupt the configuration. Once again, systems with emergent causal structures seem to naturally avoid situations with ill-defined or pathological causal behavior. Whether one attempts to push causality to its limits by constructing configurations that are causally pathological, or to blur it by superposing distinct geometries, the system appears to resist it.

Part of the motivation for exploring these scenarios lies in the desire to better understand the interplay between gravitational theories and the quantum realm. While it is common to assume that a theory of quantum gravity should resolve the singularities of General Relativity, it is far less conventional to consider whether gravity might, in turn, offer insights into foundational issues in quantum mechanics. This is tied to the broader observation that most approaches to quantum gravity focus on quantizing the gravitational field, yet few explore the possibility of adapting the rules of quantum theory to the general relativistic framework. In Penrose’s words, this would involve ``gravitizing'' quantum mechanics, a crucial step in reconciling the two theories. 

Our modest contribution in this thesis has been to illustrate, within the concrete framework of analogue gravity, how causality plays a fundamental role in the formulation of physical theories. We have shown that when a system is configured in such a way that its causal structure becomes ill-defined, it tends to evolve away from that configuration toward one with a well-behaved and consistent causal structure. In fact, we interpret this as suggesting that successfully reconciling quantum rules with the gravitational realm must involve imposing constraints on the causality of the system. While some degree of causal indeterminacy may arise at the quantum level, we argue that huge departures from a well-defined causality should be dynamically suppressed.

\subsection*{Building gravitational theories}

To gain deeper insight into the conditions under which a system might admit an effective description governed by a version of Einstein's equations, given that it is not the case for the majority of known analogue systems, we have revisited General Relativity from a constructive perspective. Specifically, in Chapter~\ref{Ch3:SelfCoupling} we have re-examined its formulation as a consistent theory of a self-interacting massless spin-2 field propagating on a fixed Minkowski background.

We have focused on the off-shell approach to reconstructing General Relativity, which builds the full nonlinear theory starting from the Fierz-Pauli action for a linear spin-2 field. Through a reverse-engineering exercise, we have demonstrated that any Diff-invariant theory, including higher-order theories of gravity, can be obtained from its linearization via the bootstrapping procedure. From a constructive approach, at each step of this iterative process, ambiguities naturally emerge and must be addressed. Contrary to some claims in the literature, we have shown that these ambiguities are physically meaningful: different choices in resolving them can lead to inequivalent nonlinear completions of the same linear theory. A notable example is offered by Lovelock theories (in spacetime dimensions greater than four), whose linearized dynamics always reduce to the Fierz-Pauli form and can all be obtained through bootstrapping. Each of them arises from different choices made during the procedure.

Second, we have highlighted that much of the analysis in the literature is contaminated by the implicit expectation that the end result of the bootstrapping procedure must be General Relativity, or more generally a background-independent theory. However, we have shown that this does not need to be the case. By starting from the WTDiff theory, an alternative description of a massless spin-2 field in Minkowski spacetime, we demonstrated that it is possible to reconstruct Unimodular Gravity as a consistent nonlinear completion. In this case, the graviton couples order by order to the traceless part of the energy-momentum tensor. While our method was limited to a reverse-engineering approach, it effectively illustrates that consistent nonlinear completions of massless spin-2 fields can lead to background-dependent theories. Whether a broader class of such background-dependent completions exists to linear theories remains an open question worthy of further investigation.

Unimodular Gravity was already known, but we believe that most of the results regarding its relationship to General Relativity have remained either unclear or sparsely distributed across the literature. For this reason, we found it worthwhile to undertake a critical review and perform a careful, systematic comparison between the two theories and fill the gaps in the literature when needed. This was the content of Chapter~\ref{Ch4:UG}. While various claims have suggested potential distinctions between Unimodular Gravity and General Relativity, our analysis shows that, apart from the nature of the cosmological constant, the two theories are equivalent. In Unimodular Gravity, the cosmological constant arises as an integration constant rather than as a coupling parameter. As a result, the theory possesses an additional global degree of freedom.

Our analysis extends to higher-derivative theories of gravity and arbitrary matter couplings, provided they respect the symmetries of the theory, and applies at the classical, semiclassical, and perturbative quantum levels. We have also revisited the conformal factor problem in Euclidean gravity, showing that in Unimodular Gravity its nature changes, manifesting instead as a longitudinal diffeomorphism. In all cases considered, whether involving the Einstein–Hilbert term, higher-derivative generalizations, or metric-affine frameworks, we consistently find that the only distinction from General Relativity lies in the cosmological constant.   

Building on our analysis of the theory and its comparison with General Relativity, we have also revisited the standard arguments often cited to support the claim that General Relativity emerges as the low-energy limit of string theory. We have shown that, in fact, it remains a matter of choice whether this low-energy description is framed within a Diff-invariant (General Relativity-like) or a WTDiff-invariant (Unimodular Gravity-like) theory. This observation already prompts a broader reconsideration: whether a theory that reproduces General Relativity-like dynamics, such as one where the metric emerges from more fundamental degrees of freedom, must necessarily be described by a Diff-invariant framework.

\subsection*{Alternative to black holes}

Asymptotically flat black hole geometries in General Relativity exhibit a remarkable simplicity. In fact, a common interpretation of the no-hair theorems suggests that all solutions to the vacuum nonlinear field equations with these properties fall within the two-parameter Kerr family. This is particularly striking given the nonlinear and intricate nature of the Einstein vacuum equations.

As an attempt to disentangle some of the subtleties behind the existing theorems, we have focused on the physical content underlying Israel’s theorem. We deconstructed the theorem in its simplest setting, adding the further simplifying assumption of axisymmetry. The full analysis that we carried out in Chapter~\ref{Ch5:Nohair} has uncovered aspects of General Relativity that, to the best of our knowledge, have not been clearly explained elsewhere.   

A key assumption underlying the theorems is that the geometry satisfies the vacuum Einstein equations throughout all the external region. This assumption limits the applicability of these results to astrophysical contexts, as black holes are never truly isolated. Moreover, it raises questions about the physical relevance of the vacuum assumption, specifically whether matter in equilibrium in the external region could introduce additional parameters in the solution. We have shown that the essential physical insight behind the no-hair theorems is that, for any gravitational environment, there is only one possible geometric shape for the horizon which makes it a nonsingular surface. This implies that the horizon adjusts from spherical symmetry to match the structure of the surrounding gravitational environment, but such environment does not ``grow hair'' on the black hole.  

We further explored how the presence of a horizon constrains the metric. We showed that if a horizon is replaced by an object parametrically close to it, such as a maximum redshift surface (which becomes an infinite redshift surface in the black hole limit), the geometry will approach that of a black hole as long as the curvature invariants remain bounded. This analysis demonstrates that integrating Einstein's equations from infinity to a high redshift surface, along with specific curvature bounds, restricts the allowed geometries. In the infinite redshift limit, the only nonsingular solution that arises from this integration is the Schwarzschild black hole, provided the curvature invariants remain finite.

In Chapter~\ref{Ch6:ToroidalBHs}, we constructed the first 4D black hole geometries with toroidal event horizons that are nonsingular in the external region. It is a natural follow-up to the previous results, built upon the same tools developed there. Along with a detailed construction of these geometries, we explained why previous attempts led to singularities in the external region. Additionally, we conducted a thorough analysis of energy condition violations, which are required to have exotic horizons, demonstrating that, in the specific case considered, all pointwise energy conditions are violated. 

Finally, in Chapter~\ref{Ch7:Buchdahl}, we turned to a new question: what types of compact objects can arise in General Relativity that are not black holes but still mimic most of their properties? We showed that under generic assumptions for the energy-momentum tensor, such as energy conditions, causality constraints, e.g., finite sound speeds, and similar physically motivated criteria, objects cannot be made arbitrarily compact. We revisited two classic theorems concerning fluid spheres that match onto an external Schwarzschild geometry. The first, Buchdahl’s theorem, establishes a bound on compactness under the assumptions of isotropy and monotonically decreasing energy density. We demonstrated using a simple toy model that relaxing isotropy allows objects to become arbitrarily compact, provided no energy or causality conditions are imposed. Once those are enforced, new bounds on compactness emerge, which we have reviewed in detail.

Regarding the assumption of a monotonically decreasing energy-density profile, we revisited Bondi’s theorem, which shows that a new compactness bound arises if one only assumes nonnegative energy densities. We constructed a toy model of a thick shell that asymptotically approaches Bondi’s bound without introducing distributional sources, offering a clearer physical interpretation. Additionally, we presented a second model involving negative energy densities, capable of reaching arbitrarily close to the black hole compactness limit, just as Bondi originally argued should be possible. Interestingly, this second model shares many features with recently found semiclassical stars, which are exact solutions of the self-consistent semiclassical gravity equations in the Polyakov approximation for spherically symmetric perfect fluid configurations.

\newpage

\section*{\textit{Coda.} Some personal thoughts}

In conclusion, I shall now offer some personal thoughts on the work presented in this thesis. These final pages should be taken as a summary of the main lessons that I have learned during this thesis together with some of the questions that this thesis has opened and I would like to address in the future. All of them fit within the framework of gravitational physics and its interplay with the quantum realm. This should be taken as a personal note, which is specially biased toward my interests at the moment of writing. 

There are two well-known key aspects of General Relativity that I have reconsidered after this thesis, since some of the results in the thesis shed new light on them. First and foremost, General Relativity is remarkably robust. By this, I mean that small deformations of the theory, such as modifications to its dynamics through higher-curvature terms, semiclassical corrections, or other extensions, tend to lead to pathologies, and hence General Relativity is often preferred over these deformations. Even though General Relativity harbors certain pathologies, such as singularities arising from gravitational collapse, it remains a well-behaved theory in the sense that these issues are typically hidden behind event horizons. Furthermore, General Relativity is stable, showing no signs of pathological behavior like runaway solutions or ghost modes when perturbing around background geometries.

In contrast, most of the deformations of General Relativity when not understood as EFT corrections, tend to introduce instabilities. For example, adding higher-order curvature terms or adopting a metric-affine formulation often leads to Ostrogradsky instabilities due to higher-order equations of motion, or to the emergence of singular solutions. Unless such modifications are treated perturbatively within an EFT framework, with General Relativity being the leading-order behavior, they frequently result in pathological dynamics. 

The second key property is that General Relativity is surprisingly simple in the following sense. Despite being a nonlinear theory, its space of solutions displays a remarkable degree of simplicity. A prime example are the stationary black hole solutions: all of them are uniquely described by the biparametric Kerr family. This is the standard reading of the no-hair theorems, and to date, no counterexamples have been found. Out of the vast space of possible vacuum solutions, only a bi-parametric family (with these parameters representing the mass and angular momentum) is realized. In this sense, vacuum black holes are strikingly simple objects, especially when compared to something as familiar as a star in Newtonian gravity, which can in principle require an infinite set of multipole moments to describe its gravitational field, as all of them are independent. 

This simplicity extends to dynamical aspects as well. For instance, one can compare Einstein's equations to another famous set of nonlinear equations: the Navier-Stokes equations. While General Relativity has a well-posed initial value problem, the analogue for Navier-Stokes equations remains unresolved and is one of the Clay Millennium Problems. A heuristic explanation for General Relativity's comparative simplicity is that gravitational waves propagate strictly at the speed of light, without dispersing. As a result, they do not form shock waves, a property that distinguishes them from other nonlinear wave phenomena. 

Regarding the robustness of General Relativity, this thesis has questioned whether such robustness is uniquely singled out from a constructive perspective. In particular, from the viewpoint of the bootstrapping approach used to reconstruct nonlinear interacting theories from their linear limits. We have seen that any diffeomorphism-invariant theory can, in principle, be obtained through such a procedure. Moreover, even when restricting to theories that propagate only the two degrees of freedom associated with the graviton, we are not limited to the standard Fierz-Pauli description. An alternative is provided by the WTDiff theory, which also describes a massless spin-2 field in flat spacetime. Remarkably, WTDiff can be bootstrapped to yield Unimodular Gravity as its nonlinear completion.

The only difference found between General Relativity and Unimodular Gravity, or more generally, Diff-invariant theories and their WTDiff-invariant counterparts, is the character of the cosmological constant. In Diff-invariant theories it appears as a coupling constant and in WTDiff invariant theories as an integration constant. At the classical level, I do not believe that this difference offers any significant new insight. First of all, the cosmological constant can be safely ignored for most of the gravitational physics that does not involve cosmological scales. And even at cosmological scales, gravitational theories cannot be directly tested, as cosmological observations inherently rely on assumptions about the underlying geometry, making them model-dependent ~\cite{Ellis2012}. From this perspective, whether the cosmological constant is an integration constant or a coupling constant is irrelevant. Independently of that, one would need to fix it for the specific model (we mean the assumptions made on the geometry and the matter content, e.g., homogeneity and isotropy, perfect fluid, etc.) and gravitational theory (we mean the specific dynamics, i.e., the equations for the gravitational field). 

It has also been argued in the literature that the distinction between General Relativity and WTDiff-invariant theories could prove advantageous when quantum corrections are taken into account, particularly because the cosmological constant in WTDiff theories does not suffer from a naturalness problem. I have critically examined this line of reasoning and believe it should be treated with caution. As a guiding principle for theory building, I find it slightly sloppy and insufficiently sharp to be useful, especially in the case of the cosmological constant, whose value is ultimately fixed by model-dependent observations.  From my perspective, the real interest in WTDiff-invariant theories lies in their potential to offer fresh perspectives for constructing gravitational theories beyond General Relativity and Unimodular Gravity. For example, since the cosmological constant in Unimodular Gravity behaves as a dynamical variable, albeit with a trivial dynamics, it would be worthwhile to explore extensions where it acquires nontrivial dynamics.

Moreover, WTDiff-invariant theories, as formulated in this thesis, incorporate a background structure, a fixed volume form, that, while breaking full diffeomorphism invariance, remains dynamically harmless. It does not lead to any pathologies or actually introduce dynamics that deviates from the General Relativity one. Admittedly, I found this somewhat disappointing, as my initial motivation for studying these theories was the hope of uncovering meaningful dynamical differences from General Relativity. Nonetheless, I believe the analysis remains valuable, as it demonstrates that the presence of background structures in a theory does not necessarily compromise its consistency or physical viability.

I believe that many of the theoretical prejudices against background-dependent theories are rooted in historical developments. Since General Relativity was constructed getting rid of the background structures from the outset, it has often been tacitly adopted as an epistemological principle in model building, favoring background independence as a guiding criterion. However, the analysis presented in this thesis demonstrates that the presence of background structures does not necessarily lead to inconsistencies or pathologies, and thus, they should not be dismissed based on this principle alone. 

In fact, as suggested by the first two chapters of this thesis, background structures can help in circumventing potential problems that background independent theories may have. In such theories, causality is determined dynamically, which opens the door to causally pathological solutions, particularly if appropriate constraints on the matter content are not enforced. For instance, in the first chapter of the thesis we showed that background structures might help in precisely avoiding these situations. Furthermore, at the quantum level, our results from the second chapter precisely suggests that, although an instance of a blurred causality is expected at the quantum level, the system abhors any strong deviation from a fixed causality. From this perspective, background structures that impose a fixed causal framework can be advantageous, helping to sidestep these potential issues.

Another clear example of background-dependent theories where the background structure is not only harmless but essential is provided by the ghost-free massive gravity theories developed by de Rham, Gabadadze, and Tolley~\cite{deRham2010}. In these models, a fixed reference metric is required to construct the interaction terms, yet this background does not introduce any pathological behavior. Inspired by this, it would be interesting to explore whether a similar construction could be adapted to WTDiff-invariant theories. While such an extension would likely involve additional fields and increased complexity, it may offer a path toward generating a dynamical cosmological constant within this framework.

On another front, the analysis of no-hair theorems presented in this thesis has led to two main insights. First, the validity of no-hair theorems does not rely on the actual presence of an event horizon. Any object approaching horizon formation, provided that we constrain its curvature to remain sub-Planckian, exhibits a structure that is parametrically close to its black hole counterpart. Second, the presence of an external gravitational environment does not alter this conclusion: while the black hole geometry may adapt to the environment, no additional parameters are introduced. In this sense, the essential ingredients that constrain the multipole structure of the configuration are the existence of arbitrarily high redshift surfaces together with an asymptotically flat region.

The no-hair theorems are dynamical, as they rely on Einstein's equations, but they only apply to stationary configurations. As such, they do not provide any insights on how these stationary states are approached dynamically. Moreover, they inherently exclude certain pathological geometries, such as the Curzon solution, with suitable assumptions. I believe that further exploration of no-hair theorems from the perspective adopted in this thesis, particularly in more general settings like the fully nonaxisymmetric static case and, specially, the stationary and axisymmetric case, could also unveil new or already known but not deeply explored physically interesting classes of geometries that merit study in their own right.

My main interest in these types of solutions stems from the possibility that they might represent the idealized end-states of certain dynamical processes. For example, I suspect that the Curzon solution could correspond to an extremely anisotropic gravitational collapse, in which matter becomes infinitely smashed in one direction tending to generate an infinitely thin and diluted disk of matter. At the ring of this disk is where the singularity would be located. This is specially relevant, as from a dynamical perspective, our understanding of gravitational collapse in highly nonspheroidal configurations remains limited, especially in scenarios that fall far outside the scope of the hoop conjecture~\cite{Klauder1972}. 

In summary, many aspects of General Relativity remain poorly understood, and several important problems are yet to be resolved. Exploring frameworks beyond General Relativity offers a valuable opportunity not only to address these challenges, but also to gain new insights into the theory itself.


\cleardoublepage

\setcounter{secnumdepth}{2}
\appendix

\chapter{Fundamentals of quantum mechanics and gravitation}
\label{AppA}

\fancyhead[LE,RO]{\thepage}
\fancyhead[LO,RE]{Fundamentals of quantum mechanics and gravitation}


This appendix serves as a complement to Chapter~\ref{Ch2:AnalogueSuperp}. The original motivation for many authors to explore gravitationally inspired models for quantum state reduction stemmed from the foundational problems of quantum mechanics. Our analysis, viewed through the lens of analogue and emergent gravity, also provides new insights into these issues, as discussed in Chapter~\ref{Ch2:AnalogueSuperp}. We specifically focused on comparing our work with Penrose’s ideas because they are presented in a form that is closer to our model. However, it is worth noting that, in spirit, many other models share similarities with Penrose’s approach. In preparing the article in which the chapter is based~\cite{Barcelo2021c}, we conducted an extensive review of the literature, attempting to gather and summarize the key features of these works, as well as to describe standard approaches to addressing quantum mechanics foundational problems without invoking gravity. Since it is difficult to find a clear and organized summary of the relevant literature on this topic, we have chosen to include it here in this appendix.

\section{The quantum puzzle}
\label{QuantumPuzzle}

Quantum mechanics, as described by just a dynamical wavefunction, faces two related problems. The first one is how to accommodate the apparent absence of superpositions of macroscopic systems. As the formalism per se does not suggest a division between microscopic and macroscopic systems, the same rules for superposition should apply to both realms. The second problem, is how to account for the specific although probabilistic results of actual acts of measurement in the laboratories, i.e., how to account for the apparent intrinsic statistical result of individual acts of measurement. The formalism of quantum mechanics just provides us with a statistics for the outcome of a huge number of experiments repeated under identical conditions. Although a pragmatic attitude towards interpreting the origin of these probabilities can be taken, it is quite undesirable and problematic from a more realistic point of view. The reason for this is that such a pragmatic approach directly assumes that no reality or meaning can be given to the observations. Nowadays, such a ``statistical'' interpretation is not problematic from an epistemological point of view but our opinion is that it would hardly be a complete description of the physical world. However, other interpretations of quantum mechanics pretend to answer this question in order to, first, be able to assign an ontological nature to the systems underlying observations and, second, open the door to modifications of the theory with a potential observational impact which might be obscure or counterintuitive within this pragmatic view. We can catalog these alternative best known ways of solving the conundrum as (see for instance the review~\cite{Bassi2013}):
\begin{itemize} 
\item
\textbf{Copenhagen interpretation}: It postulates a (imprecise) division between macroscopic and microscopic entities following classical and quantum evolution laws respectively \cite{Bohr1928}. It incorporates two distinct evolution laws for quantum systems \cite{vonneumann1955}: the unitary evolution, when no external measurement is taken, and the non-unitary evolution (state reduction or collapse), when one measures a quantum property of the system using a classical apparatus. This law has a probabilistic nature. In practice these assumptions have proved to work extremely well. However, many researchers have tried to provide a deeper understanding of this separation. Furthermore, from a physical point of view the division itself seems rather unnatural and somehow an ad hoc imposition: for the theory to be predictive we would additionally need an exhaustive classification of the systems according to their classical or quantum behavior. This suggests that a more conceptually satisfying underlying explanation might exist which effectively agrees with the Copenhagen interpretation. 

\item
\textbf{Decoherence and many worlds}: 
Many scientist believe that the disappearance of correlations in the way towards the macroscopic world is caused by a phenomenon often called environmental decoherence \cite{Zeh1970,Zurek1981,Zurek1982}. This phenomenon can be described within standard linear quantum mechanics and it is a consequence of the fact that we never observe the entire world but restrict our observations to partial open subsystems, with the rest of the world acting as an environment. Interactions of the system with the environment, which are always present, tend to suppress the correlations of the system with itself in favor of correlations of the system with the environment. Then, the effect of partial tracing over the environment degrees of freedom, typically difficult to monitor, leads to a reduction of the state of the open system much like the standard collapse of the wave function in the Copenhagen interpretation.

In fact, from the viewpoint of environmental decoherence the non-unitary collapse of the wave function when measuring a quantum property with a classical (macroscopic) apparatus is just an effective description of a very efficient (almost instantaneous) process of decoherence by the environment. From this viewpoint, it is also reasonable to expect that the lifetime of quantum superpositions would diminish progressively as a system becomes more and more macroscopic, since their interactions with the environment are usually enhanced by the complexity of such object. Here, the words {\em macroscopic} and {\em complexity} are used in an intuitive, nontechnical manner. From a mathematical point of view, this is an explanation of how we can pass from a quantum probability distribution (a general density matrix) to a classical probability distribution (a density matrix diagonal in some basis, the so-called pointer basis) which has no trace of quantum correlations. The existence of macroscopic superpositions is not fundamentally forbidden, only highly suppressed.

However, this mechanism by itself, does not solve the second quantum puzzle we have mentioned. The final description of a measurement process is a probabilistic mixture of all the possible outcomes. In fact there is nothing fundamental forbidding that the lost correlations are recovered in the far future. For finite systems, such a recovery is inevitable: recurrences always occur, although the timescales at which they occur (the Poincar\'e recurrence time) might be too large for typical systems~\cite{Bocchieri1957}. Again, this is something that we do not see but we do not know whether the reason is fundamental or just practical. Accepting the absence of any fundamental evolution law other than the unitary evolution leads naturally to a many-world interpretation of quantum mechanics, if one insists on taking a realist approach and neglects the possibility of having hidden variables.

The many-worlds interpretation put forward by Everett~\cite{Everett1957,Wheeler1957} assumes that only the unitary evolution really exists. Results of a measurement different from the one we actually obtain, live in different coexistent worlds we do not have access to. This interpretation displays several problems, for instance, being the evolution deterministic it is not clear how Born's probability rule is recovered for repeated experiments under identical conditions~\cite{Bassi2013}. For more potentially unpleasant philosophical consequences, see~\cite{Bacciagaluppi2001}.

The combination of the decoherence mechanism and the many-worlds interpretation applied to the already reduced quantum states leads to a conceptual framework which can be currently considered as the ``establishment view'', paraphrasing~\cite{Bassi2013}. However, it is important to remark that this viewpoint is far from being sufficiently verified and free of conceptual puzzles~\cite{Bacciagaluppi2007}. 

\item
\textbf{Bohmian mechanics}: In this theoretical framework, originally conceived by de Broglie and later further pursued and formalized by Bohm~\cite{Bohm1952,Bohm1995}, the wavefunction is an abstract physical entity that guides the trajectories of real classical particles or fields, which are treated as in classical statistical mechanics. Measurement is a purely classical process, and the collapse of the wave function after measurement only encodes a reselection of the statistical properties of the particles in the ensemble. Bohmian mechanics is experimentally equivalent to the Copenhagen interpretation (expect perhaps in situations of nonequilibrium \cite{Valentini1991,Valentini1991b}). Its interest resides in the fact that it is a proof of principle that an ontological interpretation of quantum mechanics is possible. 

\item
\textbf{Modified quantum-mechanical theories}: Even acknowledging the success of the standard quantum mechanical formalism, since the very birth of quantum mechanics there has been a minoritarian but steady stream of works trying to find modified quantum mechanical frameworks free from some of the puzzles of the standard formulation.  
On the one hand, there are many proposals prescribing nonlinear generalizations of quantum mechanics which could combine in a single dynamical law both the unitary and the non-unitary evolution~\cite{Garcia-Moreno2019,Morgan2020,Barcelo2012,Oppenheim2018,Oppenheim2020}. On the other hand, there are more or less vague attempts to add to the standard formalism a fundamental source of decoherence. These approaches attempt to solve the first of the two problems we have presented by introducing a modified dynamics which forbids superposition of macroscopic objects. Some of them try to introduce nonlinearities in the Schr\"{o}dinger equation, in order to have a single equation describing the classical and quantum regimes. For a recent review of these ideas see~\cite{Paredes2019}. Other proposals try to introduce some kind of stochastic dynamics in order to model the processes of localization of the wave function, being the proposals of Ghirardi-Rimini-Weber~\cite{Ghirardi1986} one of the most famous examples.
There are other proposals which associate the fundamental loss of coherence to the omnipresence of gravity~\cite{Karolyhazy1966,Komar1969,Karolyhazy1974,Diosi1986,Coleman1988,Diosi1989,Gisin1989,Penrose1989,Ghirardi1990b,Penrose1992,Percival1995,Pearle1996,Penrose1996}. These approaches tend to suggest that both the classical and quantum dynamics are only effective approximations of an universal dynamics. In other words, both kinds of dynamics seem to correspond to certain limits of a more general theory.

A strong criticism to these kind of fundamental modifications of quantum mechanics was presented in~\cite{Banks1984}, suggesting that dynamical laws which map pure state to mixed state are not viable, since they either violate causality or energy conservation. However, in~\cite{Unruh1995}, the authors argue that evolution laws transforming pure states into mixed states can be chosen in such a way that causality and energy-momentum conservation hold at all scales, exemplifying it with some non-Markovian models. Furthermore, they point out that it is possible to choose a subset of the Markovian models analyzed in~\cite{Banks1984} avoiding causality problems and such that all the possible deviations from the ordinary dynamics, including the energy conservation violations, are confined to a given sector of the Hilbert space, avoiding conflicts with experimental results. In~\cite{Unruh2017}, the authors argue that the belief that energy conservation violation is inevitable with non-unitary evolution laws comes from the misbelief that any environmental decoherence requires energy transfer among the system and the environment. They emphasize that it is possible to build quantum mechanical systems, like the ones considered in~\cite{Egusquiza1999,Unruh2012}, which interacts with an environment in a way such that the evolution erases the quantum coherences of the system and exact energy conservation holds.
\end{itemize}

\section{Gravity's role in quantum state reduction}
\label{GravityRole}

As we have already mentioned, there has been a steady trend trying to see whether the standard formalism of quantum mechanic fails in some regime~\cite{Wigner1962,Bialyanicki1976,Pearle1976,Ellis1983,Pearle1989,Ghirardi1986,Ghirardi1990,Bassi2013,Feldmann2012,Bassi2016}. 
Furthermore, there have also been many attempts to relate this violations of quantum mechanics with a gravitational origin \cite{Karolyhazy1966,Komar1969,Karolyhazy1974,Diosi1986,Coleman1988,Diosi1989,Gisin1989,Penrose1989,Ghirardi1990b,Penrose1992,Percival1995,Pearle1996,Penrose1996,Frenkel2002,Hu2003,Giulini2011,Adler2014,Hu2014,Penrose2014,Sharma2014,Bera2015,Singh2015,Donadi2020}.
Here we briefly review some of these ideas.

\subsection{Karolyhazy's and Di\'osi's ideas}
In 1966 Karolyhazy suggested that the absence of a clear quantum behavior in macroscopic bodies could be (at least partially) caused by the fluctuations in spacetime itself~\cite{Karolyhazy1966}. Our interpretation of what he was saying is the following (at the risk of going further than he intended).

We start from the viewpoint that spacetime is not a fundamental structure, but a more phenomenological structure that serves to describe macroscopic behaviors of rulers and clocks. Then, it is reasonable to think that the actual quantum nature of microscopic rulers and clocks would transfer some of its fuzziness into the very definition of spacetime. But then, when a quantum system moves under the influence of a fuzzy spacetime, its effect is to prevent the wave function of macroscopic systems to behave too much quantum mechanically. For instance, it would prevent the wave function of a macroscopic body to widely spread. In turn, the close-to-classical behavior imprinted in the macroscopic matter, would make the spacetime, controlled by this matter, to behave also close to classically. 

Years later, Di\'osi proposed similar ideas~\cite{Diosi1989}. He built on top of another proposal for reduction of macroscopic fluctuations due to Ghirardi, Rimini and Weber (GRW). The GRW model~\cite{Ghirardi1986} is just a modification of standard quantum mechanics in order to account for the reduction of state. In fact, the only difference with standard quantum mechanics appears in the dynamical law. Whereas in the quantum case the evolution is governed by a uniparametric family of unitary operators $U_t$, in this case sudden jumps of the wavefunction that tend to spatially localize the system are introduced. The sudden jumps introduce spontaneous localization processes at scales $L$, and they are modeled by a Poissonian distribution with a certain frequency $\lambda$. These two parameters, $L$ and $\lambda$, are the only free parameters of the GRW model. 

For concreteness, let us consider the case of just one nonrelativistic particle. The master equation for its density matrix is given by 
\begin{equation}
    \frac{\dd}{\dd t} \rho(t) = - i [H,\rho(t)] - T[\rho(t)],
\end{equation}
$\rho$ representing as usual the density matrix of the system, $H$ its Hamiltonian (the first part accounts for the unitary evolution, that is, standard quantum mechanics), and the last term models the spontaneous localization process. The operator $T$, in the position basis, can be written as 
\begin{equation}
    \matrixel{x}{T[\rho(t)]}{y}= \lambda \left( 1 - \exp \left[ - (x - y)^2/4 L^2 \right] \right)  \matrixel{x}{\rho(t)}{y}.
    \label{GRWdensmat}
\end{equation}
Thus, we see that the effect of the new term is to suppress off-diagonal elements of the density matrix in the position basis, i.e., delocalized states. An operator $T_i$ would appear for each particle in the case we are considering many-particle systems. As showed in~\cite{Ghirardi1986}, for typical Hamiltonians the center-of-mass degrees of freedom decouple from the internal ones. The result of such a decoupling in this limit is that the center-of-mass follows an equation of motion of the type (\ref{GRWdensmat}) while the internal degrees of freedom follow an ordinary Schr\"{o}dinger equation of motion. Furthermore, the parameter $\lambda$ in (\ref{GRWdensmat}) is replaced with $N \lambda$, being $N$ the number of particles of the system. Thus, the model is able to account qualitatively for the absence of macroscopic superpositions while keeping the microscopic dynamics approximately untouched~\cite{Bassi2013}. Actually, it means that the more \emph{complex} and \emph{macroscopic} a system is, the faster its quantum correlations tend to decay. In this context, the measure of how complex and macroscopic a system is can be associated with the number of particles that compose it. The experimental bounds that have been put to the parameters of the model until now can be found in~\cite{Bassi2016,Feldmann2012} 

The GRW proposal constitutes a representative of some models that attempt to provide a unified framework for the observed microscopic quantum dynamics, and the absence of macroscopic superpositions. They typically introduce stochastic differential equations to account for both phenomena. In particular, Di\'osi~\cite{Diosi1989} tried to assign a gravitational origin to the localization processes. He introduced the QMUDL (Quantum Mechanics with Universal Density Localization) model, where instead of the wave-function (as in the GRW model) it is the probability density (and thus the matter density) that undergoes localization, following Karolyhazy's ideas~\cite{Karolyhazy1966}. Furthermore, the parameters characterizing the length scale and frequency of these localization processes are dimensionally fixed in terms of the Newton constant (realizing, again, the ideas of Karolyhazy). In fact, they just leave one pure number as a free parameter of the theory. 

Di\'osi's model was able to phenomenologically account for the possible role that gravity might play in the quantum state reduction. His main achievement was to take Karolyhazy's vague ideas, concerning gravity and quantum mechanics, and capture them in a simple model which differs from standard quantum mechanics. Penrose's ideas, described in Chapter~\ref{Ch2:AnalogueSuperp}, involve different arguments explaining why and how gravity might be a source of decoherence, but with similar phenomenological consequences. In a sense, Di\'osi's arguments are much more phenomenological in nature. 

A step further in the formulation of these ideas was taken in the past years. New models have been proposed where gravity not only acts as a source of decoherence for the quantum system, but also the backreaction of the quantum system on the gravitational field is accounted in the Newtonian regime. These models rely on the result of~\cite{Diosi1998}, where it was shown that it is possible to give a phenomenological description for the interaction of a classical variable with a quantum degree of freedom modeling it as a continuous measurement, at the expense of necessarily introducing fluctuations on the classical variables. Two proposals which are built on top of these principles are the Kafri-Taylor-Milburn~\cite{Kafri2014} model and the Tilloy-Di\'osi model~\cite{Tilloy2017}. The former model applies only to two particles moving in one dimension where the Newtonian potential can be linearized, while the latter is general within the regime of applicability of Newtonian gravity. A recent analysis~\cite{Gaona2021}, where some natural extensions of the Kafri-Taylor-Milburn model are introduced to include an arbitrary number of particles, shows that all these generalizations are straightforwardly ruled out experimentally or internally inconsistent. They conclude that just the Tilloy-Di\'osi model is a viable description of a quantum mechanical system interacting with a classical gravitational field through this continuous measurement process. Furthermore, at the phenomenological level, it reduces to the original proposal of Di\'osi and the one of Penrose for the gravitationally induced reduction of state~\cite{Tilloy2017}.

\subsection{Quantum mechanics with stochastic time}

We saw that Penrose arguments for the gravity induced reduction of state in quantum mechanics rely on the ill-posedness of the concept of time translation on this superposition of spacetimes. This is the ultimate reason for the suggested instability of the superposition of macroscopic distributions of matter, and the existence of a characteristic decay time for them. Thus, we can assume that, forgetting about the gravitational fields, i.e., tracing over them in some sense, we would end up in a quantum mechanics formulation in terms of some kind of stochastic time that fluctuates. In fact, such a formulation of quantum mechanics with a stochastic time was presented in~\cite{Garay1999,Egusquiza1999} and we follow it here.

If we make superpositions of slightly different distributions of matter, we expect the time variable to have small fluctuations. As a consequence, the laws of quantum mechanics will be slightly affected. In other words, we expect that some of the fuzziness of matter is transferred to the spacetime itself and, from a quantum mechanical perspective, to our choice of time to perform evolution. In order to model these features, we will define the ideal time $s$ in terms of which the Schr\"{o}dinger equation is written, as $s=t + \delta(t)$, being $t$ a generic time (it could be, for example, the proper time of some observer or the coordinate time of some of the spacetimes we are superposing) and $\delta(t)$ a stochastic variable that represents small fluctuations due to the uncertainty in our choice of time. This uncertainty models our inability to choose one privileged time or, in other words, the impossibility to identify pointwise both spacetimes.

With this, the Schr\"{o}dinger equation for our generic density matrix can be written as follows 
\begin{equation}
    i \partial_s \rho(s)= [H,\rho(s)],
\end{equation}
which can be rewritten in terms of the variable $t$ as
\begin{equation}
    i \partial_t \rho (t +  \delta(t)) = [H, \rho(t +  \delta (t))] (1 + \alpha(t)),
\end{equation}
being $\alpha(t) = \frac{\dd \delta(t)}{\dd t}$. This quantity can be interpreted as the error due to our choice of time and we assume that there exists some probability distribution functional $\mathcal{P}[\alpha(t)]$ for it. Moreover, we will assume it to be stationary since we will be concerned with superpositions of stationary spacetimes. In addition, we will assume that there is no systematic rate of increase (or decrease) of the fluctuations for the situations we are considering and thus $\expval{\alpha(t)} = 0$\footnote{Notice that this error could be straightforwardly corrected.}.

Moving to the interaction picture ($\rho \longrightarrow \rho^I = e^{i H t} \rho e^{-i H t}$) and integrating in $t$, we can reach the following integro-differential equation for the density matrix 
\begin{equation}
    i \partial_t \rho^I (t +  \delta(t) ) = \alpha(t) [H,\rho(t)] - i \int ^t_0 \dd t' \alpha(t') \alpha(t) [H,[H,\rho^I(t+ \delta(t)]].
\end{equation}
Assuming that time fluctuations are small enough, we can perform a power expansion of $\rho$ and neglect terms of order greater than $\alpha^3$. Moreover, the density matrix we have access to is the averaged density matrix $\expval{\rho(t +  \delta(t))}= \rho_{\textrm{eff}}(t)$ over the space of possible fluctuations. Averaging the expression above, using the fact that $\mathcal{P}[\alpha(t)]$ is stationary, and taking into account the considerations above, we obtain the following expression in the interaction picture
\begin{equation}
    i \partial_t \rho_{\textrm{eff}}^I(t) = - i \int ^t_0 \dd \tau c(\tau) [H,[H,\rho_{\textrm{eff}}^I(t-\tau)]] + \order{\expval{\alpha(t)^3}},
\end{equation}
where we have introduced the time correlation function for the fluctuations $c(t-t') = \expval{\alpha(t) \alpha(t')}$. The assumption of small correlations in the fluctuations allows us to neglect the $\order{\expval{\alpha(t)^3}}$ terms, since huge ones would lead to highly nonlocal effects that could be observed in principle. 

We will assume that the characteristic time of correlations of the fluctuations are much smaller than the characteristic time of evolution of the system. Then, we expect the density matrix to not evolve significantly within a correlation time and thus we can replace $\rho_{\textrm{eff}}^I(t-\tau)$ with $\rho_{\textrm{eff}}^I(t)$ and take the limit of integration to infinity. This approximation, often called the Markov approximation, allows to obtain a coarse grained density matrix~\cite{Breuer2002}, i.e., a density matrix which does not capture fluctuations at time-scales shorter than the characteristic correlation time of the stochastic variable. This approximation can be done in two steps. The first one is replacing $\rho_{\textrm{eff}}^I(t-\tau)$ with $\rho_{\textrm{eff}}^I(t)$, which means assuming the evolution of the density matrix is much slower than the typical correlation time of the fluctuations, something ensured by choosing a weak enough coupling~\cite{Rivas2012}. The resulting equation of this first step is often called the Redfield equation~\cite{Redfield1957}. The second step in this approximation is to replace the upper limit of integration by $\infty$, something that can be done as long as the time correlation function $c(\tau)$ decays fast enough at large $\tau$. 

Since we are concerned with stationary superpositions of times, it seems plausible to assume that memory effects in our system will have little impact. Moving back to the Schr\"{o}dinger picture and putting altogether these approximations, the density matrix dynamics is encoded in the following master equation 

\begin{equation}
    i \partial_t \rho_{\textrm{eff}} = [H,\rho_{\textrm{eff}}] - i \xi [H,[H,\rho_{\textrm{eff}}]].
\end{equation}
$\xi$ is a parameter with units of time resulting from the integration of the correlation function $c(t)$ and it controls the characteristic timescale of decay for the correlations. Clearly, the first term is the usual unitary part while the second term, which is diffusive but dissipationless, is responsible for the reduction of state, as an effective residue of the stochastic time we need to use to describe our system. This means that the pointer basis, i.e., the basis in which the correlations of the density matrix decay exponentially is the energy basis, in agreement with Penrose's heuristics. Explicitly, assuming that the spectrum of $H$ is discrete, we can write the evolution of the density matrix in the basis of eigenstates of $H$. We label those states as $\ket{n,g_n}$, with $g_n$ accounting for the possible degeneracies of the spectrum and being $H \ket{n,g_n} = E_n \ket{n,g_n}$ their defining property.
\begin{equation}
\matrixel{m,g_m}{\rho_{\textrm{eff}} (t)}{n,g_n} = \matrixel{m,g_m}{\rho_{\textrm{eff}} (0)}{n,g_n} e^{-i (E_n-E_m) t} e^{-(E_n-E_m)^2 \xi t}.
\end{equation}
Thus, the pointer basis, i.e., the basis in which the density matrix becomes asymptotically diagonal, coincides with the set of energy eigenvectors. The greater the energy difference between two eigenstates, the faster they undergo decoherence, with a characteristic decay timescale $\tau_{nm} = \xi^{-1} (E_n-E_m)^{-2}$.


\chapter{Bootstrapping gravitational theories}
\label{AppB}

\fancyhead[LE,RO]{\thepage}
\fancyhead[LO,RE]{Bootstrapping gravitational theories}


Consider a generic action functional $S[\mathrm{Q}^I]$ depending on a certain family of spacetime fields that we denote collectively as $\{\mathrm{Q}^I (x) \}$. These fields will be the metric (or vielbein), the contorsion and disformation tensors and the matter fields. The indices $(I,J...)$ represent a placeholder for all the fields living in the manifold and their internal, spacetime and frame indices. Moreover, let $\{\bar{\mathrm{Q}}^I\}$ be a generic background configuration. Now we evaluate the action in a configuration that deviates from such a background, $\mathrm{Q}^I = \bar{\mathrm{Q}}^I + \lambda \mathrm{q}^I$, where $\lambda$ is a dimensionless bookkeeping expansion parameter. The total action can then be rearranged as a series of the form:
\begin{align}
    S[\mathrm{Q}] = \sum_{n = 0}^{\infty} \lambda^n S^{(n)} [\bar{\mathrm{Q}},  \mathrm{q}],
\end{align}
where the partial actions $S^{(n)}$ are given by
\begin{align}
    S^{(n)} [\bar{\mathrm{Q}},  \mathrm{q}] = \frac{1}{n!} \frac{\dd^n}{\dd \lambda^n} S [\bar{\mathrm{Q}} + \lambda  \mathrm{q}] \bigg\rvert_{\lambda  = 0}.
    \label{Eq:Partial_Actions_Gen}
\end{align}

\section{Generating formula}
\label{App:General_Fields_genformula}
The derivatives with respect to $\lambda$ in \eqref{Eq:Partial_Actions_Gen} can be replaced by the standard functional derivative operator with respect to the background fields
\begin{align}
    \frac{\dd}{\dd \lambda} S[\bar{\mathrm{Q}} + \lambda  \mathrm{q}] = \int \dd^{D+1} x \  \mathrm{q}^I (x) \frac{\delta}{\delta \bar{\mathrm{Q}}^{I} (x) } S[\bar{\mathrm{Q}} + \lambda  \mathrm{q}].
\end{align}
Here, it is important to highlight again that there is a sum in $I$ that covers all the fields of the theory and all their indices. Furthermore, we have dropped surface integrals that arise when integrating by parts. These terms are neglected since we are assuming the perturbations $ \mathrm{q}^I$ of all the fields to have compact support.\footnote{
    This condition can be relaxed to rapidly enough decaying fields at infinity, instead of compact support.}
Furthermore, the $n$-th derivative can be expressed as:
\begin{align}
    \frac{\dd^n}{\dd \lambda^n} S[\bar{\mathrm{Q}} + \lambda  \mathrm{q}] = \left[ \int \dd^{D+1} x\  \mathrm{q}^I(x) \frac{\delta}{\delta \bar{\mathrm{Q}}^{I} (x) }\right]^n S[\bar{\mathrm{Q}} + \lambda  \mathrm{q}]
    \label{Eq:Partial_Derivatives_Gen}
\end{align}
where we are using the simplified notation \eqref{Eq:Shorcut_Notation} for the operator in the right hand side. Actually, it is possible to express all the terms $S^{(n)}$ for $n>2$ in terms of derivatives of $S^{(2)}[\bar{\mathrm{Q}},  \mathrm{q}]$. This is clear from Eqs.~\eqref{Eq:Partial_Actions_Gen} and~\eqref{Eq:Partial_Derivatives_Gen}, since we can combine them to find out that
\begin{align}
    S^{(n)} [\bar{\mathrm{Q}},  \mathrm{q}] &= \frac{1}{n!} \frac{\dd^n}{\dd \lambda^n} S [\bar{\mathrm{Q}} + \lambda  \mathrm{q}] \bigg\rvert_{\lambda  = 0}\nonumber\\
    &= \frac{1}{n!} \left[ \int \dd^{D+1} x\  \mathrm{q}^I(x) \frac{\delta}{\delta \bar{\mathrm{Q}}^{I} (x) }\right]^n S[\bar{\mathrm{Q}} + \lambda  \mathrm{q}] \bigg\rvert_{\lambda  = 0}\nonumber\\
   &= \frac{1}{n!} \left[ \int \dd^{D+1} x\  \mathrm{q}^I(x) \frac{\delta}{\delta \bar{\mathrm{Q}}^{I} (x) }\right]^{n-2} \left(\left[ \int \dd^{D+1} x\  \mathrm{q}^I(x) \frac{\delta}{\delta \bar{\mathrm{Q}}^{I} (x) }\right]^2 S[\bar{\mathrm{Q}} + \lambda  \mathrm{q}] \bigg\rvert_{\lambda  = 0}\right)\nonumber\\
    &= \frac{2}{n!} \left[ \int \dd^{D+1} x\  \mathrm{q}^I(x) \frac{\delta}{\delta \bar{\mathrm{Q}}^{I} (x) }\right]^{n-2} S^{(2)} [\bar{\mathrm{Q}},  \mathrm{q}]\, .
\label{Eq:Quadratic_Actions}
\end{align}
We have used that the background $\bar{\mathrm{Q}}$ is generic, namely no information is lost when going from $S[\mathrm{Q}]$ to $S[\bar{\mathrm{Q}}]$. Having said this, notice that, once we move all the dynamics from $\mathrm{Q}^I$ to $ \mathrm{q}^I$, we have that $S^{(2)}$ is all we need to build the whole action for $ \mathrm{q}^I$: $S^{(0)}$ is independent of $ \mathrm{q}^{I}$, and hence, it is irrelevant for the classical dynamics, and $S^{(1)}$ vanishes for a background $\bar{\mathrm{Q}}^{I}$ that is solution of the theory, i.e., a configuration such that
\begin{equation}
    \frac{\delta S[\mathrm{Q}]}{\delta \mathrm{Q}^I} \bigg\rvert_{\mathrm{Q}^I=\bar{\mathrm{Q}}^I} = 0\,.
\end{equation}
Notice that it was crucial for this that the two fields $\bar{\mathrm{Q}},\mathrm{q}$ correspond to the splitting of a single field $\mathrm{Q} = \bar{\mathrm{Q}} + \lambda \mathrm{q} $. 

\section{A functional identity}
\label{App:General_Fields_funidentity}

Now we proceed to show that each order in the expansion $S^{(n)}$ corresponds to the coupling of the $ \mathrm{q}^I$ field to the variation of $S^{(n-1)}$ with respect to the background of that field $\bar{\mathrm{Q}}^{I}$. Indeed, the following identity holds
\begin{align}
    \frac{\delta S^{(n)}   [\bar{\mathrm{Q}}, \mathrm{q}]}{\delta  \mathrm{q}^{I}} = \frac{\delta S^{(n-1)} [\bar{\mathrm{Q}}, \mathrm{q}]}{\delta \bar{\mathrm{Q}}^{I}}.
    \label{Eq:Key_Bootstrap_Gen}
\end{align}

\paragraph*{\textbf{Proof 1.}}
To show this identity we will follow~\cite{Butcher2009}. Let us first consider an arbitrary variation with respect to $\mathrm{q}^I$ or $\bar{\mathrm{Q}}^I$ and let us call it $X^I$. We have the following identity
\begin{align}
    \int \dd^{D+1} x\ X^{I} (x) \frac{\delta S^{(n)}[\bar{\mathrm{Q}},\mathrm{q}]}{\delta \mathrm{q}^I (x)} = \int \dd^{D+1} x\ X^{I} (x) \frac{1}{n!} \frac{\delta }{\delta \mathrm{q}^I (x) } \left(\frac{\dd^n}{\dd \lambda^n} S[\bar{\mathrm{Q}} + \lambda \mathrm{q}] \bigg\rvert_{\lambda  = 0}\right).
\end{align}
We can rewrite the last term as a derivative with respect to a new parameter $\sigma$.
\begin{align}
    \frac{1}{n!} \frac{\partial}{\partial \sigma } \frac{\partial^n}{\partial \lambda^n} S[\bar{\mathrm{Q}} + \lambda (\mathrm{q} + \sigma X) ] \bigg\rvert_{\lambda  = 0, \sigma = 0},
\end{align}
where clearly the derivatives are taken first and the substitution $\lambda = \sigma = 0$ is made at the end. Interchanging the order of derivatives
\begin{align}
    \frac{1}{n!} \frac{\partial^n}{\partial \lambda^n} \frac{\partial}{\partial \sigma } S[\bar{\mathrm{Q}} + \lambda (\mathrm{q} + \sigma X) ] \bigg\rvert_{\lambda  = 0, \sigma = 0}.
\end{align}
We can introduce a new variable $\tau = \lambda \sigma $, so that $\frac{1}{\lambda } \frac{\partial}{\partial \sigma} = \frac{\partial}{\partial \tau}$ at fixed $\lambda$. Hence, the expression above can be rewritten as
\begin{align}
    \frac{1}{n!} \frac{\partial^n}{\partial \lambda^n} \left(  \lambda \frac{\partial}{\partial \tau } S[\bar{\mathrm{Q}} + \lambda \mathrm{q} + \tau X ]\right) \bigg\rvert_{\lambda  = 0, \tau = 0} .
\end{align}
Evaluating the derivative with respect to $\lambda$ we find
\begin{align}
    & \frac{1}{n!} \left(\lambda \frac{\partial^n}{\partial \lambda^n}   \frac{\partial}{\partial \tau } S[\bar{\mathrm{Q}} + \lambda \mathrm{q} + \tau X]  + n \frac{\partial^{n-1}}{\partial \lambda^{n-1}} \frac{\partial}{\partial \tau} S [\bar{\mathrm{Q}} + \lambda \mathrm{q} + \tau X]\right) \bigg\rvert_{\lambda  = 0, \tau = 0} \nonumber \\
    &\qquad\qquad\qquad = \frac{1}{(n-1)!} \frac{\partial^{n-1}}{\partial \lambda^{n-1}} \frac{\partial}{\partial \tau} S [\bar{\mathrm{Q}} + \lambda \mathrm{q} + \tau X] \bigg\rvert_{\lambda  = 0, \tau = 0}.
\end{align}
Using the definition of $S_n[\bar{\mathrm{Q}},\mathrm{q}]$ introduced in Eq.~\eqref{Eq:Partial_Actions_Gen}, we find
\begin{align}
    \frac{\partial}{\partial \tau} S^{(n-1)} [\bar{\mathrm{Q}} + \tau X, \mathrm{q}] \bigg\rvert_{\tau = 0} = \int \dd^{D+1} x\ X^I(x) \frac{\delta}{\delta \bar{\mathrm{Q}}^{I} (x)} S^{(n-1)} [\bar{\mathrm{Q}}, \mathrm{q}],
\end{align}
concluding our proof of Eq.~\eqref{Eq:Key_Bootstrap_Gen}. We recall that in all these steps we have ignored boundary terms as mentioned above.

\paragraph*{\textbf{Proof 2.}}
A perhaps more transparent derivation of this result can be given as follows. Let us begin with the definition of $S^{(n)}[\bar{\mathrm{Q}},\mathrm{q}]$
\begin{align}
    S^{(n)} [\bar{\mathrm{Q}},\mathrm{q}] = \frac{1}{n!} \left[ \int \dd^{D+1} x\ \mathrm{q}^{I} (x) \frac{\delta}{\delta \bar{\mathrm{Q}}^I (x)} \right]^n S[\bar{\mathrm{Q}} + \lambda \mathrm{q}] \bigg\rvert_{\lambda  = 0} .
\end{align}
Now, let us compute the variation with respect to one of the fields $\mathrm{q}^J(y)$. For that purpose, we recall that the $n$ variations with respect to the background field can be rearranged with the help of the following notation: 
\begin{align}
   \left[ \int \dd^{D+1} x\  \mathrm{q}^I(x) \frac{\delta}{\delta \bar{\mathrm{Q}}^{I} (x) }\right]^n := \int \dd^{D+1} x_1\  \mathrm{q}^{I_1}(x_1) \frac{\delta}{\delta \bar{\mathrm{Q}}^{I_{1}}(x_1) } \ldots \int \dd^{D+1} x_n\  \mathrm{q}^{I_n}(x_n) \frac{\delta}{\delta \bar{\mathrm{Q}}^{I_{n}}(x_n) }\,.
   \label{Eq:Shorcut_Notation}
\end{align}
Hence if we perform a variation with respect to $\mathrm{q}^J (x)$, it simply hits the product of the $\mathrm{q}^I(x)$ terms. This gives
\begin{align}
   & \frac{\delta}{\delta \mathrm{q}^{J}(y)} \int \dd^{D+1} x_1 \ldots \int \dd^{D+1} x_n \  \mathrm{q}^{I_1}(x_1) \ldots  \mathrm{q}^{I_n}(x_n) \frac{\delta}{\delta \bar{\mathrm{Q}}^{I_1}(x_1) } \ldots \frac{\delta}{\delta \bar{\mathrm{Q}}^{I_{n}}(x_n) } S [\bar{\mathrm{Q}} + \lambda \mathrm{q}]  \bigg\rvert_{\lambda  = 0}  \nonumber \\
   & \qquad = n \int \dd^{D+1} z\ \delta^{{D+1} } (y - z) \frac{\delta}{\delta \bar{\mathrm{Q}}^J (z)} \int \dd^{D} x_1 \ldots \int \dd^{D} x_{n-1} \  \mathrm{q}^{I_1}(x_1) \ldots  \mathrm{q}^{I_{n-1}}(x_{n-1}) \nonumber \\
   & \qquad \qquad \frac{\delta}{\delta \bar{\mathrm{Q}}^{I_1}(x_1) } \ldots \frac{\delta}{\delta \bar{\mathrm{Q}}^{I_{n-1}}(x_{n-1}) } S [\bar{\mathrm{Q}} + \lambda \mathrm{q}]  \bigg\rvert_{\lambda  = 0} .
\end{align}
The last part is the partial action $S^{(n-1)} [\bar{\mathrm{Q}},\mathrm{q}]$, up to a multiplicative factorial $(n-1)!$. This, together with the prefactor $n$ gives $n (n-1)! = n!$, canceling the $n!$ above. Also, the $\delta$-distribution can be used to rearrange everything simply as a derivative with respect to $\bar{\mathrm{Q}}^J (y)$.
\begin{align}
    & \int \dd^{D+1} z\ \delta (y - z) \frac{\delta}{\delta \bar{\mathrm{Q}}^J (z)} \left[ \int \dd^{D+1} x\ \mathrm{q}^{I} (x) \frac{\delta}{\delta \bar{\mathrm{Q}}^I (x)} \right]^{n-1} S[\bar{\mathrm{Q}} + \lambda \mathrm{q}] \bigg\rvert_{\lambda  = 0} \nonumber\\
    & \qquad = \int \dd^{D+1} z \delta(y - z) \frac{\delta}{\delta \bar{\mathrm{Q}}^J (z)} S^{(n-1)} [\bar{\mathrm{Q}},\mathrm{q}] \nonumber\\
    & \qquad = \frac{\delta}{\delta \bar{\mathrm{Q}}^J (y)} S^{(n-1)} [\bar{\mathrm{Q}},\mathrm{q}].
\end{align}
Hence, we find again expression~\eqref{Eq:Key_Bootstrap_Gen}.


\chapter{Unimodular gravity and background structures}
\label{AppC}

\fancyhead[LE,RO]{\thepage}
\fancyhead[LO,RE]{Unimodular gravity and background structures}


This appendix supplements Chapter~\ref{Ch4:UG} with additional analyses, background context, and technical details.

\section{Coordinate changes and active diffeomorphisms}
\label{App:Diffs}

We have found convenient to clarify the difference between passive and active diffeomorphisms. Although in GR this distinction is not needed because the theory is background independent, in background dependent theories the distinction is important. 

\subsection{Passive and active diffeomorphisms}
\label{App:PassivevsActive}
On the one hand, \emph{passive diffeomorphisms} correspond to general coordinate transformations. These transformations only relabel the points of the manifold with new coordinates. For a given tensor field, a coordinate transformation affects both to its components and its functional dependency on the coordinates. To be precise, a coordinate transformation $x^\mu\to 
 x'{}^\mu=f^\mu(x)$ modifies a tensor field $T^{\mu...}{}_{\nu...}$ according to the standard law
 \begin{equation}
     T^{\mu...}{}_{\nu...} (x) =  (J^{-1})^\mu{}_\rho\big|_{f(x)}...\, J^\sigma{}_\nu\big|_x...\, T'{}^{\rho...}{}_{\sigma...}(f(x)),\qquad J^\sigma{}_\mu\big|_x := \frac{\partial f^\sigma}{\partial x^\mu}(x)\,.
 \end{equation}
At the level of the action it is also important to keep in mind the transformation law of the product of  $\dd^{D+1} x$ and a volume scalar density. They transform in opposite ways so their product is invariant:
\begin{equation}
    \mathfrak{s}(x) \dd^{D+1} x = \Big(\mathfrak{s}'(f(x)) \abs{J(x)}\Big)\Big( \dd^{D+1} x' \abs{J(x)}^{-1} \Big) = \mathfrak{s}'(f(x)) \dd^{D+1} x' \,,\label{eq:GCTsddx}
\end{equation}
where $J := \det(J^\sigma{}_\mu)$.

On the other hand, \emph{active diffeomorphisms} truly act on the points of the manifold. They map points of the manifold to points of the manifold. Thus, the tensor fields must transform so that they take the same value at the same points. In practice one does the substitution:\footnote{
    Infinitesimally, the effect of an active diffeomorphism is described by the Lie derivative $\delta_{\xi} T^{\mu...}{}_{\nu...} = \left( \mathcal{L}_{\xi} T \right)^{\mu...}{}_{\nu...}$. In particular, for the metric field, we have the following infinitesimal transformation $
    \delta_{\xi} g_{\mu \nu} = \left( \mathcal{L}_{\xi} g  \right)_{\mu \nu} = \nabla_{\mu} \xi_{\nu} + \nabla_{\nu} \xi_{\mu}.$
}
\begin{equation}
    T^{\mu...}{}_{\nu...} (x) \quad\xrightarrow{\mathrm{Diff}} \quad  (J^{-1})^\mu{}_\sigma\big|_{f(x)}... \,J^\rho{}_\nu\big|_x...\,T^{\sigma...}{}_{\rho...}(f(x))\,, \label{eq:activediff}
\end{equation}
where now $x^\mu \to x'{}^\mu=f^\mu(x)$. The functions $f^\mu$ should be seeing as the functions that give the new coordinates of the considered point after the transformation. Of course, an additional factor with a power of the determinant $\abs{J}$ must be added to \eqref{eq:activediff} in the case of a tensor density. By default, the functional expressions of all the fields are assumed to be affected by these transformations. 

In general, any theory written in a covariant way, i.e., in abstract tensor notation, is invariant under both passive and active diffeomorphisms that hit all the of the tensorial quantities \emph{simultaneously}. The situation changes dramatically when we consider \emph{background structures}. We say that a tensor density $\Omega^{\mu...}{}_{\nu...}$ is a background structure if it is not affected by active diffeomorphisms. More precisely, such a field transforms (ignoring indices and density weights) as
\begin{equation}
    \Omega^{\mu...}{}_{\nu...} (x) \quad\xrightarrow{\mathrm{Diff}} \quad  \,\Omega^{\mu...}{}_{\nu...}(f(x))\,.\label{eq:bgstructransf}
\end{equation}
When there are background structures, the theory distinguishes between passive diffeomorphisms (which always hit all the tensor fields in the theory) and active diffeomorphisms. However, the theory may not be invariant under passive diffeomorphisms from the beginning since the equations might be written in a non-explicitly tensorial way.

\subsection{Enforcing passive diffeomorphisms}
\label{App:Rep_Inv}

It is always possible to make a theory invariant under passive diffeomorphisms or, if it is an action involving no spatial dependence, time-reparametrization invariant. For many purposes, it is not useful since it usually involves adding explicitly background structures to the equations, instead of using such structures at our disposal to simplify them. However, there are two purposes for which it could be useful. First of all, it helps to make background structures in our theory explicit. Writing an action in a coordinate-invariant way requires to express all background structures in arbitrary coordinates. Second, it introduces an additional, albeit spurious, gauge symmetry that can be advantageous. In some calculations, choosing a gauge where the background structure remains manifest can significantly simplify the analysis.

Let us illustrate this with an example. Consider the nonrelativistic particle that is given by the standard action
\begin{align}
   S [x] = \frac{1}{2} \int \dd t \left[\left( \frac{\dd x}{\dd t}\right)^2-V(x)\right].
   \label{Eq:free_particle}
\end{align}
This action is not invariant under time-reparametrizations, as it can be straightforwardly checked. Under a time reparametrization $t \rightarrow t(\tau) $, we have that the action changes as 
\begin{align}
   S [x]  = \int \dd \tau \frac{\dd t}{\dd \tau} \left[\frac{1}{2} \frac{1}{ ( \dd t / \dd \tau ) ^2}  \left( \frac{\dd x}{\dd \tau}\right)^2  -V(x)\right].
   \label{Eq:Change_Coords_Newton}
\end{align}
However, we can make it explicitly time-reparametrization invariant by introducing a fixed (nondynamical) one-form $\boldsymbol\omega=\omega(\tau) \dd \tau$ as 
\begin{align}
    S [x; \boldsymbol\omega] =  \int \dd \tau \omega(\tau)\left[ \frac{1}{2}\frac{ 1}{\omega(\tau)^2}\left(\frac{\dd x}{\dd \tau}\right)^2-V(x) \right]. 
\end{align} 
This action is manifestly invariant under time-reparametrizations since the function $\omega(\tau)$ is introduced in such a way that its transformation law compensates the combination of the transformation law of the integration variable and the function $x(\tau)$. Let us see this explicitly. In Eq.~\eqref{Eq:Change_Coords_Newton}, if we replace the integration element $ \dd t$ by the one-form $\omega(\tau) \dd \tau$, we automatically make the second term time reparametrization invariant. In the case of the first term, we need to add a factor $\omega^{-2}$ to compensate the change coming from the $\dd x /\dd \tau$ piece of the action. In this way, we see that the action with the $\omega$ function is now time-reparametrization invariant. If we choose a time parameter such that $\omega(\tau) = 1$, we find the action~\eqref{Eq:free_particle} we began with. The original action was written down taking the Newtonian absolute time as the time parameter and, hence, the function $\omega(\tau)$ describes such absolute time function in an arbitrary parametrization.

As we advanced, this formulation of the nonrelativistic particle makes the background structure (in this case an absolute time function) explicit, at the expense of introducing additional gauge symmetries in our description. Although for the nonrelativistic particle it is far from clear how this reformulation could be useful, in UG it is much easier to work with the background structures manifest. In particular, to understand the structure of the theory.

\section{Alternative formulations of Unimodular Gravity}
\label{App:UG_Formulations}

In this Appendix, we collect common alternative formulations of UG that are used in the literature and discuss the advantages of the approach that we have taken in the main text, where we work with an unconstrained metric. In Subsection~\ref{App:UG_constrained} we discuss the formulation of UG as a constrained system. In Subsection~\ref{App:HT} we discuss Henneaux and Teitelboim formulation of UG with an additional vector field.

\subsection{Unimodular gravity: constrained Einstein-Hilbert action}
\label{App:UG_constrained}
UG is characterized by having a fixed volume form $\boldsymbol{\omega}$ which will be used to fix the determinant of the metric. This means that we want the volume form defined via the metric $\boldsymbol{g}$ to be fixed as
\begin{equation}
    \sqrt{|g|} = \omega,
    \label{fixedvolumeform}
\end{equation}
where $\omega$ is the tensor density associated with the volume form $\boldsymbol{\omega}$. UG is typically claimed to be GR with the additional constraint $\sqrt{|g|} = 1$. However, this is obviously an identity that can hold just in a particular coordinate system or a local chart: the left-hand side is a tensor density while the right-hand side is a constant. 

Dynamically it is possible to implement this constraint adding a Lagrange multiplier $\lambda$ to the action as
\begin{equation}
S [\boldsymbol{g}] = \frac{1}{2 \kappa_{D}^2} \int \dd^{D+1} x \sqrt{|g|} R(\boldsymbol{g}) + \int \dd^{D+1} x \lambda \left( \sqrt{|g|} - \omega \right) + S_{\textrm{matter}}(\boldsymbol{g}, \Phi).
\end{equation}
Having fixed the determinant of the metric ensures that we take traceless variations of the metric since $g_{\mu \nu} \delta g^{\mu \nu} = 0$. This means that we end up with the traceless part of Einstein equations
\begin{equation}
R_{\mu \nu} (\boldsymbol{g}) - \frac{1}{D+1}  g_{\mu \nu} R(\boldsymbol{g}) = \kappa_{D}^2 \left( T_{\mu \nu} (\boldsymbol{g}) - \frac{1}{D+1}  g_{\mu \nu} T(\boldsymbol{g}) \right). 
\label{tracelesseinstein}
\end{equation}
Using the same arguments used in Section~\ref{Subsec:Einstein_Hilbert}, we can end up finding the equivalence of this formulation with the unconstrained formulation introduced in the text. 

\subsection{Unimodular gravity \`a la Henneaux-Teitelboim}
\label{App:HT}

There is another way of formulating UG introducing an additional vector density $V^{\mu}$, as originally done by Henneaux and Teitelboim~\cite{Henneaux1989}. We take the following action with $\lambda$ playing again the role of a Lagrange multiplier 
\begin{equation}
    S [\boldsymbol{g},\boldsymbol{V}]= \frac{1}{2 \kappa_D^2} \int \dd ^{D+1} x \sqrt{|g|} R (\boldsymbol{g}) + \int \dd^{D+1} x \lambda \left( \sqrt{|g|} - \partial_{\mu} V^{\mu} \right) + S_{\textrm{matter}}(\boldsymbol{g}, \Phi).
    \label{henneauxteitelboim}
\end{equation}
The equations of motion are the Einstein equations with the role of the cosmological constant being played by $\lambda$ 
\begin{equation}
R_{\mu \nu} (\boldsymbol{g})- \frac{1}{2} g_{\mu \nu} R(\boldsymbol{g}) - \frac{\lambda}{2} g_{\mu \nu} =  \kappa_D^2 T_{\mu \nu} (\boldsymbol{g}),
\end{equation}
with the constraint that it must be constant
\begin{equation}
\nabla_{\mu} \lambda = 0 \quad\rightarrow\quad \lambda = -2 \Lambda,
\end{equation}
where we have chosen in the last line the constant field to be written in terms of the constant $\Lambda$. It plays the role of the cosmological constant. Finally, we have the following equation obtained by varying with respect to $\lambda$ 
\begin{equation}
\partial_{\mu} V^{\mu} = \sqrt{|g|}.
\end{equation}
In addition to diffeomorphism invariance, there is a new gauge symmetry associated with the transformations that leave $g_{\mu \nu}$ and $\lambda$ invariant while act on $V^{\mu}$ as 
\begin{equation}
V^{\mu} \rightarrow V^{\mu} + \epsilon^{\mu}, \qquad \textrm{with} \quad \partial_{\mu} \epsilon^{\mu} = 0.
\end{equation}
This formulation is closer to the one that we use in the main text, in the sense that the metric itself is unconstrained. However, this is achieved at the expense of introducing additional vector field $V^{\mu}$. Within the components of $V^{\mu}$, just $V^0$ has a physical meaning, the other three are pure gauge needed to write fully covariant equations. Actually, we can see that the role played by the component $V^0$ is, somehow, the role of the fixed volume form introduced in the previous section. To see this, we just need to notice that if we introduce the following function 
\begin{equation}
T(t) = \int \dd^Dx V^0,
\end{equation}
we will have that 
\begin{equation}
T(t_2) - T(t_1) = \int_{t_1}^{t_2} \dd x^0 \int \dd^D x \partial_{\mu} V^{\mu} = \int_{t_1}^{t_2} \dd x^0 \int \dd^D x \sqrt{|g|}.
\end{equation}
In fact, $T(t)$ is a time function conjugate to the cosmological constant~\cite{Henneaux1989} and clearly controls the volume element of the spacetime. In that sense, it is the manifestation in this formalism of having a fixed volume form as a background structure. The description presented here, can also be done in terms of a fully antisymmetric $D$-dimensional form taking the dual of the tensorial objects introduced here~\cite{Henneaux1989}. 

\section{Examples of illegal gauge fixings}
\label{Sec:Illegal_gauge}

We refer to \emph{illegal gauge fixings} as those cases where imposing the gauge condition at the level of the action does not yield the same results as imposing it directly on the equations of motion. Below, we present two illustrative examples. It is important to emphasize that we are not suggesting such gauges cannot be used within the equations of motion themselves.

\subsection{Electrodynamics in the Coulomb gauge with matter content}

Take Maxwell's action coupled to a conserved current $j^{\mu}$
\begin{align}
    S =   \int \dd^{D+1} x \left[ - \frac{1}{4} F^{\mu \nu} F_{\mu \nu} + A_{\mu} j^{\mu} \right].
\end{align}
Its equations of motion can be directly computed and one finds:
\begin{align}
    \Box A^{\mu} - \partial^{\mu} ( \partial_{\nu} A^{\nu} ) = j^{\mu}. 
\end{align}
Now we can always perform a gauge transformation such that $A_0 = 0$. To see this, under a gauge transformation we have: 
\begin{align}
    A_{\mu} \rightarrow A'_{\mu} = A_{\mu} + \partial_{\mu} \alpha,
\end{align}
and we can always choose an $\alpha$ such that $A'_0 = 0$, for example:
\begin{align}
    \alpha(t,\boldsymbol{x}) = - \int_{t_0}^t \dd t' A_0(t', \boldsymbol{x}),
\end{align}
does the job. In this gauge, $A^{\mu}$ is a purely spatial vector that we denote as $\boldsymbol{A}$ and the equations of motion reduce to:
\begin{align}
    & \partial_t \left( \nabla \cdot \boldsymbol{A} \right) = j^0, \label{eq:Gausslaw}\\
    & (- \partial_t^2 + \nabla^2 ) \boldsymbol{A} - \nabla ( \nabla \cdot \boldsymbol{A} ) = \boldsymbol{j}.
\end{align}
The same is not obtained if we fix the gauge at the level of the action though. It is in this sense that the gauge fixing is ``illegal''. To see this, upon substituting $A_0 = 0$ in the action we find
\begin{align}
    S\big|_{A^0=0} = \int \dd^{D+1} x \left[\frac{1}{2} ( \partial_t \boldsymbol{A} )^2 - \frac{1}{2} ( \nabla \times \boldsymbol{A} )^2 + \boldsymbol{j} \cdot \boldsymbol{A} \right].
\end{align}
If we vary this action, we notice that all the information related to charge conservation (the Gauss law \eqref{eq:Gausslaw}) is lost, and we only get the second equation for $\boldsymbol{A}$: 
\begin{align}
    (- \partial_t^2 + \nabla^2 ) \boldsymbol{A} - \nabla ( \nabla \cdot \boldsymbol{A} ) = \boldsymbol{j}. 
\end{align}
Thus, this gauge fixing cannot be performed at the level of the action in order to reduce the number of variables since we lose information about the existence of a conserved charge. It should be possible to still fix this gauge at the level of the action and implement a variational principle by enforcing Gauss law somehow with additional terms, but we will not pursue that approach here. 

\subsection{Stueckelberging Proca and the Lorenz gauge}
Let us consider now a massive spin-1 theory:
\begin{align}
    S_{\text{Proca}} = \int \dd^{D+1} x \left[ - \frac{1}{4} F_{\mu \nu} F^{\mu \nu} - \frac{1}{2} m^2A_{\mu}A^{\mu} \right]. 
\end{align}
We introduce a Stueckelberg field by making the replacement $A_{\mu} \rightarrow A_{\mu} + \partial_{\mu} \varphi$, giving the following action: 
\begin{align}
    S_{\text{Proca-St}} = \int \dd^{D+1} x \left[ - \frac{1}{4} F_{\mu \nu} F^{\mu \nu} - \frac{1}{2} m^2 A_{\mu} A^{\mu} - m^2 A^{\mu} \partial_{\mu} \varphi - \frac{1}{2} m^2 \partial_{\mu} \varphi \partial^{\mu} \varphi \right]. 
\end{align}
This action is invariant under the following gauge transformation that hits the two fields
\begin{align}
    A_\mu &\to A^\prime_\mu=A_\mu +\partial_{\mu}\alpha\, \\
    \varphi &\to \varphi^\prime=\varphi -\alpha\,.
\end{align}
The equations of motion of the theory are:
\begin{align}
    & \Box \varphi + \partial_{\mu} A^{\mu} = 0, \\
    & \partial_{\mu} F^{\mu \nu} - m^2 A^{\nu} - m^2 \partial^{\nu} \varphi = 0. 
\end{align} 
We now can fix the Lorenz gauge $\partial_{\mu} A^{\mu} = 0$, since we can always find a gauge transformation that leads to it. To see this, given a field configuration $A_{\mu}$, upon a gauge transformation we have that the divergence of the transformed field is:
\begin{align}
    \partial_{\mu} A^{ \prime \mu} =  \partial_{\mu} A^{\mu} + \Box \alpha. 
\end{align}
Hence, we can always find an $\alpha$ obeying the following equation
\begin{align}
    \Box \alpha = - \partial_{\mu} A^{\mu}. 
\end{align}
The situation is completely different if we try to implement this gauge fixing at the level of the action. The whole coupling between the $A^{\mu}$ and the Stueckelberg field occurs through a term that vanishes in the Lorenz gauge. To see it explicitly, we can take the action and perform an integration by parts to reach
\begin{align}
     S_{\text{Proca-St}} = \int \dd^{D+1} x \left[ - \frac{1}{4} F_{\mu \nu} F^{\mu \nu} - \frac{1}{2} m^2 A_{\mu} A^{\mu} + m^2  \partial_{\mu} A^{\mu} \varphi - \frac{1}{2} m^2 \partial_{\mu} \varphi \partial^{\mu} \varphi \right]. 
\end{align}
If we fix the Lorenz gauge, we lose the coupling between $A^{\mu}$ and $\varphi$, reaching the following action
\begin{align}
    S_{\text{Proca-St}}\big|_{\partial_{\mu} A^{\mu} =0} = \int \dd^{D+1} x \left[ - \frac{1}{4} F_{\mu \nu} F^{\mu \nu} - \frac{1}{2} m^2 A_{\mu} A^{\mu}  - \frac{1}{2} m^2 \partial_{\mu} \varphi \partial^{\mu} \varphi \right],
\end{align}
and its equations of motion are clearly the decoupled equations for the vector $A^{\mu}$ and the scalar $\varphi$: 
\begin{align}
    & \Box \varphi = 0, \\
    & \partial_{\mu} F^{\mu \nu} - m^2 A^{\nu}  = 0. 
\end{align} 
We conclude that again, this gauge fixing is illegal since it cannot be performed at the level of the action.

\section{Scattering amplitudes}
\label{App:Scattering}

We denote the scattering amplitudes in QFT involving $n$-particles as $A \left( p_1^{h_1}, ...,p_n^{h_n} \right)$, where $\{ p_i,h_i\}_{i = 1, \ldots, n}$ represent the momentum and the helicity of each of the particles. Generic QFT computations based on Feynman diagrams techniques do not focus on directly computing the scattering amplitudes but instead compute the off-shell object $A^{\mu_1 \ldots \mu_n} \left( p_1, ...,p_n \right)$, from which the amplitudes are obtained through the relation
\begin{align}
    A \left( p_1^{h_1}, ...,p_n^{h_n} \right) = \epsilon_{\mu_1}^{h_1} \cdots \epsilon^{h_n}_{\mu_n} A^{\mu_1 \ldots \mu_n} \left( p_1, ...,p_n \right),  
\end{align}
where we have contracted the off-shell amplitudes with the polarization vectors to find the on-shell amplitude with a given helicity. It is convenient to note that the scattering amplitude involving $n$ external particles in  four spacetimes dimensions has the mass dimension $[E]^{4-n}$. We focus on massless particles and hence the momentum variables $p^{\mu} = (p^0,p^1,p^2,p^3)$ are constrained to obey $p^2 = 0$. The helicities then simply correspond to $h = \pm \mathfrak{s}$, with $\mathfrak{s}$ the spin of the particles that we are considering. It is convenient to introduce the matrix $P_{\alpha \dot{\alpha}}$ which is expressed as
\begin{align}
    P_{\alpha \dot{\alpha}} = p_{\mu} ( \sigma^{\mu} )_{\alpha \dot{\alpha}},
\end{align}
where we have introduced the four-matrices $\sigma^{\mu} = \left( \sigma^0, \sigma^1, \sigma^2 , \sigma^3 \right)$, with $\sigma^0$ the identity matrix and $\sigma^i$, $i = 1,2,3$ the Pauli matrices. With this, the matrix components are
\begin{align}
    P_{\alpha \dot{\alpha}} = \left( \begin{array}{cc}
    p_0 + p_3 & p_1 + i p_2 \\
    p_1 - i p_2 & p_0 - p_3
    \end{array}
    \right).
\end{align}
The massless condition $p^2 = 0$ translates into the condition that the determinant of the matrix needs to vanish, i.e., $\det P_{\alpha \dot{\alpha}} = 0$. Notice that any $SL(2,\mathbb{C})$ transformation preserves the determinant. Furthermore, since it vanishes, we can actually decompose the matrix into two spinors
\begin{align}
    P_{\alpha \dot{\alpha}} = \lambda_{\alpha} \Tilde{\lambda}_{\dot{\alpha}}. 
    \label{Eq:FactorSpinors}
\end{align}
For real momenta the two spinors would be related through complex conjugation. However, for our purposes it is convenient to work with complex momenta. Thus, in general, for complex momenta we have two copies of the group $SL(2,\mathbb{C})$ acting independently on the two spinors $\lambda$ and $\Tilde{\lambda}$. They correspond to the $(1/2,0)$ and $(0,1/2)$ representations of $SL(2,\mathbb{C})$, respectively. The product of two spinors is computed by contracting their components with the Levi-Civita symbol $\epsilon$. Consider two spinors associated with momenta $p_1$ and $p_2$, $\lambda^{(1)}, \Tilde{\lambda}^{(1)}$ and $\lambda^{(2)}, \Tilde{\lambda}^{(2)}$, then it is convenient to introduce the following notation for their products
\begin{align}
     \langle 12 \rangle & :=  \epsilon^{\alpha \beta} \lambda^{(1)}_{\alpha} \lambda^{(2)}_{\beta} , \\
    [12]  & := \epsilon^{\dot{\alpha} \dot{\beta}} \lambda^{(1)}_{\dot{\alpha}} \lambda^{(2)}_{\dot{\beta}}.
\end{align}
In general, product of momenta can be expressed in this notation, for instance, we have 
\begin{align}
        p_i \cdot p_j = \expval{ij} [ij].
\end{align}
Notice that the factorization from Eq.~\eqref{Eq:FactorSpinors} in terms of the two spinors is not unique, since we can always perform the transformations
\begin{align}
    \lambda \rightarrow t \lambda, \qquad \Tilde{\lambda} \rightarrow t^{-1} \Tilde{\lambda},
\end{align}
that leave invariant the amplitude. They correspond to the little group scaling, which are the transformations of the Lorentz group that leave invariant the momenta of the particle. A scattering amplitude undergoes the following transformation
\begin{align}
        A^{h_1 ... h_n} \left( \lambda^{(1)}, \Tilde{\lambda}^{(1)}, ..., t\lambda^{(i)}, t^{-1} \Tilde{\lambda}^{(i)},..., \lambda^{(n)}, \Tilde{\lambda}^{(n)}\right)=\\
        t^{-2h_i} A^{h_1 ... h_n} \left( \lambda^{(1)}, \Tilde{\lambda}^{(1)}, ..., \lambda^{(i)}, \Tilde{\lambda}^{(i)},..., \lambda^{(n)}, \Tilde{\lambda}^{(n)}\right)
\end{align}
under a little group scaling. This actually fixes the scattering amplitudes for 3 external particles. For amplitudes involving three external particles with two equal helicities and one opposite we have:
\begin{align}
    A(1^{-},2^{-},3^{+}) = g \left(  \frac{\expval{12}^3}{\expval{13}\expval{23}} \right)^{\mathfrak{s}}, \quad 
    A\left( 1^{+},2^{+},3^{-}\right) = g \left(  \frac{\left[12\right]^3}{\left[ 13 \right] \left[ 23 \right]} \right)^{\mathfrak{s}}. 
\end{align}
Thus, we see that for $\mathfrak{s} = 1$, the coupling constant is dimensionless and for $\mathfrak{s} = 2$, it carries dimensions of $[E]^{-1}$ and it can be identified with $M_P^{-1}$ in four spacetime dimensions. If these terms were arising from a Lagrangian, they would necessarily correspond to terms with one derivative for spin-1 fields, i.e., $ \sim g A A \partial A $, and terms with two derivatives for the spin-2 field, i.e. $\sim M_P^{-1} h \partial h \partial h$, which arises from the Einstein-Hilbert action $\sqrt{-g} R $. 

The scattering amplitude with equal helicities lead to the following expression
\begin{align}
    A(1^{-},2^{-},3^{-}) = g' \left( \expval{12} \expval{23} \expval{13} \right)^{\mathfrak{s}}, \quad A\left( 1^{+},2^{+},3^{+}\right) = g' \left(  \left[ 1 2 \right] \left[ 2 3 \right] \left[ 1 3 \right] \right)^{\mathfrak{s}}. 
\end{align}
In this case, both of the coupling constants turn out be dimensionful and arise from higher-dimensional operators in the theory. For the case $\mathfrak{s} = 1$, we would actually have $[g'] = [E]^{-2}$. Furthermore, it would need to arise from a term with three derivatives of the $A$ field, i.e., a term of form $ \sim (\partial A)^3$ in the action. Regarding the gravitational case $\mathfrak{s} = 2$, we would have that the coupling constant dimensions are $[g'] = [E]^{-5}$ and actually corresponds to a term involving six derivatives of the fields, i.e. a term of the form $ \sim (\partial^2 h )^3$, which necessarily arises from cubic terms in curvature $\sqrt{-g} R^3$. 

From these expressions we automatically conclude that a single spin-1 particle cannot interact with itself at leading order. The reason is that the exchange of any of the two particles in the amplitude $A(1^{-},2^{-},3^{+})$, leads to a violation of Bose symmetry, since we would pick a minus factor. If we consider several species of interacting particles which we label with latin indices $a,b,\ldots$, the coupling constant would instead be replaced by 
\begin{align}
        A(1^{a -},2^{b -},3^{c +}) = g f^{abc} \frac{\expval{12}^3}{\expval{13}\expval{23}}, 
\end{align}
and the tensor $f^{abc}$ would need to be fully antisymmetric to make the amplitude symmetric under the exchange of any two particles. Furthermore, by working out the four-particle amplitude in terms of the three-particle one, as discussed in Section~\ref{Subsec:Constructibility}, and applying similar arguments, one finds that the tensor $f^{abc}$ must satisfy
\begin{align}
    f^{abc} f^{cde} + f^{bdc} f^{cae} + f^{dac} f^{cbe} = 0, 
\end{align}
which is precisely the Jacobi identity. 

For spin-2 fields, this analysis simply leads to the conclusion that interactions between massless spin-2 particles can always be redefined in a way that effectively diagonalizes and decouples them.


\chapter{Static and axisymmetric metrics and multipole expansions}
\label{AppD}

\fancyhead[LE,RO]{\thepage}
\fancyhead[LO,RE]{Static and axisymmetric metrics and multipole expansions}


\section{Curzon metric and higher multipoles}
\label{App:Curzon}

\subsection{Curzon metric}

Consider the Curzon metric~\cite{Curzon1925}, which can be expressed in terms of the Weyl line-element~\eqref{Eq:Metric} where the functions $U(r,z)$ and $V(r,z)$ are given by
\begin{align}
    & U(r,z) = - \frac{M}{\sqrt{r^2+z^2}}, \\
    & V(r,z) = - \frac{M^2 r^2}{2(r^2 + z^2)^2}, 
\end{align}
with $M$ a real parameter. We need to distinguish the situation in which $M$ is positive and the situation in which $M$ is negative. It is convenient to change coordinates to spherical coordinates
\begin{align}
    & r = R \sin \theta, \\
    & z = R \cos \theta,
\end{align}
in terms of which the metric reads:
\begin{align}
    \dd s^2 = - e^{2 U} \dd t^2 +e^{-2U} \left[ e^{2V} \left( \dd R^2 + R^2 \dd  \theta ^2 \right) + R^2 \sin^2 \theta \dd \varphi^2 \right],
\label{Eq:Line-Element- }
\end{align}
with 
\begin{align}
    & U(R) = - \frac{M}{R}, \\
    & V(R,\theta) = - \frac{M^2 \sin^2 \theta }{2 R^2}.  
\end{align}
The area of the constant $R$ spacelike surfaces (for a fixed time $t$) is given by the expression: 
\begin{align}
    A(R) = \int \dd \Omega \sqrt{g^{(2)}}
     = R^2 e^{2M/R} \int_0^{\pi} \dd \theta \int^{2 \pi}_0 \dd \varphi \sin \theta e^{-M^2 \sin^2 \theta/ (2R^2)},
\end{align}
which upon performing the integral over $\varphi$ and rearranging  terms reads
\begin{align}
    A(R) = 2 \pi R^2 e^{2M/R - M^2/(2 R^2)} \int_{-1}^1 \dd (\cos \theta) e^{M^2 \cos^2 \theta / (2R^2)}.
\end{align}
We can absorb $M$ into the radial coordinate by performing the change $R \rightarrow  |M| R$, since it only corresponds to a choice of units. Also by parity arguments we can multiply by 2 and integrate between $0$ and $1$ to obtain
\begin{align}
    & A(R)/M^2 = 4 \pi R^2 e^{2/R - 1/(2R^2)}  \int_{0}^1 \dd u e^{ u^2 / (2R^2)} \quad M >0, \\
    & A(R)/M^2 = 4 \pi R^2 e^{-2/R - 1/(2R^2)}  \int_{0}^1 \dd u e^{ u^2 / (2R^2)} \quad M<0.
\end{align}
We will also use the Kretschmann scalar of the metric
\begin{align}
    \mathcal{K} = & - 8 M^2 e^{\frac{2 M \left(M \sin ^2(\theta )-2 R\right)}{R^2}} \nonumber  
    \\ & \times \frac{3 R^2 \left(M^2 (\cos (2 \theta )-3)+4 M R-2 R^2\right)-2 M^3 \sin ^2(\theta ) (M-3 R)}{R^{10}}.
\end{align}

\paragraph*{\textbf{Positive M.}}
The limit $R \rightarrow 0$ is highly directional for the curvature. As we approach $R=0$ through a direction of fixed $\theta=\theta_0$, we have that the  
\begin{align}
    & \lim_{R \rightarrow 0} \mathcal{K}(R,\theta_0) = \infty, \ \forall \theta_0 \neq 0, \pi, \\
    & \lim_{R \rightarrow 0} \mathcal{K} (R,0) = \lim_{R \rightarrow 0} \mathcal{K} (R,\pi) = 0.
\end{align}
It is quite surprising that as we approach the (at least coordinate) singularity located at $z = r = 0$, the Kretschmann scalar may vanish or not, depending on the direction through which we approach it. This is an archetypal example of directional singularity. We have depicted in Fig.~\ref{Fig:Curvatures_Redshift}, the equipotential surfaces and the surfaces of constant Kretschmann scalar for the sake of comparison with their Schwarzschild counterpart. 

\begin{figure}
\begin{center}
\includegraphics[width=0.75 \textwidth]{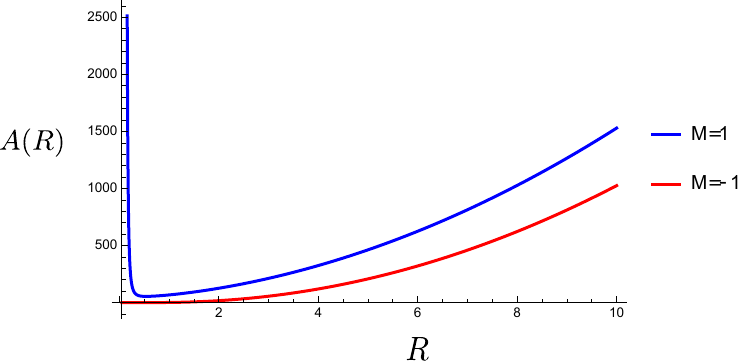}
\caption{We are depicting here the area function as a function of $R$ for $M=1$ in blue and $M=-1$ in red. For $M=1$ the area monotonically decreases from $R=0$ to a minimum at $R \simeq 0.538905$, and then becomes monotonically increasing. For $M = -1$ it is always a monotonically increasing function that vanishes at the origin.}
\label{Fig:AreaFunction}
\end{center}
\end{figure}

Furthermore, we can take a look at the area function of the constant $R$ surfaces. As we move from infinity to lower values of $R$, the area monotonically decreases until we hit a minimum at $R \simeq 0.538905M$, see Fig~\ref{Fig:AreaFunction}. From that point on, we have that the area of the surfaces increases without a bound as we approach $R \to 0$, where the area becomes infinite~\cite{Stachel1968}.

Regarding the surface gravity, it can be obtained as
\begin{align}
    \kappa^2 = - \frac{1}{2} \lim (\nabla_\mu \chi_\nu) (\nabla^\mu \chi^\nu),
\end{align}
where the limit represents that we are evaluating the object on the horizon. For the line element in Eq.~\eqref{Eq:Metric}, it is shown by Geroch and Hartle, see Eq.(3.4) from~\cite{Geroch1982}, that the surface gravity admits the following expression as the horizon is approached:
\begin{align}
    \kappa^2(r,z) = \lim e^{4 U (r,z) - 2 V(r,z)} \left[ \left(\partial_r U(r,z) \right)^2 + \left( \partial_z U(r,z) \right)^2 \right]. 
\end{align}
If we plug in the functions for the Curzon metric, we find that the leading order behavior at the origin along an arbitrary direction $r = R \sin \theta$ and $z = R \cos \theta$ is given by
\begin{align}
    \kappa^2 (\theta) = \frac{M^2 }{R^4}\exp{\frac{M \big(M [-\cos (2 \theta ])+M-8 R\big)}{2 R^2}} + \text{subdominant terms}.  
\end{align}
Taking the square root and taking into account that the term in the exponential is such that as long as $\cos \theta \neq 1$ the argument is positive, we find that:
\begin{align}
    \kappa (\theta) \rightarrow + \infty \qquad \theta \neq 0, \pi.
\end{align}
If we approach the point through along the $z$-axis, i.e. along $r = 0$,  which we recall is the direction along which the Kretschmann remains finite, we find that the surface gravity is given by:
\begin{align}
    \kappa_{\text{z-axis}} = \lim _{z \rightarrow 0} \frac{M e^{- {2 M}/{\left| z\right| }}}{z^2} = 0.
\end{align}
This could have been anticipated since the Kretschmann scalar remains finite and includes a term involving derivatives of $\kappa$ in directions transverse to the horizon.

\paragraph*{\textbf{Negative M.}}
This case is much simpler and it qualitatively represents the same situation that we have in the Schwarzschild metric with negative mass parameter. Namely, $R=0$ represents a naked timelike singularity that cannot be hit in finite proper time. Furthermore, the area of the surfaces in this case vanishes as $R \rightarrow 0$ and the curvature invariants become infinite along any direction. Let us see this in detail.

\begin{figure}[H]
\begin{center}
\includegraphics[width=0.45 \textwidth]{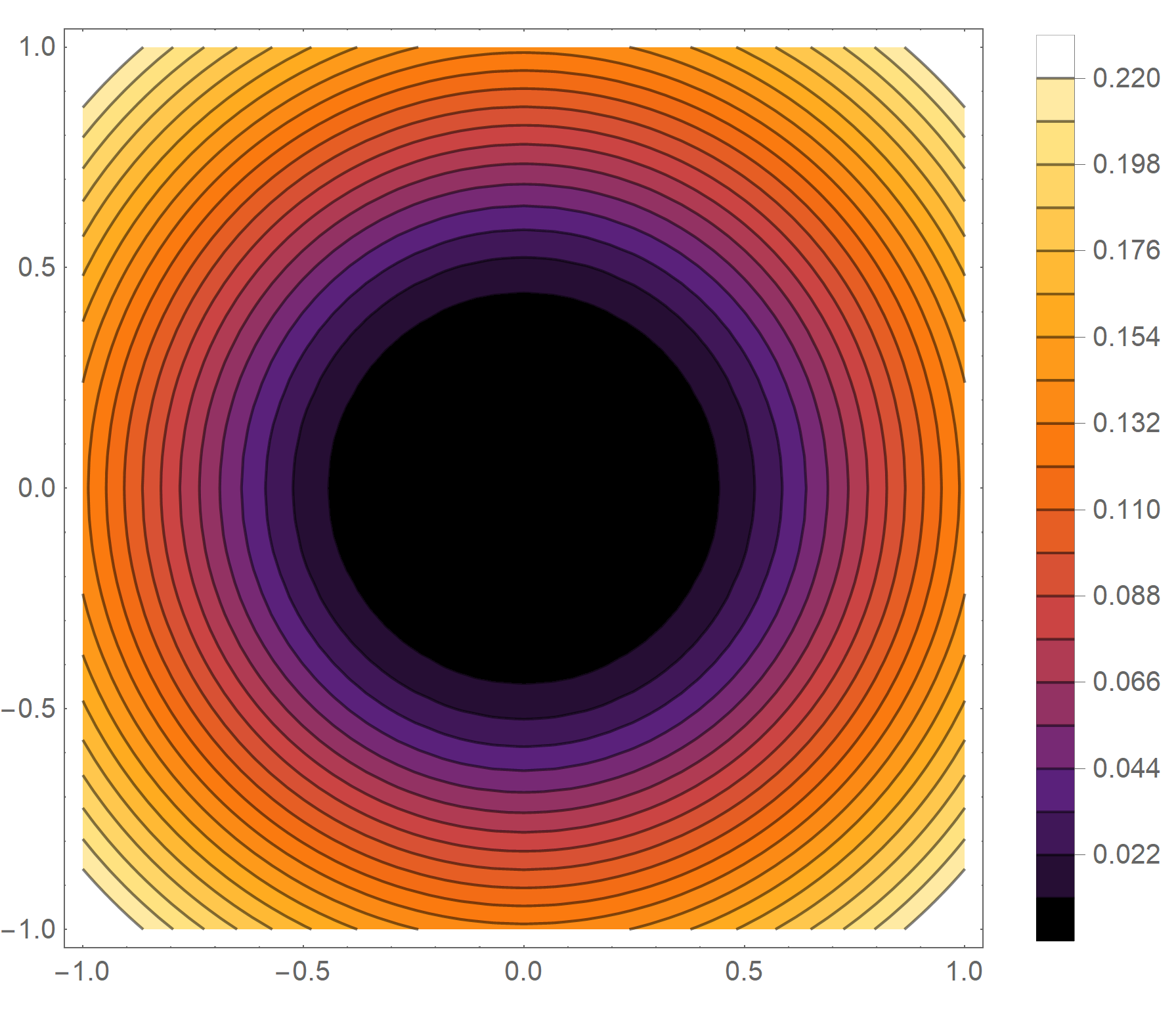}
\includegraphics[width=0.45 \textwidth]{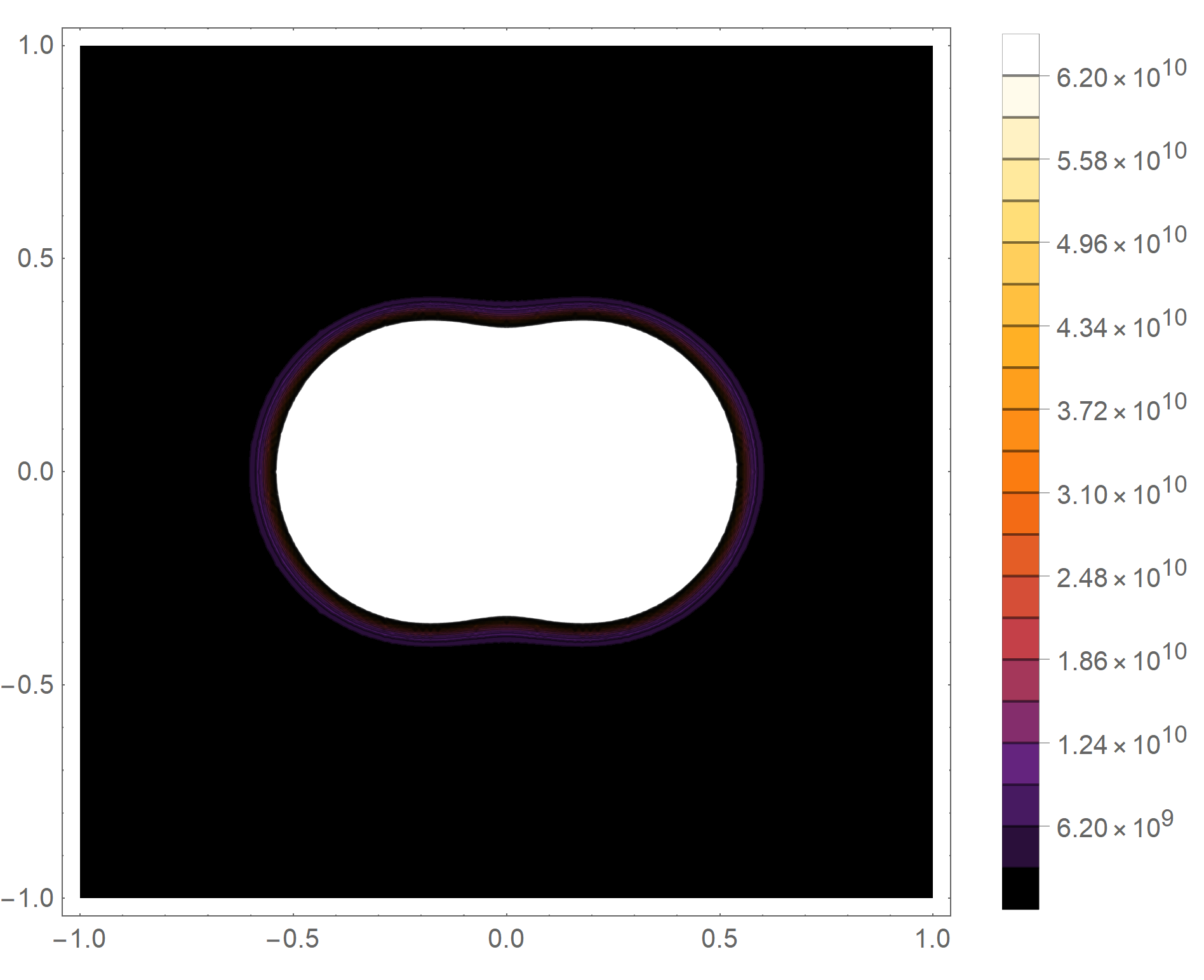}
\caption{We show the constant redshift surfaces (left) and the constant Kretschmann scalar surfaces (right) on a fixed-$\varphi$ plane for the Curzon solution with negative mass, \mbox{$M = -1$.}}
\label{Fig:Curvatures_Redshift_Negative}
\end{center}
\end{figure}

First of all, it is easy to see that  the limit
\begin{align}
    \lim_{R \rightarrow 0} \mathcal{K} (R, \theta_0),
\end{align}
is unbounded independently of the direction along which we take it. In the case of positive mass, the prefactor $e^{2M(M \sin^2 \theta_0 - 2R)/R^2}$ was divergent along every direction, except for the directions $\theta_0 = 0, \pi$, for which the first term in the argument of exponential vanishes and hence, the $e^{-4M/R}$ term dominates over the $1/R^{10}$ factor. However, in this case, the negative mass causes the exponential to diverge regardless of the value of $\theta_0$, leading to a divergence in the Kretschmann scalar. Similarly, the redshift function also diverges as $R \to 0$, as as illustrated in Fig.~\ref{Fig:Curvatures_Redshift_Negative}. 
The area function is also qualitatively different in this case. It is a monotone increasing function of $R$, approaching $0$ as $R \to 0$, as it can be seen in Fig.~\ref{Fig:AreaFunction}.

The singularity is such that it cannot be hit by any timelike geodesic nor any observer with finite integrated acceleration along its trajectory. We show it following the discussion in~\cite{Chakrabarti1983}. An observer will be described by a trajectory that is a timelike curve $\gamma$ (not necessarily a geodesic), with $\xi^\mu$ its unit tangent vector pointing toward the future. In general, a trajectory will have nonvanishing acceleration $a^\mu = \xi^\nu\nabla_\nu \xi^\mu$ unless it is a geodesic. We will consider that a trajectory is reasonable for an observer if the integrated acceleration along the trajectory is not infinite, namely: 
\begin{align}
    \int_{\gamma} a < \infty,
\end{align}
with $a = \sqrt{a^\mu a^\nu g_{\mu \nu}}$. This condition is equivalent to the statement that there exists a function $m$ bounded from below such that 
\begin{align}
    J^\mu = - \xi^\nu \nabla_\nu \left( m \xi^\mu \right),
\end{align}
is always a timelike or null future oriented vector. We can show this expanding its expression
\begin{align}
    J^\mu = - \xi^\mu \xi^\nu \nabla_\nu m - m \xi^\nu \nabla_\nu \xi^\mu = - \xi^\mu \xi^\nu \nabla_\nu m - m a^\mu. 
\end{align}
We observe that $a^\mu \xi_\mu  = 0$, as it can straightforwardly be checked by taking the derivative $\xi^\mu \nabla_\mu$ in $\xi^\mu \xi_\mu = -1$. We now compute
\begin{align}
    g_{\mu \nu} J^\mu J^\nu = \left( \xi^\sigma \nabla_\sigma m \right)^2 - m^2 a^2,
\end{align}
and demand $g_{\mu \nu} J^\mu J^\nu \leq 0$, so we get that
\begin{align}
    a \leq - \xi^\mu \nabla_\mu \log m. 
\end{align}
An integration of this, demonstrates the statement about the equivalence between the existence of $m$ and the finite acceleration. The spacetime that we are considering is static, hence we have the timelike Killing vector field $t^{\mu}$ and it satisfies the Killing equation $\nabla_{(\mu} t_{\nu)}=0$. As a consequence of this, we can identify the following quantity with the effective energy
\begin{align}
    E = -\xi_\mu t^\mu. 
\end{align}
In principle, $\gamma$ is not a geodesic, and hence $E$ is not conserved. However, we can prove the identity 
\begin{align}
    | \xi^\mu \nabla_\mu E | \leq a E.
\end{align}
Expanding the derivative of $E$ we have $\xi^\mu \nabla_\mu E = -a_\mu t^\mu - \xi^\mu \xi^\nu \nabla_{\mu} t_{\nu}$. The last term vanishes since it is the Killing equation contracted with a symmetric tensor. We have then:
\begin{align}
    -a_\mu t^\mu = a_\mu t_\nu h^{\mu \nu}, 
\end{align}
where we have introduced the projector onto the space orthogonal to the trajectory $h^{\mu \nu} = g^{\mu \nu} + \xi^\mu \xi^\nu$, and use the fact that the acceleration is always orthogonal to the vector $\xi^\mu$. Now we can apply the Cauchy-Schwarz inequality to find
\begin{align}
    a_\mu t_\nu h^{\mu \nu} \leq \left( a_\mu a_\nu h^{\mu \nu} \right)^{1/2} \left( t_\mu t_\nu h^{\mu \nu}\right)^{1/2} = a \left( t^\mu t_\mu + E^2 \right) \leq a E,
\end{align}
since in our conventions the timelike character of $t^\mu$ means that $t^\mu t_\mu \leq 0$. This leads to $\abs{\xi^\mu \nabla_\mu E} \leq aE$. If we integrate this inequality now along the curve, we conclude that $E$ needs to be finite along the curve also. We can use the triangular inequality applied to the definition of the energy leads to
\begin{align}
    E = - \xi^\mu t_\mu \geq \left( - t^\mu t_\mu \right)^{1/2} \left( - \xi^\mu \xi_\mu \right)^{1/2} = \left( - t^\mu t_\mu \right)^{1/2}. 
\end{align}
Applying this to our spacetime we have that $E$, which needs to be bounded (and hence there exists an upper bound $\Tilde{E} \in \mathbb{R}$ for it), satisfies:
\begin{align}
   \Tilde{E} \geq  E \geq  (- t^\mu t_\mu)^{1/2}  = e^{ {\abs{m}}/{R}}.
\end{align}
With this, we conclude that it is not possible to reach arbitrarily small values of the $R$ coordinate. Hence, no ``decent'' timelike trajectory can reach the naked singularity located at $R = 0$. 
\subsection{Higher-multipole configurations}
We can consider the solution generated by an arbitrary multipole located at the origin:
\begin{align*}
    U^{(\ell)} = - \frac{M^{(\ell)}}{(r^2 + z^2)^{\frac{\ell+1}{2}}} P_{\ell} \left( \frac{z}{\sqrt{r^2+z^2}} \right),
\end{align*}
and we can now perform the integration of the equations for $V$ in order to find the following:
\begin{align}
    V^{(\ell)} = - \frac{(\ell+1) (M^{(\ell)})^2}{2 (r^2+z^2)^{\ell+2}} \left[ P_{\ell}^{2} \left( \frac{z}{\sqrt{z^2 + r^2}} \right) - P_{\ell+1}^{2} \left( \frac{z}{\sqrt{z^2 + r^2}} \right) \right]. 
\end{align}
Actually, it is possible to perform the integral even for an arbitrary superposition of multipole configurations:
\begin{align*}
    U(r,z) = - \sum_{\ell=0}^{\infty} \frac{M^{(\ell)}}{(r^2 + z^2)^{\frac{\ell+1}{2}}} P_{\ell} \left( \frac{z}{\sqrt{r^2+z^2}}\right),
\end{align*}
with the function $V$ reading:
\begin{align}
    V(r,z) = & - \sum_{\ell, m =0}^{\infty} \frac{M^{(\ell)} M^{(m)} (\ell + 1) (m + 1)}{(\ell + m + 2) (r^2 + z^2)^{\ell + m + 2/2}} \\ 
    \times & \left[ P_{\ell} \left( \frac{z}{\sqrt{z^2 + r^2}} \right) P_{m} \left( \frac{z}{\sqrt{z^2 + r^2}} \right) - P_{\ell+1} \left( \frac{z}{\sqrt{z^2 + r^2}} \right) P_{m+1} \left( \frac{z}{\sqrt{z^2 + r^2}} \right) \right]. 
\end{align}
Computing the Kretschmann scalar for different multipole configurations reveals a behavior qualitatively similar to the Curzon metric: some directions approaching the origin $r = z = 0$ lead to a curvature singularity, while others do not.

\section{Bach-Weyl rings}
\label{App:Rings}

We can consider the solution to the Poisson equation associated with a density profile of the form 
\begin{align}
    \rho(\boldsymbol{x}) = \frac{M_{\text{ring}}}{2 \pi}  \delta (z) \delta( r - r_{\text{ring}}), 
\end{align}
which represents a ring of radius $r_{\text{ring}}$ located at the $z = 0$ plane and constant density. The solution can be expressed in terms of the Green function explicitly as 
\begin{align}
    U( \boldsymbol{x}) = -  \int \dd^3 \boldsymbol{x'} \frac{\rho ( \boldsymbol{x'})}{\abs{\boldsymbol{x} - \boldsymbol{x'}}}.
    \label{Eq:FundamSol}
\end{align}
It is possible to analytically perform this integral and express it in terms of the elliptic function $K$~\cite{Abramowitz1964} as
\begin{align}
    U_{\text{ring}}(r,z) = - \frac{2 M_{\text{ring}} }{\pi \sqrt{z^2 + (r_{\text{ring}}+r)^2 }} K \left(  \frac{4r_{\text{ring}}r}{z^2 + (r_{\text{ring}}+r)^2} \right). 
    \label{Eq:RingPotential}
\end{align}
Everywhere except for the ring where matter is located, this function is a harmonic function, and hence it is solution to Laplace equation. Thus, we can take it to generate a solution of Eqs.~\eqref{Eq:DrV}-\eqref{Eq:DzV}. It leads to the so-called Bach-Weyl ring solution~\cite{Bach2012} for which $V$ admits an analytic expression:
\begin{align}
    & V_{\text{ring}}(r,z) =  \frac{ M_{\text{ring}}^2  }{4 \pi^2 r r_{\text{ring}}} \left( \frac{4 r r_{\text{ring}}}{z^2 + (r_{\text{ring}}+r)^2} \right)^2 \Bigg[ - \left[ K \left( \frac{4rr_{\text{ring}}}{z^2 + (r+r_{\text{ring}})^2} \right) \right]^2 \nonumber \\
    & + 4 \frac{z^2 + (r-r_{\text{ring}})^2}{z^2 + (r+r_{\text{ring}})^2}  K \left( \frac{4rr_{\text{ring}}}{z^2  + (r+r_{\text{ring}})^2} \right)  K' \left( \frac{4rr_{\text{ring}}}{z^2 + (r+r_{\text{ring}})^2}  \right)  \nonumber \\
    & + 4 \frac{4rr_{\text{ring}}}{z^2 + (r+r_{\text{ring}})^2} \frac{z^2 + (r-r_{\text{ring}})^2}{z^2 + (r+r_{\text{ring}})^2}  \left[ K' \left( \frac{4rr_{\text{ring}}}{z^2 + (r+r_{\text{ring}})^2}  \right)  \right]^2 \Bigg] \nonumber \\
    & +  \frac{ M_{\text{ring}}^2 }{4 \pi^2 r_{\text{ring}}^2}  \left( \frac{4rr_{\text{ring}}}{z^2 + (r+r_{\text{ring}})^2} \right)^2 \Bigg[ - \left[ K \left( \frac{4rr_{\text{ring}}}{z^2 + (r+r_{\text{ring}})^2} \right) \right]^2 \nonumber \\
    & + 4 \frac{z^2 + (r-r_{\text{ring}})^2}{z^2 + (r+r_{\text{ring}})^2} K \left( \frac{4rr_{\text{ring}}}{z^2 + (r+r_{\text{ring}})^2} \right)  K' \left( \frac{4rr_{\text{ring}}}{z^2 + (r+r_{\text{ring}})^2} \right) \nonumber \\
    & - 8 \frac{z^2 + (r-r_{\text{ring}})^2}{z^2 + (r+r_{\text{ring}})^2} \frac{z^2 + r^2 + r_{\text{ring}}^2}{z^2 + (r+r_{\text{ring}})^2} \left[ K'  \left( \frac{4rr_{\text{ring}}}{z^2 + (r+r_{\text{ring}})^2} \right) \right]^2 \Bigg].
\end{align}
This function $V_{\text{ring}}(r,z)$ obeys all the requirements of regularity of the metric: it   approaches zero quadratically in $r$ as we approach the $z$-axis, while it goes smoothly to zero at spatial infinity. This solution can be simply translated in the $z$ axis by making the shift $z \rightarrow z - z_0$ above. 

We can now combine this with the Schwarzschild solution in order to find a new solution representing a black hole distorted by the gravitational field of a ring of matter, namely:
\begin{align}
    U(r,z;z_0) = U_S(r,z) + U_{\text{ring}}(r,z-z_0).
\end{align}
We have plotted the corresponding derivative of $V_{\text{comb}}$, namely 
\begin{align}
    \partial_r V = r \left[  \left( \partial_r  U \right)^2 - \left( \partial_z  U \right)^2  \right], 
\end{align}
in Fig.~\ref{Fig:Saturns_Integrands} for the ring lying on the same plane as the black hole $z_0 =0 $ and a value $z_0 = 5$. 

\section{Convergence of multipole expansions}
\label{App:Multipolar}

As discussed in the main text, Weyl coordinates are not well-suited for describing spherically symmetric configurations. In fact, when performing a multipole expansion in these coordinates, the Weyl multipole expansion, a spherically symmetric configuration yields an infinite series of terms. Moreover, this expansion only converges outside a finite region surrounding the black hole horizon, which in Weyl coordinates lies at $r = 0$ and $z \in (-M,M)$. We demonstrate this behavior here. To develop some intuition about the convergence of multipole expansions, we consider first a simple electromagnetic toy model: two point charges separated by a finite distance.

\subsection{Electromagnetic example}
\begin{figure}
\begin{center}
\includegraphics[width=0.45 \textwidth]{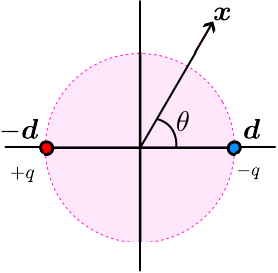}
\caption{We pictorially represent the radius of convergence of the multipole expansion of the potential generated by two equal charges with opposite signs.}
\label{Fig:Charges}
\end{center}
\end{figure} 

The full potential of two equal charges with opposite signs located at $\pm \boldsymbol{d}$, respectively, see Fig.~\ref{Fig:Charges} for a pictorial representation, is given by the sum of the Coulomb potential generated by both of them
\begin{align}
    U = \frac{q}{4 \pi \abs{\boldsymbol{x} - \boldsymbol{d}}} - \frac{q}{4 \pi \abs{\boldsymbol{x} + \boldsymbol{d}}}.
\end{align}
We can now expand the denominators at large values of $r$ compared to their separation, and we find the following expression:
\begin{align}
    \frac{1}{\abs{\boldsymbol{x} + \boldsymbol{d}}} = \frac{1}{\sqrt{r^2 + d ^2 - 2 d r \cos \theta}} = \frac{1}{r} \frac{1}{\sqrt{1- 2 \frac{d}{r} \cos \theta + \frac{d^2}{r^2}}}.  
\end{align}
Similarly we have for the other term 
\begin{align}
    \frac{1}{\abs{\boldsymbol{x} + \boldsymbol{d}}} = \frac{1}{\sqrt{r^2 + d ^2 + 2 d r \cos \theta}} = \frac{1}{r} \frac{1}{\sqrt{1+ 2 \frac{d}{r} \cos \theta + \frac{d^2}{r^2}}},
\end{align}
which is identical upon the exchange $d \to -d$. Recalling the generating formula for the Legendre polynomials~\cite{Lebedev1972}
\begin{align}
    \frac{1}{\sqrt{1 - 2 x t + t^2 }} = \sum_{n=0}^{\infty} P_n(x) t^n,
    \label{Eq:LegendreGenerating}
\end{align}
where $P_n(x)$ represents the Legendre polynomial of degree $n$, we can write the following expansions:
\begin{align}
    \frac{1}{\abs{\boldsymbol{x} + \boldsymbol{d}}} = & \frac{1}{r} \sum_{n=0}^{\infty} P_n ( \cos \theta) \left( \frac{d}{r} \right)^n, \\
    \frac{1}{\abs{\boldsymbol{x} - \boldsymbol{d}}} = &  \frac{1}{r} \sum_{n=0}^{\infty} P_n ( \cos \theta) \left( - \frac{d}{r} \right)^n. 
\end{align}
Combining both expansions, we find the following asymptotic expression for the potential 
\begin{align}
    U(r,\theta) = \frac{q}{2 \pi r} \sum_{n = 0}^{\infty} P_{2n+1} \left( \cos \theta \right) \left( \frac{d}{r} \right)^{2n + 1}. 
\end{align}
We now want to study the radius of convergence of this series. For such purpose, we notice that there is one special direction, $\theta = \pi /2$ along which all the terms in the series identically vanish since
\begin{align}
    P_{2n+1} \left( 0 \right) = 0, \quad \forall n \in \mathbb{Z}^{+} \cup \{ 0 \}.
\end{align}
Thus, from the series we have that 
\begin{align}
    U(r, \frac{\pi}{2}) = 0, \quad \forall r \in \mathbb{R} - \{ 0 \}.
\end{align}
This is precisely what we also find from the exact potential: the plane $z = 0$ lies exactly between two point charges of equal magnitude but opposite sign, so the potential must necessarily vanish there. At first glance, this might suggest that the convergence of the series is highly directional. However, we can show that this behavior is simply an artifact of the highly symmetric configuration we’re considering. In fact, the series converges outside a disk centered at the origin of coordinates, except for the region inside the disk intersecting with the $\theta = \pi /2$ plane, where it also converges. 

First of all, we can realize that the series has the form
\begin{align}
    U (r, \theta) = \frac{q}{2 \pi r} \sum_{n = 0}^{\infty} a_n,
\end{align}
where the coefficients $a_n$ are given by
\begin{align}
    a_n = P_{n} \left( \cos \theta \right) \left( \frac{d}{r} \right)^{2n+1}. 
\end{align}
The radius of convergence $\mathfrak{r}$ can be directly obtained applying the root test, according to which
\begin{align}
    \lim_{n \to \infty} \sup \abs{a_n}^{\frac{1}{n}} = \mathfrak{r}^{-1}.  
\end{align}
With the asymptotic expansion for $P_n \left( \cos \theta \right)$~\cite{Lebedev1972},
\begin{align}
    P_n \left( \cos \theta  \right) = \sqrt{ \frac{2}{\pi n \sin \theta}} \cos \left[ \left(n + \frac{1}{2}\right) \theta - \frac{\pi}{4} \right] + \order{\frac{1}{n^{3/2}}}
\end{align}
we have that the $\abs{a_n}^{\frac{1}{n}}$ at large $n$ are of the form
\begin{align}
    \abs{a_n}^{\frac{1}{n}} = \abs{\frac{d}{r}}^{2 + \frac{1}{n}} \left( \abs{\frac{2}{\pi \sin \theta}}\right)^{\frac{1}{n}} n^{- \frac{1}{2n}} \abs{\cos \left[ \left(n +  \frac{1}{2} \right) - \frac{\pi}{4} \right]} ^{\frac{1}{n}} + \text{subleading terms},
\end{align}
where the subleading terms can be safely ignored when taking the $n \to \infty$ limit. Computing the supremum and taking the limit $n \to \infty$ we find the series converges for   $ r > d, \quad \forall \theta \neq \pi/2$. This is pictorially represented in Fig.~\ref{Fig:Charges}. 
\subsection{Schwarzschild expansion in Weyl coordinates}
\begin{figure}
\begin{center}
\includegraphics[width=0.45 \textwidth]{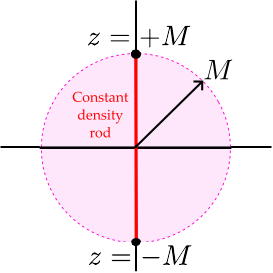}
\caption{We pictorially represent the radius of convergence of the Weyl multipole expansion of the Schwarzschild potential function $U$ in Weyl coordinates.}
\label{Fig:Schwarzschild_Multipoles}
\end{center}
\end{figure} 

We can now consider the Schwarzschild solution, which is given by the potential in Eq.~\eqref{Eq:Schwarzschild}. However, for the purpose of performing its Weyl multipole expansion, it is more convenient to consider its integral representation as the convolution of the Green function with the constant-density segment (equal to $1/2$) lying on $r = 0$ and $z \in [-M,M]$:
\begin{align}
    U (r,z) = - \frac{1}{2} \int_{-M}^{M} \frac{dz'}{\sqrt{r^2 + z^2 + z^{\prime 2}} - 2 z z^{\prime}}. 
\end{align}
Changing to spherical coordinates $(r,z) \to (R \sin \theta, R \cos \theta) $, we have
\begin{align}
    U (R,\theta) = - \frac{1}{2R} \int_{-M}^{M} \frac{d z^{\prime}}{\sqrt{1 + \frac{z^{\prime 2}}{R^2} - 2 \frac{z^{\prime}}{R} \cos \theta}}.
\end{align}
Again, we can identify the generating function of the Legendre polynomials, see Eq.~\eqref{Eq:LegendreGenerating}, from which it follows:
\begin{align}
    U(R, \theta) = -  \frac{1}{2R} \int_{-M}^{M} dz' \sum_{n=0}^{\infty} P_n \left( \cos \theta \right) \left( \frac{z'}{R} \right)^n.
\end{align}
By commuting the sum and the integral and evaluating the latter, we arrive at the following expansion of the Schwarzschild solution in Weyl coordinates:
\begin{align}
    U(R,\theta) = - \frac{M}{R} \sum_{n=0}^{\infty}  \frac{P_{2n} \left( \cos \theta \right)}{2n+1} \left( \frac{M}{R} \right) ^{2n}.
\end{align}
As discussed in the main text, we conclude that Weyl coordinates are not well-suited for spherically symmetric configurations, since the Schwarzschild solution is described as an infinite series in the multipole expansion. 

Furthermore, to analyze the radius of convergence of the series, we again apply the root test to the coefficients $a_n$ which in this case take the form
\begin{align}
    a_n = \frac{P_{2n} \left( \cos \theta \right)}{2n+1} \left( \frac{M}{R} \right) ^{2n}.
\end{align}
From this, we find that the series converges for $R>M$. This implies that the Weyl multipole expansion is not well-suited to describe a neighborhood of radius $M$ around the black hole horizon in Weyl coordinates. See Fig.~\ref{Fig:Schwarzschild_Multipoles} for a pictorial representation.


\chapter{Toroidal black holes}
\label{AppF}

\fancyhead[LE,RO]{\thepage}
\fancyhead[LO,RE]{Toroidal black holes}


The geometry constructed in Chapter~\ref{Ch6:ToroidalBHs} is built through a cut-and-paste procedure. The Israel junction conditions specify the criteria that must be met for the resulting geometry to be well-defined and to support a distributional energy-momentum tensor.

The first of these conditions imposes the continuity of the metric across any matching surface. In our case, there are two such surfaces where a discontinuity could in principle arise: the surface $\Sext$, which matches the external flat region $\Omega_3$ with the interpolating region $\Omega_2$, and the surface $\Sint$, which matches the interpolating region $\Omega_2$ with the internal toroidal black hole region $\Omega_1$. In both cases, the continuity of the metric can be ensured.

The second condition relates the jump in the extrinsic curvature across the shell with the distributional energy-momentum tensor $S^{i}_{\ j}$,
\begin{align}
    8\pi S^{i}_{\ j} = - \left( \left[ \left[ K^{i}_{\ j} \right] \right] -  \left[ \left[ K \right] \right] \delta^i_{\ j} \right),
\end{align}
where we have introduced the notation $\left[ \left[ \mathcal{O} \right] \right] = \mathcal{O}^{+} - \mathcal{O}^{-}$ to denote the jump of any quantity across one of these surfaces. Normal directions are always taken to point outward—namely, from $\Omega_3$ to $\Omega_2$, and from $\Omega_2$ to $\Omega_1$.

\section{Thin shell at the external surface}
\label{Sec:Sext}
Let us go first with the shell that is located in $\Sext$. For that purpose, we take first the parametrization of the torus in terms of the coordinates $y^i = (\Text, \aext, \bext)$:
\begin{align}
    & T ( \Text) = \Text, \\
    & X (\aext, \bext) = \cos (\aext) \Phi(\bext), \\
    & Y (\aext, \bext) = \sin (\aext) \Phi(\bext), \\
    & Z (\aext, \bext) = b  \sin (\bext)\,, 
\end{align}
where we have introduced the convenient abbreviation:
\begin{equation}
    \Phi(\beta) := \Rext + b \cos(\beta)\,.
    \label{Eq:Phidef}
\end{equation}
In terms of these coordinates, we can compute the following basis forms in $\Sext$
\begin{align}
    e^{\mu}{}_{i} = \frac{\partial X^\mu}{\partial y^i},  
\end{align}
which leads to
\begin{align}
    e^T{}_i \dd y^i& = \dd \Text, \\
   e^X{}_i \dd y^i&= - \sin (\aext) \Phi(\bext) \dd \aext - b \cos (\aext) \sin (\bext) \dd \bext, \\
    e^Y{}_i \dd y^i& = \cos (\aext) \Phi(\bext) \dd \aext - b \sin (\aext) \sin (\bext) \dd \bext, \\
   e^Z{}_i \dd y^i& = b \cos (\bext) \dd \bext.
\end{align}
The induced metric on the surface reads:
\begin{align}\label{eq:h+Region1}
    (\dd s_+^2)_{\Omega_1} = - \dd \Text^2 +\Phi(\bext)^2 \dd \aext^2 + b^2 \dd \bext^2\,. 
\end{align}
We can now compute the unit normal vector to the surface:
\begin{align}
    (n_+^\mu)_{\Omega_1} = \frac{1}{b} \left[ \left(1- \frac{\Rext}{\sqrt{X^2+Y^2}}  \right)X  \delta^{\mu}{}_{X} + \left(1- \frac{\Rext}{\sqrt{X^2+Y^2}}  \right)Y \delta^{\mu}{}_{Y} + Z \delta^{\mu}{}_{Z} \right],
\end{align}
Furthermore, with it we can now compute the extrinsic curvature of the surface $\Sext$, as seeing from outside region $\Omega_1$:
\begin{align}\label{eq:KijOmega1}
    (K^+_{ij})_{\Omega_1} = e^\mu{}_i e^\nu{}_j(\nabla_\mu n_{+\nu})_{\Omega_1} .
\end{align}
Given that the spacetime is flat and it is expressed in Minkowski coordinates, we have that the Christoffel symbols vanish and we can substitute $\nabla\to \partial$ in \eqref{eq:KijOmega1}. We have the following nonvanishing components:
\begin{align}
    & (K^+_{\aext \aext})_{\Omega_1} =  \Phi(\bext) \cos(\bext) , \\
    & (K^+_{\bext \bext})_{\Omega_1} = b . 
\end{align}
The trace reads:
\begin{align}
    (K^+)_{\Omega_1} = \frac{1}{b}+\frac{\cos(\bext)}{\Phi(\bext)} \,.
\end{align}
We can now repeat the exercise from the inside region $\Omega_2$. The surface, which corresponds to $ \ell = \Rext$, is parametrized by $(\tau, \alpha, \beta)$
\begin{align}
    (\dd s_+^2)_{\Omega_2} = - \dd \Text^2 + \Phi(\bext)^2 \dd \aext^2 + b^2 \dd \bext^2.
    \label{Eq:CurvedTorus}
\end{align}
In this case, the vector normal to the surface is
\begin{align}
    (n^a_+)_{\Omega_2} = \delta^a{}_r, 
\end{align}
and we can again compute the nonvanishing components of the extrinsic curvature to find:
\begin{align}
    (K^+_{\Text\Text})_{\Omega_2} &=  -\frac{1}{2} H'(\Rext),\\
    (K^+_{\aext \aext})_{\Omega_2} &= \frac{1}{2} \partial_\ell \mathcal{F}(\Rext, \beta_+),
\end{align}
and the trace reads:
\begin{align}
    (K^+)_{\Omega_2} = \frac{1}{2}\left[H'(\Rext) + \frac{\partial_\ell\mathcal{F}(\Rext,\bext)}{\Phi(\bext)^2}\right] \,.
\end{align}
The only nontrivial discontinuities in the extrinsic curvatures at $\Sext$ are given by:
\begin{align}
    [[ K^{+}_{\Text\Text} ]] &= \frac{1}{2} H'(\Rext), \\
    [[ K^{+}_{\aext\aext} ]] &= \Phi(\bext)\cos(\bext) -\frac{1}{2} \partial_\ell \mathcal{F}(\Rext, \beta_+) , \\ 
    [[ K^{+}_{\bext\bext}]] &= b, \\
    [[K^{+}]] &= \frac{1}{b}+\frac{\cos(\bext)}{\Phi(\bext)}-\frac{1}{2} H'(\Rext)-\frac{1}{2}\frac{\partial_\ell\mathcal{F}(\Rext,\bext)}{\Phi(\bext)^2},
\end{align}
and hence the nontrivial components of the energy-momentum tensor read: 
\begin{align}
    & 8 \pi S^+_{\Text\Text} = -\frac{1}{b}- \frac{\cos(\bext)}{\Rext + b \cos (\bext)} + \frac{\partial_\ell \mathcal{F}(\Rext,\bext)}{2(\Rext + b \cos (\bext))^2}, \\
    & 8 \pi S^+_{\aext\aext} =  (\Rext + b \cos (\bext))^2\left(\frac{1}{b}-\frac{1}{2}H'(\Rext)\right), \\
    & 8 \pi S^+_{\bext\bext} = \left[\frac{\cos(\bext)}{\Rext + b \cos (\bext)}-\frac{1}{2}H'(\Rext)-\frac{\partial_\ell \mathcal{F}(\Rext,\bext)}{2(\Rext + b \cos (\bext))^2}\right]b^2\,. 
\end{align}
where we have inserted $\Phi(\bext)$ back using Eq.~\eqref{Eq:Phidef}.

\section{Thin shell at the internal surface}
\label{Sec:Sint}
Now we repeat the exercise with the shell located at $\Sint$. From the outside region $\Omega_2$ we take $ \ell = \Rext$ and take the parametrization $(\tau, \alpha, \beta)$ to get the following induced metric on the surface:
\begin{align}\label{eq:ds-Om2}
   (\dd s_-^2)_{\Omega_2} = -\dd  \Tint^2 + \frac{m^2}{\pi^2} \dd  \aint^2 + b^2 \dd  \bint^2,
\end{align}
and the extrinsic curvature is again
\begin{align}
    (K^-_{\Tint\Tint})_{\Omega_3} &= -\frac{1}{2} H'(\Rint) \,,\\
    (K^-_{\aint\aint})_{\Omega_3} &= \frac{1}{2} \partial_\ell\mathcal{F}(\Rint,\bint)\,. 
\end{align}
The trace in this case reads:
\begin{align}
    (K^-)_{\Omega_2} =  \frac{\pi^2}{2m^2} \partial_\ell\mathcal{F}(\Rint,\bint) +\frac{1}{2} H'(\Rint)\,.
\end{align}
From the inside region $\Omega_3$, $\Sint$ corresponds with the hypersurface with $ \ell = \Rint$. If we parametrize it making the identification $\left(t = m \Tint / \Rint, z=\aint, \varphi=\bint \right)$ we reach the following expression for the induced metric:
\begin{align}\label{eq:h-Region3}
    (\dd s_-^2)_{\Omega_3} = -  \dd \Tint^2 + \frac{m^2}{\pi^2} \dd \aint^2 + m^2 \dd  \bint^2\,.
\end{align}
Continuity with \eqref{eq:ds-Om2} requires:
\begin{equation}
    b^2 = m^2\,,
\end{equation}
which can be used to eliminate the parameter $m$. The extrinsic curvature from the inside can be determined by realizing that also the normal vector to the surface is
\begin{align}
    (n_-^\mu)_{\Omega_3} = \delta^\mu{}_{\ell}, 
\end{align}
and we find that there is only one nonvanishing component of the extrinsic curvature:
\begin{align}
    (K^-_{\Tint\Tint})_{\Omega_3} = - \frac{1}{\Rint},
\end{align}
with the trace being 
\begin{align}
    (K^-)_{\Omega_3} = \frac{1}{\Rint}. 
\end{align}
The nonvanishing jumps in the extrinsic curvatures at $\Sint$  are given by:
\begin{align}
    & [[K^{-}_{\Tint\Tint}]] = -\frac{1}{2} H'(\Rint)+\frac{1}{\Rint}, \\
    & [[ K^{-}_{\aint\aint} ]] =  \frac{1}{2}\partial_\ell\mathcal{F}(\Rint,\bint) , \\ 
    & [[K^{-}]] = \frac{\pi^2}{2b^2} \partial_\ell\mathcal{F}(\Rint,\bint) +\frac{1}{2} H'(\Rint) -\frac{1}{\Rint},
\end{align}
leading to the following nontrivial components for the distributional energy-momentum tensor at $\Rint$ (after substituting $\Phi$):
\begin{align}
    8 \pi S^-_{\Tint\Tint} & = -\frac{\pi^2}{2b^2}\partial_\ell\mathcal{F}(\Rint,\bint) , \\
    8 \pi S^-_{\aint\aint} & = \frac{b^2}{\pi^2}\left[\frac{1}{2}H'(\Rint) -\frac{1}{\Rint}\right] , \\
    S^-_{\bint\bint} & =\pi^2 S^-_{\aint\aint} - b^2 S^-_{\Tint\Tint}  . 
\end{align}


\chapter{Spherically symmetric configurations}
\label{AppE}

\fancyhead[LE,RO]{\thepage}
\fancyhead[LO,RE]{Spherically symmetric configurations}


\section{Fluid spheres in GR}
\label{App:FluidSpheres}

Let us consider the line element in Eq.~\eqref{Eq:LineElement2}, and the solution corresponding to a perfect fluid with constant density profile, namely:
\begin{align}
    \rho(r) = \begin{cases}
      \rho_0,  & r \leq R\\
      0, & r>R
    \end{cases}. 
\end{align}
For such density profile, the mass function $m(r)$ takes the form
\begin{align}
    m(r) = \begin{cases}
      \frac{4}{3} \pi r^3 \rho_0,  & r \leq R\\
      \frac{4}{3} \pi R^3 \rho_0, & r>R
    \end{cases},
\end{align}
as a straightforward integration demonstrates, and we have the relation $M = (4/3) \pi R^3 \rho_0$ among the parameters $\left( M,R,\rho_0 \right)$. The pressure $p(r)$ and the redshift $\Phi(r)$ also admit explicit expressions given by
\begin{align}
    p(r) = \rho_0 \left[ \frac{(1-2M/R)^{1/2} - (1-2Mr^2/R^3)^{1/2}}{(1-2Mr^2/R^3)^{1/2} - 3 ( 1 - 2M/R)^{1/2} }\right]
    \label{Eq:ConstantDens_Pressure}
\end{align}
and
\begin{align}
    \Phi(r) = \log \left[ \frac{3}{2} \left( 1 - 2 M /R \right)^{1/2} - \frac{1}{2} \left( 1 - 2 Mr^2/R^3\right)^{1/2} \right].
    \label{Eq:ConstantDens_Redshift}
\end{align}
for $r\leq R$. Everywhere else, the pressure is zero, and the redshift function matches its Schwarzschild counterpart, given in Eq.~\eqref{Eq:SchwMetric}. The value of the pressure at the core $r=0$ is given by
\begin{align}
    p_c = p (0) = \rho_0 \left[ \frac{1 - (1-2M/R)^{1/2}}{3 (1-2M/R)^{1/2} - 1 } \right],
\end{align}
which clearly blows up when $R \rightarrow 9 M/4$, i.e., as Buchdahl's limit is approached. Furthermore, it is possible to think about this limit in an alternative way as the maximum mass that a uniform-density star can attain, namely:
\begin{align}
    M_{\text{Buchdahl}} < \frac{4}{9 (3 \pi)^{1/2}} \rho_0^{-1/2}. 
\end{align} 
Finally, we note that even though the density is not a continuous function because of the jump that it displays at $ r =R$, $h$ is still continuous, which is what is required for the proof of Buchdahl's theorem. In fact, the constant density profile can be understood as the limit of a one-parameter family of smooth $C^{\infty}$ functions $\rho_{\ell}(r)$, such as:
\begin{align}
    \rho_{\ell} (r) =  \begin{cases}
      \rho_0\, e^{ \frac{4 \ell^2}{R^2} + \frac{\ell^2}{r (r-R)}  }, &  r < R \\
      0 & r > R, 
    \end{cases},  
    \label{Eq:RegDensity}
\end{align}
which in the limit $\ell \rightarrow 0^+ $ reduces to a Heaviside function with density $\rho_0$ in the interior $r\in (0, R)$ and vanishing density in the external region. 

\section{Fully anisotropic shell matching spherically symmetric spacetimes}
\label{App:AnisotropicShell}
Let us consider a static and spherically symmetric spacetime that is foliated in spheres of proper radius $r$. Consider a line element of the form:
\begin{align}
    \dd s^2 = -f(r) \dd t^2 + h(r) \dd r^2 +r^2 \dd \Omega_2.
\end{align}
We can consider that the function $f(r)$ and $h(r)$ are piecewise defined, namely that we have:
\begin{align}
    f(r) = \begin{cases}
      f_{-} (r),  & r < R \\
      f_{+} (r),  & r > R 
    \end{cases}, \qquad
        h(r) = \begin{cases}
      h_{-} (r),  & r < R \\
      h_{+} (r),  & r > R 
    \end{cases},
\end{align}
The first of Israel junction conditions states that the metric needs to be continuous and hence the induced metric on a surface cannot have a discontinuity. The induced metric for the surface  $r=R$, whose components we denote as $h_{ij}$, can be computed from the outside and inside metrics to obtain:
\begin{align}
    & \dd s^2_{+} = - f_{+}(R) \left(\frac{\dd t_{+}}{\dd \tau} \right)^2 \dd \tau^2 + R^2 \dd \Omega_{+,2} ^2, \\
    & \dd s^2_{-} = - f_{-}(R) \left(\frac{\dd t_{-}}{\dd \tau} \right)^2 \dd \tau^2 + R^2 \dd \Omega_{-,2} ^2,
\end{align}
where $\tau$ represents the proper time for observers at constant $r = R$ and $\theta$. Continuity of the metric ensures that we can take an orthonormal basis on the surface such that $\Theta^{\mu} \partial_\mu= \partial_{\theta^{+}} = \partial_{\theta^{-}}$, \mbox{$\Phi^\mu \partial_\mu= \partial_{\varphi^{+}} = \partial_{\varphi^{-}}$}, and 
\begin{align}
    f_{+}(R) \left( \frac{\dd t_{+}}{\dd \tau} \right)^2 = f_{-} (R) \left( \frac{\dd t_{-}}{\dd \tau} \right)^2,
\end{align}
which translates into the condition $\sqrt{f_{+}(R)} t_{+} = \sqrt{f_{-} (R)} t_{-}$, choosing the same origin for the time coordinate. In other words, the induced metric is given by
\begin{align}
    \dd s^2 = - \dd \tau^2 + R^2 \dd \Omega^2,
\end{align}
since we can identify the angular coordinates, and the proper time along the shell is adapted to the redshift functions from the inside and the outside. 

We now need $n_{\mu}$, the unit normal vector to the surface which is given by
\begin{align}
    & n^{+}_{\mu} = \sqrt{h_{+}(r)} \delta^{r}_{\ \mu} ,
    & n^{-}_{\mu} = \sqrt{h_{-}(r)}\delta^{r}_{\ \mu}. 
\end{align}
To compute the extrinsic curvature, we need $\nabla_{\mu} n_{\nu}$, and we can extract the different components of the extrinsic curvature through
\begin{align}
    K_{\tau \tau} & = u^\mu u^\nu \nabla_\mu n_\nu , \\ 
    K_{\theta \theta} & = \Theta^\mu \Theta^\nu \nabla_\mu n_\nu , \\
    K_{\Phi \Phi} & = \Phi^{\mu} \Phi^{\nu} \nabla_\mu n_\nu. 
\end{align}
Computed from the outside the extrinsic curvature gives:
\begin{align}
    & K^{+}_{\tau \tau} = - \frac{1}{2} \frac{f'_{+}(R)}{f_{+}(R) \sqrt{h_{+}(R)}}, \\ 
    & K^{+}_{\theta \theta} =   \frac{R}{\sqrt{h_{+}(R)}}, \\
    & K^{+}_{\Phi \Phi} =  \frac{R \sin^2 \theta }{\sqrt{h_{+}(R)}} , \\
    & K^{+} = K^{+}_{ij}h^{ij} = \frac{1}{2} \frac{f'_{+}(R)}{f_{+}(R) \sqrt{h_{+}(R)}} + 2 \frac{R}{\sqrt{h_{+}(R)}}, 
\end{align}
whereas from the inside we simply need to replace $f_{+}$ and $h_{+}$ by $f_{-}$ and $h_{-}$ respectively:
\begin{align}
    & K^{-}_{\tau \tau} = - \frac{1}{2} \frac{f'_{-}(R)}{f_{-}(R) \sqrt{h_{-}(R)}}, \\ 
    & K^{-}_{\theta \theta} =   \frac{R}{\sqrt{h_{-}(R)}}, \\
    & K^{-}_{\Phi \Phi} =  \frac{R \sin^2 \theta }{\sqrt{h_{-}(R)}} , \\
    & K^{-} = K^{-}_{ij}h^{ij} = \frac{1}{2} \frac{f'_{-}(R)}{f_{-}(R) \sqrt{h_{-}(R)}} + 2 \frac{R}{\sqrt{h_{-}(R)}}, 
\end{align}
The second junction condition dictates that the jump in the extrinsic curvature is directly proportional to the distributional energy-momentum tensor. Explicitly, we have:
\begin{align}
    8  \pi S^{i}_{\ j} = - \left( \left[ \left[ K^{i}_{\ j} \right] \right] -  \left[ \left[ K\right] \right] \delta^i_{\ j} \right),
\end{align}
where we have introduced the notation $\left[ \left[ \mathcal{O} \right] \right] = \mathcal{O}^{+} - \mathcal{O}^{-}$, representing precisely the jump in the extrinsic curvatures. We can compute it to find:
\begin{align}
    & \left[ \left[ K^{\tau}_{\ \tau}\right] \right] =  \frac{1}{2} \frac{f'_{+}(R)}{f_{+}(R) \sqrt{h_{+}(R)}} -  \frac{1}{2} \frac{f'_{-}(R)}{f_{-}(R) \sqrt{h_{-}(R)}}, \\ 
    & \left[ \left[ K^{\theta}_{\ \theta}\right] \right] =  \frac{1}{R \sqrt{h_{+}(R)}} - \frac{1}{R \sqrt{h_{-}(R)}} , \\
    & \left[ \left[ K^{\Phi}_{\ \Phi}\right] \right] =  \frac{1}{R \sqrt{h_{+}(R)}} - \frac{1}{R \sqrt{h_{-}(R)}}, \\
    & \left[ \left[ K \right] \right] = \frac{1}{2} \left( \frac{f'_{+}(R)}{f_{+}(R) \sqrt{h_{+}(R)}} - \frac{f'_{-}(R)}{f_{-}(R) \sqrt{h_{-}(R)}} \right) + \frac{2}{R} \left[ \frac{1}{ \sqrt{h_{+} (R)}} - \frac{1}{\sqrt{ h_{-} (R)}} \right]. 
\end{align}
From this expression we can determine $S^{i}{}_{j}$. For our purpose, it is interesting that we can express the tensor as that of a perfect fluid,
\begin{align}
    S_{ij} = \sigma u_i u_j + \tilde{p}_t \left( h_{ij} + u_i u_j \right),
\end{align}
where $\sigma$ and $\tilde{p}_t$ represent the surface energy density and the surface tangential pressure. We have $S^{\tau}_{\ \tau} = - \sigma$ and $S^{\theta}_{\ \theta} = \tilde{p}_{t}$. They are given by
\begin{align}
     \sigma & = \frac{1}{4 \pi R} \left( \frac{1}{\sqrt{h_{-}(R)}} - \frac{1}{\sqrt{h_{+}(R)}} \right), \\
     \tilde{p}_t & = \frac{1}{8 \pi R} \left[ \frac{1}{ \sqrt{h_{+} (R)}} - \frac{1}{\sqrt{ h_{-} (R)}} \right] + \frac{1}{16 \pi} \left( \frac{f'_{+}(R)}{f_{+}(R) \sqrt{h_{+}(R)}} - \frac{f'_{-}(R)}{f_{-}(R) \sqrt{h_{-}(R)}} \right).
    \label{Eq:DensitiesApp}
\end{align}
In general, whenever $\tilde{p}_t \neq 0$, the presence of a thin-shell automatically introduces anisotropic pressures into play. However, in the special cases where $\tilde{p}_t = 0$, the energy-momentum tensor has only one nonvanishing component, the energy density, thus restoring isotropy. The models proposed by Bondi in~\cite{Bondi1964} precisely fit in this category.


\newpage

\fancyhead[LE,RO]{\thepage}
\fancyhead[LO,RE]{Bibliography}

\begingroup
  \renewcommand{\cleardoublepage}{}
  \renewcommand{\clearpage}{}
  \printbibliography
\endgroup

\includepdf[
  pages=-,
  pagecommand={\thispagestyle{empty}},
  fitpaper=true
]{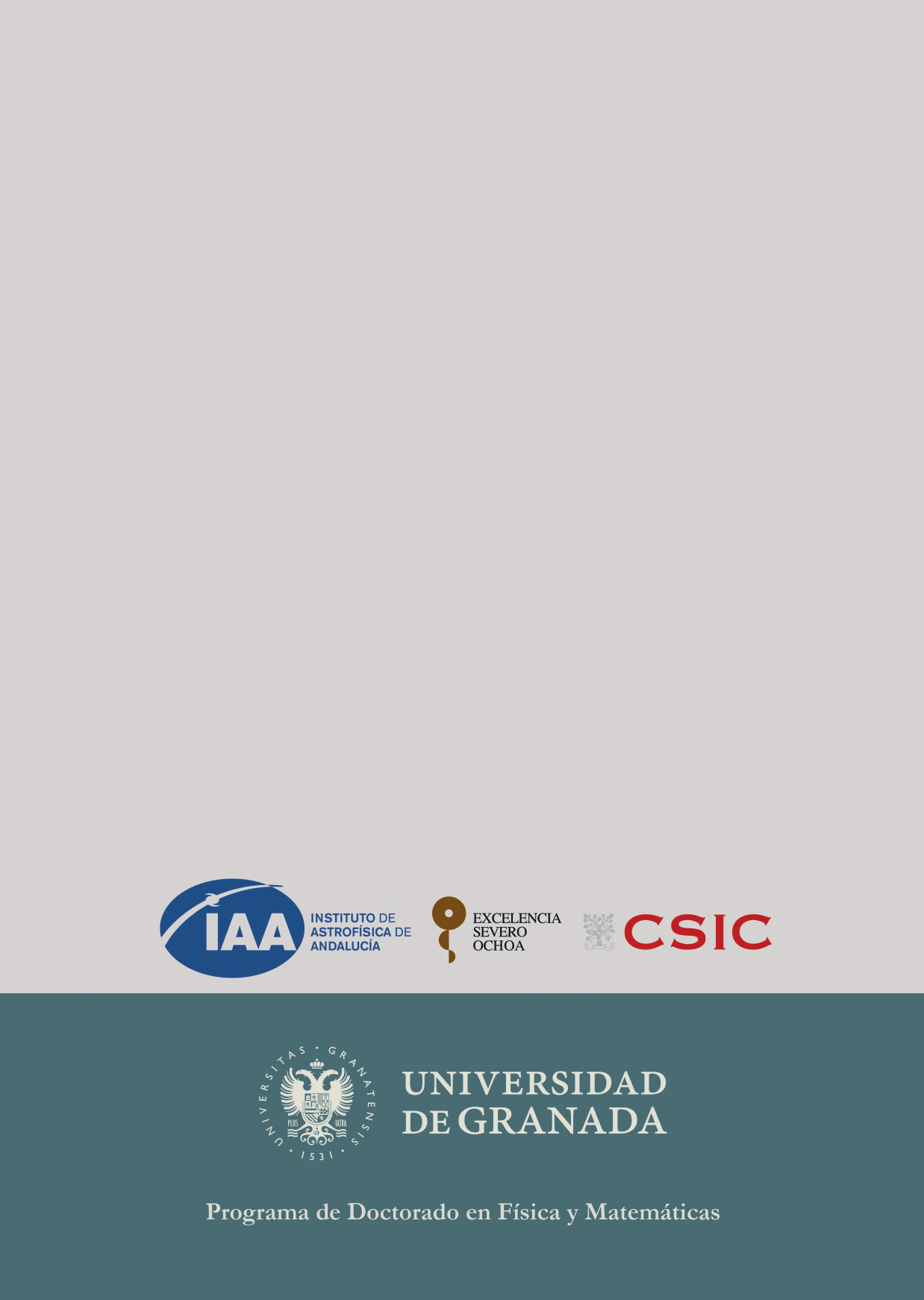}

\end{document}